\documentclass[a4paper,11pt,twoside]{article}

\usepackage{etex}
\usepackage{a4wide,bm,epsfig,booktabs}
\usepackage{setspace}
\usepackage{subfigure}
\usepackage{placeins}  
\usepackage{caption}
\usepackage{tikz}

\usepackage[ngerman, english]{babel}
 
\usepackage{relsize}    
\usepackage{graphicx}    
\usepackage{cancel}      
\usepackage{slashed}  
\usepackage{verbatim}   
\usepackage{hyperref}
\usepackage{amsmath,enumerate} 
\usepackage[ntheorem]{defoqua}

\usepackage{enumerate}     
\usepackage{stmaryrd}  
\usepackage{cite} 
 
\usepackage[flushmargin,hang]{footmisc}

\usepackage{fancyheadings}    
\usepackage[arrow, matrix,curve]{xy}

\let\origenumerate\enumerate
\def\enumerate{\origenumerate\itemsep0pt}
\let\origitemize\itemize
\def\itemize{\origitemize\itemsep0pt}

\newcommand{\tsd}{\!\upharpoonleft\! \upharpoonright\!}
\newcommand{\tdd}{\!\upharpoonleft\! \downharpoonright\!}

\newcommand{\itspace} {}%{\vspace{-1.2ex}}
\newcommand{\itspacec} {}%{\vspace{-1ex}}
\newcommand{\itspacecc} {}%{\vspace{-1ex}}

\newcommand{\he} {\hspace{1pt}}

\newcommand{\LM} {L_{[\mm]}}
\newcommand{\LMS} {L_{[\mm']}}
\newcommand{\LSM} {L'_{[\mm]}}
\newcommand{\LSMS} {L'_{[\mm']}}

\newcommand{\III} {\mathcal{I}}
\newcommand{\JJJ} {\mathcal{J}}

\usepackage
[%xindy, 
toc,          %Einträge im Inhaltsverzeichnis 
]      %im Inhaltsverzeichnis auf section-Ebene erscheinen 
{glossaries} 

\usepackage{glossary-longragged}

\makeglossaries

\newcommand{\EE} {\mathrm{E}}

\newcommand{\ssigma} {s}

\newcommand{\TOHO} {\mathcal{T}_{\mathrm{H}}}

\newcommand{\madgx} {{\mG_x^{\Ad'}}}

\newcommand{\m} {\mathfrak{m}}
\newcommand{\M} {\mathbb{M}}

\newcommand{\headl}[3] {
\newglossaryentry{#1}{   
  name=$\textbf{#2}$\hspace*{-2000pt}\vspace*{6pt},  
  description={}
  , sort=#3 , nonumberlist
}
\glsadd{#1}
}

\headl{PFB}{Principal Fibre Bundles:}{A}
\headl{GRA}{Groups and Actions:}{B}
\headl{PSPAL}{Projective Spaces and Actions on Lie Algebras:}{C}
\headl{SSG}{Special Symmetry Groups and Maps:}{E}
\headl{SSGC}{Sets of Curves:}{I}
\headl{REL}{Relations:}{J}
\headl{EXTGA}{Extension of Group Actions:}{L}
\headl{QCSP}{Quantum Configuration Spaces:}{N}
\headl{HOMPA}{Homomorphisms of Paths:}{U}
\headl{HOMPPPA}{Normalized Radon Measures:}{W}
\headl{BOCF}{Standard Homogeneous Isotropic LQC:}{Y}
\headl{ABBBABA}{Abbreviations:}{Z}
\newcommand{\RB} {\mathbb{R}_{\mathrm{Bohr}}}
\newglossaryentry{RB}{  
  name=$\mathbb{R}_{\mathrm{Bohr}}$, 
  description={The Bohr compactification of $\RR$. (homeomorphic to $\Spec(\CAP(\RR))$)}
  , sort=YYYYY
} 

\newcommand{\muB} {\mu_{\mathrm{Bohr}}}
\newglossaryentry{muB}{  
  name=$\mu_{\mathrm{Bohr}}$, 
  description={Haar measure on $\RB$.}
  ,sort=YYYYYYY 
}

\newcommand{\Mm} {\M}

\newcommand{\grF} {\mathrm{P_\mathrm{FN}}}
\newcommand{\grl} {\mathrm{P_{\lin,0}}}

\newcommand{\xsimorb} {{\mathfrak{G}_x}}
\newcommand{\xsimorbb} {\mathfrak{G}}
\newglossaryentry{XSIMORB}{ 
  name=$\xsimorb$, 
  description={Equivalence classes in $\mathfrak{g}\backslash \mg_x$ w.r.t.\ $\xsim$.} %$\Ad\colon G_x\times \mg\rightarrow \mg$.}
  , sort=JJJJJJJ
} 

\newglossaryentry{KAPPA}{  
  name=$\kappa$, 
  description={The map $\kappa\colon \A\rightarrow \HOM$. Bijection if $\Pa$ is independent.}
  ,sort=UUUUU 
}

\newcommand{\wt}[1] {\widetilde{#1}} 
\newcommand{\ovl}[1]{\overline{#1}} 
\newcommand{\cp} {\circ}
\newcommand{\Borel}{\mathfrak{B}}
\newcommand{\dd} {\mathrm{d}}
\newcommand{\J} {\mathfrak{J}}
\newcommand{\JJ} {\mathfrak{J}} 
\newcommand{\dom}{\operatorname{\mathrm{dom}}}
\newcommand{\im}{\mathrm{im}}

\newcommand{\Con} {\mathcal{A}}
\newglossaryentry{Con}{  
  name=$\Con$, 
  description={Set of smooth connections on the principal fibre bundle $P$.}
  , sort=AAAAAAAAAAAA
}

\newcommand{\hx} {\hat{x}}
\newcommand{\Homeo} {\mathrm{Homeo}}
\newcommand{\diff} {\Delta}
\newglossaryentry{DIFF}{  
  name=$\diff$, 
  description={Difference of two points in the same fibre of a principal fibre bundle.}
  ,sort=AAAAAAAA 
}
\newglossaryentry{PSIX}{  
  name=$\psi_x$, 
  description={The map $P\rightarrow S$, $p\mapsto \diff(\nu_x,p)$ for $\{\nu_x\}_{x\in M}$ a fixed family with $\nu_x\in F_x$ for all $x\in M$.}
  , sort=AAAAAAAAA 
}

\newcommand{\me} {\mathbb{1}}

\newcommand{\mLAS} {\mu_{\mathfrak{g}}}                    
\newglossaryentry{mLAS}{  
  name=$\mu_{\mathfrak{g}}$, 
  description={Normalized Radon measure on $\IHOMLAS$.}
  ,sort=WWWWWWW
}

\newcommand{\mAL} {\mu_{\mathrm{AL}}}             
\newglossaryentry{mAL}{  
  name=$\mAL$, 
  description={Ashtekar-Lewandowski measure on $\HOMW$.}
  ,sort=WWW 
} 

\newcommand{\mFNS} {\mu_{\mathrm{FN}}}
\newglossaryentry{MFNS}{  
  name=$\mFNS$, 
  description={Reduced Ashtekar-Lewandowski measure on $\IHOMFNS$.}
  ,sort=WWWWW
} 
\newcommand{\muH} {\mu_{\mathrm{H}}}
\newcommand{\DG} {\Gamma}
\newcommand{\GR} {\Gamma}
\newcommand{\res} {\mathfrak{r}}

\newglossaryentry{NU}{  
  name=$\nu$, 
  description={Fixed family $\{\nu_x\}_{x\in M}\subseteq P$ with $\nu_x\in F_x$ for all $x\in M$.} 
  ,sort=AAAAAA
}

\newcommand{\Ck} {C^k}
\newcommand{\Pa} {\mathcal{P}}
\newglossaryentry{Pa}{   
  name=$\Pa$, 
  description={Set of $\Ck$-paths in the base manifold $M$.}
  , sort=III 
} 

\newglossaryentry{PALPHA}{  
  name=$\Pa_\alpha$, 
  description={Indexed set of $\Ck$-paths in the base manifold $M$.}
  , sort=IIII 
}

\newcommand{\Pacs} {\Pa_\mathrm{CNL}}
\newglossaryentry{PACS}{   
  name=$\Pacs$, 
  description={Set of continuously but not Lie algebra generated curves.}
  , sort=IIIIIIIIIIIII
} 

\newcommand{\Pac} {\Pa_\mathrm{C}}
\newglossaryentry{PAC}{   
  name=$\Pac$, 
  description={Set of continuously generated curves. $\Pac=\Pags\sqcup \Pacs$}
  , sort=IIIIIIIIIIIIII  
} 
\newcommand{\Pas} {\mathcal{P}}

\newcommand{\PaC} {\mathfrak{P}}
\newglossaryentry{PaC}{   
  name=$\PaC$, 
  description={$\Cstar$-algebra of cylindrical functions that correspond to the set of curves $\Pa$.}
  , sort=NNN  
}

\newcommand{\PACALPHA} {\PaC_\alpha}

\newglossaryentry{PACALPHA}{  
  name=$\PACALPHA$, 
  description={$\Cstar$-algebra of cylindrical functions that corresponds to $\Pa_\alpha$.}
  ,sort=NNNN
}

\newcommand{\Pax} {{\Pa_0}}
\newcommand{\Pay} {{\Pa_\alpha}}

\newcommand{\PaCx} {{\PaC_0}}
\newcommand{\PaCy} {{\PaC_\alpha}}

\newcommand{\s} {{\vec{s}}}
\newcommand{\g} {{\vec{g}}}

\newcommand{\wg} {{\pm\g_\iota}}

\newcommand{\ms} {\mathfrak{s}}
\newcommand{\mg} {\mathfrak{g}}

\newcommand{\dttB}[2] {{\textstyle\frac{\dd}{\dd#1}}\big|_{#1=#2}}
\newcommand{\dt}[2] {\textstyle\frac{\dd}{\dd#1}\big|_{#1=#2}}
\newcommand{\w} {\omega}
\newcommand{\qw} {\ovl{\omega}}
\newcommand{\Add}[1] {\Ad_{#1}}
\newglossaryentry{ADD}{  
  name=$\Ad$, 
  description={The left action $\Ad\colon G\times \mg \rightarrow \mg$, $(g,\g)\mapsto \Add{g}(\g)$.}
  , sort=D
}

\newcommand{\Co}[1] {\alpha_{#1}}
\newglossaryentry{CONJ}{  
  name=$\alpha_g$, 
  description={Conjugation in $G$ by $g\in G$.}
  ,sort=BBBBB  
}

\newglossaryentry{FX}{  
  name=$F_x$, 
  description={Fibre $\pi^{-1}(x)$ over $x\in M$ in $P$.}
  ,sort=AAAAA
}

\newcommand{\CC}[1] {C^{#1}}
\newcommand{\parall}[2] {\mathcal{P}_{#1}^{#2}}
\newglossaryentry{PATRA}{  
  name=$\parall{\gamma}{\w}$, 
  description={Parallel transport $\parall{\gamma}{\w}\colon F_{\gamma(a)}\rightarrow F_{\gamma(c)}$ along $\gamma\colon [a,b]\rightarrow M$ w.r.t.\ $\w$.}
  , sort=AAAAAAAAAAAAAA
}
\newcommand{\RR} {\mathbb{R}}
\newcommand{\CCC} {\mathbb{C}}
\newcommand{\TT} {\mathrm{T}}
\newcommand{\ZZ} {\mathbb{Z}}
\newcommand{\NN} {\mathbb{N}}
\newcommand{\Q}{\mathbb{Q}}
\newcommand{\NNge} {{\mathbb{N}_{>0}}}

\newcommand{\SOD} {{\mathrm{SO}(3)}}

\newcommand{\SU} {{\mathrm{SU}(2)}}
\newcommand{\su} {{\mathfrak{su}(2)}}
\newcommand{\uberl} {\varrho}
\newcommand{\uberll}[2] {#1(#2)}

\newcommand{\Aut} {\mathrm{Aut}}

\newcommand{\GT}{\mathcal{G}}
\newcommand{\Spec}{\mathrm{Spec}}
\newcommand{\Cstar} {C^*}
\newcommand{\Cinf} {C_0}
\newcommand{\Cb} {B}

\newcommand{\GB} {G_{\mathrm{Bohr}}}
\newcommand{\iB} {i_{\mathrm{B}}}
\newcommand{\CAP} {C_{\mathrm{AP}}}
\newglossaryentry{CAP}{  
  name=$\CAP(G)$, 
  description={Almost periodic functions on the LCA group $G$.}
  ,sort=YYY
}

\newcommand{\aA} {\mathfrak{A}}
\newglossaryentry{aA}{  
  name=$\aA$, 
  description={$\cw$-invariant $\Cstar$-subalgebra of the bounded functions on some set $X$.}
  ,sort=LLLLL
}
\newcommand{\bB} {\mathfrak{B}} 
\newcommand{\gG} {\mathfrak{G}}
\newcommand{\rR} {\mathfrak{R}} 
\newcommand{\dD} {\cC}
\newcommand{\uU} {\mathfrak{K}}
\newcommand{\Gen} {\mathfrak{G}}

\newcommand{\X} {\ovl{X}}
\newcommand{\x} {\ovl{x}}
\newcommand{\z} {\ovl{z}}
\newcommand{\ux} {\underline{x}}

\newcommand{\wti} {\:\widetilde{\times}\:}

\newglossaryentry{Q}{  
  name=$Q$, 
  description={The Lie group $G\times S$.}
  ,sort=BBBBBB 
} 
\newglossaryentry{QP}{   
  name=$Q_p$, 
  description={The stabilizer of $Q$ w.r.t.\ $\THA$.}
  ,sort=BBBBBBBB
} 

\newcommand{\THA} {\Xi}       % WangTheta
\newglossaryentry{THA}{  
  name=$\THA$, 
  description={Action $\THA\colon ((g,s),p)\mapsto \Phi(g,p)\cdot s$ of $Q$ on $P$.}
  , sort=BBBBBBB
}
\newcommand{\qrep} {\rho}  
\newglossaryentry{QREP}{  
  name=$\rho$, 
  description={In Section \ref{CHarinvconn} the representation $\rho\colon Q \rightarrow \Aut(\mathfrak{s})$, $(g,s)\mapsto \Add{s}$.}
  , sort=BBBBBBBBB
}

\newcommand{\wm} {\varphi}       % induced on  base manifold
\newglossaryentry{WM}{  
  name=$\wm$, 
  description={Action induced by $\Phi$ on the base manifold $M$.}
  , sort=BBBBBBBBBBB
}

\newcommand{\cw} {\theta}     % induced on  set of connections
\newglossaryentry{CW}{  
  name=$\cw$, 
  description={In the bundle context, action induced by $\Phi$ on the set $\Con$ of smooth connections.}
  , sort=BBBBBBBBBBBBB   
}

\newcommand{\ah} {\vartheta}        %Antihomomorphism
\newcommand{\Gg}{\Phi}

\newcommand{\fiba} {\phi}       %Buendel-Faser compensations wirkung
\newglossaryentry{FIBAP}{  
  name=$\fiba_p$, 
  description={Homomorphisms $\fiba_p\colon G_{\pi(p)}\rightarrow S$ with $\Phi(h,p)=p\cdot \fiba_p(h)$ for all $h\in G_{\pi(p)}$.}
  , sort=AAAAAAAAAA 
}

\newcommand{\Pe} {\Phi_E}
\newcommand{\Pii} {\Phi_{SP}}
\newcommand{\Ph} {\Phi_H} 

\newglossaryentry{VR}{  
  name=$\varrho$,  
  description={The universal covering map $\varrho\colon \SU\rightarrow \SOD$.}
  , sort=EEE
}

\newcommand{\Ge} {{G_E}} 
\newglossaryentry{GE}{  
  name=$\Ge$, 
  description={The semi direct product $\mathbb{R}^3 \rtimes_\uberl \SU$.}
  ,sort=EEEEE
}
\newcommand{\Gi} {{G_{SP}}}
\newglossaryentry{GI}{  
  name=$\Gi$, 
  description={$\SU$ acting on $\RR^3\times \SU$ via $(\sigma,(x,s))\mapsto (\varrho(\sigma)(x),\sigma\cdot s)$.}
  ,sort=EEEEEE 
}
\newcommand{\Gh} {{G_H}}
\newglossaryentry{GH}{  
  name=$\Gh$, 
  description={$\RR^3$ acting via translations in the first factor of the principal bundle $\RR^3\times \SU$.}
  ,sort=EEEEEEE
}
\newcommand{\Gee} {\mathbb{R}^3 \rtimes_\uberl \SU}

\newcommand{\GLNC} {{\mathrm{GL}(n,\mathbb{C})}}

\newcommand{\A} {\ovl{\Con}}                 %% quant A
\newglossaryentry{AQ}{  
  name=$\A$, 
  description={Spectrum of $\PaC$. Space of generalized connections. }
  , sort=NNNNNN   
}

\newcommand{\AR} {{\Con_\red}}               %% A red
\newglossaryentry{AR}{  
  name=$\AR$, 
  description={Set of $\Phi$-invariant connections.  }
  ,sort=BBBB  
}

\newcommand{\ARQ} {{\ovl{\Con_\red}}}        %% A red quant
\newglossaryentry{ARQ}{  
  name=$\ARQ$, 
  description={The space $\XRQ$ for $\X=\A$ and $\aA=\PaC$.}
  ,sort=NNNNNNNNNNN 
} 
\newcommand{\ARRQ} {}  %% A red restr quant
\DeclareRobustCommand{\ARRQ} {{\underline{\Con_\red}}} 
\newglossaryentry{ARRQ}{  
  name=$\ARRQ$, 
  description={The space $\XRRQ$ for $\XR=\AR$ and $\aA=\PaC$.}
  ,sort=NNNNNNN   
}

\newcommand{\AQR} {{\A_\red}}                %% A quant red
\newglossaryentry{AQR}{  
  name=$\AQR$, 
  description={The space $\XQR$ for $\X=\A$ and $\aA=\PaC$.}
  ,sort=NNNNNNNNNNNNNNN
}

\newcommand{\ARQw} {{\ovl{\Con_{\red,\w}}}}  
\newcommand{\ARRQw} {{\underline{\Con_{\red\:\w}}}}  %% A red restr quant %rput pstricks
\newcommand{\AQRw} {{\A_{\red,\w}}}
\newcommand{\AQRL} {{\A_{\red,\lin}}}

\newcommand{\RedGauge} {\Hom_{\red,\GAG}(\Pa,\IsoF)}
\newcommand{\RedGaugew} {\Hom_{\red,\GAG}(\Paw,\IsoF)}

\newcommand{\GAG} {{\mathcal{G}}}

\newcommand{\AQRFNS} {\ovl{\Con}_{\red,\mathrm{FN}}}
\newcommand{\AQRFS} {\ovl{\Con}_{\red,\mathrm{FS}}}

\newcommand{\ZX} {{X}} 
\newcommand{\QZX} {\X}
\newcommand{\QSX} {\XNR}
\newcommand{\SX} {{Y}}

\newcommand{\IOTAY} {\iota_{\ZX}}

\newglossaryentry{IOTAY}{   
  name=$\IOTAY$, 
  description={Canonical inclusion $\iota_X\colon \ZX \hookrightarrow\Hom(\aA,\mathbb{C})$, $x\mapsto [f\mapsto f(x)]$ for $\aA\subseteq B(\ZX)$ a $\Cstar$-subalgebra.}
  ,sort=LLL  
}

\newcommand{\YNR} {{\overline{X_\upsilon}}}       %%Y nü reduziert
\newcommand{\XNR} {{\underline{X^\upsilon}}}      %%X nü reduziert

\newcommand{\specw} {\Theta} %Lifted action on some spectrum
\newglossaryentry{SPECW}{  
  name=$\specw$, 
  description={Extension of $\cw\colon G\times X\rightarrow X$ to the spectrum of a $\Cstar$-algebra of the bounded functions on $X$.}
  , sort=LLLLL 
} %

\newglossaryentry{CCW}{  
  name=$\cw$, 
  description={In the general context, a left action $\cw\colon G\times X\rightarrow X$.}
  , sort=LLLL   
}

\newcommand{\XR} {{X_\red}}                  %%X red
\newglossaryentry{XR}{  
  name=$\XR$,   
  description={Set of invariant (classical) elements, i.e., $\XR=\{x\in X\:|\: \cw(g,x)=x\text{ for all }g\in G\}$.}
  ,sort=LLLLL 
} 

\newcommand{\XQR} {{\X_\red}}                %% quant X red
\newglossaryentry{XQR}{   
  name=$\XQR$, 
  description={Set of invariant spectral (quantum) elements.}
  ,sort=LLLLLLLLLLL  
}

\newcommand{\XRQ} {{\overline{X_\red}}}           %%X red quant
\newglossaryentry{XRQ}{   
  name=$\XRQ$, 
  description={Closure of $\iota_X(X_\aA\cap \XR)$ in $\X$.}
  ,sort=LLLLLLLLL 
}

\newcommand{\XRRQ} {}     %%X red restr quant
\DeclareRobustCommand{\XRRQ} {{\underline{X_\red}}} 
\newglossaryentry{XRRQ}{  
  name=$\XRRQ$, 
  description={Spectrum of the restriction $\Cstar$-algebra $\rR=\ovl{\aA|_{\XR}}\subseteq B(X_\red)$.}
  ,sort=LLLLLLL 
}

\newcommand{\kk} {\vec{m}}
\newcommand{\os} {\hspace{1pt}}
\newcommand{\xii} {\vec{\eta}}
\newcommand{\Span} {\mathrm{span}}
\newcommand{\PMS} {(P,\pi,M,S)}
\newglossaryentry{PMS}{   
  name=$\PMS$,  
  description={Principal fibre bundle with total space $P$, base manifold $M$ and structure group $S$.}
  , sort=AAAA 
}

\newcommand{\GPHI} {(G,\Phi)}
\newglossaryentry{GPHI}{   
  name=$\GPHI$,  
  description={Lie group of automorphism of $\PMS$.}
  , sort=BBB    
}

\newcommand{\hvect} {\raisebox{1.1pt}{$($}\vspace{1pt}\raisebox{-1.1pt}{$\vec{0}_{\mathfrak{g}},\vec{h}$}\raisebox{1.1pt}{$)$}}

\newcommand{\vg} {\widetilde{\gamma}}
\newcommand{\vw} {\widetilde{\w}}
\newcommand{\vp} {\widetilde{p}}

\newcommand{\TRHOM} {\Hom(\Pa,S)}

\newcommand{\HOM} {\Hom(\Pa,\IsoF)}
\newcommand{\HOMLAS} {\Hom(\Pags,\IsoF)}
\newcommand{\HOMW} {\Hom(\Paw,\IsoF)}
\newcommand{\IHOM} {\Hom_\red(\Pa,\IsoF)}

\newcommand{\IHOMW} {\Hom_\red(\Paw,\IsoF)}
\newcommand{\HOMInd}[1] {\Hom(\Pa_{#1},\IsoF)}
\newcommand{\IHOMInd}[1] {\Hom_\red(\Pa_{#1},\IsoF)}
\newcommand{\IHOMLA} {\Hom_\red(\Pag,\IsoF)}
\newcommand{\IHOMLAS} {\Hom_\red\!\big(\Pags,\IsoF\big)}

\newcommand{\IHOMLAI} {\Hom_\red(\Pagw,\IsoF)}
\newcommand{\IHOMLL} {\Hom_\red(\Pal,\IsoF)}

\newcommand{\IHOMFNS} {\Hom_\red(\Pafns,\IsoF)}

\newcommand{\HOMFNS} {\Hom(\Pafns,\IsoF)}

\newcommand{\ITRHOM} {\Hom_\red(\Pa,S)}

\newcommand{\ITRHOMW} {\Hom_\red(\Paw,\SU)}
\newcommand{\ITRHOML} {\Hom_\red(\Pal,\SU)}

\newcommand{\Iso} {\mathrm{Mor}}
\newcommand{\IsoF} {\mathrm{Mor_F}}
\newcommand{\AF} {\Iso_{\mathrm{F}}}
\newcommand{\homm} {\varepsilon}
\newcommand{\hommm} {\epsilon}

\newcommand{\simpr} {\sim} 

\newcommand{\cpsim} {\sim_\cp}
\newglossaryentry{CPSIM}{  
  name=$\cpsim$, 
  description={$\gamma_1\cpsim \gamma_2$ iff $\exists$ open intervals $I_i\subseteq \dom[\gamma_i]$ for $i=1,2$ such that $\gamma_1(I_1)=\gamma_2(I_2)$.}
  , sort=JJJ
}
 
\newcommand{\xsim} {\sim_x}
\newglossaryentry{XSIM}{  
  name=$\xsim$, 
  description={$\vec{g}\xsim\vec{g}'$ for $\g,\g'\in \mg\backslash \mg_x$ iff there is $g\in G$ such that $\wm_g\cp \gamma^x_{\vec{g}} \cpsim \gamma^x_{\vec{g}'}$.}
  , sort =JJJJJ
}

\newcommand{\csim} {\sim_\Con}
\newcommand{\psim} {\sim_{\mathrm{par}}}
\newcommand{\isim} {\sim_{\im}}
\newcommand{\segsim} {\sim_{\delta}}
\newcommand{\segsims} {\sim_{\delta'}}

\newcommand{\lin} {\mathrm{l}}
\newcommand{\mc} {\mathrm{c}}

\newcommand{\zd} {k} 

\newcommand{\Paf} {{\Pa_{\mathrm{F}}}}
\newglossaryentry{PAF}{  
  name=$\Paf$, 
  description={Set of free curves. $\Paf= \Pafns\sqcup\Pafs$}
  , sort=IIIIIIIIIIIIIIIIIIIII
}

\newcommand{\Pafs} {{\Pa_{\mathrm{FS}}}}
\newglossaryentry{PAFS}{  
  name=$\Pafs$, 
  description={Set of free symmetric curves.}
  , sort=IIIIIIIIIIIIIIIIIII
}

\newcommand{\Pafns} {{\Pa_{\mathrm{FN}}}}
\newglossaryentry{PAFNS}{  
  name=$\Pafns$, 
  description={Set of free non-symmetric curves.}
  , sort=IIIIIIIIIIIIIIIII
}

\newcommand{\gc}[4] {\gamma_{\vec{#1},\vec{#2}}^{#3,#4}}

\newcommand{\Paw} {{\Pa_\w}}
\newglossaryentry{PAW}{  
  name=$\Paw$, 
  description={Set of embedded analytic curves in $M$.}
  , sort=IIIII  
}

\newcommand{\Pall} {{\Pa_\lin}} 
\newcommand{\Pal} {{\Pa_\lin^\sim}} 
\newcommand{\Paln} {{\Pa^\sim_{\lin,0}}} 
\newcommand{\Pacirc} {{\Pa_\mc}}

\newcommand{\ARQInd}[1] {{\ovl{A_{\red,#1}}}}

\newcommand{\ARRQInd}[1] {{\underline{\Con_{\red\: #1}}}} 

\newcommand{\arrqalpha} {}
\DeclareRobustCommand{\arrqalpha} {{\underline{\Con_{\red\: \alpha}}}} 

\newcommand{\AQRInd}[1] {{\A_{\red,#1}}}
\newcommand{\AInd}[1] {{\A_{#1}}}

\newglossaryentry{KAPPAALPHA}{  
  name=$\kappa_\alpha$, 
  description={The map $\kappa$ from Definition \ref{def:indepref}.\ref{def:mapkappa} that corresponds to $\Pa_\alpha$.} 
  ,sort=UUUUUUU 
}

\newglossaryentry{KAPPALAI}{  
  name=$\kappa_{\mg'}$, 
  description={The map $\kappa$ from Definition \ref{def:indepref}.\ref{def:mapkappa} that corresponds to $\Pagw$.} 
  ,sort=UUUUUUUUU 
}

\newglossaryentry{AALPHA}{  
  name=$\AInd{\alpha}$, 
  description={The quantum space that corresponds to $\Pa_\alpha$.} 
  , sort=NNNNNNN  
}
\newglossaryentry{AQRALPHA}{  
  name=$\AQRInd{\alpha}$, 
  description={The quantum-reduced configuration space that corresponds to $\Pa_\alpha$.}
  , sort=NNNNNNNNNNNNNNNNN  
}

\newglossaryentry{ARQALPHA}{   
  name=$\ARQInd{\alpha}$, 
  description={Closure of $\im[\iota_\Con]$ in $\A$.}
  , sort=NNNNNNNNNNNNN 
}
\newglossaryentry{ARRQALPHA}{  
  name=$\arrqalpha$, 
  description={The quantized reduced classical space that corresponds to $\Pa_\alpha$.}
  ,sort=NNNNNNNNN   
}

\newcommand{\AQRLAI} {{\A_{\mg',\red}}}
\newglossaryentry{AQRLAI}{  
  name=$\AQRLAI$, 
  description={The quantum-reduced configuration space that corresponds to $\Pagw$.}
  ,sort=NNNNNNNNNNNNNNNNNNN 
} 
\newcommand{\ARQLAI} {\ovl{\Con_{\mg',\red}}}
\newglossaryentry{ARQLAI}{  
  name=$\ARQLAI$, 
  description={The quantized reduced space that corresponds to $\Pagw$.}
  ,sort=OOOOOOOOOOOOOOOOO
}

\newcommand{\AQRLA} {{\A_{\red,\mg}}}

\newcommand{\ARQLL} {\ovl{\Con_{\red,\lin}}}

\newcommand{\ARRQLL} {}
\DeclareRobustCommand{\ARRQLL} {{\underline{\Con_{\red\:\lin}}}} 
\newcommand{\kla} {\eta}

\newcommand{\mG} {\mathfrak{G}}
\newcommand{\oo} {\mathfrak{o}}
\newcommand{\vc}{{\vec{c}}}
\newcommand{\Pags} {{\Pa^\sim_\mg}}
\newglossaryentry{PAGS}{  
  name=$\Pags$, 
  description={Set of curves equivalent to some Lie algebra generated curve.}
  , sort=IIIIIIIIIIII 
}

\newcommand{\Pagw} {{\Pa_{\mg'}}}
\newglossaryentry{PAGW}{  
  name=$\Pagw$, 
  description={$\Phi$-invariant subset of $\Paw$ being closed under decomposition and  $\Pag\subseteq \Pagw$.}
  , sort=IIIIIIIIII 
} 

\newcommand{\Pag} {{\Pa_\mg}}
\newglossaryentry{PAG}{  
  name=$\Pag$, 
  description={Set of Lie algebra generated curves.}
  , sort=IIIIIIIII 
}

\newglossaryentry{HOM}{  
  name=$\HOM$, 
  description={Homomorphisms of paths in $\Pa$.}
  ,sort=UUU 
}

\newglossaryentry{HOMALPHA}{  
  name=$\HOMInd{\alpha}$, 
  description={Homomorphisms of paths in $\Pa_\alpha$.}
  ,sort=UUUU} 

\newglossaryentry{HOMLAI}{  
  name=$\HOMInd{\mg'}$, 
  description={Homomorphisms of paths in $\Pa_{\mg'}$. %For $\Pa_\alpha$ independent the image of $\kappa_\alpha$.
  }
  ,sort=UUUUUUUUUUUUUUU
}

\newglossaryentry{IHOM}{  
  name=$\IHOM$, 
  description={ For $\Pa$ independent the set of invariant homomorphisms, i.e., $\kappa\big(\AQR\big)$.}
  ,sort=UUUUUUUUUUUUUUUUU
}

\newglossaryentry{IHOMALPHA}{  
  name=$\IHOMInd{\alpha}$, 
  description={ For $\Pa_\alpha$ independent the set of invariant homomorphisms, i.e., $\kappa_\alpha\big(\AQR\big)$.}
  ,sort=UUUUUUUUUUUUUUUUUUU
}

\newglossaryentry{IHOMLAI}{  
  name=$\IHOMLAI$, 
  description={The set of invariant homomorphisms in $\Pagw$, i.e., $\kappa_{\mg'}(\AQR)$.}
  ,sort=UUUUUUUUUUUUUUUUUUUUU
}

\newglossaryentry{ADDSTRICH}{
  name=$\Ad'$, 
  description={For $x\in M$ fixed, the left action $\Ad'\colon G_x\times \pr_\mg(\mg\backslash\mg_x)\rightarrow \pr_\mg(\mg\backslash\mg_x)$ induced by $\Ad$.} %$\Ad\colon G_x\times \mg\rightarrow \mg$.}
  ,sort=DDD
}    

\newglossaryentry{ADSTRGXSTAB}{  
  name=$G^x_{[\g]}$, 
  description={Stabilizer $G^x_{[\g]}\subseteq G_x$ of $[\g]$ w.r.t.\ $\Ad'$.} %$\Ad\colon G_x\times \mg\rightarrow \mg$.}
  , sort=DDDDD
} 

\newcommand{\Sp} {\mathrm{P}}
\newcommand{\gag}{{\gamma_{\g}^x}}
\newcommand{\adif} {\rho}
\newcommand{\pip} {\ovl{\pi}}
\newcommand{\Eq} {\mathfrak{M}}

\newglossaryentry{EQP}{  
  name=$\Eq_p$, 
  description={For $p\in P$ the set of $\Ad_{G_{\pi(p)}}^p$-equivariant morphisms $\Psi\colon \mg\backslash \mg_{\pi(p)} \times \RR_{>0}\rightarrow S$.} %$\Ad\colon G_x\times \mg\rightarrow \mg$.}
  ,sort=DDDDDDDDDDD 
} 

\newcommand{\Per} {\mathfrak{N}}
\newcommand{\mm} {{\vec{m}}}  
\newcommand{\nn}{{\vec{n}}}
\newcommand{\vv}{\vec{v}}
\newcommand{\ww}{\vec{w}}
\newcommand{\rr}{{\vec{r}}}

\newglossaryentry{SPMSU}{  
  name=$\Sp\ms$, 
  description={Projective space that corresponds to $\su$.}
  ,sort=CCCCC
}  

\newglossaryentry{PRMSU}{  
  name=$\pr_{\ms}$,  
  description={Projection $\su \rightarrow \Sp\ms$.}
  , sort=CCCCCCCCC
}

\newglossaryentry{SPMG}{  
  name=$\Sp\mg$, 
  description={Projective space that corresponds to $\mg$.}
  , sort=CCC
}  

\newglossaryentry{PRMG}{  
  name=$\pr_\mg$, 
  description={Projection $\mg \rightarrow \Sp\mg$.}
  , sort=CCCCCCC 
} 

\newcommand{\MK}{{\mathfrak{K}}}
\newcommand{\ML}{{\mathfrak{L}}}
\newcommand{\KMAX}{{\MK_{\mathrm{m}}}}
\newcommand{\Mult} {\mathfrak{F}}

\newglossaryentry{GGAMMA}{  
  name=$G_\gamma$,
  description={Stabilizer of the curve $\gamma\in \Paw$, i.e., $G_\gamma=\{g\in G \:|\: \wm_g\cp \gamma \psim \gamma\}$.}
  ,sort=JJJJJJJJJJJJ
} 

\newglossaryentry{HGAMMADELTA}{ 
  name=$H_{\gamma,\delta}$,
  description={Elements $g$ of $G$ for which $\gamma\cpsim \wm_g\cp \delta$ holds.}
  ,sort=JJJJJJJJJJ 
}   

\newcommand{\MPD} {\Mult_{p,\delta}}
\newglossaryentry{FPDELTA}{ 
  name=$\MPD$,
  description={Maps $\RR\backslash \{0\}\rightarrow S$ with certain morphism and invariance properties.}
  ,sort=DDDDDDDDDDDDD 
}

\newglossaryentry{CSIM}{ 
  name=$\csim$, 
  description={$\gamma\csim \gamma'$ 
    iff $\parall{\gamma}{\w}=\parall{\gamma'}{\w}$ for all $\w\in \Con$.}
  , sort=JJJJJJJJJ
}

\newglossaryentry{PSIM}{ 
  name=$\psim$, 
  description={$\gamma_1\psim \gamma_2$ iff $\gamma_1=\gamma_2\cp \adif|_{\dom[\gamma_1]}$ for $\adif$ an analytic diffeomorphism with $\dot\adif>0$.}
  , sort=JJJJJJJJ
}

\newglossaryentry{SEGSIM}{ 
  name=$\sim_\gamma$, 
  description={$h\sim_\gamma h'$ for $h,h'\in G$ and $\delta\in \Paw$ iff $h^{-1}h'\in G_\gamma$.}
  , sort=JJJJJJJJJJJJJ
}
\newcommand{\qRR} {}
\DeclareRobustCommand{\qRR} {{\underline{\RR}}} 
\newglossaryentry{qRR}{ 
  name=$\qRR$, 
  description={The classically reduced space $\ARRQw$.}
  , sort=YYYYYYYYY
} 

\newglossaryentry{LQG}{
  name=$\text{LQG}$,
  description={Loop Quantum Gravity.}
  , sort=ZZZ
}

\newglossaryentry{LQC}{ 
  name=$\text{LQC}$,
  description={Loop Quantum Cosmology.}
  , sort=ZZZZZ
}

\newglossaryentry{Tor1}{ 
  name=$H_{\s}$,
  description={The maximal torus in $\SU$ given by $H_{\s}=\{\exp(t\s)\:|\: t\in \RR\}$.}
  , sort=EEEEEEEEEEEEE
}
\newglossaryentry{Tor2}{ 
  name=$H_{\vv}$,
  description={The maximal torus in $\SU$ given by $H_{\vv}=\{\exp(t\murs(\vv))\:|\: t\in \RR\}$.}
  , sort=EEEEEEEEEEEEEEE
}

\newglossaryentry{Tor3}{ 
  name=$H_{x}$,
  description={The maximal torus in $\SU$ given by $H_{x}=\{\exp(t\murs(x))\:|\: t\in \RR\}$.}
  , sort=EEEEEEEEEEEEEEEEE
}
\newglossaryentry{Tor4}{ 
  name=$H_{v}$,
  description={The maximal torus in $\SU$ given by $H_{v}=\{\exp(t\murs(v))\:|\: t\in \RR\}$.}
  , sort=EEEEEEEEEEEEEEEEEEE
}

\newcommand{\oRR} {Y}
\newcommand{\qR} {{\ovl{\RR}}}
\newglossaryentry{qR}{ 
  name=$\qR$,
  description={The space $\RR\sqcup \RB$ homeomorphic to $\qRR$.}
  , sort=YYYYYYYYYYY
}

\newcommand{\qRY} {{\qR_Y}}
\newcommand{\BRq}{\Borel\raisebox{0.2ex}{$($}\raisebox{-0.1ex}{$\qR$}\raisebox{0.2ex}{$)$}}
\newcommand{\Lzw}[2] {L^{2}\!\left(\hspace{1pt}#1,#2\right)}

\newcommand{\Multl} {\Lambda'}
\newcommand{\Transl} {\Sigma'}
\newcommand{\Multw} {\Lambda}
\newcommand{\Transw} {\Sigma}
\newcommand{\TTransw} {\ovl{\Sigma}}

\newcommand{\RPLUS} {{\footnotesize $\Sigma_\RR$}}

\newcommand{\cC} {\mathfrak{C}}
\newglossaryentry{cC}{  
  name=$\cC$, 
  description={Cylindrical functions.}
}

\newcommand{\leqZ} {\leq_{\mathbb{Z}}}
\newcommand{\pc} {\beta_c}
\newcommand{\prfl}[2] {\im[#1]\sqcup S^{|#2|}}
\newcommand{\T} {\mathcal{T}} 
\newcommand{\pih}{\widehat{\pi}}
\newcommand{\chih}{\widehat{\chi}}
\newcommand{\ZN}{\mathbb{Z}\backslash\{0\}}
\newcommand{\OO}{\mathcal{O}}

\newcommand{\BTK} {\raisebox{0pt}{$\Borel\raisebox{1pt}{$\big($}\raisebox{-0ex}{$S^{|L|}$}\raisebox{1pt}{$\big)$}$}}
\newcommand{\mmu} {\wt{\mu}}
\newcommand{\pillstr} {\big(\raisebox{-0.1ex}{$\pi^{L'}_{L}$}\big)}
\newcommand{\Ts}{S^1_{\mathrm{s}}}
\newcommand{\mus}{\mu_{\mathrm{s}}}
\newcommand{\muL}{\lambda}
\newcommand{\Hil} {\mathcal{H}}
\newcommand{\HHH} {\mathrm{H}}
\newglossaryentry{HHH}{  
  name=$\HHH$, 
  description={The set of homeomorphisms $\adif \colon (0,1)\rightarrow \RR$.}
  ,sort=YYYYYYYYYYYYY
}

\newcommand{\F} {\mathrm{F}}
\newcommand{\NB} {0_{\mathrm{Bohr}}}

\newcommand{\murs} {\mathfrak{z}}

\newglossaryentry{MURS}{  
  name=$\murs$, 
  description={Canonical identification $\murs\colon \RR^3\rightarrow \su$, $\vec{e}_i\mapsto \tau_i$ for $i=1,2,3$.}
  ,sort=EEEEEEEEEEE
}

\begin{document}

\begin{titlepage}
  %% \sffamily
  %% \raggedleft
  \begin{figure} 
    \subfigure{\includegraphics[width=0.5\textwidth]{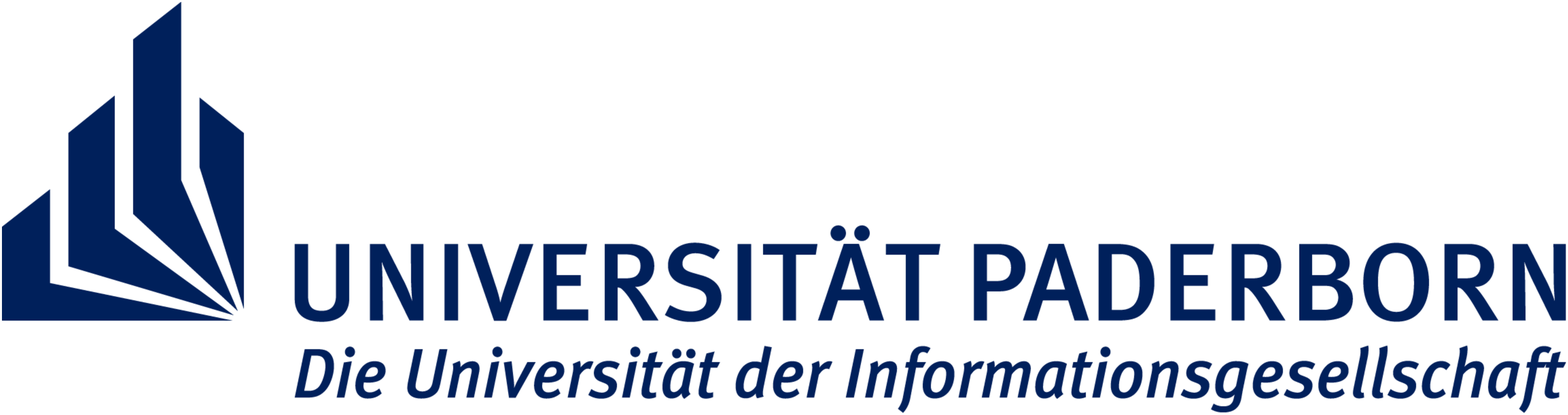}}
    \hfill 
    \subfigure{\includegraphics[width=0.18\textwidth]{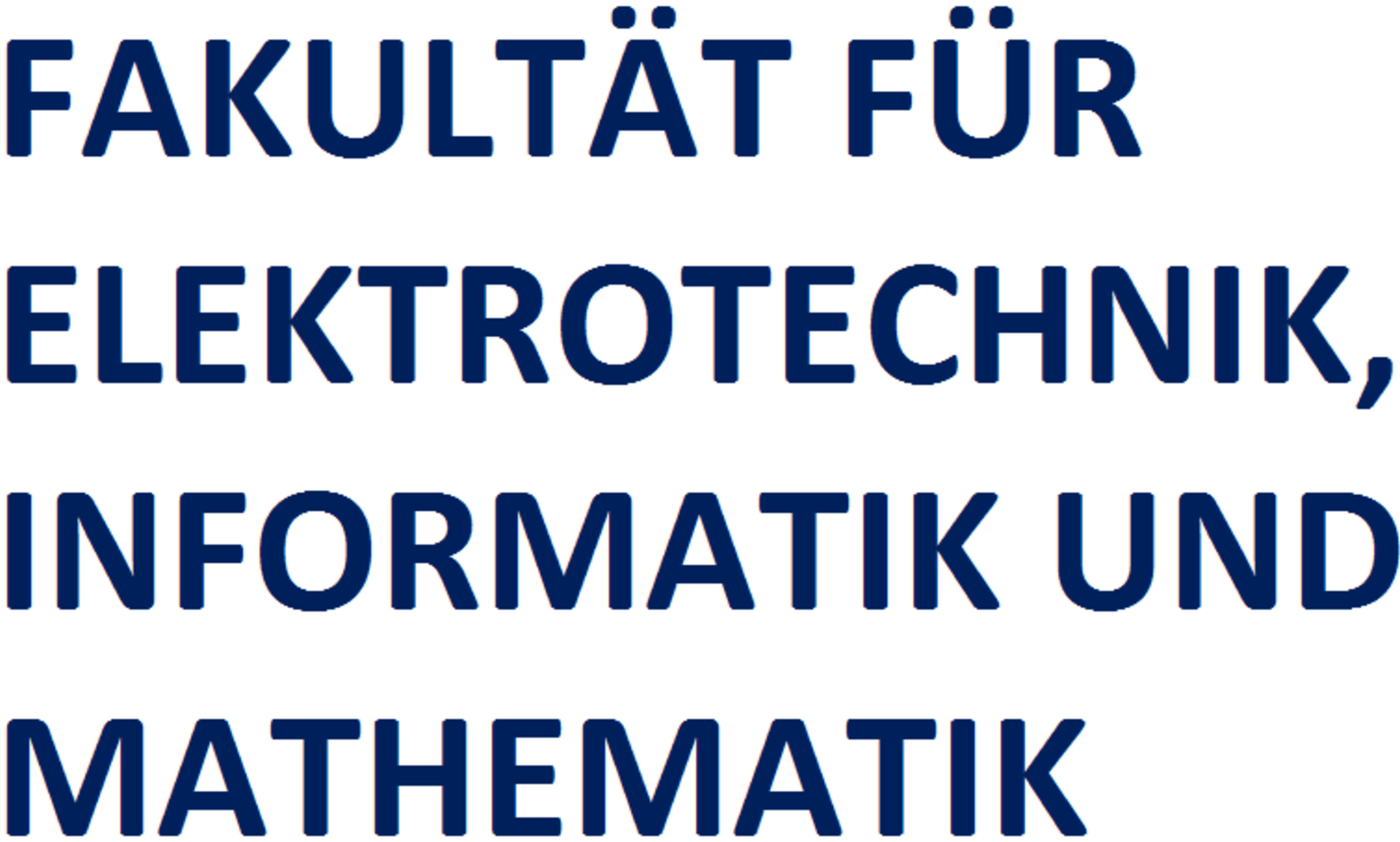}} 
  \end{figure}
  \phantom{fggf}

  \vspace{20pt}
  \centering

  {\bfseries\Huge Invariant Connections and %on Principal Fibre Bundles 
    Symmetry Reduction in Loop Quantum Gravity}\\

  \vspace{45pt}

  {\Large Von der Fakult\"at f\"ur Elektrotechnik, Informatik und Mathematik
    der\\[2pt]
    Universit\"at Paderborn} \\[23pt]
  {\large zur Erlangung des akademischen Grades}\\[23pt]
  {\Large doctor rerum naturalium}\\[4pt]
  {\Large (Dr. rer. nat.)}

  \vspace{40pt}

  {\large genehmigte Dissertation}\\[10pt]
  {\large von}\\[15pt]
  {\Large Dipl.-Phys. Maximilian Hanusch}

 % \vspace{4pt}
  \vspace{29pt}
  {\Large
    \begin{tabbing}
      \qquad\:\: Erster Gutachter:\qquad\qquad\qquad\:\: \= Prof. Dr. Christian Fleischhack\\[1pt]
      \qquad\:\: Zweiter Gutachter:\> Prof. Dr. Joachim Hilgert\\[1pt]
      \qquad\:\: Dritter Gutachter:\> Prof. Dr. Hanno Sahlmann\\[21pt]
      \qquad\:\: Tag der m\"undlichen Pr\"ufung: \> 16.12.2014
    \end{tabbing}}
  \vspace{6pt}
  {\Large Paderborn 2014}\\[15pt]
%  {\Large Diss. EIM-M/xxx}
  % \end{titlepag
\end{titlepage}

\newpage

{\selectlanguage{ngerman}
{\small
\begin{center}
{\bf Zusammenfassung}
\end{center}
\vspace{5pt}
\noindent
  Ziel dieser Arbeit ist die Entwicklung allgemeiner Techniken und Konzepte, die eine mathematisch fundierte Symmetriereduktion von (Quanten-)Eichfeldtheorien erm\"oglichen. Das Haupt\-augenmerk liegt hierbei auf dem Gebiet der Schleifenquantengravitation, im Speziellen auf der Reduktion des Quantenkonfi\-gurationsraumes und der Konstruktion von normierten Radonma{\ss}en auf den resultierenden reduzierten Quantenr\"aumen. 
  F\"ur die Reduktion bieten sich hierbei prinzipiell zwei M\"oglichkeiten. Zum einen kann man den reduzierten klassischen Konfigurationsraum quantisieren (RQ-Reduktion). Zum anderen kann man versuchen, die Symmetrie (genauer: die Gruppenwirkung) vom klassischen auf den Quantenraum zu heben, um dort die Reduktion direkt auf Quantenniveau durchzuf\"uhren (QR-Reduktion). Bislang wurde in der Schleifenquantengravitation nur der RQ-Zugang verfolgt. Da uns der QR-Zugang aus physikalisch-konzeptionellen Gr\"unden jedoch als ad\"aquater erscheint, haben wir ihn in dieser Arbeit systematisch entwickelt. Insbesondere haben wir die jeweiligen RQ- und QR-Konfigurationsr\"aume miteinander verglichen. 
      
Im ersten Teil dieser Arbeit legen wir die mathematischen Grundlagen f\"ur das QR-Reduktions\-prinzip. Hierf\"ur zentral ist die   
Fortsetzung von Gruppen\-wirkungen  $\cw\colon G\times X\rightarrow X$ auf Mengen $X$ zu Gruppen\-wirkungen $\specw\colon G\times \Spec(\aA)\rightarrow \Spec(\aA)$
auf Spektren von $\Cstar$-Algebren $\aA\subseteq B(X)$. In Analogie zur klassischen Situation, in welcher der reduzierte Raum aus allen invarianten Zusammenh\"angen $\AR$ auf dem zugrunde\-liegenden Hauptfaserb\"undel besteht, ist der quanten-reduzierte Konfigurationsraum definiert durch
\begin{align*}
	\A_\red:=\{\qw\in \A\: | \: \specw(g,\qw)=\qw\quad \forall\: g\in G\},
\end{align*}
wobei $\A$ den Quantenkonfigurationsraum der vollen Theorie bezeichnet. 
Dem  gegen\"uber steht die in der Schleifenquantengravitation  \"ubliche RQ-Strategie, bei der das Spektrum einer separierenden $\Cstar$-Algebra von beschr\"ankten Funktionen auf $\AR$ berechnet wird. 
  Wir zeigen, dass der QR-Raum $\A_\red$ stets den entsprechenden RQ-Raum enth\"alt und dass diese Inklusion \"ublicherweise sogar echt ist. Insbesondere kommutieren Reduktion und Quantisierung in diesem Kontext im Allgemeinen also nicht. 
  
  Im zweiten Teil dieser Arbeit konstruieren wir normierte Radonma{\ss}e sowohl auf den \mbox{RQ-} als auch auf den QR-reduzierten R\"aumen.
 	F\"ur letzteren Fall zeigen wir, dass der quanten-reduzierte Konfigurationsraum unter recht milden Voraussetzungen in ein Produkt vier leichter handhabbarer Konfigurations\-r\"aume zerf\"allt. Auf zweien dieser R\"aume werden normierte Radonma{\ss}e konstruiert. Im Falle einer freien und eigentlichen Gruppenwirkung (und der Strukturgruppe $\SU$) liefert dies sogar ein normiertes Radonma{\ss} auf dem vollen QR-Raum, da hier die \"ubrigen Faktoren nicht auftreten. 
  Im Rahmen der urspr\"unglichen Reduktion auf klassischem Niveau widmen wir uns dagegen dem Spezialfall der homogen-isotropen Schleifenquantenkosmologie. 
  Hier untersuchen wir die ma{\ss}theoretische Eigenschaften des kosmologischen Quantenkonfigurationsraumes $\RR\sqcup \RB$, der durch Quantisierung des reduzierten klassischen Konfigurationsraumes $\RR$ entsteht.  
  Wir zeigen, dass Invarianz unter der kanonischen Fortsetzung $\Transw\colon \RR\times (\RR\sqcup \RB) \rightarrow \RR\sqcup \RB$ der additiven Gruppenwirkung  \RPLUS$\colon \RR \times \RR \rightarrow \RR$ bereits das Haarma{\ss} auf dem $\RB$-Anteil sowie das Nullma{\ss} auf dem $\RR$-Anteil fixiert. Damit erhalten wir denselben
  kinematischen Hilbertraum wie in der Standardschleifenquantenkosmologie. 
  Im Anschluss konstruieren wir weitere normierte Radonma{\ss}e auf $\RR\sqcup \RB$ und vergleichen die resultierenden $L^2$-Hilbertr\"aume miteinander. 
  Auf den in dieser Arbeit konstruierten Hilbertr\"aumen sollen in weiterf\"uhrenden Projekten entsprechend reduzierte Observablenalgebren dargestellt werden.  
 
 Im letzten Abschnitt beweisen wir ein allgemeines Charakterisierungstheorem f\"ur invariante Zusammenh\"ange auf Hauptfaserb\"undeln. Dieses verallgemeinert die beiden klassischen Resultate von Wang sowie Harnad, Shnider und Vinet. Die besondere St\"arke dieser Verallgemeinerung liegt hierbei in ihrer breiten Anwendbarkeit, die es sogar sehr oft erlaubt, R\"aume invarianter Zusammenh\"ange explizit zu bestimmen. Dies ist zum einen wichtig f\"ur die Symmetriereduktion im urspr\"unglichen Sinne und zum anderen f\"ur die Untersuchung der Inklusionsrelationen zwischen den jeweiligen quantisierten reduzierten bzw. reduzierten quantisierten Konfigurationsr\"aumen.
 
}
  } 
  \clearpage
  {\small
  \begin{center}
{\bf Abstract}
\end{center}
\vspace{5pt}
\selectlanguage{english}
\noindent   
The intention of this dissertation is to provide general tools and concepts that allow to perform a mathematically substantiated symmetry reduction in (quantum) gauge field theories. 
	Here, the main focus is on the framework of loop quantum gravity (LQG) where we concentrate on the reduction of the quantum configuration space and the construction of normalized Radon measures on the reduced spaces. For the reduction part, one basically has  the following two possibilities. First, one can quantize the reduced classical configuration space (RQ reduction). Second, one can try to lift the symmetry (i.e., the group action) from the classical to the quantum space in order to perform a symmetry reduction directly on quantum level (QR reduction). Since in LQG only the first approach has been studied so far, we now systematically develop the second approach. In course of that, we attack the important question whether in this context quantization and reduction do commute or not.
	
	In the first part of this thesis we develop the mathematical backbone of the QR-reduction concept. This is based on the extension of group actions $\cw\colon G\times X\rightarrow X$ on sets $X$ to group actions $\specw\colon G\times \Spec(\aA)\rightarrow \Spec(\aA)$ 
 on spectra of $\Cstar$-algebras $\aA\subseteq B(X)$. The QR-configuration space then is formed by the set 
\begin{align*}
	\A_\red:=\{\qw\in \A\: | \: \specw(g,\qw)=\qw\quad \forall\: g\in G\}
\end{align*}
	with $\A$ the quantum configuration space of full LQG. This is in analogy to the classical situation where the reduced classical space $\AR$ is formed by such smooth connections that are invariant under the whole symmetry group. In contrast to that, within the traditional RQ approach one quantizes the space $\AR$ by calculating the spectrum of a separating $\Cstar$-algebra of bounded functions thereon. We show that $\A_\red$ always contains this quantized reduced classical space and that this inclusion is usually even proper. Hence, in this context, quantization and reduction in general do not commute.
	
	In the second part of this dissertation, we construct normalized Radon measures on both the RQ and the QR configuration spaces.  
	We show that, for sufficiently nice group actions, the space $\A_\red$ can be written as a product of (at most) four QR configurations spaces, each corresponding to a certain symmetry type of curves in the base manifold of the underlying bundle. We construct normalized Radon measures on two of these spaces which, together, give rise to a normalized Radon measure on $\A_\red$ whenever the action induced on the base manifold is proper and free. 
Within the traditional approach we concentrate on homogeneous isotropic loop quantum cosmology. Here, we investigate the measure theoretical aspects of the cosmological quantum configuration space $\RR\sqcup \RB$. This arises from quantizing the reduced classical configuration space $\RR$, parametrizing the set of invariant smooth connections in this case. We show that invariance under the canonical extension $\Transw\colon \RR\times (\RR\sqcup \RB) \rightarrow \RR\sqcup \RB$ of the additive group action \RPLUS$\colon \RR \times \RR \rightarrow \RR$ already singles out the normalized Radon measure 
	\begin{align*}
		\mu\colon  A_\RR\sqcup A_{\mathrm{Bohr}} \mapsto \muB(A_{\mathrm{Bohr}} )% \qquad \text{for}\qquad A_\RR\sqcup A_{\mathrm{Bohr}} \in \Borel(\RR)\sqcup \Borel(\RB) =\Borel(\RR\sqcup \RB)
	\end{align*}
	for $A_\RR\sqcup A_{\mathrm{Bohr}} \in \Borel(\RR)\sqcup \Borel(\RB)=\Borel(\RR\sqcup \RB)$ and $\muB$ the Haar measure on $\RB$. 
This means, we end up with the same kinematical Hilbert a space as we have in standard homogeneous isotropic LQC. Finally, we construct further normalized Radon measures on $\RR\sqcup \RB$ and investigate the corresponding $L^2$-Hilbert spaces. It is then part of future projects to establish representations of respective reduced observable algebras on the kinematical Hilbert spaces constructed in this work. 	
	
In the last part, we prove a general characterization theorem for invariant connections on principal fibre bundles which extends the classical result of Wang. We consider several special cases of the general theorem including the result of Harnad, Shnider and Vinet. Our theorem turns out to be predestined for calculating sets of invariant connections explicitly, and, thus, is an important tool for performing symmetry reduction in the traditional way. In addition to that, also for the investigations of the inclusion relations between quantized reduced classical and the respective quantum-reduced configuration space it is usually crucial to determine the respective sets of invariant connections explicitly.
}

\newpage
\section*{Acknowledgements}
The author thanks Christian Fleischhack for numerous discussions
and many helpful comments on several drafts of this dissertation. He thanks Johannes Aastrup, Martin Laubinger and Benjamin Schwarz for continued support, in particular, at the beginning of his doctoral studies. 
He is grateful for discussions with various members of the math faculty of the University of Paderborn. In particular, with Joachim Hilgert, Bernhard Kr\"otz, Alexander Schmeding and Andreas Schmied. He thanks Gerd Rudolph for general discussions and comments on a first draft of the paper \cite{InvConn}, and the anonymous SIGMA referees for several helpful comments and suggestions. He also thanks  Benjamin Bahr and Alexander Stottmeister for providing him with physical background 
as well as Jonathan Engle, Christian Fleischhack, Hanno Sahlmann and Stefan Waldmann for supporting him on his academic path. 
 Finally, he expresses his deep gratitude to all the beloved people who have encouraged him during his time in Paderborn both close and from distance. 
 
 The author has been supported by the Emmy-Noether-Programm of
the Deutsche Forschungsgemeinschaft under grant FL~622/1-1.
\newpage
\tableofcontents 

\newpage

\section{Introduction} 

\subsection{Quantum Gravity}
One of the most challenging problems of modern physics is the embedding of quantum mechanics and general relativity into a superordinated (and mathematically substantiated) physical theory. Such a unified description is expected to play a role whenever massive objects are concentrated on small spaces being the case, e.g., for big bang scenarios or black holes. There, one would wish such a theory to resolve the singularities that appear when one describes these phenomena  
in the classical framework of general relativity. 

Serious difficulties in combining quantum mechanics with general relativity arise from the conceptual differences of these two theories and from the lack of physical experiments hinting to some kind of quantum gravity effects. Such effects, however, one would expect from a theory unifying gravitational and quantum nature of matter. 
So, at this point, unification can only be done on a purely theoretical level, and here one basically has the choice between the following two strategies. First, one can try to construct a completely new theory which, in the appropriate physical limits, reproduces general relativity and quantum mechanics. Second, one can try to quantize general relativity directly, hoping to end up with a unified theory or, to be more realistic, to get some hints on how such a theory should look like. 
Following these philosophies, promising candidates are string theory, a pertubative approach to the construction of a superordinated theory, and the loop quantum gravity approach we will follow in this thesis.   

\subsubsection{Loop Quantum Gravity}
Being a non-pertubative and background independent approach, loop quantum gravity (\gls{LQG}) \cite{BackLA, Thiemann} 
seems to be appropriate for understanding quantum gravity effects near classical singularities where curvature is by no means small. Indeed, within this context it was possible to derive the Bekenstein-Hawking area law for a large class of black holes. \cite{blackbeken, Doma}  
Now, being a canonical quantization of gravity, LQG is based on a splitting of space-time into time and space, entailing that the four-dimensional covariance of general relativity is no longer manifest. \cite{Thiemann, Hanno} In particular, the 4-dimensional diffeomorphism constraint splits up into a spatial and the Hamiltonian constraint, the latter one defining the dynamics of LQG. 
Unfortunately, the quantization of the Hamiltonian of the full theory turns out to be difficult and is, at this point, not completely understood. \cite{Thiemann0} Here, symmetry reduced versions of LQG like loop quantum cosmology \cite{Bojolivrev} can help to better understand this quantization for the full theory. 
In addition to that,
 mathematical developments like the 
 reduction concept to be developed in this work carry over to a bigger class of gauge field theories, so that LQG is not an isolated field of research but an approach to quantum gravity whose developments enhance other areas of theoretical physics and even mathematics. 
\subsubsection{Challenges}
Although symmetry reduced versions of loop quantum gravity exists, no conceptually satisfying reduction concept has been developed so far. Indeed, 
to this point reduction has been done on a rather intuitional level, so that the connection to the full theory was usually not manifest but had to be established in a laborious way. Standard homogeneous isotropic loop quantum cosmology (\gls{LQC}) may serve as a prime example for this. However, since symmetric situations in nature  occur,  for a physical theory one would expect to have a reasonable reduction concept which allows to get rid of superfluous degrees of freedom. 

Now, being an Ashtekar approach to quantum field theories, LQG is based on functional integrals. So, besides the definition of a reduced quantum configuration space, a reduction theory here should include the construction of corresponding Radon measures which (as for the full theory) allow to integrate over field configurations.  
	In addition to that, appropriately reduced holonomy-flux algebras (and Hamiltonians) have to be represented on the respective kinematical Hilbert spaces of square integrable functions. Then, states on the reduced algebras have to be embedded into respective symmetric sectors of LQG. \cite{BojoKa} 
Once such a concept has been established, the consideration of highly symmetric systems might allow to make verifiable predictions that can help to advance the full theory.

\subsection{Mathematical Context}
The basic mathematical objects studied in this thesis are spaces of connections on principal fibre bundles, spectra of $\Cstar$-algebras of bounded functions and normalized Radon measures. Since we are investigating the problem of symmetry reduction, also left actions of (Lie) groups will play an important role. 

\subsubsection{Invariant and Generalized Connections}
In its simplest form, the configuration space\footnote{Indeed, actually the quotient $\Con\slash \GAG$ of $\Con$ w.r.t.\ the set $\GAG$ of gauge transformations of $P$ 
  is considered as physically relevant configuration space. However, to keep it simple, in this work we concentrate on the space $\Con$ and its quantum analogue $\A$ (see below). Here, the main reason is that we rather expect technical than conceptual difficulties when carrying over the developments of this work to the ``up to gauge''-case, i.e., to the quotient $\A\slash \ovl{\mathcal{G}}$ of $\A$ w.r.t.\ the compact group $\ovl{\mathcal{G}}$ of generalized gauge transformations. See, e.g., the outlook section for some more details.} 
 of a classical gauge field theory is formed by the set $\Con$ of smooth connections on a principal fibre bundle $(P,\pi,M,S)$.\footnote{In the LQG approach we usually have $P=\Sigma\times \SU$ for a 3-dimensional Cauchy surface $\Sigma$ of a space-time 
$M$.} 
Symmetries are realized by Lie groups of automorphisms $(G,\Phi)$, and symmetry reduction 
 just means to calculate the set 
	\begin{align}
	\label{eq:invconnPhi} 
 	\AR:=\{\w\in \Con\:|\: \Phi_g^*\w=\w\text{ for all }g\in G\} 
 	\end{align} 
 of smooth connections invariant under pullbacks by all 
symmetry group elements. \cite{Wang, HarSni} 
 
Then, in the Ashtekar approach to quantum gauge field theories, the quantum configuration space $\ovl{\Con}$ of the full theory is formed by the spectrum of a separating $C^*$-algebra $\PaC$ of bounded functions on $\Con$. 
Here, the $C^*$-algebra of cylindrical functions $\PaC\subseteq B(\Con)$ is  generated by matrix entries\footnote{With respect to some  faithful matrix representation of the structure group $S$.} of parallel transports along the elements of a fixed set $\Pa$ of curves in the base manifold of $P$. Here, the main reason for switching from $\Con$ to $\A$ is that (in contrast to $\Con$) there exists a natural measure on $\A$, the Ashtekar-Lewandowski one. This measure allows to integrate over field configurations and defines the $L^2$ Hilbert space on which the unique  representation \cite{UniqLew, ChUn} of the holonomy-flux algebra is realized.  
The separation property of $\PaC$ here guarantees that the space $\Con$ is canonically embedded\footnote{As we have not chosen any topology on $\Con$, this just means that $\iota_\Con$ is injective.} via  
\begin{align*}
\iota_\Con \colon \Con\rightarrow \A,\quad \w \mapsto [f\mapsto f(\w)].
\end{align*}
 Since 
$\iota_\Con(A)\subseteq\A$ is even dense \cite{Rendall}, the space $\A$ can be seen as some kind of compactification of $\Con$ (provided that the structure group is compact, as then $\PaC$ is unital).

\subsubsection{Symmetry Reduction}
Traditionally, symmetry reduction in LQG has been
 done by calculating the spectrum of a $\Cstar$-algebra of the form $\ovl{\PaC'|_{\AR}}\subseteq B(\AR)$. This is just the closure of
	\begin{align*} 
 	\PaC'|_{\AR}:=\left\{f|_\AR\:\big|\: f\in \PaC'\right\}
	\end{align*} 
	in $B(\AR)$, where $\PaC'$ denotes the $\Cstar$-algebra of cylindrical functions that corresponds to some set $\Pa'$ of curves in $M$. 
Choosing $\Pa\neq \Pa'$ then usually offers the problem that the reduced space cannot be naturally embedded into $\A$. Indeed, this is exactly the case for homogeneous isotropic LQC ($P=\RR^3\times \SU$) where originally the set of linear curves was used for $\Pa'$. \cite{Brunnhack} In particular, in view of the embedding strategy for states proposed in \cite{BojoKa} this is disadvantageous. To fix this problem, in \cite{ChrisSymmLQG} the space $\ARQ=\Spec\big(\ovl{\PaC|_{\AR}}\big)$ was introduced.  Indeed, $\ARQ$ is naturally embedded in $\A$, even homeomorphic to the closure of $\iota_\Con(\AR)$ in $\A$.

Now, $\ARQ$ arises from a quantization of the reduced classical space $\AR$ and not from a reduction of the quantum space $\A$. 
So, in this thesis we will follow the second, conceptual more satisfying approach which has not been investigated so far.  
This means that we will perform a symmetry reduction directly on the quantum level just by extending the left action 
		$\cw\colon G\times \Con \rightarrow \Con$, 
			$(g,\w)\mapsto \Phi_{g^{-1}}^*\w$
	which determines   
	 \begin{align*}
	\AR=\{\w\in \Con\:|\: \cw(g,\w)=\w\:\: \forall\:g\in G \}
	\end{align*}
	to a left action $\specw\colon G\times \A\rightarrow \A$. This will provide us with the quantum-reduced configuration space 
	\begin{align*}
	\AQR:=\{\ovl{\w}\in \A\:|\: \specw(g,\ovl{\w})=\ovl{\w}\:\: \forall\: g\in G \}
	\end{align*}
	which, as we will see, always contains the quantized reduced classical space $\ARQ$. In several situations, this inclusion even turns out to be proper so that in our context quantization and reduction usually  do not commute. This is also the main reason why it is much easier to construct measures on $\AQR$ than on $\ARQ$.\footnote{For the case of homogeneous isotropic LQC we will construct measures on $\ARQ$ by hand, see Section \ref{sec:HomIsoCo}.}  Indeed, for 
	$\AQR$ so-called modification techniques are available, which only work with limitations for $\ARQ$. This is just because they do not leave this space 
	invariant.  
To get a rough idea what modification of a generalized connection here  means, it is important to know that for compact and connected structure groups (the typical LQG case) one can identify $\A$ with the space of homomorphisms $\TRHOM$. Such homomorphisms map curves in $\Pa$ to structure group elements, and 
 modification then just means to change the values of such a map along some fixed curves in a specific way. The same techniques then will also allow to construct normalized Radon measures by means of projective structures on $\AQR$.

\subsubsection{Projective Structures and Normalized Radon Measures}
A practicable way to construct normalized Radon measures on compact Hausdorff spaces (and the usual way to do it for quantum configuration spaces in LQG) 
is to identify the space $X$ of interest as a projective limit of a reasonable family of compact Hausdorff spaces $\{X_\alpha\}_{\alpha\in I}$.   
 Here, reasonable means that each of these spaces carries a natural normalized Radon measure, and that all these measures are consistent in the way described below.  
Here, the basic idea is that for $X$ high-dimensional in a certain sense, each $X_\alpha$ catches finitely many degrees of freedom of this space.  

Now, to be a projective limit basically means that\footnote{This definition differs slightly from the standard one \cite{ProjTechAL}, but is equivalent to it.} 
\begin{enumerate}
\item
\label{akakakaka}
There exist continuous surjective projection maps $\pi_\alpha\colon X\rightarrow X_\alpha$ which separate the points in $X$
\item
\label{bkakakaka}
	For each two $\alpha_1,\alpha_2\in I$ ($I$ is a directed set) with $\alpha_1\leq \alpha_2$ there exists a transition map $\pi^{\alpha_2}_{\alpha_1}\colon X_{\alpha_2}\rightarrow X_{\alpha_1}$ with $\pi^{\alpha_2}_{\alpha_1}\cp \pi_{\alpha_2}=\pi_{\alpha_1}$. 
 \end{enumerate}
  A Riesz-Markov argument shows that the normalized Radon measures on $X$ are in bijection with the consistent families $\{\mu_\alpha\}_{\alpha\in I}$ of normalized Radon measures $\mu_\alpha$ on $X_\alpha$. Here, consistency just means that for $\alpha_1,\alpha_2\in I$ with $\alpha_1\leq \alpha_2$ the equality $\pi^{\alpha_2}_{\alpha_1}(\mu_{\alpha_2})=\mu_{\alpha_1}$ holds.
  
Then, for $\A\cong \TRHOM$ one usually chooses $I$ to consist of certain finite tuples $\alpha=(\gamma_1,\dots,\gamma_k)$ of curves $\gamma_1,\dots,\gamma_k\in \Pa$, and defines 
\begin{align*}
\pi_\alpha(\homm):=(\homm(\gamma_1),\dots,\homm(\gamma_k))\in S^k
\end{align*} 
for $S^k$ the $k$-fold product of the structure group. Choosing these tuples appropriately then ensures that $\pi_\alpha$ is surjective, and that the Haar measure on $S$ can be used to define a respective consistent family of normalized Radon measures.
Now, we will see that under the identification $\A\cong \TRHOM$ the quantum-reduced space $\AQR\subseteq \A$ corresponds to a subset $\ITRHOM \subseteq \TRHOM$ of homomorphisms that fulfil certain invariance properties. These properties give non-trivial restrictions to the images of the maps $\pi_\alpha$, so that we will be forced to adapt the whole projective structure in order to obtain a non-trivial measure on $\ITRHOM$.
For the spaces $\ARQ$ the situation is even more difficult, just because the subsets $\pi_\alpha\big(\ARQ\big)\subseteq \pi_\alpha\big(\AQR\big)$ are usually much more complicated.

\subsection{Aims and Organization}
Although this dissertation is rather motivated by the framework of LQG, its main goal is to  provide general tools and concepts that allow to perform a mathematically rigorous symmetry reduction also in other (quantum) gauge field theories. Here, we will focus on the reduction of the quantum configuration space and the definition of reasonable Radon measures thereon. 
We also attack the question whether, in our context, quantization and reduction commute or not. In course of this, we will prove a general characterization theorem for invariant smooth connections on principal fibre bundles which generalizes the classical results of Wang \cite{Wang} and Harnad, Shnider and Vinet \cite{HarSni}. 
 The definition of representations of respective reduced holonomy-flux algebras on the constructed $L^2$-Hilbert spaces 
 is left as a future task.

This work is organized as follows:
\begingroup
\setlength{\leftmargini}{12pt}
\begin{itemize}
\item
  The preliminaries in Section \ref{sec:prel} contain the notations, as well as the basic definitions, conventions and facts concerning principal fibre bundles, projective structures and Radon measures.  
\item
  In the first part of Section \ref{sec:specmathback}, 
  we present the LQC relevant cases which  will serve as prime examples during this work. In addition to that, we discuss the properties of $\SU$ being relevant for our later calculations. In the second part, we will collect the facts on spectra of $\Cstar$-algebras of bounded functions which build the mathematical backbone of the reduction concept introduced in \cite{ChrisSymmLQG} and that one, we will develop in Section \ref{sec:SPecExtGr}.  
  In the last part, we highlight the most crucial properties of the Bohr compactification of a locally compact abelian group. This is of relevance because the Bohr compactification $\RB$ of $\RR$ plays a key role in standard homogeneous isotropic LQC (see Section \ref{sec:HomIsoCo}), and also in the constructions in Subsection \ref{sec:ConSp}. We close Section \ref{sec:specmathback} with a characterization of the continuous abelian group structures on spectra of certain unital $\Cstar$-algebras.
\item
  In Section \ref{sec:SPecExtGr}, we are going to lift Lie groups $(G,\Phi)$ of automorphisms of principal fibre bundles $P$ to spectra $\A$ of $\Cstar$-algebras of cylindrical functions $\PaC\subseteq B(\Con)$. Here, $\Con$ denotes the set of smooth connections on the respective bundle and $B(\Con)$ the set of bounded functions on $\Con$.  
  This will provide us with the notion of an invariant generalized connection, the quantum analogue of an invariant classical (smooth) one.
  
  In the first step, we will use the concept of a $\Cstar$-dynamical system in order to extend an action $\text{\gls{CCW}}\colon G\times X\rightarrow X$ (of a group $G$ on a set $X$) 
  to an action $\specw\colon G\times \Spec(\aA)\rightarrow \Spec(\aA)$   on the spectrum of a $\Cstar$-subalgebra $\aA\subseteq B(X)$. 
  Then, we adapt this to the case where $X$ equals the set of smooth connections on a principal fibre bundle. Here, it will be crucial that 
  the set of paths $\Pa$, used for the definition of $\PaC$, is invariant in the sense that 
	\begin{align*}
	\gamma\in \Pa\qquad \Longrightarrow \qquad \wm_g\cp \gamma \in \Pa\qquad\forall\: g\in G
	\end{align*}  
   with $\wm\colon G\times M\rightarrow M$ the action induced by $\Phi$ on the base manifold $M$ of $P$. 
   Finally, we will consider the case where the structure group $S$ of $P$ is compact, and where the set of curves has the additional property of independence.\footnote{This is the case, e.g.\, if $\Pa$ is the set of embedded analytic curves and $S$ is connected.} In this situation, it will be possible to identify the quantum-reduced configuration spaces $\AQR$ with spaces of so-called invariant homomorphisms. This will later be important for the investigations of the inclusion relations between $\ARQ$ and $\AQR$, as well as the construction of measures on $\AQR$. At the end of Section \ref{sec:SPecExtGr}, we will show that the spaces $\ARQ$ and $\AQR$ are usually of measure zero w.r.t.\ the Ashtekar-Lewandowski measure (LQG standard measure) on $\A$.
\item
 In Section \ref{susec:LieALgGenC}, we will develop modification techniques for invariant homomorphisms in the event that the induced action $\wm$ on $M$ is analytic and pointwise
  proper. 
In the first part, we collect the relevant facts and definitions concerning analytic and Lie algebra generated curves, whereby 
 $\gamma$ is called Lie algebra generated iff it is (up to parametrization) of the form $\gamma\colon t\mapsto \wm_x(\exp(t\cdot\g))$ for some $x\in M$ and $\g\in \mg\backslash \mg_x$. Here, $\mg_x$ denotes the Lie algebra of the $\wm$-stabilizer $G_x$ of $x$.

In Subsection \ref{sec:ModifLAGC}, we will modify invariant homomorphisms along such Lie algebra generated curves, and in Subsection \ref{sec:inclrel} we apply this in order to show that the inclusion $\ARQ\subseteq \AQR$ is proper in several situations. In particular, we conclude that quantization and reduction do not commute in \mbox{(semi-)homogeneous} LQC. For homogeneous isotropic LQC this will be shown in Section \ref{sec:HomIsoCo}. 

In the last part of Section \ref{susec:LieALgGenC}, we will prove an analogue of the modification result from Subsection \ref{sec:ModifLAGC}, but now for free 
curves. These are embedded analytic curves $\gamma$ containing a subcurve $\delta$ (free segment) for which $\wm_g\cp \delta =\delta$ holds whenever $\im[\wm_g\cp \delta] \cap \im[\delta]$ is infinite for some $g\in G$. We will show that, under the condition that $\wm$ is analytic and pointwise proper, each free curve $\gamma$ is discretely generated by the symmetry group.\footnote{This means that we find a (maximal) free segment $\delta$ of $\gamma$ such that $\gamma$ admits a decomposition into finitely many subcurves, each being (up to parametrization) an initial or final segment of $\wm_g\cp \delta$ for some $g\in G$. Here, each of these subcurves which is not an initial or final segment of $\gamma$, then even equals (up to parametrization) the full segment $\wm_g\cp\delta$ for the respective $g\in G$.}
 Moreover, we will see that if each $\wm$-stabilizer is even a normal subgroup and $\wm$ is in addition transitive or proper, then each embedded analytic curve is either free or (up to parametrizations) Lie algebra generated. 
For instance, this is the case in \mbox{(semi-)homogeneous} LQC, where it will us to construct a normalized Radon measure on $\AQR$ if the structure group is $\SU$, see Section \ref{sec:MOQRCS}. 
  \item
  	In Section \ref{sec:MOQRCS}, we will construct normalized Radon measures on certain quantum-reduced configuration spaces for the case that $\wm$ is analytic and pointwise proper. Here, the main idea is to split up the set $\Paw$ of embedded analytic curves into suitable subsets $\Pa_\alpha$, $\alpha\in I$, each being closed under decomposition and inversion of its elements. In fact, then (Subsection \ref{subsub:InvHoms})
  	\begin{align*}
  		\AQRw\cong\prod_{\alpha\in I}\AQRInd{\alpha}
  	\end{align*}
  	holds and, provided that we have constructed normalized Radon measures on each of the spaces $\AQRInd{\alpha}$, we obtain a normalized Radon measure on $\AQRw$, just by taking the Radon product one. 		
  	For instance, if $\wm$ is proper and free, we have $\AQRw\cong\AQRInd{\mg}\times \AQRInd{\mathrm{F}}$ where $\AQRInd{\mg}$ corresponds to the set of Lie algebra generated and $\AQRInd{\mathrm{F}}$ to the set of the free curves.  
	So, in this case it suffices to construct normalized Radon measures on these two spaces, which is exactly the content of Section \ref{sec:MOQRCS}.
	 
	In fact, in the first part of this section, 
	we will construct a normalized Radon measure on $\AQRFNS$ for the case that $S$ is compact and connected. Here, $\AQRFNS$ corresponds to the set of such  free curves whose stabilizer\footnote{As we will see, this is a well-behaving quantity if $\wm$ is analytic and pointwise proper.} is trivial, whereby 
	in the above situation 
	 $\AQRFNS=\AQRInd{\mathrm{F}}$ holds, just because 
there 
	$\wm$ was assumed to be free. 
	
	In the second part of Section \ref{sec:MOQRCS}, we will construct a normalized Radon measure $\mLAS$ on  $\AQRInd{\mg}$,  exemplarily, for the most LQC relevant case that $S=\SU$ (and for each $n$-torus). Here, we will require some additional conditions on the $\wm$-stabilizers, which appear to hold, e.g.\, in spherically symmetric, \mbox{(semi-)homogeneous} and homogeneous isotropic LQC. 
	Then, if $\wm$ is in addition transitive (such as in homogeneous and homogeneous isotropic LQC), we have the reasonable kinematical Hilbert space $L^2(\AQRLA,\mLAS)$. Indeed, in the transitive case $\AInd{\mg}$, hence $\AQRLA$ is a physically meaningful candidates for   a quantum(-reduced) configuration space because 
	  there $\iota\colon \Con \rightarrow \AInd{\mg}$ is injective as, in this situation, the cylindrical functions that correspond to $\Pags$ separate the points in $\Con$.  
  \item
  	In Section \ref{sec:HomIsoCo}, we will focus on homogeneous isotropic LQC.  We show that quantization and reduction do not commute and investigate the measure theoretical aspects of the classically reduced quantum configuration space $\ARQ$. This space corresponds to the set of  embedded analytic curves in $M=\RR^3$ and is  
  	homeomorphic to the compact Hausdorff space\footnote{The topology on $\RR\sqcup \RB$ is quite tricky. Details will be given in Section \ref{sec:HomIsoCo}.} $\RR\sqcup \RB$. \cite{ChrisSymmLQG} 
  	In contrast to the LQC standard approach (where the reduced quantum space is defined by all linear curves in $\RR^3$ and is homeomorphic to the compact abelian group $\RB$) on $\RR\sqcup \RB$ no Haar measure is available.
  	This will be shown in the first part of Section \ref{sec:HomIsoCo}, where we prove that no continuous group structure can exist on this space. 	
Then, changing the focus from Haar to normalized Radon measures, we will show that $\muB$ is the unique normalized Radon measure which is invariant under  
  	 the canonical extension  $\Transl\colon \RR\times \RB \rightarrow \RB$ of the additive group action \RPLUS$\colon \RR \times \AR \rightarrow \AR$ of $\RR$ on $\AR\cong \RR$. 	 
 	Moreover, we will prove that the same invariance condition  singles out a normalized Radon measure on 
  	 $\ARQ\cong \RR\sqcup \RB$. This measure even 
  	defines the same kinematical Hilbert space $\Lzw{\RB}{\muB}$ as we have in standard LQC, supporting this approach from the mathematical side.  
  	In the last part of Section \ref{sec:HomIsoCo}, we will use projective structures in order to construct further normalized Radon measures on $\RR\sqcup \RB$ and, finally, compare the respective Hilbert spaces of square integrable functions thereon.
      \item 
  	In Section \ref{CHarinvconn}, we will prove a characterization theorem for invariant connection on principal fibre bundles which generalizes the classical results of Wang \cite{Wang} and Harnad, Shnider and Vinet \cite{HarSni}. We consider several special situations such as (almost) fibre transitivity (Case \ref{scase:slicegleichredcluster} and \ref{th:wang}), Lie groups of gauge transformations (Case \ref{scase:GaugeTransf}), trivial bundles (Case \ref{scase:trivbundle}) and the gauge fixing situation (Case \ref{scase:OneSlice}). Along the way, we give applications to loop quantum gravity. In particular, we will calculate the \mbox{(semi-)homogeneous} and spherically symmetric  connections already introduced in Example \ref{ex:LQC} and which we will use in Subsection \ref{sec:inclrel} in order to show that the inclusion $\ARQ\subseteq \AQR$ is proper in \mbox{(semi-)homogeneous} and spherically symmetric  LQC. We also show that the set of invariant connections depends crucially on the explicit lift of an action $\wm\colon G\times M\rightarrow M$ to $P$, see Remark \ref{rem:liftuntersch}.
\end{itemize}
\endgroup
\noindent
Each of the sections \ref{sec:SPecExtGr} -- \ref{CHarinvconn} closes with a short summary of its most relevant results.

\section{Preliminaries}
\label{sec:prel}
In this brief section we fix the notations and collect some facts and definitions  which are more or less standard, but crucial for this thesis.
\subsection{Notations}
\label{sec:notations}
Manifolds are always assumed to be smooth or analytic. If $M, N$ are manifolds and $f\colon M\rightarrow N$ differentiable, then $\dd f\colon TM\rightarrow TN$ denotes the differential map between their tangent bundles. The map $f$ is said to be an immersion iff for each $x\in M$ the restriction $\dd_xf:=\dd f|_{T_xM}\colon T_xM\rightarrow T_{f(x)}N$ is injective. Elements of tangent spaces usually are written with arrows, such as $\vv_x\in T_xM$. Here, we subscript the base point $x$ whenever it helps to clarify the calculations. In particular, in Section \ref{CHarinvconn} this will be helpful to keep the track of the calculations.  

Let $V$ be a finite dimensional vector space. A $V$-valued 1-form $\w$ on the manifold $N$ is a smooth map $\w\colon TN\rightarrow V$ whose restriction $\w_y:=\w|_{T_yN}$ is linear for all $y\in N$. The pullback of $\w$ by $f$ is the $V$-valued 1-form $f^*\w\colon TM\rightarrow V$, $\vec{v}_x\rightarrow \w_{f(x)}(\dd_xf(\vec{v}_x))$. If it is clear which tangent space $\vv$ belongs to, we usually do not subscript $\w$ by the base point, e.g., we write $\w(\vv_x)$ instead of $\w_x(\vv_x)$.

Let $G$ be a Lie group and $\mathfrak{g}$ its Lie algebra. For $g\in G$ we define the corresponding conjugation map by $\text{\gls{CONJ}}\colon G\rightarrow G$, $h\mapsto g h g^{-1}$. Its differential $\dd_e\Co{g}\colon \mathfrak{g}\rightarrow \mathfrak{g}$ at the unit element $e\in G$ is denoted by $\Add{g}$, and by \gls{ADD} we will denote the left action $\Ad\colon G\times \mg\rightarrow \mg$. 

Let $\Psi$ be a (left) action of the Lie group $G$ on the manifold $M$. For $g\in G$ and $x\in M$ we define 
$\Psi_g\colon M\rightarrow M$, $y\mapsto \Psi(g,y)$ and $\Psi_x\colon G\rightarrow M$, $h\mapsto\Psi(h,x)$, respectively.  
If it is clear which action is meant, 
we will often write $L_g$ instead of $\Psi_g$ as well as $g\cdot x$ or $g x$ instead of $\Psi_g(x)$. 
For $\vec{g}\in \mathfrak{g}$ and $x\in M$ the map 
	\begin{align}
	\label{eq:fundvf}
		\wt{g}(x):=\dttB{t}{0}\: \Psi_x(\exp(t\vec{g}))
	\end{align}
 is called the \emph{fundamental vector field w.r.t.\ $\vec{g}$}.
The Lie subgroup $G_x:=\left\{g\in G\: \big| \: g\cdot x=x\right\}$ is called the \emph{stabilizer} of $x\in M$ (w.r.t.\ $\Psi$), and its Lie algebra $\mathfrak{g}_x$ equals $\ker[\dd_x\Psi]$, see e.g.\ \cite{DuisKolk}.
The \emph{orbit} of $x$ under $G$ is the set $Gx:=\im[\Psi_x]$. $\Psi$ is said to be \emph{transitive} iff $Gx=M$ holds for one (and then each) $x\in M$. The action $\Psi$ is called \emph{proper at $x$} iff for each net
$\{g_\alpha\}_{\alpha\in I}\subseteq G$ the convergence of  
$\{\Psi(g_\alpha,x)\}_{\alpha\in I}\subseteq M$ implies the existence of a convergent subnet\footnote{This is a net $\{h_\beta\}_{\beta\in J}\subseteq G$ together with a map $\iota\colon J\rightarrow I$ such that $h_\beta=g_{\iota(\beta)}$ for all $\beta\in J$. Moreover, for each $\alpha\in I$ we find $\beta \in J$ such that $\iota(\beta')\geq \alpha$ holds for all $\beta'\geq \beta$.} of $\{g_\alpha\}_{\alpha\in I}$. This is equivalent to require that $\wm_x^{-1}(K)\subseteq G$ is compact whenever $K\subseteq M$ is compact, i.e., that $\wm_x$ is a proper map. Then, $\wm$ is called \emph{pointwise proper} iff it is proper at $x$ for all $x\in M$.  
Finally, $\Psi$ is said to be \emph{proper} iff the convergences of $\{\Psi(g_\alpha,x_\alpha)\}_{\alpha\in I}\subseteq M$ and $\{x_\alpha\}_{\alpha\in I}\subseteq M$ imply the existence of a convergent subnet of $\{g_\alpha\}_{\alpha\in I}$. Analogous conventions hold for right actions.

A curve is a continuous map $\gamma\colon D\rightarrow M$. Here $D\subseteq \mathbb{R}$ is an interval, i.e., a set of the form $(a,b]$, $[a,b)$ or $[a,b]$ with $a<b$. The curve $\gamma$ is said to be of class\footnote{We allow $k\in \{\mathbb{N}_{\geq 1},\infty,\omega\}$, where $\omega$ means analytic.} $\CC{k}$ iff there is a $\Ck$-curve (in the sense of $\CC{k}$ maps between manifolds) $\gamma'\colon I\rightarrow M$ such that $\gamma'|_{D}=\gamma$. Here, $I$ is an open interval that contains $D$, and $\gamma'$ is called an extension of $\gamma$ in this case. The same conventions hold for diffeomorphisms $\adif\colon D\rightarrow D'\subseteq\RR$.
The $\Ck$-curve $\gamma\colon D\rightarrow M$ is called an embedding iff we find an extension $\gamma'\colon I\rightarrow M$ which is an injective immersion and a homeomorphism onto its image equipped with the relative topology. If $k=\w$, we say that $\gamma$ is an embedded analytic curve. 
A curve $\gamma$ is called \emph{piecewise} (embedded) $\CC{k}$ or (embedded) \emph{$\CC{k}$-path} iff there are real numbers $a=\tau_0 <\ldots <\tau_k=b$ such that for each $0\leq i\leq k-1$ the restriction $\gamma|_{[\tau_i,\tau_{i+1}]}$ is an (embedded) curve of class $\CC{k}$. For the case that $\gamma$ is $C^k$ for some $k\in \NN_{\geq 1}$ or if $t$ is not contained in the interior of $D$, we will define the tangent vector $\dot\gamma(t) \in T_{\gamma(t)}M$ 
in the canonical way.\footnote{Recall that there occur some technical difficulties if one just treats $M$ as $\CC{k}$-manifold. This is because only for $k=\infty$ the algebraic definition of a tangent vector coincides with the geometric one. So, for $k\neq \infty$, i.e, if $\gamma$ is a $C^k$-curve in the smooth manifold $M$, we use some smooth chart $(U,\phi)$ around $\gamma(t)$ in order to obtain a smooth curve $\delta$ through $\gamma(t)$ with $\dttB{s}{t}(\phi\cp\delta)(t)=\dttB{s}{t}(\phi\cp\gamma)(t)$. Then, we define the tangent vector of $\gamma$ at $t$ to be the equivalence class $[\delta] \in T_{\gamma}(t)M$. Now, for the case that $t$ is not contained in the interior of $D$, we just use an extension of $\gamma$ in order to define $\dot\gamma(t)$.} 
In the following, $I$ and $K$ will usually denote open and compact intervals, respectively, whereby $I$ also will occur as index set if it is not in conflict with our notations.

If $W$ is a set and $U\subseteq W$ a subset, then $U^c=W\backslash U$ denotes the complement of $U$ in $W$. 
If $W$ is a topological space, by $\ovl{U}$ we usually denote the closure of $U$ in $W$. A different convention holds for subsets $\bB\subseteq B(Z)$ of $\Cstar$-algebras of bounded functions. Here, $\ovl{\bB}$ denotes the $\Cstar$-subalgebra of $B(Z)$ which is generated by $\bB$, see also Convention \ref{conv:Boundedfunc}.

If $Y$ is a locally compact Hausdorff space, then $\Cinf(Y)$ denotes the set of complex-valued, continuous functions on $Y$ that \emph{vanish at infinity}. This is that for each $f\in \Cinf(Y)$ and $\epsilon >0$ there is a compact subset $K_\epsilon\subseteq Y$ such that $|f|\leq \epsilon$ on $Y\backslash K_\epsilon$. If $\aA$ is a Banach algebra, then $\Spec(\aA)$ denotes the set of all non-zero, multiplicative, $\mathbb{C}$-valued functionals on $\aA$. 
Then, by $\GT \colon \aA \rightarrow \Cinf(\Spec(\aA))$, $a\mapsto \left[\hat{a}\colon f\mapsto f(a)\right]$ we will denote the Gelfand transformation.
Recall that the Gelfand-Naimark theorem states that $\GT$ is an isometric $^*$-isomorphism if $\aA$ is a $\Cstar$-algebra.

\subsection{Principal Fibre Bundles}
\label{subsec:InvConn}
Let $\pi\colon P\rightarrow M$ be a smooth map between the manifolds $P$ and $M$, and denote by $\text{\gls{FX}}:=\pi^{-1}(x)\subseteq P$ the fibre over $x\in M$ in $P$. Moreover, let $S$ be a Lie group that acts via $R\colon P\times S\rightarrow P$ from the right on $P$. 
If there is an open covering $\{U_\alpha\}_{\alpha\in I}$ of $M$ and a family $\{\phi_\alpha\}_{\alpha\in I}$ of diffeomorphisms
$\phi_\alpha\colon \pi^{-1}(U_\alpha)\rightarrow U_\alpha\times S$ with
\vspace{-3pt}
\begin{align}
  \label{eq:bundlemaps}
  \phi_\alpha(p\cdot s)=\big(\pi(p),[\pr_2\cp\phi_\alpha](p)\cdot s\big) \qquad\quad \forall\:p\in \pi^{-1}(U_\alpha),\forall\: s\in S,
\end{align}
\newline
\vspace{-28pt}
\newline
then \gls{PMS} is called \emph{principal fibre bundle} with total space $P$, projection map $\pi$, base manifold $M$ and structure group $S$. Here, $\pr_2$ denotes the projection onto the second factor $S$. It follows from \eqref{eq:bundlemaps} that $\pi$ is surjective, and that:
\begingroup
\setlength{\leftmargini}{20pt}
\begin{itemize}
\item
  \itspacecc
  $R_s(F_x)\subseteq F_x$ for all $x\in M$ and all $s\in S$.
\item
  \itspacec
  If $x\in M$ and $p,p'\in F_x$, then $p'=p\cdot s$ for a unique element $s\in S$.
\end{itemize}  
\endgroup
\itspacec
\noindent
Then, for $p,p'\in F_x$ contained in the same fibre, we will denote by $\text{\gls{DIFF}}(p,p')\in S$ the unique element for which $p'=p\cdot s$ holds.

The subspace $Tv_pP:=\ker[d_p\pi]\subseteq T_pP$ is called \emph{vertical tangent space} at $p\in P$ and
\vspace{-3pt}
\begin{align*}
  \wt{s}(p):=\dttB{t}{0}\: p\cdot \exp(t\vec{s})\in Tv_pP\qquad\quad \forall\: p\in P
\end{align*}
\newline
\vspace{-30pt} 
\newline
denotes the fundamental vector field of $\vec{s}$ w.r.t.\ the right action of $S$ on $P$. Recall that the map $\mathfrak{s}\ni\vec{s}\rightarrow \wt{s}(p)\in Tv_pP$ is a vector space isomorphism for all $p\in P$.

Complementary to that, a (smooth) \emph{connection} $\w$ is an $\mathfrak{s}$-valued 1-form on $P$ with  
\begingroup
\setlength{\leftmargini}{20pt}  
\begin{itemize}
\item
  \itspacecc
  $R_s^*\w= \Add{s^{-1}}\cp\: \w$ for all $s\in S$,
\item
  \itspacec
  $\w_p(\wt{s}(p))=\vec{s}$ for  all $\vec{s}\in \mathfrak{s}$. 
\end{itemize}
\endgroup
\itspacecc
\noindent
The subspace $Th_pP:=\ker[\w_p]\subseteq T_pP$ is called the \emph{horizontal tangent space} at $p$ (w.r.t.\ $\w$). We have $\dd R_s(Th_pP)=Th_{p\cdot s}P$ for all $s\in S$ and one can show that $T_pP= Tv_pP\oplus Th_pP$ holds for all $p\in P$. The set of smooth connections on $P$ is denoted by \gls{Con} in the following.

\subsubsection{Parallel Transports}
Let $\gamma\colon [a,b]\rightarrow M$ be a $\CC{1}$-curve in $M$ and $\w$ a connection on $P$. Then, for each $p\in F_{\gamma(a)}$ there is a unique $\CC{1}$-curve $\gamma_p^\w\colon [a,b]\rightarrow P$ with 
 $\pi\cp\gamma_p^\w=\gamma$, $\gamma_p^\w(a)=p$ as well as $\dot\gamma_p^\w(t)\in Th_{\gamma_p^\w(t)}P$, i.e., 
  $\w_{\gamma_p^\w(t)}(\dot\gamma_p^\w(t))=0$ 
for all $t\in [a,b]$. \cite{KobNomiz} 
This curve is called \emph{horizontal lift} of $\gamma$ w.r.t.\ $\omega$ in $p$ and 
\begin{align*}
	\text{\gls{PATRA}}\colon F_{\gamma(a)}&\rightarrow F_{\gamma(b)}\\
p&\mapsto \gamma_p^\w(b)
\end{align*}
is called \emph{parallel transport} along $\gamma$ w.r.t.\ $\omega$. The map $\parall{\gamma}{\omega}$ is a morphism, i.e., $\parall{\gamma}{\omega}(p\cdot s)=\parall{\gamma}{\omega}(p)\cdot s$ holds for all $p\in F_{\pi(p)}$ and all $s\in S$.
For a $\CC{1}$-path $\gamma$ one defines the parallel transport by $\parall{\gamma}{\omega}:= \parall{\gamma_0}{\omega}\cp\dots\cp\parall{\gamma_{k-1}}{\omega}$ where $\gamma_i$ are $\CC{1}$-curves with $\gamma_i=\gamma|_{[\tau_i,\tau_{i+1}]}$ for $0\leq i\leq k-1$ and $a=\tau_0<\ldots< \tau_k=b$. It is straightforward to see that this definition is independent of the explicit decomposition of $\gamma$.

\subsubsection{Automorphisms and Invariant Connections}
\label{subsub:invconn}
A diffeomorphism $\kappa\colon P\rightarrow P$ is said to be an \emph{automorphism} iff
$\kappa(p\cdot s)=\kappa(p)\cdot s$ holds for all $p\in P$ and all $s\in S$. It is straightforward to see that an $\mathfrak{s}$-valued 1-form $\w$ on $P$ is a connection iff this is true for the pullback $\kappa^*\w$. A \emph{Lie group \gls{GPHI} of automorphisms of $P$} is a Lie group $G$ together with a left action $\Phi$ of $G$ on $P$
such that the map $\Phi_g$ is an automorphism for each $g\in G$. This is equivalent to say that $\Phi(g,p\cdot s)=\Phi(g,p)\cdot s$ holds for all $p\in P$, $g\in G$ and all $s\in S$. In this situation, we will often write $gps$ instead of $(g\cdot p)\cdot s=g\cdot(p\cdot s)$. Each such left action $\Phi$ gives rise to three further important left actions:
\begingroup
\setlength{\leftmargini}{20pt}
\begin{itemize}
\item
  \itspacecc
  The action $\wm$ induced on the base manifold is defined by
  \begin{align}
    \label{eq:INDA} 
    \begin{split}
      \text{\gls{WM}}\colon \quad G\times M &\rightarrow M\\
      (g,m)&\mapsto (\pi\cp\Phi)(g, p_m)
    \end{split}
  \end{align}
  where $p_m\in F_m$ is arbitrary. Then, $\wm$ is smooth because for $s_0\colon U_0\rightarrow P$ a smooth local section with $U_0\subseteq M$ open we have $\wm|_{G\times U_0}=\Phi(\cdot,s_0)$. Then, $\Phi$ is called \emph{fibre transitive} iff $\wm$ is transitive.
\item
  \itspacec
  We equip $\text{\gls{Q}}=G\times S$ with the canonical Lie group structure and define \cite{Wang} 
  \begin{align}
    \label{eq:THETA}
    \begin{split}
      \text{\gls{THA}}\colon \qquad Q \times P&\rightarrow P\\
      ((g,s),p)&\mapsto \Phi\left(g, p\cdot s^{-1}\right).
    \end{split} 
  \end{align}
\item
  \itspacec
  The action $\cw$ induced on the set $\Con$ of smooth connections is defined by 
  \begin{align}
    \label{eq:connact}
    \begin{split}
      \text{\gls{CW}}\colon\quad G\times \Con&\rightarrow \Con\\
      (g,\w)&\mapsto \Phi_{g^{-1}}^*\w.
    \end{split} 
  \end{align}
\end{itemize}
\endgroup
\vspace{4pt}
\begin{definition}[Invariant Connection]
  \label{def:Invconn}
  A connection $\w$ is called $\Phi$-invariant iff $\Phi_g^*\w=\w$ holds for all $g\in G$. 
\end{definition}
This definition is equivalent to require that for each $p\in P$ and $g\in G$ the differential $\dd_pL_g$ induces an isomorphism between the horizontal tangent spaces $Th_pP$ and $Th_{gp}P$. In literature sometimes this condition is used to define $\Phi$-invariance of connections. 

We conclude this subsection with the following straightforward facts, see also \cite{Wang}:
\begingroup
\setlength{\leftmargini}{20pt}
\begin{itemize}
\item
  \itspacec
  Consider the representation $\text{\gls{QREP}}\colon Q \rightarrow \Aut(\mathfrak{s})$, $(g,s)\mapsto \Add{s}$. Then it is straightforward to see that each $\Phi$-invariant connection $\w$ is of type $\rho$, i.e., $\w$ is an $\mathfrak{s}$-valued 1-form on $P$ with $L_{q}^*\w=\rho(q)\cp \w$ for all $q\in Q$.
\item
  \itspace
  An $\mathfrak{s}$-valued 1-form $\w$ on $P$ with $\w(\wt{s}(p))=\vec{s}$ for all $\vec{s}\in \mathfrak{s}$ is a $\Phi$-invariant connection iff it is of type $\rho$. 
\item
  \itspace
  Let \gls{QP} denote the stabilizer of $p\in P$ w.r.t.\ $\THA$ and $G_{\pi(p)}$ the stabilizer of $\pi(p)$ w.r.t.\ $\wm$. Then
  $G_{\pi(p)}=\left\{h\in G\: | \:L_h\colon F_p\rightarrow F_p \right\}$ and we have the Lie group homomorphism 
\begin{align}
\label{eq:phip}  
  \text{\gls{FIBAP}}\colon G_{\pi(p)}\rightarrow S\quad \text{by requiring that}\quad \Phi(h,p)=p\cdot\fiba_p(h)\quad \text{for all}\quad h\in G_{\pi(p)}.
\end{align}  
 If $\mathfrak{q}_p$ and $\mathfrak{g}_{\pi(p)}$ denote the Lie algebras of $Q_p$ and $G_{\pi(p)}$, respectively, then
  \vspace{-6pt}
  \begin{align}
    \label{eq:staoQ}  
    Q_p=\{(h,\fiba_p(h))\:|\:h\in G_{\pi(p)}\}\qquad\text{and}\qquad \mathfrak{q}_p=\big\{\big(\hspace{1pt}\raisebox{-1pt}{$\vec{h}$},\dd_e\fiba_p\big(\hspace{1pt}\raisebox{-1pt}{$\vec{h}$}\hspace{1.5pt}\big)\big)\:\big|\:\raisebox{-1pt}{$\vec{h}$}\in \mathfrak{g}_{\pi(p)}\big\}.
  \end{align}
\end{itemize}
\endgroup

\subsection{Projective Structures and Radon Measures}
\label{subsec:ProjStruc}
In this subsection, we will collect the necessary facts on projective structures and Radon measures. Here, our conventions concerning Radon measures are the same as in \cite{Elstrodt}, see Definition \ref{def:Mass}. 
We start with the following non-standard definition of a projective limit:
\begin{definition}[Projective Limit]
  \label{def:ProjLim}
  Let $\{X_\alpha\}_{\alpha\in I}$ be a family of compact Hausdorff spaces where $(I,\leq)$ is a directed set. Recall that this means that $\leq$ is a reflexive and transitive relation on $I$, and that for each two $\alpha,\alpha'\in I$ we find some $\alpha''\in I$ with $\alpha,\alpha'\leq \alpha''$. A compact Hausdorff space $X$ is called projective limit of the family $\{X_\alpha\}_{\alpha\in I}$ iff
  \begingroup
  \setlength{\leftmargini}{20pt} 
  \begin{enumerate}
  \item
    \label{def:ProjLim1}
    \itspacec
    For each $\alpha\in I$ there is a continuous, surjective map $\pi_\alpha\colon X\rightarrow X_\alpha$. 
  \item
    \label{def:ProjLim2}
    \itspacec 
    For $\alpha_1,\alpha_2 \in I$ with $\alpha_1\leq \alpha_2$ there is a continuous map $\pi^{\alpha_2}_{\alpha_1}\colon X_{\alpha_2}\rightarrow X_{\alpha_1}$ for which $\pi^{\alpha_2}_{\alpha_1}\cp \pi_{\alpha_2}=\pi_{\alpha_1}$ holds.  
      
    It follows that each of these maps is surjective and that $\pi^{\alpha_2}_{\alpha_1}\cp\pi^{\alpha_3}_{\alpha_2}=\pi^{\alpha_3}_{\alpha_1}$ holds if  $\alpha_1\leq \alpha_2\leq\alpha_3$ for $\alpha_1,\alpha_2,\alpha_3\in I$.
  \item
    \label{def:ProjLim3}
    \itspacec
    If $x,y\in X$ with $x\neq y$, then there is some $\alpha\in I$ with $\pi_\alpha(x)\neq \pi_\alpha(y)$.
  \end{enumerate}
  \endgroup
\end{definition}
It is proven in Lemma \ref{lemma:equivalence} that the above definition of a projective limit is equivalent to the usual definition \cite{ProjTechAL}  as a subset
 \begin{align*}  
    \widehat{X}=\left\{\hx \in \textstyle\prod_{\alpha\in I}X_\alpha\: \:\big|\:\: \pi_{\alpha_1}^{\alpha_2}(x_{\alpha_2})=x_{\alpha_1}\:\: \forall\:\alpha_2\geq \alpha_1\right\}
  \end{align*} 
 of the Tychonoff product $\prod_{\alpha \in I}X_\alpha$. In particular, each two projective limits of a fixed family of compact Hausdorff spaces are homeomorphic if the same transition maps are used. 

Anyhow, in this thesis the main reason for writing a compact Hausdorff $X$ as a projective limit is due to Lemma \ref{lemma:normRM}. This states that we obtain a normalized Radon measures on $X$ if we define a consistent family (see next definition) of normalized Radon measures on the spaces $X_\alpha$. Using the standard Tychonoff product approach, here we always had to take care of the identification of $X$ with the space $\widehat{X}$.
For this reason, Definition \ref{def:ProjLim} is much more convenient for the purpose to construct normalized Radon measures on $X$.
\begin{Definition}[Borel, Radon Measures]
  \label{def:Mass}
  \begingroup
  \setlength{\leftmargini}{20pt}
  \begin{enumerate}
  \item
    \label{def:Mass1}
    A Borel measure $\mu$ on a Hausdorff space $Y$ is a locally finite\footnote{This means that for each $y\in Y$ we find $U\subseteq Y$ open with $y\in U$ and $\mu(U)<\infty$.} measure $\mu\colon \mathfrak{B}(Y)\rightarrow [0,\infty]$, where $\mathfrak{B}(Y)$ denotes the Borel $\sigma$-algebra of $Y$. It  is said to be normalized if $\|\mu\|:=\mu(Y)=1$.
  \item
    \label{def:Mass2}
    \itspace
    A Borel measure $\mu$ is called inner regular iff for each $A\in \mathfrak{B}(Y)$ we have 
	\begin{align*}    
    	\mu(A)=\sup\{\mu(K):K \text{ is  compact and }K\subseteq A\}.
    \end{align*} 
  \item
    \label{def:Mass3}
    \itspace
    A Radon measure $\mu$ is an inner regular Borel measure. It is called finite if     
    $\mu(Y)<\infty$ holds. Recall that each finite Radon measure is outer regular, i.e., for each $A\in \Borel(Y)$ we have
    \begin{align*} 
    	\mu(A)=\inf\{\mu(U):U \text{ is  open and }A\subseteq U\}.
    \end{align*}
  \item
    \itspace
    Assume that we are in the situation of Definition \ref{def:ProjLim} and $\{\mu_\alpha\}_{\alpha\in I}$ is a family of Radon measures $\mu_\alpha\colon \mathfrak{B}(X_\alpha)\rightarrow [0,\infty]$. Then, $\{\mu_\alpha\}_{\alpha\in I}$ is called consistent iff $\mu_{\alpha_1}$ equals the push forward measure $\pi^{\alpha_2}_{\alpha_1}(\mu_2)$ whenever $\alpha_1\leq \alpha_2$ for $\alpha_1,\alpha_2\in I$. 
  \end{enumerate}
  \endgroup
\end{Definition}

\begin{lemma}
  \label{lemma:normRM}
  \begin{enumerate}
  \item
    \label{lemma:normRM1}
    Let $\mu$ be a finite Radon measure on $X$ and $f\colon X\rightarrow Y$ a continuous map. Then, the push forward measure $f(\mu)$ is a finite Radon measure on $Y$.
  \item
    \label{lemma:normRM2}
    Let $X$ and $\{X_\alpha\}_{\alpha\in I}$ be as in Definition \ref{def:ProjLim}. Then, the normalized Radon measures on $X$ are in bijection with the consistent families of normalized Radon measures on $\{X_\alpha\}_{\alpha\in I}$.
  \end{enumerate}
\begin{proof}
  \begin{enumerate}
  \item
    Since $\mu$ is finite, $f(\mu)$ is a finite Borel measure. Then, inner regularity of $f(\mu)$ is straightforward from inner regularity of $\mu$.  
  \item
    See Lemma \ref{lemma:ConstMeas}. 
  \end{enumerate}
\end{proof} 
\end{lemma}
In Subsection \ref{sec:ConSp}, we will construct measures on Tychonoff products $X=X_1\times \dots\times X_k$ from measures $\mu_1,\dots,\mu_k$ on topological spaces $X_1,\dots,X_k$ that are not second countable. In this case, the Borel $\sigma$-algebra $\Borel(X)$ is usually larger than the product $\sigma$-algebra $\Borel(X_1)\otimes \dots\otimes \Borel(X_k)$. For this reason, we cannot use the standard product measure approach for $ \sigma$-finite measures (see, e.g., \cite[Chap.V, \S 1]{Elstrodt}). Now, if the spaces $X_i$ are locally compact Hausdorff and $\mu_i$ are finite Radon measures, then we have the notion of the Radon product measure, being defined on $\Borel(X)$. Since such Radon product measures seem to occur in standard literature usually only by means of existence statements without a concrete definition, we decided to provide such a definition at this point. This is just to clarify what exactly we mean by Radon product measures in the following. 
\begin{lemdef}[Radon Product Measure]
  \label{def:ProductMa}
  For a topological space $Z$ we denote by $C_c(Z)$ the set of continuous functions having compact support in $Z$.
  \begin{enumerate}
  \item
    \label{def:ProductMa1}
    Let $\mu$ and $\nu$ be normalized Radon measures on the locally compact Hausdorff spaces $X$ and $Y$, and denote by $\III$ and $\JJJ$ the corresponding normalized\footnote{We have $\|\III\|=\mu(X)=1$ as well as $\|\JJJ\|=\mu(Y)=1$ by 2.8 Satz in \cite[Chap.VIII, \S 2]{Elstrodt}.} positive linear functionals on $C_c(X)$ and $C_c(Y)$, respectively. 
    \begingroup
	\setlength{\leftmarginii}{15pt}
    \begin{itemize} 
    \item
    By  Theorem (13.2) in \cite{Hewitt} we obtain a well-defined positive linear functional 
    \begin{align*}
    	\III\times \JJJ\colon C_c(X\times Y)\rightarrow \mathbb{C}
	\end{align*}    
     just by  
    \begin{align}
      \label{eq:wellfunc}
      (\III\times \JJJ) (f):=\III\left[x\mapsto \JJJ(f(x,\cdot))\right]=\JJJ\left[y\mapsto \III(f(\cdot,y))\right]\qquad \forall\: f\in C_c(X\times Y).
    \end{align}
    In particular, this means that $f_y\colon x\mapsto \III(f(\cdot,y))\in C_c(X)$ and $f_x\colon y\mapsto \JJJ(f(x,\cdot))\in C_c(Y)$. Observe that  $\III\times \JJJ$ is normalized because: 
    \begingroup
	\setlength{\leftmarginiii}{13pt}
	\begin{itemize}
	\item[$\triangleright$]
	\vspace{5pt}    
    $\|\III\times \JJJ\|\leq 1$ since for each $f\in C_c(X\times Y)$ we have  
    \begin{align*}
    	|(\III\times \JJJ)(f)|\stackrel{\eqref{eq:wellfunc}}{\leq}\|\III\|\|x\mapsto \JJJ(f(x,\cdot))\|_\infty\leq \|\III\|\|\JJJ\|\sup_{x\in X}\|f(x,\cdot)\|_\infty \leq \|f\|_\infty
    \end{align*}
    \item[$\triangleright$]
    $\|\III\times \JJJ\|\geq 1-\epsilon$ for each $\epsilon\in (0,1)$ since
    we find $f_1\in C_c(X)$ and $f_2\in C_c(Y)$ with 
	\begin{align*}    
    |\III(f_1)|\geq\sqrt{1-\epsilon}\cdot \|f_1\|_\infty\qquad \text{ as well as}\qquad |\III(f_2)|\geq \sqrt{1-\epsilon}\cdot  \|f_2\|_\infty,
	\end{align*}    
    so that for $f_1\otimes f_2 \colon (x,y)\mapsto f_1(x)\cdot f_2(y)$ we have 
    \begin{align*}
    	|(\III\times \JJJ)(f_1\otimes f_2)|\stackrel{\eqref{eq:wellfunc}}{=}|\III(f_1)\cdot \JJJ(f_2)|\geq |1-\epsilon|\cdot \|f_1\|_\infty\cdot \|f_2\|_\infty=|1-\epsilon|\cdot \|f_1\otimes f_2\|_{\infty}.
    \end{align*}
    \end{itemize}
    \endgroup
	\item    
    Let $\mu\times \nu$ denote the respective normalized Radon measure on $\Borel(X\times Y)$ defined by the Riesz-Markov theorem in the form 2.5 Satz in \cite[Chap.VIII, \S 2]{Elstrodt}. 
    	 Then, by \eqref{eq:wellfunc} for all $f\in C_c(X\times Y)$ Fubini's formula holds
    \begin{align*} 
      \int_{X\times Y}f\: \dd (\mu\times \nu) = \int_X\left(\int_Yf(x,y)\:\dd\nu(y)\right)\dd\mu(x)=\int_Y\left(\int_Xf(x,y)\:\dd\mu(x)\right)\dd\nu(y).
    \end{align*}
    \end{itemize}
    \endgroup
  \item
    \label{def:ProductMa2}
    For $i=1,\dots,n$ let $\mu_i$ be a normalized Radon measure on the locally compact Hausdorff space $X_i$, and let $\III_i$ denote the corresponding normalized positive linear functional on $C_c(X_i)$. Then, it is straightforward from Part \ref{def:ProductMa1}) that the linear functional 
    \begin{align}
      \label{eq:wellfuncs}
      \III_n(f):=\III_{\sigma(1)}\left[x_{\sigma(1)}\mapsto \III_{\sigma(2)}\left[\cdots  x_{\sigma(n-1)}\mapsto \III_{\sigma(n)}(x_{\sigma(n-1)}\mapsto f(x_1\dots,x_n) )\cdots \right]\right] 
    \end{align}
    for $f\in C_c(X_1\times{\dots} \times X_n)$ is well defined, normalized, positive and independent of $\sigma \in S_n$. In particular, the corresponding Radon measure  $\mu_n$ on $\Borel(X_1\times {\dots} \times X_n)$ is normalized, and
    due to \eqref{eq:wellfuncs} a respective Fubini's formula holds.
  \item
    \label{def:ProductMa3}
    Let $\{X_\iota\}_{\iota\in I}$ be a family of compact Hausdorff spaces and $\{\mu_\iota\}_{\iota \in I}$ a family of respective  normalized Radon measures. We consider the compact Hausdorff space $X:=\prod_{\iota\in I}X_\iota$ and define a normalized Radon measure $\mu_I$ on $X$ as follows.
    
    Let $\J$ denote the set of all finite tuples $J=(\iota_1,\dots,\iota_k)$ of mutually different elements of $I$. Define $X_J:=X_{\iota_1}\times {\dots} \times X_{\iota_k}$, $\mu_J:=\mu_{\iota_1}\times {\dots} \times \mu_{\iota_k}$ and the respective projection map by
    \begin{align*} 
    	\pi_J\colon X\rightarrow X_J,\:\:
    	\prod_{\iota\in I}x_\iota \mapsto (x_ {\iota_1},\dots,x_ {\iota_k}).
	\end{align*}    	
    We write $J\leq J'$ for $J,J'\in \JJ$ with $J=(\iota_1,\dots,\iota_k)$ and $J'=(\iota'_1,\dots,\iota'_{k'})$ iff there exists an injection $\sigma\colon \{\iota_1,\dots,\iota_k\}\rightarrow \{\iota'_1,\dots,\iota'_{k'}\}$. Finally, we consider the transition maps 
     \begin{align*}   	
    \pi^{J'}_J\colon X_{J'}\rightarrow X_J,\:\:\big(x_{\iota'_1},\dots,x_{\iota'_{k'}}\big)\mapsto \big(x_{\sigma(\iota_1)},\dots,x_{\sigma(\iota_{k})}\big). 
     \end{align*}   	
    Obviously, $(\J,\leq)$ is a directed set, and $X$ a projective limit of $\{X_J\}_{J\in \JJ}$. Moreover, it is straightforward from \eqref{eq:wellfuncs} that $\{\mu_J\}_{J\in \JJ}$ is a respective consistent family of normalized Radon measures. We define $\mu_I$ to be the corresponding normalized Radon measure on $X$ provided by Lemma \ref{lemma:normRM}.
    \hspace*{\fill} $\lozenge$
  \end{enumerate}
\end{lemdef}
\begin{remark}[Fubini]
\label{rem:unednlichfubini}
	In the situation of Lemma and Definition \ref{def:ProductMa}.\ref{def:ProductMa3} assume that $I=I_1\sqcup I_2$ for $I_1,I_2\neq \emptyset$. Moreover, let $\mu_1$ and $\mu_2$ denote the corresponding Radon product measures on $X_1=\prod_{\iota\in I_1} X_\iota$ and $X_2=\prod_{\iota\in I_2}X_\iota$, respectively. Then, it is easy to see that $\mu=\mu_1\times \mu_2$ holds. 
	
	In fact, by continuity of the corresponding linear functionals $\III,\III_1,\III_2$ one only has to check that $\III=\III_1\times \III_2$ holds on the dense $^*$-subalgebra $\mathrm{Cyl}(X)\subseteq C(X)$ consisting of such continuous function which can be written in the form $f\cp \pi_J$ with $f\in C(X_J)$ for some $J\in \J$, see also proof of Lemma \ref{lemma:normRM}.\ref{lemma:normRM2}. This, however, is straightforward from the definitions. 
\end{remark}
\section{Special Mathematical Background}
\label{sec:specmathback}
This section essentially collects some background material we could not find in the standard literature in this form. 

In the first part, we will fix some conventions concerning certain symmetric situations that appear in loop quantum gravity (\gls{LQG}) and will serve as prime examples during this work. In particular, we will specify sets of invariant connections that belong to these situations. Since the corresponding calculations are rather technical than illustrative, we have decided to shift them to Section \ref{CHarinvconn} and the Appendix, and
only to provide the relevant information at this point. 
The necessary techniques  
will be developed in Section \ref{CHarinvconn}, where we attack the characterization problem of invariant connections in full generality. 
In the second part, we will prove some straightforward facts on maps on spectra and then apply them to the case where the $\Cstar$-algebra consists of certain bounded functions on a set. This is important for the investigations in Section \ref{sec:SPecExtGr} where we consider  $\Cstar$-subalgebras of bounded functions on sets of smooth connections.
In the last part, we will give a brief introduction into the Bohr compactification of a locally compact abelian group. This will be used 
to characterize the abelian continuous group structures on spectra of $\Cstar$-subalgebras of bounded functions by families of quasi-characters, see Definition \ref{def:quasichar}. 
\subsection{Loop Quantum Gravity Case}
Usually, in loop quantum gravity (LQG) the structure group is $\SU$, and the principal fibre bundle is of the form $P=\Sigma \times \SU$ for a 3-dimensional manifold (Cauchy surface) $\Sigma$. Although the results of this dissertation apply to a much larger class of principal bundles and structure groups, in the LQG relevant examples of this work the bundle will be just of the form $\RR^3\times \SU$. We now collect some facts, notations and conventions concerning the Lie group $\SU$, the elements of $\RR^n$ and sets of curves in $\RR^3$. 
\begin{convention}
  \label{conv:sutwo1}
  \begingroup
  \setlength{\leftmargini}{20pt}
  \begin{enumerate}
  \item
  \label{conv:sutwo111}
    By $\text{\gls{VR}}\colon \SU \rightarrow \SOD$ we denote the universal covering map.  
    Then, for $\ssigma\in \SU$ we have
    \begin{align}
      \label{eq:covm}
      \ssigma(x):=\varrho(\ssigma)(x)=\murs^{-1}(\Add{\ssigma}(\murs(x)))
    \end{align}
    for the linear isomorphism $\text{\gls{MURS}}\colon \RR^3 \rightarrow \mathfrak{su}(2)$, $\sum_{i=1}^3 x^i \vec{e}_i\mapsto \sum_{i=1}^3 x^i \tau_i$ with
    \begin{align*}
      \tau_1:=\begin{pmatrix} 0 & -\I  \\ -\I & 0  \end{pmatrix}\qquad\qquad \tau_2:=\begin{pmatrix} 0 & -1  \\ 1 & 0  \end{pmatrix}\qquad\qquad \tau_3:=\begin{pmatrix} -\I & 0  \\ 0 & \I  \end{pmatrix}.
    \end{align*}
    Although this identification of $\RR^3$ with $\su$ is standard, we decided to introduce the map $\murs$ at this point. The main reason is that in our applications we will need this identification permanently, as it simplifies the formulas and calculations drastically, see, e.g., \eqref{eq:rotinvconn}, \eqref{eq:homis} and the Appendices \ref{subsec:InvIsoHomWang} and \ref{subsec:IsotrConn}.
    The alternative would be to calculate in coordinates which we want to avoid whenever it is possible.
    
    Recall that\footnote{Observe that $\exp$ is surjective because $\SU$ is compact and connected.} each $\ssigma\in \SU$ can be written in the form 
    \begin{align}
      \label{eq:expSU2}
      \ssigma= \cos(\alpha/2)\cdot \me + \sin(\alpha/2)\cdot \murs(\vec{n})=\exp\big(\alpha/2\cdot\murs(\vec{n})\big)
    \end{align} 
    for some $\vec{n}\in \RR^3$ with $\|\vec{n}\|=1$ and $\alpha\in [0,2\pi)$. In this case $\uberl(\ssigma)$ rotates a point $x$ by the angle $\alpha$ w.r.t.\ the axis $\vec{n}$.
    \item
    	In the sequel, $\RR^n$ will occur as vector space (tangent space), as base manifold of a principal fibre bundle, and as a symmetry group. In the first case we write its elements with arrows such as $\vv$, and the same will be done if an element of $\RR^n$ occurs as a normal vector, rotation axis or as the traversing direction of a linear curve, see next point. In the second case, i.e., if $\RR^n$ is considered as a base manifold, we usually write $x$, and in the last case we will write $v$. Nevertheless, in all three situations we will make free use of the vector space operations in $\RR^n$. For instance, if $G=\RR^n$ acts via addition on $M=\RR^n$, we will write $v+x$. If $G=\RR_{\neq 0}$ acts via multiplication on $M=\RR^n$, we will write $\lambda \cdot x$. In the same way, if $n=3$, we also apply the above map $\murs$ to all these elements, i.e., $\murs(\vv)$, $\murs(x)$ and $\murs(v)$.
\item 
 \label{conv:sutwo2}
    Whenever $M=\Sigma=\RR^3$, the sets $\Pall,\Pal,\Paln,\Pacirc$ are defined as follows:
    \begingroup
    \setlength{\leftmargini}{20pt}
    \begin{enumerate}
    \item
      \label{conv:lincirccurves1} 
      According to the notations introduced in the previous point,  
      by $\Pall$ we will denote the set of curves of the form 
      $x+\gamma_{\vec{v},l}$ for $x, \vec{v}\in \mathbb{R}^3$ with $\|\vec{v}\|=1$ and 
      $\gamma_{\vec{v},l}\colon [0,l]\rightarrow \mathbb{R}$, $t\mapsto t \vec{v}$. 
      
      By $\Pal$ we will denote the set of all linear curves, i.e., the set of all embedded analytic curves with $\im[\gamma]$ contained in a line through some fixed point $x\in \RR^3$. 
      Then, $\Pal$ consists of all embedded analytic curves $\gamma$ equivalent to an element of $\Pall$, i.e.,
	\begin{align*}      
      	\gamma = \delta\cp \adif\quad\text{for some}\quad \delta\in \Pall\quad\text{and}\quad \adif \colon I\rightarrow \RR\quad\text{an analytic diffeomorphism}.
     \end{align*}
     Here, $I$ denotes some open interval which contains $\dom[\gamma]$.
     
     Finally, by $\Paln\subseteq \Pal$ we will denote the subset of all linear curves traversing through the origin.
    \item
      \label{conv:lincirccurves2}
      Let $\Pacirc$ consist of all circular curves, i.e., all curves of the form 
      \begin{align*}
        \gc{n}{r}{x}{\tau} \colon [0,\tau]&\rightarrow \RR^3\\
        t&\mapsto x + \cos(t)\:\vec{r} + \sin(t) \:\vec{n}\times \vec{r}  
      \end{align*}
      for $\vec{n},\vec{r},x\in \RR^3$ with $\|\vec{n}\|=1$ as well as $0< \tau< 2\pi$.
    \end{enumerate}
    \endgroup  	
    	
    \item
    For $\s\in \su$ and $\vv,x,v \in \RR^3$ we define the corresponding maximal tori in $\SU$ by 
	\begin{align}  
	\label{eq:tori}  
	\text{\gls{Tor1}}:=\{\exp(t\s)\:|\: t\in \RR\}\qquad \text{\gls{Tor2}}:=H_{\murs(\vv)}\qquad \text{\gls{Tor3}}:=H_{\murs(x)}\qquad \text{\gls{Tor4}}:=H_{\murs(v)}.
    \end{align}
   Hence, we suppress the map $\murs$ in the last three cases.  
\hspace*{\fill}$\lozenge$
  \end{enumerate}
  \endgroup
\end{convention}
In Section \ref{sec:MOQRCS} we will be concerned with certain equivariant maps to $\SU$.  
Inter alia, the next lemma then will be essential for calculating spaces of such maps.
\begin{lemma}
  \label{lemma:torus}
  \begin{enumerate}
  \item
    \label{lemma:torus1}
    If $s s'=s's$ for $s,s'\in \SU$, then $s,s'\in H_{\vec{n}}$ for some $\vec{n}\in\RR^3\backslash \{0\}$. This torus is uniquely determined if $s\neq \pm \me$ or $s'\neq \pm \me$. 
  \item
    \label{lemma:torus2}
    Let $s,s'\in H_{\vec{n}}$ be different from $\pm\me$. If $s'=\alpha_h(s)$ holds for some $h\in \SU$, then we have the following two possibilities:
     \begingroup
    \setlength{\leftmarginii}{17pt}
    \begin{enumerate}
    \item[a)]
      \vspace{-7pt}
      $s=s'$ and $h\in H_{\vec{n}}$ 
    \item[b)]
    	$s'=s^{-1}$ and $h=\exp\left(\textstyle\frac{\pi}{2}\cdot\murs(\vec{m})\right)$ for some $\mm\in \RR^3$ orthogonal to $\vec{n}$ with $\|\mm\|=1$ and uniquely determined up to a sign. 	
    	
    	In particular, $\alpha_h(s_0)=s_0^{-1}$ holds for $\pm \me\neq s_0\in H_{\vec{n}'}$ iff $\langle\vec{m},\vec{n}'\rangle=0$.
    \end{enumerate}
    \endgroup
  \end{enumerate} 
  \begin{proof}
    \begin{enumerate}
    \item
      This follows by straightforward calculation involving \eqref{eq:expSU2}, or from the general theory of compact and connected Lie groups as $\{-\me,\me\}$ are the irregular elements of $\SU$. 
    \item
      We write $s=\exp(t\cdot\murs(\vec{n}))$ and $s'=\exp(t'\murs(\vec{n}))$ for unique reals $t,t'\in [0,2\pi)$, cf.\ \eqref{eq:expSU2}. 
      Then
      \begin{align}
        \label{eq:bla}
        \exp\big(t\cdot \murs(\varrho(h)(\vec{n}))\big)=\alpha_h(s)=s'=\exp(t'\murs(\vec{n})). 
      \end{align}
      If $\varrho(h)(\vec{n})=\vec{n}$, then $t=t'$ and $s=s'$, hence, $h\in H_{\vec{n}}$ by Part \ref{lemma:torus1}), so that case \textit{a)} holds.  

      If $\varrho(h)(\vec{n})\neq\vec{n}$ for $\vec{v}:=\varrho(h)(\vec{n})$, then from \eqref{eq:bla}, \eqref{eq:expSU2} and $\|\vec{v}\|=1$ we obtain
      \begin{align*}
        \cos(t)\cdot \me + \sin(t)\cdot \murs(\vec{v})=\cos(t')\cdot \me + \sin(t')\cdot \murs(\vec{n}).
      \end{align*}
      Then $\cos(t)=\cos(t')$, hence $\sin(t)=\pm \sin(t')$, i.e., $\vec{n}=-\vec{v}$ because $\vv\neq \vec{n}$ and since $\sin(t),\sin(t')\neq 0$ as $s,s'\neq \pm \me$. 
      Consequently, $\varrho(h)$ is a rotation by $\pi$ around an axis $\vec{m}$ orthogonal to $\vec{n}$. So, if $s_0\in H_{\vec{n}'}$ with $\langle\vec{m},\vec{n}'\rangle=0$, then 
	\begin{align*}
		\alpha_h(s_0)=\exp\!\big(t_0\cdot \murs(\varrho(h)(\vec{n}' ))\big)=\exp(-t_0\cdot \murs(\vec{n}'))=s_0^{-1}.
	\end{align*}	      
	Finally, that this equality only holds for such $s_0\neq \pm \me$ with $s_0\in H_{\vec{n}'}$ for $\langle\vec{m},\vec{n}'\rangle=0$, is clear from the fact that $\mm$  is unique up to a sign.
    \end{enumerate}
  \end{proof}
\end{lemma}
The next example collects the most relevant loop quantum cosmology (\gls{LQC}) cases to be discussed in this work. The proofs concerning invariant connections can be found in the Appendix \ref{subsec:InvIsoHomWang} and Section \ref{CHarinvconn}. The reason for providing these results already at this point is that we will need them during the following sections. 
\begin{example}[Loop Quantum Cosmology]
  \label{ex:LQC}
  Let $P=\RR^3\times \SU$. We consider the following symmetry groups and actions:\footnote{Observe that $\Pe(g,p)=g\cdot_\varrho p$ for $\cdot_\varrho$ the group multiplication in $\Gee$.}
  $$
  \begin{tabular}{rlrll}
    \gls{GE}\!\!\!\!\!&$:=\Gee$  \qquad\qquad &  \quad\qquad $\Pe((v,\sigma),(x,s))$\!\!\!\!\!&$:=(v+\text{\gls{VR}}(\sigma)(x),\sigma s)$\\
    \gls{GI}\!\!\!\!\!&$:=\SU$   \qquad\qquad &  \quad\qquad $\Pii(\sigma,(x,s))$\!\!\!\!\!&$:=(\varrho(\sigma)(x),\sigma s)$\\
    \gls{GH}\!\!\!\!\!&$:=\RR^3$ \qquad\qquad &  \quad\qquad $\Ph(v,(x,s))$\!\!\!\!\!&$:=(v+x,s)$
  \end{tabular}$$
  For each of these actions the corresponding induced action is pointwise proper. In the second case, this is clear from compactness of $\SU$, and in the last case this holds  because $\wm$ is just the addition in $\RR^3$. In the first case ,we have $\Ge=P$ so that even $\Pe$ is (pointwise) proper as topological groups always act properly on themselves. 
  \par
  \begingroup
  \leftskip=8pt
  \vspace{8pt}
  \noindent
  {\bf\textit{Homogeneous Isotropic LQC:}}
  \vspace{2pt}
  
  \noindent
  The Lie group $\Ge$ resembles the Euclidean group $\RR^3 \rtimes \SOD$ in the sense that 
  the action $\wm$ induced by $\Pe$ on $\RR^3$ gives rise to the same orbits as the canonical action of the euclidean group does. 
  The $\Pe$-invariant connections are of the form
  \vspace{-3pt}
  \begin{align}
    \label{eq:homis}
    \w^c(\vec{v}_x,\vec{\sigma}_s)= c \Add{s^{-1}}[\murs(\vec{v}_x)]+s^{-1}\vec{\sigma}_s \qquad \forall\:(\vec{v}_x,\vec{\sigma}_s)\in T_{(x,s)}P,
  \end{align}  
  where $c$ runs over $\mathbb{R}$. This follows from Wang's original theorem \cite{Wang} and the explicit calculations can be found in Appendix \ref{subsec:InvIsoHomWang}.
  
  \vspace{8pt}
  \noindent
  {\bf\textit{Spherically Symmetric LQC:}}
  \vspace{2pt}
  
  \noindent
  It is proven in Example \ref{bsp:Rotats} that the spherically symmetric connections, i.e., the $\Pii$-invariant ones are of the form
  \begin{align}
    \label{eq:rotinvconnn}
    \begin{split}
      \w^{abc}(\vec{v}_x,\vec{\sigma}_s):= \Add{s^{-1}}\!\big[&a(x)\hspace{1pt}\murs(\vec{v}_x)+ b(x)[\hspace{1pt}\murs(x),\murs(\vec{v}_x)]
      +c(x)[\:\murs(x),[\hspace{1pt}\murs(x),\murs(\vec{v}_x)]]\big]+ s^{-1}\vec{\sigma}_s
    \end{split}
  \end{align}
  for $(\vec{v}_x,\vec{\sigma}_s)\in T_{(x,s)}P$. Here, $a,b,c\colon \mathbb{R}^3\rightarrow \mathbb{R}$ are rotation invariant maps that can be written in the form
  $a(x)=f\big(\|x\|^2\big)$, $b(x)=g\big(\|x\|^2\big)$, $c(x)=h\big(\|x\|^2\big)$
  for smooth functions $f,g,h\colon (-\epsilon,\infty)\rightarrow \RR$ and some $\epsilon>0$. 

  \vspace{8pt}
  \noindent
  {\bf\textit{(Semi-)Homogeneous LQC:}}
  \vspace{2pt}
  
  \noindent
  This is a special case of Example \ref{ex:transinv}, where by semi-homogeneity we mean that $G=G_{SH}$ is a two-dimensional linear subspace of $\RR^3$ and the corresponding action is $\Phi_{SH}(v,(x,s)):=(v+x,s)$.   
  Here, for $W$ an algebraic complement of $G$ in $\RR^3$, the $\Phi_{SH}$-invariant connections are parametrized by the smooth maps $\psi \colon \mg\times TW\rightarrow \su$ whose restrictions $\psi|_{\mg\times T_xW}$ are linear for all $x\in W$, hence by the smooth maps $\psi \colon \RR^2\times T\RR\rightarrow \su$ whose restrictions $\psi|_{\RR^2\times T_t\RR}$ are linear for all $t\in \RR$.  
  
  In the homogeneous case, i.e., if $\Gh=\RR^3$, the $\Ph$-invariant connections are in bijection with the linear maps $L\colon \RR^3\rightarrow \su$, see also Example \ref{ex:eukl}. \hspace*{\fill}$\lozenge$
  
  \endgroup
  \par
\end{example}

\subsection{Bounded Functions and Maps on Spectra}
\label{subsec:GrAcOnSp}
In the first part, we relate automorphisms of abelian $\Cstar$-algebras with homeomorphisms of their spectra. In the second part, we will apply this to $\Cstar$-algebras of bounded functions, being the relevant quantities in the framework of loop quantum gravity. We collect some basic facts on denseness \cite{Rendall,ChrisSymmLQG} of a set $X$ in the spectrum of a $\Cstar$-subalgebra of bounded functions thereon as well as on embeddability of the spectrum of a restriction $\Cstar$-algebra into the spectrum of the full one. In particular, this will provide us with the mathematical backbone of traditional reduction approach in the framework of Loop Quantum Gravity. \cite{ChrisSymmLQG} 

\subsubsection{Maps on Spectra}
\begin{definition}
	For a Banach algebra $\aA$, let $\|\cdot\|_\aA$ denote the corresponding norm. For $a\in \aA$ denote by $\|\cdot\|_a\colon \aA'\rightarrow \RR_{\geq 0}$ the seminorm $\|\chi\|_a:=|\chi(a)|$ for $\aA'$ the topological dual of $\aA$. 
\end{definition}
\begin{Lemma} 
  \label{lemma:homzuspec}
  \begin{enumerate}
  \item
    \label{lemma:homzuspec1}
    If $\lambda\colon \aA\rightarrow \bB$ is a homomorphism of abelian Banach algebras $\aA$ and $\mathfrak{B}$, then the map
    \begin{align*}
    \ovl{\lambda}\colon \Spec(\bB)\rightarrow \Spec(\aA),\quad
    \chi \mapsto \chi \cp \lambda
    \end{align*}
    is continuous if it is well defined. In particular, this is the case if $\lambda$ is surjective or unital.\footnote{More precisely, if $\aA$ and $\bB$ are unital and $\lambda(1_\aA)=1_\bB$.}
  \item
    \label{lemma:homzuspec2}
    If $\aA$ is an abelian $\Cstar$-algebra, then $\eta\colon \Aut(\aA)\rightarrow \Homeo(\Spec(\aA))$, $\lambda\mapsto\ovl{\lambda}$
    is a group antiisomorphism.
  \end{enumerate}
  \begin{proof}
    \begin{enumerate}
    \item
      The image of $\ovl{\lambda}$ consists of homomorphisms. So, the only case in which
      well-definedness fails is when $\ovl{\lambda}(\chi)=0$ for some $\chi\in \Spec(\bB)$. 
      If $\lambda$ is unital, then $\ovl{\lambda}(\chi)(1_{\aA})=(\chi\cp\lambda)(1_{\aA})=\chi(1_\bB)=1\neq 0$ for all $\chi\in \Spec(\bB)$, so that $\ovl{\lambda}$ is well defined in this case. If $\lambda$ is surjective and $\chi\in \Spec(\bB)$, then $\chi(b)\neq 0$ for some $b\in \bB$ and we find $a\in\aA$ with $\lambda(a)=b$. Hence, $\ovl{\lambda}(\chi)(a)=\chi(b)\neq 0$ for all $\chi \in \Spec(\bB)$ also in this case. For continuity let $\Spec(\bB)\supseteq\{\chi_\alpha\}_{\alpha\in I}\rightarrow \chi$ be a converging net. Then for $a\in \aA$ and $\epsilon> 0$ we find $\alpha_\epsilon\in I$ such that $\|\chi_\alpha-\chi\|_{\lambda(a)}\leq \epsilon$ for all $\alpha\geq \alpha_\epsilon$. But for $\alpha\geq \alpha_\epsilon$ we have
      \begin{align*} 
      	\big\|\ovl{\lambda}(\chi_\alpha)-\ovl{\lambda}(\chi)\big\|_a=|\chi_\alpha(\lambda(a))-\chi(\lambda(a))|=\|\chi_\alpha-\chi\|_{\lambda(a)}\leq \epsilon
	  \end{align*}      	
      	 so that
      $\Spec(\aA)\supseteq\{\ovl{\lambda}(\chi_\alpha)\}_{\alpha\in I}\rightarrow \ovl{\lambda}(\chi)$ shows continuity of $\ovl{\lambda}$.
    \item
      Each $\lambda\in \Aut(\aA)$ is a surjective homomorphism so that the image of $\eta$ consists of well defined and continuous maps. Then $\eta$ is an antihomomorphism as 
      \begin{align*}
        \eta(\lambda\cp \lambda')(\chi)=(\chi \cp \lambda)\cp \lambda'=\eta(\lambda')(\chi\cp \lambda)=\eta(\lambda')(\eta(\lambda)(\chi)).
      \end{align*}
      Now, $\lambda^{-1}\in \Aut(\aA)$ exists so that $\eta(\lambda)^{-1}=\eta(\lambda^{-1})$ is continuous as well. This means that the image of $\eta$ consists of homeomorphisms and shows well-definedness of this map. To verify injectivity assume that $\eta(\lambda)=\eta(\lambda')$ for $\lambda,\lambda'\in \Aut(\aA)$. Then for 
      $a\in\aA$ and $\chi\in \Spec(\aA)$ we have
      \begin{align*}
        \GT(\lambda(a))(\chi)&=(\chi\cp\lambda)(a)=\eta(\lambda)(\chi)(a)\\
        &=\eta(\lambda')(\chi)(a)=(\chi\cp\lambda')(a)
        =\GT(\lambda'(a))(\chi)
      \end{align*}
      so that $\GT(\lambda(a))=\GT(\lambda'(a))$. Then injectivity of $\GT$ implies $\lambda(a)=\lambda'(a)$ for all $a\in \aA$, hence $\lambda=\lambda'$. For surjectivity of $\eta$ define 
      \begin{align}
        \label{eq:tautau}  
        \begin{split} 
          \tau\colon\Homeo(\Spec(\aA))&\rightarrow\Aut(\aA)\\
          h&\mapsto \left[a\mapsto \GT^{-1}\left[\GT(a)\cp h\right]\right].
        \end{split}
      \end{align}
      This is well defined because $\GT(a)\cp h\in \mathrm{C}_0(\Spec(\aA))$ for each $a\in \aA$ so that $\tau(h)\colon \aA\rightarrow \aA$ is well defined for all $h\in \Homeo(\Spec(\aA))$. Moreover, $\tau(h)$ is a homomorphism since $\GT$ and $\GT^{-1}$ are. 
      Finally, 
      \begin{align*}
        \left[\tau(h^{-1})\cp \tau(h)\right](a)&=\GT^{-1}\left[\GT(\tau(h)(a))\cp h^{-1}\right]
        =\GT^{-1}\left[\GT(a)\cp h \cp h^{-1}\right]=a
      \end{align*}
      so that $\tau(h)\in \Aut(\aA)$.
      Then for $\chi\in \Spec(\aA)$ and all $a\in \aA$ we have
      \begin{align*}
        \eta(\tau(h))(\chi)(a)&=\ovl{\tau(h)}(\chi)(a)=\chi(\tau(h)(a))
        =\chi\big(\GT^{-1}[\GT(a)\cp h]\big)\\
        &=[\GT(a)\cp h](\chi)=\GT(a)(h(\chi))=h(\chi)(a),
      \end{align*}
      hence $\eta\cp \tau = \id_{\Homeo(\Spec(\aA))}$, which shows surjectivity of $\eta$.
    \end{enumerate}
  \end{proof}
\end{Lemma}
\subsubsection{Bounded Functions}
\label{subsec:boundedfun}
For a set $\ZX$ let $\Cb(\ZX):=\{f\colon \ZX\rightarrow \mathbb{C}\:|\:\|f\|_{\infty}< \infty\}$ denote the set of bounded, complex-valued functions on $\ZX$. Then, $\Cb(\ZX)$ is an abelian $\Cstar$-algebra w.r.t.\ the supremum norm $\|\cdot\|_\infty$.
\begin{convention}
  \label{conv:Boundedfunc}
  Let $\aA\subseteq \Cb(\ZX)$ be some fixed $\Cstar$-subalgebra and $\upsilon\colon \SX\rightarrow \ZX$ a map with $\SX$ some further set.
  \begingroup
  \setlength{\leftmargini}{20pt}
  \begin{itemize}
  \item
    \label{conv:Boundedfunc1}
    \itspacec
    Let $Z$ be a set, $\bB\subseteq B(Z)$ a subset and $\Gen$ the $^*$-algebra generated by $\bB$. Then, by $\ovl{\bB}$ we will denote the closure of $\mathfrak{G}$ in $B(Z)$.
  \item
    \label{conv:Boundedfunc2}
    \itspace 
    The spectrum of $\aA\subseteq B(\ZX)$ is denoted by $\QZX$ in the following. This is motivated by the first part of the next lemma. Since $\ZX$ is not assumed to carry any topology, this will not be in  conflict with the notation concerning closures of subsets of topological spaces introduced in  Subsection \ref{sec:notations}.
  \item
    \label{conv:Boundedfunc3}
    \itspacec
    The pullback of $\aA$ by $\upsilon$ is the $^*$-algebra $\upsilon^*(\aA):=\{f\cp \upsilon\:|\: f\in \aA\}\subseteq B(\SX)$. The closure $\rR_\upsilon:=\ovl{\upsilon^*(\aA)}\subseteq B(\SX)$ is called restriction of $\aA$ w.r.t.\ $\upsilon$. Its spectrum is denoted by $\XNR$ in the following.\footnote{In view of the second point, it would be logical to denote $\Spec(\rR_\upsilon)$ by $\ovl{Y^\upsilon}$. However, we will write $\XNR$ because the set $Y$ is already encoded in the map $\upsilon$. Moreover, our notation suggests that we are dealing with the spectrum of a \emph{restriction} $\Cstar$-algebra. Then, if $\aA$ is unital, we have $\XNR\cong \YNR$ by Lemma \ref{lemma:dicht}.\ref{lemma:dicht3}, where the latter space (defined in the last point of this Convention \ref{conv:Boundedfunc}.) is a closed subset of $\X$. For this reason, it makes sense to use the letter $X$ for both spaces.}
  \item  
    \label{conv:Boundedfunc4}
    \itspace
    Let $\ZX_\aA$ denote the set of all $x\in \ZX$ for which the map
    \begin{align*} 
      \text{\gls{IOTAY}}\colon \ZX &\rightarrow \mathrm{Hom}(\aA,\mathbb{C})\\
      x&\mapsto [f\mapsto f(x)]
    \end{align*} 
    is non-zero, i.e., $\ZX_\aA=\{x\in \ZX\: |\:\exists\: f\in \aA : f(x)\neq 0\}$. Hence, $x\in \ZX_\aA$ iff $\iota_\ZX(x)\in \Spec(\aA)$. 
  \item
    \label{conv:Boundedfunc5}
    \itspacec
    The set $\YNR\subseteq \QZX$ is defined to be the closure 
	\begin{align*} 
	\YNR:=\ovl{\iota_\ZX\big(\ZX_\aA\cap\upsilon(\SX)\big)}\subseteq \Spec(\aA)
	\end{align*}	
	 of $\iota_\ZX\big(\ZX_\aA\cap\upsilon(\SX)\big)$ in $\Spec(\aA)$.  
    The third part of the next lemma then shows that 
	\begin{align*}    
    	\YNR\cong \XNR=\Spec(\rR_\upsilon)
	\end{align*}    	
    	 holds if $\aA$ is unital.\hspace*{\fill}$\lozenge$
  \end{itemize}
  \endgroup
\end{convention}
The first part of the following lemma is a slight variation of Proposition 2.1 in \cite{ChrisSymmLQG} which, in turn, originates from \cite{Rendall}. The second part can also be derived from Corollary 2.19 in \cite{ChrisSymmLQG}.
\begin{lemma} 
  \label{lemma:dicht}
  \begin{enumerate}
  \item
    \label{lemma:dicht1}
    If $\ZX$ is a set and $\aA\subseteq \Cb(\ZX)$ a $\Cstar$-subalgebra, then $\iota_X(\ZX_\aA)\subseteq \Spec(\aA)$ is dense, i.e., $\QZX=\ovl{\iota_\ZX(\ZX_\aA)}$. The map $\iota_\ZX$ is injective iff $\aA$ separates the points in $\ZX_\aA$.
  \item
    \label{lemma:dicht2}
    Let $\aA$ be unital, $\SX$ a set and $\upsilon\colon \SX \rightarrow \ZX$ a map.  
    Then, $\ovl{\upsilon^*}\colon \XNR\rightarrow \QZX$ is
    the unique continuous map which extends $\upsilon$ in the sense that the following diagram is commutative:
    \begin{center}
      \makebox[0pt]{
        \begin{xy}
          \xymatrix{
            \XNR \ar@{->}[r]^-{\ovl{\upsilon^*}}   &  \QZX  \\
            \SX\ar@{.>}[u]^{\iota_{\SX}}\ar@{->}[r]^-{\upsilon} & \ZX  \ar@{.>}[u]^{\iota_\ZX}
          }
        \end{xy}
      }
    \end{center}    
    The map $\ovl{\upsilon^*}$ is an embedding.\footnote{This means that $\ovl{\upsilon^*}$ is a homeomorphism to its image equipped with the relative topology.}
  \item
    \label{lemma:dicht3}
    In the situation of Part \ref{lemma:dicht2}) we have 
	\begin{align*}
		\im\!\left[\ovl{\upsilon^*}\right]=\ovl{\upsilon^*}(\XNR)\stackrel{!}{=}\ovl{\iota_X(\upsilon(Y))}=\YNR,
	\end{align*}     
    i.e., $\XNR\cong \YNR$ by the embedding property of $\ovl{\upsilon^*}$.
  \item
    \label{lemma:dicht4}
    If $\rho \colon \ZX\rightarrow \ZX$ is a map, then
    $\aA\subseteq \rho^*(\aA)$ implies $\rho(\ZX_\aA)\subseteq \ZX_\aA$. 
  \end{enumerate}
  \begin{proof}
    \begin{enumerate}
    \item
      Assume there is $\chi\in U:=\Spec(\aA)\backslash \ovl{\iota_\ZX(\ZX_\aA)}$. Then $U$ is an open neighbourhood of $\chi$. Since the space $\Spec(\aA)$ is locally compact Hausdorff, by Urysohn's lemma we find a continuous function $\hat{f}\colon \Spec(\aA)\rightarrow [0,1]$ such that $\hat{f}(\chi)=1$ and $\hat{f}$ has compact support contained in $U$. 
      Then $\hat{f}\in \Cinf(\Spec(\aA))$ so that $\hat{f}=\GT(f)$ for some $f\in \aA$. Consequently, $f(x)=\iota_\ZX(x)(f)=\hat{f}(\iota_\ZX(x))=0$ for all $x\in \ZX_\aA$ by construction, and $f(x)=0$ for all $x\in \ZX\backslash \ZX_\aA$ by definition of $\ZX_\aA$. Consequently, $f=0$ and $\hat{f}=\GT(f)=0$, which contradicts $\hat{f}(\chi)=1$. 
      The injectivity statement is immediate from the definitions.
    \item
      Since $\aA$ and $\rR_\upsilon$ are unital, we have $\ZX_\aA=\ZX$ and $\SX_{\rR_\upsilon}=\SX$.  
      Then $\upsilon^*\colon \aA \rightarrow \rR_\upsilon$ is a unital algebra homomorphism, so that by Lemma \ref{lemma:homzuspec}.\ref{lemma:homzuspec1} the map $\ovl{\upsilon^*}\colon \Spec(\rR_\upsilon)\rightarrow \Spec(\aA)$ is continuous and well defined.\footnote{Observe that there is no ad hoc reason for $\upsilon^*(\aA)$ to be closed, i.e., $\upsilon^*$ is not necessarily surjective. For this reason we are forced to assume unitality in order to guarantee well-definedness of $\ovl{\upsilon^*}$.} Obviously, $\ovl{\upsilon^*}\cp \iota_\SX = \iota_\ZX\cp \upsilon$ so that uniqueness follows from denseness of $\im[\iota_\SX]$ in $\XNR=\Spec(\rR_\upsilon)$ and continuity of $\ovl{\upsilon^*}$. 
      
      Now, assume that $\ovl{\upsilon^*}(\chi)=\ovl{\upsilon^*}(\chi')$ for $\chi,\chi'\in \Spec(\rR_\upsilon)$. Then
      \begin{align*}
        \chi(\upsilon^*(f))=\ovl{\upsilon^*}(\chi)(f)=\ovl{\upsilon^*}(\chi')(f)=\chi'(\upsilon^*(f))\qquad\forall\:f\in \aA,
      \end{align*}
      hence
      $\chi|_{\upsilon^*(\aA)}=\chi'|_{\upsilon^*(\aA)}$. Since $\upsilon^*(\aA)\subseteq\rR_\upsilon$ is dense and $\chi,\chi'$ are continuous, $\chi=\chi'$ follows.
      Now, since $\XNR$ is compact and $\im\!\left[\ovl{\upsilon^*}\right]$ is a Hausdorff space,  the bijective continuous map $\ovl{\upsilon^*}\colon \XNR\rightarrow \im\!\left[\ovl{\upsilon^*}\right]$ is a homeomorphism. 
    \item
      By \ref{lemma:dicht2}) we have $\ovl{\upsilon^*}(\iota_\SX(\SX))=\iota_\ZX(\upsilon(\SX)) $, hence
      \begin{align*}
        \ovl{\upsilon^*}\!\left(\QSX\right)\stackrel{\ref{lemma:dicht1})}{=}\ovl{\upsilon^*}\Big(\ovl{\iota_\SX(\SX)}\Big)\subseteq \ovl{\ovl{\upsilon^*}(\iota_\SX(\SX))}=\ovl{\iota_\ZX(\upsilon(\SX))}.
      \end{align*}
      Here, for the inclusion in the third step we have used continuity of $\ovl{\upsilon^*}$. This shows 
      $ \ovl{\upsilon^*}\!\left(\QSX\right)\subseteq \ovl{\iota_\ZX(\upsilon(\SX))}$. For the converse inclusion we calculate 
      \begin{align*}
        \iota_\ZX(\upsilon(\SX))\subseteq\ovl{\upsilon^*}\left(\ovl{\iota_\SX(\SX)}\right)\stackrel{\ref{lemma:dicht1})}{=}\ovl{\upsilon^*}\left(\QSX\right),
      \end{align*}
      hence $\ovl{\iota_\ZX(\upsilon(\SX))}\subseteq  \ovl{\upsilon^*}\!\left(\QSX\right)$ 
      since $ \ovl{\upsilon^*}\!\left(\QSX\right)$ 
      is compact.
    \item
      If $x\in \ZX_\aA$, then $f(x)\neq 0$ for some $f\in \aA$. Since by assumption we have $f=g\cp \rho$ for some $g\in \aA$, we obtain $g(\rho(x))\neq 0$, hence $\rho(x)\in \ZX_\aA$.
    \end{enumerate}
  \end{proof}
\end{lemma}

\subsection{The Bohr Compactification}
\label{subsec:Bohrcomp}
We start with some basic definitions and facts, cf.\ \cite{RudinFourier}. Then we show that the Bohr compactification of a locally compact abelian group $G$ equals the spectrum of the almost periodic functions $\CAP(G)$ on $G$. Finally, we characterize the continuous abelian group structures on the spectrum of a unital $\Cstar$-subalgebra of the bounded functions on a set. 
\subsubsection{Basic Definitions}
\label{sss:Basicdefs}
The statements not proven here can be found, e.g., in \cite{RudinFourier}:

\vspace{5pt}
\noindent
If $G$ is an LCA group, then the dual group $\DG$ of $G$ is the set of continuous homomorphisms $\chi\colon G\rightarrow S^1$, for $S^1\subseteq \mathbb{C}$ the unit circle, endowed with the group structure
\begin{align*}
  (\chi*\chi')(g):=\chi(g)\cdot\chi'(g),\qquad\quad \chi^{-1}(g):=\ovl{\chi(g)},\qquad\quad 1_\DG\colon g\mapsto 1 \qquad\quad \forall\:g\in G.
\end{align*}
The elements of $\DG$ always separate the points in $G$ and if $G$ is compact, then $\int_{G}\chi\: \dd \muH=0$ iff $\DG\ni \chi\neq 1$ for $\muH$ the Haar measure on $G$. 
The group $\DG$ becomes an LCA group when equipped with the topology generated by the sets 
\begin{align*}
	B_{K,\epsilon}(\chi):=\{\chi'\in \DG\:|\: |\chi(g)-\chi'(g)|< \epsilon\text{ for all }g\in K\}.
\end{align*}
Here $K\subseteq G$ is compact, $\chi\in \DG$ and $\epsilon>0$. If $\widehat{\DG}$ denotes the dual of $\DG$, Pontryagin duality states that the map $j\colon G\rightarrow \widehat{\DG}$, $j(g)\colon \chi\mapsto\chi(g)$ is an isomorphism and a homeomorphism.

Now, if we equip $\DG$ with the 
discrete topology, then we obtain a further LCA group $\DG_d$. The Bohr compactification $\GB$ of $G$ is defined to be the dual of $\DG_d$. $\GB$ is compact since for duals of discrete LCA groups this is always the case. Moreover, we have $\widehat{\DG}\subseteq \GB$ because $\GB$ equals the set of all homomorphisms $\psi\colon \DG\rightarrow S^1$ whereas $\widehat{\DG}$ consists of the continuous (w.r.t.\ the topology on $\DG$, not w.r.t.\ the discrete topology on $\DG_d$) ones.
One can show that the map $i_{\mathrm{B}}\colon G\rightarrow \GB$, defined as $j$ above, is a continuous isomorphism to the dense subgroup $\iB(G)\subseteq \GB$. Since $\DG_d$ is discrete, each compact set is finite. Consequently, the topology on $\GB$ is generated by the sets 
\begin{align}
  \label{eq:bohrtop}
  B_{\chi,\epsilon}(\psi):=\{\psi '\in \GB\:|\: |\psi(\chi)-\psi'(\chi)|< \epsilon\}\qquad\text{for}\quad\epsilon>0 \quad\text{and}\quad\chi\in \DG_d.
\end{align}
\begin{lemma}
  \label{lemma:Bohriso}
  Let $\text{\gls{CAP}}\subseteq B(G)$ denote the $\Cstar$-algebra generated by the elements of $\DG$. Then, the restriction map $\res\colon \Spec(\CAP(G))\rightarrow \GB$, $\psi\mapsto \psi|_{\DG}$ is a homeomorphism. 
  \begin{proof}
    Obviously, $\res$ is well defined and continuous. Moreover, it is injective because $\CAP(G)$ is generated by $\DG$. 
    So, the crucial part is to show surjectivity of $\res$. For this, let $\widehat{\psi}\in\GB$ and define $\psi$ on the $^*$-span $\Gen$ of $\DG$ in $\CAP(G)$ by
     \begin{align*}
    \psi(f):=\textstyle\sum_{i=1}^k \beta_i\cdot \widehat{\psi}(\chi_i) \qquad\text{for}\qquad \Gen\ni f=\sum_{i=1}^k \beta_i \cdot\chi_i.
     \end{align*}
    This map is well defined and linear because $\DG\subseteq \CAP(G)$ is a linearly independent subset. In fact, let $\Gamma_B$ denote the dual group of $\GB$ and $j_\Gamma\colon \Gamma_d \rightarrow  \Gamma_B$ the canonical map $j_\Gamma(\chi)\colon \psi \mapsto \psi(\chi)$ defined as the map $j$ above. Then $j_\Gamma(\Gamma)\subseteq C(\GB)$  
   is contained in the dual group of $\GB$, so that 
     $\int_{\GB}j_\Gamma(\chi)\: \dd \muH=0$ iff $\chi\neq 1$ for $\muH$ the Haar measure on $\GB$ and $\chi \in \DG$. Now, it is straightforward to see that
	\begin{align*}    
     \psi(\ovl{f}\cdot f')=\ovl{\psi(f)}\cdot\psi(f')\qquad\text{holds for all}\qquad f,f'\in \Gen. 
	\end{align*}     
     Then $\psi$ extends (by linearity and continuity) to a well-defined element\footnote{Multiplicativity on $\CAP(G)$ follows from continuity and multiplicativity on $\Gen$.} of $\Spec(\CAP(G))$ if we can show that $\psi$ is continuous on $\Gen$. For this, let $\Gen\ni f=\sum_{i=1}^k \beta_i \cdot\chi_i$ be as above and $\epsilon>0$. We choose $g\in G$ such that $\big|\widehat{\psi}(\chi_i)-\iB(g)(\chi_i)\big|\leq \frac{\epsilon}{k\max(|\beta_1|,\dots,|\beta_k|)}$ for $1\leq i\leq k$ and obtain
    \begin{align*}
      |\psi(f)|&\leq \big|\textstyle\sum_{i=1}^k\beta_i\cdot \big[ \widehat{\psi}(\chi_i)-\iB(g)(\chi_i)\big]\big|+ \big|\textstyle\sum_{i=1}^k \beta_i\cdot \iB(g)(\chi_i)\big|\\
      & \leq \max(|\beta_1|,\dots,|\beta_k|)\cdot\textstyle\sum_{i=1}^k \frac{\epsilon}{k\max(|\beta_1|,\dots,|\beta_k|)} + \|f\|_\infty,
    \end{align*}
    hence $|\psi(f)|\leq \|f\|_\infty +\epsilon$ for all $\epsilon>0$. This shows $|\psi(f)|\leq \|f\|_\infty$ for all $f\in \CAP(G)$, hence continuity of $\psi$.
  \end{proof}
\end{lemma}
So, in the following we tacitly identify $\GB$ with $\Spec(\CAP(G))$ where we carry over the group structure and the corresponding Haar measure $\muH$ from $\GB$ to $\Spec(\CAP(G))$, i.e., we define 
\begin{align*}
  \psi_1 + \psi_2 := \res^{-1}(\res(\psi_1)+ \res(\psi_2))\qquad \psi^{-1}:=\res^{-1}\big(\raisebox{-0pt}{$\ovl{\res(\psi)}$}\big)\qquad e:=\res^{-1}(e)\qquad \muB:=\res^{-1}(\muH).
\end{align*}
A more concrete description of this group structure is given in the proof of Proposition \ref{prop:Specgroup}. There we show that abelian group structures on spectra of unital $\Cstar$-subalgebras of bounded functions can be encoded by families of quasi-characters on the underlying set, see Definition \ref{def:quasichar}.  
\begin{lemconv}[Bohr Compactification of $\RR$]
\label{lemconv:RBMOD}
\begin{enumerate}
\item  
\label{lemconv:RBMOD1}
 If $G=\RR$, then $\GR$ consists exactly of the functions of the form
  $\chi_l\colon x\mapsto e^{\I l x}$ for $l\in \RR$. \cite{RudinFourier}. Since we do not consider Bohr compactifications of other locally compact abelian groups in the following, by \gls{muB} we will always denote the Haar measure on $\RB$. 
 \item
    \label{prop:Bohrmod24} 
    Let $\psi \in \RB$, $L=\{l_\alpha\}_{\alpha \in I}\subseteq \RR$ a collection of $\Q$-independent reals, and $L^\perp\subseteq \mathbb{R}$ a subset for which $L \sqcup L^\perp$ is a $\mathbb{Q}$-base of $\RR$. Then, for $\{q_\alpha\}_{\alpha\in I}\subseteq  \Q$ and $\{s_\alpha\}_{\alpha\in I}\subseteq S^1$ we find $\psi'\in \RB$ with 
\begin{align*}    
 \psi'(\chi_{q_\alpha\cdot l_\alpha})=s_\alpha\quad  \forall\:\alpha\in I\qquad\quad\text{and}\qquad\quad \psi'(\chi_{l})=\psi(\chi_{l})\quad\forall\: l \in \Span_{\mathbb{Q}}(L^\perp). 
\end{align*}
   In fact, we choose $\{x_\alpha\}_{\alpha\in I}\subseteq \RR$ such that $\chi_{q_\alpha\cdot l_\alpha}(x_\alpha)=s_\alpha$ holds for all $\alpha\in I$, and define
    \begin{align*}
      \zeta\left(\chi_{q\cdot l_\alpha}\right):=\chi_{q\cdot l_\alpha}(x_\alpha)\: \text{ if }\:\alpha\in I \qquad\quad\text{as well as}\qquad\quad \zeta\left(\chi_{q\cdot l}\right):=\psi(\chi_{q\cdot l})\: \text{ if }\:l\in L^\perp
    \end{align*}
    for all $q\in \mathbb{Q}$. 
    Then, for $l\in \RR$ arbitrary we have a unique representation of the form 
	\begin{align*}    
    	l=\textstyle\sum_{i=1}^k q_i\: l_{\alpha_i} + \textstyle\sum_{j=1}^{k'} q_j'\: l'_j .
	\end{align*} 
	with $l'_j \in L^\perp$ and $q_i,q'_j \in \Q$ for $1\leq i\leq k$, $1\leq j\leq k'$. Here, we define
    \begin{align*}
      \psi'\!\left(\chi_{l}\right):=\textstyle\prod_{i=1}^k\zeta\big(\raisebox{1pt}{$\chi_{q_i\cdot l_{\alpha_i}}$}\big) \cdot \textstyle\prod_{j=1}^q\zeta\big(\raisebox{1pt}{$\chi_{q'_j\cdot l'_j}$}\big) 
    \end{align*}
    as well as $\psi(1):=1$. Then, it is straightforward to see that $\psi'\colon \Gamma\rightarrow S^1$ is a homomorphism with the desired properties.
    \item
    \label{prop:Bohrmod21}
    Let $\Per$ denote the set of maps $\phi\colon \RR_{>0}\rightarrow [0,2\pi)$ with
    \begin{align*}	
      \phi(l+l')=\phi(l)+\phi(l')\:\bmod \: 2\pi\qquad \forall\:l,l'\in \RR_{>0}.
    \end{align*}
   Then $\RB\cong \Per$.
   
   In fact, if $\phi\in \Per$, then $\psi\colon \chi_l\mapsto \e^{\I \sign(l) \phi(|l|)}$ for $l\neq 0$ and $\psi(1):=1$ is a well-defined element of $\RB$ because
      \begin{align*}
        \psi(\chi_{l}\cdot\chi_{l'})
        =\e^{\I \sign(l+l')\: \phi(|l+l'|)}
        =\e^{\I \sign(l)\:\phi(|l|)} \e^{\I \sign(l')\:\phi(|l'|)}
        =\psi(\chi_{l})\cdot\psi(\chi_{l'}).
      \end{align*}
      Here, the second equality is clear for $\sign(l)=\sign(l')$ and 
      follows from 
      \begin{align*}
        \phi(l)=\phi(l-l'+l')=\phi(l-l')+\phi(l') +2\pi n \quad\text{for}\quad l>l'>0
      \end{align*}
      in the other cases.    
      Now, if $\psi\in \RB$, then $\psi(\chi_l)\in S^1$, hence $\psi(\chi_l)=\e^{\I \phi'(l)}$ for $\phi'(l)\in [0,2\pi)$ uniquely determined. This defines a map $\phi'\colon \RR \rightarrow [0,2\pi)$ whose restriction $\phi:=\phi'|_{\RR_{>0}}$ has the desired properties.
\item
 \label{rem:RBOHR}
  In the following, by the Bohr compactification 
  \gls{RB}  
  of $\RR$ we will understand both 
  \begingroup
  \setlength{\leftmarginii}{15pt}
  \begin{itemize}  
  \item
  In Subsection \ref{sec:ConSp}, the dual group of $\GR_d$ (see Convention \ref{conv:sammel}).
  \item
  	In Section \ref{sec:HomIsoCo}, the spectrum of the $\Cstar$-algebra $\CAP(\RR)$ generated by the set $\GR$ of continuous characters on $\RR$.
  \end{itemize}
  \endgroup
	\noindent
	In both cases we will refer to the modifications result in Part \ref{prop:Bohrmod24} which obviously also applies to the $\Spec(\CAP(\RR))$ case.  
\end{enumerate}
\end{lemconv}
\subsubsection{Abelian Group Structures and Quasi-Characters}
\begin{definition}
  \label{def:quasichar}
  Let $X$ be a set and $\aA\subseteq B(X)$ some unital $C^*$-subalgebra. A subset $\uU\subseteq \aA$ with\footnote{Recall Convention \ref{conv:Boundedfunc} for the closure of a subset of a $\Cstar$-subalgebra of bounded functions.} $\ovl{\uU}=\aA$ is called family of quasi-characters on $X$ iff 
  \begingroup
  \setlength{\leftmargini}{15pt}
  \begin{enumerate}
  \item
    \label{def:quasichar1}
    $\im[f]\subseteq S^1$ for all $f\in \uU$.
  \item
    \label{def:quasichar2}
    $\uU$ is closed under pointwise multiplication and complex conjugation.
  \item
    \label{def:quasichar3}
    The elements of $\uU$ are linearly independent.
  \item
    \label{def:quasichar4}
    For $x,y\in X$ there is a net $\{z_\alpha\}_{\alpha\in I}\subseteq X$ such that $f(x)f(y)=\lim_\alpha f(z_\alpha)$ for all $f\in \uU$. Moreover, there is a net $\{e_\alpha\}_{\alpha\in J}$ such that $\lim_\alpha f(e_\alpha)=1$ for all $f\in \uU$.
  \end{enumerate}
  \endgroup
  \noindent
  If $X$ carries an abelian group structure, then $f\in \aA$ is said to be a character iff $\im[f]\subseteq S^1$, and 
	\begin{align*}
		 f(x+ y)=f(x)\cdot f(y)\qquad\text{as well as}\qquad f(x^{-1})=\ovl{f}(x)
	\end{align*}  
  holds for all $x,y\in X$.
\end{definition}

\begin{proposition}
  \label{prop:Specgroup}
  \begin{enumerate}
  \item
  Let $X$ be a set and $\aA\subseteq B(X)$ a unital $C^*$-algebra. Then, the families of quasi-characters are in bijection with the continuous abelian group structures on $\X$. 
  \item
  If $X$ carries an abelian group structure, then a continuous abelian group structure on $\X$ is compatible in the sense that
  \begin{align*} 
  	\iota_X(x)+  \iota_X(y)=\iota_X(x + y)\qquad \forall\:x,y\in X_\aA
  \end{align*}  	
  	iff the respective family $\uU$ consists of characters.  
  \end{enumerate}
  \end{proposition}
   The technical details of the proof can be found in Proposition \ref{prop:SpecgroupA}.
  \begin{proof}
  	Each family of quasi-characters $\uU$ gives rise to a continuous abelian group structure on $\X$ just by 
    \begin{align*}
      (\psi_1+\psi_2)(f):= \textstyle\sum_{i=1}^n \beta_i\:\psi_1(f_{\alpha_i})\psi_2(f_{\alpha_i})\qquad \psi^{-1}(f):=\textstyle\sum_{i=1}^n \beta_i\:\ovl{\psi_1(f_{\alpha_i})}\qquad e(f):=\sum_{i=1}^n\beta_i
    \end{align*}  	
     for $\psi_1,\psi_2,\psi \in \X$ and $f:=\textstyle\sum_{i=1}^n\beta_i f_{\alpha_i}$ with $f_{\alpha_i}\in \uU$, $\beta_i\in \mathbb{C}$ for $1\leq i\leq n$. 
     Here, one has to show that the above maps extend by continuity to $\aA$ being the closure of the $^*$-algebra generated by $\uU$. 
  	Conversely, if $\X$ carries a continuous abelian group structure and $\Gamma$ denotes its dual group, then $\uU:=\GT^{-1}(\Gamma)$ is the desired family of quasi-characters for $\GT\colon \aA\rightarrow C(\X)$ the Gelfand transform.
  \end{proof}
Of course, if $G$ is a compact abelian group, then $\DG\subseteq \CAP(G)$ is the family of (quasi-)characters that corresponds to the continuous group structure on $\GB\cong \Spec(\CAP(G))$.

\section{Spectral Extensions of Group Actions}
\label{sec:SPecExtGr}
In this section, we develop a reduction concept which, applied to the framework of loop quantum gravity, allows to perform a symmetry reduction directly on the quantum level. 
We will consider the general situation where the symmetry 
is represented by a Lie group of automorphisms of the principal fibre bundle of interest. The basic idea 
then is to lift this symmetry  
  to a left action on the quantum configuration space of LQG. In analogy to the classical situation, where the reduced configuration space is formed by the set $\AR$ of invariant connections on $P$, the quantum-reduced configuration space then will be formed by such elements which are invariant under the whole symmetry group. In contrast to that, reduction in LQG traditionally means to quantize the reduced classical space $\AR$, and we will see that the resulting space is always contained in our quantum-reduced one. Moreover, in the next section we will show that this inclusion is  even proper in several situations, so that in this context quantization and reduction usually do not commute.

In a first step, we will use the concept of a $\Cstar$-dynamical system in order to extend a left action $\cw\colon G\times X\rightarrow X$ of a group $G$ on a set $X$ to the spectrum of a $\Cstar$-subalgebra $\aA\subseteq B(X)$.
Then, we adapt this to the case where $X$ equals the set of smooth connections on a principal fibre bundle, and where $\aA$ is generated by parallel transports along 
a distinguished set of curves in its base manifold.  
Finally, we consider the case where the structure group is compact, and where the set of curves has an additional  independence property. In this situation, it will be possible to identify the quantum configuration space of LQG with a space of homomorphisms of paths. The quantum-reduced configuration space then  will be formed by the invariant ones. In the next section, this will allow us to investigate the inclusion relations between quantized reduced classical and respective quantum-reduced configuration spaces in much more detail.

We now start with some general statements concerning group actions on spectra of abelian $\Cstar$-algebras.
\subsection{Group Actions on Spectra}
\label{subsec:graonsp} 
Recall \cite{0922.46050} that a $\Cstar$-dynamical system is a triple $(\aA,G,\ah)$ consisting of a $\Cstar$-algebra $\aA$, a group $G$ and an antihomomorphism $\ah\colon G\rightarrow \Aut(\aA)$. If $G$ is a topological group, then $\ah$ is said to be continuous iff for each $a\in \aA$ the map $\ah(\cdot)(f)\colon G\rightarrow \aA$, $g\mapsto \ah(g)(a)$ is continuous. In \cite{0922.46050} it is shown that each $\Cstar$-dynamical system with $G$ locally compact\footnote{The proof there also works if $G$ is an arbitrary topological group.} and $\ah$ continuous, gives rise to a continuous action $\specw$ of $G$ on $\Spec(\aA)$.
In the next lemma we discuss this assignment for the abelian case. In the first part, we will drop the continuity assumptions and show that the assignment $\ah \mapsto \specw$ is bijective.  
In the second part, we then investigate the continuity properties of the respective maps. For instance, if
$\aA$ is unital, it turns out that $\specw$ is continuous iff $\ah$ is. This will provide us with a necessary condition for continuity of $\specw$.  
\begin{lemma} 
  \label{lemma:leftactionsvsautos}
  Let $\aA$ be an abelian $\Cstar$-algebra. 
  \begin{enumerate}
  \item
    \label{lemma:leftactionsvsautos1}
    The $\Cstar$-dynamical systems $(\aA,G,\ah)$ are in bijection with the left actions $\specw\colon G\times \Spec(\aA)\rightarrow \Spec(\aA)$ for which $\specw_g$ is continuous for all $g\in G$.
  \item
    \label{lemma:leftactionsvsautos2}
    If $G$ is a topological group, then continuity 
    of $\ah$ implies continuity of $\specw$. The converse implication holds if $\aA$ is unital.
  \end{enumerate}
  \end{lemma}
  Recall that $\specw_g\colon \Spec(\aA)\rightarrow \Spec(\aA)$ is defined by $\specw_g(\chi):=\specw(g,\chi)$.
  \begin{proof}
    \begin{enumerate}
    \item
      Let $\ah\colon G\rightarrow \Aut(\aA)$ be given and define the corresponding left action by $\specw(g,\chi):=\eta(\ah(g))(\chi)$. Then, $\specw_g$ is well defined and continuous by Lemma \ref{lemma:homzuspec}.\ref{lemma:homzuspec2}.
      Moreover, since $\eta$ and $\ah$ are antihomomorphisms, the left action property follows from
      \begin{align*}
        \specw(gh,\chi)&=\eta(\ah(gh))(\chi)=\eta(\ah(h)\cp\ah(g))(\chi)\\ 
        &=\eta(\ah(g))\left(\eta(\ah(h))(\chi)\right)
        =\specw(g,\eta(\ah(h))(\chi))=\specw(g,\specw(h,\chi)).
      \end{align*}
      Conversely, if $\specw\colon G\times \Spec(\aA)\rightarrow \Spec(\aA)$ is a left action, then $\specw_g\in \mathrm{Homeo}(\Spec(\aA))$ for each $g\in G$ so that $\ah(g):=\tau(\specw_g)$ is an element of $\Aut(\aA)$. Here, $\tau\colon \mathrm{Homeo}(\Spec(\aA))\rightarrow \Aut(\aA)$ denotes the map \eqref{eq:tautau}, which is just the inverse of $\eta$. Since $\eta$ is an antiisomorphism, the same is true for $\tau=\eta^{-1}$, so that
      \begin{align*}
      \ah(gh)=\tau(\specw_{gh})=\tau(\specw_g\cp \specw_h)
      =\tau(\specw_h)\cp \tau(\specw_g)=\ah(h)\cp \ah(g). 
      \end{align*}
    \item
      Assume that $\ah$ is continuous and let $G\times \Spec(\aA)\supseteq \{(g_\alpha,\chi_\alpha)\}_{\alpha\in I}\rightarrow (g,\chi)\in G\times \Spec(\aA)$ be a converging net. Then $\{g_\alpha\}_{\alpha\in I}\rightarrow g$ and $\{\chi_\alpha\}_{\alpha\in I}\rightarrow \chi$ are converging nets as well. By continuity of $\ah$ for each $a\in \aA$ and each $\epsilon>0$ we find $\alpha_\epsilon\in I$ such that $\|\ah(g)(a)-\ah(g_\alpha)(a)\|_\aA\leq \frac{\epsilon}{2}$ for all $\alpha \geq \alpha_\epsilon$. Then, since $\|\chi_\alpha\|_{\mathrm{op}}\leq 1$ for all $\alpha\in I$, we obtain 
      \begin{align*}
        |\chi_\alpha\big(\ah(g)(a)-\ah(g_\alpha)(a)\big)|\leq \|\ah(g)(a)-\ah(g_\alpha)(a)\|_\aA\leq\frac{\epsilon}{2}\qquad \forall\:\alpha \geq \alpha_\epsilon.
      \end{align*}
      Moreover, since $\{\chi_\alpha\}_{\alpha\in I}\rightarrow \chi$, we find $\alpha'_{\epsilon}\in I$ with 
      $\|\chi-\chi_\alpha\|_{\ah(g)(a)}\leq \frac{\epsilon}{2}$
      for all $\alpha\geq \alpha'_\epsilon$.
      Then for $\alpha\in I$ with $\alpha\geq \alpha_\epsilon,\alpha'_\epsilon$ we have
      \begin{align*}
        \|\specw(g,\chi)-\specw(g_\alpha,\chi_\alpha)\|_a
        &=\|\eta(\ah(g))(\chi)-\eta(\ah(g_\alpha))(\chi_\alpha)\|_{a}\\
        &=|\chi(\ah(g)(a))-\chi_\alpha(\ah(g_\alpha)(a))|\\
        &\leq \|\chi-\chi_\alpha\|_{\ah(g)(a)} + |\chi_\alpha\big(\ah(g)(a)-\ah(g_\alpha)(a)\big)|\leq \epsilon,
      \end{align*}
      which shows the first part.
      
      Now, let $\aA$ be unital. For fixed $a\in \aA$ we consider the continuous function
      \begin{align*}
        h\left((g,\chi),(g',\chi')\right):= (\GT(a)\cp \specw)(g,\chi)-(\GT(a)\cp \specw)(g',\chi'), 
      \end{align*}
      where $(g,\chi),(g',\chi')\in G\times \Spec(\aA)$. Then $h^{-1}(B_\epsilon(0))$ is open and contains $((g,\chi),(g,\chi))$ for all $(g,\chi)\in G\times \Spec(\aA)$. Let $g\in G$ be fixed. Then for each $\chi\in \Spec(\aA)$ we find open neighbourhoods $B_\chi\subseteq G$, $U_\chi \subseteq \Spec(\aA)$ of $g$ and $\chi$, respectively, such that 
	\begin{align*}      
      B_\chi\times U_\chi \times B_\chi\times U_\chi\subseteq h^{-1}(B_\epsilon(0))
     \end{align*}
     holds.   
      By compactness of $\Spec(\aA)$  
      there are $\chi_1,\dots, \chi_n\in \Spec(\aA)$ such that the corresponding sets $U_{\chi_j}$ cover $\Spec(\aA)$. Then $B_g:= B_{\chi_1} \cap \dots \cap B_{\chi_n}$ is an open neighbourhood of $g$ and we obtain
      \begin{align*}
        \left|(\GT(a)\cp\specw)(g,\chi)-(\GT(a)\cp\specw)(h,\chi)\right|< \epsilon \qquad \forall\: \chi \in \Spec(\aA),  h\in B_g.
      \end{align*}
      Consequently, $\left\|\GT(a)\cp\specw_g -\GT(a)\cp\specw_h\right\|_{\infty}\leq \epsilon$, so that
      \begin{align*}
        \|\ah(g)(a)-\ah(h)(a)\|_{\aA}&= \|\tau(\specw_g)(a)-\tau(\specw_h)(a)\|_{\aA}\\
        &=\|\GT^{-1}[\GT(a)\cp\specw_g]-\GT^{-1}[\GT(a)\cp\specw_g]\|_{\aA}\\
        &=\|\GT^{-1}[\GT(a)\cp\specw_g-\GT(a)\cp\specw_g]\|_{\aA}\\
        &=\left\|\GT(a)\cp\specw_g-\GT(a)\cp\specw_h\right\|_{\infty}\\
        &\leq \epsilon
      \end{align*}
      for all $h\in B_g$. This shows continuity of $\ah(\cdot)(a)$ at $g\in G$. 
    \end{enumerate}
  \end{proof}

\begin{definition}[$\specw$-Invariance]
  \label{def:ixschlange}
  Let $\cw$ be a left action of the group $G$ on the set $X$ and $\aA\subseteq \Cb(X)$.
  \begin{enumerate}
  \item
    \label{def:ixschlange1}
    $\aA$ is called $\cw$-invariant iff $\cw_g^*(\aA)\subseteq \aA$ for all $g\in G$, i.e., if $(\aA,G,\ah)$ is a $\Cstar$-dynamical system with $\ah(g)(f):=\cw_g^*(f)$ for all $g\in G$ and all $f\in \aA$.
  \item
    \label{def:ixschlange2}
    We define the set 
	\begin{align*}    
    	\text{\gls{XR}}:=\left\{x\in X\:|\:\cw(g,x)=x\text{ for all } g\in G \right\}
	\end{align*}    
     of invariant elements, and denote the spectrum of the $\Cstar$-algebra $\ovl{\aA|_{\XR}}$ by \gls{XRRQ}.  
  \item
    \label{def:ixschlange3} 
    We define      
    	$\text{\gls{XRQ}}:=
    	\ovl{\iota_X(X_\aA\cap \XR)}$ to be the closure of $\iota_X(X_\aA\cap \XR)$  in $\X$.
  \end{enumerate}
\end{definition}
\begin{remark}
	Let $Y:=\XR$ and $\upsilon \colon Y\rightarrow X$ the canonical inclusion map. Moreover, let $\XNR$, $\YNR$ denote the corresponding spaces introduced in Convention \ref{conv:Boundedfunc}. Then
	\begingroup
	\setlength{\leftmargini}{15pt}
	\begin{itemize}
	\item
		$\XRRQ=\XNR$ because $\ovl{\aA|_{\XR}}=\rR_\upsilon=\ovl{\upsilon^*(\aA)}$,
	\item
		$\XRQ=\YNR$ because $\YNR=\ovl{\iota_X(X_\aA\cap \upsilon(\XR))}=
    	\ovl{\iota_X(X_\aA\cap \XR)}$.		
		\hspace*{\fill}$\lozenge$
	\end{itemize}
	\endgroup
\end{remark}
\begin{Proposition} 
  \label{prop:autspec} 
  Let $\cw$ be a left action of the group $G$ on the set $X$ and $\text{\gls{aA}}\subseteq \Cb(X)$ a $\cw$-invariant 
  $\Cstar$-algebra.  
  \begin{enumerate}
  \item
    \label{prop:autspec1}
    There is a unique left action $\text{\gls{SPECW}}\colon G\times \X \rightarrow \X$
    such that:
    \begin{enumerate}
    \item[$\mathrm{(a)}$]
      $\specw_g$ is continuous for all $g\in G$,
    \item[$\mathrm{(b)}$]
      $\specw$ extends $\cw$ in the sense that on $X_\aA$ we have
      \begin{align}
        \label{eq:extens}
        \specw_g\cp\iota_X=\iota_X\cp\cw_g\qquad\forall\: g\in G.
      \end{align}
    \end{enumerate}
    $\specw$ is explicitly given by $\specw\big(g,\x\big)= \x\cp \cw_g^*$. 
  \item
    \label{prop:autspec2}
    If $G$ is a topological group, then $\specw$ is continuous if $\cw^*_\bullet f\colon G\rightarrow \aA$, $g \mapsto f\cp \cw_g$ is continuous for each $f\in \aA$. The converse implication holds if $\aA$ is unital.
  \item
    \label{prop:autspec3}
    The set of invariant elements 
	\begin{align*}    
    	\XQR:=\left\{\x \in \X\:\big|\: \specw(g,\x)=\x\text{ for all }g\in G\right\}
	\end{align*}    
     is closed in $\X$, and we have $\XRQ\subseteq \XQR$.
  \item
    \label{prop:autspec4}
    If $\aA$ is unital, then  
    $\XRRQ\cong \XRQ$ via $\ovl{i_{\XR}^*}\colon \XRRQ\rightarrow \XRQ\subseteq \X$. 
   The following diagram is commutative
     \begin{center}
      \makebox[0pt]{
        \begin{xy}
          \xymatrix{
            \XRRQ\: \ar@{->}[r]^-{\ovl{i_{\XR}^*}}_-{\cong}   & \: \XRQ\:  \ar@{->}[r]^-{\subseteq}   & \:\XQR\: \ar@{->}[r]^-{\subseteq} & \:\X \\
          \XR\: \ar@{.>}[u]^{\iota_{\XR}} \ar@{^{(}->}[r]^-{i_{\XR}}& \:i_{\XR}(\XR)\:  \ar@{.>}[u]^{\iota_X} \ar@{->}[rr]^{\subseteq} & & \: X \ar@{.>}[u]^{\iota_{X}}.    
          }
        \end{xy}
      }
    \end{center}
  \end{enumerate}
  \begin{proof}
    \begin{enumerate}
    \item
      First observe that \eqref{eq:extens} makes sense because $\cw_{g^{-1}}^*(\aA)\subseteq \aA$ implies $\aA = \cw_{g}^*(\cw_{g^{-1}}^*(\aA))\subseteq \cw_{g}^*(\aA)$, hence $\cw_g\colon X_\aA\rightarrow X_\aA$ by Lemma \ref{lemma:dicht}.\ref{lemma:dicht4}.
      For uniqueness, let $\specw$ and $\specw'$ be two such extensions of $\cw$. 
      Then, by \eqref{eq:extens} we have
      $\specw'_g|_{\iota_X(X_\aA)}=\specw_g|_{\iota_X(X_\aA)}$ for all $g\in G$ so that $\specw'_g=\specw_g$ by $\mathrm{(a)}$  and denseness of $\iota_X(X_\aA)$ in $\Spec(\aA)$. For existence, consider the $\Cstar$-dynamical system $(\aA,G,\ah)$ for $\ah(g):= \cw_g^*$.
      Then, Lemma \ref{lemma:leftactionsvsautos}.\ref{lemma:leftactionsvsautos1} provides us with a corresponding left action $\specw\colon G\times \Spec(\aA)\rightarrow \Spec(\aA)$ such that $\specw_g$ is continuous for all $g\in G$. This action is  
      given by
      \begin{align*}
        \specw(g,\ovl{x})=\eta(\ah(g))(\ovl{x})=\ovl{x}\cp \ah(g)=\ovl{x}\cp \cw^*_g,
      \end{align*}
      so that for $x\in X_\aA$ and all $f\in \aA$ we have
      \begin{align*}
        \specw_g(\iota_X(x))(f)
        =\iota_X(x)(\cw_g^*f)=f(\cw_g(x))=(\iota_X\cp\cw_g)(x)(f).
      \end{align*}
    \item
      We have $\cw_\bullet^*f=\ah(\cdot)(f)$, so that the continuity statement is clear from Lemma \ref{lemma:leftactionsvsautos}.\ref{lemma:leftactionsvsautos2}.
    \item
      Let $\XQR\supseteq \{\x_\alpha\}_{\alpha\in I}\rightarrow \x\in \X$ be a converging net. Then for all $g\in G$ it follows from continuity of $\specw_g$ that 
      $\specw(g,\x)=\specw\left(g,\lim_\alpha\x_\alpha\right)=\lim_\alpha\specw(g,\x_\alpha)=\lim_\alpha \x_\alpha=\x$, 
      hence closedness of $\XQR$. Moreover, $\iota_X(\XR\cap X_\aA)\subseteq \XQR$ by \eqref{eq:extens} so that $\ovl{\iota_X(\XR\cap X_\aA)}\subseteq \XQR$.
    \item
      This follows from the parts \ref{lemma:dicht2}), \ref{lemma:dicht3}) of Lemma \ref{lemma:dicht} if we define $\upsilon:=i_{\XR}$.
    \end{enumerate}
  \end{proof}
\end{Proposition}
\begin{remark}
\begingroup
\setlength{\leftmargini}{20pt}
\begin{itemize}
\item
In the following sections, we will see that the details of the inclusion relations between the sets $\XRQ$ and $\XQR$ usually are far from being trivial.
\item
If $\aA$ is unital, then Part \ref{prop:autspec1}) can also be derived from Corollary 2.19 in \cite{ChrisSymmLQG} by extending each $\specw_g^*\colon \aA\rightarrow \aA$ uniquely to a continuous map $\specw_g\colon \Spec(\aA)\rightarrow \Spec(\aA)$. In fact, it follows from the uniqueness property of these maps that $\specw_{gh}=\specw_g\cp \specw_h$ holds for all $g,h\in G$. Then, $\specw\colon (g,\ovl{x})\mapsto \specw_g(\ovl{x})$ is a well-defined group action with the properties from Proposition \ref{prop:autspec}.\ref{prop:autspec1}.  
\item
In Subsection \ref{sec:InvGenConnes} we will use Proposition \ref{prop:autspec} in order to perform a symmetry reduction on quantum level in the framework of LQG. Then, in Section \ref{sec:HomIsoCo}, we will use this proposition in order to derive some uniqueness statements concerning normalized Radon measures on cosmological quantum configuration spaces occurring in LQG, see Proposition \ref{lemma:bohrmassdichttrans} and  Corollary \ref{cor:eindbohr}.
\end{itemize}
\endgroup
\end{remark}
\subsection{Invariant and Generalized Connections}
\label{sec:InvGenConnes}
We now adapt the previous subsection to the situation where the action $\cw$ comes from a Lie group of automorphisms $(G,\Phi)$ on a principal fibre bundle \gls{PMS}. This means that $\cw$ is given by \eqref{eq:connact}, i.e.,  acts on the set $\Con$ of smooth connections on $P$ by 
\begin{align*}
      \text{\gls{CW}}\colon G\times \Con\rightarrow \Con,\quad
      (g,\w)\mapsto \Phi_{g^{-1}}^*\w.
  \end{align*}
  Moreover, the $\Cstar$-algebra $\text{\gls{aA}}\subseteq B(\Con)$ is generated by parallel transports along suitable curves in $M$ in this case. 
  
More precisely, let \gls{Pa} be a fixed set of $\Ck$-paths in the base manifold $M$ and $\text{\gls{NU}}=\{\nu_x\}_{x\in M}\subseteq P$ a fixed family of elements with $\nu_x\in F_x$ for all $x\in M$. By $\cC$ we will denote the set of all bounded functions on $\Con$ which are of the form 
\begin{align}
\label{eq:deltadef}
	h_\gamma^\nu\colon \w\mapsto f\cp \text{\gls{DIFF}}\left(\nu_{\gamma(b)},\text{\gls{PATRA}}\big(\nu_{\gamma(a)}\big)\right)
\end{align} 
	 for $\gamma\in \Pa$ with $\dom[\gamma]=[a,b]$ and $f\in C_0(S)$. Recall that $\parall{\gamma}{\w}\colon F_{\gamma(a)}\rightarrow F_{\gamma(b)}$ denotes the parallel transport along $\gamma$ w.r.t.\ $\w$ as well as $\diff(p,p')\in S$ the unique element with $p'=p\cdot \diff(p,p')$ for $p,p'$ contained in the same fibre over $M$.  
	 The corresponding unital $\Cstar$-algebra of cylindrical functions is defined by $\text{\gls{PaC}}:=\ovl{\cC}\subseteq B(\Con)$. According to Convention \ref{conv:Boundedfunc}, here $\ovl{\cC}$ denotes the closure of the $^*$-algebra generated by $\cC$  in $B(\Con)$.
This definition is independent of the choice of $\nu$, as for $\nu'=\{\nu'_x\}_{x\in M}\subseteq P$ another such family and $\Pa\ni \gamma\colon [a,b]\rightarrow M$ we have  
\begin{align} 
  \label{eq:indep}
  \diff\big(\nu'_{\gamma(b)},\parall{\gamma}{\w}\big(\nu'_{\gamma(a)}\big)\big)=  
  \Delta\big(\nu'_{\gamma(b)},\nu_{\gamma(b)}\big)\cdot\diff\big(\nu_{\gamma(b)},\parall{\gamma}{\w}\big(\nu_{\gamma(a)}\big)\big)\cdot\Delta\big(\nu_{\gamma(a)},\nu'_{\gamma(a)}\big).
\end{align}
\begin{convention} 
  \label{hgamma} 
  \begingroup
  \setlength{\leftmargini}{20pt}
  \begin{itemize}
  \item
    In the following, let $\text{\gls{PSIX}}(p):=\diff(\nu_x,p)$ for all $p\in F_x$ and all $x\in M$, and define
    \begin{align*}
      h_\gamma^\nu(\w):=\diff\left(\nu_{\gamma(b)},\parall{\gamma}{\w}\big(\nu_{\gamma(a)}\big)\right)=\psi_{\gamma(b)}\left(\parall{\gamma}{\w}\big(\nu_{\gamma(a)}\big)\right)\qquad\forall \:\w\in \Con. 
    \end{align*}
  \item
    If $S$ is compact, then there exists \cite{BroeD} a faithful matrix representation $\rho\colon S\rightarrow \GLNC$ of $S$, which we will fix in the following.  Then, since the matrix entries of $\rho$ separate the points in $S$, the functions 
	\begin{equation}
	\label{eq:gniij}    
    	\:[h_\gamma^\nu]_{ij}:=\rho_{ij}\cp h_\gamma^\nu, 
    \end{equation}    
    together with the unit, generate $\PaC$.
  \item
    We will denote the spectrum of $\PaC$ by \gls{AQ}. If it is necessary to avoid confusion, we will write \gls{AALPHA} with an index $\alpha$ referring to the involved set of curves \gls{PALPHA}. The respective $\Cstar$-algebra of cylindrical functions then is denoted by \gls{PACALPHA}.
  \end{itemize}
  \endgroup
\end{convention}
The elements $\ovl{\w}\in \A$ are called generalized connections and form the quantum configuration space of LQG. Here, the main reason for replacing $\Con$ by $\A$ is that the latter space is compact and can be equipped with a normalized Radon measure in a canonical way, see Subsection \ref{subsub:InvHoms}. Moreover, in the relevant cases $\Con$ is canonically embedded\footnote{Here, this just means that $\iota_\Con$ is injective.} via the map $\iota_\Con$ (see Convention \ref{conv:Boundedfunc}) as the next lemma shows. More concretely, it suffices, e.g., that each tangent vector of $M$ is realized as a final tangent vector of a curve in $\Pa$ for which each final segment is an element of $\Pa$ as well.  Thus, if $M=\RR^3$, it suffices that $\Pa$ contains all linear curves:
\begin{lemma}
  \label{lemma:separating}
  The map $\iota_\Con\colon \Con \rightarrow \A$ is injective if for each $\vec{v}\in TM$ there is $\gamma\in \Pa$ and $s\in\dom[\gamma]=[a,b]$, such that $\dot\gamma(s)=\vec{v}$ and $\gamma|_{[a,t]}\in \Pa$ for all $t\in (a,s]$ or $\gamma|_{[t,b]}\in \Pa$ for all $t\in [s,b)$. 
  \begin{proof}
    This is a straightforward generalization of Appendix A in \cite{ChrisSymmLQG},  see Lemma \ref{lemma:separatingA}.
  \end{proof} 
\end{lemma}
According to Definition \ref{def:ixschlange}.\ref{def:ixschlange1} we have
\begin{align*}
  \text{\gls{AR}}=\{\w\in\Con\:|\: \cw(g,\w)=\w\text{ for all }g\in G\}= \{\w\in\Con\:|\: \Phi_g^*\w=\w\text{ for all }g\in G\},
\end{align*}
so that $\AR$ equals the set of $\Phi$-invariant connections on $P$, see Definition \ref{def:Invconn}.
Then, traditionally, in LQG the space $\text{\gls{ARRQ}}:=\Spec\big(\raisebox{0pt}{$\ovl{\PaC_\alpha|_{\AR}}$}\big)$ (cf.\ Definition \ref{def:ixschlange}.\ref{def:ixschlange2})
is used as reduced quantum configuration space, see, e.g., \cite{MathStrucLQG}. Here, the index $\alpha$ hints to the fact that usually not the same set of curves as for the full theory is used to define $\PaC_\alpha$. 

To summarize, we have made the following specifications:
\begin{align*}
	\cw	 \qquad \text{--} \qquad &\text{left action } \cw\colon G\times \Con\rightarrow \Con,\:(g,\w)\mapsto \Phi_{g^{-1}}^*\w\text{ for }(G,\Phi)\\[-6pt]
								&\text{a Lie group of automorphisms of }P\\ 
	\aA \qquad \text{--} \qquad  &\Cstar\text{-algebra }\PaC\text{ generated by the parallel transport functions }h_\gamma^\nu\text{ for }\gamma\\[-6pt] 
								 &\text{contained in the fixed set of $\Ck$-paths $\Pa$ in $M$.}\\%[4pt]
    \XR \qquad \text{--} \qquad  &\text{set $\AR$ of invariant connections on $P$.}\\[2pt]
    \XRRQ \qquad \text{--} \qquad  &\text{spectrum $\ARRQ=\Spec\big(\ovl{\PaC|_{\AR}}\big)$.}\\[2pt]
    \XRQ \qquad \text{--} \qquad  &\text{closure $\ARQ=\ovl{\iota_\Con(\Con_\PaC \cap \AR)}\subseteq \A$.}			 
\end{align*}
 Now, in order to perform a reduction on quantum level, i.e, to obtain a well-defined extension $\specw$ of $\cw$ to $\A$ providing us with the space $\AQR$ of invariant generalized connections, we first have to investigate under which assumptions the $\Cstar$-algebra $\PaC$ is  $\cw$-invariant. It turns out, to be sufficient that $\Pa$ fulfils the following invariance property. 
\begin{definition}
  \label{def:InvPfade} 
  A set $\Pa$ of $\CC{k}$-paths in $M$ is said to be $\Phi$-invariant iff $\wm_g \cp \gamma\in \Pa$ holds for all $g\in G$ and all $\gamma \in \Pa$. Here, \gls{WM} denotes the left action induced by $\Phi$ on $M$, see \eqref{eq:INDA}. \hspace*{\fill}$\lozenge$ 
\end{definition}
  Of course, if $\Pa$ is a collection of $\CC{k}$-paths in $M$ and $\wm$ is of class $\CC{k}$, the set $\langle \Pa\rangle$ of all compositions $\wm_g\cp \gamma$ with $g\in G$ and $\gamma \in \Pa$ is $\Phi$-invariant and consists of $\CC{k}$-paths as well. 
\begin{lemma}
  \label{lemma:CylSpecActionvorb}
  If $\Pa$ is $\Phi$-invariant, then $\PaC$ is $\cw$-invariant.
  \end{lemma}
  \begin{proof}
   	This is just a straightforward consequence of the fact that if $\gamma_p^\w$ is the horizontal lift of $\gamma\colon [a,b]\rightarrow M$ in $p\in F_{\gamma(a)}$ w.r.t.\ the connection $\w$, then $\Phi_g\cp \gamma_p^\w$ is the horizontal lift of $\wm_g\cp \gamma$ in the starting point $\Phi_g(p)$ w.r.t.\ the connection $\cw(g,\w)$.  The technical details can be found in Lemma \ref{lemma:CylSpecActionvorbA}.
  \end{proof}
We now are able transfer Proposition \ref{prop:autspec} to generalized connections, where we only consider the case where $S$ is compact, i.e., where $\PaC$ is unital. This is basically because for our further considerations we will need that $S$ is compact anyway. For instance, we will identify $\A$ with a space of homomorphisms of paths, and for this compactness will be necessary. In addition to that, the homeomorphism $\ARRQ\cong \ARQ$ which we have (see next corollary) if $S$ is compact will be crucial for the investigations of the inclusion relations between $\ARRQ$ and $\AQR$. 
\begin{corollary}
  \label{cor:CylSpecAction}
  Let $\Pa$ be $\Phi$-invariant and $S$ be compact. Then 
	\begin{align*}
		  \specw\colon G\times \A&\rightarrow \A, \quad
		  	 (g,\ovl{\w})\mapsto \ovl{\w}\cp \cw_g^*
	\end{align*}	  
  is the unique left action such that: 
    \begingroup
  	\setlength{\leftmargini}{20pt}
   \begin{itemize}
    \item
      $\specw_g$ is continuous for all $g\in G$,
    \item
      $\specw$ extends $\cw$ in the sense that on $\Con$ we have
      \begin{align*}
        \specw_g\cp\iota_\Con=\iota_\Con\cp\cw_g\qquad\forall\: g\in G.
      \end{align*}
    \end{itemize}
    \endgroup	
\noindent  
  The quantum-reduced space $\text{\gls{AQR}}=\left\{\w \in \A\:\big|\: \specw(g,\w)=\w\text{ for all }g\in G\right\}$ is compact and the following diagram is commutative:
   \begin{align}
    \label{eq:inclusionsdiag}
    \begin{split}
      \makebox[0pt]{
        \begin{xy}
          \xymatrix{
            \text{\gls{ARRQ}}\: \ar@{->}[r]^-{\ovl{i_{\AR}^*}}_-{\cong}   &  \:\text{\gls{ARQ}}\: \ar@{->}[r]^-{\subseteq}   & \:\text{\gls{AQR}}\:\ar@{->}[r]^-{\subseteq} &\:\text{\gls{AQ}} \\
           \ar@{.>}[u]^{\iota_{\AR}} \text{\gls{AR}} \: \ar@{^{(}->}[r]^-{i_{\AR}} & \:i_{\AR}(\AR)\:  \ar@{.>}[u]^{\iota_\Con} \ar@{^{(}->}[rr]^{\subseteq}  &  &\:\:\text{\gls{Con}} \ar@{.>}[u]^{\iota_{\Con}}.    
          }
        \end{xy}
      }
    \end{split}
  \end{align}
  Here, $\ovl{i_{\AR}^*}\colon \ARRQ\rightarrow \ARQ$ is a homeomorphism.
  The action $\specw$ is continuous iff $\theta_\bullet^*f\colon G\rightarrow \PaC$, $g \mapsto f\cp \theta_{g^{-1}}^*$ is continuous for all $f$ of the form $\rho_{ij}\cp h_\gamma^\nu$.
  \begin{proof}
    By Lemma \ref{lemma:CylSpecActionvorb} we have $\cw_g^*(\PaC)\subseteq \PaC$ for all $g\in G$. Moreover,
    $\PaC$ is unital and $\cw_\bullet^*f\colon G\rightarrow\PaC$, $g\mapsto \cw_g^*f$, is continuous for all $f\in \PaC$ iff\footnote{For this, observe that $\cw_\bullet^*1=1$ and that $\cw_\bullet^*[\lambda f+ \ovl{\mu g}\hspace{1pt}]=\lambda\: \cw_\bullet^*f+ \mu\: \ovl{\cw_\bullet^*g}$ for $\lambda,\mu\in \mathbb{C}$ and $f,g\in \PaC$.}  this is the case for all generators $f=\rho_{ij}\cp h^\nu_\gamma$.  
    Consequently, the claim follows from Proposition \ref{prop:autspec}.
  \end{proof}
\end{corollary}
\begin{remark}
\label{rem:subspt}
The elements of $\AQR$ are called invariant generalized connections and the space $\AQR$ will be equipped with its canonical subspace topology in the following.. 
\end{remark}
The next example shows that the action $\specw$ is usually not continuous in the standard cosmological applications. 

\begin{example}[Discontinuous Spectral Action]
  \label{ex:Eukl}
  Let $P=\RR^3 \times \SU$, $\Ge=\Gee$ and $\Pe$  
  be as in Example \ref{ex:LQC}, define $\nu:=\{(x,\me)\}_{x\in \mathbb{R}^3}$ and let $\Pa$ contain all linear curves in $\mathbb{R}^3$. The following arguments also apply to the homogeneous case, i.e., where the group is $\Gh=\RR^3$ and $\Ph$ is as in the same Example.

	\vspace{6pt}  
  \noindent
  We show that for the linear curve $\gamma_0\colon [0,1]\rightarrow \RR^3$, $t\mapsto t\vec{e}_1$ and $f:=\rho_{11}\cp h_{\gamma_0}^\nu$ the map $\theta_\bullet^*f\colon G\rightarrow \PaC$, $g\mapsto f\cp \cw_g$ is not continuous at $g=e\in \Ge$. For this, we define 
    \begingroup
  	\setlength{\leftmargini}{20pt}
	\begin{itemize}
	\item
		smooth connections $\w^r$ for $r\in \RR$ and
	\item
		group elements $g_\lambda \in \Ge$ for $\lambda\in \RR$ with $g_\lambda \rightarrow e\in \Ge$ for $\lambda \rightarrow 0$,
	\end{itemize} 
	\endgroup  
  \noindent
 such that for each $\lambda >0$ we find $r>0$ with $|\theta_{e}^*f(\w^r)- (\theta_{g_\lambda}^*f )(\w^r)|=1$, i.e., $\|\theta_{e}^*f- (\theta_{g_\lambda}^*f )\|\geq 1$. From this it is clear that $\theta_\bullet^*f\colon G\rightarrow \PaC$ is not continuous at $g=e\in \Ge$, so that $\specw$ is not continuous by Corollary \ref{cor:CylSpecAction}.  
	\begingroup
  	\setlength{\leftmargini}{25pt}
	\begin{enumerate}
	\item[(a)]
		  For $r\in \RR$ we define the connection $\omega^r$ by the right invariant geometric distribution specified by the following smooth sections $\mathcal{E}^r_i\colon P \rightarrow TP$ for $1\leq i\leq 3$:
  \begin{align*}
  	\mathcal{E}^r_1(x,s)&:=(\vec{e}_1,r x_2\:\tau_2\cdot s)\in T_{(x,s)}P \qquad\text{ for }\qquad x=(x_1,x_2,x_3)\in \RR^3
  	\\
  	\mathcal{E}^r_{i}(x,s)&:=(\vec{e}_{i},\vec{0})\hspace{109.5pt}\text{for }\qquad i=2,3.
  \end{align*} 
    Here, $\tau_2\cdot s:=\dd_\me R_s(\tau_2)\in T_s\SU$.
	\item[(b)]
		   Let $\gamma_y\colon[0,1]\rightarrow \mathbb{R}^3$ denote the linear curve that starts on the $\vec{e}_2$-axis at $y$ and traverses in $\vec{e}_1$-direction with constant velocity $\vec{e}_1$, i.e., $\gamma_y(t)=y\cdot\vec{e}_2 + t\cdot\vec{e}_1$.
  Then, its horizontal lift $\gamma_y^{r}\colon [0,1]\rightarrow \mathbb{R}^3\times \SU$ w.r.t.\ $\w^r$ in $(y\cdot \vec{e}_2,\me)$ is given by 
  $\gamma_y^{r}(t)=\left(\gamma_y(t),\exp(try\cdot \tau_2)\right)$ because
  $\pi \cp \gamma_y^{r}=\gamma_y$ and
  \begin{align*}
    \dot\gamma_y^{r}(t)=(\vec{e}_1,ry\tau_2\cdot \exp(try\tau_2))=\mathcal{E}^r_1\left(\gamma_y(t),\exp(try\cdot \tau_2)\right)=\mathcal{E}^r_1\left(\gamma_y^{r}(t)\right).
  \end{align*}
  	\item[(c)]
  	By the choice of $\nu$ we have 
  \begin{align*}
    h_{\gamma_y}^\nu(\w^r)=\pr_2\cp \gamma_y^{r}(1)=\exp(ry\cdot \tau_2)\stackrel{\eqref{eq:expSU2}}{=}\begin{pmatrix} \cos(ry) & -\sin(ry)  \\ \sin(ry) & \cos(ry)\end{pmatrix}.
  \end{align*} 
  Then, for $g=(-\lambda \vec{e}_2,\me)$ we obtain\footnote{See, e.g., proof of Lemma \ref{lemma:CylSpecActionvorb}, i.e., \eqref{eq:trafogenerators} in Lemma \ref{lemma:CylSpecActionvorbA}, where $\gamma'=\wm_{g^{-1}}\cp\gamma$.}  
  $\cw_g^*h^\nu_{\gamma_y}=h^\nu_{\gamma_{\lambda +y}}$, hence 
  \begin{align}
    \label{eq:Absch}
    \begin{split}
      \big\|\cw_{e}^*\big[\rho_{11}\cp h^\nu_{\gamma_0}\big]-\cw_{g}^*\big[\rho_{11}\cp h^\nu_{\gamma_0}\big]\big\|_{\infty}
      &\geq \big|\left(\rho_{11}\cp h^\nu_{\gamma_0}-\cw_g^*\big[\rho_{11}\cp h^\nu_{\gamma_0}\big]\right)(\w^r)\big|\\
      &=\big|\big(\rho_{11}\cp h^\nu_{\gamma_0}-\rho_{11}\cp h^\nu_{\gamma_{\lambda }}\big)(\w^r)\big|\\
      &=|1-\cos(\lambda r)|.
    \end{split}
  \end{align}
  This expression equals $1$ for $r=\pi/(2\lambda)$ and implies discontinuity of the action $\specw$ as we have explained above.
    \hspace*{\fill} {$\lozenge$} 
	\end{enumerate} 
	\endgroup  
\end{example}

\subsection{Homomorphisms of Paths and Invariance}
\label{subsec:InvHoms}
To this point, we have seen that a Lie group of automorphisms $(G,\Phi)$ on $P$ and a set of $\Ck$-paths $\Pa$ provides us with the quantized reduced classical spaces $\ARRQ\cong\ARQ$. Under the condition that $\Pa$ is $\Phi$-invariant, we even obtain the quantum-reduced configuration space $\AQR$.
  
 In order to investigate the inclusion properties between the spaces $\AQR$ and $\ARQ$, and to construct reasonable measures thereon, 
 we now are going to identify the elements of $\A$ with homomorphisms of paths, so-called generalized parallel transports. These are maps assigning to a curve $\gamma\in \Pa$ an equivariant mapping $F_{\gamma(a)}\rightarrow F_{\gamma(b)}$ where $\dom[\gamma]=[a,b]$. This identification will be possible under the assumption that $S$ is compact and that the set $\Pa$ has some additional independence property. Under this assignment $\AQR$ then occurs as a subspace of homomorphisms being invariant in a natural sense. 
The big advantage of considering $\A$ as a space of homomorphisms is 
the possibility to apply geometrical techniques like modification of (in this case invariant) homomorphisms to be developed in Section \ref{susec:LieALgGenC}. These modifications are crucial for investigations concerning the inclusion relations between $\ARQ$ and $\AQR$ (cf.\ Subsection \ref{sec:inclrel}). Indeed, they will allow us to  construct invariant elements that cannot be contained in $\ARQ$ or, more precisely, that cannot be approximated by classical (smooth) invariant connections. 
In addition to that,  
modification turns out to be a key tool for the construction 
of measures on $\AQR$, cf.\ Remark \ref{rem:euklrem} or Section \ref{sec:MOQRCS}.       

For the rest of this section, let $\PMS$ be a principal fibre bundle with compact structure group $S$, and $\nu=\{\nu_x\}_{x\in M}\subseteq P$ a family with $\nu_x\in F_x$ for all $x\in M$. Finally, recall the map $\psi_x(p)=\diff(\nu_x,p)$ for $p\in F_x$ from Convention \ref{hgamma}.

\subsubsection{Homomorphisms of Paths}
\label{sec:hompaths}
We start our considerations with a short 
introduction into homomorphisms of paths, and then highlight their relation to the space $\A$. 
\begin{definition}
\begin{enumerate}
\item
  Let $\gamma\colon [a,b]\rightarrow M$ be a curve.
  \begingroup
  \setlength{\leftmarginii}{15pt}
  \begin{itemize}
  \item
    \vspace{-4pt}
    The inverse of $\gamma$ is defined by $\gamma^{-1}\colon [a,b]\ni t\mapsto \gamma(b+a-t)$.  
  \item
    \vspace{2pt}
    A decomposition of $\gamma$ is a family of curves $\{\gamma_i\}_{0\leq i\leq k-1}$ 
    with $\gamma|_{[\tau_i,\tau_{i+1}]}=\gamma_i$ for $0\leq i\leq k-1$ and real numbers $a=\tau_0<{\dots}<\tau_k=b$. 
  \end{itemize}
  \endgroup
  \noindent 
  \item
  A set  of $\Ck$-paths $\Pa$ is said to be stable under decomposition and inversion iff
\begingroup
  \setlength{\leftmarginii}{15pt}
  \begin{itemize}
  \item
      \vspace{-4pt}
    $\gamma\in \Pa$ implies $\gamma^{-1}\in \Pa$,
  	\item
  	    \vspace{2pt}
	for each decomposition $\{\gamma_i\}_{0\leq i\leq k-1}$ of $\gamma$ we have $\gamma_i\in \Pa$ for all $0\leq i\leq k-1$. 
  \end{itemize}
  \endgroup  
  \end{enumerate}
\end{definition}
The space $\HOM$ of homomorphisms of paths (generalized parallel transports) is defined as follows.
\begin{definition}
  \label{def:hompaths}
  \begingroup
  \setlength{\leftmargini}{20pt}
  \begin{itemize}
  \item
    For $x,x'\in M$ let $\Iso(x,x')$ denote the set of equivariant mappings $\phi \colon F_x\rightarrow F_{x'}$, and define 
	\begin{align*}    
    	\AF:=\bigsqcup_{x,x'\in M}\Iso(x,x').
    \end{align*}
   Of course, here equivariance means that $\phi \cp R_s=R_s\cp \phi$ holds for all $s\in S$.
  \item
    \itspacecc
    Define the equivalence relation\footnote{$\gamma_1$ and $\gamma_2$ are called holonomy equivalent in this case.} \gls{CSIM} on $\Pa$ by $\gamma\csim \gamma'$ 
    iff $\parall{\gamma}{\w}=\parall{\gamma'}{\w}$ holds for all $\w\in \Con$. Observe that by definition we have that $\gamma\not\csim \gamma'$ if the start and end points of $\gamma$ and $\gamma'$ do not coincide. 
  \item
    \itspacecc
    Let \gls{HOM} denote the set of all maps $\homm\colon \Pa \rightarrow \AF$ such that for $\gamma\in \Pa$ with $\dom[\gamma]=[a,b]$ we have:
    \begin{enumerate}
      \itspace
    \item[\textrm{(a)}]
      $\homm(\gamma)\in \Iso(\gamma(a),\gamma(b))$
      and $\homm(\gamma)=\id_{F_{\gamma(a)}}$ if $\gamma$ is constant,
    \item[\textrm{(b)}]
      $\homm(\gamma)= \homm(\gamma_{k-1})\cp \dots\cp \homm(\gamma_{0})$ if $\{\gamma_i\}_{0\leq i\leq k-1}\subseteq \Pa$ is a decomposition of $\gamma$,
    \item[\textrm{(c)}]
      $\homm(\gamma^{-1})=\homm(\gamma)^{-1}$,
    \item[\textrm{(d)}]
      $\homm(\gamma)=\homm(\gamma')$ if 
      $\gamma\csim \gamma'$.
    \end{enumerate}
  \end{itemize}
  \endgroup
\end{definition}
\begin{remark}
\label{rem:homomorphbed}
\begin{enumerate}
\item
Obviously, $\homm\in  \HOM$ mimics the algebraic properties of the map $\gamma\mapsto \parall{\gamma}{\w}$ for $\w\in \Con$. Consequently, $\Con$ can canonically be identified with a subset of $\HOM$. Then, one may ask whether there is a natural topology on $\HOM$ for which this subset is even dense.\footnote{This means that for each $\homm\in \HOM$ we find a net $\{\w_\alpha\}_{\alpha\in I}\subseteq \Con$ with $\homm(\gamma)=\lim_\alpha \parall{\gamma}{\w_\alpha}$ for each $\gamma\in \Pa$ in a reasonable sense.} We will see that for compact $S$ the above identification of $\Con$ with a subset of $\HOM$ extends to the spectrum $\A$ in a canonical way, i.e., $\A$ can be identified with a subset of $\HOM$ as well. If the set $\Pa$ has the additional property of independence, see Definition \ref{def:indepref}, this identification even turns out to be bijective, so that in this case $\A\cong \HOM$ holds. Hence, carrying over the topology on $\A$ to $\HOM$, $\Con$ can be considered as a dense subset of $\HOM$ just because $ \iota_\Con(\Con)$ is dense in $\A$. A concrete description of this topology is given in    
Definition \ref{def:indepref}.\ref{conv:muetc}.
\item
\label{rem:euklrem5}
Our definition of $\HOM$ differs from the traditional one \cite{Ashtekar2008} in the following two points:

	{\bf Decompositions instead of concatenations:}
	
	\noindent
 We require $\homm$ to be compatible w.r.t.\ decompositions and not w.r.t.\  concatenations of paths. This avoids unnecessary technicalities as, in the following sections, it will allow us to restrict to embedded analytic curves instead of considering all the piecewise ones. For this observe that both sets give rise to the same $\Cstar$-algebra of cylindrical functions. Hence, define the same quantum configuration space $\A$. 
  In particular, we get rid of unnecessary redundancies such as the occurrence of curves 
$\gamma\colon [0,t]\rightarrow M$ with\footnote{In other words, $\gamma$ is holonomical equivalent to a concatenation of a curve with its inverse.} 
\begin{align}
\label{eq:stichstr}  
  \gamma|_{[s,t]}\csim \big[\gamma|_{[0,s]}\big]^{-1}\text{ for some }s\in (0,t),
\end{align}  
   i.e., of curves 
   holonomy equivalent to the constant curve $[0,1]\ni t\mapsto \gamma(0)$. Without going to much into detail, we state that 
   if $\Pa$ consists of embedded analytic curves and $S$ is connected, then (cf.\ Lemma \ref{lemma:BasicAnalyt}.\ref{lemma:BasicAnalyt2}): 
   
   $\gamma_1 \csim \gamma_2$ for $\gamma_1,\gamma_2\in \Pa$ with $\dom[\gamma_i]=[a_i,b_i]$ for $i=1,2$
   {\bf iff} $\gamma_1\isim \gamma_2$, i.e., 
\begin{align*}  
 	 \im[\gamma_1]=\im[\gamma_2]\qquad \text{and}\qquad \gamma_1(a_1)=\gamma_2(a_2),\: \gamma_1(b_1)=\gamma_2(b_2) 
\end{align*}
   {\bf iff} 
  we find an analytic diffeomorphism $\adif\colon \dom[\gamma_1] \rightarrow \dom[\gamma_2]$ with $\gamma_1=\gamma_2\cp \rho$ (we write $\gamma_1 \text{\gls{PSIM}} \gamma_2$ in this case). 
   
In the piecewise embedded analytic situation things are more complicated as there $\csim$ and $\isim$ cannot longer coincide.
 Indeed, let $\delta \colon [0,1]\rightarrow M$ be piecewise embedded analytic with $\delta(0)=\delta(1)$ and $\parall{\delta}{\w} \neq \id_{F_{\delta(0)}}$ for some $\w\in \Con$.\footnote{Apply, e.g., Proposition A.1 in \cite{ParallTranspInWebs} in order to construct such a connection.}
\vspace{-25pt} 
 
   \hspace{140pt} 
	\begin{tikzpicture}
	\draw[->,line width=1pt] (0,0) .. controls (0,2.5) and (-2.5,0.5) .. (-0.1,0);
	\draw (-1.5,1) node {\(\delta\)};
	\draw[->,line width=1pt] (0,0) .. controls (0,2.5) and (2.5,0.5) .. (0.1,0);
		\draw (1.5,1) node {\(\delta'\)};
	\filldraw[black] (0,0) circle (1.7pt);
	\end{tikzpicture}

\vspace{9pt}
  We choose a curve $\gamma\colon[0,2]\rightarrow M$ with $\gamma|_{[0,1]}=[\gamma|_{[1,2]}]^{-1}=\delta$ and obviously have $\gamma\isim \delta$, but $\parall{\gamma}{\w}=\id_{F_{\delta(0)}}\neq \parall{\delta}{\w}$. 
Moreover, even if we first identified piecewise embedded analytic curves which only differ by an insertion or a rejection of a segment $\gamma$ fulfilling \eqref{eq:stichstr}, we also would need non-commutativity of the structure group if we want to replace $\csim$ by some piecewise version of $\psim$. Indeed, let 
$\delta$ and $\delta'$ be piecewise embedded analytic as sketched in the picture above. Then, for $\gamma_1,\gamma_2$ 
piecewise embedded analytic such that $\gamma_1$ first runs through $\delta$, then through $\delta'$, and $\gamma_2$ does it the other way round,
   we cannot even find a continuous map $\adif \colon \dom[\gamma_1]\rightarrow \dom[\gamma_2]$ such that $\gamma_1=\gamma_2\cp \adif$ holds. Consequently, $\gamma_1\nsim_{\mathrm{par}} \gamma_2$, but we have $\gamma_1\csim \gamma_2$ for abelian $S$ because then 
   \begin{align*}   
   	\parall{\gamma_1}{\w}=\parall{\delta}{\w}\cp \parall{\delta'}{\w}=\parall{\delta'}{\w}\cp \parall{\delta}{\w}=\parall{\gamma_2}{\w}.
	\end{align*}   

	\vspace{10pt}
	{\bf Values in $\boldsymbol{\IsoF}$ instead in the structure group $\boldsymbol{S}$:} 
	
	\noindent
	If $\nu=\{\nu_x\}_{x\in M}$ is a choice of elements $\nu_x \in F_x$ as described in the beginning of this subsection, then $\HOM$ can be identified with the set $\TRHOM$ of all maps $\hommm\colon \Pa \rightarrow S$ that fulfil 
    the algebraic properties \textrm{(b)} -- \textrm{(d)} from Definition \ref{def:hompaths}. The corresponding bijection
    \begin{align*} 
    	\Omega_\nu\colon \HOM\rightarrow \TRHOM
	\end{align*}    	
    	 is just given by
    \begin{align}
    \label{eq:ON}
      \Omega_\nu(\homm)(\gamma)=(\psi_{\nu_{\gamma(b)}}\cp  \homm(\gamma))(\nu_{\gamma(a)})\qquad \text{for}\qquad \dom[\gamma]=[a,b].
    \end{align}
    Its inverse is
    \begin{align*}
    	\Omega_\nu^{-1}(\hommm)(\gamma)(p):=\nu_{\gamma(b)}\cdot \hommm(\gamma)\cdot \psi_{\nu_{\gamma(a)}}(p)\qquad \forall\:p\in F_{\gamma(a)}.
    	\end{align*}
    	 
    This identification is especially convenient if $P$ is trivial, i.e., if $P=M\times S$. indeed, then we have the canonical choice $\nu_x:=(x,e)$ for all $x\in M$ whereby $\psi_x=\pr_2$. 
    So, for such a trivial bundle we define
    \begin{align}
    \label{eq:tricBHoms}
      \Omega:=\Omega_\nu\qquad\quad\text{and have}\qquad\quad \TRHOM=\Omega(\HOM),  
    \end{align}
    Hence, $\Omega(\homm)(\gamma)=\pr_2(\homm(\gamma)(\gamma(a),e))$ for $\gamma\in \Pa$ with $\dom[\gamma]=[a,b]$.
    
    In the following, we only refer to the spaces  $\TRHOM$ in specific applications. This is because the general formulas (and the proofs) are
     usually less technical if we work with the spaces $\HOM$.\footnote{The reader who might not believe is encouraged to compare the formulas \eqref{eq:kappadef} and \eqref{eq:algpro}. Moreover, he might write down \eqref{eq:invprop} (and the respective proof) in terms of the space $\Hom(\Pa,S)$ for an arbitrary choice $\{\nu_x\}_{x\in M}$, and then does the same for the projection maps $\pi_p$ we will introduce in Lemma and Definition \ref{def:topo}.\ref{def:topo3}.} In addition to that, we do not need to refer to any choice $\{\nu_x\}_{x\in M}$ in this case.    
\hspace*{\fill}$\lozenge$
\end{enumerate}
\end{remark}
We now come to the definitions we need to identify the spaces $\A$ and $\HOM$.
\begin{definition}
  \label{def:indepref}
  \begin{enumerate}
  \item
   \label{conv:muetc}
   We define the topology $\TOHO$ on $\HOM$ to be generated by the sets of the form 
    \begin{align}   
    \label{eq:opensets}  
    U^{p_1,\dots,p_k}_{\gamma_1,\dots,\gamma_k}(\homm):=\left\{\homm'\in \HOM\:|\: \homm'(\gamma_i)(p_i)\in  \homm(\gamma_i)(p_i)\cdot U\quad\forall \: 1\leq i\leq k \right\}. 
  \end{align}
  Here $U$ denotes a neighbourhood of $e\in S$, $\homm\in \HOM$, $k\in \mathbb{N}_{>0}$ and $\Pa \ni \gamma_i\colon [a_i,b_i] \rightarrow M$, $p_i \in F_{\gamma_i(a_i)}$ for $1\leq i\leq k$. 
  \item
  \label{refnbff}
  We define independence of a set of curves $\Pa$ as follows.
  \begingroup
  \setlength{\leftmarginii}{14pt}
  \begin{itemize}
  \item
  \vspace{-8pt}
    A refinement of a finite subset $\{\gamma_1,\dots,\gamma_l\}\subseteq \Pa$ is a finite collection $\{\delta_1,\dots,\delta_n\}\subseteq \Pa$ such that for each path $\gamma_j$ we find a decomposition $\{(\gamma_j)_i\}_{1\leq i\leq k_j}$ such that each subcurve $(\gamma_j)_i$ is equivalent to one of the paths $\delta_r$ or $\delta_r^{-1}$ for $1\leq r\leq n$.
  \item
    \itspacecc
    A family $\{\delta_1,\dots,\delta_n\}\subseteq \Pa$ is said to be independent\footnote{Our notion of independence coincides with that of weakly independence introduced in \cite{Ashtekar2008}.} iff for each collection $\{s_1,\dots,s_n\}\subseteq S$  
    there is $\w\in \Con$ such that (recall \eqref{eq:deltadef}) $h^\nu_{\delta_i}(\w)= s_i$ for all $1\leq i\leq n$. Due to \eqref{eq:indep} this definition does not depend on the explicit choice of $\nu$.
  \item
    \itspacecc
    $\Pa$ is said to be independent iff each finite collection $\{\gamma_1,\dots,\gamma_l\}\subseteq \Pa$
    admits an independent refinement.\footnote{In particular, $\Pa$ cannot contain constant curves.}
  \end{itemize}
  \endgroup
   \item
  \label{def:mapkappa}
  	Let $S$ be compact. For $\qw\in \A$ and $\gamma\in \Pa$ with $\dom[\gamma]=[a,b]$ we let
	\begin{align}  
		\label{eq:dfdcccccf}
  		\parall{\gamma}{\qw}(p):=\lim_\alpha \parall{\gamma}{\w_\alpha}(p) \qquad \forall\: p\in F_{\gamma(a)}
	\end{align}   	
  	for $\Con\supseteq \{\w_\alpha\}_{\alpha\in I}\rightarrow \qw$ an approximating net, and define 
  \begin{align}
    \begin{split}
      \label{eq:kappadef}
      \text{\gls{KAPPA}}\colon \ovl{\Con} &\rightarrow \HOM\\
      \ovl{\w}&\mapsto \left[\gamma \mapsto  \parall{\gamma}{\qw}\right].
    \end{split}
  \end{align}			
  \end{enumerate}
\end{definition}
\begin{remark}
Obviously, \eqref{eq:dfdcccccf} just means to define the generalized parallel transport function that corresponds to $\ovl{\w}$ as a limit of the classical parallel transports w.r.t.\ the elements $\w_\alpha$ approximating $\ovl{\w}$. This is the coordinate free description of the map $\kappa$. A more concrete formula involving a choice of $\{\nu_x\}_{x\in M}$ as well as a choice of a faithful matrix representation $\rho$ of $S$ is provided in the next lemma adapting the standard results \cite{Ashtekar2008} to our framework and shows that \eqref{eq:kappadef}, hence \eqref{eq:dfdcccccf} is well defined. 
\end{remark}
\begin{lemma}
\label{lemma:homeotopo}
Let $S$ be compact and $\rho$ a faithful matrix representation of $S$. Moreover, let $\{\nu_x\}_{x\in M}\subseteq P$ be a family of elements with $\nu_x\in F_x$ for all $x\in M$. 
\begin{enumerate}
	\item
	\label{lemma:SpeczuHommm}
	The map \eqref{eq:kappadef} is well defined and injective. It is surjective if $\Pa$ is independent. Moreover, the following formula holds for $\gamma\in \Pa$ with $\dom[\gamma]=[a,b]$ and $[h_\gamma^\nu]_{ij}=\rho_{ij}\cp h_\gamma^\nu$ as in \eqref{eq:gniij}:
	\begin{align}      
      \label{eq:algpro}
        \kappa(\ovl{\w})(\gamma)(p)=\nu_{\gamma(b)}\cdot \rho^{-1}\left(\left(\ovl{\w}\left([h^\nu_\gamma]_{ij}\right)\right)_{ij}\right)\cdot \psi_{\gamma(a)}(p)\qquad \forall \: p\in F_{\gamma(a)}.
    \end{align}
	\item
	\label{it:ddd}
	If $\Pa$ is independent, then $\TOHO$ is the unique topology making $\kappa$ a homeomorphism.
	\end{enumerate}
\end{lemma}
  \begin{proof}
  \begin{enumerate}
  \item
    $\rho$ is a homeomorphism to $B:=\im[\rho]\subseteq \GLNC$ since $S$ is compact and $B$ is Hausdorff. 
    Then, by compactness of $B$ we have  
    \begin{align*}
    \lim_\alpha\big(\iota_\Con(w_\alpha)([h^\nu_\gamma]_{ij})\big)_{ij}=(\ovl{\w}([h^\nu_\gamma]_{ij})_{ij}\in B 
    \end{align*}
    For this, observe that $I$ is directed and that by the definition of the Gelfand topology on $\A$ for each $\epsilon>0$ and all $1\leq i,j\leq n$ we find $\alpha_{ij}\in I$ with 
    \begin{align*}
         \big\|\qw\left([h^\nu_\gamma]_{ij}\right)-[h^\nu_\gamma]_{ij}(\w_\alpha)\big\|<\epsilon\qquad \forall\: \alpha\geq  \alpha_{ij}.
    \end{align*} 
    So, using continuity of $\rho^{-1}$ and that of the right multiplication in the bundle $P$, we obtain
    \begin{align*}      
      \begin{split}
        \kappa(\ovl{\w})(p)&=
        \parall{\gamma}{\qw}(p)=\lim_\alpha \parall{\gamma}{\w_\alpha}(p)\\
       & =\lim_\alpha \nu_{\gamma(b)}\cdot \psi_{\gamma(b)}\left(\parall{\gamma}{\w_\alpha}(\nu_{\gamma(a)})\right)\cdot \psi_{\gamma(a)}(p)\\  
        &=\lim_\alpha \nu_{\gamma(b)}\cdot h_\gamma^\nu(\w_\alpha)\cdot \psi_{\gamma(a)}(p)\\
        &=\lim_\alpha\nu_{\gamma(b)}\cdot \rho^{-1}\left(\left(\iota_\Con(\w_\alpha)\left([h^\nu_\gamma]_{ij}\right)\right)_{ij}\right)\cdot \psi_{\gamma(a)}(p)\\
        &=\nu_{\gamma(b)}\cdot \rho^{-1}\left(\left(\ovl{\w}\left([h^\nu_\gamma]_{ij}\right)\right)_{ij}\right)\cdot \psi_{\gamma(a)}(p),
      \end{split}
    \end{align*}
    hence \eqref{eq:algpro}. 
    In particular, this shows that the limit $\lim_\alpha \parall{\gamma}{\w_\alpha}(p)$ exists and is independent of the choice of the net $\{\w_\alpha\}_{\alpha\in I}$. 
    Using the same continuity arguments, we conclude that $\kappa(\ovl{\w})\in \HOM$ from $\kappa\left(\im[\iota_\Con]\right)\subseteq \HOM$. 
    Let $\dD\subseteq \PaC$ denote the dense unital $^*$-subalgebra of $\PaC$ generated by the functions $[h^\nu_\gamma]_{ij}$.  
    If $\kappa(\ovl{\w}_1)=\kappa(\ovl{\w}_2)$ for $\ovl{\w}_1,\ovl{\w}_2\in \Spec(\PaC)$, then $\ovl{\w}_1|_\dD=\ovl{\w}_2|_\dD$ by \eqref{eq:algpro}, hence $\ovl{\w}_1=\ovl{\w}_2$ by continuity of this maps. This shows injectivity of $\kappa$.

   Now, if $\Pa$ is independent and $\homm\in \HOM$, we define the corresponding preimage $\qw_\homm\in \A$ as follows. Let $[h^\nu_\gamma]_{ij}$ denote the complex conjugate of the function $[h^\nu_\gamma]_{ij}$ and define
	\begin{align*}    
    	\qw_\homm(1):=1\qquad\quad
    \qw_\homm([h^\nu_\gamma]_{ij}):=\rho_{ij}\cp \psi_{\nu_{\gamma(b)}}\cp\homm(\gamma)(\nu_{\gamma(a)}) \quad\qquad \qw_\homm([h^\nu_\gamma]^*_{ij}):=\ovl{\qw_\homm([h^\nu_{\gamma}]_{ji})}. 
    \end{align*}
    For $f\in \dD$ choose a representation as a sum of products of the form 
	\begin{align*}    
    	F=\sum_l \lambda_l\cdot h_{l,1}\cdot {\dots}\cdot h_{l,n(l)}
    \end{align*} 
    where $\lambda_l \in \mathbb{C}$ and each
    $h^l_{k}$ equals $1$, a generator $[h^\nu_\gamma]_{ij}$ or the complex conjugate $[h^\nu_\gamma]^*_{ij}$ of a generator. We assign to $f$ the value
	\begin{align*}    
    	\qw_\homm(F):=\sum_l\lambda_l\cdot \qw_\homm(h_{l,1})\cdot {\dots}\cdot \qw_\homm(h_{l,n(l)}).
    \end{align*}
    Obviously, $\qw_\homm$ is a $^*$-homomorphism if it is well defined and, it extends (by linearity) to an element of $\Spec(\PaC)$ if it is continuous. For well-definedness assume that $F_1$,$F_2$ are two representations of $f$,
    and denote by $[\gamma_1],\dots, [\gamma_m]$ the equivalence (holonomy equivalence) classes of the paths that occur in both expressions. 
    Let $\{\delta_1,\dots,\delta_n\}\subseteq \Pa$ be a refinement
    of $\{\gamma_1,\dots,\gamma_m\}$ and $\w\in\Con$ with\footnote{For simplicity, assume that $\dom[\delta_r]=[0,1]$ for $1\leq r\leq n$.} $h^\nu_{\delta_r}(\w)=\psi_{\nu_{\delta_r(1)}}\homm(\delta_r)(\nu_{\delta_r(0)})$ for $1\leq r\leq m$. Then it follows from the algebraic properties of parallel transports and $\homm$ that 
    \begin{align*}    
      \qw_\homm(F_1)=F_1(\w)=f(\w)=F_2(\w)=\qw_\homm(F_2).
    \end{align*}    	
    This shows well-definedness, and continuity now follows from    
    	$|\qw_\homm(f)|=|f(\w)|\leq \|f\|_\infty$.
    By construction we have $\kappa(\qw_\homm)=\homm$, which shows surjectivity of $\kappa$.
    \item
	The space $\HOM$ equipped with $\TOHO$ is obviously Hausdorff. So, since $\A$ is compact and $\kappa$ is bijective, we only have to show that $\kappa$ is continuous. For this, let $\A \supseteq\{\ovl{\w}_\alpha\}_{\alpha\in I}\rightarrow \qw \in\A$ be a converging net, $\Pa\ni\gamma\colon [a,b]\rightarrow M$, $p\in F_{\gamma(a)}$ and $U\subseteq S$ a neighbourhood of $e\in S$. Then, by \eqref{eq:algpro} we have 
	 \begin{align*}
		F_{\gamma(b)}\ni \kappa(\ovl{\w}_\alpha)(\gamma)(p) &\stackrel{\eqref{eq:algpro}}{=} \nu_{\gamma(b)}\cdot \rho^{-1}\left(\left(\ovl{\w}_\alpha\left([h^\nu_\gamma]_{ij}\right)\right)_{ij}\right)\cdot \psi_{\gamma(a)}(p)\\
		& \hspace{3.2pt}\rightarrow \:\:\nu_{\gamma(b)}\cdot \rho^{-1}\left(\left(\ovl{\w}\left([h^\nu_\gamma]_{ij}\right)\right)_{ij}\right)\cdot \psi_{\gamma(a)}(p)
		= \kappa(\ovl{\w})(\gamma)(p)\in F_{\gamma(b)},
	\end{align*}
	so that $\{\kappa(\ovl{\w}_\alpha)(\gamma)(p)\}_{\alpha\in I}$ converges in $F_{\gamma(b)}$ to $\kappa(\ovl{\w})(\gamma)(p)\in F_{\gamma(b)}$. Consequently, we find $\alpha_0 \in I$ such that
	\begin{align*}
		\kappa(\ovl{\w}_\alpha)(\gamma)(p)\in \kappa(\ovl{\w})(\gamma)(p)\cdot U\qquad \forall\: \alpha\geq \alpha_0, 
	\end{align*}	
	hence $\kappa(\ovl{\w}_\alpha)\in U_{\gamma}^p(\kappa(\ovl{\w}))$ for all $\alpha\geq \alpha_0$. 
	Since $U_{\gamma_1,\dots,\gamma_k}^{p_1,\dots,p_k}(\homm)=U_{\gamma_1}^{p_1}(\homm)\cap \dots \cap U_{\gamma_k}^{p_k}(\homm)$ and $I$ is directed, the claim follows.
    \end{enumerate}
\end{proof}

\subsubsection{Invariant Homomorphisms}
\label{subsub:InvHoms}
We now use the identification of $\A$ with $\HOM$ in order to obtain a more concrete description of the space $\AQR$. 
\begin{definition}[Invariant Homomorphisms]
\label{def:invhomm}
Let $\Pa$ be $\Phi$-invariant and independent.
We define the space of $\Phi$-invariant homomorphisms by  
$\text{\gls{IHOM}}:=\kappa\big(\AQR\big)$ and equip it with the subspace topology w.r.t.\ $\HOM$.\footnote{Since $\kappa$ is a homeomorphisms, this is just the topology carried over from $\AQR$ by $\kappa$, cf. Remark \ref{rem:subspt}.}
\end{definition}
\begin{lemma}
  \label{lemma:ShapeInvHomms}
  Let $\Pa$ be $\Phi$-invariant and independent and $\homm\in \HOM$.
  Then $\homm\in \IHOM$ iff 
  \begin{align}
    \label{eq:invprop}
    \homm(\wm_g\cp \gamma)(\Phi_g(p))=\Phi_g( \homm(\gamma)(p))\qquad \forall\: g\in G,\forall\:\gamma\in \Pa,\forall\:p\in F_{\gamma(a)},\;
  \end{align}
  where $\dom[\gamma]=[a,b]$. 
  \begin{proof}
  Observe that $\homm\in \IHOM$ iff $\kappa(\Theta_g(\kappa^{-1}(\homm)))=\homm$ 
   for all $g\in G$, i.e., iff
	\begin{align*}  
		\kappa(\Theta_g(\kappa^{-1}(\homm)))(\gamma)(p)=\homm(\gamma)(p)\qquad \forall\: \gamma\in \Pa,\:\forall\:g\in G.
  	\end{align*}  
    So, let $\{\w_\alpha\}_{\alpha\in I}\subseteq \Con$ be a net with $\{\iota_\Con(\w_\alpha)\}_{\alpha\in I}\rightarrow \kappa^{-1}(\homm)$.
    Then from
    \begingroup
    \setlength{\leftmargini}{20pt}
    \begin{itemize}
    \item[a)]
      continuity of $\specw_g$ by Corollary \ref{cor:CylSpecAction},
    \item[b)]
      $\specw_g\cp \iota_\Con = \iota_\Con \cp \cw_g$ by Corollary \ref{cor:CylSpecAction},
    \item[c)]
     $\parall{\gamma}{\cw(g,\w)}(p)=\Phi_g \!\left(\parall{g^{-1}\gamma}{\w}(\Phi_{g^{-1}}(p))\right)$ for $g^{-1}\gamma:=\wm_{g^{-1}}\cp \gamma$,\footnote{See, e.g., \eqref{eq:patra} in the proof of Lemma \ref{lemma:CylSpecActionvorb}.}  
    \item[d)]
      continuity of $\Phi$ and
    \item[e)]
      the definition of $\kappa$
    \end{itemize}
    \endgroup 
    \noindent
    we obtain  
    that 
    \begin{align}
      \label{eq:InvHomTau}
      \begin{split}
        \kappa\big((\specw_g\cp \kappa^{-1})(\homm) \big)(\gamma)(p) &\stackrel{\mathrm{a)}}{=}\kappa\left(\lim_\alpha \big(\Theta_g\cp \iota_\Con\big)(\w_\alpha)\right)(\gamma)(p)\\%\nonumber\\
        &\stackrel{\mathrm{b)}}{=}\kappa\left(\lim_\alpha (\iota_\Con\cp\cw_g)(\w_\alpha)\right)(\gamma)(p)\stackrel{\mathrm{e)}}{=}\lim_\alpha \parall{\gamma}{\cw_g(\w_\alpha)}(p)\\
        &\stackrel{\mathrm{c)}}{=}\lim_\alpha \Phi_g\!\left(\parall{g^{-1}\gamma}{\w_\alpha}\left(\Phi_{g^{-1}}(p)\right)\right)\stackrel{\mathrm{d)}}{=}\Phi_g\!\left(\lim_\alpha\parall{g^{-1}\gamma}{\w_\alpha}\left(\Phi_{g^{-1}}(p)\right)\right)\\%\nonumber\\
        &\stackrel{\mathrm{e)}}{=}\Phi_g\Big(\homm\!\left(g^{-1}\gamma\right)\left(\Phi_{g^{-1}}(p)\right)\!\Big)
      \end{split}
    \end{align}
    Replacing $g$ by $g^{-1}$, we see that $\homm\in \IHOM$ iff for all $g\in G$, $\gamma\in \Pa$ and all $p\in F_{\gamma(a)}$ we have $\Phi_g(\homm(\gamma)(p))=\homm(\wm_g\cp \gamma)(\Phi_g(p))$.
  \end{proof}
\end{lemma}
In the following let $\Pa$ be $\Phi$-invariant and independent.
\begin{remark}[Invariance up to Gauge Transformations]
\label{rem:homaction}
    \begin{enumerate}
    \item
    \label{rem:euklrem4}
    \vspace{6pt}
    Instead of $\AQR\cong\IHOM$, one also can consider the space $\RedGauge$ of generalized connections that are invariant up to gauge transformations. This is the set of elements $\homm\in \HOM$ which fulfil that for each $g\in G$ there is a generalized gauge transformation\footnote{This just means that $\pi \cp\sigma =\pi$ and $\sigma(p\cdot s)=\sigma(p)\cdot s$ for all $\in S$.} $\sigma\colon P\rightarrow P$ with $\specw_g(\homm)=\sigma^*(\homm)$, i.e., 
    \begin{align*}
      \specw_g(\homm)(\gamma)(p)=(\sigma \cp \homm(\gamma))(\sigma^{-1}(p))\qquad \forall\: \gamma\in \Pa, \forall\:p\in F_{\gamma(a)}
    \end{align*}
    with $\dom[\gamma]=[a,b]$. Obviously, $\AQR\cong \IHOM\subseteq \RedGauge$, and usually $\RedGauge$ would be seen as the more physical space. However, in this thesis we want to concentrate on the spaces $\AQR$ and $\ARQ$ just because we rather expect technical than conceptual difficulties when carrying over the current developments to the ``up to gauge'' case.
     \item 
    \label{rem:euklrem55}
In Remark \ref{rem:homomorphbed}.\ref{rem:euklrem5} we have seen that each family $\nu=\{\nu_x\}_{x\in M}\subseteq P$ of elements with $\nu_x\in F_x$ for all $x\in M$ allows to identify the spaces $\HOM$ and $\TRHOM$ via the bijection 
	\begin{align*}    
    	\Omega_\nu\colon \HOM\rightarrow \TRHOM 
	\end{align*} 
	defined by \eqref{eq:ON}.   	
	Then, composing $\Omega_\nu$ with the bijection $\kappa \colon \A\rightarrow \HOM$ from Definition \ref{def:indepref}.\ref{def:mapkappa}, we obtain from \eqref{eq:ON} and \eqref{eq:algpro} that
    \begin{align}
      \label{eq:hilfseq}
      \begin{split}
        (\Omega_\nu\cp \kappa)(\ovl{\w})(\gamma)&=\rho^{-1}\left(\left(\ovl{\w}([h_\gamma^\nu]_{ij})\right)_{ij}\right)
      \end{split}
    \end{align}    
    for $\rho$ is a faithful matrix representation of $S$. 
	\item    
    For $P$ a trivial principal fibre bundle we extend \eqref{eq:tricBHoms} to
 \begin{align*}
      \Omega:=\Omega_\nu\quad\:\: \TRHOM=\Omega(\HOM) \quad\:\: \ITRHOM:=\Omega(\IHOM) 
    \end{align*}
    for $\nu$ the canonical choice $\nu_x=(x,e)$ for all $x\in M$. Recall that then (see Remark \ref{rem:homomorphbed}.\ref{rem:euklrem5})
	\begin{align}
	\label{eq:trivp}    
    	\Omega(\homm)(\gamma)=\pr_2(\homm(\gamma)(\gamma(a),e))\qquad\text{for $\gamma\in \Pa$ with $\dom[\gamma]=[a,b]$ holds}.
    \end{align}
   The spaces $\ITRHOM$ will only occur in concrete examples in the following.\hspace*{\fill}$\lozenge$
    \end{enumerate}
\end{remark}
    Fixing $\nu$, one can always try to evaluate the invariance condition  \eqref{eq:invprop} in terms of the spaces $\TRHOM$. However, if $\Phi$ is too complicated, even an appropriate choice of $\nu$ will give rise to cumbersome conditions. This is one of the reasons why we prefer to deal with the space $\HOM$ instead of $\TRHOM$ in this work. 
 	Nevertheless, the next example shows that in our LQC prime examples (Example \ref{ex:LQC}) the action $\Phi$ is simple enough to give rise to very natural invariance conditions in terms of the space $\Hom(\Pa,\SU)$. Basically, this is the case because for each of the groups $\Ge$, $\Gi$, $\Gh$ we have that 
 	$\pr_2(g\cdot (x,\me))$ does not depend on the base point $x$, i.e., the effect of a group element on the fibres is constant in a certain sense. 
\begin{example}[Loop Quantum Cosmology]
    \label{rem:euklrem6}
    Let $P=\RR^3\times \SU$, and $G:=\Ge$, $\Phi:=\Pe$ be defined as in Example \ref{ex:LQC}. Moreover, assume that $\Pa$ is independent, $\Pe$-invariant and closed under decomposition and inversion of its elements.
       \begingroup
	\setlength{\leftmargini}{20pt} 
    \begin{itemize}
    \item
    \label{rem:euklrem61}
    We have $g^{-1}=(-\uberll{\sigma^{-1}}{v},\sigma^{-1})$ for $g=(v,\sigma)\in \Ge$. 	
   Moreover, if $\gamma\in \Pa$ with $\dom[\gamma]=[a,b]$ and $\homm\in \IHOM$, then \eqref{eq:invprop} and \eqref{eq:trivp} give
	\begin{align*}    
    	\Omega(\homm)(\wm_g\cp\gamma)&  \stackrel{\eqref{eq:trivp}}{=} \pr_2(\homm(\wm_g\cp \gamma)((\wm_g(\gamma(a)),\me)))\stackrel{\eqref{eq:invprop}}{=}(\pr_2\cp\Phi_g)\big(\homm(\gamma)(\Phi_{g^{-1}}((\wm_g(\gamma(a)),\me)))\big).
    \end{align*}
   The left hand side equals $\Omega(\homm)(v+\sigma(\gamma))$ and for the right hand side we obtain
	\begin{align*}    
    	(\pr_2\cp\Phi_g)\big(\homm(\gamma)(\Phi_{g^{-1}}((\wm_g(\gamma(a)),\me)))\big)&=(\pr_2\cp\Phi_g)\big(\homm(\gamma)\big(\big(\gamma(a),\sigma^{-1}\big)\big)\\
    	&=\sigma\cdot \pr_2\big(\homm(\gamma)\big(\big(\gamma(a),\sigma^{-1}\big)\big)\\
    	&=\sigma\cdot \pr_2\big(\homm(\gamma)\big(\big(\gamma(a),\me\big)\big)\cdot \sigma^{-1}\\
    	&= \alpha_\sigma(\Omega(\homm)(\gamma)).
    \end{align*}
    It follows that $\hommm \in \Hom_\red(\Pa,\SU)$ iff  
    \begin{align}
      \label{eq:InvGenConnRel}
      \hommm(v+\uberll{\sigma}{\gamma)}=(\Co{\sigma}\cp \hommm)(\gamma)\qquad \forall\:(v,\sigma)\in \Ge, \forall\:\gamma \in \Pa
    \end{align}
    holds.
    In particular, this means that the value of $\hommm$ is independent of the starting point of the path $\gamma$. In the same way, we see that for the spherically symmetric and \mbox{(semi-)homogeneous} cases we have
    \begin{align}
      \begin{array}{lrclcl} 
        \label{eq:algrels}
        \Gi \colon  & \hommm(\uberll{\sigma}{\gamma})\!\!\!&=&\!\!\!\Co{\sigma}\cp \hommm(\gamma) && \forall\:\sigma\in \SU,\forall\:\gamma \in \Pa, \\
        \Gh\colon&  \hommm(v+\gamma)\!\!\! & =&\!\!\! \hommm(\gamma)&& \forall\:v\in \RR^3,\forall\:\gamma \in \Pa \\
        G_{SH}\colon&  \hommm(L(v)+\gamma)\!\!\! & =&\!\!\! \hommm(\gamma)&& \forall\:v\in \RR^2\subseteq \RR^3,\forall\:\gamma \in \Pa 
      \end{array}
    \end{align} 
    for $\Gi$, $\Gh$ and $G_{SH}$ (semi-homogeneous) defined as in Example \ref{ex:LQC}, and $L\colon \RR^2 \rightarrow \RR^3$ an injective linear map. 
    \item
    \label{rem:euklrem62}
	It already follows from \eqref{eq:InvGenConnRel} that for $\Ge$ we have $\hommm(x+\gamma_{\vv,l})\in H_{\vv}$ for all $\hommm\in \Hom_\red(\Pa,\SU)$ and the straight lines 
	$x+ \gamma_{\vv,l}\in \Pal$ defined as in Convention \ref{conv:sutwo1}.\ref{conv:sutwo2}.
	
	In fact, obviously $\sigma(\gamma_{\vv,l})=\gamma_{\vv,l}$ holds for each $\sigma\in H_{\vec{v}}$ just because $\sigma$ rotates around the velocity vector $\vv$ of the linear curve $\gamma_{\vv,l}$. Then, \eqref{eq:InvGenConnRel} shows
	\begin{align*} 
    	\hommm(x+\gamma_{\vv,l})=\hommm(x+\sigma(\gamma_{\vv,l}))\stackrel{\eqref{eq:InvGenConnRel}}{=}\Co{\sigma}\cp \hommm(\gamma_{\vv,l})\stackrel{\eqref{eq:InvGenConnRel}}{=}\Co{\sigma}\cp \hommm(x+\gamma_{\vv,l})\qquad \forall\: \sigma\in H_{\vv},
    \end{align*}
    so that, choosing $\sigma\neq \pm\me$, Lemma \ref{lemma:torus}.\ref{lemma:torus1} implies that $\hommm(x+ \gamma_{\vv,l})\in H_{\vv}$.	
   Obviously, then for $\Gi$ the same statement holds if $x\in \Span_\RR(\vv)$. \hspace*{\fill}$\lozenge$  
   \end{itemize}
   \endgroup
\end{example}
The next remark collects some straightforward consequences of the case where $\Pa$ naturally decomposes into independent, $\Phi$-invariant subsets being closed under decomposition and inversion of their elements. For this, we will need
\begin{convention}
  \label{conv:invhomm}
  If the set of curves \gls{PALPHA} carries the index $\alpha$, then (in reference to Convention \ref{hgamma}) 
  we denote the respective quantum spaces by 
	\begin{align*}  
  	\text{\gls{ARRQALPHA}}\qquad\qquad\quad \text{\gls{ARQALPHA}} \qquad\qquad\quad \text{\gls{AQRALPHA}}\qquad\qquad\quad \text{\gls{AALPHA}}.
  \end{align*}
  Moreover, we denote the corresponding bijection from Definition \ref{def:indepref}.\ref{def:mapkappa} by 
	\begin{align*}   
   	\text{\gls{KAPPAALPHA}}\colon \A_\alpha\rightarrow \text{\gls{HOMALPHA}}\qquad\text{and define}\qquad\text{\gls{IHOMALPHA}}:=\kappa_\alpha\big(\AQRInd{\alpha}\big). 
	\end{align*}
\end{convention}
\begin{remark}[Restricting Homomorphisms]
\label{rem:restriction}
    Let $\Pay \subseteq \Pax$ both be independent, $\Phi$-invariant and closed under decomposition and inversion. Then we have the following commutative diagram: 
      \begin{center}
        \makebox[0pt]{
          \begin{xy}
            \xymatrix{
              &\Spec\big(\raisebox{0pt}{$\ovl{\PaCx|_{\AR}}$}\big)=\hspace{-25pt} &\ARRQInd{0}\ar@{->}[r]^-{\ovl{i^{*}_{\AR}}}_{\cong}\ar@{->}[d]_-{\ovl{\upsilon_{\alpha 0}}} & \ARQInd{0}\ar@{->}[r]^-{\subseteq}& \AQRInd{0} \ar@{->}[d]^-{\res'_{0\alpha}} \ar@{->}[r]^-{\subseteq} &  \AInd{0}\ar@{->}[d]^-{\res'_{0\alpha}}  \ar@{->}[r]^-{\kappa_0}_-{\cong} & \HOMInd{0}\ar@{->}[d]^-{\res_{0\alpha}}\\
              &\Spec\big(\raisebox{0pt}{$\ovl{\PaCy|_{\AR}}$}\big)=\hspace{-25pt}&\ARRQInd{\alpha}\ar@{->}[r]^-{\ovl{i^*_{\AR}}}_{\cong} & \ARQInd{\alpha}\ar@{->}[r]^-{\subseteq}& \AQRInd{\alpha}\ar@{->}[r]^-{\subseteq}&\AInd{\alpha}\ar@{->}[r]^-{\kappa_\alpha}_-{\cong}& \HOMInd{\alpha},
            }
          \end{xy}
        }
      \end{center}
    where we have omitted the indices $0$ and $\alpha$ at the maps $\ovl{i^*_{\AR}}$. We have used the following maps:
	\vspace{5pt}    
    
    \begin{tabular}{rrlll}
		$\upsilon_{\alpha 0}\colon$&\hspace{-12pt}$\ovl{\PaCy|_{\AR}}$ &\hspace{-8pt} $\hookrightarrow$ & \hspace{-8pt} $\ovl{\PaCx|_{\AR}}$ \quad & \hspace{-12pt} \text{--}\:\: The canonical inclusion. \\
		$\ovl{\upsilon_{\alpha0}}\colon$&\hspace{-12pt}$\ARRQInd{0}$ &\hspace{-8pt} $\rightarrow$ & \hspace{-8pt} $\ARRQInd{\alpha}$\quad & \hspace{-12pt} \text{--}\:\: The corresponding map from Lemma \ref{lemma:homzuspec}.\ref{lemma:homzuspec1}. \\
		$\res_{0\alpha}\colon$&\hspace{-12pt}$\homm$ & \hspace{-8pt} $\mapsto$ & \hspace{-8pt} $\homm|_{\Pay}$\quad & \hspace{-12pt} \text{--}\:\: The restriction of the elements of $\HOMInd{0}$ to $\Pay$.
\end{tabular}
	\vspace{5pt}  
	 
\noindent
Moreover, $\res'_{0\alpha}:= \kappa_\alpha^{-1}\cp \res_{0\alpha} \cp \kappa_0$ is  
 $\res'_{0\alpha}$ is continuous because $\kappa_0$ is a homeomorphism by Lemma \ref{lemma:homeotopo}.\ref{it:ddd},  and since $\res_{0\alpha}$ is continuous just by the definition of the topologies on $\HOMInd{0}$ and $\HOMInd{\alpha}$. 
Commutativity of the above diagram now follows from the definitions of $\kappa_0$ and $\kappa_\alpha$, \eqref{eq:inclusionsdiag}, as well as the fact that $\ovl{\upsilon_{\alpha0}}$ just assigns to $\psi\in \ARRQInd{0}$ the restriction $\psi|_{\ovl{\PaCy|_{\AR}}}$. Indeed, the last two points give
	\begin{align*}
	 \iota_{\Con}\cp i_{\AR}\:\stackrel{\star}{=}\:\ovl{i^*_\AR}\cp \iota_{\AR}\qquad\text{and} \qquad \iota^\alpha_{\AR}\:\stackrel{\star\star}{=}\:\ovl{\upsilon_{\alpha0}}\cp \iota^0_{\AR}, 
	\end{align*}
	respectively, and then commutativity is 
easily checked on the dense subset $\iota_{\AR}(\AR)\subseteq \ARRQInd{0}$. For this, let $\Pay\ni \gamma\colon [a,b]\rightarrow M$ and $p\in F_{\gamma(a)}$. Then, for $\w\in \AR$ we have
	\begin{align*} 
    	\Big(\res_{0\alpha}\cp \kappa_0\cp \ovl{i^*_\AR}\Big)&\Big(\iota^0_{\AR}(\w)\Big)(\gamma)(p)\\
    	& \stackrel{\eqref{eq:inclusionsdiag}}{=}\kappa_0\big(\big(\iota^0_\Con\cp i_{\AR}\big)(\w)\big)(\gamma)(p)\stackrel{\eqref{eq:kappadef}}{=}\parall{\gamma}{i_{\AR}(\w)}(p)
	\stackrel{\eqref{eq:kappadef}}{=}\kappa_\alpha\big(\big(\iota^\alpha_\Con\cp i_{\AR}\big)(\w)\big)(\gamma)(p)\\
	& %\hspace{3pt}=\:\hspace{1pt}
	\stackrel{\star}{=}
	\kappa_\alpha\big(\big(\ovl{i^*_\AR}\cp\iota^\alpha_{\AR}\big)(\w)\big)(\gamma)(p)
	\stackrel{\star\star}{=}\kappa_\alpha\big(\big(\ovl{i^*_\AR}\cp \ovl{\upsilon_{\alpha0}} \cp\iota^0_{\AR}\big)(\w)\big)(\gamma)(p)\\
	&\hspace{3pt}=\:\hspace{1pt}
	\big(\kappa_\alpha\cp \ovl{i^*_\AR}\cp \ovl{\upsilon_{\alpha0}} \big)\big(\iota^0_{\AR}(\w)\big)(\gamma)(p).		
%    		
%    		& \hspace{-3.5pt}\stackrel{\eqref{eq:algpro}}{=}\nu_{\gamma(b)}\cdot \rho^{-1}\left(\left(\ovl{i^*_\AR}(\psi)\left([h^\nu_\gamma]_{ij}\right)\right)_{ij}\right)\cdot \psi_{\gamma(a)}(p)\\
%    		& = \nu_{\gamma(b)}\cdot \rho^{-1}\left(\left(\psi\!\left([h^\nu_\gamma]_{ij}\big|_{\AR}\right)\right)_{ij}\right)\cdot \psi_{\gamma(a)}(p)\\
%    		&= \nu_{\gamma(b)}\cdot \rho^{-1}\left(\left(\ovl{\upsilon_{0\alpha}}(\psi)\!\left([h^\nu_\gamma]_{ij}\big|_{\AR}\right)\right)_{ij}\right)\cdot \psi_{\gamma(a)}(p)\\
%    		& \hspace{-3.5pt}\stackrel{\eqref{eq:algproo}}{=}\nu_{\gamma(b)}\cdot \rho^{-1}\left(\left(\ovl{i^*_\AR}(\ovl{\upsilon_{0\alpha}}(\psi))\left([h^\nu_\gamma]_{ij}\right)\right)_{ij}\right)\cdot \psi_{\gamma(a)}(p)\\
%    		%&=\kappa_\alpha\Big(\ovl{i^*_\AR}(\ovl{\upsilon_{0\alpha}}(\psi))\Big)(\gamma)(p)\\
%    		&=\Big( \kappa_\alpha \cp  \ovl{i^*_\AR} \cp \ovl{\upsilon_{0\alpha}}\Big)(\psi)(\gamma)(p).
    \end{align*}
    Here, the superscripts $0$ and $\alpha$ hint to the fact that $\iota_\Con^0$ and $\iota_\Con^\alpha$ map into different spectra. 
 \hspace*{\fill}$\lozenge$
\end{remark}
Remark \ref{rem:restriction} provides us with the following
\begin{proposition}
    \label{rem:euklrem2b}
   	Assume that we have $\Pax=\bigsqcup_{\alpha\in I}\Pa_\alpha$ for non-empty sets $\Pa_\alpha$ which are independent, $\Phi$-invariant and closed under decomposition and inversion. Then, $\IHOMInd{0}\cong \prod_{\alpha\in I}\IHOMInd{\alpha}$  	
%   	$\AQRInd{0}\cong \prod_{\alpha\in I}\AQRInd{\alpha}$ 
   	via the map
   	\begin{align*}
   	 \Xi^0_{I}\colon \IHOMInd{0}&\rightarrow \prod_{\alpha\in I}\IHOMInd{\alpha},\\
   	  \qw &\mapsto \prod_\alpha \res_{0\alpha}(\qw). 
	\end{align*} 
	In particular, if each $\IHOMInd{\alpha}$ carries a normalized Radon measures $\mu_\alpha$, then we obtain a normalized Radon measure $\mu$ on $\IHOMInd{0}$ just by 
\begin{align*}    
    \mu:=\big(\Xi^0_{I}\big)^{-1}\!\left(\textstyle\prod_{\alpha\in I}\mu_\alpha\right)
\end{align*}    
     for $\prod_{\alpha\in I}\mu_\alpha$ the Radon product measure on $\prod_{\alpha\in I}\IHOMInd{\alpha}$ from Lemma and Definition \ref{def:ProductMa}.\ref{def:ProductMa3}.
\end{proposition}	
\begin{proof}  	 
   	 The map $\Xi^0_{I}$ is obviously continuous and injective. Moreover, it is surjective because if $\homm_\alpha\in \IHOMInd{\alpha}$ for all $\alpha\in I$, 
    then $\homm(\gamma):= \homm_\alpha(\gamma)$ for $\gamma\in \Pa_\alpha$
    is a well-defined element of $\IHOMInd{0}$, just by the properties of the sets $\Pa_\alpha$. 
\end{proof}
We close this section with some investigations concerning  
the measure theoretical aspects of the reduced spaces $\ARQ$ and $\AQR$. 
The final remark then contains an outlook of the next sections. 

Assume that $M$ is analytic, $\wm$ is an analytic action, $S$ is compact and connected and \gls{PAW} the set of embedded analytic curves in $M$. Moreover, assume that $\Pa\subseteq \Paw$ is $\Phi$-invariant\footnote{Since $\wm$ is analytic, one can always choose $\Pa=\Paw$.} and closed under decomposition and inversion of its elements. As we will see in Lemma \ref{lemma:BasicAnalyt}.\ref{lemma:BasicAnalyt3}, then $\Pa$ is independent. We recall the 
Ashtekar-Lewandowski measure\footnote{This will be a special case of the construction in Subsection \ref{sec:FreeM}, see  Definition \ref{def:AshLewx}.} \cite{ProjTechAL} \gls{mAL} 
on $\A$  
being characterized by the following property:
\vspace{5pt} 

\noindent
Let $\alpha=(\gamma_1,\dots,\gamma_k)$ for $\gamma_1,\dots,\gamma_k \in  \Pa_\w$ be a finite subset such that\footnote{We write $\gamma \text{\gls{CPSIM}} \gamma'$ iff there are open intervals $I\subseteq \dom[\gamma]$ and $I'\subseteq \dom[\gamma']$ such that $\gamma(I)=\gamma'(I')$ holds, see Definition \ref{def:analytLieAlgBD}.\ref{def:analytLieAlgBD1}.} 
% \begin{align*}
$\gamma_i \nsim_\cp \gamma_j$ holds for all $1\leq i\neq j\leq k$, and where 
$\dom[\gamma_i]=[a_i,b_i]$ for $i=1,\dots,k$.
Then, the push forward of $\mAL$ by the map 
\begin{align}
\label{projm}
\begin{split}	
\pi_\alpha\colon \A&\rightarrow S^k\\
  \ovl{\w}&\mapsto \big(\psi_{\gamma_1(b_1)}\big(\kappa(\ovl{\w})(\gamma_1)(\nu_{\gamma_1(a_1)})\big),\dots,\psi_{\gamma_k(b_k)}\big(\kappa(\ovl{\w})(\gamma_k)(\nu_{\gamma_k(a_k)})\big)\big)
  \end{split}
\end{align}
equals the Haar measure on $S^{|\alpha|}$.
\vspace{5pt} 

\noindent
A closer look at the invariance property \eqref{eq:invprop} gives
\begin{lemma}
  \label{cor:AshtLew}
  If $\dim[S]\geq 1$ and $\gamma\nsim_\cp \wm_g\cp\gamma$ holds for some $\gamma\in \Paw$ and $g\in G$, then the Borel sets $\AQR, \ARQ\subseteq \A$ are of measure zero w.r.t. $\mAL$. 
  \begin{proof}
    The subsets $\AQR, \ARQ$ are Borel sets as they are compact by Corollary \ref{cor:CylSpecAction}. Since by the same Corollary we have $\ARQ\subseteq \AQR$, it suffices to show that $\mAL(\AQR)=0$. Now,
    \begin{align*}
      \mAL\!\left(\AQR\right)\leq \mAL\!\left(\pi_\alpha^{-1}\big(\pi_\alpha\big(\AQR\big)\big)\right)=\mu_\alpha(\pi_\alpha\big(\AQR\big))\qquad \forall\:\alpha\in I,
    \end{align*}    
    and for $\alpha:=(\gamma,\wm_g\cp \gamma)$ we find $d,b\in S$ such that $B:=\pi_\alpha\big(\AQR\big)=\{(s,c\cdot s\cdot d)\:|\: s\in S\}$ holds. 
    
    In fact, for $\qw\in \AQR$ and $\homm:=\kappa(\qw)$ we have
    \begin{align}
      \label{eq:invpi}
      \begin{split}
    	\pi_{\wm_g\cp \gamma}(\qw)&=\psi_{\wm_g(\gamma(b))}\big(\homm(\wm_g\cp\gamma)(\nu_{\wm_g(\gamma(a))})\big)\stackrel{\eqref{eq:invprop}}{=}\psi_{\wm_g(\gamma(b))}\big(\Phi_g\cp \homm(\gamma)\big(\Phi_{g^{-1}}(\nu_{\wm_g(\gamma(a))})\big)\big)\\
    	&=\underbrace{\psi_{\wm_g(\gamma(b))}(\Phi_g(\nu_{\gamma(b)}))}_{c}\: \cdot\: \psi_{\nu_{\gamma(b)}}(\homm(\gamma)(\nu_{\gamma(a)}))\cdot \underbrace{\psi_{\gamma(a)}\big(\Phi_{g^{-1}}(\nu_{\wm_g(\gamma(a))})\big)}_{d},
      \end{split}
    \end{align}
    which equals $c \cdot \pi_\gamma(\qw)\cdot d$.
    
    Now, since $\dim[S]\geq 1$, we find $\{s_n\}_{n\in\NN}\subseteq S$ with $s_n\neq s_m$ for $n\neq m$, hence 
	\begin{align*}
	B\cdot (e,s_n) \cap B\cdot (e,s_m)=\emptyset\qquad  \text{ for } n\neq m.
	\end{align*}	    
     Then, if $\mu_\alpha(B)> 0$, $\sigma$-additivity and translation invariance of $\mu_\alpha$ would imply that
	\begin{align*}
	 \mu_\alpha(S\times S)\geq \sum_{n\in \NN}\mu_\alpha(B\cdot (e,s_n))=\sum_{n\in \NN}\mu_\alpha(B)=\infty
	\end{align*}	  
	holds. This, however, contradicts that $\mu_\alpha$ is normalized.  
  \end{proof}
\end{lemma}
We conclude this section with the following remark, providing an outlook of the next sections. 
\begin{remark}
  \label{rem:euklrem}
  \begin{enumerate}
    \item
    \label{rem:euklrem0}
    If $\wm$ is analytic and pointwise proper, then
    $\AQR, \ARQ\subseteq \A$ are of measure zero w.r.t. $\mAL$ whenever $\dim[S]\geq 1$ and $G_x\neq \{e\}$ for some $x\in M$. 
    
    In fact, we will see in Lemma \ref{lemma:sim}.\ref{lemma:sim2} that for such an action $\wm$ and $\g\in \mg\backslash\mg_x$ for $x\in M$ we always find $l>0$ such that the restriction of the curve $\delta \colon t\mapsto \wm_x(\exp(t\cdot\g))$ to $[0,l]$ is embedded analytic. Then, for $g:=\exp(l/3\cdot \g)$ and $\gamma:=\delta|_{[0,l/3]}$ the requirement $\gamma \nsim_\cp \wm_g\cp \gamma$ of Corollary \ref{cor:AshtLew} is obviously fulfilled.
  \item
    \label{rem:euklrem1}
    Corollary \ref{cor:AshtLew} already shows that the invariance properties of the elements of $\AQR$ give non-trivial restrictions to the images of the projection maps $\pi_\alpha$. In particular, this is the case if the curves  forming the index $\alpha=(\gamma_1,\dots,\gamma_k)$ are related in the sense that $\wm_g\cp\gamma_i\csim \gamma_j$ holds for some $g\in G$ and some $1\leq i\neq j\leq k$. 
    
    Now, non-trivial restrictions can also occur if we have $\alpha=\gamma$ for a single curve $\gamma\in \Pa$, namely if $\gamma$ is invariant under some symmetry group element $g\in G\backslash\{e\}$, i.e., $\phi_g\cp\gamma =\gamma$. Indeed, then by \eqref{eq:invpi} we have $\pi_\gamma(\qw)=\pi_{\wm_g\cp \gamma}(\qw)=c\cdot \pi_\gamma(\qw)\cdot d$, and for $S$ compact and connected and the case that $c=d^{-1}\neq \pm \me$, such a relation can already restrict $\im[\pi_\gamma]$ to a maximal torus as Lemma \ref{lemma:torus} shows.\footnote{See also the second point in Example \ref{rem:euklrem6}} So, when constructing measures on the spaces $\ARQ$ and $\AQR$ by means of such projection maps, one has to take these invariance properties into account.  
    \item 
    	As we will see in Subsection \ref{sec:inclrel}, the inclusion $\ARQ\subseteq \AQR$ is usually proper. Consequently, there are further highly non-trivial restrictions to the image of $\pi_\alpha$ when restricting to $\ARQ$. For this reason, it seems to be very hard to provide a general notion of a reduced measure on these spaces. Anyhow,
    for the case of homogeneous isotropic LQC, in Section \ref{sec:HomIsoCo}, we will discuss the measure theoretical aspects of the space $\ARQ$ in detail. 
    \item
    \label{rem:dsdfdf}
For the space $\AQR$ we will investigate (next two sections) the case where $\Pa= \Paw$ holds and where $\wm$ is analytic and pointwise proper. This will allow us to follow the lines of Proposition \ref{rem:euklrem2b}, as, in this case, $\Paw$ splits up into 
free and continuously generated curves. We denote the respective sets by $\Paf$ and $\Pac$. The set $\Paf$ consists of all $\gamma\in \Paw$ which contain a subcurve $\delta$ (free segment) such that
\begin{align*}
	\wm_g\cp \delta \cpsim\delta \:\:\:\text{ for }\:\:\: g\in G  \qquad\Longrightarrow \qquad \wm_g\cp \delta = \delta, 
\end{align*}
and then $\Pac$ is just its complement in $\Paw$. We will see in Proposition \ref{prop:freeseg}.\ref{prop:freeseg1} that each $\gamma\in \Paf$ is discretely generated by the symmetry group, i.e., is build up finitely many translates of initial and final segments\footnote{Here, translates of initial and final segments of $\delta$ can only occur as initial and final segments of $\gamma$.} of one of its maximal free segments $\delta$. Now,
	\begingroup
\setlength{\leftmarginii}{15pt} 
\begin{itemize}
	\item
	\vspace{-3pt}
		If $\wm$ is transitive or proper and admits only stabilizers which are normal subgroups, then $\Pac$ equals the set $\Pags$ of Lie algebra generated curves, i.e., of embedded analytic curves equivalent to a curve of the form $[0,l]\ni t\mapsto \wm(\exp(t\cdot\g),x)$ for some $x\in M$, $\g\in \mg\backslash \mg_x$ and $l>0$. Then, in special situations (such as if $\wm$ acts free), we will be able to construct a normalized Radon measure on $\IHOMLAS$ for the case that $S=\SU$. This will be done in Subsection \ref{sec:ConSp}.
	\item
		Let $\Pafns\subseteq \Paf$ denote the subset of non-symmetric curves, i.e., of curves for which $\wm_g\cp \gamma \neq \gamma$ holds for all $g\in G\backslash\{e\}$.\footnote{In other words, the stabilizer of $\gamma$ (a well-behaving and important quantity in the situation where $\wm$ is analytic and pointwise proper) is trivial.} 
		We will see in Subsection \ref{sec:FreeM} that $\HOMFNS$ carries a natural normalized Radon measure whenever the structure group of the bundle  is compact and connected. This measure even specializes to the Ashtekar-Lewandowski measure on $\HOMW$ if $G=\{e\}$.  This is because, according to our definitions (see Definition \ref{def:freeSeg}.\ref{def:freeSeg32}), then $\Pafns=\Paw$ holds.
	\item
		If $\wm$ is proper and free, then $\Pac=\Pags$ and $\Paf=\Pafns$ holds. Here, both sets are independent, $\Phi$-invariant and closed under decomposition and inversion of their elements. So, for $S=\SU$ we will obtain a normalized Radon measure on
		\begin{align*}
			\IHOM\cong\IHOMLAS \times  \IHOMFNS
		\end{align*}  
		just by taking the Radon product of the measures from the first two points, cf.\ Proposition \ref{rem:euklrem2b}. 
\end{itemize}
\endgroup
\item 
\label{it:sdsdds}
The set $\Pags$ will be of particular interest because:
\begingroup
\setlength{\leftmarginii}{15pt} 
\begin{itemize}
\item
	\vspace{-3pt}
	Lie algebra generated curves turn out to be very useful for investigations concerning the inclusion relations between the spaces $\ARQ$ and $\AQR$, as they often allow to decide the inclusion problem on the level of the Lie algebras of the symmetry and the structure group, see Subsection \ref{sec:inclrel}. 
 \item
     Whenever $\wm$ acts transitively, the cylindrical functions that correspond to $\Pags$ separate the points in $\Con$, so that 
      $\iota\colon \Con \rightarrow \AInd{\mg}$ is injective. This is clear from surjectivity of $\dd_e\wm_x \colon \mg \rightarrow T_xM$ and Lemma \ref{lemma:separating}. So, in this case not only $\AInd{\mg}\cong \HOMLAS$ but also $\AQRLA\cong \IHOMLAS$ is a physically meaningful candidate for a (reduced) quantum configuration space. Indeed, there are approaches to LQG which only use linear curves to define $\A$ \cite{JonEng3}. Moreover, originally only linear curves were used to define the quantum configuration space of LQC and also in the reduction paper \cite{BojoHomoCosmo} only Lie algebra generated curves were taken into account.\footnote{In this rather heuristic paper the author restricts to the transitive situation with $G_x=\{e\}$, and quantizes (unspecified) sets containing $\AR$ by using parallel transport functions along Lie algebra generated curves.} 
\end{itemize}   
\endgroup
  \end{enumerate} 
\end{remark}
\subsection{Summary}
\label{concl:Actionlift}
\begin{enumerate}
\item 
  \label{concl:Actionlift1}
  In the first part of this section we have seen that it is always possible to extend a left action 
	\begin{align*}
	\cw\colon G\times X\rightarrow X
	\end{align*}	  
   of a group $G$ on a set $X$ uniquely to a left action 
   	\begin{align*}
	\specw\colon G\times \X\rightarrow \X
	\end{align*}	
   on the spectrum $\X$ of a $\Cstar$-algebra $\aA\subseteq B(X)$. The action $\specw$ is always continuous in $\X$, and continuous if $G\ni g\mapsto \cw_g^*f\in \aA$ is continuous for all $f\in \aA$. For $\aA$ unital, here even the iff statement holds. 
   
   Continuity of $\specw_g$ and the extension property\footnote{Recall that $X_\aA$ just denotes the set of elements $x\in X$ for which $\aA\rightarrow \mathbb{C}$, $f\mapsto f(x)$ is not the zero functional.} 
	\begin{align*}
		\specw_g\cp \iota_X|_{X_\aA}=\iota_X\cp \cw_g|_{X_\aA}\qquad \forall\: g\in G
	\end{align*}  
   always imply that $\XRQ\subseteq \XQR$ holds. Here, $\XRQ$ denotes the closure in $\X$ of  
\begin{align*}
	\XR\cap X_\aA=\{x\in X_\aA\:|\: \cw(g,x)=x\quad\forall\: g\in G\}
\end{align*}  
    and $\XQR=\{\x\in \X\:|\: \specw(g,\x)=\x\quad\forall\: g\in G\}$ is the closed subset of $\specw$-invariant elements.  
\item 
  \label{concl:Actionlift2}
  In the second part, we have applied the first one to the quantum gauge field situation  
  where $X$ equals the set $\Con$ of smooth connections on a principal fibre bundle $\PMS$, and $\cw$ comes from a Lie group $(G,\Phi)$ of automorphisms of $P$, i.e., 
	\begin{align*}  
  	\cw(g,\w)=\Phi_{g^{-1}}^*\w\qquad\forall\: g\in G. 
	\end{align*}  
  In this situation $\aA$ is the $\Cstar$-algebra $\PaC$ of cylindrical functions that corresponds to any set $\Pa$ of $\CC{k}$-paths in $M$ which is invariant under pullbacks by the translations $\wm_g$ ($\Pa$ is $\Phi$-invariant). Here, $\wm$ denotes the action induced by $\Phi$ on $M$. In particular, we have seen that the set $\ARQ$ of quantized invariant (classical) connections is always contained in the compact set $\AQR$ of invariant generalized (quantum) connections.
\item 
  \label{concl:Actionlift3}
  In the last part of this section, we have identified the quantum spaces $\A$ with spaces $\HOM$ of homomorphisms of paths. We have seen that this is always possible if the set $\Pa$ is independent.\footnote{This is the case, e.g., if $M$ is analytic, $S$ is compact and connected, and $\Pa$ is the set $\Paw$ of embedded analytic curves in $M$, cf.\ Lemma \ref{lemma:BasicAnalyt}.\ref{lemma:BasicAnalyt3}.} Under the further assumption that $\Pa$ is $\Phi$-invariant, this has allowed us to identify $\AQR$ with the subspace $\IHOM\subseteq \HOM$ of invariant homomorphisms. 
  
  Moreover, using invariance, we have shown that in the analytic (LQG) situation, i.e., $M$ and $\wm$ are analytic, $\wm$ is pointwise proper and $\Pa=\Paw$, the sets $\ARQ$ and $\AQR$ are of measure zero w.r.t.\ the Ashtekar-Lewandowski measure $\mAL$ (standard measure in LQG) on $\A$ provided that $\dim[S]\geq 1$, $G_x\neq \{e\}$ holds for some $x\in M$, and $S$ is compact and connected. 
  Moreover, we have illustrated that defining a normalized Radon measures on $\AQR\cong \IHOMW$ can be done by decomposing $\Paw$ into elements continuously,  
  and into elements discretely generated (see Remark \ref{rem:euklrem}.\ref{rem:dsdfdf}) by the symmetry group. This decomposition will be discussed in much more detail in the next section. 
\end{enumerate}

\section{Modification of Invariant Homomorphisms}
\label{susec:LieALgGenC}
Let $\PMS$ be a principal fibre bundle and $(G,\Phi)$ a Lie group of automorphisms thereon. In the previous section, we have introduced the space $\AQR$ of invariant generalized connections that corresponds to a $\Phi$-invariant set $\Pa$ of $\CC{k}$-curves in $M$. For the case that $\Pa$ is in addition independent and $S$ is compact, we have identified  $\AQR$ with the subset $\IHOM \subseteq \HOM$ of homomorphisms having additional invariance properties.  
We now switch to the analytic situation, i.e., we consider embedded analytic curves and actions $\Phi$ for which the induced action $\wm$ on $M$ is analytic and pointwise proper. We will prove certain modification results for invariant homomorphisms which are crucial for both our investigations of the inclusion relations between $\ARQ$ and $\AQR$, and the construction of normalized Radon measures on $\AQR$ in Section \ref{sec:MOQRCS}. In analogy to Lemma and Convention \ref{lemconv:RBMOD}.\ref{prop:Bohrmod24}, modification here just means to change the value of a given invariant homomorphism on some distinguished set of curves in a specific way.

In the following, $\Paw$ will always denote the set of embedded analytic curves in $M$, and $S$ is assumed to be compact and connected. Recall that compactness of $S$ guarantees well-definedness of the map $\kappa\colon \A\rightarrow \HOM$, $\qw\mapsto\big[\gamma \mapsto \parall{\gamma}{\qw}\hspace{1pt}\big]$. Connectedness is crucial because:
\begingroup
\setlength{\leftmargini}{20pt}
\begin{itemize}
\item
	As we will see in Lemma \ref{lemma:BasicAnalyt}.\ref{lemma:BasicAnalyt3}, a subset $\Pa\subseteq \Paw$ closed under decomposition and inversion of it elements is independent if $S$ is connected. Recall that this ensures that $\kappa$ is bijective, i.e., a homeomorphism w.r.t.\ the topology $\TOHO$ from Definition \ref{def:indepref}.\ref{conv:muetc}.
\item
	If $\dim[S]\geq 1$, the equivalence relation $\csim$ equals the (elsewise coarser) equivalence relation $\psim$ that will occur in the fundamental statements of this section.
	Here, $\gamma_1\psim\gamma_2$ just means that both curves coincide up to parametrization, i.e., 
	 \begin{align*}  
  		\gamma_1\text{\gls{PSIM}}\gamma_2 \:\: \Longleftrightarrow \:\: \gamma_1=\gamma_2\cp \adif|_{\dom[\gamma_1]}\: \text{ for }\:  \adif \colon I\rightarrow \RR \:\text{ analytic diffeomorpism with }\dot\adif>0.
  	\end{align*}
\end{itemize}
\endgroup
\noindent
In Subsection \ref{subsec:LaGC}, we will collect the relevant facts and definitions concerning analytic and Lie algebra generated curves. In Subsection \ref{sec:ModifLAGC}, we modify invariant homomorphisms along such Lie algebra generated curves, and in Subsection \ref{sec:inclrel} we will apply this in order to obtain some general conditions which allow to decide whether the inclusion 
 $\ARQ\subseteq\AQR$ 
 is proper. Basically, this will be done by constructing elements of $\AQR$ that cannot be approximated by classical (smooth) invariant connections. 
In particular, we conclude that quantization and reduction do not commute in \mbox{(semi-)homogeneous}  loop quantum cosmology. For homogeneous isotropic LQC, this will be shown in Section \ref{sec:HomIsoCo}. Finally, in the last part of this section, we will prove an analogue of the modification result from Subsection \ref{sec:ModifLAGC}, now for free 
curves, cf.\ Remark \ref{rem:euklrem}.\ref{rem:dsdfdf}. We first show that each analytic embedded curve which contains a free segment 
is discretely generated 
 by the symmetry group. Then, we use this in order to modify homomorphism along such free segments.

\subsection{Analytic and Lie Algebra Generated Curves}
\label{subsec:LaGC}
This subsection basically collects the properties of analytic and Lie algebra generated curves that will be  relevant for our later considerations. We will start with some elementary facts on analytic curves. Then, 
we highlight the most important properties of the Lie algebra generated ones, in particular, for the case that the induced action $\wm$ is analytic and pointwise proper.
\subsubsection{Basic Properties of Analytic Curves}
We start with a lemma collecting some standard properties of (embedded) analytic curves. For this, we will need 
\begin{definition}
  \label{remdef:eqrels}
  Let $\gamma_1,\gamma_2\in \Paw$ with $\dom[\gamma_i]=[a_i,b_i]$ for $i=1,2$. We define the equivalence relations $\psim$ and $\isim$ on $\Paw$ by 
	\begin{align*}  
  		\gamma_1\hspace{2.5pt}\text{\gls{PSIM}}\hspace{1pt}\gamma_2 \quad \Longleftrightarrow \quad &
  		%&\gamma_1=\gamma_2\cp \adif \text{ for }  \adif \colon \dom[\gamma_1]\rightarrow \dom[\gamma_2] \text{ an analytic diffeomorphism.}\footnotemark\\[5pt]    
%
\gamma_1=\gamma_2\cp \adif|_{\dom[\gamma_1]}\: \text{ for }\:  \adif \colon I\rightarrow \RR \:\text{ analytic diffeomorpism with }\dot\adif>0.\\  		
  		\gamma_1\isim\gamma_2  \quad \Longleftrightarrow \quad & \im[\gamma_1]=\im[\gamma_2]\:  \text{ and }\: \gamma_1(a_1)=\gamma_2(a_2)\: \text{ as well as }\: \gamma_1(b_1)=\gamma_2(b_2).
  	\end{align*} 
  	Recall that this means that $I\subseteq \RR$ is open and that $\dom[\gamma_1]\subseteq I$ holds. 
\end{definition}
\begin{convention}
\label{conv:extensaccum}
\begingroup
\setlength{\leftmargini}{20pt}
\begin{itemize}
\item
	In the sequel, by an accumulation point of a topological space $X$ we will understand an element $x\in X$ for which we find a net $\{x_\alpha\}_{\alpha\in I}\subseteq X\backslash\{x\}$ with $\lim_\alpha x_\alpha=x$.  
\item
	According to the notations subsection, domains of curves 
	are always assumed to be intervals (non-empty interior). 
\item
	An analytic (immersive) curve $\gamma$ is called extendible iff we find an analytic (immersive) curve $\delta$ with open domain, such that $\gamma=\delta|_{\dom[\gamma]}$ holds and $\dom[\gamma]\subseteq \dom[\delta]$ is properly contained. 
	It   	
	is called maximal (or inextendible) iff no extension exists. The same conventions hold for analytic diffeomorphisms $\adif\colon I\rightarrow \RR$. (Observe that the domain of a maximal analytic (immersive) curve is necessarily open.)
\end{itemize}
\endgroup
\end{convention}

\begin{lemma}
  \label{lemma:BasicAnalyt}
  \begin{enumerate}
  \item
    \label{lemma:BasicAnalyt1}
	 Let $\gamma_i\colon (a_i,b_i)\rightarrow M$ be an analytic embedding for $i=1,2$ and $x$ an accumulation point of $\im[\gamma_1]\cap \im[\gamma_2]$ (w.r.t.\ the subspace topology inherited from $M$). Then $\gamma_1(I_1)=\gamma_2(I_2)$ for open intervals $I_i\subseteq (a_i,b_i)$ with $x\in  \gamma_i(I_i)$ for $i=1,2$.
 \item
    \label{maximalextension}
    	Each (immersive) analytic curve has a maximal extension. 
  \item 
    \label{lemma:BasicAnalyt2}
    We have 
	\begin{align*}
		\gamma_1\isim \gamma_2\quad \Longleftrightarrow\quad \gamma_1\psim \gamma_2 \quad 
		%\text{and} \quad \gamma_1\psim \gamma_2\quad 
		\Longrightarrow \quad \gamma_1\csim \gamma_2. 
	\end{align*}     
   If $\dim[S]\geq 1$, then 
	\begin{align*}
		\gamma_1\isim \gamma_2\quad \Longleftrightarrow\quad \gamma_1\psim \gamma_2 \quad \Longleftrightarrow\quad \gamma_1\csim \gamma_2.
	\end{align*}   
  \item 
    \label{lemma:BasicAnalyt4}
    Let $\delta \colon [0,k] \rightarrow M$ and $\delta' \colon [0,k'] \rightarrow M$ be analytic embeddings. 
   % More precisely,
    \begingroup
    \setlength{\leftmarginii}{20pt}
    \begin{enumerate}
    \item
    \label{dddd}
    	If $\delta$ and $\delta'$ share an initial segment, i.e.,
    	\vspace{-16pt}
	\begin{align*}
	\delta|_{[0,s]}\isim \delta'|_{[0,s']} \qquad \text{for}\qquad  s\in (0,k],\: s'\in (0,k'],
%    	 \delta([k_1,k_1+t])=\delta'([k_1',k'_1+t'])\qquad \text{for some}\qquad t,t'>0,
%
		\hspace{98.25pt}
		\begin{tikzpicture}
		\filldraw[black] (0,0) circle (2pt);	
		\draw[->,line width=0.7pt,color=red] (0.25,0) .. controls (1,0) and (1,0.5) .. (0.5,0.5);
	    \draw[->,line width=1.5pt] (0,0) -- (1.5,0);
		\draw (1.3,0.3) node {\(\delta\)};
		\draw[color=red] (0.25,0.5) node {\(\delta'\)};
		\draw[-,line width=1pt] (0.37,0) -- (0.37,0.15);
		\end{tikzpicture}
	\end{align*}    	 
    then either 
    \begingroup
    \setlength{\leftmarginiii}{15pt}
	\begin{itemize}
	\item    
			\vspace{-5pt} 
    \hspace{18pt}$\delta \psim \delta'|_{[0,t']}$\quad for\quad $t'\in (0,k']$ \qquad or\hspace*{\fill}	 
    	\begin{tikzpicture}
    	\filldraw[white] (2,0) circle (2pt);
    	\filldraw[black] (0,0) circle (2pt);
		\draw[->,line width=1.05pt,color=red] (0,0) -- (2,0);	
	    \draw[->,line width=1.5pt] (0,0) -- (1.5,0);
		\draw (1,0.2) node {\(\delta\)};
		\draw[color=red] (1.8,0.2) node {\(\delta'\)};	
		\end{tikzpicture}
	\item	
	\vspace{-5pt} 
    $\delta|_{[0,t]}\psim \delta'$\quad\hspace{17pt} for\quad $t\phantom{'}\in (0,k)$.\hspace*{\fill}	 
    	\begin{tikzpicture}
    	\filldraw[white] (2,0) circle (2pt);
    	\filldraw[black] (0,0) circle (2pt);	
		\draw[->,line width=1.1pt,color=red] (0,0) -- (1,0);
	    \draw[->,line width=1.5pt] (0,0) -- (1.5,0);
		\draw[color=red] (0.7,0.2) node {\(\delta'\)};
		\draw (1.7,0.2) node {\(\delta\)};	
		\end{tikzpicture}
     \end{itemize}
     \endgroup
    \item	 
    \vspace{5pt} 
	If $\delta$ and $\delta'^{-1}$ share an initial segment, i.e.,
	   	\vspace{-15pt}	    
	\begin{align*}
	\delta|_{[0,s]}\isim \big[\delta'|_{[s',k']}\big]^{-1} \qquad \text{for}\qquad  s\in (s,k],\: s'\in[0,k'),
%    	 \delta([k_1,k_1+t])=\delta'([k_1',k'_1+t'])\qquad \text{for some}\qquad t,t'>0,
%
	\hspace{73pt}
	\begin{tikzpicture}
	%    \draw[->,line width=1.1pt, color=white] (0,0) -- (2,0);
	%	\filldraw[white] (2,0) circle (2pt);
		%\filldraw[black] (0,0) circle (1pt);
		\filldraw[red] (0.5,0.5) circle (1.5pt);			
		\draw[-,line width=0.6pt,color=red] (0.25,0) .. controls (1,0) and (1,0.5) .. (0.5,0.5);
		\draw[->,line width=1.5pt] (0,0) -- (1.5,0);
		\draw[->,line width=1.05pt] (0.25,0) -- (0,0);
		\draw (1.3,0.3) node {\(\delta\)};
		\draw[color=red] (0.25,0.5) node {\(\delta'\)};
		\draw[-,line width=1.3pt] (0.4,0) -- (0.4,0.15);
		\end{tikzpicture}
	\end{align*}    	 
    then either 
    \begingroup
    \setlength{\leftmarginiii}{15pt}
	\begin{itemize}
	\item    
		\vspace{-5pt} 	
    \hspace{18pt}$\delta \psim \big[\delta'|_{[t',k']}\big]^{-1}$\quad for\quad $t'\in [0,k')$ \qquad or\hspace*{\fill}	 
    	\begin{tikzpicture}
    	\filldraw[white] (2,0) circle (2pt);
    	%\filldraw[black] (0,0) circle (2pt);
    	\filldraw[red] (2,0) circle (1.5pt);
    	\draw[->,line width=1.0pt,color=red] (2,0) -- (0,0);	
	    \draw[->,line width=1.5pt] (0,0) -- (1.5,0);
		\draw (1,0.2) node {\(\delta\)};
		\draw[color=red] (1.8,0.25) node {\(\delta'\)};	
		\end{tikzpicture}
	\item	
		\vspace{-5pt}
    $\delta|_{[0,t]}\psim \delta'^{-1}$\quad\hspace{29.5pt} for\quad $t\phantom{'}\in (0,k)$.\hspace*{\fill}	 
    	\begin{tikzpicture}
    	\filldraw[white] (2,0) circle (2pt);
	   	\filldraw[red] (1,0) circle (1.5pt);	
	   	\draw[->,line width=1.1pt,color=red] (1,0) -- (0,0);
	    \draw[->,line width=1.5pt] (0,0) -- (1.5,0);
		\draw[color=red] (0.7,0.2) node {\(\delta'\)};
		\draw (1.7,0.2) node {\(\delta\)};	
		\end{tikzpicture}
     \end{itemize}
     \endgroup
    \end{enumerate}
    \endgroup    
    \item
    \label{Basanalyt}
    	Let $\gamma_i\colon [0,k_i]\rightarrow M$ be an analytic embedding for $i=1,2$. Then, for $i=1,2$ the set $\gamma_i^{-1}(\im[\gamma_1]\cap\im[\gamma_2])$ is the disjoint union of finitely many isolated points and $0\leq m\leq 2$ disjoint compact   
intervals $\{L_i^p\}_{1\leq p\leq m}$ with 
$\gamma_1|_{L_1^p}\isim \big[\gamma_2|_{L_2^p}\big]^{\pm 1}$ for $1\leq p\leq m$. 
    	
   If $m=1$, then either 
    \begingroup
    \setlength{\leftmarginii}{15pt}
	\begin{itemize} 
	\item  
	\vspace{-5pt} 	
  		$L_i^m=[0,k_i]$\:\: for some \:\:$i\in \{1,2\}$\qquad or\hspace*{\fill}	 
    	\begin{tikzpicture}
   	\filldraw[white] (2,0) circle (2pt);	
		\draw[<->,line width=1.0pt,color=red] (0.5,0) -- (1.2,0);
		\draw[<->,line width=1.5pt] (0,0) -- (2,0);
		\draw[color=red] (0.85,0.2) node {\(\gamma_i\)};
		\draw (1.7,0.2) node {\(\gamma_j\)};	
		\end{tikzpicture}
	\item 
		\vspace{-5pt}
		$L_i^m$ is of the form $[l_i,k_i]$ or $[0,l_i]$ with $0<l_i<k_i$ for $i=1,2$. \hspace*{\fill}	 	 
    	\begin{tikzpicture}
    	\filldraw[white] (2,0) circle (2pt);	
		\draw[<->,line width=1.0pt,color=red] (0.9,0) -- (2,0);
		\draw[<->,line width=1.5pt] (0,0) -- (1.5,0);
		\draw (0.5,0.2) node {\(\gamma_j\)};
		\draw[color=red] (1.8,0.25) node {\(\gamma_i\)};	
		\end{tikzpicture}
	\end{itemize}
	\endgroup
	  
	If $m=2$, then for $i=1,2$ we have%\hspace*{\fill} 
	\vspace{-10pt}	
	\begin{align*}	
		 \quad\:\: L_i^p=[0,l^p_i]\:\:\text{ and }\:\: L_i^q=[l^q_i,k_i]\:\:\text{ for some }\:\:  l_i^p\neq l_i^q,\: 1\leq p\neq q\leq 2.
		\hspace{50pt}
		\begin{tikzpicture}
    	\filldraw[white] (2,0) circle (2pt);
%    	%\filldraw[black] (0,0) circle (2pt);
%    	\filldraw[black] (2,0) circle (1.5pt);	
		\draw[<-,line width=1.0pt,color=red] (1.15,0) -- (1.6,0);
		\draw[->,line width=1.0pt,color=red] (0.4,0) -- (0.85,0);
		\draw[-,line width=1pt,color=red] (1.6,0) .. controls (1.8,0) and (2,0.5) 		.. (1,0.5);
		\draw[-,line width=1pt,color=red] (0.4,0) .. controls (0.2,0) and (0,0.5)
		.. (1,0.5);
%		\draw[-,line width=1.1pt,color=red] (0.8,0.5) -- (1.2,0.5);
	    \draw[<->,line width=1.5pt] (0.4,0) -- (1.6,0);
		\draw (1,0.2) node {\(\gamma_j\)};
		\draw[color=red] (0,0.4) node {\(\gamma_i\)};	
		\end{tikzpicture} 
	\end{align*} 
    \item
    \label{lemma:BasicAnalyt3}
    If $S$ is connected with $\dim[S]\geq 1$, and 
    $\Pas\subseteq \Paw$ is closed under decomposition and inversion of its elements, then $\Pas$ is independent.
  \end{enumerate}
  \end{lemma}
  \begin{proof}
	\begin{enumerate}
		\item 
			 Let $x=\gamma_i(\tau_i)$ with $\tau_i\in (a_i,b_i)$ for $i=1,2$. Moreover, let $(U,\phi)$ be an analytic submanifold chart of $\im[\gamma_1]$ which is centered at $x$ and maps $\im[\gamma_1]\cap U$ into the $x$-axis. Choose $\epsilon>0$ such that $\gamma_2(B_{\epsilon}(\tau_2))\subseteq U$ and consider the analytic functions $f_{\zd}:=\phi^{\zd}\cp\gamma_2|_{B_{\epsilon}(\tau_2)}$ for $k=2,\dots,\dim[M]$. Then $\tau_2$ is an accumulation point of zeroes of $f_{\zd}$, so that $f_{\zd}=0$ by analyticity of $f_{\zd}$. This shows $\phi(\gamma_2(B_{\epsilon}(\tau_2))\subseteq \phi(U\cap\im[\gamma_1])$, hence $\gamma_2(B_{\epsilon}(\tau_2))\subseteq \im[\gamma_1]$. 
    The claim then holds for $I_2:=B_{\epsilon}(\tau_2)$ and $I_1:=\gamma_1^{-1}( \gamma_2(I_2))$.
        	\item 
    	Let $\gamma\colon D\rightarrow M$ be an analytic curve, denote by $\EE$ the set of all extensions of $\gamma$ with open domain and define its maximal extension
    	\begin{align*}
    		\gamma_0\colon I:= \textstyle\bigcup_{\delta\in \EE}\dom[\delta]\rightarrow M
    	\end{align*}
    	by $\gamma_0(x):=\delta(x)$ for $\delta\in \EE$ with $x\in \dom[\delta]$. Then, $\gamma_0$ is well defined because for $\delta'\in \EE$ with $x\in \dom[\delta']$ we necessarily have $\delta'(x)=\delta(x)$. In fact, since $\delta$ and $\delta'$ coincide on $D$, by analyticity they coincide on $\dom[\delta]\cap\dom[\delta']\ni x$.
 	If $\gamma$ is in addition immersive, we restrict $\gamma_0$ to the maximal (necessarily open) interval containing $D$ on which $\gamma_0$ is an immersion.   
		\item
		  The implication $\psim\: \Rightarrow\:  \isim$  is obvious and $\psim\: \Rightarrow\:  \csim$  is clear because parallel transports are invariant under reparametrization by orientation preserving diffeomorphisms.
        
      Now, assume that $\gamma_1\sim_{\im} \gamma_2$ and let $\gamma_i'\colon (a_i',b_i')\rightarrow M$ be an analytic embedding with $[a_i,b_i]\subseteq (a_i',b_i')$ and $\gamma_i'|_{[a_i,b_i]}=\gamma_i$ for $i=1,2$. Then $\gamma_i(a_i)$ and $\gamma_i(b_i)$ are accumulation points of $\im[\gamma'_1]\cap\im[\gamma'_2]$, so that
      by Part \ref{lemma:analytCurvesIndepetc1}) we can arrange that $N:=\im[\gamma_1']= \im[\gamma_2']$ just by shrinking the intervals $(a'_i,b_i')$ in a suitable way. Now, $N$ is a real analytic submanifold in the topological sense and the maps $\gamma_i'$ are diffeomorphisms w.r.t.\ its analytic structure. 
      Consequently, $\adif:=\gamma_2'^{-1}\cp \gamma_1'$ is the desired diffeomorphism. 
      
	Now, let $\dim[S]>1$ and $\nu=\{\nu_x\}_{x\in M}$ a choice of elements with $\nu_x\in F_x$ for all $x\in M$. If $\gamma_1\csim \gamma_2$, then $\gamma_1(a_1)=\gamma_2(a_2)$ and $\gamma_1(b_1)=\gamma_2(b_2)$ by definition, and we have to show that  
      $\im[\gamma_1]= \im[\gamma_2]$. Now, if this is not true, then we find $\tau_1\in (a_1,b_1)$ or $\tau_2\in (a_2,b_2)$ such that $\gamma_1(\tau_1)\notin \im[\gamma_2]$ or $\gamma_2(\tau_2)\notin \im[\gamma_1]$, respectively. If $\gamma_1(\tau_1)\notin \im[\gamma_2]$, by compactness of $\im[\gamma_2]$ we find a neighbourhood $U$ of $\gamma_1(\tau_1)$ in $M$ with  $U\cap \im[\gamma_2]=\emptyset$. Let $\w\in \Con$ be fixed and $s:=h_{\gamma_2}^\nu(\w)$. If we choose $s'\neq s$, then by Proposition A.1 in \cite{ParallTranspInWebs} we find $\w'\in\Con$ such that $\w'$ equals $\w$ outside $U$ and $h_{\gamma_1}^\nu(\w')=s'$. But, $h_{\gamma_2}^\nu(\w)=h_{\gamma_2}^\nu(\w')$ since $\im[\gamma_2]\subseteq M\backslash U$, so that 
	\begin{align*}      
      h_{\gamma_2}^\nu(\w')=h_{\gamma_2}^\nu(\w)=s\neq s'= h_{\gamma_1}^\nu(\w')
    \end{align*}
    contradicts that $\gamma_1\csim \gamma_2$.
    \item  
  It suffices to show \textit{(a)} because  \textit{(b)} follows from \textit{(a)} if we replace $\delta'$ by $\delta'^{-1}$. Let
  \begin{align*}      
    t:=\sup\big\{s\in [0,k]\:\big|\:\exists\: s'\in [0,k'] \colon \delta|_{[0,s]}\isim \delta'|_{[0,s']} \big\}.
  \end{align*}
  By assumption we have $t>0$, so that $\delta|_{[0,t]}\isim \delta'|_{[0,t']}$ for some $t'\in (0,k']$ by continuity, hence $\delta|_{[0,t]}\psim \delta'|_{[0,t']}$ by Part \ref{lemma:BasicAnalyt2}).
  \begingroup
  \setlength{\leftmarginii}{17pt}
  \begin{itemize}
  \item
	If $t=k$, then $\delta\psim \delta'|_{[0,t']}$ and we have done (the same for $t'=k'$).
\item
	If $t<k$ and 
	$t'<k'$, then $\delta(t)=\delta'(t')$ is an accumulation point of 
	\begin{align*}	
		\delta((0,t+\epsilon))\cap \delta'((0,t'+\epsilon'))\quad \text{for} \quad\epsilon,\epsilon'>0 \quad\text{suitable small}.
	\end{align*}
	Then, $\delta|_{[0,t_0]}\isim \delta'|_{[0,t'_0]}$ for some $t_0>t$, $t'_0> t'$ by Part \ref{lemma:BasicAnalyt1}), 
 	  contradicting the choice of $t$. 
  \end{itemize}	 
  \endgroup 
  \item
      The statement is clear if $T:=\im[\gamma_1]\cap \im[\gamma_2]$ is finite. In the other case there exists an  
      accumulation point of $T$, just by compactness of $T$. Let 
$\gamma_i'\colon (a_i',b_i')\rightarrow M$ be an extension of $\gamma_i$ for $i=1,2$.  
       Part \ref{lemma:analytCurvesIndepetc1}) shows that we find  $[s_i,t_i]\subseteq [0,k_i]$ for $i=1,2$ with $\gamma_1|_{[s_1,t_1]}\isim [\gamma_2|_{[s_2,t_2]}]^{\pm 1}$. The claim now follows by repeated application of Part \ref{lemma:BasicAnalyt4}). 
       
       In fact, replacing one of the curves by its inverse if necessary, we can assume that $\gamma_1|_{[s_1,t_1]}\isim \gamma_2|_{[s_2,t_2]}$ holds. By Part \ref{lemma:BasicAnalyt4}) we can assume\footnote{The case $t_2=k_2$ follows analogously.} that $t_1=k_1$, i.e., $\gamma_1|_{[s_1,k_1]}\isim \gamma_2|_{[s_2,t_2]}$, and  then the same part shows that one of the following two cases holds:
       \begin{align*}
           \gamma_1 &\isim \gamma_2|_{[r_2,t_2]}\quad \text{ for }\quad 0\leq r_2\leq s_2 \qquad \text{or}\\
		\gamma_1|_{[r_1,k_1]} &\isim \gamma_2|_{[0,t_2]}\quad\hspace{3.8pt} \text{ for }\quad0<r_1\leq s_1. 
       \end{align*} 
	   In the first case the claim is clear, and  in the second one it is clear if 
       $T':=\gamma_1([0,r_1])\cap \gamma_2([t_2,k_2])$ if finite.
       In the other case, $T'$ admits an accumulation point just by compactness.      
    Then, applying Part \ref{lemma:analytCurvesIndepetc1}) we find $[x_1,y_1]\subseteq [0,r_1)$ and $[x_2,y_2]\subseteq (t_2,k_2]$ with 
	\begin{align*}    
    	\gamma_1|_{[x_1,y_1]}\isim \big[\gamma_2|_{[x_2,y_2]}\big]^{\pm 1}.
    	\hspace{40pt}
    			\begin{tikzpicture}	
		\draw[<-,line width=1.0pt,color=red] (1.1,0) -- (1.6,0);
		\draw[-,line width=1pt,color=red] (1.6,0) .. controls (2,0.6) and (0.3,0.6) .. (0.7,0);
		\draw[->,line width=1pt,color=red] (0.7,0) .. controls (0.9,0) and (0.9,0) .. (0.95,0.28);
			    \draw[<->,line width=1.5pt] (0.3,0) -- (1.6,0);
		\draw (1.27,0.22) node {\(+\)};
		\draw[color=red] (1.9,0.25) node {\(\gamma_2\)};
		\draw (0.2,0.2) node {\(\gamma_1\)};	
		\end{tikzpicture} 
		\qquad\quad
    	\begin{tikzpicture}
		\draw[<-,line width=1.0pt,color=red] (1,0) -- (1.6,0);
		\draw[-,line width=1pt,color=red] (1.6,0) .. controls (1.9,0.3) and (1,0.7) .. (0.9,0.3);
		\draw[-,line width=1pt,color=red] (0.9,0.3) .. controls (0.9,0) and (0.8,0) .. (0.7,0);
		\draw[->,line width=1pt,color=red] (0.7,0) .. controls (0.6,0) and (0.6,0.2) .. (0.6,0.4);
		\draw[<->,line width=1.5pt] (0.3,0) -- (1.6,0);
		\draw (1.25,0.22) node {\(-\)};
		\draw[color=red] (1.85,0.3) node {\(\gamma_2\)};
		\draw (0.2,0.2) node {\(\gamma_1\)};	
		\end{tikzpicture}     
	\end{align*} 
	Combining Part \ref{lemma:BasicAnalyt4}) with injectivity of $\gamma_1$ and $\gamma_2$, we see that $\gamma_1|_{[x_1,y_1]}\isim [\gamma_2|_{[x_2,y_2]}]^{- 1}$ is not possible, and that 
	$\gamma_1|_{[0,r_1']}\isim \gamma_2|_{[t_2',k_2]}$ for $r_1'\in (0,r_1)$ and $t'_2\in (t_2,k_2)$.    
 \item
    	The full proof can be found in Lemma \ref{lemma:analytCurvesIndepetc}.\ref{lemma:analytCurvesIndepetc5}. 
    	Basically, one applies Part \ref{Basanalyt}) in order to show that each finite collection $\{\gamma_1,\dots,\gamma_k\}$ admits a refinement $\{\delta_1,\dots,\delta_m\}$ such that $\im[\delta_i]\cap \im[\delta_j]$ is finite for $i\neq j$. The rest then follows from compactness of $\im[\delta_i]$ and Proposition A.1 in \cite{ParallTranspInWebs}.	 
\end{enumerate}	
  \end{proof}
\subsubsection{Basic Properties of Lie Algebra Generated Curves}
In this subsection, we collect the relevant properties of Lie algebra generated curves. 
First, we provide the basic definitions and highlight their most important properties. Then, we show that for $\wm$ analytic and pointwise proper the set $\Pags$ of Lie algebra generated curves as well as its complement in $\Paw$ both are closed under decomposition and inversion of their elements. This will be extended to an decomposition of $\Paw$ into four natural subsets (invariant under inversions and decompositions as well) in Corollary \ref{rem:freinichtLiealg}. Recall that, as we have shown in Proposition \ref{rem:euklrem2b}, such a decomposition gives rise to a respective factorization of the space $\AQRw$.
\begin{definition}
  \label{def:analytLieAlgBD}
  \begin{enumerate}
  \item
    \label{def:analytLieAlgBD1}
    We say that the two curves $\gamma_1,\gamma_2$ in the manifold $M$ share an open segment and write $\gamma_1\text{\gls{CPSIM}} \gamma_2$ iff we find open intervals $I_i\subseteq \dom[\gamma_i]$ for $i=1,2$ with $\gamma_1(I_1)=\gamma_2(I_2)$.
  \item 
    \label{def:analytLieAlgBD2}
    For $x\in M$ and $\vec{g}\in \mathfrak{g}\backslash\mathfrak{g}_x$ we define the curve
	\begin{align*}    
    	\gamma^x_{\vec{g}}\colon \RR &\rightarrow M\\
    	t&\mapsto \wm(\exp(t\cdot\vec{g}),x).
	\end{align*}    	
  \item
    \label{def:analytLieAlgBD3}
    \itspacec
    Let $x\in M$ and $\vec{g},\vec{g}'\in \mathfrak{g}\backslash\mathfrak{g}_x$. We say that $\vec{g}$ and $\vec{g}'$ are related and write $\vec{g}\text{\gls{XSIM}}\vec{g}'$ iff there is some $g\in G$ for which $\gamma^x_{\vec{g}} \cpsim \wm_g\cp\gamma^x_{\vec{g}'}$ holds. 
     \item
    \label{def:analytLieAlgBD4}
    For $x\in M$ and $\vec{g}\in \mathfrak{g}\backslash\mathfrak{g}_x$ we define the subgroup $G_{[\g]}^x\subseteq G_x$ by 
    \begin{align}
    	\text{\gls{ADSTRGXSTAB}}:=\{h\in G_x\:|\: \Ad_h(\g)\in \spann_\RR(\g)\}.
    \end{align}
    Here, the brackets in the subscript $[\g]$ refer to the fact that $G_{[\g]}^x=G_{[\g']}^x$ holds if $\g'= \lambda \cdot\g$ for some $\lambda\neq 0$, see also Remark and Definition \ref{rem:ppropercurve}.\ref{rem:ppropercurve1}. 
  \end{enumerate}
\end{definition}

\begin{remdef}
\label{rem:ppropercurve}
\begingroup
\setlength{\leftmargini}{23pt}
\begin{enumerate}
\item
\label{rem:ppropercurve0}
In the following, we will tacitly  use that 
\begin{align*}
\wm_h\cp \gamma_\g^x(t)&=\wm\big(h\cdot \exp(t\cdot \g)\cdot h^{-1},x\big)\\
&=\wm\big(\exp\big(t\cdot \Ad_h(\g)\big),x\big)\\
&=\gamma_{\Ad_h(\g)}^x(t)
\end{align*}
for all $t\in \RR$ and all $h\in G_x$.
\item
\label{rem:ppropercurve1}
	Let \gls{SPMG} denote the projective space\footnote{This is the set $\mg\backslash\{0\}$ modulo the equivalence relation $\g'\simpr  \g$\quad $\Longleftrightarrow$\quad $\g'\in \spann_\RR(\g)$.} that corresponds to $\mg$, and %$\ms$, respectively,  
    	let $\text{\gls{PRMG}}\colon  \mg\rightarrow \Sp\mg$ denote
    the corresponding projection map. Moreover, for $\g\in \mg\backslash\mg_x$ let $[\g]$ be the class of $\g$ in $\pr_\mg(\mg\backslash \mg_x)$, i.e., $[\g]:=\pr_\mg(\g)$.
     
    Then, the subgroup $G_{[\g]}^x$ from Definition \ref{def:analytLieAlgBD}.\ref{def:analytLieAlgBD4} can also be characterized as follows. 
     Since $\Ad(h)$ is linear for each $h\in G_x$, the map  
	\begin{align*}       
       \text{\gls{ADDSTRICH}}\colon G_x\times \pr_\mg(\mg\backslash \mg_x)&\rightarrow \pr_\mg(\mg\backslash \mg_x),\quad
       (h,[\g])\mapsto \big[\!\Ad_h(\g)\big]
	\end{align*}       
        is a well-defined left action, and then $\text{\gls{ADSTRGXSTAB}}\subseteq G_x$ is just the stabilizer of $[\g]$ w.r.t.\ $\Ad'$. The relevance of this action and its orbits will be discussed in Remark \ref{lemremmgohnermgx}. 
 \item
 \label{rem:ppropercurve2}
 The importance of the group $G_{[\g]}^x$ comes from the condition in the next Lemma \ref{lemma:sim}.\ref{lemma:sim3} which, for the special case that $V=\spann_\RR(\g)$ is one-dimensional and $\wm$ is analytic, can more naturally be written in the form 
 (see Definition \ref{def:stable} and Lemma and Remark \ref{rem:dfggfg}.\ref{rem:dfggfg1})
     \begin{align}
     \label{eq:staby}
	\gamma_{\g}^x|_{[0,l]} \psim \gamma_{\pm\Ad_h(\g)}^x|_{[0,l']}\quad\text{for }\quad h\in G_x \qquad\Longrightarrow \qquad h\in G_{[\g]}^x,
	\end{align}
	whereby $l,l'>0$ are such that both curves are embedded analytic.\footnote{Due to Lemma \ref{lemma:sim}.\ref{lemma:sim2} such reals always exist, see also Part \ref{rem:ppropercurve4}) of this remark.}    
	As we will see in the next subsection, this is the key condition (besides analyticity and pointwise properness of $\wm$) which will make it possible to modify invariant homomorphisms along Lie algebra generated  curves that correspond to $\g\in \mg\backslash \mg_x$. Basically, this will be done by replacing the last factor in the general parallel transport formula \eqref{eq:trivpar} (for invariant connections along Lie algebra generated curves) by certain equivariant mappings $\Psi\colon \spann_\RR(\g)\rightarrow S$. The point here is that if the invariant homomorphism comes from an invariant connection, the left hand side of \eqref{eq:staby} already implies that the value of this homomorphism on both curves coincide, just because parallel transports are invariant under reparametrizations. In the purely algebraical setting, however, we will need \eqref{eq:staby} in order to guarantee this. 
\item 
\label{rem:ppropercurve3}
Observe that $\gamma^x_{\vec{g}}$ is analytic if $\wm$ is analytic because the exponential map of $G$ is analytic. Part \ref{lemma:sim5} of the next lemma then states that if $\wm$ is in addition pointwise proper, each Lie algebra generated curve $\gag$ is maximal in the following sense:

An analytic immersion $\gamma$ sharing an open segment with $\gag$, i.e., $\gamma\cpsim \gag$ is already a subcurve of $\gag$. This will be important for our modifications in Subsection \ref{sec:ModifLAGC} and  
 is usually not true if $\wm$ is not pointwise proper:

Indeed, let $\wm\colon \RR_{>0}\times \RR^n \rightarrow \RR^n$, $(\lambda,y)\mapsto \lambda\cdot y$, $0\neq x\in\RR^n$ and $\g=1\in T_1\RR_{>0}$. Then $\gag(t)=\e^{t}\cdot x$ since the exponential map $\exp\colon \mg \rightarrow \RR_{>0}$ is just given by $\lambda\mapsto \e^{\lambda}$. However, obviously $\gamma\colon \RR \rightarrow \RR$, $t\mapsto t\cdot x$ is an immersion with $\gamma\cpsim \gag$ but $\im\big[\gag\big]=\RR_{>0}\cdot x$ is properly contained in $\im[\gamma]=\RR \cdot x$.\hspace*{\fill}$\dagger$

In addition to that, pointwise properness will guarantee that $\xsim$ even defines an equivalence relation on $\mg\backslash \mg_x$ as we will see in the last Part of the next lemma. 
\item
\label{rem:ppropercurve4}
 The second part of Lemma \ref{lemma:sim} shows that Lie algebra generated curves are always immersions and that for each $\g\in \mg\backslash \mg_x$ there is a unique number $\tau_\g\in \RR_{>0}\sqcup{ \infty}$ such that $\gag|_{[a,a+l]}$ is embedded iff $l<\tau_\g$. It is a very useful observation that this already implies that $\Ad_h(\g)=\pm \g$ holds for each $h\in G_{[\g]}^x$ provided that $\wm$ is pointwise proper.

 	In  fact, if $\lambda\neq \pm 1$, then $\lambda\neq 0$ because $\Ad_{h^{-1}}\cp\Ad_h=\id_\mg$. In particular, replacing $h$ by $h^{-1}$, we can assume that $|\lambda|<1$ holds. Since $\Ad_h$ is linear, scaling $\g$ if necessary, by the above statements, we also can assume that $y:=\wm(\exp(\g), x)\neq x$. Now, the sequence defined by 
	\begin{align*}
		\qquad\qquad\qquad\quad\wm(h^n,y)&=\wm(h^n\cdot \exp(\g)\cdot h^{-n},x)\qquad\qquad\qquad\qquad\qquad\qquad (h\in G^x_{[\g]}\subseteq G_x)\\
		&=\wm(\exp(\Ad_{h^{n}}(\g)),x)=\wm(\exp(\lambda^n\cdot \g),x)
	\end{align*}		
	has the same limit as $\{\wm(h^n,x)\}_{n\in \NN}$, namely $x$. Then, pointwise properness of $\wm$ implies that $g\cdot y =g \cdot x$ holds for some $g\in G$. Hence, $y=x$, which contradicts the choice of $y$.  
\hspace*{\fill}$\lozenge$
\end{enumerate}
\endgroup
\end{remdef}
\begin{lemma}[Lie Algebra Generated Curves]
  \label{lemma:sim}  
  Let $x\in M$ and $\vec{g},\vec{g}'\in \mathfrak{g}\backslash \mathfrak{g}_x$. 
  \begin{enumerate}
  \item
    \label{lemma:sim1}  
    The horizontal lift of $\gamma:=\wm_{g'}\cp \gamma^x_{\vec{g}}|_{[0,l]}$  
    w.r.t.\ the invariant connection $\w$ in the point $p\in F_x$ is given by (recall that $\wt{g}$ denotes the fundamental vector field \eqref{eq:fundvf} which corresponds to $\g$)
    \begin{align}	
      \label{eq:trivpar}
      \wt{\gamma}(t):=g'\cdot \exp(t\cdot\vec{g})\cdot p\cdot \exp(-t \cdot \w(\wt{g}(p)))\qquad \forall\: t\in [0,l].
    \end{align}
  \item
    \label{lemma:sim2} 
    The curve $\gamma:=\gamma^x_{\vec{g}}$ is an immersion. 
    \begingroup
	\setlength{\leftmarginii}{17pt}
    \begin{itemize}
    \vspace{-4pt}
    \item
    	If $\gamma$ is injective, then $\gamma|_{[a,b]}$ is embedded for all $[a,b]$ with $a<b$.
    \item
    \vspace{3pt}
    	If $\gamma$ is not injective, then it is cyclic in the sense that there is $\tau\in \RR_{>0}$ uniquely determined, such that 
	\begin{align*}
	\gamma(t)=\gamma(t')\qquad\Longleftrightarrow \qquad t=t' + n\tau \quad\text{for some}\quad n\in \mathbb{Z}.
	\end{align*}	    	 
    	Hence, the curves $\gamma|_{[a,b]}$ are embedded iff $b\in (a,a+\tau)$ for $a,b\in \RR$.
    \end{itemize}
    \endgroup     
  \item
    \label{lemma:sim3}
    Let $V\subseteq \mg$ be an $\Add{G_x}$-invariant linear subspace with $V\cap \mg_x =\{0\}$. 
   Then,\footnote{Here, $\psim$ is defined as in the analytic case, whereby the respective diffeomorphism $\adif$ is assumed to be smooth instead of analytic.}
    \begin{align*}
	\gamma_{\g'}^x|_{[0,l']} \psim \wm_h\cp \gamma_{\g}^x|_{[0,l]}\quad\text{for }\quad h\in G_x,\: \g,\g'\in V \qquad\Longrightarrow \qquad \textstyle\frac{l}{l'} \vec{g}=\Add{h^{-1}}(\g').
	%\textstyle\frac{l}{l'} \vec{g}'= \Add{h^{-1}}(\g),
	\end{align*}
  \item
    \label{lemma:sim4} 
    If $\vec{g}\xsim \vec{g}'$, then 
    $\lambda \vec{g}-\Add{h^{-1}}(\vec{g}')\in \mathfrak{g}_x$
    for some $h\in G_x$ and $\lambda\in \RR_{\neq 0}$. In particular, if $G_x=\{e\}$, then $\vec{g}\nsim_x \vec{g}'$ whenever $\g,\g'$ are linearly independent.
  \item
    \label{lemma:sim5} 
    Let $\wm$ be analytic and pointwise proper.
        \begingroup
	\setlength{\leftmarginii}{15pt}
	\begin{enumerate}
		\item
		\label{sim5a}
		\vspace{-4pt}
			If $\gamma\colon D \rightarrow M$ is an analytic immersion with $\gamma\cpsim\wm_g \cp \gag$, then $\gamma=\wm_g \cp\gag\cp\adif$ for some analytic diffeomorphism $\adif\colon I\rightarrow \RR$ ($I$ open).
			\vspace{2pt}
		\item
		\label{sim5b}
		     If 
 $\delta\colon K \rightarrow M$ is an embedded analytic curve with $\delta \cpsim \wm_g\cp\gamma_{\g}^x$, then 
    $\delta \psim \wm_g\cp\gamma_{\pm\g}^x|_{K}$ 
    for $K \subseteq \RR$ a compact interval.
	\end{enumerate}	 
	\endgroup   
	   \item
    \label{lemma:sim444}
    	If $\wm$ is analytic and pointwise proper, then $\xsim$ defines an equivalence relation on $\mg\backslash \mg_x$.
  \end{enumerate}
  \end{lemma} 
 	Obviously, claim $\textit{(b)}$ in Part \ref{lemma:sim5} is immediate from claim $\textit{(a)}$ in Part \ref{lemma:sim5}, and it also can be proven by the techniques from Lemma \ref{lemma:BasicAnalyt}.\ref{lemma:BasicAnalyt4}. We here decided to show the more general statement as it makes some of our further proofs more elegant, and because the only additional notion which we need is that of a maximal extension of an analytic (immersive) curve. At this point, the author expresses his gratitude to Christian Fleischhack for advising him to introduce maximal extensions in order to prove the more general statement.   
  \begin{proof}
    \begin{enumerate}
    \item
      Let $\xi:= \w(\wt{g}(p))$. 
      Then, $\pi(\wt{\gamma}(t))=\wm_{g'}\cp\wm(\exp(t\vec{g}),x)=\gamma(t)$ and 
      \begin{align*}
        \w\big(\raisebox{-1pt}{$\dot{\wt{\gamma}}(t)$}\big)%= \w\left(\dd_e\Phi_{\exp(t\vec{g})\cdot p\cdot\exp(-t \xi)}(\vec{g}) -\wt{\xi}(\exp(t\vec{g})\cdot p\cdot\exp(-t\xi)) \right)
        &\stackrel{\eqref{eq:trivpar}}{=} \w\Big((\dd R_{\exp(-t\xi)}\cp \dd L_{\exp(t \vec{g})})(\wt{g}(p))\Big)
        - \w\left(\wt{\xi}\big(R_{\exp(-t\xi)}\big(\Phi(\exp(t\vec{g}), p)\big)\big)\right)\\
        &\:\hspace{1pt}= (\Add{\exp(t\xi)}\cp\: \w)(\wt{g}(p)) - \xi=\Add{\exp(t\xi)}(\xi) -\xi =0,
      \end{align*}
      where we have used $\Phi$-invariance of $\w$ in the first two steps.
    \item
      First observe that $\gamma$ is an immersion because\footnote{Observe that $\Ad_g\colon \mg_x \rightarrow \mg_y$ is an isomorphism for $y=\wm(g,x)$. In fact, if $\Ad_{{g}^{-1}}(\g)\in \mg_x$ for $\g\in \mg$, then $\g=\Ad_{g}(\vc)$ for $\vc\in \mg_x$, hence $\wm(\exp(t\g),y)=\wm_{g}\cp \wm(\exp\big(t\vc), x)=y$ for all $t\in \RR$, i.e., $\g\in \mg_y$.} 
      \begin{align*}
        \dot \gamma (t)=\dd_e\wm_{\wm(\exp(t\vec{g}), x)}(\vec{g})=0\qquad \Longleftrightarrow\qquad  \vec{g}\in \mathfrak{g}_{\wm(\exp(t\vec{g}), x)}=\Add{\exp(t\vec{g})}(\mathfrak{g}_x).
      \end{align*} 
      In fact, then $\dot \gamma (t)=0$ implies $\vec{g}=\Add{\exp(t\vec{g})}\big(\vec{h}\big)$ for some $\vec{h}\in \mathfrak{g}_x$, hence $\vec{h}=\Add{\exp(-t\vec{g})}(\vec{g})=\vec{g}$, which contradicts the choice of $\vec{g}$. 
      
      If $\gamma$ is injective and $[a,b]\subseteq \RR$, then for $\epsilon>0$ the curve $\gamma|_{[a-\epsilon,b+\epsilon]}$ is an embedding just by compactness of $[a-\epsilon,b+\epsilon]$. Of course, $\gamma|_{(a-\epsilon,a+\epsilon)}$ then is an embedding as well. 

      Now, assume that $\gamma$ is not injective and observe that   
      \begin{align}
        \label{eq:cycl}
        \gamma(t')=\gamma(t)\qquad \Longleftrightarrow\qquad\gamma(t'+a)=\gamma(t+a)\text{ for all }a\in \RR .
      \end{align}
      Let $\tau$ denote the infimum of $\{t\in \RR_{>0}\:|\: \gamma(0)=\gamma(t)\}\subseteq \RR$. 
       Then $\gamma(\tau)=\gamma(0)$ by continuity of $\gamma$, and $\tau>0$ because $\gamma$ is locally injective as it is an immersion. 
      So, if $\gamma(t')=\gamma(t)$ for $t,t'\in \RR$, then $\gamma(0)=\gamma(t-t')$ by \eqref{eq:cycl}, and we have
      $0\leq (t-t')+n \tau \leq \tau$ for some $n\in \mathbb{Z}$. Then, minimality of $\tau$ shows that $t-t' + n\tau \in \{0,\tau\}$, hence $t-t'=n' \tau$ for some $n'\in \mathbb{Z}$.
    \item 
      By assumption, we find a diffeomorphism $\adif\colon  I'\rightarrow \RR$ with $\gamma_{\g'}^x|_{[0,l']}=\wm_h\cp \gamma_{\g}^x\cp \adif|_{[0,l']}$ where $\adif(0)=0$ and $\adif(l')=l$. Then  
%      This means that for
      \begin{align*}
      G_x\ni H(t):=\exp(-t\cdot \mathrm{Ad}_{h^{-1}}(\g'))\cdot \exp(\rho(t)\cdot \g) \qquad \forall\: t\in [0,l']
      \end{align*}
      and $H(0)=e$.
      Hence, $\mg_x\ni\dot H(0)=\dot\rho(0)\cdot\g -\Add{h^{-1}}(\g')$, so that by $\Add{G_{x}}$-invariance of $V$ and since  $\g,\g'\in V$, $V\cap \mg_x=\{0\}$, we have
      \begin{align}
        \label{eq:vor}
      	\Add{h^{-1}}(\g')=\dot\rho(0)\cdot\g.
      \end{align}
      Then 
        $H(t)=\exp(-t \dot\adif(0)\cdot\g)\cdot \exp(\adif(t)\cdot\g)=\exp([\adif(t)-t \dot\adif(0)]\cdot \g)$,
      hence
      \begin{align*} 
      	\mg_x\ni \dd L_{H(t)^{-1}}\dot H(t)=[\dot\adif(t)-\dot\adif(0)]\cdot\g.
	\end{align*}     
      This shows $\dot\adif(t)=\dot\adif(0)$, hence $\adif(t)=\lambda t$ for $\lambda=\dot\adif(0)$. Then $\lambda=\frac{l}{l'}$ because $l=\adif(l')=\lambda l'$, 
      so that the rest is clear from \eqref{eq:vor}.    
    \item
      By assumption, we find $g\in G$ and $I,I' \subseteq \RR$ open intervals with       
     	$\wm_g\cp \gamma^x_{\vec{g}}(I)=\gamma^x_{\vec{g}'}(I')$.       
      Since both curves are embeddings, $\adif\colon I'\rightarrow I$ defined by 
      \begin{align*}	
        \adif:=\left(\wm_g\cp \gamma^x_{\vec{g}}\big|_{I}\right)^{-1}\cp \gamma^x_{\vec{g}'}|_{I'}
      \end{align*}		
      is a diffeomorphism %$\adif\colon (t'_0,t'_1)\rightarrow (t_0,t_1)$ 
      for which we can assume\footnote{In fact, elsewise we replace $\g$ by $-\g$. For this, observe that $\gamma_\g^x|^{-1}_{[0,l]}= \wm_{\exp(l\g)}\cp\gamma_{-\g}^x|_{[0,l]}$.} that $\dot\adif>0$.
	Now,  
	\begin{align*}
		G_x\ni \exp(-t'\g')\cdot g\cdot \exp(\adif(t')\:\g)\qquad\forall\:t'\in I'
	\end{align*}        
	and for $t_0'\in I'$ fixed we define 
     $G_x\ni h:=\exp(-t'_0\vec{g}')\cdot g\cdot\exp(\adif(t'_0)\:\vec{g})$. 
      Then, for 
      \begin{align*}
        \delta(t'):=h^{-1}\cdot\exp(-t'\g')\cdot g\cdot\exp(\adif(t')\:\g)\qquad\forall\:t'\in I'
      \end{align*}	
      we have $\delta(t'_0)=e$,
      and a simple calculation shows that 
      \begin{align*}
      		\mathfrak{g}_x\ni\dot\delta(t'_0)= \lambda\vec{g}-\Add{h^{-1}}(\vec{g}')\qquad\text{ for }\qquad\lambda := \dot\adif(t'_0)\neq 0.     
      \end{align*}
    \item
It suffices to consider the case where $g=e$ because $\wm$ is analytic, and $\gamma$ and $\wm_g\cp \gamma_{\g}^x$ share an open segment iff $\wm_{g^{-1}}\cp \gamma$, $\gamma_{\g}^x$ do so. 
  Then, substituting $\g$ by $-\g$ if necessary we find an analytic diffeomorphism $\adif \colon I\rightarrow \RR$ with $\gamma|_I=\gag\cp \adif$ and $\dot\adif>0$, see also Part \ref{lemma:sim4}). Let $\gamma$ be maximal immersive and denote by $\adif'\colon I'\rightarrow \RR$ the maximal immersive extension of $\adif$. Then $\gamma|_{I'}=\gag \cp \adif'$ because $\gamma|_{I'}$ and $\gag \cp \adif'$ coincide on $I$. We now have to show that $I'=\dom[\gamma]$. For this, let $I'=(r',s')$, $\dom[\gamma]=(r,s)$ and assume that $s'<s$.
  
	We choose a monotonous increasing sequence $\{s'_n\}_{n\in \mathbb{N}}\subseteq I'$ with $\lim_n s'_n= s'$:	
  \begingroup
  \setlength{\leftmarginii}{17pt}
	\begin{itemize}
	\item
	  	\vspace{-3pt}
		If $\lim_n \adif(s'_n)=t$ exists, then $\gamma(s')$ is an accumulation point of $\im[\gamma]\cap \im[\gag]$, so that 
		Lemma \ref{lemma:BasicAnalyt}.\ref{lemma:BasicAnalyt1} provides us with 
		open intervals $J'\ni s'$ and $J\ni t$, 
		 and an analytic diffeomorphism $\adif_0\colon J'\rightarrow J$ with $\gamma|_{J'}=\gag\cp \adif_0$. Then, glueing together $\adif'$ and $\adif_0$ provides us with a contradiction to maximality of $\adif'$.
  	\item 	
  		\vspace{2pt}
  	 	If $\lim_n \adif(s'_n) =\infty$, let\footnote{We define $\tau_\g:=\infty$ iff $\gag$ is injective, see also next definition.} $0<d<\tau_\g$ and choose $s'_n$ in such a way that 
		\begin{align*}
		\adif(s'_n)=\adif(s'_0)+n\cdot d\qquad \forall\: n\in \NN_{\geq 1}.
	\end{align*}		  	 	
  	 Then, for $x_1:=\wm_{\exp(s'_0\cdot \g)}(x)$, $x_2:=\wm_{\exp(d\cdot \g)}(x_1)\neq x_1$ and $g_n:=\exp(n d\cdot \g)$ we have
  	 \begin{align*}
  	 	\lim_n \wm(g_n,x_1)=\gamma(s')=\lim_n \wm(g_n,x_2).
  	 \end{align*}
  	 Hence, $\wm(g,x_1)=\wm(g,x_2)$ for some $g\in G$ by pointwise properness of $\wm$. This implies $x_1=x_2$ and contradicts the choices.
  	\end{itemize}
  	Consequently, $s'=s$, and in the same way we see that $r=r'$ holds.
  	\endgroup 
    \item
		This is a straightforward consequence of Part \ref{lemma:sim5}. In fact, if $\g\xsim\g'$ and $\g'\xsim \g''$ with
	\begin{align*}		
		 \wm_g\cp \gag \cpsim \gamma_{g'}^x \qquad\quad \text{as well as}\qquad\quad \gamma_{g'}^x \cpsim \wm_{g''}\cp \gamma_{\g''}^x,
	\end{align*}
		then for $K'$ a compact interval by Part \ref{lemma:sim5} we find open intervals $K$ and $K''$ such that 
	\begin{align*}		
		\wm_g\cp \gamma^x_{\pm\g}|_{K}\psim \gamma^x_{\g'}|_{K'}\psim \wm_{g''}\cp\gamma^x_{\pm\g''}|_{K''}
	\end{align*}		
		holds, hence $\g\xsim \g''$.  
  \end{enumerate}
  \end{proof} 
So, for the case that $\wm$ is analytic and pointwise proper, the above lemma provides us with the following notions:
\begin{definition}
  \label{def:liegeneq}
  Let $\wm$ be analytic and pointwise proper. 
  \begin{enumerate}
  \item
    \label{def:liegeneq144}
  For $\g\in\mg \backslash \mg_x$, we denote by $\tau_\g\in \RR_{>0}\sqcup {\infty}$ the period of $\gag$ introduced in Lemma \ref{lemma:sim}.\ref{lemma:sim2}, where we define $\tau_\g:=\infty$ if $\gag$ is injective. Recall that $\gag|_{[a,b]}$ is an embedding iff $b-a<\tau_\g$. 
  \item
    \label{def:liegeneq23}
  	By \gls{XSIMORB} we will denote the set $(\mg\backslash \mg_x) \slash_{\xsim}$ of equivalence classes  w.r.t.\ $\xsim$.
  \item  
    \label{def:liegeneq1}
    By $\text{\gls{PAG}}\subseteq \Paw$ we denote the set of all embedded analytic curves of the form $\wm_g \cp \gag|_K$ for $x\in M$, $\g\in \mg\backslash \mg_x$, $g\in G$ and $K\subseteq \RR$ compact.\footnote{Of course, for $K=[k_0,k_1]$ this means that $k_1-k_0<\tau_\g$ for $\tau_\g$ the period of $\gag$.} 
   \item  	
    \label{def:liegeneq2}  	
    By $\text{\gls{PAGS}} \subseteq \Paw$ we will denote the set of all curves equivalent ($\psim$) to a curve in $\Pag$. By the second part of Lemma \ref{lemma:sim}.\ref{lemma:sim5}, this is exactly the set of all $\gamma\in \Paw$ with $\gamma \cpsim \delta$ for some $\delta\in \Pag$.
    \item
    \label{def:liegeneq11}
    	For $x\in M$, $g\in G$ and $y:=\wm_g(x)$, we define $\wt{\Add{g}}\colon \xsimorbb_x\rightarrow \xsimorbb_y$, $[\g]\mapsto [\Add{g}(\g)]$. This map is well defined because 
	\begin{align*}    	
    	\g\xsim \g'&\quad\Longrightarrow \quad \wm_{g_0}\cp \gag \cpsim \gamma^x_{\g'}\:\text{ for some }\: g_0\in G_x\\
    	& \quad\Longrightarrow \quad \wm_{g\cdot g_0\cdot g^{-1}}\cp \gamma^{y}_{\Ad_g(\g)} \cpsim \gamma^y_{\Ad_g(\g')}\\
    	  &\quad\Longrightarrow \quad \Ad_{g}(\g)\sim_y \Ad_{g}(\g').
    \end{align*}
    Then, $\wt{\Ad}_{g}$ is even bijective as its inverse is just given by $\wt{\Ad}_{g^{-1}}\colon \xsimorbb_y\rightarrow \xsimorbb_x$.
  \end{enumerate}
\end{definition}
We close our considerations by stating that in the analytic and pointwise proper case $\Pag, \Pags$ and $\Paw\backslash \Pags$ are closed under decomposition and inversion of their elements. In particular, by Proposition \ref{rem:euklrem2b} we have the splitting\footnote{Provided, of course, that $\Pag$ and $\Paw\backslash \Pags$ are not empty. Elsewise, the respective factor just has to be dropped.} $\AQRw\cong \AQRInd{\mg}\times \AQRInd{\mg^c}$ where the latter factor is the quantum-reduced space which corresponds to the set $\Paw\backslash \Pags$. 
\begin{corollary}
  \label{cor:decompo}
  Let $\wm$ be analytic and pointwise proper. 
  Then, $\Pag, \Pags$ and $\Paw\backslash \Pags$ are closed under decomposition and inversion of their elements.
  \begin{proof}
    For $\Pag$ and $\Pags$ the statement is clear from the definitions and 
    \begin{align}
      \label{eq:hom}
    %  \gag|_{[a,b]}\psim \wm_{\exp(a \g)}\cp \gamma_{\g}^x|_{[0,b-a]} %\qquad\quad\text{and}\qquad\quad 
    \big[\wm_g\cp\gag|_{[a,b]}\big]^{-1}= \wm_{g }\cp \gamma_{-\g}^{y}|_{[a,b]}\qquad\text{for}\qquad y:=\wm(\exp([a+b]\cdot\g),x).
    \end{align}
    Now, if $\gamma$ is not equivalent to an element of $\Pag$, i.e., if $\gamma\in \Paw\backslash \Pags$,  then 
    the same must be true for $\gamma^{-1}$, just because  $\Pags$ is closed under inversions and
	\begin{align*}
		\gamma\psim \delta \qquad \Longleftrightarrow \qquad\gamma^{-1}\psim \delta^{-1}.
	\end{align*}
    Moreover, by Lemma \ref{lemma:sim}.\ref{lemma:sim5}
    there cannot exist a subcurve $\gamma'=\gamma|_K$ of $\gamma$ which is equivalent to an element of $\Pag$, i.e., which is contained in $\Pags$. Consequently, $\Paw\backslash \Pags$ is closed under decompositions and inversions as well.
  \end{proof}
\end{corollary}

\subsection{Modifications along Lie Algebra Generated Curves}
\label{sec:ModifLAGC}
In the sequel, let $\wm$ be always analytic and pointwise proper and $\dim[S]\geq 1$.\footnote{Recall that this ensures that the equivalence relations $\csim$ and $\psim$ coincide, see Lemma \ref{lemma:BasicAnalyt}.\ref{lemma:BasicAnalyt2}.} Moreover, let $\Pagw\subseteq \Paw$ be a $\Phi$-invariant subset which is closed under decompositions and inversions and that contains all Lie algebra generated curves, i.e., the set $\Pag$ from Definition \ref{def:liegeneq}.\ref{def:liegeneq1}. Then, according to Convention \ref{conv:invhomm}, we will denote by $\ARQLAI$ the respective quantized reduced classical configuration space, and  
  	by $\text{\gls{IHOMLAI}}\cong\text{\gls{KAPPALAI}}\big(\text{\gls{AQRLAI}}\big)$ the respective quantum-reduced one.  

We now are going to modify invariant homomorphisms, i.e., elements of $\IHOMLAI$ along Lie algebra generated curves. Here, we will make use of the fact that for each $p\in F_x$ the action $\Phi$ provides us with a canonical lift of $\gag$, namely $t\mapsto \Phi_{\exp(t\g)}(p)$. Basically, then we will mimic \eqref{eq:trivpar} by replacing the last factor 
	\begin{align*}
		\Psi_\w \colon \g \mapsto \exp(-\w(\wt{g}(p)))
	\end{align*}
	by maps  
		$\Psi\colon V\rightarrow S$ (with $V$ a special linear subspace of $\mg$) 
	 having the correct equivariance property. For this, observe that by invariance of $\w$  
	 we have 
\begin{align*} 	 
	 \Psi_\w \cp\Ad_h =\alpha_{\phi_p(h)}\cp\:\Psi_\w\qquad\forall\:h\in G_{\pi(p)}
	 \end{align*}
	  with $\text{\gls{FIBAP}}\colon G_{\pi(p)}\rightarrow S$  the unique Lie group homomorphisms which fulfils (cf.\ \eqref{eq:phip})
	\begin{align*}
		 \Phi(h,p)=p\cdot \fiba_p(h)\qquad \forall\:h\in G_{\pi(p)}. 
	\end{align*}	
	For instance, 
	%\begin{definition}	
	let  $V\subseteq \mg$ be an $\Add{G_{x}}$-invariant linear subspace with $V\cap \mg_{x}=\{0\}$. Then,  a linear map $L\colon V\rightarrow \ms$ is called $\Add{G_{x}}^p$-equivariant iff 
	\begin{align*}    
    	\Ad_{\fiba_p(h)}\cp \: L = L\cp \Ad_{h}\qquad \forall\: h\in G_{x},
    \end{align*}
   and in the second part of Proposition \ref{th:invhomm} we will modify invariant homomorphism by means of such maps. This will be helpful, in particular, for our investigations concerning the inclusion relations between the spaces $\ARQLAI$ and $\AQRLAI$ in Subsection \ref{sec:inclrel}. However, for our investigations in Subsection \ref{sec:ConSp} (construction of measures), we will need a slightly different version of this:	
\begin{definition}
\label{def:eqmaps}
A map $\Psi\colon \spann_\RR(\g) \rightarrow S$ with
    \begin{align}
    \label{eq:equii}	
      % \label{eq:homeig}
      \Psi((\lambda+\mu)\cdot\g)=\Psi(\lambda\cdot \g)\cdot \Psi(\mu\cdot \g)\qquad \forall\:\lambda,\mu\in \RR 
      %\quad\qquad\text{as well as} \quad\qquad \Psi(\g,l+l')=\Psi(\g,l)\cdot\Psi(\g,l')
     % %\qquad \forall\:l,l'\in \RR_{>0},\forall\: \lambda\neq 0
    \end{align}   
     is said to be $\Add{G_{[\g]}}^p$-equivariant iff 
	\begin{align} 
	   \label{eq:equiii}    
     	\alpha_{\fiba_p(h)}\cp\:\Psi=\Psi\cp \Ad_h\qquad\forall\:h\in G^x_{[\g]}
	\end{align}     	
	holds. \hspace*{\fill}$\lozenge$
\end{definition}
	 Then, in order to obtain well-defined homomorphisms when modifying by means of such maps, we will be forced to require $\g$ to have the additional property of stability. This generalizes the condition in Lemma \ref{lemma:sim}.\ref{lemma:sim3} which automatically holds for $\Add{G_{x}}$-invariant linear subspaces as discussed above.  
	\begin{definition}[Independent and Complete Families]
	\label{def:stable}
	 \begin{enumerate}
 	\item	
 		\label{def:stable2} 
 		An element $\g\in \mg \backslash\mg_x$ is called \emph{stable} iff 
	\begin{align}
	\label{eq:stab1}
		\gamma^x_{\g}|_{[0,l]} \psim \gamma^x_{\pm\Ad_h(\g)}|_{[0,l']}\quad\text{for}\quad h\in G_x\qquad \Longrightarrow\qquad h\in G_{[\g]}^x.%\frac{l}{l'}\g=\Ad_{h^{-1}}(\g').
	\end{align}	
	\item
	 \label{eq:iindepg3}    
    \itspacec
    A family of stable elements $\{\g_{\alpha}\}_{\alpha\in I_x}\subseteq \mg\backslash \mg_x$ is said to be \emph{independent} iff $\g_\alpha \nsim_x \g_\beta$ holds for all $\alpha,\beta \in I_x$ with $\alpha\neq \beta$. It is called \emph{complete} iff for each $\mathfrak{r}\in \xsimorb$ we have $\g_\alpha \in \mathfrak{r}$ for some $\alpha\in I_x$.\footnote{Recall that $\xsimorb$ denotes the classes in $\mg\backslash \mg_x$ w.r.t.\ $\xsim$.} 
	 \end{enumerate}
	\end{definition}
	\begin{lemrem}[Stability]
	\label{rem:dfggfg}
	\begin{enumerate}
	\item
		\label{rem:dfggfg0}
	It already follows from stability \eqref{eq:stab1} of $\g \in \mg\backslash \mg_x$ (and analyticity) that 
	\begin{align*}
		 \pm\Ad_h(\g)=\g\qquad\text{and}\qquad l'=l\qquad \text{holds},
	\end{align*}		
	hence $\gamma^x_{\g} = \gamma^x_{\pm\Ad_h(\g)}$.

	 In fact, by Remark and Definition \ref{rem:ppropercurve}.\ref{rem:ppropercurve4}, we have $\Ad_h(\g)=q\cdot \g$ for $q\in\{-1,1\}$, hence $\gamma^x_{\g}|_{[0,l]} = \gamma^x_{\pm q\cdot\g}\cp \adif|_{[0,l]}$ for $\adif\colon [0,l]\rightarrow [0,l']$ a diffeomorphism with $\dot\adif>0$. Then,
	\begin{align*}	
		H(t):=\exp(-t\cdot \g)\cdot \exp(\pm q \adif(t)\cdot \g)\in G_x \qquad \forall\: t\in [0,l],
	\end{align*}	
		 so that $\pm q=1$ and $\dot\adif=1$ because
	\begin{align*}
	\mg_x\ni \dd L_{H(t)^{-1}}\dot H(t)=[\pm q\dot\adif(t)-1]\cdot \g\qquad \forall \: t\in [0,l].	
	\end{align*} 
	This shows  $\pm\Ad_h(\g)=\pm q\cdot \g=\g$ as well as $\adif(t)=t$, hence $l'=\adif(l)=l$. 
	\item
		\label{rem:dfggfg1}
		As already mentioned in Remark and Definition  \ref{rem:ppropercurve}.\ref{rem:ppropercurve2}, Condition \eqref{eq:stab1} generalizes 	 
	 the property from Lemma \ref{lemma:sim}.\ref{lemma:sim3} in a specific way.
	  	  
	 More concretely, $\g_0\in \mg\backslash \mg_x$ is stable iff
	\begin{align}
		\label{eq:stab2}
		\gamma^x_{\g'}|_{[0,l']} \psim \gamma^x_{\Ad_h(\g)}|_{[0,l]}\quad\text{for}\quad h\in G_x,\:\g,\g'\in [\g_0]\qquad \Longrightarrow\qquad \frac{l}{l'}\g=\Ad_{h^{-1}}(\g').
	\end{align}
In fact, obviously \eqref{eq:stab2} implies \eqref{eq:stab1}. Now, if \eqref{eq:stab1} holds for $\g_0$, and if we have    
		\begin{align}
		\label{eq:stab2gfgf}
		\gamma^x_{\g'}|_{[0,k']} \psim \gamma^x_{\Ad_h(\g)}|_{[0,k]}\quad\text{for}\quad h\in G_x,\:\g,\g'\in [\g_0],
	\end{align}  
	then $\g=q\lambda\cdot \g_0$, $\g'=q'\lambda'\cdot \g_0$ for some $\lambda,\lambda'>0$ and $q,q'\in \{-1,1\}$, hence 
			\begin{align*}
		\gamma^x_{q'\cdot \g_0}\big|_{[0,\lambda'k']} \psim \gamma^x_{q\cdot \Ad_h(\g_0)}\big|_{[0,\lambda k]}.
	\end{align*}
	Then, the group property of $G^x_{[\g_0]}$ and \eqref{eq:stab1} show that $h\in G^x_{[\g_0]}$ if $q'=1$ or $q=1$. Moreover, if $q,q'=-1$, by analyticity of the involved curves we have  
	\begin{align*}
		\gamma^x_{ \g_0}|_{[0,r']} \psim \gamma^x_{ \Ad_h(\g_0)}|_{[0,r]}
	\end{align*}
	for some $r',r>0$, so that \eqref{eq:stab1} shows $h\in G^x_{[\g_0]}$ as well. Consequently, 
	\begin{align*}	
	\Ad_h(\g)=\pm \g=\lambda \cdot \g'\qquad\text{for some}\qquad \lambda\neq 0, 
	\end{align*}	
	and then the arguments from Part \ref{rem:dfggfg0}) applied to \eqref{eq:stab2gfgf} show the right hand side of \eqref{eq:stab2}. 	
	\item
	\label{rem:dfggfg2}	 
	 It is clear from \eqref{eq:stab2} and Lemma \ref{lemma:sim}.\ref{lemma:sim3} that $\g$ is stable whenever $\spann_\RR(\g)$ is contained in an $\Ad_{G_x}$-invariant linear subspace $V$ of $\mg$ with $V\cap \mg_x=\{0\}$. For instance, this is always true if $G_x=\{e\}$ holds. 
	\item 
		Let 
	 $L\colon V\rightarrow \ms$ be an $\Ad_{G_x}^p$-equivariant linear map and $0\neq \g\in V$. Then, it is an interesting observation that\footnote{Here, the minus sign is just to show the resemblance to \eqref{eq:trivpar}.} 
	\begin{align*}	
	\Psi\colon \spann_\RR(\g)\rightarrow S,\quad \g\mapsto \exp(- L(\g))
	\end{align*}
	 is $\Add{G_{[\g]}}^p$-equivariant. Indeed, examples for such $\Ad_{G_x}^p$-equivariant linear maps $L\colon \mg\rightarrow \ms$ are  given by $(\Phi_p^*\w)|_{\mg}$ for $\w$ a $\Phi$-invariant connection on $P$. 
	 \item
	 	In Proposition \ref{th:invhomm}, we will modify elements $\homm\in \IHOMLAI$, first, by specifying their values on the curves $\gamma_{\g_0}^x|_{[0,l]}$. Second, by extending them to the curves $\wm_g\cp \gamma_{\g_0}^x|_{[0,l]}$ in such a way that the resulting homomorphism $\homm'$ is invariant.   
	Then, in order to guarantee that $\homm'$ is well defined, we will need condition \eqref{eq:stab1}, i.e., stability of $\g$. This condition guarantees that $\homm'$ takes the same values on each two curves being related as on the left hand side of \eqref{eq:stab2}. Basically, this is just because the equivariances of the involved maps then will cancel out each other in the correct way, see \eqref{eq:clacu0}. 
		 \item
	 	\label{rem:dfggfg44}
	 	The stability property of an element $\g\in \mg\backslash\mg_x$ usually has to be checked by hand. 
	 	
	 	Indeed, in Subsection \ref{sec:ConSp} (for the case that $S=\SU$) we will discuss the situation where each $\wm$-orbit $\m$ admits an independent and complete family $\{\g_\alpha\}_{\alpha \in I_x}\subseteq \mg\backslash\mg_x$ (of stable elements) for some $x\in\m$. We will show that then 
$\IHOMLAS$ is homeomorphic to a Tychonoff product of compact Hausdorff spaces on each of which a natural Radon measure exists. 
This will allow us to a define a normalized Radon measure on $\IHOMLAS$ just by taking the Radon product one. In particular, this will be possible for the cases of \mbox{(semi)-homogeneous}, spherically symmetric and homogeneous isotropic loop quantum cosmology as for these situations we will show by hand that such independent and complete families of stable elements exist. 
	 \hspace*{\fill}$\lozenge$
	 \end{enumerate}
%	 \endgroup 
	 \end{lemrem} 
The next lemma collects some elementary properties of independent and complete families that will become relevant for our considerations in Subsection \ref{sec:ConSp}. 
Recall that there we use such families in order to parametrize the space $\IHOMLAS$. The last point of the next lemma then will show injectivity of this parametrization. 
The first two points are just to switch between such different ones.   
\begin{lemma}
  \label{lemma:completee}
  Let $y=\wm_g(x)$ for $x\in M$ and $g\in G$. Moreover, let 
 $\{\g_{\alpha}\}_{\alpha\in I_x}\subseteq \mg\backslash \mg_x$ and $\{\g'_{\beta}\}_{\beta\in I_y}\subseteq \mg\backslash \mg_y$ both be independent and complete. 
  \begin{enumerate}
    \item
    \label{lemma:completee1}
    There is a unique bijection $\tau_{xy}\colon I_x\rightarrow I_y$ with\footnote{Recall Definition \ref{def:liegeneq}.\ref{def:liegeneq11} for the definition of $\wt{\Ad}$.}    
    $\wt{\Ad}_g([\g_\alpha])=[\g_{\tau_{xy}(\alpha)}]$ for all $\alpha\in I_x$.
  \item
  	\label{it:completee1}
  	Let $\wt{\Ad}_g([\g])=[\g']$ for $[\g] \in \xsimorb$ and $[\g']\in \xsimorbb_y$.   
	Then, we find $g_0\in G$ and an analytic diffeomorphism $\adif\colon \RR\rightarrow \RR$ with $\adif(0)=0$ and $\adif(\tau_\g)=\tau_{\g'}$, such that
	 \begin{align}
      \label{eq:simcon}
      \gamma^y_{\g'}= \wm_{g_0}\cp \gamma^x_{\pm \g}\cp \adif. 
    \end{align} 
    In particular, if $\g\xsim \g'$ for $\g,\g'\in \mg\backslash \mg_x$, then we find $h\in G_x$ with
    \begin{align}
      \label{eq:simconfestfuss}
      \gamma^x_{\g'}= \wm_h\cp \gamma^x_{\pm \g}\cp \adif =\gamma^x_{\pm \Ad_h(\g)}\cp \adif. 
    \end{align} 
   \item
    \label{lemma:homzueq33} 
    The values of $\homm\in \IHOMLAI$ on the curves 
	\begin{align*}
		    \wm_g\cp \gamma^y_{\g'}|_{[0,l']}\qquad \text{for all }g\in G,\:\g'\in \mg\backslash\mg_y\text{ and } l'\in (0,\tau_{\g'})
	\end{align*}	    
      are completely determined by its values on the curves $\gamma^x_{\g_\alpha}|_{[0,l]}$ for all $\alpha\in I_x$ and $l\in (0,\tau_{\g_\alpha})$. 	
  \end{enumerate}
  \begin{proof}
    \begin{enumerate}
    \item
    	This is clear from bijectivity of $\wt{\Ad}_g$.
    \item   
	Since $\Ad_g(\g)\xsim \g'$,  we find $g'\in G$ with $\gamma_{\Ad_g(\g)}^y\cpsim \wm_{g'}\cp \gamma^y_{\g'}$. Hence, 
    \begin{align*}
    	\wm_g\cp \gamma_{\g}^x=\gamma_{\Ad_g(\g)}^y\cpsim \wm_{g'}\cp \gamma^y_{\g'}\qquad&\Longrightarrow\qquad
    	  \wm_{g^{-1} g'}\cp \gamma^y_{\g'}\cpsim\gag \\ 
    	  \qquad&\Longrightarrow\qquad \wm_{g^{-1} g'}\cp\gamma^y_{\g'}= \gamma^x_{\pm\g}\cp \adif\hspace{50pt} (\text{Lemma } \ref{lemma:sim}.\ref{lemma:sim5})
    	  \\ \qquad&\Longrightarrow\qquad \gamma^y_{\g'}= \wm_{g'^{-1} g}\cp\gamma^x_{\pm\g}\cp \adif
    	  %\\ 
    	  %\qquad&\Longrightarrow\qquad \gamma^y_{\g'}|_{[0,l]}\psim \wm_{g'^{-1} g\cdot  \exp(\pm \adif(0)\cdot \g)}\cp\gag|_{[0,\adif(l)]}   	
    \end{align*}
	for $\adif\colon \RR\rightarrow \RR$ an analytic diffeomorphism.      
        \item
        By \eqref{eq:simcon} we find $\alpha\in I_x$ with $\wm_g\cp \gamma^y_{\g'}|_{[0,l]}\psim \wm_{g\cdot g_0}\cp \gamma_{\pm\g_\alpha}^x|_{[0,\adif(l)]}$, hence
		\begin{align*}
		\wm_g\cp \gamma^y_{\g'}|_{[0,l]}&\psim \wm_{g\cdot g_0}\cp \gamma_{\g_\alpha}^x|_{[0,\adif(l)]}\qquad\quad \text{or}\\
		\wm_g\cp \gamma^y_{\g'}|_{[0,l]}&\psim \wm_{g\cdot g_0}\cp \gamma_{-\g_\alpha}^x|_{[0,\adif(l)]}\psim \wm_{g\cdot g_0}\cp \wm_{\exp(-\adif(l)\cdot \g_\alpha)}\cp \big[\gamma_{\g_\alpha}^x|_{[0,\adif(l)]}\big]^{-1}.
		\end{align*}		
		The claim now follows from the invariance and homomorphism properties of $\homm$.
    \end{enumerate}
  \end{proof}
\end{lemma}
Before we come to the desired modification result, we want to show how the action  
\begin{align*}
\text{\gls{ADDSTRICH}}\colon G_x\times \pr_\mg(\mg\backslash \mg_x)\rightarrow \pr_\mg(\mg\backslash \mg_x),\quad
       (h,[\g])\mapsto \big[\!\Ad_h(\g)\big]
\end{align*}
	introduced in Remark and Definition \ref{rem:ppropercurve}.\ref{rem:ppropercurve1} can help to find independent and complete families as discussed above. 
\begin{remark}[$\Ad'$-Orbits]
\label{lemremmgohnermgx}
 Let $\madgx:=\pr_\mg(\mathfrak{g}\backslash \mg_x)\slash G_x$ denote the set of $\Ad'$-orbits in $\pr_\mg(\mathfrak{g}\backslash \mg_x)$ and choose 
	$\oo \in \madgx$ as well as $[\g],[\g']\in \oo$. Then, $\g'=\lambda \Ad_h(\g)$ for some $\lambda\neq 0$ and $h\in G_x$. Hence, 
	\begin{align}
	\label{eq:compladorb}
		 \gamma^x_{\g'}=\gamma^x_{\lambda \Ad_h(\g)} = \wm_h\cp \gamma_{\lambda \g}^x \cpsim \wm_h\cp \gamma_{\g}^x,  
	\end{align}
	which shows that $\g\xsim\g'$. 	
	So, in order to obtain a complete and independent family $\{\g_\alpha\}_{\alpha\in I}\subseteq \mg\backslash \mg_x$ as in Definition \ref{def:stable}.\ref{eq:iindepg3}, one can proceed as follows:
	  \begingroup
  \setlength{\leftmargini}{15pt}
	\begin{itemize}
		\item[--]
		\vspace{-2pt}
			First, one identifies two orbits $\oo,\oo' \in \madgx$ iff there are $[\g]\in \oo$ and $[\g']\in \oo'$ with $\g\xsim \g'$.
		\item[--]
		Second, if possible, one chooses a stable element $[\g_\alpha]$ in each of the ``remaining'' orbits.
	\end{itemize}
	\endgroup
\end{remark}
The first part of the next proposition shows, how one can modify invariant homomorphisms along Lie algebra generated curves that  correspond to stable elements $\g\in \mg\backslash \mg_x$. The second part is adapted to the requirements of the next subsection, where we will use it in order to construct invariant generalized connections that cannot be approximated by the classical (smooth) invariant ones. Hence, cannot be contained in $\ARQLAI$. 
\begin{proposition}
  \label{th:invhomm}
  Let $\wm$ be analytic and pointwise proper, assume that $\IHOMLAI\neq \emptyset$, fix $p\in P$ and define $x:=\pi(p)$. 
  \begin{enumerate}  
  \item
    \label{th:invhomm1}
    Let $\{\g_\alpha\}_{\alpha \in I_x}\subseteq \mg\backslash\mg_x$ be independent, and $\{\Psi_\alpha\}_{\alpha\in I_x}$ a family of $\Add{G_{[\g_\alpha]}}^p$-equivariant maps $\Psi_\alpha\colon \spann_\RR(\g_\alpha)\rightarrow S$. 
  
   For $\alpha\in I_x$, we define 
     \begin{align}
    \label{eq:ihdef}
    \Upsilon_{\pm, \alpha,l}\colon F_x\ni p'\mapsto
%      \wt{\homm}\left(\gamma^{x}_{\pm\g_\alpha}|_{[0,l]}\right)(p')
      \Phi_{p}\left(\exp(\pm l\cdot \g_\alpha)\right)\cdot  \Psi_\alpha(\pm l\cdot \g_\alpha) \cdot \Delta(p,p').
    \end{align}
    Then, for each $\homm'\in \IHOMLAI$ the map
    \begin{align}
      \label{eq:homdeff}
      % \label{eq:AbaenderHom}
      \homm(\gamma)(p'):= 
      \begin{cases} 
        (\Phi_g\cp \Upsilon_{\pm, \alpha,l})(\Phi_{g^{-1}}(p')) &\mbox{if } \gamma \csim \wm_g\cp \gamma_{\pm\g_\alpha}^x|_{[0,l]} \:\text{ for }\:\alpha \in I_x,\:g\in G\\
        \homm'(\gamma)(p') & \mbox{else}
      \end{cases}
    \end{align}
    is a well-defined element of $\IHOMLAI$. 
  \item
    \label{th:invhomm2}
    Let $\{V_\alpha\}_{\alpha\in I}$ be non-trivial $\Add{G_{x}}$-invariant linear subspaces of $\mg$ for which the sums $V_\alpha \oplus \mg_x$ (for all $\alpha \in I$) and $V_\alpha\oplus V_\beta \oplus \mg_x$ (for $\alpha\neq \beta$) are direct. Moreover, for each $\alpha\in I$ let $L_\alpha\colon V_\alpha \rightarrow \ms$ be a non-trivial $\Add{G_{x}}$-equivariant linear map.
     
    For $\g_\alpha \in V_\alpha\backslash\{0\}$, let 
    \begin{align}
    \label{eq:defmodh}
    	\Upsilon_{\g_\alpha,l}\colon F_x\ni p'\mapsto \Phi_{p}\left(\exp(l\cdot \g_\alpha)\right)\cdot  \exp\left(-l L_\alpha(\g_\alpha)\right) \cdot \Delta(p,p').
    \end{align}
    Then, for each $\homm'\in \IHOMLAI$ the map
    \begin{align}
      % \label{eq:homdef}
      % \label{eq:AbaenderHom}
            \label{eq:homdef}
      \homm(\gamma)(p'):= 
      \begin{cases} 
        (\Phi_g\cp \Upsilon_{\g_\alpha,l})(\Phi_{g^{-1}}(p')) &\mbox{if } \gamma \csim \wm_g\cp \gamma_{\g_\alpha}^x|_{[0,l]} \text{ for }\g \in V_\alpha, g\in G\\
        \homm'(\gamma)(p') & \mbox{else}
      \end{cases}
    \end{align}
    is a well-defined element of $\IHOMLAI$.
  \end{enumerate}    
  \begin{proof}
    \begin{enumerate}
    \item
      The crucial part is well-definedness, the rest are just straightforward calculations. For well-definedness, let $\gamma\in \Pagw$, $\alpha,\beta \in I_x$, $\g=\pm\g_\alpha$, $\g'=\pm\g_\beta$ and $g,g'\in G$ with 
		\begin{align*}      
      	\gamma \csim \wm_{g}\cp \gamma_{\g}^x|_{[0,l]}\qquad\qquad\text{and}\qquad\qquad \gamma \csim\wm_{g'}\cp \gamma_{\g'}^x|_{[0,l']}.
		\end{align*}      	 
      Then, $\wm_{g}\cp \gamma_{\g}^x|_{[0,l]}  \csim\wm_{g'}\cp \gamma_{\g'}^x|_{[0,l']}$, and we have to verify that $\homm(\wm_g\cp \gamma_{\g}^x|_{[0,l]})=\homm(\wm_{g'}\cp \gamma_{\g'}^x|_{[0,l']})$ holds. Applying the definitions, we see that this is equivalent to show that 
	  \begin{align*}      
     \homm(\gamma_{\g'}^x|_{[0,l']})= \homm(\wm_{g'^{-1} g}\cp \gamma_{\g}^x|_{[0,l]})
      \end{align*}
      holds. Since $g'^{-1}g\in G_x$ and $\gamma_{\g'}^x|_{[0,l']}\csim \wm_{g'^{-1}g}\cp  \gamma_{\g}^x|_{[0,l]}$,
      it suffices to show that 
	  \begin{align*}      
      	\gamma_{\g'}^x|_{[0,l']}\csim \wm_h\cp \gamma_{\g}^x|_{[0,l]} \text{ for }h\in G_x \qquad \Longrightarrow\qquad \homm(\gamma_{\g'}^x|_{[0,l']})=\homm(\wm_h\cp \gamma_{\g}^x|_{[0,l]}). 
      \end{align*}
      However, in the above situation we have $\g_\alpha\xsim \g_\beta$, hence $\alpha=\beta$ by independence of $\{\g_\alpha\}_{\alpha\in I_x}$.  
      Then $\g,\g'\in [\g_\alpha]$, so that by stability of $\g_\alpha$ and \eqref{eq:stab2} 
      we have 
      $\lambda \g=\Ad_{h^{-1}}(\g')$ for  
      $\lambda=\frac{l}{l'}$,      
      hence 
      \begin{align}
        \label{eq:clacu0}
        \begin{split}
          \homm(\wm_h\cp \gamma_{\g}^x|_{[0,l]})(p') &=\Phi_p(h\cdot\exp(l'\lambda\hspace{1pt} \g))\cdot \Psi_\alpha( l\cdot \g)\cdot \Delta(p,\Phi_{h^{-1}}(p'))\\
          &=\Phi(h\cdot h^{-1}\exp(l'\g')\cdot h,  p)\cdot\Psi_\alpha(l  \cdot \g)\cdot \fiba_p(h)^{-1}\cdot \Delta(p,p')\\
          &=\Phi(\exp(l' \g'),p)\cdot\Psi_\alpha\big(l\Ad_{h}(\g)\big)\cdot  \Delta(p,p')\\
          &=\Phi(\exp(l' \g'),p)\cdot \Psi_\alpha(l'\g')\cdot  \Delta(p,p')\\
          &= \homm\big(\gamma_{\g'}^x|_{[0,l']}\big)(p').
        \end{split}
      \end{align}
      Now, it is immediate from the definitions that 
      $\homm(\gamma)(p'\cdot s)=\homm(\gamma)(p')\cdot s$ for all $s\in S$ and that
     \begin{align*} 
      \homm(\wm_g\cp \gamma)(\Phi_g(p))=(\Phi_g\cp\homm)(\gamma)(p)\qquad \forall\: g\in G. 
	\end{align*}      
      So, in order to show  
      that $\homm\in \IHOMLAI$ it remains to show the homomorphism properties of $\homm$. Here, it suffices to verify these properties for such curves that are equivalent to one of the curves $\wm_g\cp\gamma^x_{\g}|_{[0,l]}$ for $\g=\pm\g_\alpha$ with $\alpha\in I_x$. In fact, if $\gamma$ is not equivalent to one of these curves, then by Corollary \ref{cor:decompo} the same is true for its inverse and all its subcurves $\gamma|_{K'}$ for $K'\subseteq \dom[\gamma]$ so that the homomorphism properties here are clear from $\homm'\in \IHOMLAI$.  
      
      Now, since 
      \begin{align*}
      \wm_g\cp 
        % \label{eq:hom}
        \gag|_{[a,b]}\csim\wm_{g\cdot \exp(a \g)}\cp \gamma_{\g}^x|_{[0,b-a]} \qquad\text{as well as}\qquad \big[\gag|_{[a,b]}\big]^{-1}\csim\gamma_{-\g}^x|_{[-b,-a]}
      \end{align*}
      for $\gamma\csim \wm_g\cp\gamma^x_{\g}|_{[0,l]}$ we have $\gamma^{-1} \csim \wm_{g\cdot \exp(l\vec{g})}\cp \gamma_{-\g}^x|_{[0,l]}$. Hence, for $h:=g\cdot \exp(l\cdot\vec{g})$ and $\g=\pm \g_\alpha$ we obtain
      \begin{align*}
        \homm\big(\gamma^{-1})(q)%&=\Phi_{g'}\cp\homm\left(\wm_{{g'}^{-1}}\cp \gamma^{-1}\right)(\Phi_{g^{-1}}(q))\\
		&=(\Phi_{h}\cp\Upsilon_{\mp,\alpha,l})(\Phi_{h^{-1}}(q))\\   
%        &=(\Phi_{h}\cp\wt{\homm})\left( \gamma^x_{-\vec{g}}|_{[0,l]}\right)(\Phi_{h^{-1}}(q))\\
        %
        %
        &=\Phi_h\cp \Phi_{p}\left(\exp(-l\cdot\vec{g})\right)\cdot \Psi_\alpha^{-1}(l \g)\cdot \Delta(p,\left(\Phi_{h^{-1}}(q)\right)).
      \end{align*} 
      Then, for $q:=\homm(\gamma)(p')=(\Phi_g\cp\Upsilon_{\pm,\alpha,l})(\Phi_{g^{-1}}(p'))$ we have
      \begin{align*}
        \Delta\left(p,\Phi_{h^{-1}}(q)\right)&=\Delta\left(p,(\Phi_{\exp(-l\cdot\g)}\cp\Upsilon_{\pm,\alpha,l})(\Phi_{g^{-1}}(p'))\right) \\
        &=\Delta\left(p,p\cdot\Psi_\alpha(l \cdot\g)\cdot  \Delta(p,\Phi_{g^{-1}}(p'))\right)\\
        &=\Psi_\alpha(l \cdot\g)\cdot  \Delta\left(p,\Phi_{g^{-1}}(p')\right).
      \end{align*}
      Since $\Phi_h\cp \Phi_{p}\left(\exp(-l\vec{g})\right)=\Phi\left(g\cdot \exp(l\g)\cdot \exp(-l\g) ,p\right)=\Phi(g,p)$, we get
      \begin{align*}
        \homm\big(\gamma^{-1}\big)\big(\homm(\gamma)\big(p'\big)\big)&= \Phi_g(p)\cdot \Delta\left(p,\Phi_{g^{-1}}(p')\right)\\
        &=\Phi\left(g,p\cdot \Delta\left(p,\Phi_{g^{-1}}(p')\right)\right)=\Phi(g, \Phi(g^{-1},p'))=p',
      \end{align*}
      hence $\homm\big(\gamma^{-1}\big)\cp \homm(\gamma)=\id_{F_{\gamma(0)}}$.  
      
      Finally, let $\gamma\csim \wm_g\cp \gamma_{\g}^x|_{[0,l]}$ with $\dom[\gamma]=[a,b]$ and $s\in (a,b)$. We choose $l'\in (0,l)$ with $\gamma(s)=\wm_g\cp \gamma_{\g}^x(l')$ and define $\gamma_1:=\gamma|_{[a,s]}$ as well as $\gamma_2:=\gamma|_{[s,b]}$. Then 
      \begin{align*}
      \gamma_1& \csim \wm_g\cp \gamma_{\g}^x|_{[0,l']}\hspace{73pt}\qquad\text{and}\\ 
        \gamma_2& \csim \wm_g\cp \gamma_{\g}^x|_{[l',l]}\csim \wm_{h}\cp \gamma_{\g}^x|_{[0,l-l']}\quad \text{for}\quad h:=g\cdot \exp(l'\vec{g}).
      \end{align*}
      Now, $\Phi_{h^{-1}}(\homm(\gamma_1)(p'))=p\cdot \Psi_\alpha(l' \g)\cdot \Delta(p,\Phi_{g^{-1}}(p'))$ and
      \begin{align*}
        \homm(\gamma_2)(q)&=(\Phi_h\cp\Upsilon_{\pm,\alpha,l-l'})(\Phi_{h^{-1}}(q))\\
        &=\Phi\left(g\cdot \exp(l'\g)\cdot \exp([l-l']\hspace{1pt} \g), p\right)\cdot \Psi_\alpha([l-l']\hspace{1pt} \g)\cdot \Delta(p,\Phi_{h^{-1}}(q))\\
        &=\Phi\left(g\cdot \exp(l\cdot\g), p\right)\cdot \Psi_\alpha([l-l']\hspace{1pt}  \g)\cdot \Delta(p,\Phi_{h^{-1}}(q)),
      \end{align*}
      so that
      $\homm(\gamma_2)\cp\homm(\gamma_1)(p')=\Phi\left(g\cdot \exp(l\cdot\g), p\right)\cdot \Psi_\alpha(l \g)\cdot \Delta(p,\Phi_{g^{-1}}(p'))
      =\homm(\gamma)(p')$.
    \item 
      The crucial part is well-definedness, as the rest follows analogously to Part \ref{th:invhomm1}). Now, as in Part \ref{th:invhomm1}) it suffices to show that for $\g\in V_\alpha$, $\g'\in V_\beta$, $h\in G_x$
	\begin{align*}      
       \gamma_{\g'}^x|_{[0,l']}\csim \wm_h\cp\gamma_{\g}^x|_{[0,l]}\qquad\Longrightarrow \qquad \homm(\gamma_{\g'}^x|_{[0,l']})=\homm(\wm_h\cp\gamma_{\g}^x|_{[0,l]}).
    \end{align*}
    Now, in the above situation we have $\alpha=\beta$ just by Lemma \ref{lemma:sim}.\ref{lemma:sim4} and the direct sum property $V_\alpha \oplus V_\beta \oplus \mg_x$. By Part \ref{lemma:sim3}) of the same lemma we even have $\textstyle\frac{l}{l'} \g=\Ad_{h^{-1}}(\g')$, so that for 
      \begin{align*}    
    	\Psi_\alpha(l\cdot \g):=\exp(-l L_\alpha(\g))\qquad \forall\:l>0,\:\forall\: \g\in V_\alpha\text{ and } \alpha \in I
      \end{align*}    
      the calculation \eqref{eq:clacu0} shows the claim.  	
    \end{enumerate}
  \end{proof}
\end{proposition}

\subsection{Inclusion Relations}
\label{sec:inclrel}
In this brief subsection, we will use the modification results from the previous part, in order to derive some general conditions which allow to decide whether 
the inclusion
 $\ARQLAI\subseteq\AQRLAI$ 
 is proper. This will be done in Proposition \ref{prop:incl}, where we construct elements of $\IHOMLAI\cong \AQRLAI$ that cannot be approximated by the elements of $\iota_\Con(\AR)\subseteq \AQRLAI$, i.e., by classical (smooth) invariant connections. In the first part of this proposition, we will provide a criterion which can be applied whenever the set $\AR$ of invariant connections is explicitly know. 
 Then, in the last two parts of the same proposition, we will basically use that due to formula \eqref{eq:trivpar} (and linearity of the involved maps) the parallel transports along Lie algebra generated curves which correspond to linearly dependent Lie algebra elements are related in a certain way. In particular, this will allow us to show that quantization and reduction do not commute in \mbox{(semi-)homogeneous} LQC as well. 

In the following, we still assume that $S$ is compact and connected with $\dim[S]\geq 1$, and that $\wm$ is analytic and pointwise proper. We fix $p\in P$, define $x:=\pi(p)$, and let $V, V_1,V_2,V_3$ be non-trivial $\Add{G_x}$-invariant linear subspaces of $\mg$ such that $V_i\oplus V_j\oplus \mg_x$ is direct for all $1\leq i\neq j\leq 3$ and 
\begin{align*}
	V\cap \mg_x=\{0\}\qquad V_1\cap \mg_x=\{0\} \qquad V_2\cap \mg_x=\{0\}\qquad V_3\cap \mg_x=\{0\}.
\end{align*} 
 Then, by $L,L_1,L_2,L_3\neq 0$ we will denote respective non-trivial $\Add{G_x}^p$-equivariant linear maps.\footnote{Observe that
  each linear map $L\colon V\rightarrow \ms$ is $\Add{G_x}^p$-equivariant if $S$ is commutative.}  
\begin{proposition} 
  \label{prop:incl}
  If 
  $\IHOMLAI\neq \emptyset$,  
  then 
  we have $\ARQLAI\subsetneq \AQRLAI$ if:   
  \begin{enumerate}
  \item 
    \label{prop:incl1}
    We find $\g\in \mg\backslash\mg_x$ stable and $\Psi\colon \spann_\RR(\g)\rightarrow S$ an $\Add{G_{[\g]}}^p$-equivariant map, such that 
    \begin{align}
    \label{eq:cccccccc}
     \Psi(\g)\notin\ovl{C}\qquad\quad\text{for}\qquad\quad C:=\bigcup_{\w\in \AR}\exp\left(\w\!\left(\hspace{0.2pt}-\dd_e\Phi_p(\g)\right)\right).
    \end{align} 
  \item
    \label{prop:incl2}
    We have $V\oplus \mg_x \subsetneq \mg$ and $S=S^1$.   
  \item
    \label{prop:incl3}
    We have $S=\SU$ and there are $\g_i\in V_i$ for $i=1,2,3$, such that $L_1(\g_1),L_2(\g_2),L_3(\g_3)$ are linearly independent and $\g_3\in\spann_\RR(\g_1,\g_2)$.
  \end{enumerate}
  \begin{proof}
    \begin{enumerate} 
    \item
      Define $\homm$ by \eqref{eq:homdeff} for $|I_x|=1$, $\g_\alpha=\g$ and $\Psi_\alpha=\Psi$, and assume that 
      $\homm \in \kappa_{\mg'}(\ARQLAI)$. Then, we find a net $\{\w_\alpha\}_{\alpha\in I}\subseteq \AR$ with $\{\kappa_{\mg'}(\iota_\Con(\w_\alpha))\}_{\alpha\in I}\rightarrow \homm$, hence $\homm(\gamma)(p)=\lim_\alpha\parall{\gamma}{\w_\alpha}(p)$ for 
$\gamma:=\gag|_{[0,1]}$. Consequently, 
      \begin{align*}
        \Phi_p(\exp(\g))\cdot \Psi(\g)&=
        \homm(\gamma)(p)= \lim_\alpha\parall{\gamma}{\w_\alpha}(p)\\
%        %\\%&        
&\hspace{-3pt}\stackrel{\eqref{eq:trivpar}}{=}\hspace{-1pt}
        \lim_\alpha \Phi_p(\exp(\g))\cdot \exp\left(\w_\alpha\!\left(-\wt{g}(p)\right)\right)\\
        &=
        \lim_\alpha \Phi_p(\exp(\g))\cdot \exp\left(\w_\alpha\!\left(-\dd_e\Phi_p(\g)\right)\right) \in \Phi_p(\exp(\g))\cdot \ovl{C},
      \end{align*}
      which contradicts the choice of $\g$. 
    \item
	We here only sketch the proof. The details can be found in Appendix \ref{app:incl}.  
	\begingroup
    \setlength{\leftmarginii}{25pt}
	\begin{itemize}
	\item[(a)]
	\vspace{-4pt}
	    We choose $0\neq \g_3\in V$, $\g_1\in \mg\backslash [V\oplus \mg_x]$  and define $\g_2:=\g_3+\g_1$. Then, for $i=1,2,3$ we let $\gamma_i:=\gamma_{\g_i}^x|_{[0,l]}$ for some $l>0$ with $l<\tau_{\g_1},\tau_{\g_2},\tau_{\g_3}$.
	\item[(b)]
	\vspace{2pt}	
		It follows that $\g_3 \nsim_x \g_1,\g_2$, so that modifying (Proposition \ref{th:invhomm}.\ref{th:invhomm2}) the homomorphism $\homm'\in \IHOMLAI$ along $\gamma_{\g_3}^x|_{[0,l]}$ does not change its values on $\gamma_{\g_i}^x|_{[0,l]}$ for $i=1,2$. 
	\item[(c)]
		\vspace{2pt}	
	We fix $\homm'\in \IHOMLAI$ and define $\homm_\mu$ by Proposition \ref{th:invhomm}.\ref{th:invhomm2} for $|I_x|=1$, $\g_\alpha:=\g_3$, $V_\alpha:=V$ and $L_\alpha:=\mu L$ for  $\mu\in \RR$, i.e., we only modify  along $\g_3$. 
		\item[(d)]
		\vspace{2pt}
		We fix $p\in F_x$. Then, for $\w\in \AR$ the value
	\begin{align*}
		\kappa_{\mg'}(\iota_\Con(\w))(\gamma_i)(p)=\parall{\gamma_i}{\w}(p)
	\end{align*}			
	 is given by \eqref{eq:trivpar} for $i=1,2,3$. In particular, the value $\kappa_{\mg'}(\iota_\Con(\w))(\gamma_3)(p)$ is related to the values $\kappa_{\mg'}(\iota_\Con(\w))(\gamma_1)(p)$ and $\kappa_{\mg'}(\iota_\Con(\w))(\gamma_2)(p)$ just because $\g_3=\g_1-\g_2$.
	 \item[(e)]
		\vspace{2pt}
		Recall the open subsets \eqref{eq:opensets} and choose $U:=\exp\hspace{1pt}(\hspace{1pt}\I (-\epsilon,\epsilon))$ for $\epsilon<\textstyle\frac{\pi}{4}$. It follows that there exists $\mu \in \RR$ such that\footnote{Here, it is important that $U^{p,p}_{\gamma_1,\gamma_2}(\homm_\mu)=U^{p,p}_{\gamma_1,\gamma_2}(\homm')$ is the same neighbourhood for all $\mu\in \RR$, just by point (b).} 
		\begin{align*}
		\kappa_{\mg'}(\iota_\Con(\w))\in U^{p,p}_{\gamma_1,\gamma_2}(\homm_\mu)\qquad\Longrightarrow \qquad \kappa_{\mg'}(\iota_\Con(\w))\notin U^{p}_{\gamma_3}(\homm_\mu),
	\end{align*}  
  	hence $\kappa_{\mg'}(\iota_\Con(\AR))\cap U^{p,p,p}_{\gamma_1,\gamma_2,\gamma_3}(\homm_\mu)=\emptyset$. Then, $\homm_\mu \notin \kappa_{\mg'}\big(\ARQ\big)=\kappa_{\mg'}\Big(\ovl{\iota_\Con(\AR)}\Big)$, hence $\AQR\ni\kappa_{\mg'}^{-1}(\homm_\mu)\notin \ARQ$, which shows the claim.  
	\end{itemize}	    
    \endgroup
    \item
    We choose any $\epsilon'\in \IHOMLAI$, being non-empty by assumption, and    define $\homm\in \IHOMLAI$ by \eqref{eq:homdef} w.r.t.\ $I_x:=\{1,2,3\}$. Moreover, we let $\gamma_i:=\gamma^x_{\g_i}|_{[0,1]}$ for $i=1,2,3$. 
    
    We choose a neighbourhood $W$ of $0$ in $\su$ such that $\exp|_{W}$ is a diffeomorphism. Then, scaling the $\g_i$, we can assume that $\s_i:=L_i(\g_i)\in W$ for $i=1,2,3$. 
      We now show that $\homm$ cannot be approximated by the elements of $\kappa_{\mg'}(\iota_\Con(\AR))$, i.e., by classical (smooth) invariant connections. 

For this, let $\|\cdot\|_\murs$ denote the euclidean norm on $\RR^3$ carried over to $\su$ by $\murs\colon \RR^3\rightarrow \su$, and choose    
      $\epsilon >0$ such that 
      \begin{align*}      	
      	B_\epsilon(\s_i):=\{\s\in \su\:|\: \|\s_i-\s\|_\murs<\epsilon\}\subseteq W\qquad \text{for }i=1,2,3.  
      \end{align*}     	
      Then, $\exp(B_\epsilon(\s_i))$ is open 
      in $\SU$, and since $\pm \me\notin \exp(B_\epsilon(\s_i))$, we have
      \begin{align*}
        U_\epsilon^{i}:=\exp^{-1}(\exp(B_\epsilon(\s_i)))=\left\{\s + 2\pi n\frac{\s}{\|\s\|}\:\bigg|\: \s\in B_\epsilon(\s_i), n\in \mathbb{Z}\right\}
      \end{align*}
      by formula \eqref{eq:expSU2}, i.e., by the periodicity property of the exponential map of $\SU$. 
      Since $B_\epsilon(\s_i)$ does not contain the origin ($\exp(0)=\me$), there is $\alpha \in (0,2\pi)$ such that\footnote{Here, $\measuredangle\left(\vec{v},\vec{w}\right)$ means the minimum of the angles between $\vec{v}$, $\vec{w}$ and $-\vec{v}$, $\vec{w}$, so that $C^i_\alpha$ is a double cone.}
      \begin{align*}
        U_\epsilon^{i}\subseteq C^i_\alpha:=\{\s \in \su\:|\: \measuredangle\left(\murs^{-1}(\s_i),\murs^{-1}(\s)\right)< \alpha)\}
      \end{align*}
      for $C_\alpha^i$ the double cone in $\su$ determined by the axis $\s_i$ and the opening angle $\alpha$. 
      Conversely, for each $\alpha\in (0,2\pi)$ we also find some $\epsilon(\alpha)> 0$ such that $U_{\epsilon(\alpha)}^{i}\subseteq C^i_\alpha$ holds. 
      
      Let $U\subseteq \SU$ be an open neighbourhood of $\me$ with $\exp(-\s_i)\cdot U \subseteq \exp(B_\epsilon(\s_i))^{-1}$ for $i=1,2,3$. 
      Then, if $\kappa_{\mg'}(\iota_\Con(\w))\in U_{\gamma_1,\gamma_2,\gamma_3}^{p,p,p}(\homm)$, by \eqref{eq:trivpar} and the definition of $\homm$ we have
      \begin{align*}
        	\exp(-\w(\wt{g}_i(p)))\in \exp(-L_i(\g_i))\cdot U= \exp(-\s_i)\cdot U\subseteq  \exp(B_\epsilon(\s_i))^{-1},
      \end{align*}
	    i.e., $\exp(\w(\wt{g}_i(p)))\in \exp(B_\epsilon(\s_i))$, hence $\w(\wt{g}_i(p))\in C^i_\alpha$ for $i=1,2,3$. Now, by the choice of $\g_3$, we have $\w(\wt{g}_3(p))\in \spann_\RR(\w(\wt{g}_1(p)),\w(\wt{g}_2(p)))$,
      being a subset of
      \begin{align*}
        C^1_\alpha+C^2_\alpha:=\{v+w\:|\: v\in C^1_\alpha,w\in C^2_\alpha\}.
      \end{align*}
      We now show that, for $\alpha$ suitable small, $\left[C^1_\alpha+C^2_\alpha\right]\cap C^3_\alpha=\{0\}$ holds. This then contradicts that $\w(\wt{g}_3(p))\in U_{\epsilon(\alpha)}^{3}\subseteq C^3_\alpha\backslash\{0\}$ and shows the claim. 
      
      Since the (pre)image of a cone of the above form under a linear isomorphism contains a cone of the above form, it suffices to consider the case where $\vec{n}_i=\vec{e}_i$ for $i=1,2,3$. Here, we have to show that the equation\footnote{The single expressions parametrize the cones around the $y$-, $z$- and $x$-axis, respectively.}
      \begin{align*}
        \begin{pmatrix}   s y \cos(\theta)\\ y \\ s y \sin(\theta) \end{pmatrix}+\begin{pmatrix}   r z \cos(\phi)\\ r z \sin(\phi)\\ z \end{pmatrix}= \begin{pmatrix}    x \\ t x \cos(\eta)\\ t x \sin(\eta) \end{pmatrix}
      \end{align*}
      has no solution for $0< r,s,t \leq \epsilon$ provided that $\epsilon$ is suitable small. But, this is clear since the determinant of
      \begin{align*} 
        \begin{pmatrix}  x\\ y \\ z \end{pmatrix}\mapsto 
	\begin{pmatrix}   
          -1 & s  \cos(\theta) & r  \cos(\phi) \\ 
          - t \cos(\eta)  & 1 &  r\sin(\phi)\\
          - t \sin(\eta)  & s \sin(\theta) &  1
        \end{pmatrix}  	 \cdot
        \begin{pmatrix}  x\\ y \\ z \end{pmatrix}
      \end{align*}
      tends to $-1$ for $\epsilon \rightarrow 0$.
    \end{enumerate}
  \end{proof}
\end{proposition}
\begin{corollary}
  \label{cor:incl}
  Let $(P,\pi,M,S)$ be a principal fibre bundle and $(G,\Phi)$ a Lie group of automorphisms of $P$. Moreover, let the induced action $\varphi$ be analytic and pointwise proper. Then, in the following situations we have $\ARQLAI\subsetneq \AQRLAI\colon$
  \begin{enumerate}
  \item
    \label{cor:incl1}
    Let $S=S^1$
    and $\dim[G]\geq2$:
    \begingroup
    \setlength{\leftmarginii}{15pt}
    \begin{itemize}
    \item
      \label{cor:incl2}
      \vspace{-4pt}
      \itspacec
      $\IHOMLAI\neq\emptyset$ and there is $x\in M$ such that $\dim[G]-\dim[G_x]\geq 2$. In addition to that, we find an $\Add{G_x}$-invariant vector $\g\in \mg\backslash \mg_x$.  
    \item
      $\varphi$ acts transitively and free.\footnote{Observe that $\wm$ then is automatically pointwise proper because $M\cong G$.}
    \end{itemize}
    \endgroup
  \item
        \label{cor:incl22}
    Let $S=\SU$, $\dim[G]\geq 2$ and $G_x=\{e\}$ for some $x\in M \colon$
    \begingroup
    \setlength{\leftmarginii}{15pt}
    \begin{itemize}
    \item
      \itspacec
            \vspace{-4pt}
      $\IHOMLAI\neq \emptyset$.  
    \item
      $\varphi$ is transitive.
    \end{itemize}
    \endgroup
  \end{enumerate}
  \begin{proof}
    \begin{enumerate}
    \item
      If $\g\in \mg\backslash \mg_x$ is $\Add{G_x}$-invariant, let $V:=\Span_\RR(\g)$ and $L\colon \lambda \g \mapsto \lambda \s$ for some fixed $\s\in \ms\backslash\{0\}$. This map is linear and, by commutativity of $S^1$, $\Add{G_x}^p$-equivariant for each $p\in \pi^{-1}(x)$. Moreover, since $\dim[G]-\dim[G_x]\geq 2$, we have $V\oplus \mg_x\subsetneq \mg$ so that the claim is clear from from Proposition \ref{prop:incl}.\ref{prop:incl2}.  

      If $G_x=\{e\}$ and $\varphi$ is transitive, then Wang's theorem \cite{Wang} (see also Case \ref{th:wang}) shows $\AR\cong \Hom_\RR(\mg,\ms)$, so that $\kappa_{\mg'}(\iota_\Con(\AR))\subseteq\IHOMLAI\neq \emptyset$. Moreover, since each $\g\in \mg\backslash\mg_x$ is $\Add{G_x}$-invariant,    
      the requirements of the first case are fulfilled. 
    \item
      We choose $\g_1,\g_2\in\mg$ linearly independent and $\g_3\in \Span_\RR(\g_1,\g_2)$ neither contained in $V_1:=\Span_\RR(\g_1)$ nor contained in $V_2:=\Span_\RR(\g_2)$. Let $V_3:=\Span_\RR(\g_3)$ and $\s_i:=\tau_i$ for $i=1,2,3$. Then the maps $L_i\colon \lambda \g_i\mapsto \lambda \s_i$ are linear and $\Add{G_x}^p$-equivariant for each $p\in \pi^{-1}(x)$ so that the claim follows from Proposition \ref{prop:incl}.\ref{prop:incl3}. The second part is clear. 
    \end{enumerate}
  \end{proof}
\end{corollary}
We close this subsection with an application of Corollary \ref{cor:incl} to loop quantum cosmology, by showing that 
$\ARQLAI\subsetneq \AQRLAI$ holds (quantization and reduction do not commute) in 
\mbox{(semi-)homogeneous} LQC.  
In Subsection \ref{subsec:QuantvsRed}, we will see that this is also true in the  
homogeneous isotropic case. 
\begin{example}[(Semi-)Homogeneous Loop Quantum Cosmology]
  \label{ex:LQCInc}
  Let $P=\RR^3\times \SU$ and $\Ph$, $\Phi_{SH}$ be defined as in Example \ref{ex:LQC}. 
  We claim that quantization and reduction do not commute in \mbox{(semi-)homogeneous} loop quantum cosmology. In fact,
      \begingroup
    \setlength{\leftmargini}{14pt}
  \begin{itemize}  
  \item
  \vspace{-4pt}
   In the homogeneous case $(\Gh,\Ph)$ this follows from the second part of Corollary \ref{cor:incl}.\ref{cor:incl22}. 
	\item
   In the semi-homogeneous case $(G_{SH},\Phi_{SH})$, the action 
  $\wm$ is pointwise proper because $\Ph$ is proper and and each linear subspace of $\RR^3$ is closed. Since $\dim[G_{HS}]\geq 2$, by the first part of Corollary \ref{cor:incl}.\ref{cor:incl2} it suffices to show that $\IHOMLAI\neq \emptyset$. But, this is clear because (as already stated in Example \ref{ex:LQC}) the set $\AR$ is in bijection with the smooth maps $\psi \colon \RR^2\times T\RR\rightarrow \su$ for which the restrictions $\psi|_{\RR^2\times T_x\RR}$ are linear for all $x\in \RR$.\hspace*{\fill}$\lozenge$
	\end{itemize}
	\endgroup  
\end{example}

\subsection{Modifications along Free Segments}
\label{sec:ModifreeSeg}
Complementary to our investigations of Lie algebra, i.e., continuously generated curves, we now are going to study the set $\Paf$ of free curves in $M$. This is the set of all embedded analytic curves that contain a segment which, in a certain sense, does not overlap with its translates by the symmetry group.\footnote{The precise definition will given below, cf.\  also Remark \ref{rem:euklrem}.\ref{rem:dsdfdf}.} We will show that each such curve is covered by finitely many translates  (of initial and final segments) of one of its maximal free segments. This will provide us with a canonical decomposition of such free curves by means of the symmetry group. In course of this, we will split up the set  $\Paw\backslash \Pags$ into three subsets $\Pacs$, $\Pafs$ and $\Pafns$ 
 being closed under inversions and decompositions as well. Recall that Proposition \ref{rem:euklrem2b} then provides us with a respective factorization 
	\begin{align}
	\label{eq:dsfs444ffff}
    \AQRw \cong  \AQRInd{\mg}\times \AQRInd{\mathrm{CNL}}   
 \times  \AQRFNS\times \AQRInd{\mathrm{FS}}.
  \end{align}
 Here, we always have $\Paf=\Pafs\sqcup \Pafns$, whereby $\Pafns$ consists of all free curves whose stabilizer is trivial,\footnote{See Definition \ref{def:freeSegg}.} so that $\Pafs$ consists of all free curves for which this is not the case. 
 
 Then, the second space $\AQRInd{\mathrm{CNL}}$ 
 is the least accessible one, just because the set $\Pacs$ of continuously but not Lie algebra generated curves is so.  
 However, we will show that $\Pacs=\emptyset$ holds, i.e.,  
that we have $\Paw=\Pags\sqcup \Paf$ whenever $\wm$ is transitive or proper and admits only stabilizers which are normal subgroups of $G$. So, if $\wm$  is in addition free, we even have $\Paw=\Pags\sqcup \Pafns$, hence 
$\AQRw \cong  \AQRInd{\mg}\times  \AQRInd{\mathrm{FN}}$, so that in this case it suffices to define normalized Radon measure on these two factors in order to define a normalized Radon measure on $\AQRw$. 

Now, in the second part of Section \ref{sec:MOQRCS}, we will use  Proposition \ref{th:invhomm}.\ref{th:invhomm1} in order to construct a normalized Radon measure $\mLAS$ on $\AQRInd{\mg}$, e.g., for the case that $\wm$ is free and $S$ equals $\SU$ or an $n$-torus. Moreover, in the last part of the present Section, we will prove an analogue of this proposition for free curves, which we then use in Subsection \ref{sec:FreeM} 
in order to construct a normalized Radon measure $\mFNS$ on $\AQRInd{\mathrm{FN}}$. So, together these two measures give rise to a normalized Radon measure 
\begin{align*}
\mLAS\times \mFNS\qquad \qquad\text{on}\qquad\qquad \AQRw \cong  \AQRInd{\mg}\times  \AQRInd{\mathrm{FN}}
\end{align*}
whenever the respective requirements hold. This will be the case, e.g.\, in \mbox{(semi-)homogeneous} LQC.
\begin{definition}
  \label{def:freeSegg}
    For $\gamma$ an analytic curve, we define the group\footnote{The group property is easily verified, and for $\gamma$ embedded analytic also clear from $\textit{(a)}$ in Lemma \ref{lemma:curvestablemma}.\ref{lemma:curvestablemma1}.} 
	\begin{align*} 
    	\text{\gls{GGAMMA}}:=\{g\in G \:|\: \wm_g\cp \gamma = \gamma\cp \adif \text{ for } \adif\colon \dom[\gamma]\rightarrow \dom[\gamma]\text{ an analytic diffeomorphism with }\dot\adif>0\}
    \end{align*}
    as well as the equivalence relation \gls{SEGSIM} on $G$ by 
	\begin{align*}
		g\sim_\gamma g'\qquad \Longleftrightarrow \qquad g^{-1}g'\in G_\gamma.
	\end{align*}	    
	Then,
      \begingroup
  \setlength{\leftmargini}{20pt}
	\begin{itemize}
	\item
		$\gamma$ is called symmetric iff $G_\gamma\neq \{e\}$.
	\item    
     $\gamma$ is called a free segment iff 
	\begin{align*}    
    	\gamma\cpsim\wm_g\cp \gamma\quad\text{for}\quad g\in G\qquad\Longrightarrow \qquad g\in G_\gamma.
    \end{align*}
    \item
    $\gamma$ is called free iff $\gamma|_K$ is a free segment for some compact interval $K\subseteq \dom[\gamma]$.
	\item
      For $\delta$ an analytic curve let 
	\begin{align*}      
      \text{\gls{HGAMMADELTA}}:=\{g\in G \:|\: \wm_g\cp \delta\cpsim \gamma \}\qquad\quad\text{as well as}\qquad\quad H_{[\gamma,\delta]}:=H_{\gamma,\delta}\hspace{1pt} \slash \!\sim_\delta.
	\end{align*} 
\end{itemize}
\endgroup
\end{definition}
The next lemma collects the relevant properties of the above quantities for the embedded analytic case. 
\begin{lemma}
  \label{lemma:curvestablemma}
  Let $\wm$ be analytic and pointwise proper, and $\Paw\ni \gamma\colon [a,b]\rightarrow M$. Moreover, let $\Paw\ni \delta\colon L\rightarrow M$ be a free segment.
\begin{enumerate}
	\item  
	\label{lemma:curvestablemma1}
  	We have $\wm_h\cp \gamma = \gamma$ for all $h\in G_\gamma$. Hence, 
		 \begin{enumerate}
		 \item[(a)]
     	\label{lemma:curvestablemma1aa}
     	\vspace{-3pt}
     		 $\gamma\cpsim\wm_g\cp \gamma \qquad \Longrightarrow \qquad\wm_g\cp \gamma = \gamma$
     	\item[(b)]
     	\label{lemma:curvestablemma1a}
     		     	\vspace{3pt}
     	%\vspace{-3pt}
     	 $G_\gamma=\bigcap_{t\in \dom[\gamma]}G_{\gamma(t)}$ is a closed subgroup of $G$,
   	 	\item[(c)] 
   	 	    \label{lemma:curvestablemma1b}
   	 	     	\vspace{2.5pt}
    	  $G_{\gamma|_D}=G_\gamma$ for each interval $D\subseteq \dom[\gamma]$.
   		\end{enumerate}
   		\item
   		\label{lemma:curvestablemma2}
   		  Assume that $\gamma$ is not free, and let $t\in [a,b]$. 
    Then, for each interval $D\subseteq [a,b]$ with $t\in K$ we find $g\in G\backslash G_{\gamma(t)}$ with $\gamma|_{K}\cpsim \wm_g\cp \gamma|_K$.
       	\item
   		\label{lemma:curvestablemma3}
     Let $\gamma$ be free and $\{h_\alpha\}_{\alpha\in I}\subseteq H_{\gamma,\delta}$ a family of representatives of $H_{[\gamma,\delta]}$. Then,
     \begin{enumerate}
     \item
     \label{it:p1}
     \vspace{-3pt}
     $H_{[\gamma,\delta]}$ is finite.
     \item
     \label{it:p2}
     \vspace{2pt}
     There exists a unique decomposition $a=k_0<k_1<\dots<k_n=b$ of $\dom[\gamma]$ into compact intervals $K_i=[k_i,k_{i+1}]$ for $0\leq i\leq n-1$, such that either 
	\begin{align}
	\label{eq:bbbb}
	\begin{split}     
     \gamma|_{K_i} &\:\:\nsim_{\cp} \:\hspace{1pt}\wm_{h_\alpha}\cp \delta \qquad\qquad\:\:\forall\:\alpha\in I \qquad\text{or}\\
     \gamma|_{K_i}&\psim \wm_{h_{\alpha_i}}\cp [\delta|_{L_i}]^{\pm 1}\quad\hspace{5.5pt}\text{for } \alpha_i\in I \text{ and } L_i\subseteq L\text{ uniquely determined}.
   \end{split}
   \end{align}  
    Here, $L_i=L=[l_1,l_2]$ if $1\leq i\leq k-2$ and $L_0,L_{n-1}$ are both of the form 
	\begin{align}
	\label{eq:oftheform}    
    	\:[m,l_2] \quad\text{ for }\quad  m\in [l_1,l_2)\qquad \qquad\text{or}\qquad\qquad [l_1,m]\quad \text{ for }\quad m\in (l_1,l_2].
    \end{align}
    \vspace{-15pt}
    \begin{align}
    \label{eq:decompo}
    \raisebox{-30pt}{
     \begin{tikzpicture}
	    \draw[-,line width=1.5pt] (-0.5,0) -- (4,0);
	    \draw[->,line width=1.5pt,dotted] (4,0) -- (6,0);
	    \draw[-,line width=1.5pt] (-0.5,-0.2) -- (-0.5,0.1);
	%   	\filldraw[black] (-0.5,0) circle (2pt);
	   	\draw (6.3,0) node {\(\gamma\)};
	    \draw[-,line width=1.25pt] (1.5,1.25) -- (2.5,1.25);
	    \draw[-,line width=1pt] (1.5,1.16) -- (1.5,1.39);
	    \draw[-,line width=1pt] (2.5,1.16) -- (2.5,1.39);	   
		\draw (2.7,1.3) node {\(\delta\)};	 
		\draw[-,line width=1.25pt,color=red] (1.515,1.35) -- (1.97,1.35); 
		\draw[color=red] (1.75,1.55) node {\(_{L_0}\)};
		\draw[-,line width=1.25pt,color=olive,dotted] (2,1.35) -- (2.48,1.35);
	    \draw[-,line width=1.25pt,dotted,color=olive] (-1,0.15) -- (-0.5,0.15);
	    \draw[-,line width=1pt] (-1,0.05) -- (-1,0.25);    
	    \draw[-,line width=1pt] (0,-0.15) -- (0,0.25);
		\draw (0,-0.3) node {\(_{k_1}\)};	
		\draw (-0.5,-0.3) node {\(_{k_0}\)};    
		\draw[color=red] (-0.25,0.35) node {\(_{L_0}\)}; 
		\draw[-,line width=1.25pt,color=red] (-0.5,0.15) -- (-0.02,0.15); 
		\draw[color=blue] (-0.25,-0.6) node {\(_{K_0}\)};
		\draw[-,line width=1.5pt,color=blue] (-0.47,0) -- (-0.02,0);
		\draw[-,line width=1pt] (0.5,-0.15) -- (0.5,0.15);  
		\draw (0.5,-0.3) node {\(_{k_2}\)}; 
		\draw[color=blue] (1.1,-0.6) node {\(_{K_2}\)};
		\draw[-,line width=1.5pt,color=blue] (0.51,0) -- (1.48,0);
		\draw[-,line width=1pt] (1.5,-0.15) -- (1.5,0.15); 
		\draw (1.5,-0.3) node {\(_{k_3}\)}; 
		\draw[-,line width=1pt] (0.485,0.15) -- (1.515,0.15); 
		\draw[-,line width=1pt] (1.9,-0.15) -- (1.9,0.15);
		\draw (1.9,-0.3) node {\(_{k_4}\)};     
		\draw[-,line width=1pt] (2.9,-0.15) -- (2.9,0.15);  
		\draw (2.9,-0.3) node {\(_{k_5}\)};
		\draw[-,line width=1pt] (1.885,0.15) -- (2.915,0.15);
		\draw[color=blue] (3.25,-0.6) node {\(_{K_5}\)};
		\draw[-,line width=1.5pt,color=blue] (2.91,0) -- (3.58,0); 
		\draw[-,line width=1pt] (3.6,-0.15) -- (3.6,0.15);
		\draw (3.6,-0.3) node {\(_{k_6}\)};
		\draw[-,line width=1pt] (3.585,0.15) -- (4,0.15);
		\draw[-,line width=1pt,dotted] (4,0.15) -- (4.6,0.15);		
%		
		%Pfeile
		\draw[->,line width=0.8pt,dotted] (2.3,1.15) .. controls (2.3,0.8) and (3.7,0.3) .. (3.8,0.3);	
		\draw (3.45,0.65) node {\(_{h_{\alpha_6}}\)};
		\draw[->,line width=0.8pt,dotted] (2.1,1.15) .. controls (2.1,0.8) and (2.5,0.4) .. (2.6,0.3);	
		\draw (2.15,0.4) node {\(_{h_{\alpha_4}}\)};		
		\draw[->,line width=0.8pt,dotted] (1.8,1.15) .. controls (1.8,0.8) and (1.4,0.4) .. (1.3,0.3);
		\draw (1,0.35) node {\(_{h_{\alpha_2}}\)};		
		\draw[->,line width=0.8pt,dotted,color=red] (1.65,1.15) .. controls (1.65,0.8) and (0.4,0.45) .. (0.08,0.3);
		\draw (0.7,0.9) node {\(_{h_{\alpha_0}}\)};			 	    
		\end{tikzpicture}}
		\end{align}
     \end{enumerate}
   	\end{enumerate}
\end{lemma}
\begin{proof}	
\begin{enumerate}
\item
	The implications $\textit{(a)}$ and $\textit{(b)}$ are clear, and the implication $\textit{(c)}$ follows from the analyticities of $\wm_h\cp \gamma$ and $\gamma$ just by
	\begin{align*}	
		h\in G_{\gamma|_D}\qquad\Longrightarrow\qquad (\wm_h\cp \gamma)|_D=\gamma|_D 
		\qquad\Longrightarrow\qquad \wm_h\cp \gamma =\gamma.
	\end{align*}
	Now, to show that $\wm_h\cp \gamma=\gamma$ holds for $h\in G_\gamma$, we assume that $\wm_h(\gamma(t_0))\neq \gamma(t_0)$ for $t_0\in (a,b)$.\footnote{Observe that the end points of $\gamma$ are necessarily fixed by $\wm_h$.} Then, we have $\wm_h(\gamma(t_0))=\gamma(t_1)$ for some $t_1\in (a,b)$, and we may assume that $t_0<t_1$ holds. Let $t_0<s_0<t_1$. Then $\wm_h(\gamma(s_0))=\gamma(s_1)$ for some $t_1< s_1$ because $\gamma^{-1}\cp \wm_{h}\cp \gamma$ is a homeomorphism, hence monotonous. Applying $\wm_h$ inductively, 
      we obtain sequences $\{t_n\}_{n\in \NN}, \{s_n\}_{n\in \NN}\subseteq (a,b)$ with $t_n <s_n< t_{n+1} < s_{n+1}$ as well as $\gamma(t_n)=\wm_{h^n}(\gamma(t_0))$ and $\gamma(s_n)=\wm_{h^n}(\gamma(s_0))$ for all  $n\in \NN$. 
      Obviously, $\lim_n s_n =\lim_n t_n\in[a,b]$ exists, so that 
	\begin{align*}      
       \lim_n \wm_{h^{n}}(\gamma(t_0))=\lim_n \gamma(t_n)=\lim_n \gamma(s_n)=\lim_n \wm_{h^{n}}(\gamma(s_0)). 
	\end{align*}       
	Pointwise properness of $\wm$ now implies $\gamma(t_0)=\gamma(s_0)$, which 
       contradicts that $t_0\neq s_0$ and injectivity of $\gamma$.
    \item
    It suffices to show the claim for $D=K$ compact. Moreover,  
    we only consider the case where $t<b$ holds because the case $a<t$ follows analogously. Then, switching from $\gamma$ to $\gamma^{-1}$ and reparametrizing if necessary, we can assume that $K=[0,r]$ with $0\leq t< r$. 
    
     Now, assume that the statement is wrong, i.e., that for each $g\in G\backslash G_{\gamma}$ with $\wm_g\cp \gamma|_{K}\cpsim \gamma|_K$ we have $g\in G_{\gamma(t)}$.\footnote{Observe that we find such a group element  since elsewise $\gamma|_K$ would be a free segment.}  
     Then, to derive a contradiction, it suffices to show that: 
   
   	 {\bf Claim:}  For $t<k_0<k<b$ we find $g\in G_{\gamma(t)}$ with $\wm_g(\gamma(k))=\gamma(k')$ for $t< k'<k_0$. 
   	 
   	 In fact, let $n_0\geq 1$ with $t+ \textstyle\frac{1}{n_0}< r$. Then, for each $n\geq n_0$ we find $g_n\in G_{\gamma(t)}$ such that 
	\begin{align*}   	 
   	 \wm_{g_n}(\gamma(r))=\gamma(t_n)\qquad \text{for}\qquad t< t_n<t+ \textstyle\frac{1}{n}.
	\end{align*}   	 
   	  Consequently, $\lim_n \wm_{g_n}(\gamma(r))=\gamma(t)=\lim_n \wm_{g_n}(\gamma(t))$, hence $\gamma(r)=\gamma(t)$ by pointwise properness of $\wm$. Since $t\neq r$, this contradicts injectivity of $\gamma$. 

      \vspace{3pt}
      {\bf Proof of the Claim:} 
      
      \vspace{-3pt}
      Since $\gamma|_{[t,k_0]}$ is not a free segment ($\gamma$ is not free), we find $g\in G\backslash G_\gamma$ with $\wm_g\cp \gamma|_{[t,k_0]} \cpsim \gamma|_{[t,k_0]}$, 
      and by assumption we have $g\in G_{\gamma(t)}$. The two possible configurations are 
      \begin{align*}  
      \hspace{40pt}
      \begin{tikzpicture}
		\draw[-,line width=1pt,color=red] (0,0) .. controls (0.1,0.6) and (0.7,0.4) .. (1,0.1);
		\draw[-,line width=1pt,color=red] (1,0.1) .. controls (1.1,0) and (1.2,0) .. (1.3,0);
		\draw[->,line width=1pt,color=red] (1.4,0) .. controls (1.5,0) and (1.7,0.2) .. (1.7,0.4);
		\draw[->,line width=1.5pt] (0,0) -- (2,0);
		\filldraw[black] (0,0) circle (2pt);
		\draw[color=red] (1.5,0.5) node {\(_{k_0}\)};
		\draw (2,-0.3) node {\(_{k_0}\)};
		\draw (-0.2,-0.02) node {\(_t\)};
		\draw (2.2,0) node {\(\gamma\)};
		\draw[color=red] (2.35,0.5) node {\(\wm_g\cp\gamma\)};	
		\end{tikzpicture}   
		\qquad\qquad\raisebox{20pt}{$\text{and}$} \qquad\qquad\quad	
    	\begin{tikzpicture}	
		\draw[-,line width=1pt,color=red] (0,0) .. controls (0,1.2) and (2,0.4) .. (1.4,0);
		\draw[->,line width=1pt,color=red] (1.1,0) .. controls (1,0) and (0.7,0) .. (0.6,0.25);
		\draw[->,line width=0.8pt,color=red] (1.1,0) .. controls (0.8,0) and (0.7,0) .. (0.5,0.45);
			    \draw[->,line width=1.5pt] (0,0) -- (2,0);
		\filldraw[black] (0,0) circle (2pt);
		\draw[color=red] (0.4,0.17) node {\(_{k_0}\)};		
		\draw[color=red] (0.72,0.43) node {\(_{k}\)};
		\draw (2,-0.3) node {\(_{k_0}\)};
		\draw (-0.2,-0.02) node {\(_t\)};
		\draw (2.25,0) node {\(\gamma.\)};
		\draw[color=red] (2,0.5) node {\(\wm_g\cp\gamma\)};	
		\end{tikzpicture} 
	\end{align*} 
      Then, by injectivity of $\gamma$ and Lemma \ref{lemma:BasicAnalyt}.\ref{lemma:BasicAnalyt4}, one of the following situations holds:   
            \begin{align*}  
      \hspace{36.5pt}
      &\raisebox{20pt}{
      \begin{tikzpicture}
	    \draw[->,line width=1pt,color=red] (0,0) -- (1.5,0);
	    \draw[->,line width=1.5pt] (0,0) -- (2,0);
		\filldraw[black] (0,0) circle (2pt);
		\draw[color=red] (1.45,0.25) node {\(_{k_0}\)};
		\draw (2,-0.3) node {\(_{k_0}\)};
		\draw (-0.2,-0.02) node {\(_t\)};
		\draw (2.2,0) node {\(\gamma\)};
		\draw[color=red] (0.6,0.25) node {\(\wm_g\cp\gamma\)};	
		\end{tikzpicture}}%
		%%%%%%%%%%%%%%%%%%%%%%%%%%%%%%%%%%%
		\qquad\qquad\qquad\raisebox{20pt}{$\text{or}$} \qquad\qquad\quad\:
		%%%%%%%%%%%%%%%%%%%%%%%%%%%%%%%%%%%	
    	\begin{tikzpicture}	
		\draw[-,line width=1pt,color=red] (0,0) .. controls (0,0.8) and (2,0.4) .. (1.4,0);
		\draw[->,line width=1pt,color=red] (1.1,0) -- (0.85,0);
		\draw[->,line width=0.8pt,color=red] (1.1,0) -- (0.4,0);
		\draw[->,line width=1.5pt] (0,0) -- (2,0);
		\filldraw[black] (0,0) circle (2pt);
		\draw[color=red] (0.9,0.25) node {\(_{k_0}\)};
		\draw[color=red] (0.5,0.23) node {\(_{k}\)};
		\draw (2,-0.3) node {\(_{k_0}\)};
		\draw (-0.2,-0.02) node {\(_t\)};
		\draw (2.25,0) node {\(\gamma.\)};
		\draw[color=red] (2,0.5) node {\(\wm_g\cp\gamma\)};	
		\end{tikzpicture} \\[-30pt]
		%%%%%%%%%%%%%%%%%%%%%%%%%%
      &\raisebox{0pt}{
      \begin{tikzpicture}
      	\draw[->,line width=1pt,color=red] (0,0) -- (2.5,0);
	    \draw[->,line width=1.5pt] (0,0) -- (2,0);
		\filldraw[black] (0,0) circle (2pt);
		\draw (1.9,0.3) node {\(_{k_0}\)};
		\draw[color=red] (2.5,-0.3) node {\(_{k_0}\)};
		\draw (-0.2,-0.02) node {\(_t\)};
		\draw[color=red] (2.7,0.25) node {\(\wm_g\cp \gamma\)};
		\draw (1,0.25) node {\(\gamma\)};	
		\end{tikzpicture}} 
	\end{align*} 
	More precisely, 
      \begingroup
      \setlength{\leftmarginii}{25pt}
      \begin{itemize}
        \item[(a)]
        In the first case, we have
		\begin{align*}        
		\gamma|_{[t,s_0]}&\psim \wm_g\cp \gamma|_{[t,k_0]} \:\text{ for some }\: s_0\in (t,k_0]\qquad\text{or}\\ 
        \gamma|_{[t,k_0]}&\psim \wm_g\cp \gamma|_{[t,s_0]} \hspace{3pt}\text{ for some }\: s_0\in (t,k_0].
        \end{align*}        
        In both situations, for $s_0=k_0$ we would have $g\in G_\gamma$, so that $s_0<k_0$ holds just because $g\notin G_\gamma$ by assumption. 
        Moreover, replacing $g$ by $g^{-1}$ if necessary, we can assume that the first relation holds. 
		Then $\wm_g(\gamma)(k_0)=\gamma(s_0)$ for some $t<s_0<k_0$, and  applying $\wm_g$ once more, we find that $\wm_g(\gamma(s_0))=\gamma(s_1)$ for some $t<s_1<s_0$ just by monotonicity of $\gamma^{-1}\cp \wm_h\cp \gamma$. Inductively, we obtain $\{s_n\}_{n\in \NN}\subseteq (t,k_0)$ with $s_{n+1}< s_n$ and $\gamma(s_n)=\wm_{g^n}(\gamma(s_0))$ for all $n\in \NN$. 
        
        Let $s:=\lim_n s_n$. Then $\gamma(s)=\lim_n \wm_{g^{n}}(\gamma(s_0))$, hence $\wm_g(\gamma(s))=\gamma(s)$ by continuity of $\wm_g$. Consequently, $\lim_n \wm_{g^n}(\gamma(s))=\gamma(s)=\im_n\wm_{g^n}(\gamma(s_0))$, so that by pointwise properness of $\wm$ we have $\gamma(s)=\gamma(s_0)$. This, however contradicts $s<s_0$ and injectivity of $\gamma$, showing that case (a) cannot occur.
      \item[(b)]
      	In the second case, we have 
      	$\wm_g\cp \gamma|_{[s,k_0]}\psim \big[\gamma|_{[s',k_0']}\big]^{-1}$ for some $t<s<k_0$ and some $t<s'<k_0'<k_0$.
	Then, Lemma \ref{lemma:BasicAnalyt}.\ref{lemma:BasicAnalyt4} and injectivity of $\wm_g\cp \gamma|_{[t,k]}$ show that 
	\begin{align*}
			\wm_g\cp \gamma|_{[s,k]}\psim \big[\gamma|_{[k',k_0']}\big]^{-1}
	\end{align*}
	 holds for some $t<k'< s'< k_0'<k_0$, hence $\wm_g(\gamma(k))=\gamma(k')$ for $1<k'<k_0$.
      \end{itemize}
      \endgroup
 \item
	 If $I$ is finite, then $\textit{(b)}$ is clear from Lemma \ref{lemma:BasicAnalyt}.\ref{Basanalyt}. However, if $I$ is infinite, then we find  $\{\alpha_{n_i}\}_{i\in \mathbb{N}}\subseteq I$ mutually different with $\gamma|_{K_{n_i}}\psim \delta^{\pm 1}$ for intervals $K_{n_i}\subseteq K$ which mutually can only share start and end points.
Let $k\neq k'$ be contained in $K_{n_0}$ 
and define sequences $\{x_i\}_{i\in \NN}, \{x'_i\}_{i\in \NN} \subseteq \dom[\gamma]$ by 
      \begin{align*}        
        x_i:=\gamma^{-1}\!\left(\wm\big(h_{\alpha_{n_i}}\cdot h^{-1}_ {\alpha_{n_0}},\gamma(k)\big)\right)\qquad\text{as well as}\qquad x'_i:=\gamma^{-1}\!\left(\wm\big(h_{\alpha_{n_i}}\cdot h^{-1}_ {\alpha_{n_0}},\gamma(k')\big)\right).
      \end{align*}
      By compactness of $\dom[\gamma]$, we can assume that $\lim_i x_i =x\in \dom[\gamma]$ exists. Then, since 
      the intervals $K_{n_i}$ can only share start and end points, it follows that $\lim_i x'_{i}=x$ holds. 
      
      \hspace{150pt}     
     \begin{tikzpicture}
     	%gamma
	    \draw[->,line width=1.5pt] (0,0) -- (4,0);
	    \filldraw[black] (0,0) circle (2pt);
	   	\draw (4.3,0) node {\(\gamma\)};
	
		\draw[-,line width=3pt] (0.5,0) -- (1.2,0);
		\draw (0.7,0.3) node {\(_{K_n}\)};  
		\draw[-,line width=1pt,color=blue] (1,-0.068) -- (1,0.4);
		\draw[color=blue] (1,0.5) node {\(_{x_n}\)};
		
		\draw[-,line width=1pt] (0.6,0.068) -- (0.6,-0.4);
		\draw (0.7,-0.5) node {\(_{x'_n}\)};
		
		\draw[-,line width=3pt] (2.45,0) -- (3.2,0); 
		\draw (3.3,0.28) node {\(_{K_{n+1}}\)};
		\draw[-,line width=1pt,color=blue] (2.6,-0.068) -- (2.6,0.4);
		\draw[color=blue] (2.6,0.5) node {\(_{x_{n+1}}\)};		 
		 
		\draw[-,line width=1pt] (3.1,0.068) -- (3.1,-0.4);
		\draw (3.1,-0.5) node {\(_{x'_{n+1}}\)};		 
		 
		\draw[-,line width=0.8pt,dashed,color=red] (0.9,-0.25) -- (2.7,-0.25);
		\draw[color=red] (1.8,-0.55) node {\(_{B_\epsilon(x)}\)};
		
		\draw[color=red] (0.85,-0.25) node {\(_(\)};
		\draw[color=red] (2.75,-0.25) node {\(_)\)};
				
		\draw[-,line width=3pt] (1.5,0) -- (2,0);  
		\draw (1.9,0.3) node {\(_{K_{m>n+1}}\)};
		\end{tikzpicture} 
		    
      Hence, for $g_i:=h_{\alpha_{n_i}}\cdot h^{-1}_ {\alpha_{n_0}}$ we have
	 \begin{align*}      
      \lim_i \wm_{g_i}(\gamma(k))=\lim_i \gamma(x_i)=\gamma(x)= \lim_i \gamma(x'_i)=\lim_i \wm_{g_i}(\gamma(k')),
      \end{align*}
      so that $\gamma(k)=\gamma(k')$ by  pointwise properness of $\wm$. This contradicts the choices and shows that $I$ is finite. 
 \end{enumerate}
\end{proof}

\begin{definition}
  \label{def:freeSeg}
  \begin{enumerate}
  \item
    \label{def:freeSeg3}
     By $\text{\gls{PAF}}\subseteq \Paw$ we will denote the set of all free embedded analytic curves.
	\item 
	 \label{def:freeSeg32}   
     We define the subsets $\Pafs,\Pafns\subseteq \Paw$ of free symmetric and free non-symmetric curves by 
    \begin{align*}    
      \text{\gls{PAFS}}:=\{\gamma\in \Paf\:|\: G_\gamma\neq\{e\}\}\quad\qquad \text{and}\qquad\qquad \text{\gls{PAFNS}}:=\{\gamma\in \Paf\:|\: G_\gamma=\{e\}\},
    \end{align*}
    respectively.
  \item
    \label{def:freeSeg4}
   The set of continuously generated curves is defined by $\text{\gls{PAC}}:=\Paw\backslash \Paf$. Then, by $\text{\gls{PACS}}:=\Pac\backslash \Pags$ we will denote the set of 
    continuously but not Lie algebra generated curves.  
  \end{enumerate} 
\end{definition}
\begin{remark}
\label{decrem}
	   It is immediate from the definition of a Lie algebra generated curve that $\Pags \cap \Paf=\emptyset$, hence $\Pags\subseteq \Paw \backslash \Paf=\Pac$ holds. Consequently, we have $\Pac=\Pags\sqcup \Pacs$ which shows that
	\begin{align}
	\label{eq:decompospw}	   
	   \Paw= \Pags\sqcup\Pacs \sqcup \Pafs\sqcup \Pafns
	 \end{align}
	 holds.
	 \hspace*{\fill} $\lozenge$
\end{remark}
The first part of (the next) Proposition \ref{prop:freeseg} shows that for $\wm$ analytic and pointwise proper, a free curve is discretely generated by the symmetry group. 
This means that for each free curve $\gamma$ we find a free segment $\delta$ such that $\gamma$ can be decomposed into subcurves, each being equivalent to a translate of an initial or final segment of $\delta$. Here, each of these subcurves which is not an initial or final segment of $\gamma$ then even equals the translate the full segment $\delta$. 
This decomposition will be inalienable for the construction\footnote{More precisely, for the separation property of the projective structure introduced there, as well as surjectivity of the involved projection maps.} of the normalized Radon measure on $\AQRInd{\mathrm{FN}}\cong \IHOMFNS$ in Subsection \ref{sec:FreeM}.

The second part of Proposition \ref{prop:freeseg} shows that $\Paw=\Pags\sqcup \Paf$ holds whenever $\wm$ is transitive and only admits  stabilizers which are normal subgroups. Obviously, then we even have $\Paw=\Pags\sqcup \Pafns$ if $\wm$ is in addition free.  
Similarly, in the third part, we will prove that each continuously generated curve is contained in a $\wm$-orbit if $\wm$ is proper.  Then, as in the second part, we will show that $\Paw=\Pags\sqcup \Paf$ holds if $\wm$ admits only normal stabilizers, hence $\Paw=\Pags\sqcup \Pafns$ if $\wm$ acts in addition freely. So, for $\wm$ non-trivial we have:
\renewcommand*{\arraystretch}{1.2}	
	\begin{center}
  \begin{tabular}{c|c|c|c|c}
    $\wm$       & $\Pags$ & $\Pacs$ & $\Pafns$ & $\Pafs$\\[3pt] \hline 
      free &  many& ? & ?& $\emptyset$\\
    transitive $+$ normal stabilizers   &  many & $\emptyset$ & ?& ?\\
      \hspace{10pt} proper $+$ normal stabilizers  &  many & $\emptyset$ & ?& ?
  \end{tabular}
  	\end{center}		
\renewcommand*{\arraystretch}{1}	
Anyhow, before we come to the proofs, we first want to give some straightforward applications of these results. 
\begin{example}[Free Curves in LQC]
  \label{eq:invelements}
  We now apply the results from Proposition \ref{prop:freeseg} to the \mbox{(semi-)homogeneous}, homogeneous isotropic and spherically symmetric LQC case.
  
  \par
  \begingroup
  \leftskip=8pt

  \vspace{8pt}

  \noindent
  {\bf\textit{(Semi-)Homogeneous LQC:}}
  \vspace{2pt}
  
  \noindent
  In the homogeneous case, $\wm$ is transitive and free, so that $\Paw=\Pags\sqcup \Pafns$ holds by Proposition \ref{prop:freeseg}.\ref{prop:freeseg2}. In the semi-homogeneous case (see Example \ref{ex:LQC}) $\wm$ is proper because the additive action of $\RR^3$ on $\RR^3$ is proper and each linear subspace of $\RR^3$ is closed. Since $\wm$ acts freely, Proposition \ref{prop:freeseg}.\ref{prop:freeseg3} shows that $\Paw=\Pags\sqcup \Pafns$  holds as well.
  \par
  \endgroup
  
  \noindent  

  \par
  \begingroup
  \leftskip=8pt

  \vspace{8pt}

  \noindent
  {\bf\textit{Homogeneous Isotropic LQC:}}
  \vspace{2pt}
  
  \noindent
  Unfortunately, $G_0=\SU \subseteq \Gee$ is not a normal subgroup, so that we cannot conclude that $\Pac=\emptyset$, i.e., $\Paw=\Pags\sqcup \Paf$ holds.  
  However, we have $\Paw=\Pags\sqcup\Pacs\sqcup \Pafns$ because $\Pafs=\emptyset$.
   
  In fact, if $\gamma\colon [0,k]\rightarrow \RR^3$ is symmetric, then we can assume that $\gamma(0)=0$ just by transitivity, and because $\wm_{g\cdot h\cdot g^{-1}}\cp (\wm_g\cp \gamma)=\wm_g\cp \gamma$ for $e\neq h\in G_\gamma$. So, let $\gamma(0)=0$ and $h\in G_\gamma$. Then $\wm_h(\gamma(0))=\gamma(0)$ shows $h\in \SU$, and since $\wm_h(\gamma(k))=\gamma(k)\neq 0$, $h$ corresponds to a rotation by some angle $\alpha\in (0,2\pi)$ around the axis (through $0$) determined by $\gamma(k)$. Since by Lemma \ref{lemma:curvestablemma}.\ref{lemma:curvestablemma1} we have $\wm_h(\gamma(t))=\gamma(t)$ for all $t\in [0,k]$, $\im[\gamma]$ is completely contained in this axis determined by $\gamma(k)$. It is now obvious\footnote{We have $\gamma \cpsim \gamma^0_{(x,0)}$ if we consider $x$ as vector in $\RR^3$, i.e., $(x,0)\in \RR^3\times \su$.} from Lemma \ref{lemma:sim}.\ref{lemma:sim5} that $\gamma\in \Pags$. This shows $\Pafs=\emptyset$.

  \vspace{8pt}
  \noindent
  {\bf \textit{Spherically Symmetric LQC:}}
  \vspace{2pt}
  
  \noindent
  By the same arguments as in the previous case, we see that a symmetric curve necessarily has to be contained in an axis through the origin. Since in the spherically symmetric situation linear curves are not Lie algebra generated, we have $\Pafs=\Paln$.\footnote{Recall (a) in Convention \ref{conv:sutwo1}.\ref{conv:sutwo2} for the definition of $\Paln$.} Moreover, each continuously generated curve has to be contained in a sphere by Proposition \ref{prop:freeseg}.\ref{prop:freeseg3}. 
  Unfortunately, we cannot conclude that $\Pacs=\emptyset$ holds, i.e.,  that we have $\Paw=\Pags\sqcup \Pafns \sqcup \Paln$ because for $x\neq 0$ the stabilizer $G_x$ equals the maximal torus $H_x\subseteq \SU$ which  is not a normal subgroup as well.    
  \hspace*{\fill}$\lozenge$ 
  \par
  \endgroup
\end{example}
We now come to the desired 
\begin{proposition}
  \label{prop:freeseg}
  Let $\wm$ be analytic and pointwise proper.
  \begin{enumerate}
  \item
    \label{prop:freeseg1}
    If $\gamma\colon [a,b]\rightarrow M$ is free, then we find a free segment $\delta\colon L\rightarrow M$ and points $a=t_0<{\dots}<t_n=b$ such that 
	\begin{align*}    
    \gamma|_{[t_i,t_{i+1}]}\psim \wm_{g_i}\cp [\delta|_{L_i}]^{\pm 1}\:\text{ for some }\: g_i\in G\qquad \forall\: 0\leq i\leq n-1.
    \end{align*} 
    Here $L_i=L=[l_1,l_2]$ for $1\leq i\leq k-2$ and $L_1,L_{n-1}$ are both of the form 
    \begin{align*}   
    	\:[m,l_2] \quad\text{ for }\quad  m\in [l_1,l_2)\qquad \qquad\text{or}\qquad\qquad [l_1,m]\quad \text{ for }\quad m\in (l_1,l_2].
    \end{align*}
    \hspace{100pt}
     \begin{tikzpicture}
     	%gamma 
	    \draw[-,line width=1.5pt] (-0.5,0) -- (4,0);
	    \draw[->,line width=1.5pt,dotted] (4,0) -- (6,0);
	    \draw[-,line width=1.5pt] (-0.5,-0.2) -- (-0.5,0.1);
	   	\draw (6.2,0) node {\(\gamma.\)};
	   	%delta
	    \draw[-,line width=1.25pt] (1.5,1.25) -- (2.5,1.25);
	    \draw[-,line width=1pt] (1.5,1.16) -- (1.5,1.39);
	    \draw[-,line width=1pt] (2.5,1.16) -- (2.5,1.39);	   
		\draw (2.7,1.3) node {\(\delta\)};	 
		\draw[-,line width=1.25pt,color=red] (1.515,1.35) -- (1.97,1.35); 
		\draw[color=red] (1.75,1.55) node {\(_{L_0}\)};
		\draw[-,line width=1.25pt,color=olive,dotted] (2,1.35) -- (2.48,1.35);
%   
	    %intervalle
	    \draw[-,line width=1.25pt,dotted,color=olive] (-1,0.15) -- (-0.5,0.15);
	    \draw[-,line width=1pt] (-1,0.05) -- (-1,0.25);   
	    
	    \draw[-,line width=1pt] (0,-0.15) -- (0,0.25);
		\draw (0,-0.3) node {\(_{k_1}\)};	
		\draw (-0.5,-0.3) node {\(_{k_0}\)};    
		\draw[color=red] (-0.25,0.35) node {\(_{L_0}\)}; 
		\draw[-,line width=1.25pt,color=red] (-0.5,0.15) -- (-0.02,0.15); 
	    
		\draw[-,line width=1pt] (1,-0.15) -- (1,0.15); 
		\draw (1,-0.3) node {\(_{k_2}\)}; 
 
		\draw[-,line width=1pt] (1.9,-0.15) -- (1.9,0.15);  
		\draw (1.9,-0.3) node {\(_{k_3}\)};
		
		\draw[-,line width=1pt] (2.8,-0.15) -- (2.8,0.15);  
		\draw (2.8,-0.3) node {\(_{k_4}\)};
		
		\draw[-,line width=1pt] (3.6,-0.15) -- (3.6,0.15);  
		\draw (3.6,-0.3) node {\(_{k_5}\)};	
		
		\draw[-,line width=1pt,dotted] (4.6,-0.15) -- (4.6,0.15);  
		\draw (4.6,-0.3) node {\(_{k_6}\)};		
		
		%Pfeile
		\draw (3.4,0.6) node {\(_{g_5}\)};
		\draw[->,line width=0.8pt,dotted] (2.4,1.15) .. controls (2.5,0.8) and (3.5,0.3) .. (3.95,0.1);	
		\draw (3,0.15) node {\(_{g_{4}}\)};
		\draw[->,line width=0.8pt,dotted] (2.27,1.15) .. controls (2.3,0.8) and (3.1,0.3) .. (3.45,0.1);
		\draw (2.3,0.15) node {\(_{g_{3}}\)};
		\draw[->,line width=0.8pt,dotted] (2.15,1.15) .. controls (2.15,0.9) and (2.4,0.3) .. (2.6,0.1);
		\draw (1.7,0.15) node {\(_{g_{2}}\)};		
		\draw[->,line width=0.8pt,dotted] (2,1.15) .. controls (2,1) and (1.5,0.4) .. (1.2,0.1);
		\draw[->,line width=0.8pt,dotted] (1.8,1.15) .. controls (1.8,0.8) and (0.7,0.2) .. (0.6,0.1);
		\draw (0.4,0.15) node {\(_{g_{1}}\)};		
		\draw[->,line width=0.8pt,dotted,color=red] (1.65,1.15) .. controls (1.65,0.8) and (0.4,0.45) .. (0.08,0.3);
		\draw (0.7,0.8) node {\(_{g_{0}}\)};			
		\end{tikzpicture}
  \item
    \label{prop:freeseg2}
    If $\wm$ is transitive and $G_x$ is a normal subgroup for (one and then each) $x\in M$, then an element of $\Paw$ is either free or contained in $\Pags$, i.e., 
    $\Paw=\Pags\sqcup \Paf$. If $\wm$ acts even free, then $\Paw=\Pags\sqcup \Pafns$ holds.
  \item
    \label{prop:freeseg3}
    If $\wm$ is proper, then each $\gamma\in \Pac$ is contained in a $\wm$-orbit. 
    If $G_x$ is in addition a normal subgroup for all $x\in M$, then $\Paw=\Pags\sqcup \Paf$ holds. 
  \end{enumerate} 
  \begin{proof}
    \begin{enumerate}
    \item
  Basically, we have to show that a suitable choice of $\delta$ fills the gaps in \eqref{eq:decompo}. 
  
  Since $\gamma$ is free, we find $K\subseteq [a,b]$ compact such that $\gamma|_{K}$ is a free segment. Let $\MK$ denote the set of all such intervals and let $\MK$ be ordered by inclusion. Then each chain $\ML$ in $\MK$ has an upper bound, namely the closure $L$ of $\bigcup_{L'\in \ML} L'$. In fact, $\delta_L:=\gamma|_L$    
  is a free segment because $\delta_L \cpsim \wm_g\cp \delta_L$ implies $\gamma|_{L'} \cpsim \wm_g\cp \gamma|_{L'}$ for some $L'\in \ML$, hence $\gamma|_{L''} \cpsim \wm_g\cp \gamma|_{L''}$ for all $\ML\ni L''\geq L'$. Consequently, $\gamma|_{L''}\psim \wm_g\cp \gamma|_{L''}$ for all $L''\geq L'$, hence $\gamma|_L\isim \wm_g\cp \gamma|_L$ by continuity. Then, $\gamma|_L\psim \wm_g\cp \gamma|_L$ by Lemma \ref{lemma:BasicAnalyt}.\ref{lemma:BasicAnalyt2}, so that $L$ is an upper bound of $\ML$. Consequently, by Zorn's lemma the set $\KMAX\subseteq \MK$ of maximal elements is non-empty. 
  
	Let $L=[l_1,l_2]\in \KMAX$, $\delta:=\gamma|_L$ and $a=k_0<\dots<k_n=b$ the respective decomposition  from Lemma \ref{lemma:curvestablemma}.\ref{lemma:curvestablemma3}. Let $\{h_\alpha\}_{\alpha \in I}\subseteq H_{[\gamma,\delta]}$ be a family of representatives and $0\leq i_1<\dots<i_m\leq n-1$ the indices for which we find $\alpha_p \in I$ with  $\gamma|_{K_{i_p}}\psim \wm_{h_{\alpha_p}}\cp [\delta|_{L_{i_p}}]^{\pm 1}$ holds, see Lemma \ref{lemma:curvestablemma}.\ref{lemma:curvestablemma3}. We now proceed in two steps:
	\begingroup
    \setlength{\leftmarginii}{22pt}
	\begin{enumerate}
	\item[\textbf{A)}]
	We first show that each $K_{i_p}$ can be assumed to be maximal, i.e., $K_{i_p}\in \KMAX$ for all $1\leq p\leq m$. 
	\item[\textbf{B)}]
	Then we show that under this assumption for each $1\leq p\leq m$ we have: 
	\begingroup
    \setlength{\leftmarginiii}{25pt}
	\begin{enumerate}
	\item[\textbf{i.)}]		
		$a<k_{i_p}$ \quad\hspace{10pt}$\Longrightarrow$\quad $\exists\:  \alpha\in I\backslash\{\alpha_p\} \colon (\wm_{h_\alpha}\cp \delta)(l_j)=\gamma(k_{i_p})$\hspace{10pt} for some $j\in \{1,2\}$.
	\item[\textbf{ii.)}]
	\vspace{3pt}	
		$k_{i_p+1}< b$ \quad$\Longrightarrow$\quad $\exists\: \alpha\in I\backslash\{\alpha_p\} \colon (\wm_{h_\alpha}\cp \delta)(l_j)=\gamma(k_{i_p+1})$ for some $j\in \{1,2\}$.
	\end{enumerate}
	\endgroup 
	\end{enumerate}
	\endgroup
	\noindent
	Since  \textbf{B)} implies that $(i_1,\dots,i_m)=(1,\dots,n-1)$,  the claim then follows.
	\vspace{2pt} 
		
	{\bf Part A:}	
	First observe that $\gamma|_{K_{i_p}}$ is a free segment for $1\leq p\leq m$ because
	\begin{align*} 
		\wm_g\cp \gamma|_{K_{i_p}} \cpsim \gamma|_{K_{i_p}}\qquad&\Longrightarrow \qquad \wm_{h_{\alpha_p}^{-1}\cdot g\cdot h_{\alpha_p}}\cp \delta \cpsim \delta\\
		 \qquad &\Longrightarrow \qquad \wm_{h_{\alpha_p}^{-1}\cdot g\cdot h_{\alpha_p}}\cp \delta =\delta\\
		 \qquad &\Longrightarrow \qquad \wm_g\cp \gamma|_{K_{i_p}}=\gamma|_{K_{i_p}},
	\end{align*}
	hence $K_{i_p}\in \MK$.	Now, it may happen that $K_{i_p}\notin \KMAX$.
	
	 \hspace{25pt}    
     \begin{tikzpicture}
     	%gamma
	    \draw[-,line width=1.5pt] (0,0) -- (4,0);
	    \draw[->,line width=1.5pt,dotted] (4,0) -- (5,0);
	   	\draw (5.3,0) node {\(\gamma\)};
	   	\draw (9.2,0) node {\(\Longrightarrow \:\:\:\:\text{ replace } L \text{ by }L':=[k_2,t]\in\KMAX\)};
 		%die k's
	 	\draw[-,line width=1.8pt,color=blue] (0,0) -- (0.9,0);
	 	\draw[-,line width=1.8pt,dotted,color=olive] (-0.4,-0) -- (0,0);	 	 			
 		
 		\draw[-,line width=1.5pt] (0,-0.2) -- (0,0.2);
		\draw (0,-0.35) node {\(_{k_0}\)}; 
		\draw[color=blue] (0.45,-0.7) node {\(_{L=K_0}\)};
			
  		\draw[-,line width=1pt] (0.9,-0.15) -- (0.9,0.15);
		\draw (0.9,-0.35) node {\(_{k_1}\)}; 
		
		\draw[-,line width=1.8pt,color=blue] (1.4,0) -- (2.2,0);
		\draw[-,line width=3pt,color=white] (2.2,0) -- (2.6,0);
	 	\draw[-,line width=1.8pt,dotted,color=olive] (2.2,-0) -- (2.6,0);
		\draw[-,line width=1pt] (1.4,-0.15) -- (1.4,0.15);
		\draw (1.4,-0.35) node {\(_{k_2}\)}; 
		\draw[color=blue] (1.8,-0.7) node {\(_{K_{i_p}=K_2}\)};
		
		\draw[-,line width=1pt] (2.2,-0.15) -- (2.2,0.15);
		\draw (2.2,-0.35) node {\(_{k_3}\)}; 
		\draw[-,line width=1pt] (2.6,-0.15) -- (2.6,0.15);
		\draw (2.6,-0.35) node {\(_{t}\)};		
		
		\draw (3.4,-0.35) node {\(_{k_4}\)}; 
		\draw[-,line width=3pt,color=white] (3.38,0) -- (3,0);
	 	\draw[-,line width=1.8pt,dotted,color=olive] (3.38,0) -- (3,0);		
		\draw[-,line width=1.8pt,color=blue] (3.4,0) -- (4.2,0);
		\draw[-,line width=1pt] (3.4,0.15) -- (4.2,0.15);
		\draw[-,line width=1pt,dotted] (3.05,0.15) -- (3.4,0.15);
		\draw[-,line width=1pt] (3.4,-0.15) -- (3.4,0.15);	
		\draw[-,line width=1pt] (4.2,-0.15) -- (4.2,0.15);	
		\draw (4.2,-0.35) node {\(_{k_5}\)};
		\draw[color=blue] (3.82,-0.7) node {\(_{K_4}\)};		
		 			
	   	\draw (1.2,0.9) node {\(_{\wm_{h_{\alpha_p}}}\)};
		\draw[->,line width=0.8pt,dotted,color=black] (0.4,0.25) .. controls (0.8,0.8) and (1.4,0.8) .. (1.8,0.25);
		\draw[-,line width=1pt] (0,0.15) -- (0.9,0.15);
		\draw[-,line width=1pt,dotted] (-0.4,0.15) -- (0,0.15);
		\draw[-,line width=1pt] (1.4,0.15) -- (2.2,0.15);
		\draw[-,line width=1pt,dotted] (2.2,0.15) -- (2.6,0.15);
		
		% K4 shift
		\draw[->,line width=0.8pt,dotted,color=black] (0.5,0.25) .. controls (1,0.7) and (3.6,0.5) .. (3.8,0.25);
		\draw (3,0.8) node {\(_{\wm_{h_{\alpha_{q}}}}\)};
		\end{tikzpicture} 
		
	In this case, let $K_{i_p}\subseteq L'\in \KMAX$ and define $\delta':=\gamma|_{L'}$. Then,
	\begin{align*}	
		\delta=\gamma|_L \psim \wm_{h_{\alpha_p}^{-1}}\cp \delta'|_{K_{i_p}}
	\end{align*}
	 so that maximality of $L$ shows that either $L=K_0$ or $L=K_{n-1}$ holds. 
	Assume that $L=K_0$ and let $a=k'_0<\dots < k'_{n'}=b$ be the decomposition of $\dom[\gamma]$ w.r.t.\ $\delta'$. Then $K_0'=K_0=L\neq L'$, so that we can assume that $L\neq K_0$ from the beginning.	
	By the same argument we also can assume that $L\neq K_{n-1}$. Then, there cannot exist $1\leq p\leq m$ with $K_{i_p}\notin \KMAX$ since this (as we have shown) would contradict that $L\neq K_0,K_{n-1}$.
	
	\vspace{6pt}  
	{\bf Part B:}	
	We only discuss the second case as the first one follows analogously. Moreover, to simplify the notations, we assume that $K_{i_p}=[k_{i_p},k_{i_p+1}]=[0,k]$ and define $K:=K_{i_p}$. 
	Recall that $\delta':=\gamma|_K$ is a maximal free segment just because $K\in \KMAX$ by {\bf Part A}. Finally,  
	let $n_0\in \mathbb{N}$ with $k+\frac{1}{n_0}\leq b$. 
	It follows that $K_n:=\big[0,k+\frac{1}{n}\big]\notin \MK$ for each $n\geq n_0$ so that we find $g_n\in G\backslash G_{\gamma|_{K_n}}= G\backslash G_{\delta'}$
		 (see $\textit{(b)}$ in Lemma \ref{lemma:curvestablemma}.\ref{lemma:curvestablemma1}) with 
			$\gamma|_{K_n}\cpsim \wm_{g_n}\cp \gamma|_{K_n}$. 
	We now proceed in two steps: 
	\begingroup
    \setlength{\leftmarginii}{17pt}
	\begin{enumerate}
	\item[\textbf{I}]
		First, we show that the claim follows if there exists $h\in G\backslash G_{\delta'}$ with $g_{n} \segsims h$ for infinitely many $n\geq n_0$. 
	\item[\textbf{II}]
		Second, we show that such an element $h$ exists.
	\end{enumerate}
	\endgroup
\vspace{2pt}  
  {\bf Step I:}  
  We can assume that $g_n\segsims h$ holds for all $n\geq n_0$, hence
	\begin{align*}
			\gamma|_{K_n}\cpsim \wm_{h}\cp \gamma|_{K_n} \qquad \forall\: n\geq n_0.
	\end{align*} 
   Then, we find sequences
  \begin{align*}   
    	K_0\supseteq \{x_m\}_{m\in \NN} \rightarrow k' \in K\qquad\text{and}\qquad K_0\supseteq \{y_m\}_{m\in\NN} \rightarrow k'' \in K
  	\end{align*}
  with $\gamma(x_m)=\wm_{h}(\gamma(y_m))$ for all $m\in \NN$, hence
  $\gamma(k')=\wm_{h}(\gamma(k''))$. Moreover, since $\gamma(k')$ is an accumulation point of   
  		$\gamma(K_0)\cap (\wm_h\cp \gamma)(K_0)$, for $\gamma'$ an extension of 
  		$\gamma$ by Lemma \ref{lemma:BasicAnalyt}.\ref{lemma:BasicAnalyt1} we find open intervals $I',I''\subseteq \dom[\gamma']$ with $k'\in I'$, $k''\in I''$ and 
	\begin{align}
	\label{eq:inters}
		  	 \gamma'(I')=\wm_h(\gamma'(I'')). %\:\text{ for }\: \gamma'\: \text{ an extension of } \:\gamma.
	\end{align}	  	 
Then, $k',k'' \in \{0,k\}$, since elsewise we would have $\delta'\cpsim \wm_h\cp \delta'$, hence $h\in G_{\delta'}$ which contradicts the choice of $h$. However, we even can assume that $k'$ and $k''$ are not both zero. 

In fact, since $\gamma|_{K}\nsim_\cp \wm_{h}\cp \gamma|_{K}$, we can arrange that $\{x_m\}_{m\in \NN}\subseteq [k,k+1/n_0]$ (then $k'=k$) or $\{y_m\}_{m\in \NN}\subseteq [k,k+1/n_0]$ (then $k''=k$). 
	Combining this with \eqref{eq:inters}, we see that there are $\epsilon,\epsilon'>0$ such that one of the following situations holds: 
	
	\vspace{10pt}
	
	\quad
 \begin{tikzpicture}
 		\draw (0.6,1.05) node {\(_{\boldsymbol{k',\: k''=k}}\)};
     	%gamma
     	\draw[-,line width=1.5pt,dotted] (0,0) -- (1,0);
	    \draw[-,line width=1.5pt] (1,0) -- (3.5,0);
	    \draw[->,line width=1.5pt,dotted] (3.5,0) -- (4,0);
	   	\draw (4.3,0) node {\(\gamma\)};
	   	\filldraw[black] (0,0) circle (2pt);
	
 		\draw[-,line width=1.5pt] (2,-0.15) -- (2,0.15);
		\draw (2,-0.4) node {\(_{k}\)}; 
	
	 	\draw[-,line width=1.5pt] (1,-0.15) -- (1,0.15);
		\draw (1,-0.4) node {\(_{0}\)};	
		
%		\draw[-,line width=1pt,dotted,color=blue] (1,-0.2) -- (2,-0.2);	
		\draw[color=blue] (1.2,0.5) node {\(_{\delta'}\)};
		
		\draw[->,line width=0.8pt, dotted,color=blue] (1,0.3) -- (1.9,0.3);		
		\draw[<-,line width=0.8pt, dotted] (2.1,0.3) -- (3,0.3);
		\draw[->,line width=0.8pt, dotted,color=red] (1.6,0.3) -- (1.9,0.3);		
		\draw[<-,line width=0.8pt, dotted,color=orange] (2.1,0.3) -- (2.4,0.3);
		
		\draw[line width=0.8pt] (2,0.3) circle (1pt);		
		\draw[->,line width=0.8pt,dotted,color=black] (1.5,0.4) .. controls (1.7,1) and (2.3,1) .. (2.5,0.4);
		\draw (2,0.6) node {\(_{\wm_h}\)};
		\draw (2,1.05) node {\(_{\text{flip}}\)};
		\draw (3.25,0.7) node {\(_{h\in G_{\delta'(k)}}\)};				
				
	 	\draw[-,line width=1.5pt] (3,-0.15) -- (3,0.15);
	 	\draw[color=orange] (2.37,0) node {\()\)};
	 	\draw[color=orange] (2.5,-0.42) node {\(_{k+\epsilon}\)};
	 	\draw[color=red] (1.63,0) node {\((\)};
	 	\draw[color=red] (1.5,-0.42) node {\(_{k-\epsilon'}\)};	
		\end{tikzpicture}
%%%%%%%%%%%%%%%%%%%%%%%%
 \begin{tikzpicture}
     	\draw (0.32,1) node {\(_{\boldsymbol{k'=k}}\)};
     	\draw (0.32,0.75) node {\(_{\boldsymbol{k''=0}}\)};
     	%gamma
	    \draw[-,line width=1.5pt,dotted] (0,0) -- (1,0);
	    \draw[-,line width=1.5pt] (1,0) -- (3.5,0);
	    \draw[->,line width=1.5pt,dotted] (3.5,0) -- (4,0);
	   	\draw (4.3,0) node {\(\gamma\)};
	   	\filldraw[black] (0,0) circle (2pt);
	
 		\draw[-,line width=1.5pt] (2,-0.15) -- (2,0.15);
		\draw (2,-0.4) node {\(_{k}\)}; 
	
	 	\draw[-,line width=1.5pt] (1,-0.15) -- (1,0.15);
		\draw (1,-0.4) node {\(_{0}\)};	
		
		\draw[color=blue] (1.2,0.5) node {\(_{\delta'}\)};
		
		\draw[->,line width=0.8pt, dotted,color=blue] (1,0.3) -- (1.95,0.3);	
		\draw[->,line width=0.8pt, dotted] (2.05,0.3) -- (3,0.3);
		\draw[-,line width=0.8pt, dotted,color=red] (1.0,0.3) -- (1.3,0.3);	
		\draw[-,line width=0.8pt, dotted,color=orange] (2.05,0.3) -- (2.3,0.3);		
		
		\draw[->,line width=0.8pt,dotted,color=black] (1.5,0.4) .. controls (1.7,1) and (2.3,1) .. (2.5,0.4);
		\draw (2,0.6) node {\(_{\wm_h}\)};
		\draw (2,1.05) node {\(_{\text{shift}}\)};				
				
	 	\draw[-,line width=1.5pt] (3,-0.15) -- (3,0.15);
	 	\draw[color=orange] (2.25,0) node {\()\)};
	 	\draw[color=orange] (2.5,-0.42) node {\(_{k+\epsilon}\)};
	 	\draw[color=red] (1.25,0) node {\()\)};
	 	\draw[color=red] (1.25,-0.42) node {\(_{\epsilon'}\)};	
		\end{tikzpicture}
%%%%%%%%%%%%%%%%%%%%%%%%
  \begin{tikzpicture}
     	\draw (0.32,1) node {\(_{\boldsymbol{k'=0}}\)};
     	\draw (0.32,0.75) node {\(_{\boldsymbol{k''=k}}\)};
     	%gamma
	    \draw[-,line width=1.5pt,dotted] (0,0) -- (1,0);
	    \draw[-,line width=1.5pt] (1,0) -- (3.5,0);
	    \draw[->,line width=1.5pt,dotted] (3.5,0) -- (4,0);
	   	\draw (4.3,0) node {\(\gamma\)};
	   	\filldraw[black] (0,0) circle (2pt);
	
 		\draw[-,line width=1.5pt] (2,-0.15) -- (2,0.15);
		\draw (2,-0.4) node {\(_{0}\)}; 
	
	 	\draw[-,line width=1.5pt] (1,-0.15) -- (1,0.15);
		\draw[->,line width=0.8pt, dotted] (1,0.3) -- (1.95,0.3);
		\draw[->,line width=0.8pt, dotted,color=blue] (2.05,0.3) -- (3,0.3);
		\draw[-,line width=0.8pt, dotted,color=red] (3.05,0.3) -- (3.3,0.3);	
		\draw[-,line width=0.8pt, dotted,color=orange] (2.05,0.3) -- (2.3,0.3);		
				
		\draw[color=blue] (2.8,0.5) node {\(_{\delta'}\)};		
		\draw[<-,line width=0.8pt,dotted,color=black] (1.5,0.4) .. controls (1.7,1) and (2.3,1) .. (2.5,0.4);
		\draw (2,0.6) node {\(_{\wm_h}\)};
		\draw (2,1.05) node {\(_{\text{shift}}\)};				
				
	 	\draw[-,line width=1.5pt] (3,-0.15) -- (3,0.15);
	 	\draw (3,-0.4) node {\(_{k}\)};
	 	\draw[color=orange] (2.25,0) node {\()\)};
	 	\draw[color=orange] (2.3,-0.42) node {\(_{\epsilon}\)};
	 	\draw[color=red] (3.25,0) node {\()\)};
	 	\draw[color=red] (3.5,-0.42) node {\(_{k+\epsilon'}\)};	
		\end{tikzpicture}
		\vspace{10pt}
		
		 \begin{tabular}{lrcrclcl}
  		\:\: &$\boldsymbol{k',\: k''=k\colon}$ &\hspace{-10pt} & $\gamma([k,k+\epsilon))$&$=$&$(\wm_h\cp\delta')((k-\epsilon',k])$ &\hspace{34pt}& $h\in H_{\gamma,\delta'}$\\
 		 &$\boldsymbol{k'=k,\: k''=0\colon}$ & & $\gamma([k,k+\epsilon))$&$=$&$(\wm_h\cp\delta')([0,\epsilon'))$ && $h\in H_{\gamma,\delta'}$\\
 		 &$\boldsymbol{k'=0,\: k''=k\colon}$ & & $\delta'([0,\epsilon))$&$=$&$(\wm_h\cp \gamma)([k,k+\epsilon'))$ && $h^{-1}\in H_{\gamma,\delta'}$.
 		\end{tabular}
  		\vspace{6pt}
  		
 		Hence, in the same order
 		\begin{align*}
 			\gamma(k_{i_p+1})\:=\:\gamma(k)&\:=\:(\wm_h\cp\delta')(k)\hspace{9.5pt}\:=\:(\wm_{h\cdot h_{\alpha_p}}\cp \delta)(l)\qquad \hspace{9.5pt}\text{for some}\quad l\in [l_1,l_2]\\
 			\gamma(k_{i_p+1})\:=\:\gamma(k)&\:=\:(\wm_h\cp\delta')(0)\hspace{10pt}\:=\:(\wm_{h\cdot h_{\alpha_p}}\cp \delta)(l)\qquad\hspace{9.5pt} \text{for some}\quad l\in [l_1,l_2]\\ 		 			\gamma(k_{i_p+1})\:=\:\gamma(k)&\:=\:(\wm_{h^{-1}}\cp\delta')(0)\:=\:(\wm_{h^{-1}\cdot h_{\alpha_p}}\cp \delta)(l)\qquad \text{for some}\quad l\in [l_1,l_2].
 		\end{align*}
 		Moreover, in the first two cases we have $h\in H_{\gamma,\delta'}$, hence 
		\begin{align*}
		h\cdot h_{\alpha_{p}}\in H_{\gamma,\delta}\qquad \Longrightarrow\qquad h\cdot h_{\alpha_{p}}\segsim h_{\alpha_q} \text{ for some }I\ni \alpha_q \neq \alpha_p.
		\end{align*}
		For this, observe that $\alpha=\alpha_p$ would imply that $h\in G_\delta$, as a straightforward calculation shows. 	
		Then, by the same arguments we see that
		 $h^{-1}\cdot h_{\alpha_p}\segsim h_{\alpha_q}$ for some $I\ni {\alpha_q} \neq \alpha_p$ in the third case. 
		 Consequently, 
		 $\textbf{B.ii.)}$ follows if we show that even $l\in \{l_1,l_2\}$ holds. This, however, follows easily from the maximality of the interval $K_{i_q}$ that corresponds to $\alpha_q$.

\vspace{2pt}  
{\bf Step II:} 

We assume that an element $h\in G\backslash G_ {\delta'}$ as in Step I does not exist. Then, replacing $\{g_n\}_{n\in \NN}$ by a subsequence we can arrange that $g_n \nsim_{\delta'} g_m$ whenever $m>n$. Moreover, we can assume that $g_n\notin H_{\gamma,\delta'}$ for all $n\geq n_0$. In fact, if for each $n\geq n_0$ we find $\wt{n}\geq n$ with  $g_{\wt{n}}\in H_{\gamma,\delta'}$, then finiteness of $I$ shows that there must be $h\in H_{\gamma,\delta'}$ with the properties from Step I. 
  Thus, we can assume that  
  \begin{align*}   
    \gamma|_{K_n}\cpsim \wm_{g_n}\cp \gamma|_{[k,k+1/n]}\qquad \forall\:n\geq n_0  \qquad \text{as well as}\qquad g_n\nsim_{\delta'} g_m\qquad\forall\:  n\neq m,
  \end{align*}
 and we now show that this is impossible. 
 For this, we write\footnote{Basically, this just means that $\gamma_1$ and $\gamma_2$ traverse into the same direction.} $\gamma_1 \tsd \gamma_2$ for $\gamma_1,\gamma_2 \in \Paw$ iff $\gamma_1|_J=\gamma_2 \cp \adif$ for $\adif\colon J\rightarrow J'$ an analytic diffeomorphism with $\dot\adif >0$. Analogously, write $\gamma_1 \tdd\gamma_2$ if $\gamma_1|_J=\gamma_2 \cp \adif$ for $\adif\colon J\rightarrow J'$ an analytic diffeomorphism with $\dot\adif <0$.
 
  We proceed in two steps: 
  \begingroup
  \setlength{\leftmarginii}{20pt}
  \begin{enumerate}
  \item[i.)]
    Let $\gamma_n:=\wm_{g_n}\cp \gamma|_{[k,k+1/n]}$ for all $n\geq n_0$ and assume that $\gamma_n \tsd\gamma_m$. 
    Then, it follows from Lemma \ref{lemma:BasicAnalyt}.\ref{lemma:BasicAnalyt4} and $g_n\nsim_{\delta'} g_m$ that either 
    \vspace{5pt}
	\begin{align*}  
  	(\wm_{g_n}\cp \gamma)(K)\subseteq \gamma([k,k+1/m])\qquad \text{or} \qquad (\wm_{g_m}\cp \gamma)(K)\subseteq \gamma([k,k+1/n])\qquad \text{holds}.
  \end{align*} 
  
  \hspace{15pt}
 \begin{tikzpicture}	
		\draw[-,line width=0.8pt] (0.8,-0.12) -- (0.8,0.12);
	 	\draw (0.8,-0.4) node {\(_{k+\frac{1}{m}}\)};
		
		\draw[-,line width=0.8pt] (1.6,-0.12) -- (1.6,0.12);
	 	\draw (1.6,-0.4) node {\(_{k}\)};
		
		\draw[-,line width=0.8pt] (2.4,-0.12) -- (2.4,0.12);
		\draw (2.4,-0.4) node {\(_{0}\)};
		
		\draw[-,line width=0.8pt,color=orange] (2.2,0.5) .. controls (1.5,0.5) and (1.4,0.7) .. (1.3,0.3);
		\draw[-,line width=0.8pt,color=orange] (1.3,0.3) .. controls (1.3,0) and (1.2,0) .. (1.1,0);
		\draw[-,line width=0.8pt,color=orange] (1.1,0) .. controls (1,0) and (1,0.2) .. (1,0.35);
		\draw[-,line width=1pt,color=orange] (1.08,0.35) -- (0.92,0.35);
	 	\draw (0.5,0.35)[color=orange] node {\(_{k+\frac{1}{n}}\)};		
		
		\draw[-,line width=0.9pt,color=orange] (1.38,0.29) -- (1.22,0.31);
	 	\draw (1.55,0.33)[color=orange] node {\(_{k}\)};		
	 	
	 	\draw[-,line width=0.9pt,color=orange] (2.2,0.42) -- (2.2,0.58);
	 	\draw[color=orange] (2.4,0.5) node {\(_{0}\)};
	 	
	 	\draw[color=orange] (1.3,0.8) node {\(\gamma_m\)};	
		 \draw[-,line width=1.2pt] (0.8,0) -- (2.4,0);
		 
		\draw (2.8,0) node {\(\gamma_n\)};	
		\end{tikzpicture}
		\quad\:\:\raisebox{20pt}{$\boldsymbol{\Longrightarrow}$}\qquad 
 \begin{tikzpicture}	
		\draw[-,line width=0.8pt] (0.8,-0.1) -- (0.8,0.12);
	 	\draw (0.8,-0.4) node {\(_{k+\frac{1}{m}}\)};
		
		\draw[-,line width=0.8pt] (1.6,-0.1) -- (1.6,0.12);
	 	\draw (1.6,-0.4) node {\(_{k}\)};
		
		\draw[-,line width=0.8pt] (2.4,-0.1) -- (2.4,0.12);
		\draw (2.4,-0.4) node {\(_{0}\)};
		
		\draw[-,line width=1.2pt,color=orange] (1.1,0.03) -- (3.3,0.03);
		\draw[-,line width=1pt,color=orange] (1.08,0.39) -- (0.92,0.39);
	 	\draw (0.5,0.39)[color=orange] node {\(_{k+\frac{1}{n}}\)};			
		\draw[-,line width=0.8pt,color=orange] (1.1,0.04) .. controls (1,0.04) and (1,0.24) .. (1,0.39);		

		\draw[-,line width=1pt,color=orange] (2.6,-0.08) -- (2.6,0.17);
	 	\draw (2.6,0.35)[color=orange] node {\(_{k}\)};		
	 	
	 	\draw[-,line width=1pt,color=orange] (3.3,-0.08) -- (3.3,0.17);
	 	\draw[color=orange] (3.3,0.35) node {\(_{0}\)};
	 		
		 \draw[-,line width=1.2pt] (0.8,0) -- (2.4,0);
		 
		\end{tikzpicture}
		\qquad \raisebox{20pt}{or}\qquad
		 \begin{tikzpicture}	
		\draw[-,line width=0.8pt] (0.8,-0.1) -- (0.8,0.12);
	 	\draw (0.8,-0.4) node {\(_{k+\frac{1}{m}}\)};
		
		\draw[-,line width=0.8pt] (1.6,-0.1) -- (1.6,0.12);
	 	\draw (1.6,-0.4) node {\(_{k}\)};
		
		\draw[-,line width=0.8pt] (2.4,-0.1) -- (2.4,0.12);
		\draw (2.4,-0.4) node {\(_{0}\)};
		
		\draw[-,line width=1.2pt,color=orange] (1.1,0.03) -- (1.5,0.03);
		\draw[-,line width=1pt,color=orange] (1.08,0.39) -- (0.92,0.39);
	 	\draw (0.5,0.39)[color=orange] node {\(_{k+\frac{1}{n}}\)};			
		\draw[-,line width=0.8pt,color=orange] (1.1,0.04) .. controls (1,0.04) and (1,0.24) .. (1,0.39);		
		
		\draw[-,line width=1pt,color=orange] (1.2,-0.08) -- (1.2,0.17);
	 	\draw (1.2,0.35)[color=orange] node {\(_{k}\)};		
	 	
	 	\draw[-,line width=1pt,color=orange] (1.5,-0.08) -- (1.5,0.17);
	 	\draw[color=orange] (1.5,0.35) node {\(_{0}\)};
	 		
		 \draw[-,line width=1.2pt] (0.8,0) -- (2.4,0);	
		\end{tikzpicture}
 	
 	In fact, elsewise we would have 
 	\begin{align*}
 		\wm_{g_n}\cp \gamma|_{K} \cpsim \wm_{g_m}\cp \gamma|_{K}&\quad \Longrightarrow\quad \delta'\cpsim \wm_{g_n^{-1} g_m}\cp \delta'
 		 \quad \Longrightarrow\quad g_n^{-1} g_m \in G_{\delta'}\\ &\quad \Longrightarrow\quad g_n\sim_{\delta'} g_m.
 	\end{align*}	
 	In particular, it cannot happen that for infinitely many $n\geq n_0$ we find $n'>n$ with  	
 		$\gamma_n \tsd \gamma_{n'}$. 
 	In fact, then for each $\epsilon>0$ we would find 
 	$n>n_0$ and $h_n\in G$ with 
     $\wm_{h_n}(\gamma(K))\subseteq \gamma([k,k+\epsilon))$.  
   Hence, we would find a sequence $\{n_i\}_{i\in \NN}\subseteq \NN$ such that $\wm_{h_{n_i}}(\gamma(0))$ and $\wm_{h_{n_i}}(\gamma(k))$ both converge to $\gamma(k)$ so that $\gamma(0)=\gamma(k)$ by pointwise properness of $\wm$. Of course, this contradicts injectivity of $\gamma$.
  \item[ii.)]
	By assumption, we have    
  	\begin{align*} 
	%\label{eq:dsfsd}   
    \hspace{-18.5pt}\gamma|_{K_{n_0}}\tdd \gamma_n %\wm_{g_{n}}\cp \gamma|_{[k,k+1/n]}
    \qquad\qquad \text{or}\qquad\qquad \gamma|_{K_{n_0}}\tsd \gamma_n %\wm_{g_{n}}\cp \gamma|_{[k,k+1/n]}
	\end{align*}    
	for infinitely many $n\geq n_0$. %, i.e., 
	
	\hspace{25pt}
        \begin{tikzpicture}	
		\draw[-,line width=1.2pt] (-0.2,-0.12) -- (-0.2,0.12);
	 	\draw (-0.2,-0.4) node {\(_{0}\)};
		
		\draw[-,line width=1.2pt] (0.8,-0.12) -- (0.8,0.12);
	 	\draw (0.8,-0.4) node {\(_{k}\)};
	 	
	 	\draw[-,line width=1.2pt] (1.4,-0.12) -- (1.4,0.12);
	 	\draw (1.6,-0.478) node {\(_{k+\frac{1}{n_0}}\)};
		
		\draw[-,line width=0.8pt,color=orange] (1.9,0.5) .. controls (1.2,0.5) and (1.1,0.7) .. (1,0.3);
		\draw[-,line width=0.8pt,color=orange] (1,0.3) .. controls (1,0) and (0.9,0) .. (0.8,0);
		\draw[-,line width=0.8pt,color=orange] (0.7,0) .. controls (0.6,0) and (0.6,0.2) .. (0.6,0.35);

		\draw[-,line width=1.1pt,color=blue] (0.68,0.35) -- (0.52,0.35);
	 	\draw (0.2,0.35)[color=blue] node {\(_{k+\frac{1}{n}}\)};		
		
		\draw[-,line width=0.9pt,color=blue] (1.08,0.29) -- (0.92,0.31);
	 	\draw (1.25,0.33)[color=blue] node {\(_{k}\)};		
	 	
	 	\draw[-,line width=0.9pt,color=blue] (1.9,0.42) -- (1.9,0.58);
	 	\draw[color=blue] (2.1,0.5) node {\(_{0}\)};
	 	
	 	\draw[color=orange] (1.2,0.8) node {\(_{\wm_{g_{n}}\: \cp\: \gamma|_{K_n}}\)};
		 \draw[->,line width=1.2pt,dotted] (2.2,0) -- (2.8,0);
		\draw (3.1,0) node {\(\gamma\)};
		\filldraw[black] (-0.7,0) circle (1.5pt);	
		\draw[-,line width=1.2pt,dotted] (-0.7,0) -- (-0.2,0);
	    \draw[-,line width=1.2pt] (-0.2,0) -- (2.2,0);
	   
	    \draw[-,line width=0.8pt, color=olive,dotted] (-0.2,-0.14) -- (0.8,-0.14);
	    \draw[color=olive] (0.4,-0.35) node {\(_{\delta'}\)};
	    
	    \draw(-0.6,0.6) node {\(\tdd\)};
		\end{tikzpicture}
		\qquad\qquad\raisebox{16pt}{or}
		\qquad\quad\qquad 
        \begin{tikzpicture}
		
		\draw[-,line width=1.2pt] (-0.2,-0.12) -- (-0.2,0.12);
	 	\draw (-0.2,-0.4) node {\(_{0}\)};
		
		\draw[-,line width=1.2pt] (0.8,-0.12) -- (0.8,0.12);
	 	\draw (0.8,-0.4) node {\(_{k}\)};
	 	
	 	\draw[-,line width=1.2pt] (1.4,-0.12) -- (1.4,0.12);
	 	\draw (1.6,-0.478) node {\(_{k+\frac{1}{n_0}}\)};
		
		\draw[-,line width=0.8pt,color=orange] (0.6,0.3) .. controls (0.5,0.7) and (0.4,0.5) .. (-0.3,0.5);
		\draw[-,line width=0.8pt,color=orange] (0.8,0) .. controls (0.7,0) and (0.6,0) .. (0.6,0.3);
		\draw[-,line width=0.8pt,color=orange] (1,0.35) .. controls (1,0) and (0.9,0) .. (0.9,0);

		\draw[-,line width=1.1pt,color=blue] (0.92,0.35) -- (1.08,0.35);
	 	\draw (1.5,0.3)[color=blue] node {\(_{k+\frac{1}{n}}\)};		
		
		\draw[-,line width=0.9pt,color=blue] (0.68,0.31) -- (0.52,0.29);%
	 	\draw (0.4,0.31)[color=blue] node {\(_{k}\)};		
	 	
	 	\draw[-,line width=0.9pt,color=blue] (-0.3,0.42) -- (-0.3,0.58);
	 	\draw[color=blue] (-0.45,0.5) node {\(_{0}\)};
	 	
	 	\draw[color=orange] (0.4,0.8) node {\(_{\wm_{g_{n}}\: \cp\: \gamma|_{K_n}}\)};
		 \draw[->,line width=1.2pt,dotted] (2.2,0) -- (2.8,0);
		\draw (3.1,0) node {\(\gamma\)};
		\filldraw[black] (-0.9,0) circle (1.5pt);	
	    \draw[-,line width=1.2pt,dotted] (-0.9,0) -- (-0.2,0);
	    \draw[-,line width=1.2pt] (-0.2,0) -- (2.2,0);
	   
	    \draw[-,line width=0.8pt, color=olive,dotted] (-0.2,-0.14) -- (0.8,-0.14);
	    \draw[color=olive] (0.4,-0.35) node {\(_{\delta'}\)};
	    
	    \draw(-0.8,0.5) node {\(\tsd\)};
		\end{tikzpicture}
	  
	  	In the first case, Lemma \ref{lemma:BasicAnalyt}.\ref{lemma:BasicAnalyt4} in combination with $g_n\notin H_{\gamma,\delta'}$ shows that 
	 		
    	\hspace{25pt}
               \begin{tikzpicture}
		
		\draw[-,line width=1.2pt] (-0.2,-0.12) -- (-0.2,0.12);
	 	\draw (-0.2,-0.4) node {\(_{0}\)};
		
		\draw[-,line width=1.2pt] (0.8,-0.12) -- (0.8,0.12);
	 	\draw (0.8,-0.4) node {\(_{k}\)};
	 	
	 	\draw[-,line width=1.2pt] (1.4,-0.12) -- (1.4,0.12);
	 	\draw (1.6,-0.478) node {\(_{k+\frac{1}{n_0}}\)};
		
		\draw[-,line width=0.8pt,color=orange] (1.9,0.5) .. controls (1.2,0.5) and (1.1,0.7) .. (1,0.3);
		\draw[-,line width=0.8pt,color=orange] (1,0.3) .. controls (1,0) and (0.9,0) .. (0.8,0);
		\draw[-,line width=0.8pt,color=orange] (0.7,0) .. controls (0.6,0) and (0.6,0.2) .. (0.6,0.35);

		\draw[-,line width=1.1pt,color=blue] (0.68,0.35) -- (0.52,0.35);
	 	\draw (0.2,0.35)[color=blue] node {\(_{k+\frac{1}{n}}\)};		
		
		\draw[-,line width=0.9pt,color=blue] (1.08,0.29) -- (0.92,0.31);
	 	\draw (1.25,0.33)[color=blue] node {\(_{k}\)};		
	 	
	 	\draw[-,line width=0.9pt,color=blue] (1.9,0.42) -- (1.9,0.58);
	 	\draw[color=blue] (2.1,0.5) node {\(_{0}\)};
	 	
	 	\draw[color=orange] (1.2,0.8) node {\(_{\wm_{g_{n}}\: \cp\: \gamma|_{K_n}}\)};
		 \draw[->,line width=1.2pt,dotted] (2.2,0) -- (2.8,0);
		\draw (3.1,0) node {\(\gamma\)};
		\filldraw[black] (-0.7,0) circle (1.5pt);	
		\draw[-,line width=1.2pt,dotted] (-0.7,0) -- (-0.2,0);
	    \draw[-,line width=1.2pt] (-0.2,0) -- (2.2,0);
	   
	    \draw[-,line width=0.8pt, color=olive,dotted] (-0.2,-0.14) -- (0.8,-0.14);
	    \draw[color=olive] (0.4,-0.35) node {\(_{\delta'}\)};
	    
		\end{tikzpicture}
		\qquad\qquad\raisebox{16pt}{$\boldsymbol{\Longrightarrow}$}\qquad\quad\qquad 
		\begin{tikzpicture}
		\draw[-,line width=1.3pt,color=orange] (1.1,0.03) --(3.4,0.03);	 	
	 	
	 	\draw[-,line width=1.2pt,dotted] (-0.3,0) -- (0.2,0);
		\draw[-,line width=1.2pt] (0.2,0) -- (2,0);
		\draw[->,line width=1.2pt,dotted] (2,0) -- (2.5,0);
		\filldraw[black] (-0.3,0) circle (1.5pt);	
	   
	 	\draw[color=orange] (2.5,0.4) node {\(_{\wm_{g_{n}}\: \cp\: \gamma|_{K_n}}\)};	   
	  
		\draw[-,line width=1.2pt] (0.2,-0.12) -- (0.2,0.12);
	 	\draw (0.2,-0.4) node {\(_{0}\)};
		
		\draw[-,line width=1.2pt] (1.2,-0.12) -- (1.2,0.12);
	 	\draw (1.2,-0.4) node {\(_{k}\)};
	 	
	 	\draw[-,line width=1.2pt] (1.8,-0.12) -- (1.8,0.12);
	 	\draw (1.8,-0.478) node {\(_{k+\frac{1}{n_0}}\)};
		
		\draw[-,line width=0.8pt,color=orange] (1.1,0.03) .. controls (1,0.03) and (1,0.2) .. (1,0.35);
		\draw[-,line width=1pt,color=blue] (1.08,0.35) -- (0.92,0.35);
	 	\draw (0.5,0.35)[color=blue] node {\(_{k+\frac{1}{n}}\)};

		\draw[-,line width=0.9pt,color=blue] (2.7,-0.12) -- (2.7,0.12);
	 	\draw (2.7,-0.4)[color=blue] node {\(_{k}\)};	

		\draw[-,line width=0.9pt,color=blue] (3.4,-0.12) -- (3.4,0.12);
	 	\draw (3.4,-0.4)[color=blue] node {\(_{0}\)};		
		\end{tikzpicture}
	\vspace{-20pt}		
		
	\begin{align*}   
	  	\gamma([k,k+1/n_0])\subseteq \im[\gamma_n]\qquad \forall n\geq n_0. 
%    (\wm_{g_n}\cp \gamma)(J)=\gamma(J')
%    %\qquad\text{for some}\qquad \epsilon> 0,
   		\end{align*}
   		
		Hence, $\gamma_{m}\tsd \gamma_n$ for all $m,n\geq n_0$, which is impossible by i.). This shows that the first case cannot occur. In the same way, we deduce that in the second case we have $\gamma(K)\subseteq \im[\gamma_n]$ for all $n\geq n_0$, hence		
		$\gamma_{m}\tsd \gamma_n$ for all $m,n\geq n_0$, contradicting i.) as well. 
  \end{enumerate}
  \endgroup
    \item
    It is straightforward to see that $\delta\in \Paw$ is not free iff  
      $\wm_g\cp \delta$ is not free for all $g\in G$.  
      Consequently, we can restrict to curves starting at a fixed point $x\in M$ in the following. 
      Moreover, since $\wm$ is transitive and $G_x$ is a closed normal subgroup, we can identify $M$ with the (analytic) Lie group $G\slash G_x$ via $\phi\colon G\slash G_x\rightarrow M$, $[g]\mapsto \wm_x(g)$. Let $L_{[g]}\colon G\slash G_x\rightarrow G\slash G_x$, $[h]\mapsto [g\cdot h]$ denote the left translation w.r.t.\ the group structure on $G\slash G_x$ and observe that 
      $L_{[g]}\cp \phi^{-1}=\phi^{-1}\cp L_g$ just because $\phi\cp L_{[g]}=L_g\cp \phi$. We define $\gamma':=\phi^{-1}\cp \gamma$.
      \begingroup
      \setlength{\leftmarginii}{20pt}
      \begin{enumerate}
      \item
      \label{it:aa}
		Then, $\gamma'$ is analytic embedded. Moreover, $\gamma'$ is not free if this is the case for $\gamma$. 
      	For this, assume that  $\gamma'|_K$ is a free segment for some $K\subseteq \dom[\gamma]$. Then $\wm_g \cp \gamma|_K \cpsim \gamma|_K$ implies 
      	\begin{align*}  
			L_{[g]}\cp \gamma'|_K=\phi^{-1}\cp L_g \cp \gamma|_K   \cpsim  \phi^{-1}\cp \gamma|_K=\gamma'|_K,
       	 % L_{[g]}\cp \gamma'|_K=L_{[g]}\cp \phi^{-1}\cp \gamma|_K=  
      	\end{align*}
     	 hence $L_{[g]}\cp \gamma'|_K = \gamma'|_K$. Then $\wm_g\cp \gamma|_K = \gamma|_K$, showing that $\gamma$ is free as well. 
      \item
      \label{it:bb}
        Assume there is $g\in G$ and an element $\g'$ of the Lie algebra of $G\slash G_x$ such that 
	\begin{align*}
		\gamma'\cpsim \gamma'_0 \colon L &\rightarrow G\slash G_x\\
			t&\mapsto L_{[g]}\cp \exp(t\g')
	\end{align*}	       
   holds for $L\subseteq \RR$ compact. Then, since the canonical projection $\pi \colon G \rightarrow G\slash G_x$ is a submersion, we find $\g\in \mg$ with $\dd_e \pi(\g)=\g'$. Hence, for all $t\in \dom[\gamma]$ we have  
      \begin{align*}    
    	\gamma=\phi \cp \gamma'\cpsim &\big[L\ni t\mapsto L_g\cp \phi(\exp(t\g'))\big]\\
    	=&\big[L\ni t \mapsto L_g\cp \phi(\pi(\exp(t\g)))\big]\\
    	=&\big[L\ni t\mapsto L_g\cp \wm_x(\exp(t\g))\big]
      \end{align*}
      because $\pi$ is a Lie group homomorphism. In combination with Lemma \ref{lemma:sim}.\ref{lemma:sim5} this implies $\gamma\in \Pags$.
	  \end{enumerate}
	  \endgroup
      So, in order to show the claim, we only have to consider the situation where $M=G$ and $g\colon [-a,a]\rightarrow G$ is an embedded analytic curve with $g(0)=e$ and $a>0$. Here, it suffices to show that there is an open interval $I\subseteq [-a,a]$ and $h\colon I\rightarrow \RR$ smooth  
      with 
      \begin{align}
        \label{eq:sdsd}
        \dot g(t)= h(t)\cdot\dd L_{g(t)} \dot g(0)\qquad \forall\: t\in I.
      \end{align}
      In fact, then for $t_0\in [i_1,i_2]\subseteq I$ fixed and $\g:=\dot g(0)$, the unique\footnote{For $t_0\in I=(i_1,i_2)$ define $v\colon [0,\min(1,i_2-t_0)]\rightarrow \mg$, $t\mapsto h(t+t_0)\cdot \g$ and apply Satz 1.10 in \cite{HelgaBaum} in order to obtain $\ovl{g}\colon [0,i_2-t_0] \rightarrow G$ uniquely determined by $\ovl{g}(0)=e$ and $\dot{\ovl{g}}(t)=\dd L_{\ovl{g}(t)}h(t+t_0)\cdot \g$. Then $\ovl{g}(t)=g(t_0)^{-1}g(t+t_0)$ because \eqref{eq:sdsd} shows that the right hand side fulfils these conditions. Consequently, $g$ is uniquely determined by \eqref{eq:sdsd}.} solution of \eqref{eq:sdsd} with  $g(t_0)=e$ is given by %curve % with $t< l_1$
      \begin{align*} 
        g(t)=g(t_0)\cdot \exp\left(\left[\int_{t_0}^t h(s)\:\dd s \right]\!\g\right)\qquad \forall\: t\in I,
      \end{align*}
      so that 
      $g \cpsim L_{g(t_0)}\cp \exp(\cdot\: \g)$. 
      
      Now, to show \eqref{eq:sdsd} let $K_n:=[-\textstyle\frac{1}{n},\textstyle\frac{1}{n}]$ and choose $g_n\in G\backslash\{e\}$ with $g|_{K_n}\cpsim L_{g_n}\cp g|_{K_n}$ for each $n\in \NNge$. This is possible because $g$ is not free.  
      Then, $\lim_n g_n =e$ as for each $n\in \NNge$ there is $x_n\in g(K_n)$ with $g_n\cdot x_n \in  g(K_n)$, hence 
	\begin{align*}
		\lim_n \:[g_n\cdot x_n], \lim_n  x_n = g(0)=e\qquad\Longrightarrow\qquad \lim_n g_n =\lim_n x_n^{-1}=e.
	\end{align*}	      
      Since $g$ is an embedding, we find a closed neighbourhood $U$ of $e$ in $G$ and $K\subseteq (-a,a)$ compact with $U\cap \im[g]=g(K)$.
By continuity\footnote{Let $V\subseteq G$ be a neighbourhood of $e$ with $V^2\subseteq U$ and choose $n_0>0$ such that $g_n\in V$ and $g(K_n)\subseteq V$ for all $n\geq n_0$.} of $g$ and  that of the group multiplication in $G$, we find $n_0>0$ such that $K_n\subseteq K$ and $g_n\cdot g(K_n)\subseteq U$ for all $n\geq n_0$.
\vspace{9pt}

	\hspace{50pt}
	  \begin{tikzpicture}   
		\draw[->,line width=2.5pt] (0,0) .. controls (1,0.8) and (2,-0.3) .. (3,1);  
		\filldraw[black] (0,0) circle (2.5pt);  		 		
  		\filldraw[white] (1.5,0.3) circle (30pt); 
  
	  	\draw[-,line width=0.8pt,color=blue] (1.7,0.32) .. controls (1.5,0.31) and (1.5,0.31) .. (1.5,0.7);
%		\draw[-,line width=0.8pt] (1.5,0.6) .. controls (1.5,0.7) and (1.5,0.8) .. (1.3,0.8);
  		\draw[-,line width=0.8pt,color=blue] (1.9,0.33) .. controls (2,0.33) and (2,0.33) .. (2,0.6);
  		
		\draw[-,line width=0.8pt,color=blue] (1.42,0.7) -- (1.58,0.7);
		\draw[-,line width=0.8pt,color=blue] (1.92,0.6) -- (2.08,0.6);  		
  		 		 
  		\draw[-,line width=1pt] (0,0) .. controls (1,0.8) and (2,-0.3) .. (3,1);     	
  		\draw[black,line width=0.8pt, dashed] (1.5,0.3) circle (30pt);  	
  		\draw (0.4,1.1) node {\(U\)};
  		\draw (3.2,1.2) node {\(g\)};	
  		\draw (0.8,0.1) node {\(_K\)};
		\draw[-,line width=1pt] (1.4,0.24) -- (1.4,0.36);

		\draw (1.4,-0.1) node {\(_{g|_{K_n}}\)}; 
		\draw[-,line width=0.6pt] (1.1,0.18) -- (1.7,0.18);
		  		
		\draw (1.1,0.3) node {\(_{[}\)};  			
		\draw (1.7,0.3) node {\(_]\)}; 
		  			
  		\draw (1.3,0.45) node {\(_e\)};
  		 			
  		\draw[color=blue] (1.55,0.95) node {\(_{\wm_{g_n}\cp\hspace{1pt}g|_{K_n}}\)}; 			
		\end{tikzpicture} 
		\qquad\qquad\raisebox{25pt}{\LARGE{$\boldsymbol{\Longrightarrow}$}}\qquad\qquad\quad
		 	  \begin{tikzpicture}   
		\draw[->,line width=2.5pt] (0,0) .. controls (1,0.8) and (2,-0.3) .. (3,1);  
		\filldraw[black] (0,0) circle (2.5pt);  		 		
  		\filldraw[white] (1.5,0.3) circle (30pt);

		\draw[-,line width=0.8pt,color=blue] (0.8,0.2) -- (0.8,0.6);
		\draw[-,line width=0.8pt,color=blue] (2.3,0.3) -- (2.3,0.6);  		
  		\draw[-,line width=0.8pt,color=blue] (0.8,0.6) -- (2.3,0.6);  
  		 		 
  		\draw[-,line width=1pt] (0,0) .. controls (1,0.8) and (2,-0.3) .. (3,1);     	
  		\draw[black,line width=0.8pt, dashed] (1.5,0.3) circle (30pt);  	
  		\draw (0.4,1.1) node {\(U\)};
  		\draw (3.2,1.2) node {\(g\)};	
  		\draw (0.8,0) node {\(_K\)};
		\draw[-,line width=1pt] (1.4,0.24) -- (1.4,0.36);

		\draw (1.4,-0.1) node {\(_{g|_{K_n}}\)}; 
		\draw[-,line width=0.6pt] (1.1,0.18) -- (1.7,0.18);
		  		
		\draw (1.1,0.3) node {\(_{[}\)};  			
		\draw (1.7,0.3) node {\(_]\)}; 
		  			
  		\draw (1.3,0.45) node {\(_e\)};
  		 			
  		\draw[color=blue] (1.55,0.85) node {\(_{\wm_{g_n}\cp\hspace{1pt}g|_{K_n}}\)}; 	
		\end{tikzpicture} 
		
		\vspace{7pt}
      Now, $g|_{K_n}\cpsim L_{g_n}\cp g|_{K_n}$ implies $g\cpsim L_{g_n}\cp g|_{K_n}$  so that Lemma \ref{lemma:BasicAnalyt}.\ref{lemma:BasicAnalyt4} provides us with compact intervals $K'\subseteq K_n$ and $K''\subseteq [-a,a]$ which are maximal w.r.t.\ the property that $g(K'')=(L_{g_n}\cp g)(K')$ holds. Since $g_n\cdot g(K_n)\subseteq U$ and $U\cap \im[g]=g(K)$, we have $K''\subseteq K$. Then, $K'=K_n$ because $K''$ neither contains $a$ nor $-a$. This shows that the curve $g_n\cdot g|_{K_n}$ is completely contained in $g|_K$.

    Consequently, we find open intervals $J_n\subseteq K$ and $I_n\subseteq K_n$ with $0\in I_n$, as well as diffeomorphisms ${\adif}_n\colon I_n\rightarrow J_n$ such that  
	\begin{align}
	\label{eq:hhhlgdfgrgf}    
    L_{g_n}\cp g|_{I_n}= g|_{J_n}\cp \adif_n\qquad\:\forall\:n\geq n_0. 
	\end{align}    
    Then $g_n=g_n\cdot g(0)=g(\adif_n(0))$, and since everything is smooth, for $t_n:=\adif_n(0)\in K$ and $\g:=\dot g(0)$ we obtain 
      \begin{align*} 
        \dd L_{g(t_n)} \g = \dd L_{g_n}\: \dot g(0)\stackrel{\eqref{eq:hhhlgdfgrgf}}{=} \dot\adif_n(0)\:  \dot g(\adif_n(0))=\dot\adif_n(0)\:  \dot g(t_n),
      \end{align*}
      hence, $\dd L_{g(t_n)^{-1}}\: \dot g(t_n) \in \Span_\RR(\g)$ for all $n\in \NNge$. 
      Let $\wt{g}\colon  K\rightarrow \mg$, $t\mapsto \dd L_{g(t)^{-1}}\: \dot g(t)$ and denote by $s$ the supremum of $\|\wt{g}(t)\|$ for $t\in K$ and $\|\cdot\|$ a fixed norm in $\mg$. Then, $s$ is finite because $\wt{g}$ is continuous. 
      
      Now, $\wt{g}$ is even analytic and intersects the image of the analytic embedded curve
	\begin{align*}
	\delta\colon [-s\slash \|\g\|,s\slash \|\g\|]\rightarrow \mg,\quad t\mapsto t\cdot \g
	\end{align*}	      
        in infinitely many points.
      Then, $\im[\delta]\cap \im[\wt{g}]$ contains an accumulation point, and it follows as in the 
       proof of Lemma \ref{lemma:BasicAnalyt}.\ref{lemma:BasicAnalyt1} that we find an  
       open interval $I\subseteq K$ for which $\wt{g}(I)\subseteq \Span_\RR(\g)$ holds. This shows that
	\begin{align*} 
	      \dd L_{g(t)^{-1}} \dot g(t)=\wt{g}(t) = h(t)\cdot \g\qquad\forall\: t\in I
    \end{align*}
     holds for some smooth map $h\colon I\rightarrow \RR$, hence \eqref{eq:sdsd}. 
    \item
   We fix $t\in (a,b)$ and choose compact neighbourhoods $K \subseteq (a,b)$ of $t$ and $U\subseteq M$ of $x:=\gamma(t)$ with $\gamma(K)=\im[\gamma]\cap U$. 
     We first show that we find $\{g'_n\}_{n\in \NN}\subseteq G\backslash G_x$ with $\wm(g'_n,x)\in \gamma(K)$ and $\lim_n g_n=e$.
    \begingroup
    \setlength{\leftmarginii}{14pt}
	\begin{itemize}     
      \item 
      \vspace{-4pt}
       Let $n_0\in \NNge$ be such that $K_n:=\big[t- \frac{1}{n},t+\frac{1}{n}\big]\subseteq K$ for all $n>n_0$ and use Lemma \ref{lemma:curvestablemma}.\ref{lemma:curvestablemma2} in order to fix some $g_n\in G\backslash G_x$ with $\wm_{g_n}\cp \gamma|_{K_n} \cpsim \gamma|_{K_n}$ for each $n\geq n_0$.       
       Then, there exist $t_n, s_n\in K_n$ with $\gamma(t_n)=\wm(g_n, \gamma(s_n))$ for all $n\geq n_0$, hence 
      \begin{align}
       \label{eq:sdfdfsfsd}
        \lim_n \wm(g_n,\gamma(s_n))=\lim_n\gamma(t_n)=\gamma(t)=x 
      \end{align}
         Then, by properness of $\wm$ we find a subnet of 
         $\{g_n\}_{n\in \NN_{\geq n_0}}$ which converges to an element $h\in G$, and since manifolds are first countable, we even can assume $\lim_n g_n =h$ from the beginning. So, by continuity of $\wm$ and \eqref{eq:sdfdfsfsd} we have   
      \begin{align*} 
      	\wm(h,x)=\wm(h,\gamma(t))=\lim_n \wm(g_n,\gamma(s_n))=x\qquad \Longrightarrow \qquad  h\in G_x,
      \end{align*}      	
      where in the second step we have used that $\lim_n \gamma(s_n)=\gamma(t)$. 
	\item
	If we can prove that $\wm(g_n,x)\in \gamma(K)$ holds for infinitely many $n\geq n_0$, we just have to replace $g_n$ by $g_n\cdot h^{-1}$ in order to get the desired sequence $\{g'_n\}_{n\in \NN}\subseteq G\backslash G_x$. 
	To this end, let $V\subseteq G$ and $W\subseteq M$ be neighbourhoods of $h$ and $x$, respectively, with $\wm(V,W)\subseteq U$. 
	\vspace{3pt}	
	
	We choose $n_0' \geq n_0$ such that $g_n\in V$ and $\gamma(K_n)\subseteq W$ for all $n\geq n_0'$. Then $g_n \cdot \gamma(K_n)\subseteq U$  for all $n\geq n_0'$, and since $\wm_{g_n}\cp \gamma|_{K_n}\cpsim \gamma|_K$, the same arguments as in Part \ref{prop:freeseg2} show that $(\wm_{g_n}\cp \gamma)(K_n)\subseteq \gamma(K)$, hence $\wm(g_n,x)\in \gamma(K)$ for all $n\geq n_0'$. \hspace*{\fill}$\dagger$
	\end{itemize}        
      \endgroup
	  Let $\mg=\mg_0\oplus \mg_x$ and $U\subseteq \mg_0$, $V\subseteq \mg_x$ be neighbourhoods of zero, such that 
	\begin{align*}
		 h\colon U\times V\rightarrow W,\quad (u,v)\mapsto \exp(u)\cdot \exp(v)
	\end{align*}	  
	  is a diffeomorphism to an open subset $W\subseteq G$. Since the differential $\dd_e f=\dd_e\wm_x|_{TU}$ of $f:=\wm_x\cp\exp|_U$ is injective, shrinking $U$ we can assume that $f$ is an analytic embedding.
	            
      We define $O:=f(U)$ and choose an analytic submanifold chart $(\phi_0,U_0)$ of $O$ around $x$. Then, we find $I\subseteq \dom[\gamma]$ open with $t\in I$ and $\gamma(I)\subseteq U_0$. Since $\lim_n g_n'= e$, we find $n'\in \NN$ such that for each $n\geq n'$ we have $g'_n\in W$, hence $g_n'=h(u_n',v_n')$ for some $(u_n',v_n')\in U\times V$. Consequently,   
	\begin{align*}       
       \wm(g_n',x)=(\wm_x\cp h)(u_n',v_n')=f(u_n')\in O\qquad \forall\:n\geq n',
	\end{align*}      
      and since $\wm(g_n',x)\in \gamma(K)$, 
      by the embedding property of $\gamma$ we find $n''\geq n'$ such that $\wm(g_n',x)\in \gamma(I)$ holds for all $n\geq n''$.
       Consequently, $\phi_0(x)$ is an accumulation point of $(\phi_0\cp \gamma)(I) \cap \phi_0(O\cap U_0)$ in $\phi_0(O\cap U_0)$. Then, by analyticity of the components of $\phi_0\cp \gamma$ we find $J\subseteq I$ open with $t\in J$ and $\gamma(J)\subseteq O\subseteq Gx$. 
       
       Now, if we choose an analytic embedding $\gamma'\colon [a-\epsilon,b+\epsilon]$ with $\gamma'|_{[a,b]}=\gamma$, then Part \ref{prop:freeseg1}) shows that $\gamma'$ is not free as well. Then, the above arguments show that for each $t\in [a,b]$ we find an open interval $J\subseteq [a-\epsilon,b+\epsilon]$ with $t\in J$ and $\gamma'(J)\subseteq G \gamma(t)$. Since finitely many of such intervals $J_1,\dots,J_k$ cover $[a,b]$, and since orbits are disjoint, $\im[\gamma]$ must be contained in $G x$. This shows the first part.

      For the second part,  
      let $L\subseteq J$ be compact, such that $\gamma(L)\subseteq O$. Then, 
\begin{align*}      
      \gamma':=\pi \cp\exp\cp  f^{-1}\cp \gamma|_L
\end{align*}      
       is an embedded analytic curve in $G\slash G_x$ which is not free. In fact, for analyticity observe that  $(V',\phi')$ with $V':=\pi(\exp(U))$ and $\phi':=(\pi\cp \exp|_U)^{-1}$ is an analytic chart of $G\slash G_x$ around $[e]$, and that $f\colon U\rightarrow O$ is an analytic diffeomorphism. Moreover, that $\gamma'$ is not free follows by the same arguments as in the proof of Part \ref{prop:freeseg2}), cf.\ (a). Consequently, $\gamma'$ is Lie algebra generated by Part \ref{prop:freeseg2}), and since $\dd_e \pi$ is a submersion and a Lie group homomorphism, we conclude that $\gamma|_L$, hence $\gamma$, is Lie algebra generated as well, cf.\ (b) in Part \ref{prop:freeseg2}). 
    \end{enumerate} 
  \end{proof}
\end{proposition}
Using the first part of the above proposition, we now obtain that each of the sets $\Pafs$, $\Pafns$, $\Pags$ and $\Pacs$ are closed under  decomposition and inversion of its elements, hence that the factorization \eqref{eq:dsfs444ffff} holds.
\begin{corollary}
  \label{rem:freinichtLiealg}
  The sets $\Pafs$, $\Pafns$, $\Pags$ and $\Pacs$ are closed under  decomposition and inversion of their elements. 
  \begin{proof}  
 	By Lemma \ref{lemma:curvestablemma}.\ref{lemma:curvestablemma1} $\Paf$ is closed under inversions, and Proposition \ref{prop:freeseg}.\ref{prop:freeseg1} shows that it is also closed under decompositions. 
    Then, Part $\textit{(b)}$ Lemma \ref{lemma:curvestablemma}.\ref{lemma:curvestablemma1} shows 
    that the sets $\Pafns$ and $\Pafs$ are closed under decomposition and inversion as well.  
    Finally, by Corollary \ref{cor:decompo} the set $\Paw\backslash \Pags=\Pacs \sqcup \Paf$ is invariant under decomposition and inversion, so that the inverse or a subcurve of an element of $\Pacs$ must be contained in $\Pacs\sqcup\Paf$. However, the  inverse of a free segment is a free segment, and freeness of a subcurve of $\gamma\in \Pacs$ already would   imply freeness of $\gamma$.    
  \end{proof}
\end{corollary}
We close this section with the following (discrete) analogue to Proposition \ref{th:invhomm}.\ref{th:invhomm1}. There, we modify  invariant homomorphisms along free segment $\delta$ by means of $G_\delta$-invariant maps:
\begin{definition}
Let $\delta\colon[k_1,k_2]\rightarrow M$ be a free segment and $p\in F_{\delta(k_1)}$. Then, by $\MPD$ we will denote the set of all maps $\Psi\colon \RR\rightarrow S$ 
 which are invariant under $G_\delta$ in the sense that $\alpha_{\fiba_p(h)}\cp \Psi=\Psi$ holds for all $h\in G_\delta$, and that fulfil
 \begin{align*}
	\Psi(\lambda-\lambda')=\Psi(\lambda)\cdot \Psi(\lambda')^{-1}  \qquad\forall\: \lambda,\lambda'\in\RR. %\text{as well as}\qquad \Psi(l+l')=\Psi(l)\cdot \Psi(l')%\qquad \text{ if }l,l',l+l'\neq 0. 
\end{align*}
\end{definition}  
\begin{remark}
		If $G_\delta=\{e\}$, examples for such invariant maps $\Psi\colon \RR \rightarrow S$ are just given by $\Psi\colon  \lambda\mapsto \exp(\lambda\cdot\s)$ for $\s\in \ms$. So, if $S$ is compact and connected ($\exp$ is surjective), then for each $s\in S$ and for each $\lambda\neq 0$ we find $\Psi \in \MPD$ with $\Psi(\lambda)=s$.
%	\item
	However, if $G_\delta$ is non-trivial, one has to decide from case to case whether such non-trivial maps exist. 
	 \hspace*{\fill} $\lozenge$
\end{remark}
\begin{proposition}
  \label{lemma:freemodify}
  Let $\Pa\subseteq \Paw$ be closed under decompositions and inversions, $\homm'\in \IHOM$, $S$ compact and connected with $\dim[S]\geq 1$,\footnote{Recall that this ensures that the equivalence relations $\csim$ and $\psim$ coincide, see Lemma \ref{lemma:BasicAnalyt}.\ref{lemma:BasicAnalyt2}.} and $\delta\colon L=[l_1,l_2]\rightarrow M$ a free segment. Moreover, let $t\in L$, $p\in F_{\delta(t)}$ and $\Psi\in \MPD$. 
	\vspace{6pt}  
  
  \noindent
  	We define $\homm \in \IHOM$ as follows. 
  For $[s_1,s_2]\subseteq L$ and $s\in L$ let 
  \begin{align*}  
  	\delta|_{[s_2,s_1]}:=[\delta|_{[s_1,s_2]}]^{-1}\qquad\text{as well as}\qquad\homm'(\delta|_{[s,s]})(q):=q
  \end{align*}
  and define  
  \begin{align*}
    \wt{\Psi}\!\left(\delta|_{[a,b]}\right)(p'):=
    \homm'\left(\delta|_{[t,b]}\right)(p)\cdot  \Psi(b-a) \cdot \diff\big(\homm'\!\left(\delta|_{[t,a]}\right)(p),p'\big)\qquad \forall\:p'\in F_{\delta(a)}.  
  \end{align*} 
 Then, for $\gamma\in \Paw$ we choose a family of representatives $\{h_\alpha\}_{\alpha\in I}$ of $H_{[\gamma,\delta]}$ and let $k_0<\dots < k_n$ denote the respective unique decomposition from Lemma \ref{lemma:curvestablemma}.\ref{lemma:curvestablemma3} of $\dom[\gamma]=[k_0,k_n]$ into compact intervals $K_i=[k_i,k_{i+1}]$ for $0\leq i \leq n-1$. Recall that then either 
	\begin{align*}
		\qquad\gamma|_{K_i}\nsim_\cp \wm_{h_\alpha}\cp \delta\qquad \forall\: \alpha\in I\qquad\qquad\text{(we define $\alpha(i):=0\notin I$ and $h_{\alpha(i)}:=e$ in this case)}
	\end{align*}	  
 or\footnote{Recall that $L_i=L$ for $1\leq i\leq n-2$ and $L_0,L_{n-1}\subseteq L$ are of the form $[l_1,l]$ or $[l,l_2]$ for $l_1<l\leq l_2$ or $l_1\leq l<l_2$, respectively.} $\gamma|_{K_i} \psim \wm_{h_{\alpha(i)}}\cp [\delta|_{L_i}]^{p_i}$ holds for $\alpha(i) \in I$, $p_i\in\{-1,1\}$  and $L_i\subseteq L$ uniquely determined. Let
  \begin{align}
  \label{eq:dkdfjs<afjfdsf}
    \wt{\homm}(\gamma|_{K_i}):= 
    \begin{cases} 
      \homm'(\gamma|_{K_i}) &\mbox{if } \alpha(i)=0,\\
      \Phi_{h_{\alpha(i)}}\cp \wt{\Psi}([\delta|_{L_i}]^{p_i})\cp \Phi_{h^{-1}_{\alpha(i)}} &\mbox{if } \alpha(i)\in I
    \end{cases}
  \end{align}
  \noindent
  as well as $\homm(\gamma):= \wt{\homm}(\gamma|_{K_0})\cp \dots \cp \wt{\homm}(\gamma|_{K_{n-1}})$.
  Then, $\homm$ is a well-defined element of $\IHOM$.
  \begin{proof}
    First assume that $\homm$ is well defined. Then $\homm(\gamma)(p'\cdot s)=\homm(\gamma)(p')\cdot s$ for $s\in S$ is immediate from the definitions. For invariance of $\homm$  let $g\in G$. Then, for $\wm_g\cp \gamma$ we can use the same index set and the same decomposition  as for $\gamma$, provided that we define $h'_{\alpha}:=g\cdot h_{\alpha}$ for all $\alpha\in I$. We obtain 
    \begin{align*}
      \homm(\wm_g\cp \gamma)\cp \Phi_g &= \wt{\homm}(\wm_g\cp \gamma|_{K_0})\cp \dots \cp \wt{\homm}(\wm_g\cp \gamma|_{K_{n-1}})\cp \Phi_g\\
      & \hspace{-3pt}\stackrel{\eqref{eq:dkdfjs<afjfdsf}}{=} \Phi_{g}\cp\wt{\homm}(\gamma|_{K_0})\cp \Phi_{g^{-1}}\cp \dots \cp \Phi_{g}\cp\wt{\homm}(\gamma|_{K_{n-1}})\cp \Phi_{g^{-1}}\cp \Phi_g
      = \Phi_g\cp \homm(\gamma). 
    \end{align*}
    Now, multiplicativity of $\homm$ is clear if $\gamma$ is splitted at some point contained in an interval $K_i$ with $\alpha(i)=0$. For the other case, it suffices to show that
    $\wt{\Psi}\left(\delta|_{[s,b]}\right)\cp \wt{\Psi}\left(\delta|_{[a,s]}\right)=\wt{\Psi}\left(\delta|_{[a,b]}\right)$ holds for all $s,a,b\in L$ with $a<s<b$ or $b<s<a$.  Now,
    \begin{align*}
      \wt{\Psi}\!\left(\delta|_{[a,s]}\right)(p')=\homm'(\delta|_{[t,s]})(p)\cdot \Psi(s-a)\cdot \diff(\homm'\left(\delta|_{[t,a]}\right)(p),p')
    \end{align*}
    and $\wt{\Psi}\!\left(\delta|_{[s,b]}\right)(q)=\homm'(\delta|_{[t,b]})(p)\cdot \Psi(b-s)\cdot \diff\!\left(\homm'\!\left(\delta|_{[t,s]}\right)(p),q\right)$, hence
    \begin{align*}
      \wt{\Psi}\!\left(\delta|_{[s,b]}\right)\big(\wt{\Psi}\!\left(\delta|_{[a,s]}\right)(p')\big)
      &=\homm'(\delta|_{[t,b]})(p)\cdot \Psi(b-s)\cdot \diff\!\left(\homm'\!\left(\delta|_{[t,s]}\right)(p),\homm'(\delta|_{[t,s]})(p)\right)\\
      &\hspace{1.8pt}\qquad\qquad\qquad\cdot \Psi(s-a)\cdot \diff\!\left(\homm'\!\left(\delta|_{[t,a]}\right)(p),p'\right)\\
      &=\homm'(\delta|_{[t,b]})(p)\cdot \Psi(b-a)\cdot \diff\!\left(\homm'\!\left(\delta|_{[t,a]}\right)(p),p'\right)\\
      &=\wt{\Psi}\left(\delta|_{[a,b]}\right)(p').
    \end{align*}
    For the inversion property observe that
    \begin{align*}
      \homm(\gamma^{-1})= \wt{\homm}\big([\gamma|_{K_{n-1}}]^{-1}\big)\cp  \dots \cp  \wt{\homm}\big([\gamma|_{K_0}]^{-1}\big).
    \end{align*}
    Then, $\homm(\gamma^{-1})\cp \homm(\gamma)=\id_{F_{\gamma(k_0)}}$ and $\homm(\gamma)\cp \homm(\gamma^{-1})=\id_{F_{\gamma(k_n)}}$ are clear if we can show that for all $0\leq i\leq n-1$ we have 
    \begin{align*}
      \wt{\homm}\big([\gamma|_{K_i}]^{-1}\big)\cp \wt{\homm}([\gamma|_{K_i}])=\id_{F_{\gamma(k_i)}}\qquad\text{and}\:\:\qquad\wt{\homm}([\gamma|_{K_i}])\cp \wt{\homm}\big([\gamma|_{K_i}]^{-1}\big)=\id_{F_{\gamma(k_{i+1})}}, 
    \end{align*}
    respectively. Again, this is clear if $\alpha(i)=0$, and in the other case we have $[\delta|_{[a,b]}]^{-1}=\delta|_{[b,a]}$, hence 
    $\wt{\Psi}\left([\delta|_{[a,b]}]^{-1}\right)(q)=\homm'(\delta|_{[t,a]})(p)\cdot \Psi(a-b)\cdot \diff\left(\homm'\!\left(\delta|_{[t,b]}\right)(p),q\right)$. Then
    \begin{align*}
      \wt{\Psi}\left([\delta|_{[a,b]}]^{-1}\right)(\wt{\Psi}\left(\delta|_{[a,b]}\right)(p'))=\homm'(\delta|_{[t,a]})(p)\cdot \diff\!\left(\homm'\!\left(\delta|_{[t,a]}\right)(p),p'\right)=p'
    \end{align*}
    as well as
    \begin{align*}
    \wt{\Psi}\left(\delta|_{[a,b]}\right)(\wt{\Psi}\left([\delta|_{[a,b]}]^{-1}\right)(p'))=\homm'(\delta|_{[t,b]})(p)\cdot \diff\!\left(\homm'\!\left(\delta|_{[t,b]}\right)(p),p'\right)=p',
    \end{align*}
   	which shows that 
	\begin{align*}
	\wt{\Psi}\left([\delta|_{[a,b]}]^{-p_i}\right)(\wt{\Psi}\left([\delta|_{[a,b]}]^{p_i}\right)(p'))=p'\qquad\text{and}\qquad\wt{\Psi}\left([\delta|_{[a,b]}]^{p_i}\right)(\wt{\Psi}\left([\delta|_{[a,b]}]^{-p_i}\right)(p'))=p'
    \end{align*}
    holds. From this, the inversion property is clear.
    \vspace{6pt}

    \noindent
    It remains to show that $\homm$ is well defined. Obviously, $\homm(\gamma)=\homm(\gamma')$ if $\gamma\csim \gamma'$, so that we only have to discuss what happens if we choose another family $\{h'_\alpha\}_{\alpha\in I'}$ of representatives of $H_{[\gamma,\delta]}$. Now, we can assume that $I=I'$, and that the decompositions of $\gamma$ into intervals $K_i$ w.r.t.\ both families coincide. In fact, this is clear from $h_\alpha \segsim h'_\alpha$  and the uniqueness of these decompositions. 
   Obviously, it suffices to show that
    \begin{align}
      \label{eq:gdfdg}
      \Phi_{h_{\alpha(i)}}\cp \wt{\Psi}([\delta|_{L_i}]^{p_i})\cp \Phi_{h^{-1}_{\alpha(i)}}=\Phi_{h'_{\alpha(i)}}\cp \wt{\Psi}([\delta|_{L_i}]^{p'_i})\cp \Phi_{{h'_{\alpha(i)}}^{\hspace{-12pt}-1}}\qquad \forall\: \alpha(i)\neq 0.
    \end{align}
    Since $h:=h^{-1}_{\alpha(i)}h'_{\alpha(i)} \in G_\delta$ we have $p_i=p_i'$. Then, if we write 
    $[\delta|_{L_i}]^{p_i}=\delta|_{[a,b]}$ for $a,b\in \RR$, we obtain
    \begin{align*}
      \big(\Phi_{h}\cp \wt{\Psi}([\delta|_{L_i}]^{p_i}) \cp \Phi_{h^{-1}}\big)(p')&=(\Phi_{h}\cp \homm')\left(\delta|_{[t,b]}\right)(p)\cdot  \Psi(b-a) \cdot \diff\big(\homm'\!\left(\delta|_{[t,a]}\right)(p),\Phi_{h^{-1}}(p')\big) \\
      &=\homm'\!\left(\wm_h\cp \delta|_{[t,b]}\right)(\Phi_{h}(p))\cdot \Psi(b-a) \cdot  \diff\big((\Phi_h\cp\homm')\!\left(\delta|_{[t,a]}\right)(p),p'\big)\\
     &=\homm'\!\left( \delta|_{[t,b]}\right)(\Phi_{h}(p))\cdot \Psi(b-a) \cdot  \diff\big(\homm'\!\left(\wm_h\cp\delta|_{[t,a]}\right)(\Phi_h(p)),p'\big)\\
       &=\homm'\!\left(\delta|_{[t,b]}\right)(p)\cdot  \fiba_p(h)\cdot \Psi(b-a) \cdot  \diff\big(\homm'\!\left(\delta|_{[t,a]}\right)(p)\cdot  \fiba_p(h),p'\big)\\   
%%%%      
      &=\homm'\!\left(\delta|_{[t,b]}\right)(p)\cdot  \fiba_p(h) \cdot \Psi(b-a) \cdot \fiba_p(h)^{-1}\cdot\diff\big(\homm'\!\left(\delta|_{[t,a]}\right)(p),p'\big)\\
      &=\homm'\!\left(\delta|_{[t,b]}\right)(p)\cdot  \alpha_{\fiba_p(h)}(\Psi(b-a)) \cdot\diff\big( \homm'\!\left(\delta|_{[t,a]}\right)(p),p'\big)\\
      &=\homm'\!\left(\delta|_{[t,b]}\right)(p)\cdot  \Psi(b-a) \cdot\diff\big(\homm'\!\left(\delta|_{[t,a]}\right)(p),p'\big)\\
      &=\wt{\Psi}([\delta|_{L_i}]^{p_i})(p'),
    \end{align*}
    hence \eqref{eq:gdfdg}. Here, in the second step, we have used \eqref{eq:invprop} for the first factor, as well as 
	\begin{align*}
		q\cdot \diff\big(q,\Phi_{h^{-1}}(p')\big)=\Phi_{h^{-1}}(p')\qquad &\Longrightarrow\qquad \Phi_h(q)\cdot \diff\big(q,\Phi_{h^{-1}}(p')\big)=p'\\
		&\Longrightarrow\qquad   \diff\big(q,\Phi_{h^{-1}}(p'\big)\big)=\diff\big(\Phi_h(q),p'\big)
	\end{align*}
	with $q,p'\in F_{\delta(a)}$ for the last one. In the third step, we have used $\wm_h\cp\delta=\delta$ in the first factor and \eqref{eq:invprop} in the last one. The fourth step is clear, and in the fifth one, we have use that
		\begin{align*}
		q\cdot \fiba_p(h)\cdot  \diff\big(q\cdot \fiba_p(h),p'\big)=p'\qquad &\Longrightarrow\qquad  \fiba_p(h)\cdot  \diff\big(q\cdot \fiba_p(h),p'\big)=\diff(q,p')\\
		&\Longrightarrow\qquad  \diff\big(q\cdot \fiba_p(h),p'\big) = \fiba_p(h)^{-1}\cdot\diff(q,p').
	\end{align*}
    Finally, in the seventh step, we have used the invariance property of $\Psi\in \MPD$. 
  \end{proof}
\end{proposition} 

\subsection{Summary}
\label{concl:modify}
In this Section, we basically have discussed the situation where the action $\wm$ induced by the symmetry\footnote{This means a Lie group of automorphisms on the principal fibre bundle $\PMS$ under consideration.} on the base manifold is analytic and pointwise proper. In the first part, 
we have modified invariant homomorphisms along Lie algebra generated curves, and in the 
second one   
we have applied this to the LQG setting in order to show that quantization and reduction do not commute in several situations. In particular, we have shown that this is the case 
in \mbox{(semi-)homogeneous}  
LQC. In the last part, we have proven that  
free curves are discretely generated by the symmetry group, and that we have the decomposition 
\begin{align*}
	\Paw=\Pags\sqcup \Pacs\sqcup \Pafns\sqcup \Pafs
\end{align*}
of $\Paw$ 
into subsets, each being closed under inversions and decomposition. Hence, the factorization
	\begin{align*}
	%\label{eq:dsfs444ffff}
    \AQRw \cong  \AQRInd{\mg}\times \AQRInd{\mathrm{CNL}}   
 \times  \AQRFNS\times \AQRInd{\mathrm{FS}}
  \end{align*} 
  by Proposition \ref{rem:euklrem2b}. 
  Moreover, we have shown that even $\Pacs=\emptyset$, hence   
$\Paw = \Pags\sqcup \Paf$ holds if $\wm$ admits only normal stabilizers and is transitive or proper. Recall that $\Paf=\Pafs\sqcup \Pafns$  denotes the set of free curves, whereby $\Pafns$ consists of such free curves whose stabilizer is trivial and $\Pafns$ of those having a non-trivial one. Consequently, 
\begin{align*}
	\Paw=\Pags\sqcup \Pafns\qquad\Longrightarrow \qquad \AQRw \cong  \AQRInd{\mg}\times  \AQRInd{\mathrm{FS}} 
\end{align*}
holds if $\wm$ is in addition free. This is the case, e.g., in  \mbox{(semi-)homogeneous} LQC and also important in view of the next section. There, we will construct normalized Radon measures on each of these two factors, providing us with the respective Radon product measures on $\AQRw$ in this case.
 For the measure on $\AQRInd{\mathrm{FS}}$, there we will use the modification result for   
 free segments which we have proven in Proposition \ref{lemma:freemodify}.

\section{Measures on Quantum-Reduced Configuration Spaces}
\label{sec:MOQRCS}
In the previous section, we have seen that $\Paw=\Pags\sqcup \Pacs\sqcup \Pafns\sqcup \Pafs$ holds  if $\wm$ analytic and pointwise proper, whereby each of the occurring sets is closed under decomposition and inversion of its elements, hence\footnote{If one the sets of curves is empty, one just has to remove the respective factor in the product \eqref{eq:dsfs444}.} 
	\begin{align}
	\label{eq:dsfs444}
    \AQRw \cong  \AQRInd{\mg}\times \AQRInd{\mathrm{CNL}}   
 \times  \AQRFNS\times \AQRInd{\mathrm{FS}}
  \end{align}
  by Proposition \ref{rem:euklrem2b}. 	
   So, in order to define a normalized Radon measure on $\AQRw$, it suffices to construct respective normalized Radon measures on each of the factors occurring on the right hand side of \eqref{eq:dsfs444}. 
 In this section, we will use the developments of the previous one in order to provide general constructions for the spaces 
\begin{align*}
	\AQRInd{\mg^\sim}\cong \IHOMLAS
	\qquad\qquad\text{and}\qquad\qquad \AQRInd{\mathrm{FN}}\cong\IHOMFNS.
\end{align*} 
Recall that these spaces correspond to the sets of Lie algebra generated and free non-symmetric curves, respectively. So, in particular, our constructions will provide us with a normalized Radon measure on $\AQRw$ whenever $S=\SU$ and $\wm$ acts properly and free. Indeed, then we have $\Paw=\Pags \sqcup\Pafs$, hence $\AQRw \cong  \AQRInd{\mg}\times \AQRInd{\mathrm{FS}}$ and all requirements of Subsection \ref{sec:ConSp} are fulfilled.

More precisely, in Subsection \ref{sec:FreeM} we will construct a normalized Radon measure on $\AQRInd{\mathrm{FN}}$ for the general case that $S$ is compact and connected. We will 
define a projective structure on $\AQRInd{\mathrm{FN}}$ which generalizes that one, used for the construction of the Ashtekar-Lewandowski measure on $\A_\w$.  
Roughly speaking, the projection maps will be defined by \eqref{projm},  whereby in the respective indices now only free segments will occur. Proposition \ref{prop:freeseg}.\ref{prop:freeseg1} then guarantees that $\AQRInd{\mathrm{FN}}$ is separated by these maps, and their surjectivity will be established by Proposition \ref{lemma:freemodify}. Then, as in the Ashtekar-Lewandowki case,  the Haar measure on $S$ can be used to define a consistent family of normalized Radon measure, providing us with a normalized Radon measure on $\IHOMFNS$. In specific circumstances, the same construction can also be used to define a normalized Radon measure on $\AQRInd{\mathrm{FS}}$. This will be demonstrated for the case of spherically symmetric LQC where the set $\Pafs$ just consists of all linear curves traversing through the origin. There, the respective projection maps can be chosen in such a way that their images are just products of maximal tori, each carrying a Haar measure by itself. However, there is no ad hoc reason why there should be such a convenient choice in the general case. So, in specific situations one has to investigate the sets $\Pafs$ and $\AQRInd{\mathrm{FS}}$ explicitly if one wants to use the developed techniques in order to construct a measures on $\AQRInd{\mathrm{FS}}$.

Now, the mentioned restriction to the structure group comes from our constructions for the space $\AQRInd{\mg^\sim}$ (Subsection \ref{sec:ConSp}). These we will exemplarily do for the most LQG-relevant case where $S=\SU$. Analogous constructions appear to be possible also for other compact and connected structure groups, in any case for the abelian ones (tori). 

Indeed, for our considerations we will assume that each $\wm$-orbit $\m=[x_\m]\subseteq M$ with $\mg\backslash \mg_{x_\m}\neq \emptyset$ (we will denote the set of all such orbits by $\Mm$) admits an independent and complete family $\{\g_\alpha\}_{\alpha\in I_\m}\subseteq \mg\backslash \mg_{x_\m}$ as in Definition \ref{def:stable}\ref{eq:iindepg3}.\footnote{Due to Lemma and Remark \ref{rem:dfggfg}.\ref{rem:dfggfg2} 
 this is always the case, e.g., if $\wm$ is free.} Then, as we will show in the first part of Subsection \ref{sec:ConSp}, $\AQRInd{\mg^\sim}$ is homeomorphic to the Tychonoff product 
 \begin{align*}
 	Y=\prod_{\m\in \Mm,\alpha\in I_\m} Y_{\m,\alpha}
 \end{align*}
 for $Y_{\m,\alpha}$ the set of $\Ad_{G_{[\g_\alpha]}}^{p_\m}$-equivariant mappings $\Psi\colon \spann_\RR(\g_\alpha)\rightarrow S$ (cf.\ Definition \ref{def:eqmaps}) equipped with a suitable topology. Here, 
 $p_\m\in F_{x_\m}$ is a fixed choice for each $\m\in \Mm$.   
  The above homeomorphism is just a straightforward consequence of Proposition \ref{th:invhomm}.\ref{th:invhomm1} where, except for compactness and connectedness, no further requirements on the structure group $S$ have to be done.   
 Then, it is the non-trivial part to calculate these factors explicitly, to define reasonable measures thereon, and, finally, to show that the measure on $\AQRInd{\mg^\sim}\cong Y$ induced by the respective  Radon product measure on $Y$ does not depend on any choices one has done. Then, following the arguments of Subsection \ref{sec:ConSp}, one immediately sees that for $S$ the n-torus, each of the above factors is either homeomorphic to the $n$-fold product of $\RB$ by %, via the map
\begin{align*}
	[\hspace{1pt}\RB\hspace{1pt}]^n&\rightarrow Y_{\m,\alpha}\\
	(\psi_1,\dots,\psi_n)&\mapsto   \big[\lambda\cdot \g_{\m,\alpha} \mapsto \big(\psi_1(\chi_\lambda),\dots,\psi_n(\chi_\lambda)\big)\big], 
\end{align*} 
or consists of the trivial map $\Psi\colon \lambda\cdot \g_\alpha \mapsto e$. This is just because by Remark \ref{rem:ppropercurve}.\ref{rem:ppropercurve4} and commutativity of $S$, equivariance of $\Psi\in Y_{\m,\alpha}$ is either a trivial condition or reads $\Psi(\g)=\psi(-\g)$, hence $\Psi(\g)=e$ for all $ \g\in \spann_\RR(\g_{\m,\alpha})$. 
Then, in both cases we have a canonical Radon measure on $Y_{\m,\alpha}$, providing us with a respective Radon product measure on $Y$. 
 
 However, for $\SU$ (and any other non-commutative structure group) equivariance gives more complicated conditions which have to be  investigated carefully. This is the content of the second part of this section.  
 In both subsections, we will discuss the \mbox{(semi-)homogeneous}, homogeneous isotropic as well as the spherically symmetric LQC case.

\subsection{The Reduced Ashtekar-Lewandowski Measure}
\label{sec:FreeM}
As already mentioned above, we now are going to equip $\AQRInd{\mathrm{FN}}\cong\IHOMFNS$ with a projective structure, which we then  use to construct a normalized Radon measure thereon. This will be done in analogy to the construction of the Ashtekar-Lewandowski measure \cite{ProjTechAL} on $\A_\w\cong \HOMW$ which we even get back 
 if $G=\{e\}$ holds. This is just because due to our definitions then $\Pafns=\Paw$, hence $\IHOMFNS=\HOMW$ holds.

In order to avoid trivialities, we will assume that $\IHOMFNS$ is not empty. Moreover, we let  
$\nu=\{\nu_x\}_{x\in M}\subseteq P$ be a fixed family with $x\in F_x$ for all $x\in M$ and $\psi_x(p):=\diff(\nu_x,p)$ for all $p\in F_x$ as in Convention \ref{hgamma}. Finally, we will assume that $S$ is compact and connected with $\dim[S]\geq 1$. 

We start our investigations with the definition of the directed set: 
\begingroup
\setlength{\leftmargini}{20pt}
\begin{itemize}
\item
  \itspace
  Let $\grF$ denote the set of all finite tuples $(\gamma_1,\dots,\gamma_k)$ with $\gamma_1,\dots,\gamma_k \in \Pafns$ free segments such that we have 
   \begin{align*}
  \wm_g\cp \gamma_i \nsim_\cp \gamma_j\qquad \forall\: g\in G,\: 1\leq i\neq j\leq k.
	\end{align*}  
\item
  \itspace
  For $(\gamma_1,\dots,\gamma_k), (\gamma'_1,\dots,\gamma'_p)\in \grF$ write $(\gamma_1,\dots,\gamma_k)\leq (\gamma'_1,\dots,\gamma'_p)$ 
  iff each $\gamma_i$ admits a decomposition $\{(\gamma_i)_j\}_{1\leq j\leq n_i}$ such that each restriction $(\gamma_j)_i$ is equivalent to one of the curves $\wm_g\cp \gamma'_j$ or its inverses for some $g\in G$ and $1\leq j\leq p$.
\end{itemize}
\endgroup 
\begin{lemma}[Directed Set]
  \label{lemma:projdefx}
    The pair $(\grF,\leq)$ is a directed set.
\end{lemma}
  \begin{proof}
      Let $(\gamma_1,\dots,\gamma_k),(\gamma'_1,\dots,\gamma'_p)\in \grF$. 
      We inductively construct an upper bound in $\grF$ as follows.
       For each $1\leq j\leq p$ let $a_j=k_0^j<\dots<k^j_{n_j}=b_j$ denote the decomposition from  Lemma \ref{lemma:curvestablemma}.\ref{lemma:curvestablemma3} of 
       $\dom[\gamma'_j]= [a_j,b_j]$ w.r.t.\ $\gamma_1\colon [l_1,l_2]\rightarrow M$ 
       into intervals $K_i^j=\big[k^j_i,k^j_{i+1}\big]$. 
       We first remove all segments $\gamma'_j|_{K_i^j}$ that are equivalent to a translate of $\gamma_1$ or its inverse. 
       \vspace{4pt}
       
       \noindent
        More precisely, let $\gamma'_{j,1},\dots, \gamma'_{j,m_j}$ 
       denote the segments $\gamma'_j|_{K_i^j}$ of $\gamma_j'$ with 
	\begin{align*}    
        \gamma'_j|_{K_i^j}\nsim_\cp \wm_g \cp \gamma_1\qquad\forall\: g\in G
	\end{align*}       
        and replace $(\gamma'_1,{\dots},\gamma'_p)$ by $(\gamma_{1,1}',\dots,\gamma'_{1,m_1},\dots,\gamma_{k,1}',\dots,\gamma'_{k,m_k})$. 
      Here, it may happen that $\gamma'_j|_{K_0}\csim \wm_g\cp [\gamma_1|_{L'}]^{\pm 1}$ or $\gamma'_j|_{K_{n-1}}\csim \wm_g\cp [\gamma_1|_{L'}]^{\pm 1}$ holds with $L'=[l_1,l]$ for $l<l_2$ or $L'=[l,l_2]$ for $l_1<l$. In this case, we split $\gamma_1$ at these (at most $2p$) points and obtain curves $\gamma_{1,1},\dots,\gamma_{1,n_1}$. 

      We now apply the same procedure to $\gamma_2$ and $(\gamma_{1,1}',\dots,\gamma'_{1,m_1},\dots,\gamma_{k,1}',\dots,\gamma'_{k,m_k})$, and inductively end up with finitely many free segments $\gamma_{1,1},\dots,\gamma_{1,q_1},\dots, \gamma_{1,p},\dots,\gamma_{1,q_p}$ and $\gamma''_1,\dots,\gamma''_m$ that fulfil
	\begin{align*}      
        \wm_g\cp \gamma_{r,s}\nsim_\cp \gamma_{r',s'}&\:\text{ for }\:(r,s)\neq (r',s') \:\text{ because }\:(\gamma_1,\dots,\gamma_k)\in \grF \qquad\text{as well as}\\
        \wm_g\cp \gamma''_{q}\nsim_\cp \gamma''_{q'}&\:\text{ for }\:q\neq q' \hspace{41.5pt}\text{ because }\:(\gamma'_1,\dots,\gamma'_p)\in \grF
    \end{align*}
     for all  $g\in G\backslash\{e\}$. For this, observe that each of these curves 
     is a subcurve of one of the free segments $\gamma_1,\dots,\gamma_k, \gamma'_1,\dots,\gamma'_p$.      
      Moreover, by construction we have
      \begin{align*}
     \gamma_{r,s}\nsim_\cp   \wm_g\cp \gamma''_{q}\qquad\forall\: g\in G\backslash\{e\},\: 1\leq r\leq k,\:1\leq s\leq q_r,\:1\leq q\leq m
      \end{align*}
       as well as  
	 \begin{align*}
      (\gamma_1,\dots,\gamma_k),(\gamma'_1,\dots,\gamma'_p)\leq (\gamma_{1,1},\dots,\gamma_{1,q_1},\dots, \gamma_{1,l},\dots,\gamma_{1,q_p},\gamma''_1,\dots,\gamma''_p).
      \end{align*}
  \end{proof}
We now are able to define the projection and transition Maps:
\begin{lemdef}[Projection and Transition Maps]
  \label{def:ProjLimALfreex}
  \begin{enumerate}
  \item
  \label{hompropers}
     For $\gamma\in \Pafns$ with $\dom[\gamma]=[a,b]$ let
	\begin{align}
	\label{eq:projmapdef}
		\pi_\gamma(\homm):=\psi_{\gamma(b)}\big(\homm(\gamma)(\nu_{\gamma(a)})\big)\qquad\forall\: \homm\in \HOMFNS. 
	\end{align}	     
    Then, it is immediate to see that for each $\homm\in \HOMFNS$ we have 
    \begin{align*}
    	\pi_{\gamma^{-1}}(\homm)=\pi_{\gamma}(\homm)^{-1}\qquad\text{as well as}\qquad \pi_\gamma(\homm)=\pi_{\gamma|_{[a,t]}}(\homm) \cdot \pi_{\gamma|_{[t,b]}}(\homm)\quad \forall\:t\in (a,b).
    \end{align*}
     Moreover, for each  $g\in G$ and $\homm\in \IHOMFNS$ we have (cf. \eqref{eq:invpi}) 
 \begin{align}
      \label{eq:invpii}
     % \begin{split}
    	\pi_{\wm_g\cp \gamma}(\homm)&=
%    	
%    	\psi_{\wm_g(\gamma(b))}\big(\homm(\wm_g\cp\gamma)(\nu_{\wm_g(\gamma(a))})\big)\stackrel{\eqref{eq:invprop}}{=}\psi_{\wm_g(\gamma(b))}\big(\Phi_g\cp \homm(\gamma)\big(\Phi_{g^{-1}}(\nu_{\wm_g(\gamma(a))})\big)\big)\\
% 
    	\underbrace{\psi_{\wm_g(\gamma(b))}(\Phi_g(\nu_{\gamma(b)}))}_{c}\: \cdot\: \psi_{\nu_{\gamma(b)}}(\homm(\gamma)(\nu_{\gamma(a)}))\cdot \underbrace{\psi_{\gamma(a)}\big(\Phi_{g^{-1}}(\nu_{\wm_g(\gamma(a))})\big)}_{d}
    	.
    %  \end{split}
    \end{align} 
  \item
    For $\alpha=(\gamma_1,\dots,\gamma_k)\in \grF$ with $\dom[\gamma_i]=[a_i,b_i]$ for $i=1,\dots,k$ we define the map $\pi_{\alpha}\colon \IHOMFNS\rightarrow X_\alpha:=S^{|\alpha|}$ by
    \begin{align}
    \label{eq:dfdssssttt}	
      \pi_\alpha(\homm):= \big(\psi_{\gamma_1(b_1)}\big(\homm(\gamma_1)(\nu_{\gamma_1(a_1)})\big),\dots,\psi_{\gamma_k(b_k)}\big(\homm(\gamma_k)(\nu_{\gamma_k(a_k)})\big)\big).
    \end{align}
    This map is surjective by (the next) Proposition \ref{lemma:consFamx}.\ref{lemma:consFamx1}.   
  \item
    Let $\alpha=(\gamma_1,\dots,\gamma_k)\leq (\gamma'_1,\dots,\gamma'_{k'})=\alpha'$ and $\{(\gamma_i)_j\}_{1\leq j\leq n_i}$ be the corresponding decomposition of $\gamma_i$ for $1\leq i\leq k$.
    \begingroup
    \setlength{\leftmarginii}{12pt}
    \begin{itemize}
    \item
   	We find $p_{ij}\in\{1,-1\}$, $1\leq m_{ij} \leq k'$ and $g_{ij}\in G$ 
    with
   	\begin{align}
   	\label{eq:fdfdfdf}
   		(\gamma_i)_j^{p_{ij}}\csim \wm_{g_{ij}}\cp \gamma'_{m_{ij}}.
   	\end{align}
    	Here, for all $1\leq i\leq k$ we have 
    $m_{ij}\neq m_{ij'}$ for $1\leq j\neq j'\leq n_i$ just by injectivity of $\gamma_i$. In addition to that, we denote by $c_{ij}$ and $d_{ij}$ the structure group elements $c$ and $d$ from \eqref{eq:invpii} that correspond to the curve $\wm_{g_{ij}}\cp \gamma'_{m_{ij}}$.
    
    Observe that the above quantities are even uniquely determined because  if \eqref{eq:fdfdfdf} in addition holds for $p'_{ij}\in\{1,-1\}$, $1\leq m'_{ij} \leq k'$ and $g'_{ij}\in G$, then 
	\begin{align*}    
		\gamma'_{m_{ij}}\csim \wm_{g_{ij}^{-1}g'_{ij}}\cp \Big[\gamma'_{m'_{ij}}\Big]^{p_{ij}\cdot p'_{ij}} \qquad&\Longrightarrow\qquad \gamma'_{m_{ij}}\cpsim \wm_{g_{ij}^{-1}g'_{ij}}\cp \gamma'_{m'_{ij}}\\[-2pt]
		&\Longrightarrow\qquad m_{ij}=m'_{ij}\\
		&\Longrightarrow\qquad g_{ij}^{-1}g'_{ij}=e\\
		&\Longrightarrow\qquad p_{ij}=p_{ij}',
    \end{align*}
    where the second step is clear from $(\gamma_1',\dots,\gamma_k')\in \grF$ and the third one from freeness of $\gamma'_{m_{ij}}$ and $G_{\gamma'_{j}}=e$. 
	\item    
    We obtain a well-defined and continuous map $\pi^{\alpha'}_\alpha\colon X_{\alpha'}\rightarrow X_\alpha$ if we define
    \begin{align}
      \label{eq:alphaalphsastrichx}
      \pi^{\alpha'}_\alpha\!\left(x_1,\dots,x_{k'}\right):=\Bigg(\prod_{j=1}^{n_1}\left(c_{1j}\cdot x_{m_{1j}}\cdot d_{1j}\right)^{p_{1j}},\dots,\prod_{j=1}^{n_k}\left(c_{kj}\cdot x_{m_{kj}}\cdot d_{kj}\right)^{p_{kj}}\Bigg).
    \end{align}
    In fact, $\pi^{\alpha'}_\alpha\cp \pi_{\alpha'}=\pi_\alpha$ is immediate from Part \ref{hompropers}), and due to surjectivity of $\pi_{\alpha'}$ (proven in Proposition \ref{lemma:consFamx}.\ref{lemma:consFamx1})) 
    this definition then cannot depend on any choices. 
       \end{itemize}
       \endgroup
	\item
		 Let $\mu_k$ denote the Haar measure on $S^{k}$ for $k\in \mathbb{N}_{\geq 1}$ and define $\mu_\alpha:=\mu_{|\alpha|}$ for $\alpha\in \grF$. Observe that $\mu_k$ equals the $k$-fold product $\mu^k$ of the Haar measure $\muH$ on $S$ (cf.\ Definition \ref{def:ProductMa}.\ref{def:ProductMa2}) as this normalized Radon measure is translation invariant. In fact, using a Riesz-Markov argument, translation invariance is straightforward from 
		 Fubini's formula.
  \end{enumerate}
\end{lemdef}
The next proposition shows that $\IHOMFNS$ is indeed a projective limit of $\{X_\alpha\}_{\alpha\in \grF}$ w.r.t.\ the above projection and transition maps. Moreover, it shows that $\{\mu_{\alpha}\}_{\alpha\in \grF}$ is a respective family of normalized Radon measures defining a normalized Radon measure on $\IHOMFNS$.  
\begin{proposition}
  \label{lemma:consFamx}
  \begin{enumerate}
  \item
    \label{lemma:consFamx1}
    The maps $\pi_\alpha\colon \IHOMFNS\rightarrow S^{|\alpha|}$ are surjective, and together they separate the elements of $\IHOMFNS$.
  \item
    \label{lemma:consFamx2}
    $\IHOMFNS$ is a projective limit of $\{X_\alpha\}_{\alpha\in \grF}$ w.r.t.\ the maps from Lemma and Definition \ref{def:ProjLimALfreex}.
  \item
    \label{lemma:consFamx3}
     $\{\mu_{\alpha}\}_{\alpha\in \grF}$ is a consistent family of normalized Radon measures w.r.t.\ $\{X_\alpha\}_{\alpha\in \grF}$.
  \end{enumerate}
  \begin{proof}
    \begin{enumerate}
    \item
      For surjectivity, let $\homm_0\in \IHOMFNS\neq \emptyset$, $\alpha=(\gamma_1,\dots,\gamma_k)\in\grF$ with $\dom[\gamma_i]=[a_i,b_i]$ and $s'_i:=\psi_{\gamma_i(b_i)}\big(\homm_0(\gamma_i)(\nu_{\gamma_i(a_i)})\big)$ for $1\leq i\leq k$. Since $S$ is compact and connected, the exponential map of $S$ is surjective. Hence,  for $(s_1,\dots,s_k)\in S^k$ we find $\s_1,\dots,\s_k\in \ms$ such that $s_i=s_i'\cdot\exp([b_i-a_i]\cdot \s_i)$ holds for all $1\leq i\leq k$. 
      
      Now, in the situation of Proposition \ref{lemma:freemodify} let $\delta:=\gamma_1$, $\Psi\colon r\mapsto \exp(r \cdot \s_1)$, $t:=a_1$ and $p:=\nu_{\gamma_1(a_1)}$. Moreover, denote by $\homm_1$ the respective modified homomorphism. Then, 
      \begin{align*}
        \pi_{\gamma_1}(\homm_1)&=\psi_{\gamma_1(b_1)}\left(\homm_1(\gamma_1)(\nu_{\gamma_1(a_1)})\right)\\
        &=\psi_{\gamma_1(b_1)}\left(\homm_0(\gamma_1)(\nu_{\gamma_1(a_1)}) \cdot\exp([b_1-a_1]\hspace{1pt} \s_1\right)\\
        &=s'_1 \cdot\exp([b_1-a_1]\hspace{1pt} \s_1) =s_1,
      \end{align*}
      and applying the same argument inductively, we obtain $\homm_k\in \IHOMFNS$ with $\pi_{\gamma_i}(\homm_k)=s_i$ for $1\leq i\leq k$, hence $\pi_\alpha(\homm_k)=(\pi_{\gamma_1}(\homm_k),\dots, \pi_{\gamma_k}(\homm_k))=(s_1,\dots,s_k)$. For this observe that modifying $\homm_i$ along $\gamma_{i+1}$ does not change its values on the curves $\gamma_j$ for $1\leq j\leq i$. This is clear because $H_{[\gamma_i,\gamma_j]}=\emptyset$ holds for $1\leq i\neq j\leq k$, so that surjectivity of $\pi_\alpha$ follows. 
      
      Now, for the separation property let $\homm,\homm'\in \IHOMFNS$ be different, i.e., $\homm(\gamma)\neq \homm'(\gamma)$ for some $\gamma\in \Pafns$. By Proposition \ref{prop:freeseg}.\ref{prop:freeseg1}, we find a maximal free segment $\delta=\gamma|_{K}$, a decomposition $t_0<\dots< t_n$ of $\dom[\gamma]=[t_0,t_n]$ and elements $g_0,\dots,g_{n-1} \in G$, such that 
	\begin{align*}
		\gamma|_{[t_i,t_{i+1}]}\csim \wm_{g_i}\cp \delta^{\pm 1}\qquad\forall\: 1\leq i\leq n-2
	\end{align*}      
	as well as 
	\begin{align*}
	\gamma|_{[t_0,t_{1}]}\csim \wm_{g_0}\cp [\delta|_{L}]^{\pm 1}\qquad \text{and}\qquad \gamma|_{[t_{n-1},t_n]}\csim \wm_{g_{n-1}}\cp [\delta|_{L'}]^{\pm 1}
	\end{align*}
  holds for $L,L'$ both of the form $[l,l_2]$ with $l_1\leq l<l_2$ or $[l_1,l]$ with $l_1<l\leq l_2$. Then, splitting $\delta$ at the respective (at most two) points $\delta(l)$, we obtain a decomposition of $\delta$ into at most three subcurves which define an index $\alpha\in \grF$ for which $\pi_\alpha(\homm)\neq \pi_\alpha(\homm')$ holds. 
    \item 
    It remains to show continuity of the maps $\pi_\alpha$. Here, it suffices to consider the case where the index $\alpha$ consists of one single free segment $\gamma$. Since the topology on $\IHOMFNS$ is the subspace topology inherited from $\HOMFNS$,\footnote{See Definition \ref{def:invhomm}.} $\pi_\gamma$  is continuous  iff
	\begin{align*}
		\pi_\gamma\cp \kappa_{\mathrm{FN}}|_{\AQRFNS}
	\end{align*}	    
    is continuous w.r.t.\ to the subspace topology on $\AQRFNS$ inherited from $\A_{\mathrm{FN}}=\Spec(\PaC_{\mathrm{FN}})$. This, however, is the case if $\pi_\gamma\cp \kappa_{\mathrm{FN}}$ is continuous w.r.t.\ the topology on $\A_{\mathrm{FN}}$. 
    Here, $\kappa_{\mathrm{FN}}\colon \A_{\mathrm{FN}}\rightarrow \HOMFNS$ denotes the respective homeomorphism from Definition \ref{def:indepref}.\ref{def:mapkappa}. However, $\pi_\gamma\cp \kappa_{\mathrm{FN}}\colon \A_{\mathrm{FN}}\rightarrow S$ is continuous by \eqref{eq:algpro}, just because 
      $\left[h_\gamma\right]_{ij}\in \PaC_{\mathrm{FN}}$. 
    \item
      We have to show that $\pi^{{\alpha'}}_{\alpha}(\mu_{{\alpha'}})=\mu_{\alpha}$ holds if $\alpha\leq {\alpha'}$, and by the Riesz-Markov theorem 
      it suffices to verify that 
	\begin{align*}
	 \int_{X_\alpha}f \:\dd\mu_\alpha=\int_{X_\alpha}f \:\dd\pi^{{\alpha'}}_{\alpha}(\mu_{{\alpha'}})\qquad\forall\: f\in C(X_\alpha). 
	\end{align*}
      Now, if $g\in C(X_{\alpha'})$, $x=(x_1,\dots,x_{k'})\in S^{k'}$ and $1\leq i\leq k'$, Fubini's formula reads
      \begin{align}
        \label{eq:Fubinix}
        \int_{X_{\alpha'}}g \:\dd\mu_{\alpha'}=  \int_{S^{k'-1}} \left(\int_S g(x) \:\dd\mu_1(x_i)\right)\dd\mu_{k'-1}(\raisebox{-0.2ex}{$x^i$}),
      \end{align}
      where $S^{k'-1}\ni x^i:=(x_1,\dots,x_{i-1},x_{i+1},\dots,x_{k'})$. Hence, for $f\in C(X_\alpha)$ we have
      \begin{align*}
      %\label{eq:hfhfhfhfhfhfhf}
        \begin{split}
          \int_{X_\alpha}f \:\dd\pi^{{\alpha'}}_{\alpha}(\mu_{{\alpha'}})&=\int_{X_{\alpha'}}\Big(f\cp \pi^{{\alpha'}}_{\alpha}\Big) \:\dd\mu_{{\alpha'}}
          =\int_{S^{k'-1}}\left(\int_S\Big(f\cp \pi^{{\alpha'}}_{\alpha}\Big)(x) \:\dd\mu_1(x_i)\right)\dd\mu_{k'-1}(\raisebox{-0.2ex}{$x^i$}).
        \end{split}
      \end{align*} 
      By the definition of $\grF$, each of the variables $x_1,\dots,x_{k'}$ occurs in exactly one of the products on the right hand side of \eqref{eq:alphaalphsastrichx}. Consequently, the components $\big[\raisebox{-0.2ex}{$\pi^{{\alpha'}}_{\alpha}$}\big]_i$ of $\pi^{\alpha'}_\alpha$ mutually depend on different variables $\ovl{x}_i=\big(x_{m_{i,1}},\dots,x_{m_{i,n(i)}}\big)$.
       
       Then, by left-, right- and inversion invariance of $\mu_1$, for $x\in S^{k'}$ we have 
      \begin{align*}
        \int_S\Big(f\cp \pi^{{\alpha'}}_{\alpha}\Big)(x) \:\dd\mu_1(x_{m_{k,1}})
        & =\int_S f\Big(\big[\raisebox{-0.2ex}{$\pi^{{\alpha'}}_{\alpha}$}\big]_1(\raisebox{-0.1ex}{$\ovl{x}_1$}),\dots,\big[\raisebox{-0.2ex}{$\pi^{{\alpha'}}_{\alpha}$}\big]_{k-1}(\raisebox{-0.1ex}{$\ovl{x}_{k-1}$}),\big[\raisebox{-0.2ex}{$\pi^{{\alpha'}}_{\alpha}$}\big]_{k}(\raisebox{-0.1ex}{$\ovl{x}_{k}$})\Big) \:\dd\mu_1(x_{m_{k,1}})
        \\
        & =\int_S f\Big(\big[\raisebox{-0.2ex}{$\pi^{{\alpha'}}_{\alpha}$}\big]_1(\raisebox{-0.1ex}{$\ovl{x}_1$}),\dots,\big[\raisebox{-0.2ex}{$\pi^{{\alpha'}}_{\alpha}$}\big]_{k-1}(\raisebox{-0.1ex}{$\ovl{x}_{k-1}$}),x_{m_{k,1}}\Big) \:\dd\mu_1(x_{m_{k,1}}).
      \end{align*}
      Then, applying the same argument inductively 
       we obtain
      \begin{equation*}
        \int_{X_\alpha}f \:\dd\pi^{{\alpha'}}_{\alpha}(\mu_{{\alpha'}}) =\int_{S^{k'}} f(x_{m_{1,1}},\dots,x_{m_{k,1}})\:\dd\mu_{\alpha'}(x)=\int_{X_\alpha} f\:\dd\mu_{\alpha},
      \end{equation*}
      where the last step follows inductively from \eqref{eq:Fubinix} and $\muH(S)=1$. 
    \end{enumerate}
  \end{proof}
\end{proposition}
\begin{lemrem}[Independence from the Choices]
By Proposition \ref{lemma:consFamx}.\ref{lemma:consFamx2}, there is a unique normalized Radon measure $\mu$ on $\IHOMFNS$ which  corresponds to the family $\{\mu_\alpha\}_{\alpha \in \grF}$ for $\mu_\alpha$ the Haar measure on $S^{|\alpha|}$. 
Now, we obtain different projection maps $\pi'_\alpha$ if we choose another family $\nu'=\{\nu'_x\}_{x\in M}\subseteq P$ with $\nu'_x\in F_x$ for all $x\in M$. This family, however, gives rise to the same measure on $\IHOMFNS$ because:  

	It follows from \eqref{eq:indep} that for each $\alpha\in \grF$ we find $d_1,b_1,\dots,d_k,b_k \in S$ such that
$\pi_\alpha(\homm)=(s_1,\dots,s_k)$ implies $\pi'_\alpha(\homm)=(d_1\cdot s_1 \cdot b_1,\dots,d_k\cdot s_k \cdot b_k)$ for all $(s_1,\dots,s_k)\in X_\alpha$.
%\item
Consequently, $\pi_\alpha'=\mathrm{d}\cdot \pi_\alpha \cdot \mathrm{b}$ holds for $\mathrm{d}:=(d_1,\dots,d_k)$ and $\mathrm{b}:=(b_1,\dots,b_k)$, so that for $B\in \Borel(X_\alpha)$ we have
\begin{align*}
	\homm\in \pi_\alpha'^{-1}(B)\quad\Longleftrightarrow \quad\pi'_\alpha(\homm)\in B
	\quad &\Longleftrightarrow \quad\pi_\alpha(\homm)\in \mathrm{d}^{-1}\cdot B\cdot\mathrm{b}^{-1}\\
	&\Longleftrightarrow \quad\homm\in \pi_\alpha^{-1}(\mathrm{d}^{-1}\cdot B\cdot\mathrm{b}^{-1}).
\end{align*} 
%\item
Hence, 
\begin{align*}
	\pi_\alpha'(\mu)(B)=\mu\left(\pi_\alpha'^{-1}(B)\right)=\mu\left(\pi_\alpha^{-1}\left(\mathrm{d}^{-1}\cdot B\cdot\mathrm{b}^{-1}\right)\right)=\mu_\alpha\left(\mathrm{d}^{-1}\cdot B\cdot\mathrm{b}^{-1}\right)=\mu_\alpha(B)
\end{align*}
by invariance of $\mu_\alpha$.\hspace*{\fill}$\lozenge$
\end{lemrem}

\begin{definition}[The Reduced Ashtekar-Lewandowski Measure]
  \label{def:AshLewx}
  The normalized Radon measure on $\IHOMFNS$ that corresponds to the consistent family of normalized Radon measures $\{\mu_\alpha\}_{\alpha\in \grF}$ is denoted by \gls{MFNS} in the following. For the special case that $G=\{e\}$, this measure is called Ashtekar-Lewandowski measure on $\HOMW$ and is denoted by \gls{mAL}.
\end{definition}

\begin{remark}[Loop Quantum Cosmology]
  \label{ex:Fullmeas}
  In the next subsection, we are going to define a normalized Radon measure $\mLAS$ on $\AQRLA\cong \IHOMLAS$, in particular, for the case of (semi-)homogeneous, spherically symmetric and homogeneous isotropic LQC. As already mentioned in Remark \ref{rem:euklrem}.\ref{it:sdsdds}, in the two transitive cases (homogeneous isotropic and homogeneous LQC) the space $\AQRLA$ is already a reasonable quantum-reduced configuration space by itself. So, there we have the reasonable kinematical Hilbert space $\Lzw{\AQRLA}{\mLAS}$.
  \begin{enumerate}
  \item
    \label{ex:Fullmeas1}
  Anyhow, in (semi-)homogeneous LQC we even have the normalized Radon measure $\mu_\red:=\mLAS \times \mFNS$ on $\AQRw\cong \IHOMW$ because $\Paw=\Pags\sqcup \Pafns$, hence 
	\begin{align*}  
	\AQRw\cong \AQRInd{\mg}\times \AQRFNS 
 % \IHOMW\cong \IHOMLAS\times \IHOMFNS
  	\end{align*} 
  	holds by Example \ref{eq:invelements}.
  \item
    \label{ex:Fullmeas2}
    In the homogeneous isotropic and spherically symmetric case we unfortunately do not know whether $\Pacs=\emptyset$ holds. In the homogeneous isotropic case, we elsewise could define $\mu_\red=\mLAS \times \mFNS$ as well, just because there $\Pafns=\emptyset$ holds  by Example \ref{eq:invelements}.
    
    In the spherically symmetric case, we additionally had to define a measure $\mu_{\mathrm{FS}}$ on
	\begin{align*}    
    \AQRFS \cong \Hom_\red(\Pafs,\IsoF)=\Hom_\red(\Paln,\IsoF)
    \end{align*}
    for $\Paln$ the set of linear curves in $\RR^3$ traversing through the origin.\footnote{See (a) in Convention \ref{conv:sutwo1}.\ref{conv:sutwo2} for the definition of $\Paln$ as well as Example \ref{eq:invelements} for the equality $\Pafs=\Paln$.} 
    This, however, is easy as we can use the same techniques as for the construction of $\mFNS$: 
	\begingroup
    \setlength{\leftmarginii}{13pt}
    \begin{itemize}
    \item 
    	 Instead of $\grF$, we consider the set $\grl$ of tuples $(\gamma_1,\dots,\gamma_k)$ with $\gamma_i\in \Paln$ and $\im[\gamma_i]\subseteq \Span_\RR(\vec{e}_1)$ for $1\leq i\leq k$. Hence, we restrict to such elements of $\Paln$ which completely traverse in the $\vec{e}_1$-axis. Obviously, each other curve in $\Pafs=\Paln$ can be obtained by rotating such a curve around a suitable axis. 
    \item	
    	 We choose $\nu_x=(x,\me)$, i.e., $\psi_x=\pr_2\colon P\ni(x,s)\mapsto s\in \SU$  for all $x\in \RR^3$, and define the projection and transition maps exactly as in Lemma and Definition \ref{def:ProjLimALfreex}. Then,
    	 \begin{align}
    	 \label{eq:dssdds}
    	 	\pi_\gamma(\homm)\stackrel{\eqref{eq:projmapdef}}{=}\psi_{\gamma(b)}\big(\homm(\gamma)(\nu_{\gamma(a)})\big)\stackrel{\eqref{eq:ON}}{=}\Omega_\nu(\homm)(\gamma)
    	 	\stackrel{\eqref{eq:tricBHoms}}{=} \Omega(\homm)(\gamma)=\hommm(\gamma)
    	 \end{align}
    	 for $\dom[\gamma]=[a,b]$ and $\hommm:=\Omega(\homm)$. Since for $\sigma\in H_{\vec{e}_1}=H_{\tau_1}$ we have $\gamma=\sigma(\gamma)$, this gives
		\begin{align*}
			\pi_\gamma(\homm)=\pi_{\sigma(\gamma)}(\homm)\stackrel{\eqref{eq:dssdds}}{=}\hommm(\sigma(\gamma))\stackrel{\eqref{eq:algrels}}{=}\alpha_\sigma(\hommm(\gamma))=\alpha_\sigma(\pi_\gamma(\homm)).
		\end{align*}
%	\item
	 Then,  Lemma \ref{lemma:torus}.\ref{lemma:torus1} shows $\im[\pi_\gamma]\subseteq H_{\tau_1}\cong S^1$,   
    and by Proposition \ref{lemma:freemodify}, we even have $\im[\pi_\alpha]\cong S^{|\alpha|}$. This follows by the same arguments as in the proof of Proposition \ref{lemma:consFamx}.\ref{lemma:consFamx1} because $\Psi_\mu\colon r\mapsto \exp(r\cdot [\mu\cdot\tau_1])$ for $\mu\in\RR$ is contained in $\Mult_{p,\gamma}$ for $p=(\gamma(a),\me)$. In fact, this is clear from $G_\gamma=H_{\tau_1}$, and that $\phi_p(\sigma)=\sigma$ holds for all $\sigma\in G_\gamma$ as for $p=(x,\me)$ we have (see Example \ref{ex:LQC} for the definiton of $\Pii\colon \SU \times P\rightarrow P$)
     \begin{align*}
     	\Pii(\sigma,p)=(x,\sigma)\qquad\Longrightarrow\qquad \Pii(\sigma,p)= p\cdot \sigma \qquad\Longrightarrow\qquad \phi_p(\sigma)=\sigma.
     \end{align*}  
      Then, using the Haar measure on $H_{\tau_1}\cong S^1$ instead of $\SU$, by the same arguments as in the $\Pafns$ case, we obtain a normalized Radon measure $\mu_{\mathrm{FS}}$ on $\Hom_\red(\Paln,\IsoF)$.
    \end{itemize}
    \endgroup
      \end{enumerate}
\end{remark} 

\subsection{Lie Algebra Generated Configuration Spaces}
\label{sec:ConSp}
In the previous part, we have constructed a normalized Radon measure on $\IHOMFNS$, whereby we have used the developments of Subsection \ref{sec:ModifreeSeg}. 
In this part, we are going to write the space $\IHOMLAS$ as a Tychonoff product of compact Hausdorff spaces which we determine exemplarily for the most LQC relevant case that $S=\SU$. Then, we define normalized Radon measures on each of these spaces, providing us with a normalized Radon measures on $\IHOMLAS$.
  
  For the first step, we will need that each $\wm$-orbit which contains some Lie algebra generated curve admits an independent and complete family of stable Lie algebra elements. We show that, under these circumstances,  $\IHOMFNS$ can be written as a Tychonoff product of compact Hausdorff spaces which are just given % which then are 
% just given 
 by the sets of equivariant maps that correspond to the elements of the above independent and complete families. Here, we will only need that the structure group is compact and connected. 
 
	In the second part, we will determine these sets of equivariant mappings explicitly for the case that $S=\SU$. We define normalized Radon measures on each of them and show that the corresponding Radon product measure gives rise to a normalized Radon measure on $\IHOMFNS$ which does not depend on any choices we have made.\footnote{For instance, the above families of stable elements.} As already sketched in the beginning of this section, these constructions also work for the abelian case and are even easier there. For non-abelian $S$ one basically has to investigate to which maximal tori equi\-variance can restrict the image of a map $\Psi\colon \spann_\RR(\g)\rightarrow S$  having the homomorphism properties \eqref{eq:equii}. Indeed, for $S$ compact and connected, multiplicativity already restricts such an image to a unique maximal torus if  $\Psi(\lambda\cdot \g)$ is regular for some $\lambda\in \RR$,\footnote{This is because $\Psi(\mu\cdot \g)\cdot \Psi(\lambda\cdot \g)=\Psi([\mu+\lambda]\cdot \g)=\Psi(\lambda\cdot \g)\cdot \Psi(\mu\cdot \g)$ for all $\mu\in \RR$.} and obviously this issue is trivial in the abelian case.  
 Finally, the set of occurring tori has to be equipped with a normalized Radon measure having certain invariance properties which  make the whole construction independent of any choices at the end. 
  Using general theory of compact and connected Lie groups, then one might obtain some generalizations of the construction we will work out in detail for $\SU$. 

So, in the next part, the structure group will be compact and connected, as well as $\SU$ in the subsequent ones.\footnote{Anyhow, since it simplifies the notations, we often will write $\ms$ instead of $\su$.} Moreover, as in the previous section, $\wm$ is always assumed to be analytic and pointwise proper.  
 By $\Mm$ we will denote the set of $\varphi$-orbits $\m$ in $M$ with $\mg\backslash \mg_x\neq \emptyset$ for one (and then each) $x\in \m$.  We fix a family $\{p_\m\}_{\m\in \Mm}\subseteq P$ of elements with $x_\m:=\pi(p_\m)\in \m\in \Mm$  for all $\m\in \Mm$, where for each such $\m$ we let $\{\g_{\m,\alpha}\}_{\alpha \in I_\m}\subseteq \mg\backslash\mg_{x_\m}$ be independent and complete. Of course, we only will consider the situation where $\IHOMLAS\neq \emptyset$ holds.\footnote{Observe that  $\IHOMLAS=\emptyset$ already implies that $\IHOMW=\emptyset$ holds just because each element of the latter space restricts to an element of the former one.}
\subsubsection{Tychonoff Products}
We now are going to write $\IHOMLAS$ as a Tychonoff product of compact Hausdorff spaces %, which will be done 
for the general case that $S$ is compact and connected. The first step towards this is performed in Lemma and Definition \ref{def:topo}, where the strategy is basically the following:
 \begingroup
 \setlength{\leftmargini}{15pt}
\begin{itemize}
\item
In the Parts \ref{def:topo1}) and \ref{def:topo3}), for $x\in M$ and $p\in F_x$, we assign to $\epsilon \in \IHOMW$ an $\Ad_{G_{x}}^p$-equivariant (in the first factor)  map $\Psi\colon \mg\backslash \mg_{x} \times \RR_{>0}\rightarrow S$ just by 
\begin{align}
\label{eq:mappodappo}
    \Psi(\g,l) := \Delta\big(\Phi_{\exp(l\cdot \g)}(p),\homm\big(\gamma_\g^{x}|_{[0,l]}\big)(p)\big).%\qquad\forall\: 0<l<\tau_\g
\end{align}
Recall that $t\mapsto\Phi(\exp(t\cdot \g),p)$ is the canonical lift of $\gag$ in $p\in F_x$ which we already have used in Proposition \ref{th:invhomm}.\ref{th:invhomm1} for modifying invariant homomorphisms along such Lie algebra generated curves $\gag$. Moreover, as usual, for $p,p'$ contained in the same fibre, here $\diff(p,p')$ denotes the unique element $s\in S$ for which $p'=p\cdot s$ holds.
\item
In Part \ref{def:topo11}), we will split up the map \eqref{eq:mappodappo} into several 
 maps $\spann_\RR(\g_\alpha)\rightarrow S$ being equivariant in the sense of Definition \ref{def:eqmaps}. Here, $\alpha$ runs over some index set $I_x$ for $\{\g_\alpha\}_{\alpha\in I_x}\subseteq \mg\backslash\mg_x$ an independent and complete family of stable elements.
\item
Then, in Part \ref{def:topo2}), we will define the relevant topologies on the occurring spaces, and in the last part we will provide the desired homeomorphism between $\IHOMLAS$ and the mentioned Tychonoff product..
\end{itemize} 
\endgroup 
\begin{lemdef}
  \label{def:topo}
  Let $p\in P$, $x:=\pi(p)$ and assume that $\mg\backslash \mg_x\neq \emptyset$ holds.
  \begingroup
  \setlength{\leftmargini}{20pt}
  \begin{enumerate}  
  \item 
    \label{def:topo1}
       By \gls{EQP} we denote the set of $\Add{G_{x}}^p$-equivariant maps  $\Psi\colon \mg\backslash \mg_{x} \times \RR_{>0}\rightarrow S$ that fulfil\footnote{The reason for introducing the factor $\RR_{>0}$ will become clear in Part \ref{def:topo3}).}  
	\begin{align} 
%	      \label{eq:Equi0}   
%    	\Psi(\lambda\cdot \g,l+l')=\Psi(\g,|\lambda| l)^{\sign(\lambda)}\cdot \Psi(\g,|\lambda| l')^{\sign(\lambda)}\qquad\forall\: \lambda\in \RR_{\neq 0},\: l,l'>0 ,
%	\end{align}    
	    \label{eq:Equi0}   
    	\Psi(\lambda\cdot \g,l)=\Psi(\g,|\lambda| l)^{\sign(\lambda)}\qquad\forall\: \lambda\in \RR_{\neq 0},\: l>0 
	\end{align} 
    and
    \begin{align}
      \label{eq:Equi}
      \Psi(\g,l+ l')=\Psi(\g,l)\cdot\Psi(\g,l')\qquad \forall\: l,l'\geq 0.
    \end{align}
    Here, equivariance means that 
	\begin{align*}    
    	\Psi(\Ad_h(\g),\cdot)=\alpha_{\fiba_p(h)}\cp \Psi(\g,\cdot) \qquad \forall\: h\in G_{x}, \forall\:\g\in \mg\backslash \mg_{x}.
    \end{align*}
      \item 
    \label{def:topo11}
    Recall the quantities we have considered in Subsection \ref{sec:ModifLAGC}. 
    Then, since we have $G_{[\g]}^x\subseteq G_x$, for each $\Psi\in \Eq_p$ and each $\g\in \mg\backslash \mg_x$ the map $\Psi_\g\colon \Span_\RR(\g)\rightarrow S$ defined by  
	\begin{align*}
		\Psi_\g(0):=\me \qquad\qquad\text{and}\qquad\qquad \Psi_\g(\lambda\cdot \g):=\Psi(\lambda\cdot\g,1)\:\text{ for }\: \lambda \in \RR_{\neq 0}
	\end{align*}  
  is $\Ad^p_{G_{[\g]}}$-equivariant in the sense of Definition \ref{def:eqmaps}. We will denote by $Y^p_\g$ the set of all such $\Ad^p_{G_{[\g]}}$-equivariant maps and define  
  \begin{align*}  
  	\res_\g^p\colon \Eq_{p}\rightarrow Y_\g^p,\quad
  	 \Psi \mapsto \Psi_{\g}.
  	\end{align*}
    \item
      \label{def:topo2}
    We equip $\Eq_p$ with the topology %$\EqT_p$ 
    generated by the sets 
    \begin{align*}   	
      U_{\g,l}(\Psi):=\{\Psi'\in \Eq_p\:|\: \Psi'(l\cdot \g)\in \Psi(l\cdot \g)\cdot U\}
    \end{align*}
    for $\Psi\in \Eq_p$, $U\subseteq S$ open, $\g\in \mg\backslash\mg_x$ and $l>0$. Similarly, we equip the spaces $Y^p_\g$ with the topologies generated by the sets 
\begin{align*}   	
      U_{\lambda}(\Psi):=\{\Psi'\in Y^p_\g\:|\: \Psi'(\lambda \cdot \g)\in \Psi(\lambda\cdot  \g)\cdot U\},
    \end{align*}    
    for $\Psi\in Y_\g^p$, $U\subseteq S$ open and $\lambda\in \RR$.  
    It is easy to see that $\res_\g^p$ is continuous w.r.t.\ these topologies.
  \item
    \label{def:topo3}
    We define $\pi_p\colon \mg\backslash \mg_{x}\times  \RR_{>0}\times \IHOMLAS \rightarrow S$ by
    \begin{align*}
    	\pi_p(\g,l,\homm) := \Delta\big(\Phi_{\exp(l\g)}(p),\homm\big(\gamma_\g^{x}|_{[0,l]}\big)(p)\big)\qquad\forall\: 0<l<\tau_\g
    \end{align*}
    as well as
    \begin{align*}    
    	\pi_p(\g,l,\homm):=\prod_{i=1}^k\pi_p(\g,l_i,\homm)\quad\text{for}\quad 0<l_1,\dots,l_k<\tau_\g \quad\text{with}\quad l=l_1+\dots+l_k.
	\end{align*}    	
	Then, straightforward calculations, cf.\ Appendix \ref{app:Bohrmodl}, similar to that in the proofs of Proposition \ref{th:invhomm}.\ref{th:invhomm1} and Proposition \ref{lemma:freemodify} show that the map 
	\begin{align}
	\label{eq:pip}
	\pip_p \colon \IHOMLAS\rightarrow \Eq_p,\quad 
	\homm\mapsto \pi_p(\cdot,\cdot,\homm)
	\end{align}
	is well defined and continuous. Consequently, $\pi_p$ is well defined as well. In particular, the topological space $Y_\g^p$ is compact. In fact, $\IHOMLAS$ is compact and $\res_\g^p\cp \pip_p$ is continuous, as well as surjective by the next corollary.
%	\item
%	\label{def:topo4}

	Obviously, we have $\pi_p=\alpha_{\Delta(p',p)}\cp\pi_{p'}$ if $\pi(p)=\pi(p')$ holds. Moreover, if $g\in G$, $s\in S$, $\g\in \mg\backslash\mg_y$ and $h\in G_{y}$ for $y:=\pi(g\cdot p)$, a straightforward calculation shows that we have, cf.\ Appendix \ref{app:Bohrmodl} 
    \begin{align}
      \label{eq:verkn}
      \pi_{g\cdot p\cdot s}(\lambda\Add{h}(\g),l,\homm)=\alpha_{s^{-1} \fiba_{g\cdot p}(h)}\cp \pi_{p}\big(\Add{g^{-1}}(\g),|\lambda| l,\homm\big)^{\sign(\lambda)}\qquad \forall\:\lambda\in \RR_{\neq 0}, \:l>0.
    \end{align}
    This equation will be relevant for our considerations in the final part of this subsection. There, we will show that the measure we  construct on $\IHOMLAS$ does not depend on any choices such as  
    the independent and complete families we have fixed in the beginning of this subsection.
      \item
      \label{def:topo5}
      To simplify the notations, let
      \begin{align*}
      Y_{\m,\alpha}:=Y_{\g_{\m,\alpha}}^{p_\m}\qquad\quad\text{as well as}\qquad\quad\res_{\m,\alpha}:=\res_{\g_{\m,\alpha}}^{p_\m}\qquad\forall\: \m\in \Mm, \forall\: \alpha\in I_\m.
      \end{align*}
       We equip $Y:=\prod_{\m\in \Mm,\alpha\in I_\m}Y_{\m,\alpha}$  with the Tychonoff topology and define 
	\begin{align*}      
      \Pi_Y\colon\IHOMLAS \rightarrow Y\qquad\text{by}\qquad \Pi_{Y}:=\prod_{\m\in \Mm}\hspace{1pt}\big[\res_{\m}\cp \pip_{p_\m}\big]
	\end{align*}       
   for the natural map $\res_{\m}:=\prod_{\alpha\in I_\m}\res_{\m,\alpha}\colon \Eq_{p_\m}\rightarrow \prod_{\alpha\in I_\m}Y_{\m,\alpha}$. 
  \end{enumerate}
  \endgroup
\end{lemdef}
The homeomorphism property of $\Pi_Y$  is a straightforward consequence of Lemma \ref{lemma:completee}.\ref{lemma:homzueq33} and Proposition \ref{th:invhomm}.\ref{th:invhomm1}, as we now show in
\begin{corollary}
	The map $\Pi_Y\colon \IHOMLAS \rightarrow Y$ is a homeomorphism.
\end{corollary}
\begin{proof}
	$\Pi_Y$ is continuous because the maps $\res_{\m,\alpha}\cp \pip_{p_\m}$ are so. Moreover, $Y$ is Hausdorff because the spaces $Y_{\m,\alpha}$ are Hausdorff. So, since $\IHOMLAS$ is compact, the claim follows if we show that $\Pi_Y$ is bijective. Now, $\Pi_Y$ is injective by Lemma \ref{lemma:completee}.\ref{lemma:homzueq33} and
	\begin{align*}	
		\big(\res_{\m,\alpha}\cp \pip_{p_\m}\big)(\homm)(\lambda\cdot \g_{\m,\alpha})&=\pip_{p_\m}(\homm)(\lambda \cdot \g_{\m,\alpha},1)\\
		&=\diff\big(\Phi_{\exp(\lambda \cdot \g_{\m,\alpha})}(p_\m),\homm\big(\gamma_{\lambda\cdot \g_{\m,\alpha}}^{x_\m}|_{[0,1]}\big)(p_\m)\big)\\
		&=\diff\big(\Phi_{\exp(\lambda \cdot \g_{\m,\alpha})}(p_\m),\homm\big(\gamma_{\g_{\m,\alpha}}^{x_\m}|_{[0,\lambda]}\big)(p_\m)\big),
	\end{align*}
	 and surjective by Proposition \ref{th:invhomm}.\ref{th:invhomm1}. In fact, if $\Psi_{\m,\alpha}\in Y_{\m,\alpha}$ for all $\m\in \Mm$ and all $\alpha\in I_{\m}$, then for each $\m\in \Mm$ we can modify an element $\homm'\in \IHOMLAS$ w.r.t.\ the family $\{\Psi_{\m,\alpha}\}_{\alpha\in I_\m}$ in order to obtain $\homm\in \IHOMLAS$ with  
	\begin{align*}
		\big(\res_{\m,\alpha}\cp \pip_{p_\m}\big)(\homm)=Y_{\m,\alpha}\qquad \forall\: \alpha\in I_\m.
	\end{align*}
	Since orbits are disjoint, i.e., $\im\big[\wm_g\cp \gamma_{\g}^{x_\m}\big]\cap \im\big[\wm_{\g'}\cp \gamma_{\g'}^{x_{\m'}}\big]=\emptyset$ if $\m\neq \m'$, it is clear that we can apply Proposition \ref{th:invhomm}.\ref{th:invhomm1} simultaneously for all $\m\in \Mm$.
\end{proof}
\subsubsection{Normalized Radon Measures}
In the previous part, we have seen that $\IHOMW$ is homeomorphic to the Tychonoff product of the compact Hausdorff spaces $Y_{\m,\alpha}$, each of them consisting of certain equivariant maps. So, in order to define a normalized Radon measure on $\IHOMW$, it suffices to define such measures on each of the factors $Y_{\m,\alpha}$. For this, we now calculate these spaces explicitly where we will restrict to the case where $S=\SU$. 

We start by recalling some facts and fixing some notations. 
\begin{convention}
  \label{conv:sammel}
  \begingroup
  \setlength{\leftmargini}{15pt}
  \begin{itemize}
  \item
 	 By $S^1$ we denote the unit circle in $\mathbb{C}$. This will not be in conflict with our notations as in the sequel no products of the structure group $S=\SU$ will occur.
  \item    
    \itspacec
    In the following, let $\DG$ denote the group of all characters on $\RR$, i.e., of all functions of the form $\chi_l\colon x\mapsto \e^{\I l x}$.    
    By the Bohr compactification $\RB$ of $\RR$ we will understand (cf.\ Lemma and Convention \ref{lemconv:RBMOD}.\ref{rem:RBOHR}) the set of unital homomorphisms $\psi\colon \DG \rightarrow  S^1$ equipped with the topology generated by the sets, cf.\ \eqref{eq:bohrtop}
    \begin{align*} 
      V_{l}(\psi):=\{\psi'\in \RB\:|\: \psi'(\chi_l)\in \psi(\chi_l)\cdot V\}
    \end{align*}
    for $\psi\in \RB$, $V\subseteq S^1$ open and $l>0$. Recall that $\RB$ is a compact abelian group when equipped with the group structure
    \begin{align*}
      \big(\psi +\psi'\big)(\chi_l):=\psi(\chi_l)\cdot \psi'(\chi_l)\qquad \psi^{-1}(\chi_l):=\psi(\chi_{-l})\qquad 1_{\mathrm{Bohr}}(\chi_l):=1
    \end{align*}
    for all $l\in \RR$. Moreover, as we have seen in Lemma and Convention \ref{lemconv:RBMOD}.\ref{prop:Bohrmod21},
    $\RB$ is parametrized the set $\Per$ of all maps $\phi\colon \RR_{>0}\rightarrow [0,2\pi)$ with
    \begin{align*}	
      \phi(l+l')=\phi(l)+\phi(l')\:\bmod \: 2\pi\qquad \forall\:l,l'\in \RR_{>0}.
    \end{align*}
  \item
    Let \gls{SPMSU} denote the projective space\footnote{This is the set $\su\backslash\{0\}$ modulo the equivalence relation $\s'\sim \s$ iff $\s'\in \spann_\RR(\s)$.} that corresponds to $\su$ as well as \gls{PRMSU} the corresponding projection map.
    \item
     Recall the map $\text{\gls{MURS}}\colon \RR^3 \rightarrow \mathfrak{su}(2)$ from Convention \ref{conv:sutwo1}.\ref{conv:sutwo111} and that $H_{\s}=\{\exp(t\cdot\s)\:|\: t\in \RR\}$.  
   \item  
    In the following, we will consider $\Sp\ms$ as an index set, and choose a fixed representative $\s_\beta$ with $\|\text{\gls{MURS}}^{-1}(\s_\beta)\|=1$ in each equivalence class $\beta\in \Sp\ms$. 
    \item 
    If $\beta =[\s\hspace{1pt}]\in \Sp\ms$, we will always mean that $\|\murs^{-1}(\s)\|=1$ holds 
    and define $H_{\beta}:=H_\s$. 
  \end{itemize}
  \endgroup
\end{convention} 
In order to determine the spaces $Y_\g^p$, 
we now have to investigate the images of the equivariant maps these spaces consist of. Due to the first part of Lemma \ref{prop:Bohrmod2}, this means to identify the maximal tori which such an equivariant map is allowed to map to. Indeed, there we show that the subset of all non-trivial maps\footnote{Observe that if $\Psi(\lambda\cdot \g)=-\me$, then $\Psi(\lambda/2\cdot \g)\neq \pm \me$.} in $Y_\g^p$ which map to the same maximal torus is parametrized by $\RB$. This means that $Y_\g^p$ is either singleton or given by the product of $\RB$ with the set off all occurring tori. The latter set will we determined in the second part of Lemma \ref{prop:Bohrmod2}. There, we show that the 
following situations can occur:
\begingroup
\setlength{\leftmargini}{20pt} 
\begin{itemize}
\item[{\bf 1)}]
	$Y_\g^p$ consists of one single element, namely the trivial map.
\item[{\bf 2)}]
	All elements of $Y_\g^p$ map into the same maximal torus.
\item[{\bf 3)}]
	Each element of $Y_\g^p$ maps into a maximal torus $H_{\vec{n}}$ with $\vec{n}$ contained in a fixed plain through the origin. Moreover, each such torus occurs. 
\item[{\bf 4)}]
	Each element of $Y_\g^p$ maps into some maximal torus, whereby all maximal tori occur.
\end{itemize}
\endgroup
\noindent
The relevant notions are provided in
\begin{definition}
\label{def:betadef}
%\begin{enumerate}
%  \item
%    \itspacec 
    Let $p\in P$, $x:=\pi(p)$ and 
    \begin{align*}
      J(\g,p):=\bigcup_{\Psi\in Y^p_\g}\beta(\Psi)\subseteq  \{0\}\sqcup\Sp\ms
    \end{align*} 
    for the quantity $\beta(\Psi)$ defined as follows: 
    \begingroup
    \setlength{\leftmargini}{15pt} 
	\begin{itemize}
	\item
	\vspace{-4pt}
		If $\Psi(\lambda\cdot\g)=\me$ for all $\lambda\in \RR$, we let $\beta(\Psi),\s_\beta(\Psi):=0\in \su$.
	\item
		\vspace{2pt}
	In the other case, it follows from 
	\begin{align*}
			\Psi(\lambda\cdot \g)\cdot \Psi(\mu\cdot \g)=\Psi([\lambda+\mu]\cdot \g)=\Psi(\mu\cdot \g)\cdot \Psi(\lambda\cdot \g)\qquad\forall\:\lambda,\mu \in \RR
	\end{align*}
	and Lemma \ref{lemma:torus}.\ref{lemma:torus1} that $\im[\Psi]\subseteq H_{\beta(\Psi)}$ holds for $\beta(\Psi)\in \Sp\ms$ uniquely determined. \hspace*{\fill}$\lozenge$
	\end{itemize}
	\endgroup	      
\end{definition}
\begin{lemma}
  \label{prop:Bohrmod2}
  Let $p\in P$, $x:=\pi(p)$ and $\g\in \mg\backslash \mg_{x}$.
  \begin{enumerate}
  \item
    \label{prop:Bohrmod22} 
    For each $\beta\in J(\g,p)$ and $\phi\in \Per$ we find $\Psi'\in Y_\g^p$ with 
	\begin{align*}    
    \Psi'(\lambda\cdot \g)=\exp(\phi(\lambda )\cdot \s_{\beta}) \qquad\forall\: \lambda > 0.
    \end{align*}
  \item
    \label{prop:Bohrmod23}
    Exactly one of the following cases holds:  
    \begingroup
    \setlength{\leftmarginii}{15pt}  
    \begin{enumerate}
    \item[{\rm \textbf{1})}]
      % \label{ccase1}
      $J(\g,p)=\{0\}$,
    \item[{\rm \textbf{2})}]
      % \label{ccase2}
      $J(\g,p)=\{0,\beta\}$ for $\beta \in \Sp\ms$ uniquely determined,
    \item[{\rm \textbf{3})}]
      % \label{ccase3}
      $J(\g,p)= \{0\}\sqcup \bigcup_{\vec{n}\in \RR^3\backslash\{0\} \colon\langle\vec{n},\vec{m}\rangle=0}\:[\hspace{1.5pt}\murs(\vec{n})\hspace{0.5pt}]$ for some $\vec{m}\in \RR^3\backslash\{0\}$,
    \item[{\rm \textbf{4})}]
      % \label{ccase4}
      $J(\g,p)=\{0\}\sqcup \Sp\ms$.
    \end{enumerate}
    \endgroup
  \end{enumerate}
  \begin{proof}
    \begin{enumerate} 
    \item
      By assumption, we find $\Psi\in Y_\g^p$ with $\beta(\Psi)=\beta$. Now, let
	\begin{align*}
		\Psi'(\lambda\cdot \g):=\exp\hspace{-1pt}\big(\hspace{-1pt}\sign(\lambda)\phi(|\lambda|)\cdot \s_\beta\big)\qquad\forall\:\lambda\in \RR.
	\end{align*}	      
	Then, $\Psi'$ fulfills \eqref{eq:equii}, just by the same arguments as in Lemma and Convention \ref{lemconv:RBMOD}.\ref{prop:Bohrmod21}, so that it remains to show \eqref{eq:equiii}.     
	If $\Psi=\me$, then $\Psi'=\me$, and we have nothing to show. In the other case, we find $\lambda>0$ with\footnote{If $\Psi(\mu \cdot \g)=-\me$, then $\Psi(|\mu| \cdot \g)=-\me$ and we choose $\lambda:=\mu/2$.} $\Psi(\lambda\cdot \g)\neq \pm \me$. Let $\Ad_h(\g)=\mu\cdot \g$ for $h\in G^x_{[\g]}$ and $\mu\in \RR$. Then, $|\mu|=1$ by Remark and Definition \ref{rem:ppropercurve}.\ref{rem:ppropercurve4}, so that
	\begin{align}
	\label{eq:dfdf}
		\alpha_{\fiba_p(h)}\big(\Psi(\lambda\cdot \g)\big)=\Psi(\lambda\mu \cdot \g)=\Psi(\lambda \cdot \g)^{\sign(\mu)}.
	\end{align}
	By Lemma \ref{lemma:torus}.\ref{lemma:torus2} we now have the following two possibilities:
	%we have the following two possibilities:
      \begin{enumerate}
      \item[1.)]
      \vspace{-4pt}
        $\mu=1$ \qquad$\Longrightarrow$\qquad  $\fiba_p(h)\in H_{\beta}$ \qquad$\Longrightarrow$
                \begin{align*}
          \alpha_{\fiba_p(h)}\cp \Psi'(\lambda\cdot \g)
          &=\alpha_{\fiba_p(h)}\cp\exp\hspace{-1pt}\big(\hspace{-1pt}\sign(\lambda)\phi(|\lambda|)\cdot\s_{\beta}\big)
          = \exp\hspace{-1pt}\big(\hspace{-1pt}\sign(\lambda)\phi(|\lambda|)\cdot\s_{\beta}\big)\\
          &=\Psi'(\lambda\cdot\g)=(\Psi'\cp \Ad_h)(\lambda\cdot\g).
        \end{align*}        
      \item[2.)]
      \vspace{2pt}  
      $\mu=-1$ \qquad$\Longrightarrow$\qquad  $\alpha_{\fiba_p(h)}(s)=s^{-1}$ for all $s\in H_{\beta}$ \qquad$\Longrightarrow$    
   		\begin{align*}
          \alpha_{\fiba_p(h)}\cp \Psi'(\lambda\cdot \g)
          &=\alpha_{\fiba_p(h)}\cp\exp\hspace{-1pt}\big(\hspace{-1pt}\sign(\lambda)\phi(|\lambda|)\cdot\s_{\beta}\big)
          = \exp\hspace{-1pt}\big(\hspace{-1pt}\sign(\lambda)\phi(|\lambda|)\cdot\s_{\beta}\big)^{-1}\\
          &=\Psi'(-\lambda\cdot \g)=(\Psi'\cp \Ad_h)(\lambda\cdot\g).
        \end{align*}
      \end{enumerate}
    \item
      Basically, this follows as in Part \ref{prop:Bohrmod22}) by repeated application of Lemma \ref{lemma:torus}.\ref{lemma:torus2} involving a case differentiation. The details of the (not complicated but long) proof can be found in Appendix \ref{app:Bohrmod}.
    \end{enumerate}
  \end{proof} 
\end{lemma}
 We now are ready to determine the spaces $Y_\g^p$, and to define normalized Radon measures thereon. Lemma and Definition \ref{def:ProductMa}.\ref{def:ProductMa3} then provides us with a normalized Radon measure $\mLAS$ on 
\begin{align*}
	\IHOMLAS\cong Y=\prod_{\m\in \Mm,\:\alpha\in I_\m}Y_{\m,\alpha},
\end{align*}
 which, in the final part of this subsection, we show to be independent of any choices. However, before we come to this, we first will give some applications to loop quantum cosmology collected in Example \ref{ex:cosmoliealgmaasse} and Remark \ref{StanLQC}. 

Now, in the first part of (the next) Lemma and Definition \ref{lemdef:projspaces}, we will define the four different spaces which $Y_\g^p$ can be homeomorphic to. 
  Except for the Hausdorff property of the defined topologies, here no difficulties will arise. 
  In the second and the third part, we define the corresponding bijections and establish their homeomorphism properties. Here, the bijections are easily defined, but for their homeomorphism property we will have some hard work to do. In the last part, we provides the measures on the spaces defined in the first part. 
\begin{lemdef}
\label{lemdef:projspaces}
    Let $p\in P$, $x:=\pi(p)$ and $\g\in \mg\backslash\mg_x$.
\begin{enumerate}
\item
\label{lemdef:projspaces1}
We define 
    \begin{align*} 
      X_{\g}^p:= 
      \begin{cases} 
        \{0_{\mathrm{Bohr}}\} 					&\mbox{if Case } {\bf 1)} \text{ holds for } J(\g,p)\qquad\textbf{Type 1},\\
        \RB 										&\mbox{if Case } {\bf 2)} \text{ holds for } J(\g,p)\qquad\textbf{Type 2},\\
        \RB\wti S^1 	&\mbox{if Case } {\bf 3)} \text{ holds for } J(\g,p)\qquad\textbf{Type 3},\\
        \RB\wti S^2 	&\mbox{if Case } {\bf 4)} \text{ holds for } J(\g,p)\qquad\textbf{Type 4}.
      \end{cases}
    \end{align*}
    Here, $S^2\subseteq\RR^3$ denotes the unit sphere and $S^1\subseteq \mathbb{C}$ the unit circle. Moreover, the products denote the quotient spaces 
	\begin{align*}
		 \RB\wti S^i:= \big[\RB\times S^i\big]\slash \sim
	\end{align*}	    
    w.r.t.\ the equivalence relation $\sim$ defined by
	\begin{align*}
		(\psi,v)\sim (\psi',v')\qquad \Longleftrightarrow \qquad (\psi',v')= (\psi^{-1},-v)\quad\text{or} \quad\psi,\psi'=0_{\mathrm{Bohr}}
	\end{align*}
	for $i=1,2$. 
	In both cases, we will denote the respective projection map by $\pr_0$. 
	
	We equip each of the above spaces with its natural topology. Hence,  
$\RB\times S^i$ with the product topology and $\RB\wti S^i$ with the respective quotient topology for $i=1,2$. Obviously, $\RB\wti S^i$ is  
compact for $i=1,2$, and the Hausdorff property is proven below.	
	\vspace{5pt}	
	
	{\bf Remark:} These definitions might seem quite artificial at a first sight. However, they have the big advantage that the canonical Radon measures on $\RB$, $S^2$ and $S^1$ (see Part \ref{lemdef:projspaces5})) can be used to define a normalized Radon measure on the above spaces in a straightforward way. In addition to that,  it should be intuitively clear already at this point that the trivial map will be assigned to the class $[(\NB,v)]$, and that the factors $S^2$ and $S^1$ will label the occurring maximal tori in the respective cases. Here, we will need the identification of $(\psi,v)$ with $(\psi^{-1},-v)$ since, due to our further definitions, $v$ and $-v$ will refer to the same maximal torus in $\SU$.
\item
\label{lemdef:projspaces3}
	For each plane\footnote{This means $\mm\in \RR^3\backslash\{0\}$ and $[\mm]=[\mm']$ if $\mm'=\lambda\cdot\mm$ for $\lambda\neq 0$.} $[\mm]:=\{\vv\in \RR^3\:|\: \langle\vv,\mm\rangle=0\}\subseteq \RR^3$, 
	we fix an angle-preserving\footnote{If $\LM,\LSM$ are two such maps, then $\LM\cp {\LSM}^{-1}\in O(2)$.} map (homeomorphism) $\LM$ between the great circle on $S^2$ cutted out by $[\mm]$ and the unit circle $S^1\subseteq \mathbb{C}$. 
	Then, if Case {\bf 3)} holds for $J(\g,p)$, i.e., if 
    \begin{align}
    \label{eq:hmnejhedrffds}
      J(\g,p)\backslash\{0\}= \bigcup_{\vec{n}\in \RR^3\backslash\{0\} \colon\langle\vec{n},\vec{m}\rangle=0}[\hspace{1.5pt}\murs(\vec{n})\hspace{0.5pt}]\quad\text{ for some } \quad\vec{m}\in \RR^3\backslash\{0\},
    \end{align} 
    then we define $\xi_{\g}^p\colon J(\g,p)\rightarrow S^1$ by (recall that due to Convention \ref{conv:sammel} $\|\murs^{-1}(\s_\beta)\|=1$ holds)
    \begin{align*}
      \xi_{\g}^p(\beta):= 
      \begin{cases} 
        v^1_0 &\mbox{if } \beta=0\\
        \LM\big(\murs^{-1}(\s_\beta)\big) & \mbox{else}
      \end{cases}
    \end{align*}
  	for $v^1_0\in S^1$ some fixed element.  
    Similarly, in Case {\bf 4)}, we define $\xi_\g^p\colon J(\g,p) \rightarrow S^2$ by 
    \begin{align*}
      \xi_{\g}^p(\beta):= 
      \begin{cases} 
        v^2_0 &\mbox{if } \beta=0\\
        \murs^{-1}(\s_\beta) & \mbox{else}
      \end{cases}
    \end{align*}
  	for $v^2_0\in S^2$ some fixed element. 
\item
\label{lemdef:projspaces4}
Using the above maps, we construct our bijection $\tau^p_\g\colon Y_\g^p\rightarrow X_\g^p$ as follows: 
\begingroup
\setlength{\leftmarginii}{15pt} 
\begin{itemize}
\item 
	\vspace{-5pt}
	For $\Psi\in Y_\g^p$ with $\beta(\Psi)\neq 0$ we find $\phi\in \Per$ uniquely determined by
	\begin{align*}
		\Psi(\lambda\cdot \g)=\exp\left(\phi(\lambda)\cdot \s_{\beta(\Psi)}\right)\qquad \forall\: \lambda\in \RR,
	\end{align*}   
	and denote by $\psi$ the corresponding element of $\RB$ from Lemma and Convention \ref{lemconv:RBMOD}.\ref{prop:Bohrmod21}. 
\item
	\vspace{2pt}	
	We define $\tau_\g^p\colon Y_\g^p\rightarrow X_\g^p$ by 
      \begin{align*}
        \tau_\g^p(\Psi):= 
        \begin{cases} 
          0_{\mathrm{Bohr}} 					    &\mbox{if Case } {\bf 1)} \text{ holds for } J(\g,p),\\
          \psi					&\mbox{if Case } {\bf 2)} \text{ holds for } J(\g,p),\\
         \big[\big(\psi,\xi_\g^p\big(\beta(\Psi)\big)\big)\big] 	
          &\mbox{if Case } {\bf 3)} \text{ or Case } {\bf 4)} \text{ holds for } J(\g,p).
        \end{cases}
      \end{align*} 	
\end{itemize}	
\endgroup	
The map $\tau_\g^p$ is a well-defined homeomorphism which is independent of the explicit choice of $\s_\beta\in \beta$ we have made in Convention \ref{conv:sammel}. In fact, the independence is immediate from the definitions, and the homeomorphism property is shown below.
 \item	
    \label{lemdef:projspaces5} 
    \itspacec
    We define normalized Radon measures $\mu_\g^p$ on the spaces $X_\g^p$ as follows. If $X_\g^p$ is of \textbf{Type 1}, there is only one possibility. If $X_\g^p$ is of \textbf{Type 2}, we define $\mu_\g^p:=\mu_{\mathrm{Bohr}}$. 
    For $X_\g^p$ of \textbf{Type 3} and \textbf{Type 4} we proceed as follows: 
    \begingroup
    \setlength{\leftmarginii}{20pt}
    \begin{itemize}
    \item[a)]
    \vspace{-2pt}
      We equip $S^1$ with the Haar measure $\mu_1$ as well as 
      $S^2$ with the canonical Radon measure $\mu_2$ induced by the Haar measure on $\SOD$.\footnote{Let $\mu_R$ denote the Haar measure on $\SOD$ and $\Omega \colon \SOD\times S^2 \rightarrow S^2$ the canonical left action. Let $\vec{n}\in S^2$ be fixed and denote by $G_{\vec{n}}$ the $\Omega$-stabilizer of $\vec{n}$. Then, $\SOD\slash G_{\vec{n}}\cong S^2$ (by the map $[g]\mapsto \Omega(g,\vec{n})$), and we equip the quotient with the push forward of $\mu_R$ by the corresponding projection map. 
        It is straightforward to see that the measure $\mu_{2}$ induced on $S^2$ by this diffeomorphism does not depend on the explicit choice of $\vec{n}\in S^2$.} 
      %%%%%%%%%%%%%%%%%%% 
      Observe that this measure is invariant under the action of $\SOD$ on $S^2$, i.e., $\mu_{2}(R_3(A))=\mu_{2}(A)$ holds for all $A\in \Borel(S^2)$ and all $R_3\in \SOD$.
    \item[b)]
      We equip $\RB\times S^1$ and $\RB\times S^2$ with the respective Radon product measures 
	\begin{align*}
		\mu_{1\times}:=\mu_{\mathrm{Bohr}}\times \mu_1\quad\qquad\text{and} \quad\qquad \mu_{2\times}:=\mu_{\mathrm{Bohr}}\times \mu_2
	\end{align*}      
       from Lemma and Definition \ref{def:ProductMa}.\ref{def:ProductMa1}.
    \item[c)]
     We define the measures on $\RB\wti S^1$ and $\RB\wti S^2$ by the push forwards of $\mu_{1\times}$ and $\mu_{2      
        \times}$ by the respective projection maps. 
    \end{itemize}
    \endgroup
\end{enumerate}
\end{lemdef}
\begin{proof}
\begin{enumerate}
\item[1)]
For $i=1,2$ we have: 
\begingroup
\setlength{\leftmarginii}{15pt} 
\begin{itemize}
\item
	If $\psi\neq 0_{\mathrm{Bohr}}$ and $v\in S^i$, then an open neighbourhood of $[(\psi,v)]$, e.g., is given by $\pr_0(U\times V)$ for $U\subseteq \RB$ an open neighbourhood of $\psi$ with $\NB\notin U$ and $V\subseteq S^i$ an open neighbourhood of $v$. This is clear from 
	\begin{align}	
	\label{eq:dsvvasdr}
	\pr_0^{-1}(\pr_0(U\times V))=U\times V \:\cup \: U^{-1}\times -V.
	\end{align}
\item
	If $U_0\subseteq \RB$ is a symmetric open neighbourhood of $0_{\mathrm{Bohr}}$, then $\pr_0(U_0\times S^i)$ is an open neighbourhood of $[(0_{\mathrm{Bohr}},v)]$ in $\RB\wti S^i$, just because 
	\begin{align}
	\label{eq:sfsfdsfsdf}
	\pr_0^{-1}\!\left(\pr_0\!\left(U_0\times S^i\right)\right)\stackrel{\eqref{eq:dsvvasdr}}{=}U_0\times S^i
	\end{align}		
	is open.
\item
	So, choosing $U_0$ and $U$ as above, such that in addition $U_0\cap U^{\pm 1}=\emptyset$ holds, we can separate $[(\NB,v_0)]$ and $[(\psi,v)]$ for $\psi \neq \NB$ and $v,v_0\in S^i$ by $\pr_0(U_0\times S^i)$ and $\pr_0(U\times V)$. In fact, by construction we even have 
	\begin{align*}
		\pr_0^{-1}(\pr_0(U_0\times S^1))\cap \pr_0^{-1}(\pr_0(U\times V))=\emptyset.
	\end{align*}		
\item
	The remaining cases follow from \eqref{eq:dsvvasdr} and the fact that for $\psi,\psi'\neq 0_{\mathrm{Bohr}}$ with $\psi\neq \psi'$ as well as $v,v'\in S^i$ with $v\neq v'$ we find neighbourhoods $U, U'$ of $\psi, \psi'$ as well as neighbourhoods $V,V'$ of $v,v'$, such that 
\begin{align*}
\emptyset&=U\cap U', U\cap U^{-1}, U'\cap {U'}^{-1}, U^{-1}\cap U'\qquad\text{as well as} \\
\emptyset&= V\cap V', V\cap -V, V'\cap -V', -V\cap V'
\end{align*}	
	holds, respectively.
\end{itemize}
\endgroup     
\item[3)]  
Obviously, $\tau_\g^p$ is well defined and injective, and  
its surjectivity follows from  Lemma \ref{prop:Bohrmod2}.\ref{prop:Bohrmod22} and Lemma and Convention \ref{lemconv:RBMOD}.\ref{prop:Bohrmod21}. So, since $Y_\g^p$ is compact and $X_\g^p$ is Hausdorff by Part \ref{lemdef:projspaces1}), we only have to prove continuity of $\tau_\g^p$ in order to show its homeomorphism property. 

{\bf Strategy:} Basically, here the difficulty is to show that the convergence of a net $Y_\g^p\supseteq \{\Psi_\iota\}_{\iota\in J}\rightarrow \Psi\in Y_\g^p$ already implies the ``convergence'' of the corresponding ``maximal tori'' $\beta(\Psi_\iota)$ to the ``maximal torus'' $\beta(\Psi)$ of the limes map. This is clear if $Y_\g^p$ is of \textbf{Type 2} (or \textbf{Type 1}) as there $\beta(\Psi_\iota)=\beta(\Psi)$ holds for all $\iota\in J$. For $Y_\g^p$ of \textbf{Type 3} or \textbf{Type 4}, we have to consider such $\lambda>0$ for which $\Psi(\lambda\cdot \g)$ is regular, i.e., different from $\pm \me$. here, we have to use the local diffeomorphism property of the map 
\begin{align*}
	 S^i\times [ (0,\pi)\sqcup (\pi,2\pi)]\ni(v,t)\mapsto \exp(t\cdot \murs(v)), 
\end{align*}
  whose local inverses give back the two possible values $(v_\lambda,t_\lambda)$ and $(-v_\lambda,2\pi-t_\lambda)$ in $[0,2\pi]$ for which   
  \begin{align*}
  	\exp( t_\lambda\cdot \murs( v_\lambda))=\Psi(\lambda\cdot \g)=\exp( (2\pi-t_\lambda)\cdot \murs( -v_\lambda))
  \end{align*}
  holds.\hspace*{\fill}$\dagger$
\begingroup
\setlength{\leftmarginii}{13pt}
\begin{itemize}
\item[$\triangleright$]
If $X_\g^p$ is of \textbf{Type 1}, we have nothing to show because $\im\big[\tau_\g^p\big]=0_{\mathrm{Bohr}}$ holds in this case.
\end{itemize}
\endgroup
 For the other types, let $Y_\g^p\supseteq \{\Psi_\iota\}_{\iota\in J}\rightarrow \Psi\in Y_\g^p$ be a converging net. 
  	We choose $\phi\in \Per$, $\{\phi_\iota\}_{\iota\in J}\subseteq \Per$ as well as $\beta\in \{0\}\sqcup\Sp\ms$, $\{\beta_\iota\}_{\iota\in J}\subseteq \{0\}\sqcup\Sp\ms$ such that for each $\lambda>0$ we have
    \begin{align}
    \label{eq:covbbb}
    \Psi(\lambda\cdot \g)&=\exp(\phi(\lambda)\cdot\s_\beta)\qquad\text{as well as}\qquad
      \Psi_\iota(\lambda\cdot \g)=\exp(\phi_\iota(\lambda)\cdot\s_{\beta_\iota})\quad \forall\: \iota\in J.
    \end{align}
   	Let $\psi\in \RB$ and $\{\psi_\iota\}_{\iota\in J}\subseteq \RB$ denote the elements of $\RB$ from Lemma and Convention \ref{prop:Bohrmod2}.\ref{prop:Bohrmod21} that correspond to $\phi$ and $\{\phi_\iota\}_{\iota\in J}$, respectively. Finally, define 
	\begin{align*}   	
   	f\colon S^2\times [0,2\pi)&\rightarrow \SU\\
   	(v,t)&\mapsto \exp(t\cdot\murs(v)), 
   	\end{align*}
    i.e., $f(v,t)=\cos(t)\cdot \me+\sin(t)\cdot\murs(v)$ by  \eqref{eq:expSU2}.
\begingroup
\setlength{\leftmarginii}{13pt}
\begin{itemize}
\item[$\triangleright$]
 If  $Y_\g^p$ is of \textbf{Type 2}, then $\beta_\iota=\beta$ holds for all $\iota\in J$. Now,   
    $g\colon S^1\rightarrow H_\beta$, $\e^{\I t}\mapsto \exp(t\cdot s_\beta)$ is a homeomorphism, and for all $\lambda\in \RR$ we have
	\begin{align*}    
     \psi(\chi_\lambda)=g^{-1}\cp \Psi(\lambda\cdot \g)\qquad\text{and}\qquad \psi_\iota(\chi_\lambda)= g^{-1}\cp \Psi_\iota(\lambda\cdot \g)\quad\forall\:\iota\in J.
     \end{align*}
    Consequently, the convergence $\{\Psi_\iota\}_{\iota\in J}\rightarrow \Psi$ implies $\{\psi_\iota\}_{\iota\in J}\rightarrow \psi$ just by the definition of the topologies on $Y_\g^p$ and $\RB$.
\end{itemize}
\endgroup   
    $\triangleright$ Let $Y_\g^p$ be of \textbf{Type 3} or of \textbf{Type 4}. 
    
    \vspace{3pt}
    We first assume that $\Psi=\me$, hence $\beta=0$: 
    \begingroup
\setlength{\leftmarginii}{15pt}
\begin{itemize}
\item[--]
    \vspace{-4pt}
    Then, $\lim_\iota \Psi_\iota(\lambda\cdot \g)=\Psi(\lambda\cdot \g)=\me$ for all $\lambda\in \RR$, so that for each such $\lambda$ and each $n\in\mathbb{N}_{>0}$ we find $\iota_0\in J$ with
	\begin{align}  
	\label{eq:jhgsd}  
     \phi_\iota(\lambda)\in [0,1/n)\sqcup (2\pi-1/n,2\pi)\qquad  \forall\:\iota\geq \iota_0.
	\end{align}     
	In fact, if $C:=[0,1/n)\sqcup (2\pi-1/n, 2\pi)$ and $U:=f\big(S^2\times C\big)$, then (cf.\ formula \eqref{eq:expSU2}) $U=\exp\big([0,1/n) \cdot \murs(S^2)\big)$  is an open neighbourhood of $\me$ in $\SU$ provided that $n$ is suitable large. Since $f^{-1}(U)=S^2\times C$, the claim is clear from equation \eqref{eq:covbbb} and the definition of $f$.	          
\item[--]
      Then, \eqref{eq:jhgsd} shows that 
      \begin{align*}
      	\lim_\iota \psi_\iota(\chi_\lambda)=\lim_\iota\e^{\I \sign(\lambda)\phi_\iota(|\lambda|)}\stackrel{\eqref{eq:jhgsd}}{=} 1\qquad\forall\: \lambda>0, 
      \end{align*}
      hence $\lim_\iota\psi_\iota = 0_{\mathrm{Bohr}}$. Consequently,
	  \begin{align*}
		\lim_\iota\tau_\g^p(\Psi_\iota)=[(0_{\mathrm{Bohr}}, v_0^i)]
	  \end{align*}	      
      for $i\in\{1,2\}$ because each neighbourhood of $[(0_{\mathrm{Bohr}},v_0^i)]$ in $\RB\wti S^i$ contains an open neighbourhood of the form $\pr_0\!  \left(U_{0}\times S^i\right)$ for $U_0$ an symmetric open neighbourhood of $0_{\mathrm{Bohr}}$ in $\RB$. 
    \item[--]
    In fact, by the definition of the quotient topology,
    $W\subseteq \RB\wti S^i$ with $[(0_{\mathrm{Bohr}},v_0^i)]\in W$ is an open neighbourhood of $[(0_{\mathrm{Bohr}},v_0^i)]$ iff $\pr_0^{-1}(W)$ is an open subset of $\RB\times S^i$, hence an open neighbourhood of $(0_{\mathrm{Bohr}},v)$ for all $v\in S^i$. Then, we have
	\begin{align*}    
    	\pr_0^{-1}(W)\supseteq \textstyle\bigcup_{\nu}U^\nu\times V^\nu 
	\end{align*}    	
    	for  open neighbourhoods $U^\nu\subseteq \RB$ of $0_{\mathrm{Bohr}}$ as well as $V^\nu\subseteq S^i$ open subsets with $S^i=\bigcup_{\nu}V^\nu$. Since $S^i$ is compact, we even have $S^i=V^{\nu_1}\cup\dots\cup V^{\nu_l}$ for finitely many indices, so that by \eqref{eq:sfsfdsfsdf} the statement holds for $U_0=U\cap U^{-1}$ with $U:=U^{\nu_1}\cap\dots\cap U^{\nu_l}$. 
\end{itemize}
\endgroup

    \vspace{3pt}
    We now assume that $\Psi(\lambda\cdot \g)\neq\me$ holds for some $\lambda>0$, hence $\beta\neq 0$: 
    \begingroup
\setlength{\leftmarginii}{13pt}
\begin{itemize}
\item[--]
    \vspace{-4pt}
    We  even can assume that $\Psi(\lambda\cdot \g)\neq\pm\me$ since elsewise we replace $\lambda$ by $\lambda/2$.  
   	Then, for $v:=\murs^{-1}(\s_\beta)$ we find open neighbourhoods 
	\begin{align*}
	V\subseteq S^2\:\text{ with }\:v\in V \qquad J\subseteq (0,\pi)\sqcup (\pi,2\pi)\:\text{ with }\:\phi(\lambda) \in J%\qquad U\subseteq \SU\:\text{ with }\: \pm\me \notin U\:\text{ and }\:\Psi(\lambda)\in U
	\end{align*}   	
	as well as $ U\subseteq \SU$ with 
	$\Psi(\lambda\cdot \g)\in U$,  
  such that $f':=f|_{V\times J}$ is a diffeomorphism to $U$. 
   	
   	In fact, since $t:=\phi(\lambda)\in (0,\pi)\sqcup (\pi,2\pi)$, i.e., $\sin(t)\neq 0$, 
   	\begin{align*}
   	0=\dd_{(t,v)} f((\Delta t,\Delta v) )=\Delta t \:[-\sin(t)\cdot\me+\cos(t)\cdot\murs(v)] +\sin(t)\cdot \murs(\Delta v)
	\end{align*}   	
   	 implies $\Delta t=0$, hence $\Delta v=0$.
	\item[--]   	
   	Let $\iota_0\in J$ be such that $\Psi_\iota(\lambda)\in U$ holds for all $\iota\geq \iota_0$. Then, for such $\iota\geq \iota_0$ we have
    \begin{align*}
      \s_{\beta_\iota}= q_\iota\cdot \big(\:\murs\cp \pr_1\cp{f'}^{-1}\big)(\Psi_\iota(\lambda\cdot \g)) \qquad\phi_\iota(\lambda)= 2\pi p_\iota + q_\iota\cdot \big(\hspace{1pt}\pr_2\cp{f'}^{-1}\big)(\Psi_\iota(\lambda\cdot \g))
    \end{align*}  
    for $q_\iota\in \{-1,1\}$ uniquely determined and $p_\iota=0$ for $q_\iota=1$ as well as $p_\iota=1$ for $q_\iota=-1$. 
    Since the occurring maps are continuous and       
    \begin{align*}
      \s_{\beta}= \big(\murs\cp \pr_1\cp{f'}^{-1}\big)(\Psi(\lambda\cdot \g)) \qquad\qquad \phi(\lambda)=\big(\pr_2\cp{f'}^{-1}\big)(\Psi(\lambda\cdot \g))
    \end{align*}
   % holds as well, 
   holds, 
    we have 
	\begin{align*}
		\s_\beta= \lim_\iota q_\iota\cdot \s_{\beta_\iota}\qquad\text{and}\qquad \phi(\lambda)=\lim_\iota  q_\iota \phi_\iota(\lambda)+ 2\pi  p_\iota.
	\end{align*}	 
    Consequently, in the \textbf{Type 3} case (analogously in the \textbf{Type 4} case) with $\mm$ as in \eqref{eq:hmnejhedrffds}, we obtain 
    \begin{align}
    \label{eq:xiconv}
    \begin{split}
      \xi_\g^p(\beta)&=\LM \big(\murs^{-1}(s_\beta)\big)=\LM\Big(\murs^{-1}\Big(\lim_\iota q_\iota\cdot s_{\beta_\iota}\Big)\Big)\\
      &=\lim_\iota q_\iota \cdot \LM \big(\murs^{-1}(s_{\beta_\iota})\big)=\lim_\iota q_\iota\cdot\xi_\g^p(\beta_\iota).
      \end{split}
    \end{align}
    \end{itemize}
   \endgroup
   \noindent   
     Then, for each further $\lambda'>0$ with $\Psi(\lambda'\cdot \g)\neq \pm \me$ we have $\s_\beta= \lim_\iota q'_\iota\cdot \s_{\beta_\iota}$ 
     for $q_\iota'\in  \{-1,1\}$ uniquely determined as well. So, since $\s_\beta= \lim_\iota q_\iota\cdot \s_{\beta_\iota}$, we find $\iota_0\in J$ such that $|q_\iota-q_\iota'|<2$ holds for all $\iota\geq \iota_0$, hence $q_i=q_i'$ for all $\iota\geq \iota_0$. 
Consequently, 
    \begin{align}
    \begin{split}
    \label{eq:psiconv}
      \psi(\chi_{\lambda'})&=\e^{\I \phi(|\lambda'|)}= \e^{\I \lim_\iota [q_\iota \phi_\iota(|\lambda'|)+2\pi p_\iota]}\\
      &=\lim_\iota\e^{\I\: [q_\iota\!\phi_\iota(|\lambda'|)+2\pi p_\iota]}=\lim_\iota\e^{\I\: q_\iota\!\phi_\iota(|\lambda'|)}=\lim_\iota \psi_\iota(\chi_{\lambda'})^{ q_\iota}.
      \end{split}
    \end{align}
  Moreover, if $\Psi(\lambda')=-\me$, then $\Psi(\lambda'/2)\neq \pm \me$ and  
	 \begin{align*}    
     \lim_\iota\psi_\iota(\chi_{\lambda'})^{q_\iota}=\big[\lim_\iota\psi_\iota(\chi_{\lambda'/2})^{q_\iota}\big]^{2}\stackrel{\eqref{eq:psiconv}}{=}\psi(\chi_{\lambda'}). 
	\end{align*}     
     Finally, if $\Psi(\lambda')=\me$, then the same arguments as in the $\Psi=\me$ case show that $\lim_\iota \psi_\iota(\lambda')= 1=\psi(\lambda')$, hence $\lim_\iota \psi_\iota(\lambda')^{q_\iota}= 1=\psi(\lambda')$ holds as well. 
     
     Consequently, $\psi(\chi_{\lambda'})=\lim_\iota \psi_\iota(\chi_{\lambda'})^{ q_\iota}$ holds for all $\lambda'>0$, so that 
     by the definition of the topology on $\RB$, we have $\lim_\iota \psi_\iota^{q_\iota}= \psi$. Hence, 
    \begin{align*}
      \lim_\iota \tau_\g^p(\Psi_\iota)&=\lim_\iota\left[\left(\psi_\iota, \xi_\g^p(\beta_\iota)\right)\right]=\lim_\iota\left[\left(\psi_\iota^{q_\iota},q_\iota \cdot\xi_\g^p(\beta_\iota)\right)\right]\\
      &=\left[\lim_\iota\left(\psi_\iota^{q_\iota},q_\iota \cdot\xi_\g^p(\beta_\iota)\right)\right]\stackrel{\eqref{eq:xiconv}}{=}\big[(\psi,\xi_\g^p(\beta))\big],
    \end{align*}
    where in the third step we have used continuity of the projection map $\pr_0$ and in the second one that $(\psi,v)\sim (\psi^{-1},-v)$ holds. 
\end{enumerate}
\end{proof}
To this point, we have identified the spaces $Y_\g^p$ with the spaces $X_\g^p$, on each of which we have defined normalized Radon measure with suitable invariance properties. So, we now are ready to identify $\IHOMLAS$ with a respective Tychonoff product, and to define a normalized Radon measure on this space. To this end, we simplify the notations by defining
 \begin{align*}
      \mu_{\m,\alpha}:=\mu_{\g_{\m,\alpha}}^{p_\m}\qquad\qquad X_{\m,\alpha}:=X_{\g_{\m,\alpha}}^{p_\m}\qquad\qquad\tau_{\m,\alpha}:=\tau_{\g_{\m,\alpha}}^{p_\m}%\qquad\forall\: \m\in \Mm, \forall\: \alpha\in I_\m.
      \end{align*}
      for all $\m\in \Mm$ and all $\alpha\in I_\m$.
\begin{definition}[The normalized Radon measure $\mLAS$]
\begingroup
\setlength{\leftmargini}{17pt}
\begin{itemize}
\item
We equip 
	\begin{align*}     
     X:=\prod_{\m\in \Mm,\alpha\in I_\m}X_{\m,\alpha}%\text{with the radon product measure }
     \end{align*}
      with the Radon product 
     (see Lemma and Definition \ref{def:ProductMa}.\ref{def:ProductMa3}) of the normalized Radon measures $\mu_{\m,\alpha}$ 
     from Lemma and Definition \ref{lemdef:projspaces}.\ref{lemdef:projspaces5}.
   \item
   By \gls{mLAS} we denote the normalized Radon measure on $\IHOMLAS$ carried over from $X$ by the homeomorphism 
	\begin{align*}
	\eta_X:=\Xi_X \cp \Pi_Y\colon \IHOMLAS\rightarrow X
	\end{align*}	   
    with
	\begin{align*}
		\Xi_X:=\prod_{\m\in \Mm,\alpha\in I_\m}\tau_{\m,\alpha}\colon Y\rightarrow X
	\end{align*}	     
	the product of the maps $\tau_{\m,\alpha}\colon Y_{\m,\alpha}\rightarrow X_{\m,\alpha}$ from Lemma and Definition \ref{lemdef:projspaces}.\ref{lemdef:projspaces4}.\hspace*{\fill}$\lozenge$
	\end{itemize}
	\endgroup
\end{definition}
Before we come to the independence of $\mLAS$ from the choices we have made, we now calculate the space $\IHOMLAS$ for our three standard LQC situations from Example \ref{ex:LQC}.
\begin{example}[Loop Quantum Cosmology]
  \label{ex:cosmoliealgmaasse}
  Assume that we are in the situation of Example \ref{ex:LQC}, where  $P=\RR^3\times \SU$. We now determine the space $X$ for the case of (semi-)homogeneous, spherically symmetric and homogeneous isotropic LQC. We start with 
  \par
  \begingroup
  \leftskip=8pt

  \vspace{8pt}
  \noindent
  {\bf\textit{(Semi)-Homogeneous LQC:}}
  \vspace{2pt}
  
  \noindent
  We even can assume that $P$ is an arbitrary $\SU$-bundle and that 
  $\wm$ acts transitively and free in homogeneous case, or just free in the semi-homogeneous one. 
    \begingroup
  \setlength{\leftmargini}{25pt}
  \begin{itemize}
  \item
  If we are in the homogeneous case, 
   $\Mm$ is singleton and  $G_x=\{e\}$ for all $\g\in \mg$. Consequently, each such $\g\in \mg\backslash \mg_x$ is stable by Lemma and Remark \ref{rem:dfggfg}.\ref{rem:dfggfg2}.  
  Let $I:=I_\m:=\Sp\mg$ and $\g_\alpha\in \alpha$ for each $\alpha\in I$. Then, $\{\g_\alpha\}_{\alpha\in I}$ is obviously complete, and independent by Lemma \ref{lemma:sim}.\ref{lemma:sim4} (see also Remark  \ref{lemremmgohnermgx}).
 It follows that
  \begin{align*} 
    \IHOMLAS\cong Y=\prod_{\alpha\in I}Y_{\m,\alpha} \cong X=\Big[\RB\wti S^2\Big]^{|\Sp\mg|}
  \end{align*}		
  because $J(\g,p)=\{0\}\sqcup \Sp\ms$ holds for all $p\in P$ and all $0\neq \g\in \mg$. This is clear from $G^x_{[\g]}=\{e\}$ since then $\Psi\colon \lambda \cdot \g \mapsto \exp(\lambda\cdot \s)$ is an element of $Y_\g^p$ for all $\s\in \su$.
\item
  In the semi-homogeneous case, i.e., if $\wm$ acts not transitively but free, the same arguments show that we have
  \begin{align*} 
    \IHOMLAS\cong \Big[\RB\wti S^2\Big]^{|\Mm\times \Sp\mg|}.
  \end{align*}
  %holds.
  \end{itemize}
  \endgroup
  
  \vspace{8pt}
  \noindent
  {\bf\textit{Spherically Symmetric LQC:}}
  \vspace{2pt}
  
  \noindent
  We now consider the second case in Example \ref{ex:LQC}. Here, 
   $\Mm$ can be parametrized by the positive $x$-axis $\vec{A}=\{\lambda \cdot\vec{e}_1 \:|\: \lambda>0\}$ as we can ignore the origin\footnote{This is in contrast to the classical (smooth) situation, where it usually makes a big difference whether one takes singular orbits (such as here the origin) into account or not.  
   Indeed, as we will see in Example \ref{bsp:Rotats} (where we  calculate the set of smooth spherically symmetric connections explicitly) this is exactly the case in spherically symmetric LQC. 
 } because its stabilizer is the whole group, i.e. $\mg\backslash \mg_0=\emptyset$. 
 	 Then, $G_{x}=H_{\tau_1}$ and $\mg_x=\Span_\RR(\tau_1)$ holds for all $x\in \vec{A}$. For each $\m \in \Mm$ we denote by $x_\m\in \m$ the unique representative contained $\vec{A}$, i.e., the only element of $\m\cap \vec{A}$. Finally, we let $p_\m:=(x_\m,e)$ for all $\m\in \Mm$. 
\vspace{6pt}  
  
  \noindent
  We claim that for each $x\in \vec{A}$ the space $\mathfrak{G}_{x}$ is parametrized by the angles in $(0,\pi/2]$:
  \begingroup
  \setlength{\leftmargini}{25pt}
  \begin{itemize}
   \item
  	For each $\g\in \mg\backslash \mg_x$,  $\im\big[\gamma_\g^x\big]$ is the circle in $\RR^3$ which arises from rotating $x$ around the axis through the origin determined by $\murs^{-1}(\g)$. In particular, if $\g'\in \mg\backslash \mg_x$ is a further element, then we either have
	\begin{align*}
		\im\big[\gamma_\g^x\big]=\im\big[\gamma_{\g'}^x\big]\qquad\Longrightarrow\qquad \g'=\pm\g
	\end{align*}	  	
  	or $\im\big[\gamma_\g^x\big]\cap\im\big[\gamma_{\g'}^x\big]$ consists of at most two points.
  \item
  	We consider the family $\{\g_{\alpha}\}_{\alpha\in (0,\pi)}\subseteq \mg\backslash\mg_x=\su\backslash \Span_\RR(\tau_1)$ of elements 
	\begin{align*}  	
  	 \g_{\alpha}:=\cos(\alpha)\cdot \tau_1 +\sin(\alpha)\cdot\tau_2\quad\text{for}\quad 0<\alpha<\pi.
  	\end{align*}
  	These elements are stable because by the first point
  	 $\gamma_{\g_\alpha}^x|_{[0,l]}\psim\gamma_{\pm\Ad_h(\g_\alpha)}^x|_{[0,l']}$ already implies that $\Ad_h(\g_\alpha)=\pm \g_\alpha$ holds.
  \item
  	Since $\Ad\colon G_{x}\times \mg\backslash \mg_x\rightarrow \mg\backslash \mg_x$ for $G_x=H_{\tau_1}$ acts on $\mg=\su\cong \RR^3$ via rotations around the $\tau_1$-axis, the above family is complete. In fact, for each $\g\in \mg\backslash \mg_x=\su\backslash \spann_\RR(\tau_1)$ we find $\alpha\in (0,\pi)$, $\lambda \neq 0$ and $h\in H_{\tau_1}$ such that $\g= \lambda\Ad_h(\g_\alpha)$, i.e., $\g\xsim \g_\alpha$ holds.
	\item
	The family $\{\g_{\alpha}\}_{\alpha\in (0,\pi)}$ is not independent because  
	\begin{align*}
	\Ad_h(\g_{\pi/2+ \epsilon})=-\g_{\pi/2- \epsilon} \qquad\Longrightarrow\qquad\g_{\pi/2+ \epsilon} \xsim \g_{\pi/2-\epsilon}\qquad\forall\: 0< \epsilon < \pi/2
	\end{align*}	
	if $h\in H_{\tau_1}$ corresponds to a rotation by the angle $\pi$. 
	\item
	So, replacing the above family by $\{\g_{\alpha}\}_{\alpha\in (0,\pi/2]}$ does not change completeness, and we even have independence. In fact, if $\alpha,\beta \in (0,\pi/2]$ with $\g_\alpha \xsim \g_\beta$, then Lemma \ref{lemma:completee}.\ref{it:completee1} shows 
  \begin{align*}
 % \label{eq:fgfgfgf}
  	\gamma_{\g_\alpha}^x|_{[0,l]}\psim  \gamma_{\pm \Ad_h(\g_\beta)}^x|_{[0,l']}\qquad\text{for some}\qquad h\in G_x.
  \end{align*}
Hence, $\g_\alpha = \pm \Ad_h(\g_\beta)$ by the first point, just because $\im\big[\gamma_{\g_\alpha}^x\big]\cap\im\big[\gamma_{\pm \Ad_h(\g_\beta)}^x\big]$ is infinite. So, since $\Ad_h$ rotates in $\su\cong \RR^3$ around the $\tau_1$-axis, it is clear by construction 
that $h=e$ and $\g_\alpha=\g_\beta$ must hold.
  \end{itemize}
  \endgroup
  \noindent
  It remains to determine the type of $J(\g_\alpha,p_\m)$ for all $\alpha\in (0,\pi/2]$ and all $\m\in \Mm$. We will show that $J(\g_\alpha,p_\m)$ is of {\bf Type 4} if $\alpha\in (0,\pi/2)$, and of {\bf Type 3} if $\alpha=\pi/2$, so that
   \begin{align*}
    \IHOMLAS\cong \Big[\RB\wti S^1\Big]^{|\RR_{>0}|}\times \Big[\RB\wti S^2\Big]^{|(0,\pi/2)\times \RR_{>0}|}
  \end{align*}
  holds. In fact, let $x\in \vec{A}$ be as above. Then, since the non-trivial elements $h\in G_{x}=H_{\tau_1}$ rotate $\g_\alpha$ in $\su\cong \RR^3$ w.r.t.\ to an angle in $(0,2\pi)$ around the $\tau_1$-axis, it is clear that $\Ad_h(\g_\alpha)$ can only be equal to $\pm \g_\alpha$ if $h=e$ or
	\begin{align*}
	\alpha=\pi/2\qquad \text{and}\qquad h=\pm\exp(\textstyle\frac{\pi}{2}\tau_1)\qquad \text{with}\qquad\Ad_h(\g_{\pi/2})=-\g_{\pi/2},
\end{align*}	  	
  	whereby $h$ corresponds to a rotation by the angle $\pi$.
  Consequently, 
    \begingroup
  \setlength{\leftmargini}{25pt}
  \begin{itemize}
	\item
	We have $G_{[\g_\alpha]}^{x}=\{e\}$ if $\alpha\in (0,\pi/2)$, so that in this case $J(\g_\alpha,p_\m)$ is of {\bf Type 4} for all $\m\in \Mm$. 
	\item
	We have	$G_{[\g_{\pi/2}]}^x=\{e,\pm\exp(\textstyle\frac{\pi}{2}\tau_1)\}$. So, since the map $\fiba_{p_\m}\colon \SU \rightarrow \SU$ equals $\id_\SU$, for each $\m\in \Mm$ and $h=\pm\exp(\textstyle\frac{\pi}{2}\tau_1)$ the equivariance of a map $\Psi\colon \spann_\RR(\g_{\pi/2})\rightarrow \SU$ just reads (for $h=e$ equivariance gives no conditions on $\im[\Psi]$)
		\begin{align*}
		%\label{equivariancebbbb}
		\Psi\big(\lambda\cdot\g_{\pi/2}\big)^{-1}=\Psi(\lambda\cdot\Ad_h(\g_{\pi/2}))=\alpha_{\fiba_{p_\m}(h)}(\Psi(\lambda\cdot\g_{\pi/2}))=\alpha_h(\Psi(\lambda\cdot\g_{\pi/2}))\qquad\forall\: \lambda\in \RR.
		\end{align*}
		 So, by Lemma \ref{lemma:torus}.\ref{lemma:torus2} it should be clear that\footnote{The maps $\Psi_{\vec{n}}\colon \lambda \cdot \g_{\pi/2}\mapsto \exp(\lambda\cdot \murs(\vec{n}))$ for $\vec{n}\in \RR^3\backslash \{0\}$ with $\langle\vec{n},\vec{e}_1\rangle=0$ are all non-trivial and $\Ad_{G_{[\g_{\pi/2}]}}^{p_\m}$-equivariant, see e.g.\ Lemma \ref{lemma:torus}.\ref{lemma:torus2}.} %(cf.\ Lema \ref{lemma:torus}.\ref{lemma:torus})  
	\begin{align*}
	J(\g_{\pi/2},p_\m)=\{0\}\sqcup \textstyle\bigcup_{\vec{n}\in \RR^3\backslash\{0\} \colon\langle\vec{n},\vec{e}_1\rangle=0}\:[\hspace{1.5pt}\murs(\vec{n})\hspace{0.5pt}]
	\end{align*}		
	 is of {\bf Type 3} for all $\m\in \Mm$.		
  \end{itemize}
  \endgroup
  \noindent

  \vspace{8pt}
  \noindent
  {\bf\textit{Homogeneous Isotropic LQC:}}  
  \vspace{2pt}
  
  \noindent
 We now consider the first case in Example \ref{ex:LQC}. Since $\wm$ is transitive, $\Mm$ is singleton, and we choose $p=(0,\me)$ so that $x=0$ as well as $\{0\}\times \SU=G_x\subseteq \Gee$ holds. %Then, the stabilizer of $x$ is $\SU$ and
 Observe that for $(\vv,\s)\in \mg= \RR^3\times\su$ and $h=(0,\sigma)\in \{0\}\times \SU=G_x$, we have 
	\begin{align}
	\begin{split}
	\label{conjuuuuu}
		 \Ad_h((\vv,\s))&=\dttB{t}{0}\alpha_h((t \vv,\exp(t \s)))\\
		 & =\dttB{t}{0} (0,\sigma)\cdot_\varrho (t \vv,\exp(t \s)) \cdot_\varrho (0,\sigma)^{-1}\\
		 				&= \dttB{t}{0} (t \varrho(\sigma)(\vv),\sigma\cdot \exp(t\cdot \s))
		 \cdot_\varrho (0,\sigma^{-1})\\
		 				&=\dttB{t}{0} (t\varrho(\sigma)(\vv),\alpha_\sigma(\exp(t\cdot \s)))\\
						&=(\varrho(\sigma)(\vv),\Ad_\sigma(\s))
	\end{split}
	\end{align}	 
 for $\Ad_\sigma$ the differential of the conjugation by $\sigma$ in $\SU$. 
\vspace{5pt} 
 
 \noindent 
 We obtain an independent and complete family of elements of 
	\begin{align*} 
 	\mg\backslash \mg_x=\big[\RR^3\times \su\big]\backslash \big[\{0\}\times \su\big]=\big[\RR^3\backslash\{0\}\big] \times \su 
	\end{align*}
	as follows. 
  We fix $\vv,\vv_{\perp}\in \RR^3\backslash\{0\}$ orthogonal to each other and normalized, and define $\s_0:=\murs(\vv)$ as well as $\s_{\perp}:=\murs(\vv_\perp)$. Moreover, let
  \begin{align*} 
    \Span_\RR(\s_0,\s_\perp)\supseteq E_\geq&:=\{\lambda_1 \s_0 +\lambda_2 \s_\perp\: |\: \lambda_1 \in \RR,\lambda_2 \geq 0\}\\
    \Span_\RR(\s_0,\s_\perp)\supseteq E_>&:=\{\lambda_1 \s_0 +\lambda_2 \s_\perp\: |\: \lambda_1 \in \RR,\lambda_2 > 0\}\\
    \Span_\RR(\s_0,\s_\perp)\supseteq E_0\hspace{2pt}&:=E_> \sqcup \{0\} 
  \end{align*}
  denote the upper and the positive upper half plane determined by $\s_0$ and $\s_\perp$, as well as 
  $E_0$ the union of the positive upper half plane with the origin. It is shown in Appendix \ref{subsec:Aclacula} that the family $\{(\vv,\s)\}_{\s\in E_0}$ is   
  independent and complete (and consists of stable elements).
	\vspace{6pt}  
  
	\noindent  
	We now calculate $J((\vv,\s),p)$ for all $\s\in E_0$. For simplicity, here we will assume that $\vv=\vec{e}_1$ and $\vv_\perp=\vec{e}_2$ holds, hence $\s_0=\tau_1$ and $\s_\perp=\tau_2$.
	\begingroup
  	\setlength{\leftmargini}{25pt} 
	\begin{itemize}
	\item
		Let $\Psi\colon \spann_\RR((\vv,\s))\rightarrow \SU$ be $\Ad^p_{G_{[(\vv,\s)]}}$-equivariant and 
	\begin{align*}
		h=(0,\sigma)\in G^x_{[(\vv,\s)]}\subseteq G_x = \{0\}\times \SU.
	\end{align*}			
		 Then, since $\fiba_{p}\colon \Gee \rightarrow \SU$ is just given by $\pr_{\SU}$, equivariance of $\Psi$ reads 
		\begin{align}
		\label{equivariancebbbb}
		\Psi(\lambda\cdot(\varrho(\sigma)(\vv),\Ad_\sigma(\s)))\stackrel{\eqref{conjuuuuu}}{=}\Psi(\Ad_h(\lambda\cdot(\vv,\s)))
		%&=\alpha_{\phi_p(\sigma)}(\Psi(\lambda\cdot(\vv,\s)))
		=\alpha_\sigma(\Psi(\lambda\cdot(\vv,\s)))\qquad\forall\:\lambda\in \RR.
		\end{align}
	\item
		If $\s=0$, then $\{0\}\times H_{\vv}\subseteq G^x_{[(\vv,0)]}$ 
		 because  
	\begin{align*}
		\Ad_h((\vv,0))\stackrel{\eqref{conjuuuuu}}{=}(\varrho(\sigma)(\vv),0)=(\vv,0)\qquad \forall\: h=(0,\sigma)\in \{0\}\times H_{\vv}.  
	\end{align*}			
Combining this with \eqref{equivariancebbbb}, we obtain $\Psi((\vv,0))=\alpha_{\sigma}(\Psi(\vv,0))$ for all $\sigma\in H_{\vv}$,
		 hence $\im[\Psi]\subseteq H_{\vv}$  
		 by Lemma \ref{lemma:torus}.\ref{lemma:torus1}. This already shows that $J((\vv,0),p)$ is either of {\bf Type 1} or of {\bf Type 2}. 
		 
		 However, $J((\vv,0),p)=\{0,[\murs(\vv)]\}=\{0,[\s_0]\}$ is of {\bf Type 2} because $\Psi\colon \lambda\cdot( \vv,0)\mapsto \exp(\lambda\cdot \murs(\vv))$ is $\Ad^p_{G_{[(\vv,0)]}}$-equivariant and non-trivial. In fact, if $h=(0,\sigma)\in G^x_{[(\vv,0)]}$ is arbitrary, then 
	\begin{align*}
		(\varrho(\sigma)(\vv),0)\stackrel{\eqref{conjuuuuu}}{=}\Ad_h((\vv,0))=(\pm \vv,0)
	\end{align*}			 
		  by Remark and Definition \ref{rem:ppropercurve}.\ref{rem:ppropercurve4}, so that $\varrho(\sigma)(\vv)=\pm \vv$ holds.  Consequently, for each $\lambda \in \RR$ we have
		\begin{align*}
			\Psi(\Ad_h(\lambda\cdot( \vv,0)))&=\Psi((\pm\lambda\cdot \vv,0))=\exp(\lambda\cdot\murs(\varrho(\sigma)(\vv)))\\
			&=\exp(\lambda\cdot\Ad_\sigma(\murs(\vv)))=\alpha_\sigma(\Psi(\lambda\cdot (\vv,0)))=\alpha_{\fiba_p(h)}(\Psi(\lambda\cdot (\vv,0))).
		\end{align*}		
	\item
		If $\s\neq 0$, then $G^x_{[(\vv,\s)]}=\{e,\pm \exp(\frac{\pi}{2} \tau_3)\}$. In fact, by Remark and Definition \ref{rem:ppropercurve}.\ref{rem:ppropercurve4} for 
	\begin{align*}
		h=(0,\sigma)\in G^x_{[(\vv,\s)]}\subseteq G_x=\{0\}\times \SU
	\end{align*}			
	we have $\Ad_h((\vv,\s))=\pm (\vv,\s)$, whereby the positive case, i.e.,  $\varrho(\sigma)(\vv)=\vv$ 
	can only occur if $\sigma=e$. This is because here we must have $\sigma\in H_{\vv}=H_{\tau_1}$, so that $\Ad_\sigma$ rotates in $\su\cong \RR^3$ around the $\tau_1$-axis. By the definition of the set $E_>$, then it is clear that $\Ad_\sigma(\s)=\s$  can only hold for $\sigma=e$. Now, if $\Ad_h((\vv,\s))=-(\vv,\s)$, then 
	$\alpha_\sigma(\exp(\lambda\cdot\s))=\exp(\lambda\cdot\Ad_\sigma(\s))=\exp(\lambda\cdot\s)^{-1}$ for all $\lambda\in \RR$, 
so that Lemma \ref{lemma:torus}.\ref{lemma:torus2} already shows that $\sigma=\pm \exp(\frac{\pi}{2} \tau_3)$ holds. This is also in line with $\varrho(\sigma)(\vv)=-\vv$, and as 
	  in the previous point it is now clear that 
	\begin{align*}
	J((\vv,\s),p)=\{0\}\sqcup \textstyle\bigcup_{\vec{n}\in \RR^3\backslash\{0\} \colon\langle\vec{n},\vec{e}_3\rangle=0}\:[\hspace{1.5pt}\murs(\vec{n})\hspace{0.5pt}]
	\end{align*}		
	 is of {\bf Type 3}.		
	\end{itemize}	
	\endgroup 
	\noindent
  It follows that 
  \begin{align}
    \label{eq:wichgle}
    \IHOMLAS\cong \RB \times  \Big[\RB \wti S^1\Big]^{|E_>|}
  \end{align}
  Here the first factor $\RB$  corresponds to $(\vv,0)\in \RR^3\times \su$ and determines the image of an invariant homomorphism on the set of linear curves $\Pal$. For this, recall that  
  each such curve is equivalent to $\wm_g\cp\gamma^x_{(\vv,0)}|_{[0,l]}$ for some $l>0$ and $g\in G$. Then,
  \begingroup
	\setlength{\leftmargini}{13pt}
	\begin{itemize}
	\item[$\triangleright$]
  Performing the constructions of this subsection for the set $\Pal$ instead of $\Pags$ provides us with the homeomorphism 
	\begin{align*}
		\eta_\lin \colon \IHOMLL\rightarrow  \RB
	\end{align*}	
	 for which $\eta_\lin(\mu_\lin)=\muB$ holds with $\mu_\lin$ the respective Radon measure on $\IHOMLL$. In particular, using Remark \ref{rem:unednlichfubini}, it is not hard to see that then $\mu_\lin$ is the push forward of $\mLAS$ by the restriction map (see Remark \ref{rem:restriction})
	\begin{align*}
			 \res_{\mg\lin}\colon \IHOMLAS\rightarrow \IHOMLL,\quad
			 	\homm \mapsto \homm|_{\Pal}.
	\end{align*}
	This has also consequences for standard homogeneous isotropic LQC being discussed in Remark \ref{StanLQC}. 
  \hspace*{\fill}$\lozenge$		
  \end{itemize}
  \endgroup
  \endgroup
  \par 
\end{example}   
\begin{remark}[Standard Homogeneous Isotropic LQC] 
  \label{StanLQC}
  \begingroup
  \setlength{\leftmargini}{20pt}
  \begin{enumerate}
  \item
  \label{StanLQC1}
    The standard quantum configuration space of homogeneous isotropic LQC \cite{MathStrucLQG} is given by $\ARRQLL$, i.e., the spectrum of the restriction $\Cstar$-algebra $\ovl{\PaC_\lin|_{\AR}}$. Here, $\PaC_\lin$ denotes the $\Cstar$-algebra of cylindrical functions that correspond to the set $\Pal$ of linear curves in $\RR^3$. By \eqref{eq:homis} we have $\AR \cong \RR$, and by Lemma \ref{lemma:separating} the map $\iota_\Con\colon \Con \rightarrow \A_\lin$ is injective because the functions in $\PaC_\lin$ separate the points in $\Con$. The latter statement means that $\A_\lin$ is a reasonable quantum configuration space because $\Con$ is naturally  embedded therein. Now, $\Pal$ is obviously $\Pe$-invariant, so that by Corollary \ref{cor:CylSpecAction} the quantum-reduced space $\AQRL\subseteq \A_\lin$ exists, is homeomorphic to $\IHOMLL$, and is physically meaningful as well. 
    
    Now, we have $\ovl{\PaC_\lin|_{\AR}}=\CAP(\RR)$, hence $\ARRQLL=\RB$. 
    In fact, for $\g:=(\vv,0)\in \RR^3 \times \su$ we have $\exp(t\cdot \g)=(t\cdot \vv,\me)$, so that by \eqref{eq:trivpar} the horizontal lift of $\gamma_l:=\gamma_\g^x|_{[0,l]}$ in $p=(x,e)$ w.r.t.\ $\w^c$ is just given by 
    \begin{align}
      \label{eq:horliftlin}
      \begin{split}
        \wt{\gamma}_l(t)&=\Pe((t\cdot\vv,\me),(x,\me))\cdot \exp(- t\cdot \w^c(\:\wt{g}((x,\me))))\\
        &=(x+t\cdot \vv,\me)\cdot \exp(-c t \cdot\murs(\vv))=(x+t\cdot\vv, \exp(-c t \cdot\murs(\vv))).
      \end{split}
    \end{align}
    Consequently, for $\nu$ the standard choice $\nu_x=(x,\me)$ for all $x\in M$, we have
    \begin{align}
      \label{eq:patralinc}
      h_\gamma^\nu(\w^c)= \exp(-c l \cdot\murs(\vv))\stackrel{\eqref{eq:expSU2}}{=} \cos(cl\|\vv\|)\cdot\me - \sin(cl\|\vv\|) \cdot\murs(\vv/\|\vv\|),
    \end{align}
    so that by \eqref{eq:expSU2} the $\Cstar$-algebra $\PaC_\lin$ is generated by the constant function $1$  and the functions $t\mapsto \sin(l t)$, $t\mapsto \cos(l t)$ for all $l\neq 0$. Hence, by the characters $\chi_l$ for $l\in \RR$. Consequently, $\PaC_\lin=\CAP(\RR)$ and $\ARRQLL=\Spec(\CAP(\RR))=\RB$.
  \item
  \label{StanLQC2}
    As we have seen in the end of Example \ref{ex:cosmoliealgmaasse}, 
	$\AQRL\cong \IHOMLL$ is homeomorphic to $\RB$ as well.     
 	A straightforward calculation now shows that the composition 
 	 	\begin{align*}
 		\eta_\lin \cp \kappa_\lin \cp \ovl{i^*_\AR}\colon \RB\cong\ARRQLL\rightarrow \RB
    \end{align*}
 is even the identity $\id_{\RB}$ on $\RB$. 
 	\begin{figure}[h]
  \begin{minipage}[h]{\textwidth}
    \begin{center}
      \makebox[0pt]{
        \begin{xy}
          \xymatrix{
           \RB \cong \ARRQLL \ar@{->}[r]^-{\ovl{i^*_\AR}}_-{\cong}   &  \: \ARQLL \: \ar@{->}[r]^-{\subseteq} & \AQRL\ar@{->}[r]_-{\cong}^-{\kappa_\lin}   &\IHOMLL \ar@{->}[r]_-{\cong}^-{\eta_\lin}& \RB.
         }
        \end{xy}}
    \end{center}
  \end{minipage} 
\end{figure}
   \FloatBarrier
    Here, 
    $\ovl{i^*_\AR}$ 
    is the identification from \eqref{eq:inclusionsdiag} and $\kappa_\lin$  
    the respective homeomorphism from Subsection \ref{subsec:InvHoms}. 
    In particular, this means that $\ARRQLL\cong \AQRL$ holds, i.e.,  that in the homogeneous isotropic case quantization and reduction commutes if one restricts to linear curves.  
    Since our construction provide us with the Haar measure $\muB$ on $\RB$, being  used for the definition of the standard LQC kinematical Hilbert space $L^2(\RB,\muB)$ \cite{MathStrucLQG}, we have reproduced this Hilbert space by performing a reduction on quantum level.
    
    Finally, observe that we now can easily embed the traditional LQC configuration space 
	\begin{align*}    
    \ARRQLL\cong \ARQLL\cong \IHOMLL \stackrel{(*)}{\cong} \ITRHOML
    \end{align*}
     into the quantum-reduced space $\IHOMW\stackrel{(**)}{\cong} \ITRHOMW$ via the map
    \begin{align*}
    \iota_\lin\colon \ITRHOML&\rightarrow \ITRHOMW,
	\end{align*}    
	defined by 
	\begin{align*}
         \iota_\lin(\hommm)(\gamma):=\homm(\gamma)\:\text{ if }\:\gamma\in \Pal\qquad\text{and}\qquad\iota_\lin(\hommm)(\gamma):=\me\:\text{ for }\:\gamma\in \Paw\backslash \Pal.
	\end{align*}
	If we use the standard choice $\nu_x=(x,\me)$ for all $x\in \RR^3$ for the identifications $(*)$ and $(**)$, on the level of the spaces $\IHOMLL$ and $\IHOMW$ this just mean to assign to $\homm\in \ITRHOML$ the element $\homm'\in \IHOMW$ with $\homm'(\gamma)=\homm(\gamma)$ for $\in \Pal$ and 
\begin{align*}	
	\homm'(\gamma)((\gamma(a),s)):=(\gamma(b),s)\qquad \forall\:s\in \SU\quad\text{and}\quad \Paw\backslash \Pal\ni \gamma\colon [a,b]\rightarrow \RR^3.
\end{align*}
	So, following this approach, there are many ways to embed $\ARRQLL$ into $\IHOMW$, whereby (at least from the mathematical point of view) using the standard choice of $\nu$ seems to be the most natural one. Its physical relevance, however, might first become clear once the dynamics of the quantum reduced theory has been successfully established.	
  \item
    Using the identification of $\IHOMLL$ with $\ITRHOML$ from Remark \ref{rem:homomorphbed}.\ref{rem:euklrem5} via the standard choice $\nu_x=(x,e)$ for all $x\in \RR^3$, one finds that the continuous group structure on $\RB\cong \ITRHOML$ corresponds to the group structure on $\ITRHOML$ defined by
    \begin{align*}
      (\hommm_1*\hommm_2)(\gamma):=\hommm_1(\gamma) \hommm_2(\gamma)\qquad\qquad \hommm^{-1}(\gamma):=\hommm(\gamma)^{-1}\qquad\qquad  e(\gamma):=\me
    \end{align*} 
    for all $\gamma\in \Pal$. 
    These operations are well defined because for each fixed $\vv\in \RR^3\backslash\{0\}$ we have\footnote{For the definition of the curves $\gamma_{\vv,l}$ See (a) in Convention \ref{conv:sutwo1}.\ref{conv:sutwo2}.} $\hommm(x+\gamma_{\vv,l})\in H_{\vv}$ for all $\hommm \in \ITRHOML$, i.e., $[\hommm_1(\gamma),\hommm_2(\gamma)]=0$ for all $\gamma\in \Pal$ and all $\hommm_1,\hommm_2 \in\ITRHOML$. This is clear from Lemma \ref{lemma:torus}.\ref{lemma:torus1} because 
    \begin{align*}
      \hommm(x+\gamma_{\vv,l})=\hommm(\gamma_{\vv,l})=\hommm(\sigma(\gamma_{\vv,l}))=\alpha_\sigma(\hommm(\gamma_{\vv,l}))=\alpha_\sigma(\hommm(x+\gamma_{\vv,l}))\qquad \forall\: \sigma\in H_{\vv}
    \end{align*}
    by \eqref{eq:algrels}. However, $[\hommm_1(\gamma),\hommm_2(\gamma)]=0$ usually\footnote{Choose, e.g., $\gamma=\gamma_\g^0$ for $\g= (\vv,\s)$ with $\s\notin \Span_\RR(\murs(\vv))$ and use that $J(\g,(0,\me))$ is of {\bf Type 3} in this case, see third part of Example \ref{ex:cosmoliealgmaasse}.} does not hold for all $\hommm_1,\hommm_2\in \ITRHOMW$ if $\gamma\notin \Pal$, so that one cannot define the same group structure, e.g., on $\ITRHOMW$.
  \end{enumerate}
  \endgroup
\end{remark}

\subsubsection{Independence from the Choices}
In this final subsection, we show that the definition of the measure $\mLAS$ does not depend on any choices we have made. We start with simplifying the notations.

We consider the index set $I:=\{[\m,\alpha]\:|\: \m\in \Mm, \alpha\in I_\m\}$ and define for each $\iota=[\m,\alpha]\in I$:\footnote{Recall Lemma and Definition \ref{def:topo} as well as Lemma and Definition \ref{lemdef:projspaces}.} 
\begin{align*}
	\g_\iota:=\g_{\m,\alpha}\qquad\qquad \Eq_\iota:=\Eq_{p_\m}\qquad\qquad Y_\iota:=Y_{\g_{\m,\alpha}}^{p_\m}\qquad\qquad  X_\iota:=X_{\g_{\m,\alpha}}^{p_\m},
\end{align*}
the measure $\mu_\iota:=\mu_{\g_{\m,\alpha}}^{p_\m}$, 
as well as the maps
\begin{align*}
\pip_\iota&:=\pip_{p_\m}\colon \IHOMLAS\rightarrow \Eq_\iota\\
\res_\iota  &:=\res_{\g_{\m,\alpha}}^{p_\m}\colon \Eq_\iota\rightarrow Y_\iota\\
\tau_\iota &:=\tau_{\g_{\m,\alpha}}^{p_\m}\colon Y_\iota\rightarrow X_\iota\\
\hspace{-90pt}\text{(if $Y_\iota$ of respective type)}\hspace{40pt} \xi_\iota&:=\xi^{p_\m}_{\g_{\m,\alpha}}\colon J(\g_{\m,\alpha},p_\m) \rightarrow S^i \quad \text{for}\quad i\in \{1,2\}\\
 %\quad \text{if}\quad J(\g_{\m,\alpha},p_\m)\quad\text{is of the respective type}\\
 \eta_\iota&:= \tau_\iota\cp\res_\iota\cp\pip_\iota\colon \IHOMLAS\rightarrow X_\iota
 %\\
 %\eta_X&:=\prod_{\iota\in I}\eta_\iota\colon \IHOMLAS\rightarrow X=%\prod_{\iota\in I}X_\iota.
\end{align*}
and $\eta=\prod_{\iota\in I}\eta_\iota\colon \IHOMLAS\rightarrow X=\prod_{\iota\in I}X_\iota$. 
Moreover, as in Lemma and Definition \ref{def:ProductMa}.\ref{def:ProductMa3}: 
\begingroup
\setlength{\leftmargini}{17pt}
\begin{itemize}
\item
	Let $\J$ denote the set of all finite tuples $J=(\iota_1,\dots,\iota_k)$ of mutually different elements of $I$.
\item
	 Define $X_J:=X_{\iota_1}\times {\dots} \times X_{\iota_k}$, $\mu_J:=\mu_{\iota_1}\times {\dots} \times \mu_{\iota_k}$, and the respective projection by
    \begin{align*} 
    	\pi_J\colon X\rightarrow X_J,\:\:
    	\textstyle\prod_{\iota\in I}x_\iota \mapsto (x_ {\iota_1},\dots,x_ {\iota_k}).
	\end{align*} 
\item
	  Write $J\leq J'$ for $J,J'\in \JJ$ with $J=(\iota_1,\dots,\iota_k)$ and $J'=(\iota'_1,\dots,\iota'_{k'})$ iff there exists an injection $\sigma\colon \{\iota_1,\dots,\iota_k\}\rightarrow \{\iota'_1,\dots,\iota'_{k'}\}$, and define the maps 
     \begin{align*}   	
    \pi^{J'}_J\colon X_{J'}\rightarrow X_J,\:\:\big(x_{\iota'_1},\dots,x_{\iota'_{k'}}\big)\mapsto \big(x_{\sigma(\iota_1)},\dots,x_{\sigma(\iota_{k})}\big). 
     \end{align*}   	
\item	
 Let $\mu_I$ denote the corresponding normalized Radon measure from Lemma \ref{lemma:normRM} on $X$.
\end{itemize}
\endgroup
\noindent 
Finally, let $\widehat{\pi}_J:=\pi_J\cp\eta$ for $J\in \JJ$. Then, 
the measure $\mLAS=\eta^{-1}(\mu_I)$ is uniquely determined by the property that 
$\widehat{\pi}_J(\mLAS)=\mu_J$ holds for all $J\in \JJ$. 
\vspace{10pt}

\noindent
We now choose a further selection $\{p'_\m\}_{\m\in \Mm}\subseteq P$ of elements with $x'_\m:=\pi(p'_\m)\in \m\in\Mm$, as well as $\{\g'_{\m,\alpha}\}_{\alpha\in I'_\m}\subseteq \mg\backslash \mg_{x'_\m}$ a respective independent and complete family for each $\m\in \Mm$. Moreover, for each plane $[\mm]$ in $\RR^3$ through $0$ we choose a further map $\LSM$ as in Lemma and Definition \ref{lemdef:projspaces}.\ref{lemdef:projspaces3}. 
Then, 
\begingroup
\setlength{\leftmargini}{17pt}
\begin{itemize}
\item
	For each $\m\in \Mm$ we fix $g_\m\in G$ with $x'_\m=\wm(g_\m,x_\m)$.
\item
	By Lemma \ref{lemma:completee}.\ref{lemma:completee1},
 we can assume that $I'_\m=I_\m$ as well as $[\g'_{\m,\alpha}]=[\Ad_{g_{\m}}(\g_{\m,\alpha})]$ 
% $g_{m,\alpha}\in G$ with $x'_\m=\wm(g_{\alpha,\m},x_{\m'})$ 
 holds for all $\m\in \Mm$ and all $\alpha\in I_\m$. We define $g_\iota:=g_{\m}$ for $\iota={[\m,\alpha]}$ and denote by $s_\iota$ the unique element $s\in \SU$ for which $p'_\m=g_\iota\cdot p_\m\cdot s$ holds.
 \item
 	Then, for $I'$ and $\g'_\iota$ defined as $I$ and $\g_\iota$ above, we can assume that $I'=I$ as well as $[\g'_\iota]=[\Ad_{g_\iota}(\g_\iota)]$ holds 
 	for all $\iota\in I$.
 	\item
 	We define the corresponding spaces $\Eq'_\iota, Y_\iota', X'_\iota, X', X'_J$ and maps $\pip'_\iota,\res'_\iota, \tau'_\iota,\xi'_\iota,\eta_\iota',\eta',\widehat{\pi}'_J$ exactly as above, and denote the respective measures by  $\mu'_\iota, \mu'_J$ and $\mLAS'$.
\end{itemize}
\endgroup 
\noindent
Then, in order to show $\mLAS=\mLAS'$, it suffices to verify that $\widehat{\pi}'_J(\mLAS)=\mu'_J$ holds for all $J\in \JJ$.\footnote{This  is immediate from the fact that by definition of $\mLAS'$, $\eta'^{-1}(\mLAS')$ is the Radon product measure on $X'$ w.r.t.\ the Radon measures $\mu'_\iota$ on $X'_\iota$.} This, however, follows if we prove that (this is done in Proposition \ref{prop:invmasss}.\ref{prop:invmasss2})
    \begin{align}
    \label{eq:dfffd}
      \mu'_{J}(\widehat{\pi}'_{J}(E))=\mu_J(\widehat{\pi}_J(E))\quad\text{for}\quad\widehat{\pi}'_{J}(E) \quad\text{measurable},
    \end{align}
    and that for $\homm,\homm'\in \IHOMLAS$, $\iota\in I$ the implication (done in Proposition \ref{prop:invmasss}.\ref{prop:invmasss1})
     \begin{align}
    \label{eq:dfffdd}
      \eta_\iota(\homm)=\eta_\iota(\homm')\qquad\Longrightarrow \qquad\eta'_\iota(\homm)=\eta'_\iota(\homm')%\qquad\forall\:\iota\in I
    \end{align}
    holds. 
In fact, then for $E:=\widehat{\pi}'^{-1}_{J}(A')$ with $A'\in \Borel(X'_J)$ we have 
    \begin{align}
    \label{eq:sdfdfhd4355}	
      \mu'_{J}(A')&=\mu'_{J}\big(\widehat{\pi}'_{J}(E)\big)\stackrel{\eqref{eq:dfffd}}{=}\mu_J\big(\widehat{\pi}_{J}(E)\big)
      =\widehat{\pi}_{J}(\mLAS)\big(\widehat{\pi}_J(E)\big)\\
      &=\mLAS\big(\widehat{\pi}_{J}^{-1}\big(\widehat{\pi}_J(E)\big)\big)=\mLAS(E)=\widehat{\pi}'_{J}(\mLAS)(A'),
    \end{align}
   	where in the fifth step we have used that $\widehat{\pi}_{J}^{-1}\big(\widehat{\pi}_J(E)\big)=E$ holds. Here, $E\subseteq\widehat{\pi}_{J}^{-1}\big(\widehat{\pi}_J(E)\big)$ is clear, and obviously we have $\homm\in \widehat{\pi}_{J}^{-1}\big(\widehat{\pi}_J(E)\big)$ iff $\widehat{\pi}_J(\homm)\in \widehat{\pi}_J(E)$. So, if we can show that the latter condition implies $\homm\in E$, then $\widehat{\pi}_{J}^{-1}\big(\widehat{\pi}_J(E)\big)= E$ follows. Now, $\widehat{\pi}_J(\homm)\in \widehat{\pi}_J(E)$ means that we find $\homm'\in E$ with  	
   	 $\kla_{\iota_i}(\homm)=\kla_{\iota_i}(\homm')$ for all $1\leq i\leq k$, hence 
	\begin{align} 
	\label{eq:szbvvuzdefg}  	 
   	 \kla'_{\iota_i}(\homm)=\kla'_{\iota_i}(\homm')\qquad\text{ for all }\:\qquad 1\leq i\leq k
   	 \end{align} 
   	 by \eqref{eq:dfffdd}. Consequently, 
    \begin{align*}	
      \widehat{\pi}'_{J}(\homm)&\hspace{3.2pt}=(\pi'_{J}\cp \kla')(\homm)=\big(\kla'_{\iota_1}(\homm),\dots,\kla'_{\iota_k}(\homm)\big)\\
      &\stackrel{\eqref{eq:szbvvuzdefg}}{=}\big(\kla'_{\iota_1}(\homm'),\dots,\kla'_{\iota_k}(\homm')\big)=\widehat{\pi}'_{J}(\homm')\in \widehat{\pi}'_{J}(E)= A',
    \end{align*}
    hence $\homm\in E$. This shows \eqref{eq:sdfdfhd4355}, so that $\widehat{\pi}'_J(\mLAS)=\mu'_J$ holds for all $J\in \JJ$, hence $\mLAS=\mLAS'$.
    
    Thus, the following proposition establishes independence of $\mLAS$ from any choices we have made.
\begin{proposition}
  \label{prop:invmasss}
  \begin{enumerate}
  \item
    \label{prop:invmasss1}
    $X_\iota$ and $X'_\iota$ are of the same type for all $\iota\in I$. Moreover, for each such $\iota$ we find a homeomorphism $\Omega\colon \RB\rightarrow \RB$ with $\Omega^{-1}(\muB)=\muB$, as well as $R_2\in O(2)$ or $R_3\in \SOD$ such that
    \begin{align}
      \label{eq:beh}
      \:\kla'_\iota(\homm)=
      \begin{cases} 
        \kla_\iota(\homm)=0_{\mathrm{Bohr}} 		&\mbox{if } X_\iota \text{ is of } \:{\bf Type 1},\\
        (\Omega \cp \kla_\iota)(\homm)       &\mbox{if } X_\iota \text{ is of } \:{\bf Type 2},\\
        \big[\big(\Omega(\psi), R_2(v)\big)\big] 	&\mbox{if } X_\iota \text{ is of } \:{\bf Type 3} \text{ and } \kla_\iota(\homm)= [(\psi, v)], \\
        \big[\big(\Omega(\psi), R_3(v)\big)\big] 	&\mbox{if } X_\iota \text{ is of } \:{\bf Type 4} \text{ and } \kla_\iota(\homm)= [(\psi, v)].
      \end{cases} 
    \end{align}
    In the {\bf Type 3} and {\bf Type 4} case  $\Omega$ is an unital homomorphism, so that the respective expressions are well defined. 
   In particular, $\eta_\iota(\homm)=\eta_\iota(\homm')$ for $\homm,\homm' \in\IHOMLAS$ implies $\eta'_\iota(\homm)=\eta'_\iota(\homm')$, showing \eqref{eq:dfffdd}.
  \item
    \label{prop:invmasss2}
	Condition \eqref{eq:dfffd} holds, i.e., we have
	\begin{align*}
	% \label{eq:dfffd}
      \mu_J(\widehat{\pi}_J(E))=\mu'_{J}(\widehat{\pi}'_{J}(E)) 
    \end{align*}
    if $\widehat{\pi}'_{J}(E)$ or $\widehat{\pi}_{J}(E)$ is measurable for $E\subseteq \IHOMLAS$. 
  \end{enumerate}
\end{proposition}
\begin{proof}
\begin{enumerate}
\item[2.)]
	This is straightforward from Riesz-Markov theorem and Fubini's formula  \eqref{eq:wellfuncs} if we can show that the statement holds for $J=\iota$, i.e., that we have $\mu_\iota(\eta_\iota(E))=\mu'_{\iota}(\eta'_\iota(E))$ if $\eta_\iota(E)$ or $\eta'_\iota(E)$ is measurable for $E\subseteq \IHOMLAS$. For this, observe that by Part \ref{prop:invmasss1}) $\eta_\iota(E)$ is measurable iff $\eta'_\iota(E)$ is so. 
	
	Now, using \eqref{eq:beh}, the statement is clear if $X_\iota$ is of \textbf{Type 1} or of \textbf{Type 2}. 	
	In the \textbf{Type 3} case we calculate 
      \begin{align*}
        \mu'_\iota(\kla'_\iota(E))&=\pr_0(\mu_{1\times})\big(\kla'_\iota(E)\big)=(\muB\times \mu_{1})\left(\pr_0^{-1}\big(\kla'_\iota(E)\right)\big)\\
        &\hspace{-3.5pt}\stackrel{\eqref{eq:beh}}{=}(\muB\times \mu_{1})\left((\Omega\times R_2) \cp \pr_0^{-1}\left(\eta_\iota(E)\right)\right)\\
        &=(\Omega^{-1}\times R_2^{-1})(\muB\times \mu_{1})\left(\pr_0^{-1}\left(\eta_\iota(E)\right)\right)\\
        &=\pr_0\!\left((\Omega^{-1}\times R_2^{-1})(\mu_{1\times})\right)\left(\eta_\iota(E)\right),
      \end{align*}
      and similarly we obtain $\mu'_\iota(\kla'_\iota(E))=\pr_0\!\left((\Omega^{-1}\times R_3^{-1})(\mu_{2\times})\right)\left(\eta_\iota(E)\right)$ in the \textbf{Type 4} case. Since $\Omega^{-1}(\muB)=\muB$, $R_2^{-1}(\mu_1)=\mu_1$ and $R_3^{-1}(\mu_2)=\mu_2$,  the Riesz-Markov theorem and Fubini's formula show that
	\begin{align*}
	 \big(\Omega^{-1}\times R_2^{-1}\big)(\mu_{1\times})=\mu_{1\times}\qquad\quad\text{and}\qquad\quad\big(\Omega^{-1}\times R_3^{-1}\big)(\mu_{2\times})=\mu_{2\times},
	\end{align*}	      
       respectively, hence the claim. 
\item[1.)]
     We proceed in two steps. 
      \vspace{4pt}
      
      \noindent
      {\bf Step 1:}
      We first assume that $\g'_\iota=\Ad_{g_\iota}(\g_\iota)$, where we have  
      \begin{align}
        \label{eq:transf}
        \begin{split}
          \pip'_\iota(\homm)(\g'_\iota,l)&=\pi_{p'_\iota}(\g'_\iota,l,\homm)=\pi_{g_\iota\cdot p_\iota\cdot s_\iota}(\Ad_{g_\iota}(\g_\iota),l,\homm)\\
          &\!\!\stackrel{\eqref{eq:verkn}}{=}\alpha_{s_\iota^{-1}}\cp \pi_{p_\iota}(\g_\iota,l,\homm)=\alpha_{s_\iota^{-1}}\cp \pip_\iota(\homm)(\g_\iota,l).       
  	\end{split}
      \end{align}
      Consequently, \eqref{eq:beh} is clear if $X_\iota$ is of \textbf{Type 1}, as then $\pip_\iota(\homm)(\g_\iota,\cdot)=\me$ holds for all $\homm\in \IHOMLAS$ so that the same is true for $\pip'_\iota(\g'_\iota,\cdot)$. 
      
      For the other types, let $\pip_\iota(\homm)(\g_\iota,\cdot)\neq \me$ for $\homm\in  \IHOMLAS$ and write 
      \begin{align}  
        \label{eq:phiB}
  	\pip_\iota(\homm)(\g_\iota,l)=\exp\big(\phi(l)\cdot\s_{\beta(\homm)}\big)\qquad\forall\:l>0      
  	\end{align} 
      for $\phi\in \Per$, and $\beta(\homm):=\beta(\Psi)$ the element from  
      Definition \ref{def:betadef} that correspond to $\Psi:=\res_\iota(\pip_\iota(\homm))\in Y_\iota$. Then, \eqref{eq:transf} shows that 
      \begin{align*}
      	\pip'_\iota(\homm)(\g'_\iota,l)=\exp\big(\phi(l)\cdot \Ad_{s^{-1}_\iota}(\s_{\beta(\homm)})\big)\qquad\forall\:l>0.
      \end{align*}
      Now, since the definition of $\tau'_\iota$ does not depend on the choice of $s_\beta\in \beta$ which we have made in Convention \ref{conv:sammel} (see also Lemma and Definition \ref{lemdef:projspaces}.\ref{lemdef:projspaces4}), we can assume that $\tau'_\iota$ is defined by using the elements $\Ad_{s^{-1}_\iota}(\s_{\beta})$ for $\beta\in \Sp\ms$. Then $\Omega=\id_{\RB}$,  
	 and since $\murs^{-1}\cp \Ad_{s^{-1}}\cp\: \murs$ is a rotation in $\RR^3$, it is clear that $X_\iota$ and $X'_\iota$ are of the same type. Then, the claim is obvious in the \textbf{Type 2} case, as there $\beta(\homm)$ is independent of $\homm\in \IHOMLAS$. In the  \textbf{Type 4} case, let $v:=\xi_\iota(\beta(\homm))=\murs^{-1}(\s_{\beta(\homm)})$. Then \eqref{eq:beh} is clear from
      \begin{align*}
  	v':&= 
        \xi'_\iota([\Ad_{s_\iota^{-1}}(\s_{\beta(\homm)})])=\murs^{-1}( \Ad_{s_\iota^{-1}}(\s_{\beta(\homm)}))\\
  	&=\big(\murs^{-1}\cp \Add{s_\iota^{-1}} \cp \:\murs \big)(\murs^{-1}(\s_{\beta(\homm)}))=R_3(v).
      \end{align*}
      Finally, in the \textbf{Type 3} case, let $\mm,\mm'\in \RR^3\backslash\{0\}$ be vectors as in \eqref{eq:hmnejhedrffds} which  correspond to $X_\iota$ and $X'_\iota$, respectively. Then, $\vec{m}'=\lambda\cdot (\murs^{-1}\cp \Add{s_\iota}\cp\: \murs)(\vec{m})=R_3(\vec{m})$ for some $\lambda\neq 0$, so that for $v:=\xi_\iota(\beta(\homm))=\LM\big(\murs^{-1}(\s_{\beta(\homm)})\big)$  we have
      \begin{align*}
  	v':&=\xi'_\iota([\Ad_{s_\iota^{-1}}(\s_{\beta(\homm)})])=\LSMS\big(\murs^{-1}(\Ad_{s_\iota^{-1}}(\s_{\beta(\homm)}))\big)
    	=\big(\LSMS\cp R_3\big)\big(\murs^{-1}(\s_{\beta(\homm)})\big)\\
        &=\big(\underbrace{\LSMS\cp \LMS^{-1}}_{\in O(2)}\cp \underbrace{\LMS\cp R_3\cp \LM^{-1}}_{\in O(2)}\big)\big(\LM\big(\murs^{-1}(\s_{\beta(\homm)})\big)\big)
        =R_2\big(v\big)
      \end{align*}	
      for $R_3:=\murs^{-1}\cp \Add{s_\iota^{-1}} \cp \:\murs\in O(3)$ and $R_2=\LSMS\cp R_3\cp \LM^{-1}\in O(2)$.
      \vspace{2pt}
      
      \noindent
      {\bf Step 2:}
      To complete the proof, we have to treat the situation where $p'_\m=p_\m$, $g_\iota=e$, $s_\iota=e$ and $x=y$ holds, in full generality. This means that we have to consider the case where we have  
      %\begin{align*}    
        $[\g'_\iota]=[\g_\iota]$,
      %\end{align*}
      i.e., $\g'_\iota \xsim \g_\iota$ but not necessarily $\g'_\iota = \g_\iota$.  
      Now, by Lemma \ref{lemma:completee}.\ref{it:completee1} we find $h\in G_x$ such that we have  
       \begin{align*}
      \gamma^x_{\g'_\iota}= \wm_{h}\cp \gamma^x_{\pm \g_\iota}\cp \adif
    \end{align*}
    for $\adif\colon \RR\rightarrow \RR$ an analytic diffeomorphism with $\adif(0)=0$ and $\adif(\tau_{\g'_\iota})=\tau_{\g_\iota}$. We define 
	 \begin{align*}    
    	s_\alpha(l):=\Delta\big(\Phi_{h^{-1}}\cp\Phi_{\exp(l\cdot \g'_\iota)}\big(\Phi_{h}(p_\m)\big),\Phi_{\exp(\pm \adif(l)\cdot\g_\iota)}\big(p_\m\big)\big)
	\end{align*}    
	as well as $s_0:=\fiba_{p_\m}(h)$, and obtain for $\homm\in \IHOMLAS$ 
      \begin{align}
        \label{eq:smult}
        \begin{split}
          \pip'_{\iota}(\homm)&(\g'_\iota,l)=\Delta\!\left(\Phi_{\exp(l\cdot \g'_\iota)}\big(p_\m\big), \homm\left(\gamma^x_{\g'_\iota}|_{[0,l]}\right)\left(p_\m\right)\right)\\
          &=\Delta\!\left(\Phi_{\exp(l\cdot \g'_\iota)}\big(p_\m\big), \homm\left(\wm_{h}\cp\gamma^x_{\wg}|_{[0,\adif(l)]}\right)\left(p_\m\right)\right)\\
          &=\Delta\!\left(\Phi_{\exp(l\cdot \g'_\iota)}\big(p_\m\big), \Phi_{h}\cp \homm\left(\gamma^x_{\wg}|_{[0,\adif(l)]}\right)\big(\Phi_{{h}^{-1}}(p_\m)\big)\right)\\
          &=\Delta\!\left(\Phi_{{h}^{-1}}\cp\Phi_{\exp(l\cdot \g'_\iota)}\big(p_\m\big),  \homm\left(\gamma^x_{\wg}|_{[0,\adif(l)]}\right)\big(\Phi_{{h}^{-1}}(p_\m)\big)\right)\\
          &=\Delta\!\left(\Phi_{{h}^{-1}}\cp\Phi_{\exp(l\cdot \g'_\iota)}\big(\Phi_{h}(p_\m)\big)\cdot \fiba_{p_\m}(h)^{-1},  \homm\left(\gamma^x_{\wg}|_{[0,\adif(l)]}\right)\left(p_\m\right)\cdot \fiba_{p_\m}(h)^{-1}\right)\\
          %%%%%%%%% 
          &=\big(\alpha_{\fiba_{p_\m}(h)}\cp \Delta\big)\!\left(\Phi_{{h}^{-1}}\cp\Phi_{\exp(l\cdot \g'_\iota)}\big(\Phi_{h}(p_\m)\big),  \homm\left(\gamma^x_{\wg}|_{[0,\adif(l)]}\right)\left(p_\m\right)\right)\\
          &=\alpha_{s_0}\left(s_\alpha(l)\cdot \Delta\!\left(\Phi_{\exp(\pm \adif(l)\cdot \g_\iota)}\big(p_\m\big), \homm\left(\gamma^x_{\wg}|_{[0,\adif(l)]}\right)\left(p_\m\right)\right)\right)\\
          % &=s_0^{-1}\cdot \Delta\left(\Phi_{\exp(l'\h)}\big(p'_\m\big), \homm\left(\gamma^y_{\h}|_{[0,l']}\right)\left(p'_\m\right)\right)\\
          &=\alpha_{s_0}(s_\alpha(l))\cdot\alpha_{s_0}\big( \pip_\iota(\homm)\big(\wg,\adif(l)\big)\big)\\
                   &=\alpha_{s_0}(s_\alpha(l))\cdot\alpha_{s_0}\big( \pip_\iota(\homm)\big(\g_\iota,\adif(l)\big)\big)^{\pm 1}.
        \end{split}
      \end{align}

      {\bf Case A:} Assume that $s_\alpha=\me$ and write, see \eqref{eq:phiB}
       \begin{align*}  
       % \label{eq:phiB}
  	\pip_\iota(\homm)(\g_\iota,l)=\exp\big(\phi(l)\cdot\s_{\beta(\homm)}\big)\qquad\forall\:l>0      
  	\end{align*} 
  	as in {\bf Step 1}. Then, \eqref{eq:smult} shows 
      \begin{align*}
        \pip'(\homm)(\g'_\iota,l)=\exp\big( \phi(\adif(l))\cdot \pm\Add{s_0}\big(\s_{\beta(\homm)}\big)\big)\qquad\forall\: 0<l<\tau_{\g}.
      \end{align*}
		Now, by the same arguments as in {\bf Step 1}, we can assume that $\tau'_\iota$ is defined by using the elements $\pm\Add{s_0}(\s_{\beta})$ for $\beta\in \Sp\ms$. Consequently,  
      \begin{align*} 
        \Omega(\psi)(\chi_l)=\psi(\chi_{\adif(l)})\qquad \forall\:\psi\in \RB, \forall\:0<l<\tau_{\g'},
      \end{align*}
      hence $\Omega(\psi)(\chi_l)=\psi(\chi_{\adif(l_1)+ \dots +\adif(l_k)})$
      if $l=l_1+\dots+l_k$ for $0<l_1,\dots,l_k<\tau_{\g'}$. 
      Then, $\Omega$ is a bijective\footnote{This is because $\adif|_{(0,\tau_{\g_\iota})}$ is bijective and $\psi \in \RB$ is uniquely determined by its values on each subset of $\DG$ of the form $\{\chi_l\:|\: l\in (0,\tau)\}$ for $\tau>0$.} and unital homomorphism which is continuous because we have
      \begin{align*}
        \|\Omega(\psi)\|_{\chi_{l}}=\|\psi\|_{\chi_{\adif(l_1)+ \dots +\adif(l_k)}}
      \end{align*}
      for the seminorm $\|\psi\|_{\chi_{l}}:=\|\psi(\chi_l)\|$, by the definition of the topology on $\RB$. Since the normalized Radon measure $\Omega(\muB)$ is translation invariant, it equals $\muB$. Hence, the same is true for $\Omega^{-1}(\muB)$. The rest now follows as in {\bf Step 1}.

      \vspace{4pt}
      {\bf Case B:} Assume that $s_\alpha\neq \me$ and let $s'_\alpha:=\alpha_{s_0}\cp s_\alpha$. We choose $\homm_0\in \IHOMLAS$ with $\pip_\iota(\homm_0)(\g_\iota,\cdot)=\me$, hence $s'_\alpha=\pip'_\iota(\homm_0)(\g'_\iota,\cdot)$ by \eqref{eq:smult}. In particular, $\im[s'_\alpha] \subseteq H_{\beta'(\homm_0)}$ for $\beta'(\homm_0)$ the element from Definition \ref{def:betadef} that correspond to $\Psi':=\res'_\iota(\pip'_\iota(\homm_0))\in Y'_\iota$.
      
	{\bf We claim that:} 
	\begin{align*}
		\im\big[\pip'_\iota(\homm)(\g'_\iota,\cdot)\big]\subseteq H_{\beta'(\homm_0)}\qquad\forall\: \homm\in \IHOMLAS,
	\end{align*}
	hence $\im[\pip_\iota(\homm)(\g_\iota,\cdot)]\subseteq \alpha_{s_0^{-1}}(H_{\beta'(\homm_0)}$) for all $\homm\in \IHOMLAS$ because $\im[s'_\alpha]\subseteq H_{\beta'(\homm_0)}$.

	{\bf Proof:}
      In fact, 
      elsewise we find $\homm_1\in \IHOMLAS$ with $0\neq \beta'(\homm_1)\neq \beta'(\homm_0)$. We choose $\pm \me\neq s_1,s_2\in H_{\beta'(\homm_0)}$ and $0<l_1,l_2< \tau_{\g'_\iota}$ $\mathbb{Z}$-independent with $s'_\alpha(l_i)=s_i$ for $i=1,2$. This is always possible because $s'_\alpha$ is continuous, $s'_\alpha(0)=\me$, and $s'_\alpha\neq\me$ as $s_\alpha\neq \me$ by assumption.
    
     Combining Proposition \ref{th:invhomm}.\ref{th:invhomm1} with Lemma \ref{prop:Bohrmod2}.\ref{prop:Bohrmod22} and the Parts \ref{prop:Bohrmod24}), \ref{prop:Bohrmod21}) of Lemma and Convention \ref{lemconv:RBMOD}, we find $\homm_1'\in \IHOMLAS$ with $\beta'(\homm'_1)=\beta'(\homm_1)$ as well as 
	\begin{align} 
	\label{eq:dfdsvc77v7676vs}    
     \pip'_\iota(\homm_1')(\g'_\iota,l_1)=\me\qquad\quad\text{and}\qquad\quad 
      \pip'_\iota(\homm_1')(\g'_\iota,l_2)=:b\neq \pm \me. 
        \end{align}
      Let $d_i:=\alpha_{s^{-1}_0}(\pip_\iota(\homm_1')\big(\g_\iota,\adif(l_i))\big)^{\pm 1}$ (confer \eqref{eq:smult}) for $i=1,2$.
      Then \eqref{eq:smult} yields 
  	\begin{align*} 
	\me\stackrel{\eqref{eq:dfdsvc77v7676vs}}{=}\pip'_\iota(\homm_1')(\g'_\iota,l_1)\stackrel{\eqref{eq:smult}}{=} s_\alpha'(l_1)\cdot \alpha_{s_0}(\pip_\iota(\homm_1')\big(\g_\iota,\adif(l_i))\big)^{\pm 1}& =s_1\cdot d_1\quad\:\:(s_1\in H_{\beta'(\homm_0)}\backslash \{-\me,\me\})\\%\quad
	&\Longrightarrow  \hspace{8.5pt} d_1\in H_{\beta'(\homm_0)}\backslash \{-\me,\me\}.
    \end{align*} 
    Hence, $\pm\me\neq b=s_2\cdot d_2 \in H_{\beta'(\homm_0)}$ because $d_1$ and $d_2$ are contained in the same maximal torus, namely $H_{\beta'(\homm_0)}$. This contradicts that $b\in H_{\beta'(\homm_1)}\backslash \{-\me,\me\}$. \hspace*{\fill}$\boldsymbol{\dagger}$
\vspace{5pt}

      Thus, for $\beta:=\big[\Ad_{s_0^{-1}}(\s_{\beta'(\homm_0)})\big]$ we have $\beta(\homm)\in \{0,\beta\}$ for all $\homm\in \IHOMLAS$ with $\beta(\homm)$ defined as in {\bf Step 1}. Similarly, we have $\beta'(\homm)\in \{0,\beta'(\homm_0)\}$ for all $\homm\in \IHOMLAS$. 
      
      Now, $Y'_\iota$ cannot be of {\bf Type 1} since this would imply that $s'_\alpha=\me$ holds. Moreover, if $Y_\iota$ would be of {\bf Type 1}, then \eqref{eq:smult} would give that $\pip'(\homm)(\g'_\iota,\cdot)=s_\alpha'$ holds for all $\homm\in \IHOMLAS$, which is only possible if $Y_\iota'$ is of {\bf Type 1}, just by Lemma \ref{prop:Bohrmod2}.\ref{prop:Bohrmod22} and Proposition \ref{th:invhomm}.\ref{th:invhomm1}. Consequently, $Y_\iota$ and $Y'_\iota$ are both of {\bf Type 2}.  
      
      So, to finish the proof, we write
      	\begin{align*}
      \begin{tabular}{rlrl}
		$\pip'_\iota(\homm_0)(\g'_\iota,l)\hspace{-9pt}$&$=\exp\big(\phi'_0(l)\cdot \s_{\beta'(\homm_0)}\big)$&for& $\phi'_0\in \Per,\forall\: 0<l<\tau_{\g'_\iota}$\\[3pt]
		$\pip'_\iota(\homm)(\g'_\iota,l)\hspace{-9pt}$&$=\exp\big(\phi'_\homm(l)\cdot \s_{\beta'(\homm)}\big)$&for& $\phi'_\homm\in \Per,\forall\: 0<l<\tau_{\g'_\iota}$\\[3pt]
		$\pip_\iota(\homm)(\g_\iota,l)\hspace{-9pt}$&$=\exp\big(\phi_\homm(l)\cdot \s_{\beta(\homm)}\big)$&for& $\phi_\homm\in \Per,\forall\: 0<l<\tau_{\g_\iota}$
	  \end{tabular}
	  \end{align*}
		whereby $\pip'_\iota(\homm_0)(\g'_\iota,l)=s'_\alpha(l)$. 
      Then, \eqref{eq:smult} and $\beta'(\homm)=\beta'(\homm_0)$ show that 
      \begin{align*}      
        \exp\big(\phi'_\homm(l)\cdot \s_{\beta'(\homm)}\big)=\exp\big(\phi'_0(l)\cdot \s_{\beta'(\homm_0)}\pm \phi_\homm(\adif(l))\cdot \s_{\beta'(\homm)}\big)\qquad\forall\:0<l<\tau_{\g'_\iota}. 
      \end{align*}       
      Hence,  
      \begin{align*}
      \Omega(\psi)(\chi_l)&= \e^{\I\hspace{1pt}\phi'_\homm(l)}=\e^{\I\hspace{1pt}[\phi'_0(l)\pm \phi_\homm(\adif(l))]}\\
      &=\e^{\I\hspace{1pt}\phi'_0(l)}\cdot \e^{\pm \I\hspace{1pt}\phi_\homm(\adif(l))}
      =\psi'_0(\chi_l)\cdot \psi\big(\chi_{\adif(l)}\big)^{\pm 1}=[\psi_0\pm \Omega_\adif(\psi)](\chi_l),
      \end{align*}
      where $\Omega_\adif(\psi)(\chi_l):=\psi\big(\chi_{\adif(l)}\big)$ 
      for all $0<l<\tau_{\g'_\iota}$. Of course, here $\psi'_0\in \RB$ denotes the element that corresponds to $\phi'_0\in \Per$. The rest now follows as in {\bf Step 1} because $\psi\mapsto \psi'_0\pm \psi$ is a homeomorphism which preserves $\muB$.
\end{enumerate}
\end{proof}

\subsection{Summary}
\label{conclmeasures}
\begin{enumerate}
\item
\label{conclmeasures1}
  In the first part of this section, we have defined the normalized Radon measure $\mFNS$ on $\AQRInd{\mathrm{FN}}$  
   for the case that $S$ is compact and connected. This measure specializes to the Ashtekar-Lewandowski measure $\mAL$ on $\AQRw$ if the symmetry group is trivial. In the second part, we have constructed the normalized Radon measure $\mLAS$ on $\AQRInd{\mg}$ for the case that $S=\SU$ and that each $\wm$-orbit $\m$ admits an independent and  complete family $\{\g_{\m,\alpha}\}_{\alpha\in I_\m}\subseteq \mg\backslash \mg_x$ for some $x\in \m$.  
  So, if in this situation additionally $\Paw=\Pags\sqcup \Pafns$ holds, we have the normalized Radon measure $\mLAS\times \mFNS$ on 
\begin{align*} 
  \AQRw\cong \AQRInd{\mg}\times \AQRFNS,
 \end{align*}
 which defines a kinematical $L^2$-Hilbert space for the corresponding reduced theory. 
  In particular, this is the case in (semi)-homogeneous LQC as we have discussed in Remark \ref{ex:Fullmeas}.\ref{ex:Fullmeas1}. Recall that for all of our constructions we have assumed that the action $\wm$ induced on $M$ is analytic and pointwise proper.
\item
As already mentioned in the beginning of this subsection, the constructions there are also possible in the abelian case ($S$ is an $n$-torus) and are even easier.\footnote{The respective spaces $Y_\g^p$ of $\Ad_{G_{[\g]}}^p$-equivariant maps 
	$\Psi\colon \spann_\RR(\g)\rightarrow S$  are either $\{e\}$ or $[\RB]^n$.} Then, for an arbitrary compact and connected structure group $S$ one has to equip the respective spaces $Y_\g^p$ 
%calculate the spaces 
of equivariant maps 
	$\Psi\colon \spann_\RR(\g)\rightarrow S$ 
with suitable Radon measures. Here, suitable means that these measure have to fulfil certain invariance properties which make the whole construction independent of any choices one has to do. For this observe that we have used fixed families of stable Lie algebra elements in order to parametrize the space $\AQRInd{\mg}$. Using the theory of compact and connected Lie groups, here one might obtain some generalizations of the constructions we have worked out for $\SU$. 
\item
 In Example \ref{ex:cosmoliealgmaasse} we have shown that the measure $\mLAS$ is also available in the spherically symmetric and in the homogeneous isotropic case. Unfortunately, there we have no measure on $\AQRw$ at this point, just because we have not determined the set $\Pacs$ (of continuously but not Lie algebra generated curves) so far. 
 
 More concretely, if this set were empty, in both cases we would have a normalized Radon measure on $\AQRw$ because:
\begingroup
\setlength{\leftmarginii}{13pt}
\begin{itemize}
\item
Then, $\AQRw\cong\AQRInd{\mg}\times \AQRFNS$ would also hold in the homogeneous isotropic case, since we have shown in Remark \ref{ex:Fullmeas}.\ref{ex:Fullmeas2} that the set ($\Pafs$ of free curves having non-trivial stabilizer) is empty there.
\item
In the spherically symmetric case we would have
 \begin{align*} 
	\AQRw\cong\AQRInd{\mg}\times \AQRFNS\times \AQRInd{\mathrm{FS}},
 \end{align*}
 where we already have constructed a normalized Radon measure on $\AQRInd{\mathrm{FS}}$ by hand, see Remark \ref{ex:Fullmeas}.\ref{ex:Fullmeas2}.
\end{itemize} 
\endgroup
\item
As already mentioned in Remark \ref{rem:euklrem}.\ref{it:sdsdds}, in the homogeneous isotropic case where $\wm$ is transitive, $\AQRLA$ is a physically meaningful quantum-reduced configuration space by itself. This is because the cylindrical functions that correspond to $\Pags$ then separate the points in $\Con$, so that 
      $\Con $ is canonically embedded via $\iota\colon \Con \rightarrow \AInd{\mg}$.  
Our constructions here provide us with the reduced kinematical Hilbert space $\Lzw{\AQRLA}{\mLAS}$, which even specializes to the standard LQC kinematical Hilbert space $\Lzw{\RB}{\muB}$ if we apply the same constructions for the subset $\Pal\subseteq \Pags$, see end of Example \ref{ex:cosmoliealgmaasse} and Remark \ref{StanLQC}.\ref{StanLQC2}.

 More precisely, we have seen that each of the maps in 

\begin{figure}[h]
  \begin{minipage}[h]{\textwidth}
    \begin{center}
      \makebox[0pt]{
        \begin{xy}
          \xymatrix{
           \RB \cong \ARRQLL \ar@{->}[r]^-{\ovl{i^*_\AR}}_-{\cong}   &  \: \ARQLL \: \ar@{->}[r]^-{\subseteq} & \AQRL\ar@{->}[r]_-{\cong}^-{\kappa_\lin}   &\IHOMLL \ar@{->}[r]_-{\cong}^-{\eta_\lin}& \RB
         }
        \end{xy}}
    \end{center}
  \end{minipage} 
\end{figure}
   \FloatBarrier
  \noindent  
is a homeomorphism and that their concatenation is just the identity on $\RB$. In particular, this means that $\ARRQLL\cong \AQRL$ holds, i.e.,  that quantization and reduction commute in homogeneous isotropic LQC if one only uses linear curves in order to define the reduced spaces. Moreover, in Remark \ref{StanLQC}.\ref{StanLQC2} we have seen that, due to this fact, the standard quantum configuration space $\ARRQLL\cong \IHOMLL$ can be embedded into the full quantum-reduced space $\AQRw\cong \IHOMW$  
just by the simple map
    %\begin{align*}
    $\iota_\lin\colon \IHOMLL\rightarrow \IHOMW$ defined by
	%\end{align*}  
	 \begin{align*}
         \iota_\lin(\homm)(\gamma):=\homm(\gamma)\:\text{ if }\:\gamma\in \Pal\qquad\text{and}\qquad\iota_\lin(\homm)(\gamma):=\me\:\text{ for }\:\gamma\in \Paw\backslash \Pal.
	\end{align*}
\end{enumerate}

\section{Homogeneous Isotropic LQC}
\label{sec:HomIsoCo}
The traditional way to do symmetry reduction in LQC is to quantize the set  
$\AR$ of 
connections invariant under the given 
symmetry group. In a first step, this means to calculate the spectrum $\ARRQInd{\alpha}$ of a separating $\Cstar$-algebra of the form $\rR=\ovl{\PaC_\alpha|_{\AR}}\subseteq B(\AR)$ \cite{MathStrucLQG,ChrisSymmLQG}.   
Here, $\PaC_\alpha$ denotes the $\Cstar$-algebra of cylindrical functions which is generated by some distinguished set 
$\Pa_\alpha$ of curves in  base manifold. 
Now, in \cite{ChrisSymmLQG} it was shown that $\ARRQInd{\alpha}$ can be compatibly embedded into the quantum configuration space\footnote{Recall that this is the spectrum of the $\Cstar$-algebra of cylindrical functions that correspond to the set $\Paw$ of embedded analytic curves in $\RR^3$.} $\A_\w$ of LQG iff 
\begin{align}
\label{eq:inkldsakhfdfd898}
\PaC_\w|_{\AR}\subseteq \PaC_\alpha|_{\AR}
\end{align}
holds. Here, compatibly means that there exist an embedding of $\ARRQInd{\alpha}$ into $\A_\w$ which extends the inclusion map $\AR\hookrightarrow \Con$ in the sense of Lemma \ref{lemma:dicht}.\ref{lemma:dicht2}. 

Now, in standard homogeneous isotropic LQC \cite{MathStrucLQG}, see also Remark \ref{StanLQC}, the set $\Pal$ of linear curves is used to define the reduced configuration space. It was shown in \cite{Brunnhack} that here \eqref{eq:inkldsakhfdfd898} does not hold, i.e., that no compatible embedding of $\ARRQInd{\lin}\cong \RB$ into $\A_\w$ exists.  
In particular, for the embedding strategy proposed for states in \cite{BojoKa} this is disadvantageous. 
So, to fix this problem, in \cite{ChrisSymmLQG} the reduced space 
$\ARRQInd{\w}$ was introduced. It was shown that $\ARRQInd{\w}$ is compatibly embedded into $\A_\w$ via the map $\ovl{i_{\AR}^*}$ 
(see, e.g., \eqref{eq:inclusionsdiag}), and that it is homeomorphic to the compact Hausdorff space $\RR\sqcup \RB$.\footnote{The topology on $\RR\sqcup \RB$ is quite tricky. Details will be given in Subsection \ref{GAM}.}

In order to define the dynamics of the reduced theory,  
it is 
reasonable to construct natural measures on the reduced quantum configuration spaces.  
Indeed, such measures usually define $L^2$-Hilbert spaces on which 
representations of respective reduced holonomy-flux algebras can be defined. 
Now, being a compact abelian group, $\ARRQLL\cong \RB$ admits the Haar measure $\muB$, which defines the standard kinematical Hilbert space $\Lzw{\RB}{\muB}$ of homogeneous isotropic LQC. 
However, for
	\begin{align*}
 	\RR\sqcup \RB\cong \text{\gls{qRR}}:=\ARRQInd{\w}
	\end{align*} 	
 	 the situation is much more difficult as there no Haar measure can exist. This will be shown in Proposition \ref{prop:noGroupStruc} where we prove that it is impossible to equip $\RR\sqcup \RB$ with a group structure continuous w.r.t.\ its Gelfand topology. Basically, this is because existence of such a structure would imply that $|\RB|=|\RR|$ holds. This, however, contradicts that for the cardinality of $\RB$ we have $|\RB|\geq |2^\RR|$. Then, changing the focus from Haar to normalized Radon measures, it is a crucial observation that $\muB$ is the unique normalized Radon measure $\mu$ on $\RB$ for which the pullbacks 
	\begin{align*} 	 
 	 {\Transl_v\!}^* \colon \Lzw{\RB}{\mu}\rightarrow \Lzw{\RB}{\mu},\quad f\mapsto f\cp \Transl_v,
 	 \end{align*}
 	 are unitary operators (even form a strongly continuous one-parameter group), cf.\ Proposition \ref{lemma:bohrmassdichttrans}. Here,
 	 $\Transl\colon \RR\times \RB \rightarrow \RB$ denotes the unique extension  (cf.\ Proposition \ref{prop:autspec}) of the additive group action \RPLUS$\colon \RR \times \AR \rightarrow \AR$	  
 	   of $\RR$ on $\AR \cong\RR$. This fact is important because the exponentiated reduced fluxes (``momentum operators'') are represented in this way. 
 	   
 	   Indeed, one now might ask whether the same condition 
singles out a normalized Radon measure on $\RR\sqcup\RB$. We will show that this is the case (Corollary \ref{cor:noext}), and that this measure is given by $\muB$ as well. In particular, the respective kinematical Hilbert space equals the standard one.
 At this point, it is important to know that the Borel $\sigma$-algebra of $\RR\sqcup \RB$ is just given by $\Borel(\RR)\sqcup \Borel(\RB)$ (the topology on $\RR\sqcup \RB$ has not such an easy decomposition), so that
\begin{align*}
	\mu(A_1\sqcup A_2):=\muB(A_2)\quad \text{for}\quad A_1\in \Borel(\RR)\quad\text{and}\quad A_2\in \Borel(\RB) 
\end{align*}
is a well-defined normalized Radon measure on $\RR\sqcup \RB$, cf.  Lemma \ref{lemma:Radon}.  In the last part, we will equip this space with a projective structure in order to construct further normalized Radon measures thereon. We then close 
this section with a brief discussion of the corresponding $L^2$-Hilbert spaces they define. 
On the way, we will show that quantization and reduction do not commute in homogeneous isotropic LQC. As in Subsection \ref{sec:inclrel}, this will be done by constructing elements of $\AQRw$ which cannot be contained in $\ARQw\cong \ARRQw\cong \RR\sqcup \RB$.

\subsection{Setting}
In the following, let $P:=\RR^3\times \SU$, $G:=\Ge=\Gee$  ($\mg=\RR^3\times \su$) and $\Phi:=\Pe$, see Example \ref{ex:LQC}. Moreover, let $\nu_x:=(x,\me)$ for all $x\in M$ and recall (see \eqref{eq:homis}) that the corresponding set $\AR$ of $\Phi$-invariant connections consists exactly of the elements of the form ($c$ runs over $\RR$)
\begin{align*}
  % \label{eq:homis} 
  \w^c_{(x,s)}(\vec{v}_x,\vec{\sigma}_x)= c \Add{s^{-1}}[\murs(\vec{v}_x)]+s^{-1}\vec{\sigma}_s \qquad \forall\:(\vec{v}_x,\vec{\sigma}_s)\in T_{(x,s)}P.
\end{align*}   
So, in the sequel we will always identify $\AR$ with $\RR$. 
 In contrast to Subsection \ref{sec:ConSp} (see Convention \ref{conv:sammel}), by $\RB$ in the following we will understand the spectrum of the $\Cstar$-algebra $\CAP(\RR)$. This has just conceptual reasons and  makes mathematically no difference as Lemma \ref{lemma:Bohriso} shows. Finally, $\rR$ will always denote the $\Cstar$-algebra (vector space direct sum) 
	\begin{align*} 
 \rR:=C_0(\RR)\oplus \CAP(\RR)\subseteq B(\RR).
\end{align*} 
 As shown in \cite{ChrisSymmLQG}, this is exactly the restriction $\Cstar$-algebra $\ovl{\PaC_\w|_{\AR}}$ (under the identification $\AR\cong \RR$), so that 
 %\begin{align*}
 $\ARRQw=\Spec(C_0(\RR)\oplus \CAP(\RR))$ holds.
 %\end{align*}

\subsection{Group Structures, Actions and some Measure Theoretical Aspects}
\label{GAM}
In this subsection, we show that the Haar measure on $\RB$ is the unique normalized Radon measure which is invariant under the spectral extension %\footnote{Cf.\ Proposition \ref{prop:autspec}.} 
$\Transl\colon \RR\times \RB \rightarrow \RB$ of the additive group action \RPLUS$\colon \RR \times \RR \rightarrow \RR$, $(v,t)\mapsto v+t$. In particular, we will identify $\muB$ as the unique normalized Radon measure $\mu$ on $\RB$ for which the maps ${\Transl_v\!}^* \colon \Lzw{\RB}{\mu}\rightarrow \Lzw{\RB}{\mu}$, $f\mapsto f\cp \Transl_v$ are unitary operators for all $v\in \RR$, see Proposition \ref{lemma:bohrmassdichttrans}. In particular, we prove that the one-parameter group $\{{\Transl_v\!}^*\}_{v\in \RR}$ then is strongly continuous. We show that the same invariance condition\footnote{Namely, that the translations w.r.t.\ the spectral extension $\Transw\colon \RR\times \qRR \rightarrow \qRR$ of {\scriptsize$\Sigma_\RR$}$\colon \RR \times \RR \rightarrow \RR$ act as unitary operators. Here, it is equivalent to require that this measure is invariant under the translations $\Transw_v$ for all $v\in \RR$.} singles out a normalized Radon measure on $\qRR$, which then even defines the same Hilbert space $\Lzw{\RB}{\muB}$ as in the traditional approach. This reinforces the standard LQC approach from the mathematical side. 
On the way, we verify that, in contrast to $\RB$, there exists no continuous group structure, hence no Haar measure on $\qRR$.

Since the group structure on $\RB$ canonically extends\footnote{We have $\iota'_\RR(x)+\iota'_\RR(y)=\iota'_\RR(x+y)$ for $\iota'_\RR\colon \RR \rightarrow\RB$ the canonical embedding from Subsection \ref{subsec:boundedfun}.} the additive group structure on $\RR$, we might start by clarifying whether there is also such an extension of the additive group structure on $\RR$ to $\qRR=\Spec(C_0(\RR)\oplus \CAP(\RR))$.  
Indeed, this would provide us with a Haar measure on $\qRR$, in particular, invariant under the action $\Transw\colon \RR \times \qRR\rightarrow \qRR$, $(v,\x)\mapsto \iota_\RR(v)+\x$. This is because $\Transw$ would be the unique extension of \RPLUS\ in this case just because it fulfils the respective conditions from Proposition \ref{prop:autspec}.\ref{prop:autspec1}. 
However, already Proposition \ref{prop:Specgroup} shows that such an extension of the additive group structure on $\RR$ cannot exist.  %Indeed, we have the following
\begin{corollary}
  \label{cor:noext}
  There is no continuous group structure on $\qRR=\Spec(\rR)$ which is compatible with the additive group structure on $\mathbb{R}$.
  \begin{proof} 
    Assume there is such a group structure. Then it is necessarily abelian as it is so on a dense subset of $\Spec(\rR)$. So, by Proposition \ref{prop:Specgroup} there is a set $\uU\subseteq \rR$ of characters on $\RR$ generating $\rR$. Since $\uU\subseteq \rR$, these characters are continuous, hence contained in $\CAP(\RR)$. For this, recall that $\CAP(\RR)$ is generated by all continuous characters on $\RR$. Since $\CAP(\RR)$ is closed, the $\Cstar$-subalgebra generated by $\uU$ must be contained in $\CAP(\RR)$. This contradicts that $\uU$ generates $\rR$ just because $C_{0}(\mathbb{R})\neq \{0\}$.   
  \end{proof}
\end{corollary}  
Anyhow, using Proposition \ref{prop:autspec} we obtain the following left actions on $\qRR$ and $\RB$.
\begin{corollary}
  \label{cor:chrisleftact}
  There are unique left actions 
  % \begin{align*}
  $\Multw\colon \RR_{\neq 0}\times \qRR \rightarrow \qRR$ and $\Transw\colon \RR \times \qRR\rightarrow \qRR$, 
  % \end{align*}
  separately continuous in $\qRR$, that extend the multiplication $\cdot \colon \RR_{\neq 0}\times \RR \rightarrow \RR$, $(\lambda,t)\mapsto \lambda \cdot t$ and the translation $+\colon \RR\times \RR \rightarrow \RR$, $(v,t)\mapsto v+ t$, respectively, in the sense of \eqref{eq:extens}. The action $\Transw$ is continuous and $\Multw$ is not so. The same statements hold for $\RB$ instead of $\qRR$ where we denote the respective actions by $\Multl$ and $\Transl$ in the following. For $\Transl$ we have
  \begin{align} 
    \label{eq:Bohrwirk}
    \Transl(v,\psi)=\iota'_\RR(v)+\psi\qquad \forall\: v\in \RR,\forall\:\psi\in \RB
  \end{align}
  with  $\iota'_\RR\colon \RR \rightarrow\RB$ the canonical embedding and $+$ the addition in $\RB$.
  \begin{proof}
    Obviously, $C_0(\RR) \oplus \CAP(\mathbb{R})$ is invariant under pullback by $\cdot_\lambda$ and $+_v$ for all $\lambda\neq 0$ and all $v\in \RR$. This is obvious for $C_0(\RR)$ and follows for $\CAP(\mathbb{R})$ from $\cdot_\lambda^*(\chi_l)=\chi_{\lambda  l}$ and $+_v^*(\chi_l)=\e^{\I l v}\cdot \chi_l$
    for all $l\in \RR$. Consequently, Proposition \ref{prop:autspec}.\ref{prop:autspec1} provides us with the unique left actions $\Multw$ and $\Transw$.
    
    Obviously $\cdot_\bullet ^*\colon \lambda \mapsto  [t\mapsto \e^{\I \lambda l t}]$ is not continuous (w.r.t.\ the supremum norm) for $l\neq 0$, so that $\Multw$ is not continuous by unitality of $\rR$ and Proposition \ref{prop:autspec}.\ref{prop:autspec2}. In contrast to that, for each $f\in \rR$ the map $+_\bullet^*(f)\colon v\mapsto [t\mapsto f(v+t)]$ is continuous, so that Proposition \ref{prop:autspec}.\ref{prop:autspec2} shows continuity of $\Transw$. Here continuity is clear if $f=\chi_l$ for some $l\in \RR$ because
    \begin{align*}
    	\|+_v^*(\chi_l)-+_{v'}^*(\chi_l)\|_\infty=\big\|\big[\e^{\I lv}-\e^{\I lv'}\big]\chi_l\big\|_\infty=\big|\e^{\I lv}-\e^{\I lv'}\big|\cdot \|\chi_l\|_\infty=\big|\e^{\I lv}-\e^{\I lv'}\big|.
    \end{align*}
    It follows for $f\in C_0(\RR)$ from 
equicontinuity of $f|_K$ for each compact subset $K\subseteq \RR$,   
 and that for each $\epsilon$ we find $K\subseteq \RR$ compact with $|f(t)|<\epsilon/ 2$ for all $t\in \RR\backslash K$. 
    
    In fact, let $K':= K + [-\delta,\delta]$ for $\delta >0$. Then for all $t\in \RR \backslash K'$ and all $v\in [-\delta, \delta]$ we have $t, v+t\in \RR\backslash K$, hence  
	\begin{align}
	\label{eq:compl}    
    	|f(t)-f(v+t)|\leq |f(t)|+|f(v+t)|< \epsilon.
    \end{align}
	Moreover, by equicontinuity of $f|_{K'}$ we find $W\subseteq \RR$ a neighbourhood of $0\in \RR$, such that    
	\begin{align}
	\label{eq:compll}
		|f(t)-f(t+v)|<\epsilon \qquad\:\forall\: t\in K',\:\forall\: v\in W.
	\end{align}
	So, combining \eqref{eq:compl} and \eqref{eq:compll} we obtain
	\begin{align*}
		|f(t)-f(t+v)|<\epsilon \qquad\:\forall\: t\in \RR,\:\forall\: v\in W\cap [-\delta,\delta],
	\end{align*}
	hence $\|\!+_v^*(f)-f \|_\infty<\epsilon$ for all $v\in W\cap [-\delta,\delta]$. This show continuity of $+^*_\bullet$ at $v=0$. Then, replacing $f$ by $+^*_{v'}(f)$, we deduce continuity of $+^*_{v'}$ for all $v'\in \RR$.
	
    Finally, \eqref{eq:Bohrwirk} holds because the left action $\RR\times \RB \rightarrow \RB$, $(v,\psi)\mapsto \iota'_\RR(v)+\psi$ fulfils the condition from Proposition \ref{prop:autspec}.\ref{prop:autspec1}, hence equals the unique left action $\Transl$.
  \end{proof}
\end{corollary} 
At the end of this subsection, the action $\Transw$ will provide us with a uniqueness statement concerning normalized Radon measures on $\qRR$. The first step toward this is performed in the following
\begin{proposition}
  \label{lemma:bohrmassdichttrans}
  \begin{enumerate}
  \item
    \label{lemma:bohrmassdichttrans1}
  If $\mu$ is a normalized Radon measure on $\RB$ with $\Transl_v(\mu)=\mu$ for all $v\in \RR$, then $\mu=\muB$. In particular, $\muB$ is the unique normalized Radon measure $\mu$ on $\RB$ for which the translations ${\Transl_v\!}^* \colon \Lzw{\RB}{\mu}\rightarrow \Lzw{\RB}{\mu}$, $f\mapsto f\cp \Transl_v$ are unitary operators. 
  \item
    \label{lemma:bohrmassdichttrans2}
  The one-parameter group $\{{\Transl_v\!}^*\}_{v\in \RR}$ of unitary operators
	\begin{align*}
	{\Transl_v\!}^* \colon \Lzw{\RB}{\muB}\rightarrow \Lzw{\RB}{\muB},\quad 	f\mapsto f\cp \Transl_v
	\end{align*}  
  is strongly continuous.
  \end{enumerate}
  \begin{proof}
  \begin{enumerate}
  \item
    We have to show that $\mu(\psi+A)=\mu(A)$ holds for all $\psi\in \RB$ and all $A\in \Borel(\RB)$. Then, by inner regularity, it suffices to show the case were $A$ is compact. 
    
    So, let $K\subseteq \RB$ be compact.
    Since $\mu$ is finite, it is outer regular. Hence, for each $n\in \NNge$ we find $U_n\subseteq \RB$ open with $K\subseteq U_n$ and $\mu(U_n)-\mu(K)\leq\textstyle\frac{1}{n}$. By continuity of the addition in $\RB$, for each $\psi'\in K$ the preimage $+^{-1}(U_n)$ contains a set $V'\times W'$ with $V',W'\subseteq \RB$ open, $\NB\in V'$ and $\psi'\in W'\subseteq U_n$. By compactness of $K$ there are finitely many such quantities $\NB\in V_1',\dots,V'_k \subseteq \RB$, $W_1',\dots,W_k'\subseteq U_n$ with $K\subseteq W_1'\cup \dots\cup W'_k=:W'$ and $V_i' + W_i' \subseteq U_n$ for $1\leq i\leq k$. 

    Let $V'\subseteq V_1'\cap\dots\cap V_k'$ be a neighbourhood of $\NB$.
    By denseness of $\iota'_\RR(\RR)$ in $\RB$ we find $x'\in \RR$ with $\psi-\iota'_\RR(x')\in V'$. Then, $\psi-\iota'_\RR(x') +K\subseteq V' +W' \subseteq U_n$, so that for each $n\in \NNge$ we obtain from \eqref{eq:Bohrwirk} that 
    \begin{align*}
      \mu(\psi +K)-\mu(K)\stackrel{\ref{eq:Bohrwirk}}{=}\mu\big(\psi-\iota'_\RR(x') +K\big) -\mu(K)\leq \mu\big(U_n\big) -\mu(K)\leq \textstyle\frac{1}{n},
      % \\
      %	&\leq|\mu\big(\psi-\iota'_\RR(x') +K\big)-\mu(U_n)|+|\mu(U_n) -\mu(K)|\\
      %	&\leq |\mu\big(\psi-\iota'_\RR(x') +K\big)-\mu(U_n)|+\textstyle\frac{1}{n}\leq \textstyle\frac{2}{n},
    \end{align*}
    hence $\mu(K)\geq \mu(\psi +K)$. Then, applying the same argument to the compact set $K':=\psi+K$ and $\psi':=-\psi$, we see that $\mu(\psi +K)=\mu(K')\geq \mu(\psi'+K')=\mu(K)$.  

    For the last statement assume that ${\Transl_v\!}^*$ is unitary for each $v\in \RR$. Then, for $K\subseteq \RB$ compact, $\chi_K$ the characteristic function of $K$ and $v\in \RR$
    we have
    \begin{align*}    
      \mu(K)&=\int_{\RB}\chi_{K}\:\dd \mu=\int_{\RB} |\chi_{K}|^2\:\dd \mu=\langle \chi_K,\chi_K \rangle 
      =\big\langle {\Transl_{v}\!}^*(\chi_K),{\Transl_{v}\!}^*(\chi_K)\big\rangle\\
      &= \int_{\RB} |\chi_{\iota'_\RR(v)+K}|^2\dd \mu =\int_{\RB}\chi_{\iota'_\RR(v)+K}\dd \mu=\mu\big(\iota'_\RR(v)+K\big).
    \end{align*} 
    By the first part this already implies the translation invariance of $\mu$.
    \item
    	We have to show that w.r.t.\ the $L^2$-norm $\|\cdot\|_2$ we have 
		\begin{align*}    	
    	\lim_{v'\rightarrow v}{\Transl_{v'}\!}^*(f)=f\qquad\forall\: f\in \Lzw{\RB}{\muB},\:\forall\: v\in \RR.
    	\end{align*}
    	Since $\muB$ is regular, $C(\RB)$ is dense in $\Lzw{\RB}{\muB}$. Moreover, since ${\Transl_{v}\!}^*(f)(\psi)=f(\iota'_\RR(v)+\psi)$, for each $f\in C(\RB)$ and each $\epsilon>0$ we find $\delta>0$ such that 
	\begin{align*}    	
    	\|{\Transl_{v}\!}^*(f)- {\Transl_{v'}\!}^*(f)\|_\infty<\epsilon \qquad\forall\: v'\in B_\delta(v). 
	\end{align*}    	
    	In fact, by 3.8 Satz in \cite{Elstrodt}, for $\epsilon>0$ we find a neighbourhood $U$ of $0_{\mathrm{Bohr}}\in \RB$ with 
    	\begin{align*}
    		\psi-\psi'\in U\qquad \Longrightarrow \qquad |f(\psi)-f(\psi')|<\epsilon.	
    	\end{align*}
    	Now, for $\delta>0$ suitable small we have $\iota'_\RR(B_\delta(0))\subseteq U$. This is just because $\iota'_\RR$ is continuous as $\CAP(\RR)$ consists of continuous functions on $\RR$, see Proposition 2.1 in \cite{ChrisSymmLQG}. Thus, for all $v'\in B_\delta(v)$ we have $\iota_\RR'(v)-\iota_\RR'(v')=\iota_\RR'(v-v')\in U$, hence 
    	\begin{align*}
    		\|{\Transl_{v}\!}^*(f)- {\Transl_{v'}\!}^*(f)\|_\infty&=\sup_{\psi\in \RB}|f(\psi + \iota_\RR'(v))-f(\psi+\iota_\RR'(v'))|\leq \epsilon%\qquad \forall\: v'\in (v-\delta,v+\delta)
    	\end{align*}
    	because $\psi+\iota'_\RR(v)-\psi+\iota'_\RR(v')=\iota'_\RR(v)-\iota'_\RR(v')\in U$.
    	
    	So, let $f\in \Lzw{\RB}{\muB}$ and $v\in \RR$ be fixed. Then, for $\epsilon>0$ by denseness of $C(\RB)$ in $\Lzw{\RB}{\muB}$ we find $f_\epsilon\in C(\RB)$ with $\|f-f_\epsilon\|_{2}\leq \frac{\epsilon}{3}$. Moreover, by the above arguments we find $\delta>0$ such that $\|{\Transl_{v}\!}^*(f_\epsilon)- {\Transl_{v'}\!}^*(f_\epsilon)\|_\infty\leq \frac{\epsilon}{3}$ for all $v'\in B_\delta(v)$. Consequently, by translation invariance of $\muB$ for all such $v'$ we have
	\begin{align*}    	
    	\|{\Transl_{v}\!}^*(f)-{\Transl_{v'}\!}^*(f)\|_2&\leq
    	\|{\Transl_{v}\!}^*(f)-{\Transl_{v}\!}^*(f_\epsilon)\|_2+\|{\Transl_{v}\!}^*(f_\epsilon)-{\Transl_{v'}\!}^*(f_\epsilon)\|_2+\|{\Transl_{v'}\!}^*(f_\epsilon)-{\Transl_{v'}\!}^*(f)\|_2 \\
    	&= \|f-f_\epsilon\|_2+\|{\Transl_{v}\!}^*(f_\epsilon)-{\Transl_{v'}\!}^*(f_\epsilon)\|_2+\|f_\epsilon-f\|_2
    	\leq \epsilon.
    \end{align*} 
    \end{enumerate}
  \end{proof}
\end{proposition}
Although Corollary \ref{cor:noext} states that no extension of the addition in $\RR$ to $\qRR$ exists, one should clarify whether there is any other continuous group structure on this space. Indeed, such a  structure would provide us with a canonical measure on $\qRR$. In addition to that we want to prove an analogue to Proposition \ref{lemma:bohrmassdichttrans}. For this and the considerations in the last subsection, it is comfortable to use the following description of $\qRR$ proven in \cite{ChrisSymmLQG}. 
\begin{lemdef}
  \label{remdefchris}
  Let $\emptyset\neq Y\subseteq \RR$ be open and $C_{0,Y}(\RR)$ the set of continuous function vanishing at infinity and outside $Y$. Then Corollary B.2 in \cite{ChrisSymmLQG} shows $C_{0,Y}(\RR)\cap \CAP(\RR)=\{0\}$ and that $\aA_Y:=C_{0,Y}(\RR)\oplus \CAP(\RR)\subseteq B(\RR)$ is closed. Moreover, if we   
  equip $Y \sqcup \RB$ with the topology generated by the sets of the following types:\cite{ChrisSymmLQG} 
  \begin{align*}
    \begin{array}{lcrclcl}
      \textbf{Type 1:} && V & \!\!\!\sqcup\!\!\! & \emptyset 
      && \text{with open $V \subseteq Y$} \\
      \textbf{Type 2:} && K^c & \!\!\!\sqcup\!\!\! & \RB
      && \text{with compact $K \subseteq Y$} \\[-1.5pt]
      \textbf{Type 3:} && f^{-1}(U) & \!\!\!\sqcup\!\!\! & \GT(f)^{-1}(U) 
      && \text{with open $U \subseteq \mathbb{C}$ and $f \in \CAP(\RR)$},
    \end{array}
  \end{align*}
  then
  Proposition 3.4 in \cite{ChrisSymmLQG} states that $\Spec(\aA_Y)\cong Y\sqcup \RB$ holds via the homeomorphism $\xi\colon Y\sqcup \RB\rightarrow \Spec(\aA_Y)$ defined by
  \begin{equation}
    \label{eq:Ksiii}
    \xi(\x) := 
    \begin{cases} 
      f\mapsto f(\x) &\mbox{if } \x\in  Y\\ 
      f_0\oplus f_{\mathrm{AP}}\mapsto \x(f_{\mathrm{AP}})  & \mbox{if } \x\in \RB.
    \end{cases}
  \end{equation}  
  Here, $f_0\in C_0(\RR)$ and $f_{\mathrm{AP}}\in \CAP(\RR)$. 
  It is straightforward to see that the subspace topologies of $Y$ and $\RB$ w.r.t.\ the above topology coincide with their usual ones. In abuse of notation\footnote{Indeed, in order to be in line with the notations in \eqref{eq:inclusionsdiag} we rather should define $\qR:=\ARQw$ and introduce another symbol for $\RR\sqcup \RB$. However, in the following we will rather deal with the space $\RR\sqcup \RB$, so that it is much more convenient to use the memorable symbol $\qR$  for this space. Besides this, the spaces $\RR\sqcup \RB$, $\ARRQw$ and $\ARQw$ are homeomorphic, and using $\qR$ for $\RR\sqcup \RB$ our notations are in line with \cite{ChrisSymmLQG}.} we define $\qRY:=Y\sqcup \RB$ as well as $\text{\gls{qR}}:=\RR\sqcup \RB$, and equip these spaces with the above topology. 
  \hspace*{\fill}{\scriptsize$\blacksquare$}
\end{lemdef}
	Except for Proposition \ref{prop:noGroupStruc}, where we show that the spaces $\qRY$ cannot be equipped with group structures continuous w.r.t.\ the above topologies, we will only deal with the space $\qR$ in the sequel. Observe that it is immediate from the above definitions that $\RR$ is a dense open subset of $\qR$.

\begin{lemrem} 
  \label{remtrnas}
  \begin{enumerate}
  \item
    \label{remtrnas1}
    In contrast to the standard LQC case, for $\qR$ the canonical embedding $\iota_\RR\colon \RR\rightarrow \qRR$ does not map $\RR$ into the $\RB$ part, but canonically onto the $\RR$ part, i.e., 
    \begin{align}
      \label{eq:rmap}
      \xi^{-1}(\iota_\RR(x))=x\in\RR\subseteq \qR. 
    \end{align}
    This is because $\xi\big(\xi^{-1}(\iota_\RR(x))\big)(f)=f(x)=\xi(x)(f)$ for all $f\in \rR$. 
  \item
    \label{remtrnas11}
    Since $\xi$ is a homeomorphism, the action $\Transw$ from Corollary \ref{cor:chrisleftact} transfers via $\xi$ to an action $\TTransw\colon \RR\times \qR\rightarrow \qR$ on $\qR$, i.e., $\TTransw(v,\x):=\xi^{-1}\big(\Transw(v,\xi(\x))\big)$. Then, for all $v\in \RR$ and $x\in\RR \subseteq \qR$ we have
    \begin{align}
      \label{eq:sigmaquer}
      \begin{split} 	
	\TTransw(v,x)&\stackrel{\eqref{eq:rmap}}{=}\xi^{-1}(\Transw(v,\iota_\RR(x)))=\xi^{-1}(\iota_\RR(v+x))=\iota_\RR(v+x)\in\RR\subseteq \qR.
      \end{split}
    \end{align}
    So, for $\x\in \RB\subseteq \qR$ and $\{x_\alpha\}_{\alpha\in I}\subseteq \RR\subseteq \qR$ a net with $\lim_\alpha x_\alpha=\x$,\footnote{Here we mean the limit in $\qR$} we have
    \begin{align}
      \label{eq:sigmaquerbohr}	
      \TTransw(v,\x)&=\lim_\alpha \TTransw(v,x_\alpha)\stackrel{\eqref{eq:sigmaquer}}{=}\lim_\alpha\iota_\RR(v+x_\alpha)=\iota'_\RR(v)+  \x\in \RB \subseteq \qR. 
      % 
      % =\xi^{-1}(\Transw(v,\xi(\x)))=\lim_\alpha \xi^{-1}(\Transw(v,\iota_\RR(x_\alpha)))=\lim_\alpha \iota_\RR(v+x_\alpha)\\
      % & =\lim_\alpha \iota_\RR(v+x_\alpha) =\iota'_\RR(v) +_\RB \x. 
    \end{align}
    Here, $\iota'_\RR\colon \RR \rightarrow \RB$ denotes the canonical embedding of $\RR$ into $\RB$, so that in the last term by $+$ we mean the addition in $\RB$.    
    The last step follows from
    
    {\bf Claim 1:} $\{\iota_\RR(v+x_\alpha)\}_{\alpha \in I}$ and $\{\iota'_\RR(v) + \iota'_\RR(x_\alpha)\}_{\alpha \in I}$ (considered as nets in $\qR$) have the same limit in $\qR$, i.e., 
    \begin{align} 
      \label{eq:limgl}
      \lim_\alpha  \iota_\RR(v+x_\alpha) = \lim_\alpha \big(\iota'_\RR(v) + \iota'_\RR(x_\alpha)\big).
    \end{align}	 

    In fact, using {\bf Claim 1}, Equation \eqref{eq:sigmaquerbohr} is clear from $\lim_\alpha \iota'_\RR(x_\alpha)=\x$. 
    This, in turn, holds because by the definition of the topology on $\qR$ for each $l\in \RR$ and each open subset $U\subseteq \mathbb{C}$ with $\x(\chi_l)\in U$ we find   
    $\alpha_0\in I$ such that $x_\alpha \in \chi^{-1}_l(U)$ for all $\alpha\geq \alpha_0$, hence $\iota'_\RR(x_\alpha)\in \GT(\chi_l)^{-1}(U)$ for all $\alpha\geq \alpha_0$. This shows $\lim_\alpha \iota'_\RR(x_\alpha)=\x$ because the Gelfand topology on $\RB$ equals the initial topology w.r.t.\ the Gelfand transforms of the functions $\chi_l$ for $l \in \mathbb{\RR}$, see e.g.\ Subsection 2.3 in \cite{ChrisSymmLQG}.
    
    {\bf Proof of Claim 1:}	 
    \begingroup
    \setlength{\leftmarginii}{15pt}
    \begin{itemize}
    \item
    For each compact $K\subseteq \RR$ we find $\alpha_0\in I$ such that $x_\alpha \in [-v+K]^c$, hence $v+x_\alpha \in K^c$ for all $\alpha\geq \alpha_0$. This is clear from $\lim_\alpha x_\alpha=\x$ because $\x\in \RB\subseteq \qR$ and $[-v+K]^c\sqcup \RB$ is an open neighbourhood of $\x$.  
    \item
    Hence, since we know that $\{\iota_\RR(v+x_\alpha)\}_{\alpha\in I}$ converges to some element of $\qR$, it must converge to some $\ovl{z}\in \RB\subseteq \qR$.
    \item 
     In order to show $\qR\supseteq \{\iota'_\RR(v) + \iota'_\RR(x_\alpha)\}_{\alpha\in I}\rightarrow \ovl{z}\in \qR$, it suffices to verify that we find $\alpha_0\in I$ with $\iota'_\RR(v) + \iota'_\RR(x_\alpha) \in \GT(\chi_l)^{-1}(U)$ for all $\alpha \geq \alpha_0$. Here $l\in \RR$ and $U\subseteq \mathbb{C}$ is open with $\z(\chi_l)\in U$. 
     In fact, then the above net converges in $\RB$ to $\ovl{z}\in \RB$, hence in $\qR$ because the subspace topology of $\RB$ w.r.t.\ the topology on $\qR$ equals its usual one. 
     \item
    Now,
	\begin{align*}    
    	\iota'_\RR(v) + \iota'_\RR(x_\alpha) \in \GT(\chi_l)^{-1}(U)\quad &\Longleftrightarrow\quad (\iota'_\RR(v) + \iota'_\RR(x_\alpha))(\chi_l) \in U\\
    		\quad &\Longleftrightarrow\quad\chi_l(v)\cdot \chi_l(x_\alpha)\in U\\
    	    \quad &\Longleftrightarrow\quad\iota_\RR(v+x_\alpha) \in \chi_l^{-1}(U),
    \end{align*}
    and we find $\alpha_0\in I$ such that $\iota_\RR(v+x_\alpha) \in \chi_l^{-1}(U)$ for all $\alpha\geq \alpha_0$. The last statement is clear from $\lim_\alpha \iota_\RR(v+x_\alpha)=\ovl{z}$  and that a base of neighbourhoods of $\z$ in $\qR$ is formed by finite intersections of sets of \textbf{Type 2} and \textbf{Type 3}.
    \hspace*{\fill}{\footnotesize$\dagger$}
    \end{itemize}
    \endgroup
  \item
    \label{lemma:nullBohr}
    Observe that $\xi(\ovl{x})(\chi_l)=1$ for all $l\in \mathbb{R}$ iff $\ovl{x}\in \{0_\mathbb{R},0_{\mathrm{Bohr}}\}$. In fact,  
    % \begin{beweis}
    %   Of course, 
    of course we have $\xi(\ovl{x})(\chi_l)=1$ for $\ovl{x}\in \{0_\mathbb{R},0_{\mathrm{Bohr}}\}$. Conversely, if $\xi(\ovl{x})(\chi_l)=1$ for all $l\in \mathbb{R}$ and
    $\ovl{x}=y\in \mathbb{R}$, then $y= 0$ because $\xi(\ovl{x})(\chi_{\pi/2y})=\I$ if $y\neq 0$. Similarly, if $\ovl{x}\in \RB$, then $\ovl{x}(\chi_l)=1=0_{\mathrm{Bohr}}(\chi_\tau)$ for all $l \in \mathbb{R}$. Hence, $\ovl{x}=0_{\mathrm{Bohr}}$ as the functions $\chi_l$ generate $\CAP(\RR)$.
    % \end{beweis}
    \hspace*{\fill}{$\lozenge$}
  \end{enumerate}
  % \endgroup
\end{lemrem}
The next proposition shows that $\qRR\cong \qR$ cannot be equipped with a group structure continuous w.r.t.\ its canonical Gelfand topology.
\begin{proposition}
  \label{prop:noGroupStruc}
  There is no continuous group structure on $\qRY$. 
  \begin{proof}
    Assume there is such a group structure with multiplication $\star$ and unit element $e$.
    In a first step we show that there is some $\ovl{x}\in \qRY$ for which the continuous\footnote{Recall that the relative topology of $\RB$ w.r.t.\ that on $\qRY$ equals its usual one.} 
    restriction
    $\star[\ovl{x},\cdot\:]|_{\RB}$ 
    takes at least one value in $\oRR$. 
    So, assume that this is not the case. Then $\star[\ovl{x},\RB]\subseteq \RB$ for each $\ovl{x} \in \qRY$ so that for $\psi\in \RB\neq \emptyset$ we obtain $e=\star\big[\psi^{-1},\psi\big]\in\RB$. It follows that $\qRY=\star\big[\qRY,e\big]\subseteq \RB$, which is impossible. 
    Consequently, we find $\ovl{x} \in \qRY$ and $\psi\in \RB$ such that $\ovl{x}\star \psi\in \oRR$.
    Then, the preimage $U$ of $Y$ under the continuous map $\star[\ovl{x},\cdot\:]|_{\RB}$  is a non-empty open subset of $\RB$.
    Since $\RB$ is compact, finitely many translates of the form $\psi+ U$ cover $\RB$, so that for the cardinality of $\RB$ we obtain
    \begin{align*}
      | \RB |&=\left|\bigcup_{i=0}^n\psi_i+U\right|\leq \left|\bigsqcup_{i=0}^n U\right| 
      =\left|\bigsqcup_{i=0}^n \star[\ovl{x},U]\right|\leq n|\mathbb{R}|=|\mathbb{R}|.
    \end{align*}
    But, this is impossible because $|\RB|> |\mathbb{R}|$.
	In fact, by Lemma \ref{lemma:Bohriso} $\RB=\Spec(\CAP(\RR))$ is in bijection with the set $\RB'$ of all (not necessarily continuous) unital homomorphisms $\Gamma\rightarrow S^1$ for $\Gamma$ the dual\footnote{The set of all continuous characters on $\RR$, i.e., $\gamma=\{\chi_l\:|\: l\in \RR\}$.} group of $\RR$. Consequently, it suffices to show that $|\RB'|> |\mathbb{R}|$.     
    
    For this, let $\{\tau_\alpha\}_{\alpha\in I}\subseteq \mathbb{R}$ be a $\mathbb{Q}$-base of $\mathbb{R}$, then\footnote{$\mathbb{R}$ equals the set $F$ of all finite subsets of $\mathbb{Q}\times I$. Then $|F|=|\mathbb{Q}\times I|$ since $\mathbb{Q}\times I$ is infinite. Similarly, one has $|\mathbb{Q}\times I|=|I|$ because $I$ is infinite.} 
    $|I|=|\mathbb{R}|$ and we obtain an injective map $\iota\colon 2^I\rightarrow \RB$ as follows.
    For $J\subseteq I$ let 
    \begin{align*}
      \delta_J(\alpha):=\left\{
	\begin{array}{ll}
          0  & \mbox{if } \alpha\in J  \\
          \frac{2\pi}{\tau_\alpha} & \mbox{if }  \alpha\notin J
	\end{array}\right. 
    \end{align*}
    and define $\psi(J)\colon \Gamma\rightarrow S^1$ by $\psi(J)(\chi_0):=1$ as well as
    \begin{align*}    	
      \psi(J)\left(\chi_{\tau}\right):=\prod_{i=1}^n \chi_{q_i \cdot \tau_{\alpha_i}}(\delta_J(\alpha_i))\quad \text{for } \tau=\sum_{i=1}^l q_i\cdot \tau_{\alpha_i} \text{ with } q_1,\dots,q_l \in \mathbb{Q}.
    \end{align*}    
    Then $\iota\colon J\mapsto \psi(J)$ is injective because
    $\psi(J)(\chi_{q\cdot \tau_\alpha})=1$ for all $q\in \mathbb{Q}$ iff $\alpha\in J$, hence %. This shows 
    $|\RB|=|\RB'|\geq |2^I|=|2^\RR|>|\mathbb{R}|$. 
  \end{proof}
\end{proposition}
So, since it is not possible to equip $\qR$, i.e., $\qRR$ with a canonical Haar measure, we now concentrate on some general properties of normalized Radon measures on $\qR$.
\begin{lemma}
  \label{lemma:Radon}
  Let $\BRq$, $\Borel(\mathbb{R})$ and $\Borel(\RB)$ denote the Borel $\sigma$-algebras of the topological spaces $\qR$, $\RR$ and $\RB$, respectively. 
  \begin{enumerate}
  \item	
    \label{lemma:Radon1}
    We have $\BRq=\Borel(\mathbb{R})\sqcup\Borel(\RB)$.
  \item
    \label{lemma:Radon2}
    If $\mu$ is a finite Radon measure on $\BRq$, then so are $\mu|_{\Borel(\mathbb{R})}$ and $\mu|_{\Borel(\RB)}$. Conversely, if $\mu_1$, $\mu_2$ are finite Radon measures on $\Borel(\mathbb{R})$ and $\Borel(\RB)$, respectively, then 
    \begin{align}
      \label{eq:RadonMeasures}	
      \mu(A):=\mu_1(A\cap \mathbb{R})+ \mu_2(A\cap \RB)\quad \text{for}\quad A\in \BRq	
    \end{align}
    is a finite Radon measure on $\BRq$. 
  \end{enumerate}
  \begin{proof}
    \begin{enumerate}
    \item
      First observe that the right hand side is a $\sigma$-algebra and remember that the relative topologies on $\mathbb{R}$ and $\RB$ w.r.t.\ the topology on $\qR$ coincide with their usual ones. So, if 
      $U\subseteq \qR$ is open, then $U \cap \mathbb{R}$, $U \cap \RB$ are open in $\mathbb{R}$ and $\RB$, respectively. Hence, $U=(U \cap \mathbb{R}) \sqcup (U \cap \RB) \subseteq \Borel(\mathbb{R})\sqcup\Borel(\RB)$, showing that $\BRq\subseteq \Borel(\mathbb{R})\sqcup\Borel(\RB)$. Conversely, if $U\subseteq \mathbb{R}$ is open, then $U$ is open in $\qR$, hence $\Borel(\mathbb{R})\subseteq \BRq$. Finally, if $A\subseteq \RB$ is closed, then $A$ is compact in $\RB$. This implies compactness of $A$ in $\qR$, so that $A\in \BRq$ for each closed $A\subseteq \RB$. Since $\Borel(\RB)$ is generated by all such closed subsets, we have $\Borel(\RB)\subseteq \BRq$ as well.
    \item
      The measures $\mu|_{\Borel(\mathbb{R})}$ and $\mu|_{\Borel(\RB)}$ are well-defined by \ref{lemma:Radon1}) and obviously finite. So, it remains to show their inner regularity. But inner regularity is clear because a subset of $\mathbb{R}$ or $\RB$ is compact iff it is so w.r.t.\ topology on $\qR$.      
      For the second statement, let $\mu$ be defined by \eqref{eq:RadonMeasures}. Then $\mu$ is a finite Borel measure by \ref{lemma:Radon1}) and its inner regularity follows by a simple $\epsilon\slash 2$ argument from the inner regularities of $\mu_1$ and $\mu_2$.
    \end{enumerate}		
  \end{proof}
\end{lemma}
So, by the above lemma each normalized Radon measure on $\qR$ can be written in the form
\begin{align}
	\label{eq:mutetc}
  \mu(A)=t\:\mu_1(A\cap \mathbb{R})+ (1-t)\:\mu_2(A\cap \RB) \qquad \forall\: A\in \BRq
\end{align}
for $t\in [0,1]$ and normalized\footnote{Of course, if $t=0$ or $t=1$, it doesn't matter which normalized Radon measure we choose on $\RR$ or $\RB$, respectively. So, in these cases we allow $\mu_1=0$ and $\mu_2=0$, respectively.}  Radon measures $\mu_1$ and $\mu_2$ on $ \Borel(\mathbb{R})$ and $\Borel(\RB)$, respectively. So, the crucial step is to fix the measures $\mu_1, \mu_2$ and the parameter $t$.
% \begin{remark}
%   \label{rem:measurefix}

For the dependence of the induced Hilbert space structure on the parameter $t$ observe that for $\mu_1,\mu_2$ fixed, $t_1,t_2\in (0,1)$ and $\mu_{t_1},\mu_{t_2}$ the respective measures defined by \eqref{eq:mutetc},
the spaces $\Lzw{\qR}{\mu_{t_1}}$ and $\Lzw{\qR}{\mu_{t_2}}$ are isometrically isomorphic.  
In fact, for $A\in \BRq$ let $\chi_A$ denote the respective characteristic function and define the map
\begin{align*}
  \phi \colon \Lzw{\qR}{\mu_{t_1}}&\rightarrow \Lzw{\qR}{\mu_{t_2}}\\
  f & \mapsto  \sqrt{\frac{t_1}{t_2}} \:\chi_{\RR}\cdot f + \sqrt{\frac{(1-t_1)}{(1-t_2)}}\:\chi_{\RB}\cdot f.
\end{align*}
Then, $\phi$ is an isometric isomorphism just by the general transformation formula, so that the parameter $t$ gives rise to at most $3$ different (not canonically isomorphic) Hilbert space structures, see also Lemma \ref{lemma:techlemma}. 
% \hspace*{\fill}$\lozenge$
% \end{remark}

Anyhow, the next corollary to Proposition \ref{lemma:bohrmassdichttrans} and Lemma and Remark \ref{remtrnas}.\ref{remtrnas11} shows that invariance of $\mu$ under the translations $\TTransw_v\colon \qR\rightarrow \qR$ already forces $t=0$ and $\mu_2=\muB$. 
\begin{corollary}
  \label{cor:eindbohr}
  \begin{enumerate}
  \item
  If $\mu$ is a normalized Radon measure on $\qR$ with $\TTransw_v(\mu)=\mu$ for all $v\in\RR$, then $\mu=\muB$, i.e., $t=0$. In particular, if one wants $\TTransw_v^* \colon \Lzw{\qR}{\mu}\rightarrow \Lzw{\qR}{\mu}$, $f\mapsto f\cp \Transw_v$ to be unitary for each $v\in \RR$, then $t$ has necessarily to be zero and $\mu_2$ has to equal $\muB$.
  \item
  	The one-parameter group $\big\{\hspace{-1pt}\TTransw_v^*\big\}_{v\in \RR}$ of unitary operators
	\begin{align*}
	\TTransw_v^* \colon \Lzw{\qR}{\muB}\rightarrow \Lzw{\qR}{\muB},\quad 	f\mapsto f\cp \TTransw_v
	\end{align*}  
  is strongly continuous.\footnote{Of course, here we mean 
	$\muB(A):=\muB(A\cap \RB)$ for all $A\in \Borel(\RR\sqcup \RB)$.} 
  \end{enumerate}
  \end{corollary}
  \begin{proof}
  \begin{enumerate}
  \item
    Assume that $t>0$ and that $\mu_1$ is finite. If $\mu(K)>0$ for $K\subseteq \RR$ compact, then $\mu_1(\RR)=\infty$ just by $\sigma$-additivity and \eqref{eq:sigmaquer}. Consequently, $\mu_1(K)=0$ for all compact subsets $K\subseteq \RR$, hence $\mu_1=0$. Now \eqref{eq:sigmaquerbohr} shows that $\mu_2$ fulfils the requirements of Proposition \ref{lemma:bohrmassdichttrans}.\ref{lemma:bohrmassdichttrans1}, so that $\mu_2=\muB$ follows. 
    Finally, unitality of the maps $\TTransw_v$ implies $\mu(\TTransw(v,K))=\mu(K)$ for each compact subset $K\subseteq \qR$ and each $v\in \RR$, hence $\TTransw_v(\mu)=\mu$ for each $v\in \RR$ by inner regularity. Consequently, the last part is clear from the first one.
    \item
    This is clear from Proposition \ref{lemma:bohrmassdichttrans}.\ref{lemma:bohrmassdichttrans2} since by \eqref{eq:sigmaquer} and \eqref{eq:sigmaquerbohr} we have 
	\begin{align*}
		\|\TTransw^*_v(f)-\TTransw^*_{v'}(f)\|_2=\|\Transl^*_v(f|_{\RB})-\Transl^*_{v'}(f|_{\RB})\|_2\qquad \forall\: f\in \Lzw{\qR}{\muB}.    
    \end{align*}
  \end{enumerate}
  \end{proof}
Even if Corollary \ref{cor:eindbohr} singles out the measure $\muB$ (and support the standard LQC approach from the mathematical side), in the following subsections (except for the next short one), we 
% Except for the next short subsection, we now 
will be concerned with the construction of a projective structure and consistent families of normalized Radon measures on $\qR$. Basically, this will be done in analogy to the construction in \cite{oai:arXiv.org:0704.2397} for the standard LQC configuration space $\RB$, and we will end up with singling out the measures of the form 
\begin{align}
  \label{eq:fammeas}
  \mu(A)=t\:\adif(\lambda)(A\cap \mathbb{R})+ (1-t)\:\muB(A\cap \RB) \qquad \forall\: A\in \BRq
\end{align}
for $t\in [0,1]$. Here, %$\muB$ is the Haar measure on $\RB$ and 
$\adif(\muL)$ denotes the push forward of the restriction of the Lebesgue measure to $\Borel((0,1))$ by a homeomorphism $\adif\colon (0,1)\rightarrow \RR$. At least the $\RB$ part of these measures then is in line with the above Corollary. 
The main intention for our investigations is rather to provide a mathematically satisfying derivation of these choices than a physically justification for this.  

However, before we start constructing measures, we first investigate the inclusion relations between the spaces $\A_\w$ and $\ARQw\cong \ARRQw= \qRR\cong \qR$.
\subsection{Quantization vs. Reduction}
\label{subsec:QuantvsRed}
As we have seen in Remark \ref{StanLQC}.\ref{StanLQC2}, $\ARRQLL\cong \AQRL$ holds, i.e., 
quantization and reduction commute if one only uses linear curves for both the full quantum configuration space $\A_\lin$ and the quantized reduced classical space $\ARRQLL$. 
In this short subsection we show that this is no longer true if one takes all embedded analytic curves into account. 
We start with the following diagram sketching the relevant spaces and their relations

\begin{figure}[h]
  \begin{minipage}[h]{\textwidth}
    \begin{center}
      \makebox[0pt]{
        \begin{xy}
          \xymatrix{
            \qR \ar@{->}[r]^-{\xi}_-{\cong} &\qRR \ar@{->}[r]^-{\ovl{i^*_\AR}}_-{\cong}   &  \: \ARQw \: \ar@{->}[r]^-{\subseteq} & \AQRw\ar@{->}[r]_-{\cong}^-{\kappa}   &\IHOMW  \ar@{->}[r]^-{\Omega}_-{\cong} & \ITRHOMW  \\ 
            & \mathbb{R} \ar@{->}[u]^-{\iota_\mathbb{R}} \ar@{->}[r]^-{\cong}    &\: \AR\: \ar@{->}[u]^-{\iota_{\AR}}, &&}
        \end{xy}}
      \caption*{\hspace{-120pt}\textbf{Embedding: }$\varsigma:=\Omega\cp \kappa \cp \ovl{i^*_\AR}\cp  \xi\colon \qR\hookrightarrow \ITRHOMW$}
    \end{center}
  \end{minipage} 
\end{figure}
\FloatBarrier

\noindent
and denote the concatenation of all these maps by $\varsigma\colon \qR\rightarrow \IHOMW$. Here, $\Omega=\Omega_\nu$ for $\nu$ the canonical choice $\nu_x=(x,\me)$ for all $x\in M$, cf.\ last two parts of Remark \ref{rem:homaction}.
To better understand what this map $\varsigma$ actually does, observe that:
    \begingroup
    \setlength{\leftmargini}{20pt}
    \begin{itemize}
    \item[a)]
  		For $\rho$ the standard representation of $\SU$, we obtain from \eqref{eq:hilfseq} that
  		\begin{align*}
  			(\Omega\cp\kappa)(\ovl{\w})(\gamma)=\big(\ovl{\w}([h_\gamma^\nu]_{ij})\big)_{ij}\qquad \forall\:\gamma\in \Paw,\:\forall\:\ovl{\w}\in \A.
  		\end{align*}     
    \item[b)]
    Applying the definitions, we find that
    \begin{align*}
     	\ovl{i^*_\AR}(\ux)([h_\gamma^\nu]_{ij})=\ux(i_\AR^*([h_\gamma^\nu]_{ij}))=\ux\big([h_\gamma^\nu]_{ij}\cp i_\AR\big)=\ux\big([h_\gamma^\nu]_{ij}|_\AR\big)\qquad \forall\: \ux\in \qRR. 
    \end{align*}
    \item[c)]
    For a linear curve $\gamma\colon [0,l]\rightarrow \RR^3$, $t\mapsto t\cdot \vv$ for $\vv\in \RR^3$ with $\|\vv\|=1$ we have
      \begin{align*}
     		[h_\gamma^\nu]_{ij}(\w^c)=\big(\pr_2\cp \parall{\gamma}{\w^c}\big)(\gamma(0),\me)= \exp(-cl \cdot\murs(\vv))\stackrel{\eqref{eq:expSU2}}{=} \cos(-cl)\cdot \me + \sin(-cl) \cdot\murs(\vv).
    \end{align*}
    Here, the first step is clear from the definitions, and for the second step one might apply \eqref{eq:trivpar} to the fact that $\gamma(t)=\wm(\exp(t\vv),0)$ is Lie algebra generated (cf.\ Definition \ref{def:analytLieAlgBD}.\ref{def:analytLieAlgBD2}), i.e., $\gamma=\gamma_{\vv}^0|_{[0,l]}$ for $0\in \RR^3$ and $\vv \in \RR^3\times \su$ the Lie algebra of $\Ge=\Gee$.    
    \end{itemize}
    \endgroup 
 \noindent
Using this, we obtain the following Corollary to Lemma and Definition \ref{remdefchris}.\ref{lemma:nullBohr}. 
\begin{corollary}
  \label{cor:nullbohr}
  If $\varsigma(\ovl{x})(\gamma)=\me$ for all $\gamma\in \Pal$,
  then $\ovl{x}\in\{0_{\mathrm{Bohr}},0_\mathbb{R}\}$. 
  \begin{proof}
  It suffices to show the claim for curves as in c) with $\vv=\vec{e}_1$ and $l>0$. Then 
    \begin{align*}
     % \label{eq:DeltavonRquer}
      \begin{split}
        \me&=\varsigma(\ovl{x})(\gamma)
        =(\Omega\cp \kappa)\Big(\ovl{i^*_\AR}(\xi(\ovl{x}))\Big)(\gamma)\\
        \!\!&\stackrel{\text{a)}}{=}\left(\Big(\ovl{i^*_\AR}(\xi(\ovl{x}))\Big)\big([h_\gamma^\nu]_{ij}\big)\right)_{ij}
        \stackrel{\text{b)}}{=}\left(\xi(\ovl{x})\left([h_\gamma^\nu]_{ij}|_\AR\right)\right)_{ij}\\    
        \!\!&\stackrel{\text{c})}{=}\begin{pmatrix} \xi(\ovl{x})(c\mapsto \cos(-l c)) & \mathrm{i}\: \xi(\ovl{x})(c\mapsto \sin(-l c))  \\ \mathrm{i}\: \xi(\ovl{x})(c\mapsto \sin(-l c)) & \xi(\ovl{x})(c\mapsto \cos(-l c)) \end{pmatrix},
      \end{split}
    \end{align*}
    hence $\xi(\ovl{x})(\chi_{l})=1$ for all $l\in \mathbb{R}$. So, the claim is clear from  
    Lemma and Definition \ref{remdefchris}.\ref{lemma:nullBohr}. 
  \end{proof}
\end{corollary}

\begin{proposition} 
  \label{lemma:propincl}
  We have $\ARQw\subsetneq \AQRw$.
 \end{proposition}
  \begin{proof}
	Let 
  $\tilde{\hommm} \colon \mathbb{R}_{>0}\times (0,2\pi)\rightarrow \mathbb{R}$ be a function with
  \begin{align*}
    \tilde{\hommm}(r,x+y)=\tilde{\hommm}(r,x)+\tilde{\hommm}(r,y) \:\bmod \: 2\pi %\quad \text{whenever}\quad r\in \mathbb{R}_{>0}\quad\text{and}\quad x,y,x+y\in (0,2\pi).
    %\qquad \forall\:x,y'\in (0,1).%     +2\pi n \qquad \text{ for some } n\in \mathbb{Z} 
  \end{align*}
  whenever $r\in \mathbb{R}_{>0}$ and $x,y,x+y\in (0,2\pi)$. 
Then, we obtain an element $\hommm \in \ITRHOMW$ 
  if we define  \big(cf.\ (b) in Convention \ref{conv:sutwo1}.\ref{conv:sutwo2} for the definition of $\Pacirc$ and $\gc{n}{r}{x}{\tau}$\big)
  \[ 
  \hommm(\gamma):= 
  \begin{cases} 
    \exp\big(\tilde{\hommm}(\|\vec{r}\|,\tau)\: \murs(\vec{n})\big) &\mbox{if } \gamma \csim \gc{n}{r}{x}{\tau}\in \Pacirc\\
    \me & \mbox{else}. 
  \end{cases}
  \] 
  Here, well-definedness follows from analyticity and Lemma \ref{lemma:sim}.\ref{lemma:sim5}.\footnote{Observe that $\gc{n}{r}{x}{\tau}\psim x + \gamma^\rr_{(0,\murs(\nn))}\big|_{[0,\tau/2]}$ for $\rr$ considered as point in $\RR^3$ and $(0,\murs(\nn)\in \RR^3\times \su=\mg$, see also beginning of the next subsection.} Moreover, invariance is straightforward to see. Here, $0_\RR$ corresponds to the choice $\wt{\hommm}=0$ ($\w_0$ is the trivial connection) but the homomorphism that corresponds to $\NB$ is not so easy to compute. However, it is easily checked that choosing different maps $\wt{\hommm}$ we obtain more than two different invariant homomorphisms with $\hommm(\gamma)=\me$ for all $\gamma\in \Pal$. Since by Corollary \ref{cor:nullbohr} that cannot all come from an element of $\qR\cong \ARQw$, the claim follows. 
\end{proof}
For the rest of this section we will be concerned with the construction of a projective structure (and consistent families of normalized Radon measures) on the space $\qR\cong\qRR$. The next subsection (except for Lemma \ref{lemma:WeierErzeuger}.\ref{lemma:WeierErzeuger2}) then serves as a motivation for our later constructions. Basically, there we discuss which problems occur if one tries to use projection maps involving the identification of $\qRR$ with a subset of $\IHOMW$. However, some of the investigations are quite technical and exhausting, and the reader not so interested in these difficulties may just take a look at the second part of Lemma \ref{lemma:WeierErzeuger} in order to proceed with Subsection \ref{subsec:ProjStrucon}.

\subsection{Motivation of the Construction}
\label{subsec:Motivation} 
Recall that the $\Cstar$-algebra $\rR=C_0(\RR)\oplus \CAP(\RR)$ is already generated by parallel transports along linear and circular curves  \cite{ChrisSymmLQG}, i.e., that the $\Cstar$-algebras $\rR=\ovl{\PaC_{\w}|_{\AR}}$ and $\rR_{\lin\mc}:=\ovl{\PaC_{\lin\mc}|_{\AR}}$ coincide. Here, $\PaC_{\lin\mc}$ denotes the
$\Cstar$-algebra of cylindrical functions that corresponds to the set of curves $\Pa_{\lin\mc}=\Pall \sqcup \Pacirc$ for $\Pall$ and $\Pacirc$ defined as in Convention \ref{conv:sutwo1}.\ref{conv:sutwo2}. This means that an element of $\qRR$ is completely determined by its values on the generators of $\PaC_{\lin\mc}$, i.e., on the matrix entries of the parallel transport functions $\RR \ni c\mapsto \parall{\gamma}{\w^c}$ for $\gamma\in \Pa_{\lin\mc}$.
It follows that, using the natural identification of $\qRR$ with a subset of $\IHOMW$ via $\varsigma'$

\begin{figure}[h]
  \begin{minipage}[h]{\textwidth}
    \begin{center}
      \makebox[0pt]{
        \begin{xy}
          \xymatrix{
            &\qRR \ar@{->}[r]^-{\ovl{i^*_\AR}}_-{\cong}   &  \: \ARQw \: \ar@{->}[r]^-{\subseteq} & \AQRw\ar@{->}[r]_-{\cong}^-{\kappa}   &\IHOMW, %\\  
         }
        \end{xy}}
      % \vspace{-3.5ex} 
      \caption*{\hspace{-120pt}\textbf{Embedding: }$\varsigma':=\kappa \cp \ovl{i^*_\AR}\colon \qRR\hookrightarrow \IHOMW$}
    \end{center}
  \end{minipage} 
\end{figure}
\FloatBarrier
\noindent
 the values of the homomorphism $\varsigma'(\ux)$ that corresponds to $\ux\in \qRR$ are completely determined by its values on the elements of $\Pa_{\lin\mc}$. Moreover, due to the invariance of $\varsigma'(\ux)$ it suffices to consider elements of $\Pa_{\lin\mc}$ which are of the form
\begin{align}
	\label{eq:kkurven}
	\gamma_l:=\gamma_{\vec{e}_1,l}\qquad\forall\: l>0\qquad\qquad\text{and}\qquad\qquad\gamma_{\tau,r}:=\gamma_{\vec{e}_3,r\vec{e}_1}^{0,\tau}\qquad\forall\: r>0,\:\forall\: \tau\in (0,2\pi),
\end{align}
i.e., we can fix a traversing direction for linear curves, and a midpoint together with a traversing plain for the circular ones.
 
Now, since all these curves are Lie algebra generated, it makes sense to use the maps $\pi_p$ (see also b) below) from Lemma and Definition \ref{def:topo} which we have introduced Subsection \ref{sec:ConSp} in order to investigate the space $\IHOMLAS$. For this observe that 
\begingroup
\setlength{\leftmargini}{20pt}
\begin{itemize}
\item[a)]
The elements of $\Pa_{\lin\mc}$ are Lie algebra generated because
\begin{align}
  \label{eq:kurven}
  x+\gamma_{\vv,l}=\gamma^x_{(\vv,0)}\big|_{[0,l]}\qquad\qquad\text{and}\qquad\qquad\gc{n}{r}{x}{\tau}\psim x + \gamma^\rr_{(0,\murs(\nn))}\big|_{[0,\tau/2]}
\end{align}
where the second equivalence is clear from the explanations to \eqref{eq:expSU2} in Convention \ref{conv:sutwo1}.\ref{conv:sutwo111}. 
\item[b)]
	 For $p\in P$, $x=\pi(p)$ and $\g\in \mg\backslash \mg_x$ we have 
\begin{align*} 
    \pi_p(\g,l,\homm) = \Delta\big(\Phi_{\exp(l\g)}(p),\homm\big(\gamma_\g^{x}|_{[0,l]}\big)(p)\big)\qquad \forall\: l<\tau_\g. 
\end{align*}
 Hence, 
 	$\pi_p(\g,l,\cdot)\colon \IHOM\rightarrow \SU$ 
 	assigns to $\homm\in \IHOM$ the difference in $F_{\gag(l)}$ between $\Phi_{\exp(l\g)}(p)$ and $\homm\big(\gamma_\g^{x}|_{[0,l]}\big)(p)\big)$. Obviously, $\pi_p(\g,l,\cdot)$ is continuous. 
 	\item[c)]
 	The big advantage of using the maps $\pi_p$ is that for $\g\in \mg\backslash \mg_x$ and $\homm\in \IHOM$ fixed, $\im[\pi_p(\g,\cdot,\homm)]$ is contained in a maximal torus in $\SU$. This is clear from Lemma \ref{lemma:torus}.\ref{lemma:torus1} and (combine \eqref{eq:Equi} with \eqref{eq:pip})
\begin{align*}
	\pi_p(\g,l+l',\homm)=\pi_p(\g,l,\homm)\cdot \pi_p(\g,l',\homm)\qquad \text{for }l,l',l+l'<\tau_\g.
\end{align*} 
\end{itemize}
\endgroup
\noindent
It follows that $\qRR$ is separated by the maps 
\begin{align}
\label{eq:projmaps}
\begin{split}
	\pi_l&\colon \ux \mapsto \pi_{(0,\me)}\big((\vec{e}_1,0),l,\varsigma'(\ux)\big)\qquad \forall\: l>0\\
	 \pi_{\tau,r}&\colon \ux \mapsto \pi_{(r\cdot \vec{e}_1,\me)}\big((0,\tau_3),\textstyle\frac{\tau}{2},\varsigma'(\ux)\big)\quad \forall\:r>0\:,\forall\: \tau\in (0,2\pi).
	 \end{split}
\end{align}
Here,  $\pi_l$ and $\pi_{\tau,r}$ correspond to the choices, see \eqref{eq:kkurven} and \eqref{eq:kurven} 
\begin{align*}
	\gamma_l=\gamma^0_{(\vec{e}_1,0)}|_{[0,l]},\: p=(0,\me)\quad\qquad\text{and}\quad\qquad  \gamma_{\tau,r}=\gamma^{r\vec{e}_1}_{(0,\tau_3)}|_{[0,\tau/2]},\: p=(r\cdot\vec{e}_1,\me),
\end{align*}
respectively.

The question whether we can use these projection maps in order to define reasonable measures on $\qRR$ then depends crucially on in which way the maximal tori (see c)) the maps
	\begin{align*}
		\pi_{\bullet}(\ux)\colon \RR_{>0}\rightarrow \SU,\: l \mapsto \pi_l(\ux)\qquad\quad\text{and}\quad\qquad \pi_{\bullet,r}(\ux)\colon (0,2\pi)\rightarrow \SU,\: \tau \mapsto \pi_{\tau,r}(\ux) 
	\end{align*}
	are mapping to depend on $\ux\in \qRR$. 
\begingroup
\setlength{\leftmargini}{15pt}
\begin{itemize}
\item
	For $\pi_\bullet$ this question is immediately answered since a straightforward calculation shows that
	\begin{align}
 	 \label{eq:patralll}
 	 \pi_l(\iota_\RR(c))&=h_{\gamma_l}^{\nu}(\w^c)
% 
 	 %(\Omega\cp \kappa)(\iota_\Con(\w_c))(\gamma_{l})\stackrel{\eqref{eq:hilfseq}}{=}(\iota_\Con(\w_c)=	%([h_{\gamma_l}^\nu)]_{ij})_{ij}=h_{\gamma_l}^\nu(\w_c)\stackrel{\eqref{eq:patralinc}}{=}    
% 	 
 	 =\begin{pmatrix} \cos(lc) &\I\sin(lc)  \\ \I\sin(lc) & \cos(lc) \end{pmatrix}
 	 =\exp(-lc \tau_1)\in H_{\tau_1}\qquad \forall\: c\in \RR,\forall\: l>0
	\end{align}
	for $\iota_\RR \colon \AR \cong \RR \rightarrow \qRR$ the canonical embedding of $\RR$ into $\Spec(\CAP(\RR)\oplus C_0(\RR))$. Since $\im[\iota_\RR]$ is dense in $\qRR$, since $\pi_l$ is continuous and since $H_{\tau_1}$ is closed, we even have $\pi_l(\qRR)=H_{\tau_1}$ for all $l>0$. So, using $\pi_l$ as projection maps, we could take the Haar measure on $H_{\tau_1}\cong S^1$ in order to define a respective consistent family of normalized Radon measures.
\item
	For $\pi_{\bullet,r}$ this dependence is more complicated. Here, a straightforward calculation shows that
  \begin{align}
    \label{eq:patrall}
    \begin{split}
      \pi_{\tau,r}(\iota_\RR(c))=\exp\big(\textstyle\frac{\tau}{2}\tau_3\big)^{-1}\cdot h_{\gamma_{\tau,r}}^\nu(\w^c)
      %&= \exp\left(\textstyle\frac{\tau}{2}\tau_3\right)^{-1}\cdot h_{\gamma_{\tau,r}}^\nu(\w_c)\\
%      &=\exp\left(\textstyle\frac{\tau}{2}\tau_3\right)^{-1}\cdot \pr_2 \cp \parall{\gamma_{\tau,r}}{\w_c}((\gamma_{\tau,r}(0),\me))\\
%      &\!\!\stackrel{\eqref{eq:trivpar}}{=}\exp\left(\textstyle\frac{\tau}{2}\tau_3\right)^{-1}\cdot \pr_2\left[\Phi_{(\gamma_{\tau,r}(0),\me)}\Big(\!\exp\left(\textstyle\frac{\tau}{2}\tau_3\right)\!\Big)\cdot \exp\Big(\!-\textstyle\frac{\tau}{2}\cdot\w_c\big(\wt{\tau_3}\big((\gamma_{\tau,r}(0),\me)\big)\big)\Big)\right]\\
%      &=\exp(-\textstyle\frac{\tau}{2}\cdot\w_c(\wt{\tau_3}((\gamma_{\tau,r}(0),\me))))\\
%      &=\exp\left(-\textstyle\frac{\tau}{2}\cdot\w_c\left(\left(\dttB{t}{0}\exp(t\tau_3)(\gamma_{\tau,r}(0)),\tau_3\right)\right)\right)\\
%      % 
%      % &=\exp\left(-\textstyle\frac{\tau}{2}\cdot\w_c\left(\left(\dttB{t}{0}\exp(t\tau_3)(\gamma_{\tau,r}(0)),\tau_3\right)\right)\right)\\
%      % 
%      &=\exp\left(-\textstyle\frac{\tau}{2}\cdot\w_c\left(\left(\dttB{t}{0}\exp(t\tau_3)(r \vec{e}_1),\tau_3\right)\right)\right)\\
%      &=\exp\left(-\textstyle\frac{\tau}{2}\cdot\w_c\left(\left(r\murs^{-1}([\tau_3,\tau_1]),\tau_3\right)\right)\right)\\
%      &=\exp\left(-\textstyle\frac{\tau}{2}\cdot\w_c\left(\left(2r\vec{e}_2,\tau_3\right)\right)\right)\\
      &=\exp\hspace{-1pt}\big(\hspace{-1pt}-\textstyle\frac{\tau}{2}\cdot[2r c\cdot\tau_2+\tau_3]\big)%\qquad \forall\: c\in \RR,\forall\: r>0,\:\forall\:\tau\in (0, 2\pi),
    \end{split}
  \end{align}
  holds for all $c\in \RR$, $r>0$ and all $\tau\in (0, 2\pi)$. Hence, $\pi_{\tau,r}(\iota_\RR(c))\in H_{[2rc\cdot\tau_2+\tau_3]}$ for all $\tau\in (0,2\pi)$, and we will see in Lemma \ref{lemma:BildCirc}.\ref{lemma:BildCirc3} that $\pi_{\tau,r}(\qRR\backslash \im[\iota_\RR])= H_{\tau_2}$ holds for all $\tau\in (0,2\pi)$ and all $r>0$. Further details of the set $\im[\pi_{\tau,r}]$ are provided in Lemma \ref{lemma:BildCirc}.
 
  Then, applying \eqref{eq:expSU2} to \eqref{eq:patrall} we immediately see that for $\pc:= \textstyle\sqrt{c^2r^2+\frac{1}{4}}$ we have %by \eqref{eq:expSU2}
\begin{align}
  \label{eq:matrixentr}
  \pi_{\tau,r}(\iota_\RR(c))\stackrel{\eqref{eq:expSU2}}{=}\begin{pmatrix} \cos(\pc \tau)+\frac{\I}{2\pc}\sin(\pc \tau) &\frac{cr}{\pc} \sin(\pc \tau)  \\ -\frac{cr}{\pc}\sin(\pc \tau) & \cos(\pc \tau)-\frac{\I}{2\pc}\sin(\pc \tau)  \end{pmatrix}.
\end{align}
\end{itemize}
\endgroup
\noindent
From this, we conclude the third part of the following
\begin{lemma}
  \label{lemma:WeierErzeuger}
  \begin{enumerate}
  \item
    \label{lemma:WeierErzeuger1}
    Let $f\in C_0(\RR)$ vanishes nowhere. Then, the functions $\{f\}\sqcup \{\chi_l\}_{l\in \mathbb{R}}$ generate a dense $^*$-subalgebra of $C_0(\mathbb{R})\oplus \CAP(\RR)$. If $f$ is in addition injective,\footnote{Recall that $\im[f]\subseteq \mathbb{C}$.} then $f$ generates a dense $^*$-subalgebra of $C_0(\RR)$. 
  \item
    \label{lemma:WeierErzeuger2}
    Each nowhere vanishing injective $f\in C_0(\RR)$ is a homeomorphism onto its image.
  \item
    \label{lemma:WeierErzeuger3}
    The matrix entries of parallel transports along all linear curves (with some fixed traversing direction) and one single circular curve already generate $C_0(\mathbb{R})\oplus \CAP(\RR)$. In particular, the projection maps $\{\pi_l\}_{l>0}$ and $\pi_{\tau,r}$ for some fixed reals $\tau,r>0$ separate the points in $\qRR$.	
  \end{enumerate}
  \begin{beweis}
    \begin{enumerate}
    \item
      Since $f(x)\neq 0$ for all $x\in \mathbb{R}$, the $^*$-algebra $\gG$ generated by $\{f \cdot\chi_l\}_{l \in \mathbb{R}}\subseteq C_0(\mathbb{R})$ separates the points in $\mathbb{R}$ and vanishes nowhere. Consequently, $\gG$ is dense in $C_0(\mathbb{R})$ by the complex Stone-Weierstrass theorem for locally compact Hausdorff spaces. Since $\CAP(\RR)$ is generated by the functions $\{\chi_l\}_{l \in \mathbb{R}}$, the first claim follows. If $f$ is in addition injective, then the $^*$-algebra generated by $f$ is dense in $C_0(\mathbb{R})$ because $f$ separates the points in $\mathbb{R}$ and vanishes nowhere.
    \item
    	Let $\RR\sqcup \{\infty\}$ denote the one point compactification of $\RR$. Then $\ovl{f}\colon \RR\sqcup \{\infty\} \rightarrow \CCC$ defined by $\ovl{f}(\infty):=0$ and $\ovl{f}|_\RR:=f$ is continuous and injective, hence a homeomorphism onto its image $\im\big[\ovl{f}\hspace{1.5pt}\big]=\im[f]\sqcup\{0\}$. Consequently, $f^{-1}=\ovl{f}^{-1}|_{\im[f]}$ is continuous as well.   
    \item
      If $l=\tau r$, then \eqref{eq:patralll} and \eqref{eq:matrixentr} show that 
      \begin{align}
      	\label{eq:funcco}  		
        f(c):=\big[\pi_{\tau,r}(\iota_\RR(c))\big]_{11}-\left[\pi_l(\iota_\RR(c))\right]_{11}=\cos(\pc \tau)+\textstyle\frac{\I}{2\pc}\sin(\pc \tau)-\cos(cr \tau)
        %\qquad \forall\:c\in \RR, 
      \end{align}
      for all $c\in \RR$, 
      hence
      $f\in C_0(\RR)$. Now, if $f(c)$ is zero, then $\sin(\pc \tau)$ must be $0$, hence $\cos(\pc \tau)$ must be $\pm 1$. Then $\cos(cr\tau)$ must be $\pm 1$ as well, showing that $c=\frac{\pi n}{r\tau}$ for some $n\in \mathbb{Z}\backslash\{0\}$. In fact, $n=0$ means $\sin(\tau/2)=0$, which contradicts that $\tau\in (0,2\pi)$. Then
      \begin{align*}  		
        \cos(\pc \tau)=\cos\left(\pi n \textstyle\sqrt{1+\textstyle\frac{\tau^2}{4 \pi^2n^2}}\right),
      \end{align*}  		
      but $n \sqrt{1+\textstyle\frac{\tau^2}{4 \pi^2n^2}}\notin \mathbb{Z}$ because $0<\tau <2\pi$. Consequently, $\cos(\pc \tau)\neq 1$, which contradicts that $\cos(\pc \tau)-\cos(cr\tau)=0$. This shows that $f$ vanishes nowhere. Now, since 
      \begin{align*}      
        h_{\gamma_{l}}^\nu(\w^c) \stackrel{\eqref{eq:patralll}}{=}\pi_{l}(\iota_\RR(c))	\qquad\text{as well as}\qquad
        h_{\gamma_{\tau,r}}^\nu(\w^c) \stackrel{\eqref{eq:patrall}}{=}\exp(\textstyle\frac{\tau}{2}\tau_3)\cdot\pi_{\tau,r}(\iota_\RR(c)),
      \end{align*}      
      $f$ is also contained in the $^*$-algebra generated by the parallel transports along $\gamma_{\tau,r}$ and $\{\gamma_l\}_{l>0}$. Since by  \eqref{eq:patralll} this algebra also contains all characters $\chi_l$,  
       the first statement follows from Part \ref{lemma:WeierErzeuger1}). 

		Now, applying the definitions we find that
		\begin{align}
		\label{eq:pilrtau}
		\begin{split}
        \pi_l(\ux)&=\left(\ux\left([h_{\gamma_l}^\nu]_{ij}|_{\AR}\right)\right)_{ij}\\
        \pi_{\tau,r}(\ux)&= \exp\left(\textstyle\frac{\tau}{2}\tau_3\right)^{-1}\cdot\big(\ux\big([h_{\gamma_{\tau,r}}^\nu]_{ij}|_\AR\big)\big)_{ij}. 
        \end{split}
      \end{align}  
      So, since we have shown that the functions $\{[h_{\gamma_l}^\nu]_{ij}|_\AR\}_{l>0}$ and $[h_{\gamma_{\tau,r}}^\nu]_{ij}|_\AR$ generate $C_0(\RR)\oplus \CAP(\RR)$, and since $\ux\in \Spec(C_0(\RR)\oplus \CAP(\RR))$, the claim follows. 
    \end{enumerate}
  \end{beweis}
\end{lemma}
\begin{remdef}
  \label{rem:bohrproj}
   \begin{enumerate}
   \item
     \label{rem:bohrproj1}
     In the following, for $k\in \NN_{\geq 1}$ by $S^k$ we understand $k$-fold product of the unit circle  $S^1$, i.e., $S^k:=\big(S^1\big)^k$. Since in the beginning of this section we have fixed the structure group of our principal fibre bundle to be $\SU$, this will not be in conflict with our notations.
     \item
    \label{rem:bohrproj2}
  As already stated in Remark \ref{StanLQC}, in standard homogeneous isotropic LQC the quantum configuration space is given by $\RB$ and homeomorphic to $\ITRHOML$.\footnote{The homeomorphism was basically due to the fact that invariance restricts the value of $\hommm\in \ITRHOML$ on a linear curve traversing into the direction $\vv$ to the maximal torus $H_{\vv}$ in $\SU$.} 
  There, a projective structure and a consistent family of normalized Radon measures can be defined as follows: \cite{oai:arXiv.org:0704.2397} 
  \begingroup
\setlength{\leftmarginii}{15pt}  
\begin{itemize}  
  \item
  One considers the set $I$ of all $\ZZ$-independent tuples $L=(l_1,\dots,l_k)$ with $l_1\dots,l_k\in \RR$, and defines $(l_1\dots,l_k) \leqZ (l'_1\dots,l'_{k'})$ iff $l_i\in \Span_\ZZ(l'_1\dots,l'_{k'})$ for all $1\leq i\leq k$. 
\item  
  One defines the projection maps by $\pi_L\colon \RB\rightarrow S^{|L|}=:X_L$, $\psi \mapsto(\psi(\chi_{l_1}),\dots, \psi(\chi_{l_k}))$ as well as the transition maps by 
  \begin{align*}	
    \pi^{L'}_L\colon S^{|L'|}&\rightarrow S^{|L|}\\
    (s_1,\dots ,s_{k'})&\mapsto \textstyle\left(\prod_{i=1}^{k'}s_i^{n_{i1}},\dots,\prod_{i=1}^{k'}s_i^{n_{ik}}\right),
  \end{align*}
  where $l_i=\sum_{j=1}^{k'}n_{ij} l'_j$ for $1\leq i\leq k$. Surjectivity of $\pi_L$ then is clear from Lemma and Convention \ref{lemconv:RBMOD}.\ref{prop:Bohrmod24}, and continuity is immediate from the definitions of the Gelfand topology on $\RB$. The remaining properties of a projective limit then are easily verified. 
\item  
  One fixes the Haar measures $\mu_{|L|}$ on $S^{|L|}=X_L$. This gives rise to a consistent family of normalized Radon measures which exactly corresponds to the Haar measure on $\RB$, i.e., $\pi_L(\muB)=\mu_{|L|}$%for $\mu_L$ the Haar measure on $S^{|L|}$
  , see, e.g., proof of Lemma \ref{lemma:Projm}.
\end{itemize}
\endgroup
  Equivalent to this is to define the projection maps (cf.\ \eqref{eq:tori} for the definition of $H_{\vec{e}_1}$) %\ref{rem:bohrproj1}) one also can define  
  \begin{align*}	
    \pi'_L\colon \ITRHOML&\rightarrow [H_{\vec{e}_1}]^{|L|}=:X'_L\\
    \homm &\mapsto (\homm(\gamma_{\vec{e}_1,l_1}),\dots,\homm(\gamma_{\vec{e}_1,l_k}))
  \end{align*}
  (and respective transition maps),
   and $\mu'_L$ the Haar measure on $[H_{\vec{e}_1}]^{|L|}\cong S^{|L|}$. \hspace*{\fill}$\lozenge$
   \end{enumerate}
\end{remdef}
The third part of Lemma \ref{lemma:WeierErzeuger} shows that for the separation property from Definition \ref{def:ProjLim}.\ref{def:ProjLim3} it suffices to consider the projection maps $\pi_l$ for all $l>0$ and $\pi_{\tau,r}$ for some fixed reals $\tau,r>0$. Let 
\begin{align*}
	\pi'_l:=\pi_l\cp \xi \qquad\qquad \text{and}\qquad\qquad	\pi'_{\tau,r}:=\pi_{\tau,r}\cp \xi.
\end{align*}
 Then, for $L\in I$ as above and 
\begin{align*} 
 	\pi''_L\colon \qR &\rightarrow [H_{\vec{e}_1}]^{|L|}\\
 		 \x&\mapsto \big(\pi'_{l_1}(\x),\dots,\pi'_{l_k}(\x)\big)
\end{align*} 	
 	 we also have $\im[\pi''_L]=[H_{\vec{e}_1}]^{|L|}$.\footnote{We even have $\pi''_L(\RB)=[H_{\vec{e}_1}]^{|L|}$ (for $\RB\subseteq \qR=\RR\sqcup \RB$). This follows from Lemma and Convention \ref{lemconv:RBMOD}.} So, using the projection maps $\pi''_L$ we can use the Haar measure on $[H_{\vec{e}_1}]^{|L|}$, and the crucial question then is  
whether such a canonical measure also exists on the image of $\pi'_{\tau,r}$. In addition to that, a suitable  directed set and corresponding transition maps have to be defined. For these reasons we now investigate the 
image of $\pi'_{\tau,r}$ in more detail. 
\begin{lemma}
  \label{lemma:BildCirc}
  Let $\tau,r>0$ be fixed.
  \begin{enumerate}
  \item
    \label{lemma:BildCirc1}
    $\pi'_{\tau,r}\big(\qR\big)$ is of measure zero w.r.t.\ the Haar measure on $\SU$.
  \item
    \label{lemma:BildCirc2}
    There is no proper Lie subgroup $H\subsetneq \SU$ that contains $\pi'_{\tau,r}\big(\qR\big)$.  
  \item     
    \label{lemma:BildCirc3}
    We have $\pi'_{\tau,r}\big(\qR\big)=\pi'_{\tau,r}\big(\RR\big)\cup H_{\tau_2}$ with
    \begin{align*}    
      \pi'_{\tau,r}(\RB)=H_{\tau_2}\qquad\quad \text{and}\qquad\quad\pi'_{\tau,r}(\RR)\:\cap\: \pi'_{\tau,r}(\RB)= \{\pm\me\}.
    \end{align*}
  \item
    \label{lemma:BildCirc4}
    Let $a_n:=\frac{\sign(n)}{r}\sqrt{\frac{n^2\pi^2}{\tau^2}-\frac{1}{4}}$ for $n\in \mathbb{Z}_{\neq 0}$ and
	\begin{align*}
		A_0:=(a_{-1},a_1)\qquad A_n:=(a_n,a_{n+1})\:\text{ for }\:n\geq 1\qquad A_n:=(a_{n-1}, a_n)\:\text{ for }\:n\leq -1.
	\end{align*}     
Then, $\pi'_{\tau,r}|_{A_n}$ is injective for all $n\in \ZZ$, 
	\begin{align*}
		 \pi'_{\tau,r}(a_{n})=\me\:\text{ iff }\:|n|\text{ is even}\qquad\qquad \pi'_{\tau,r}(a_{n})=-\me\:\text{ iff }\:|n|\text{ is odd}
	\end{align*}
 as well as $\pi'_{\tau,r}\big(A_m\big)\cap \pi'_{\tau,r}\big(A_n\big)=\emptyset$ for all $m,n\in \ZZ$ with $m\neq n$. 
    For increasing $|n|$, the sets
    \begin{align*}
		B_n:=[a_{2n},a_{2(n+1)}]\:\text{ for }\:n\geq 1\qquad\qquad B_n:=[a_{2(n-1)}, a_{2n}]\:\text{ for }\:n\leq -1
	\end{align*}  
      merge to $H_{\tau_2}$ in the following sense. For each 
    $\epsilon > 0$ we find $n_\epsilon\in \mathbb{N}_{\geq1}$ such that for $|n| \geq n_\epsilon$ we have
    \begin{align*}
      \forall\: s\in B_n:\exists\: s' \in H_{\tau_2}:\|s-s'\|_{\mathrm{op}}\leq \epsilon.
    \end{align*}
  \end{enumerate}
  \begin{proof}
    The proof can be found in Appendix \ref{sec:ProofOfLemmaImCirc}. 
  \end{proof}
\end{lemma}
The first and the second part of the above lemma already show that it is hard to equip $\im[\pi'_{\tau,r}]$ with a reasonable measure. 
In addition to that, it is difficult to define a reasonable ordering, i.e., a directed set labeling the projection spaces.  
Indeed, the first thing one might try is to define $\pi'_{\tau,r}\leq \pi'_l$ or $\pi'_l\leq \pi'_{\tau,r}$. However, then one has to define reasonable transition maps between $\im[\pi'_{\tau,r}]$ and $\im[\pi'_l]$, i.e., maps 
\begin{align*}
\TT_1\colon \im[\pi'_{\tau,r}]\rightarrow \im[\pi'_l]  \quad\qquad\text{or}\quad \qquad \TT_2\colon \im[\pi'_l]&\rightarrow \im[\pi'_{\tau,r}]
\end{align*}
with $\TT_1\cp \pi'_{\tau,r}=\pi'_l$ or $\TT_2\cp \pi'_l=\pi'_{\tau,r}$, respectively. This, however, is difficult and even impossible if $\tau r =l$: 
\begingroup
\setlength{\leftmargini}{15pt}
\begin{itemize}
\item
  \itspace
  Let $0, \frac{2\pi}{l}\in \RR\subseteq \qR$, i.e., $0=\xi^{-1}(\iota_\RR(0))$ and $\frac{2\pi}{l}=\xi^{-1}\big(\iota_\RR\big(\frac{2\pi}{l}\big)\big)$, cf.\ \eqref{eq:rmap}. Then
	\begin{align*}  
  	\pi'_l(0)=\pi_l(\iota_\RR(0))\stackrel{\eqref{eq:patralll}}{=}\me=\pi'_l\big(\textstyle\frac{2\pi}{l}\big)\qquad\text{but}\qquad \pi'_{\tau,r}(0)\stackrel{\eqref{eq:patrall}}{=}\exp(-\frac{\tau}{2}\tau_3)\neq \pi'_{\tau,r}\big(\frac{2\pi}{l}\big).
	\end{align*}  	
   Here, the inequality on the right hand side is clear because by Lemma \ref{lemma:BildCirc}.\ref{lemma:BildCirc4} we have that $\pi'_{\tau,r}(x)=\pi'_{\tau,r}(y)$ for $x\neq y$ enforces $\pi'_{\tau,r}(x)=\pm \me$. Consequently, there cannot exist a transition map $\TT_2\colon \im[\pi'_l]\rightarrow \im[\pi'_{\tau,r}]$ as then 
	\begin{align*}   
   	\pi'_{\tau,r}(0)=(\TT_2\cp \pi'_l)(0)=(\TT_2\cp \pi'_l)\big(\textstyle\frac{2\pi}{l}\big)=\pi'_{\tau,r}\big(\textstyle\frac{2\pi}{l}\big)
   \end{align*}
   would hold.
\item
  \itspace
  We have $\pi'_{\tau,r}(a_{2n})=\me$ for all $n\in \mathbb{Z}_{\neq 0}$, so that for a transition map $\TT_1\colon  \im[\pi'_{\tau,r}]\rightarrow \im[\pi'_{l}]$ we would have  
  \begin{align*}
    \TT_1(\me)=\left(\TT_1\cp\pi'_{\tau,r}\right)(a_{2n})=\pi'_l( a_{2n}) \qquad\forall \: n\in \ZZ_{\neq 0}.
  \end{align*}
  Then, for $\epsilon >0$ we find $n_\epsilon \in \mathbb{N}_{\geq 0}$ such that\footnote{It is clear that $\lim_n [l a_{2n}- 2\pi n] =0$, and that $\lim_n [l a_{2n}-  l a_{2(n+1)}]= 2\pi$. Hence, we have to show that we find $n_0\in \NN_{\geq 1}$ such that $l a_{2n}-  l a_{2(n+1)} \neq 2\pi$ for all $n\geq n_0$. Now, $l a_{2n}-  l a_{2(n+1)}=2n\pi\left[\sqrt{1-\textstyle\frac{\tau^2}{4(2n+2)^2\pi^2}}-\sqrt{1-\textstyle\frac{\tau^2}{4(2n)^2\pi^2}}\right] + 2\pi \sqrt{1-\textstyle\frac{\tau^2}{4(2n+2)^2\pi^2}}$, where the first summand tends to zero for $n\rightarrow \infty$ and is negative. Since the second summand is smaller than $2\pi$, the whole expression is smaller than $2\pi$ for $n$ suitable large.} $l a_{2n}-  l a_{2(n+1)}\in B_\epsilon(2\pi)\backslash\{2\pi\}$ for all $n\geq n_\epsilon$. But, $\pi'_{l}(a_{2n})=\pi'_{l}(\raisebox{0pt}{$a_{2(n+1)}$})$ implies $la_{2(n+1)}-la_{2n} =k_n2\pi$ for some $k_n \in \mathbb{Z}$,
  so that we get a contradiction if we choose $\epsilon< 2\pi$.
\end{itemize}
\endgroup 
\noindent
However, by Lemma \ref{lemma:WeierErzeuger}.\ref{lemma:WeierErzeuger3}) it suffices to 
take one fixed circular curve $\gamma_{\tau,r}$ into account. So, %. This means that 
we can circumvent the above transition map problem by sticking to the directed set $I$ from Remark and Definition \ref{rem:bohrproj}. More precisely, we can incorporate the map $\pi'_{\tau,r}$ into each of the projection maps as follows: 
\begin{enumerate}
\item%[1.)]
  \itspace
  For $I\ni L=(l_1,\dots,l_k)$ we can define 
	\begin{align*}
	\pi'_L\colon \qR &\rightarrow \SU^{k+1}\\
		\x&\mapsto \big(\pi'_{l_1}(\x),\dots,\pi'_{l_k}(\x),\pi'_{\tau,r}(\x)\big).
	\end{align*}  
  But, then $\im[\pi'_L]$ crucially depends on the $\mathbb{Z}$-independence of $l_1,\dots,l_k,r\tau$, as, e.g., we have $\pi_L(\RB)=[H_{\vec{e}_1}]^k \times  H_{\vec{e}_2}$ in the $\mathbb{Z}$-independent case and $\pi_L(\RB)\cong S^1$ if $l_1,\dots,l_k,r\tau$ are multiples of the same real number. For this, observe that we cannot restrict to independent tuples without adapting $\leq$. This is because for $(l',r\tau)$, $(l'',r\tau)$ independent and $(l_1,\dots,l_k)\in I$ an upper bound of $L':=l',L'':=l''$, the tuple $(l_1,\dots,l_k,r\tau)$
  does not need to be independent as well. In fact, for $l'=l-r\tau$ this cannot be true for any such $L$. 
  All this makes it difficult to find transition maps and suitable consistent families of measures for these spaces.
\item%[2.)]
  \itspace
  Basically, Lemma \ref{lemma:WeierErzeuger}.\ref{lemma:WeierErzeuger3}) is due to the fact that the $C_0(\RR)$-part 
  of the function 
  $a\colon \RR\ni c\mapsto (\pi'_{\tau,r}(c))_{11}$  (given by the \eqref{eq:funcco}) 
  vanishes nowhere. Now, we can try to find some analytic curve $\gamma$ and a projection map $\pi_\gamma\colon \qR\rightarrow \SU$ 
  such that for one of the entries $(\pi_{\gamma}(\cdot))_{ij}$, $1\leq i,j\leq 2$ the $\CAP(\RR)$-part is zero and the $C_0(\RR)$-part vanishes nowhere and is injective. Then, condition \ref{def:ProjLim3}) from Definition \ref{def:ProjLim} would hold for the projection maps $\wt{\pi}_L\colon \qR \rightarrow \im[\pi_\gamma] \sqcup [H_{\vec{e}_1}]^k$
  \begin{equation}
    \label{eq:PiLMuster}
    \wt{\pi}_L(\ovl{x}) := 
    \begin{cases} 
      \pi_\gamma(\x)  & \mbox{if } \ovl{x}\in \mathbb{R}\\
      \pi'_L(\x) &\mbox{if } \ovl{x}\in \RB
    \end{cases} 
  \end{equation}
  and we could define the transition maps and measures on $\im[\pi_\gamma]$ and $[H_{\vec{e}_1}]^k$ separately. However, even if such a curve $\gamma$ exists, it is not to be expected that it is easier to find reasonable measures on $\im[\pi_\gamma]$ than on $\im[\pi'_{\tau,r}]$.  
\end{enumerate}
In the next subsection we will follow the philosophy of the second approach. Here, we use distinguished generators of $C_0(\RR)\oplus \CAP(\RR)$ 
in order to define projective structures on $\qR$ in a more direct way. This will allow us to circumvent the image problem we have for the projection maps $\pi'_{\tau,r}$ and $\pi_\gamma$. So, the crucial part will not be to define the projective structure, but to determine the respective consistent families of normalized Radon measures. Here, the main difficulties will arise from determining the Borel $\sigma$-algebras of the projection spaces. 

\subsection{Projective Structures on $\qR$}
\label{subsec:ProjStrucon}
In this subsection, we will use the characters $\{\chi_l\}_{l\in \RR}$ and an injective nowhere vanishing $f\in C_0(\RR)$ in order to define a projective structure on $\qR$. This will be done in analogy to the definition of the projective structure on $\RB$ presented in Remark and Definition \ref{rem:bohrproj}, cf.\ \cite{oai:arXiv.org:0704.2397}. In the last part, we will use this construction in order to fix the normalized Radon measures \eqref{eq:fammeas}. 

We start with the following definitions, resetting (and collecting) some of the notations we have introduced in the previous subsection.
\begin{definition}
  \label{def:ProjLimit}
  Assume that $f\in C_0(\mathbb{R})$ is injective and $f(x)\neq 0$ for all $x\in \mathbb{R}$.
  \begin{enumerate}
  \item 
    Let $I$ denote the set of all finite tuples $L=(l_1,\dots,l_k)$ consisting of $\mathbb{Z}$-independent real numbers $l_1,\dots,l_k$. Moreover, let $|L|$ denote the length $k$ of the tuple $L$. 
  \item
    For $L,L'\in I$ define $L\leqZ L'$ iff $l_i \in \spann_{\mathbb{Z}}(l'_1,\dots,l'_{k'})$ for all $1\leq i\leq k$. 
  \item
    For $L\in I$ and $k:=|L|$ define $\pi_L\colon \qR \rightarrow \prfl{f}{L}=: X_L$ by
    \begin{equation}
      \label{eq:PiL}
      \pi_L(\ovl{x}) := 
      \begin{cases} 
        f(\ovl{x})  & \mbox{if } \ovl{x}\in \mathbb{R}\\
        \big(\ovl{x}(\chi_{l_1}),\dots,\ovl{x}(\chi_{l_k})\big) &\mbox{if } \ovl{x}\in \RB,
      \end{cases} 
    \end{equation}
    and equip $X_L$ with the final topology $\T_F$ w.r.t.\ this map. Recall that here and in the following $S^k$ just denotes the $k$-fold product of the unit circle $S^1$.
  \item
    For $L,L'\in I$ with $L\leqZ L'$ define $\pi_L^{L'}\colon X_{L'}\rightarrow X_L$ by $\pi_L^{L'}(y):=y$ if $y\in \im[f]$ and
    \begin{align}
      \label{eq:transit}
      \pi^{L'}_{L} (s_1,\dots,s_{k'}):= \left(\prod_{i=1}^{k'}{s_i}^{n^i_1} ,\dots, \prod_{i=1}^{k'}{s_i}^{n^i_{j}} \right)
      \quad\text{ if }\quad l_j=\sum_{i=1}^{k'} n^i_jl'_i
    \end{align}
    with $n_j^i\in \mathbb{Z}$ for $1\leq j\leq k=|L|$, $1\leq i\leq k'=|L'|$ and $(s_1,\dots, s_{k'})\in S^{|L'|}$.
  \end{enumerate}
\end{definition}
We now show 
that $\qR$ is indeed a projective limit of $\{X_L\}_{L\in I}$. Moreover, we determine the Borel $\sigma$-algebras of the spaces $X_L$. This will lead to an analogous decomposition of finite Radon measures as for the space $\qR$. Here, the crucial point is to show that the subspace topologies of $\im[f]$ and $S^{|L|}$ w.r.t.\ the final topology on $X_L$ are just their canonical ones.
For this, we will need the following definitions and facts:
\begingroup
\setlength{\leftmargini}{17pt}
\begin{itemize}
\item
  \itspacecc
  Let $\T_f$ and $\T_{L}$ denote the standard topologies
  on $\im[f]$ and $S^{|L|}$, respectively, i.e., 
  the subspace topology on $\im[f]$ inherited from $\mathbb{R}$ and the product topology on $S^{|L|}$.
\item
  \itspacecc
  For $L\in I$ let $\pih_L \colon \RB \rightarrow S^{|L|}$ denote the restriction of $\pi_L$ to $\RB$. 
\item    
  \itspacecc
  For $L,L'\in I$ with $L\leqZ L'$ let $\pih_L^{L'} \colon S^{|L'|} \rightarrow S^{|L|}$ denote the restriction of $\pi^{L'}_{L}$ to $S^{|L'|}$.
\item
  \itspacecc
  For $q\in \Q_{\neq 0}$ define $\chi_{l,q}:=\chi_{l/q}$ as well as $\chih_l:=\GT(\chi_l)$ for $l\in \RR$.
\item
  \itspacecc
  Since each $L\in I$ consists of 
  $\mathbb{Q}$-independent reals, we find (and fix) a subset $L^\perp\subseteq \mathbb{R}$ for which $\ovl{L}:=L\sqcup L^\perp$ is a $\mathbb{Q}$-base of $\mathbb{R}$. It is clear that, together with the constant function $1=\chi_{0}$, the functions $\left\{\chi_{l,n}\right\}_{(l,n) \in \ovl{L} \times \NN_{>0}}$ generate a dense $^*$-subalgebra 
  of $\CAP(\RR)$.   
\item
  \itspacecc
  For $p\in \mathbb{N}_{\geq 1}$ and $A\subseteq S^1$ let $\hat{p}\colon S^1\rightarrow S^1$, $s\mapsto s^p$ and define 
  \begin{align*}
    \sqrt[p]{A}:=\{s\in S^1\:|\: s^p\in A\}\qquad\text{as well as}\qquad A^p:=\{s^p\:|\: s\in A\}.
  \end{align*}   	
  If $\OO\subseteq S^1$ is open, then $\OO^p$ and $\sqrt[p]{\OO}=\hat{p}^{-1}(\OO)$ are open as well. This is because $\hat{p}$ is open (inverse function theorem) and continuous.  
\item
  \itspacecc
  For $A\subseteq S^1$ and $m\in \ZN$ we define
  \begin{equation*}
    A^{\sign(m)}:=
    \begin{cases}
      A & \mbox{if }m >0\\
      \{\ovl{z}\:|\: z\in A\} &\mbox{if } m <0.
    \end{cases}
  \end{equation*}
\end{itemize}
\endgroup
\noindent
The next lemma highlights the relevant properties of the maps $\pih_L$. 
\begin{lemma}
  \label{lemma:OpenMapp}
  Let $L=(l_1,\dots,l_k)\in I$.
  \begin{enumerate}
  \item
    \label{lemma:OpenMapp1} 
    Let $\psi \in \RB$, $q_i\in \Q$ and $s_i\in S^1$ for $1\leq i\leq k$. Then we find $\psi'\in \RB$ with 
\begin{align*}    
 \psi'(\chi_{l_i,q_i})=s_i\quad  \forall\:1\leq i\leq k\qquad\quad\text{and}\qquad\quad \psi'(\chi_{l})=\psi(\chi_{l})\quad\forall\: l \in \Span_{\mathbb{Q}}(L^\perp). 
\end{align*}
  \item
    \label{lemma:OpenMapp2}
    Let $l\in \RR$, $m_i\in \ZN$ and $\OO_i\subseteq S^1$ open for $1\leq i\leq n$. Then, for $m:=|m_1\cdot {\dots} \cdot m_n|$ and $p_i:=|\frac{m}{m_i}|$, we have
    \begin{align}
      \label{eq:WurzelUrbild}
      \bigcap_{i=1}^n \chih^{-1}_{l,m_i}(\OO_i)=\bigcap_{i=1}^n \chih^{-1}_{l,m}\left(\!\raisebox{0.0ex}{$\left[\!\sqrt[p_i]{\OO_i}\:\right]^{\sign(m_i)}\!$}\:\right)
      =\chih^{-1}_{l,m}\left(\OO\right)
    \end{align}
    for the open subset $\OO=\left[\!\sqrt[p_1]{\OO_1}\:\right]^{\sign(m_1)} \cap \dots \cap \left[\!\sqrt[p_n]{\OO_n}\:\right]^{\sign(m_n)}\subseteq S^1$.
  \item 
    \label{lemma:OpenMapp3}
    Let $A_i, B_j\subseteq S^1$ for $1\leq i\leq k$, $1\leq j\leq q$, with $B_1,\dots,B_q\neq \emptyset$. Moreover, let $h_1,\dots,h_q \in \mathbb{R}$ such that $l_1,\dots,l_k$, $h_1,\dots,h_q$ are $\mathbb{Z}$-independent. For $m_1,\dots,m_k, n_1,\dots,n_q \in \ZN$ let
    \begin{align}
      \label{eq:BaseRbohr}
      \!\!\!\!W:=\underbrace{\chih_{l_1,m_1}^{-1}(A_1)\cap\dots \cap \chih_{l_k,m_k}^{-1}(A_k)}_{U}\:\cap\: \underbrace{\chih_{h_1,n_1}^{-1}(B_1)\cap\dots \cap \chih_{h_q,n_q}^{-1}(B_q)}_{U'}.
    \end{align}
    Then $\pih_L(W)= A_1^{m_1} \times \dots \times A_k^{m_k}$.
  \item
    \label{lemma:OpenMapp4}
    The map $\pih_L$ is surjective, continuous and open.
  \end{enumerate}
  \begin{beweis}
    \begin{enumerate}
    \item
      This is clear from Lemma and Convention \ref{lemconv:RBMOD}.\ref{prop:Bohrmod24}. 
    \item
      Obviously, $\OO$ is open, and since
      the second equality in \eqref{eq:WurzelUrbild} is clear, it suffices to show that 
      \begin{align*}
        \chih^{-1}_{l,m}(A)=\chih^{-1}_{l,p\cdot |m|}\left(\big[\!\raisebox{-0.3ex}{$\sqrt[p]{A}$}\:\big]^{\sign(m)}\right)
      \end{align*}
      holds 
      for $A\subseteq S^1$, $l\in \mathbb{R}$, $p \in \mathbb{N}_{\geq 1}$ and $m\in \ZN$. 
      To show the inclusion $\supseteq$, let     
      $\psi \in \chih^{-1}_{l,p\cdot |m|}\left(\big[\!\raisebox{-0.3ex}{$\sqrt[p]{A}$}\:\big]^{\sign(m)}\right)$. Then
      \begin{align*}
        \psi(\chi_{l,p\cdot |m|}) \in \big[\!\raisebox{-0.3ex}{$\sqrt[p]{A}$}\:\big]^{\sign(m)}\qquad\Longrightarrow\qquad \psi(\chi_{l, m})=\big[\raisebox{-0.15ex}{$\psi(\chi_{l,p\cdot |m|})^{p}$}\big]^{\sign(m)}\in A.
      \end{align*}
      For the converse inclusion let $\psi \in\chih_{l,m}^{-1}(A)$. Then 
      \begin{align*}
        \big[\raisebox{-0.2ex}{$\psi(\chi_{l,p\cdot |m|})^{p}$}\big]^{\sign(m)}=\psi(\chi_{l,m})\in A \qquad\Longrightarrow\qquad \psi(\chi_{l,p\cdot |m|})\in \big[\!\raisebox{-0.15ex}{$\sqrt[p]{A}$}\:\big]^{\sign(m)}.
      \end{align*}
    \item
      We proceed in two steps:
      \begingroup
      \setlength{\leftmarginii}{15pt}
      \begin{itemize}
      \item
      	\itspace
    	We show that $\pih_L(W)=\pih_L(U)$. For this, it suffices to verify that $\pih_L(U)\subseteq \pih_L(W)$ because the converse inclusion is clear from $W\subseteq U$. 
        So, for $\psi \in U$ we have to show that $\pih_L(\psi) \in \pih_L(W)$. Since $B_j\neq \emptyset$, we find $z_j \in B_{j}$ 
        for all $1\leq j\leq q$.
        By \ref{lemma:OpenMapp1}) we find $\psi'\in \RB$ with $\psi'(\chi_{l_i,m_i})=\psi(\chi_{l_i,m_i})\in A_i$ for all $1\leq i\leq k$ and $\psi'(\chi_{h_j,n_j})=z_j\in B_j$ for all $1\leq j\leq q$. This shows $\psi'\in U\cap U'=W$, hence $\pih_L(\psi')\in \pih_L(W)$. Consequently,  
        \begin{align*}
          \pih_L(\psi)&=(\psi(\chi_{l_1,m_1}),\dots,\psi(\chi_{l_k,m_k}))
          =\big(\psi'(\chi_{l_1,m_1}),\dots,\psi'(\chi_{l_k,m_k})\big)=\pih_L(\psi')\in \pih_L(W).
        \end{align*}
      \item
        We show $\pih_L(U)=A^{m_1}_1 \times \dots \times A^{m_k}_k$. For this, it suffices to verify the inclusion $\supseteq$ as the opposite inclusion is clear from the definitions. To this end, fix $\psi\in \RB$ and let $s_i \in A_i^{m_i}$ for $1\leq i\leq k$ be chosen freely. Then, we find $z_i\in A_i$ with $z_i^{m_i}=s_i$, and  
        \ref{lemma:OpenMapp1}) provides us with some $\psi'\in \RB$ with $\psi'(\chi_{l_i,m_i})=z_i\in A_i$ for $1\leq i\leq k$. Then $\psi'\in U$ and $\psi'(\chi_{l_i})=z_i^{m_i}=s_i \in A_i^{m_i}$ for all $1\leq i\leq k$.
      \end{itemize}
      \endgroup
    \item
      Continuity of $\pih_L$ is clear from 
      \begin{align*}
        \pih_L^{-1}(A_1,\dots,A_k)= \chih_{l_1}^{-1}(A_1)\cap\dots \cap \chih_{l_k}^{-1}(A_k)\qquad\forall\:A_1,\dots A_k \subseteq S^1,
      \end{align*}
      and surjectivity is clear from Part \ref{lemma:OpenMapp3}) if we choose $A_1,\dots,A_k=S^1$.
     
      For openness observe that the $^*$-algebra generated by $1$ and $\{\chi_{l,m}\}_{(l,m) \in \ovl{L} \times \ZN}$ is dense in $\CAP(\RR)$ as it equals the $^*$-algebra generated by the all characters $\chi_l$. Then, the subsets of the form $\chi_{l,m}^{-1}(\OO)$ with $\OO\subseteq S^1$ open and $(l,m) \in \ovl{L}\times \ZN$ 	
      provide a subbasis for the topology of $\RB$.\footnote{This is because the Gelfand topology on $\RB$ equals the initial topology w.r.t.\ the Gelfand transforms of the elements of each subset $\bB\subseteq \CAP(\RR)$ that generates a dense subset $\dD$ of $\CAP(\RR)$, see e.g.\  Subsection 2.3 in \cite{ChrisSymmLQG}. Consequently, the Gelfand topology on $\RB$ equals the initial topology w.r.t.\ the functions $1$ and $\chi_{l,m}$ for $(l,m) \in \ovl{L}\times \ZN$. 
        Since the preimage of a subset of $\mathbb{C}$ under $\GT(1)$ is either empty or $\RB$, the claim is clear.} 
      So, a base of this topology  is given by all finite intersections of such subsets. Then Part \ref{lemma:OpenMapp2}) shows that, in order to obtain a base for the topology of $\RB$, it suffices to consider intersections of the form \eqref{eq:BaseRbohr} with $A_1,\dots,A_k,B_1,\dots,B_q$ open in $S^1$. For this, observe that since $\chih^{-1}_{l,m}(S^1)=\RB$, we can assume that all $l_1,\dots,l_k$ occur in each of these intersections. 
      Then, since $A_i^{m_i}$ is open if $A_i$ is open, Part \ref{lemma:OpenMapp3}) shows that $\pih_L$ is an open map. 
    \end{enumerate}
  \end{beweis}
\end{lemma}
The next lemma
highlights the crucial properties of the final topology of the spaces $X_L$. In addition to that, 
the Borel $\sigma$-algebras of these spaces are determined. 
\begin{lemma}
  \label{lemma:reltop}
  Let $L=(l_1,\dots,l_k)\in I$.
  \begin{enumerate}
  \item
    \label{lemma:reltop1}
    The subspace topologies of $\im[f]$ and $S^{|L|}$ w.r.t.\ the final topology $\T_F$ on $X_L$ are given by 
    $\T_f$ and $\T_L$, respectively. 
    For a subset $U\subseteq \im[f]$ we have $U\in \T_f$ iff $U$ is open in $X_L$.
  \item
    \label{lemma:reltop2}
    $X_L$ is a compact Hausdorff space.
  \item
    \label{lemma:reltop3}
    We have $\Borel\!\left(X_L\right)=\Borel(\im[f])\sqcup\BTK$ and 
    \begingroup
    \setlength{\leftmarginii}{20pt}
    \begin{enumerate}
    \item
      \itspacec
      If $\mu$ is a finite Radon measure on $\Borel\!\left(X_L\right)$, then $\mu|_{\Borel(\im[f])}$ and $\mu|_{\Borel(S^{|L|})}$ are finite Radon measures as well. 
    \item
      If $\mu_{f}\colon \Borel(\im[f]) \rightarrow [0,\infty)$ and $\mu_{S}\colon \BTK \rightarrow [0,\infty)$ are finite Radon measures, then 
      \begin{align}
      \label{eq:xh}
        \mu(A):=\mu_{f}(A\cap \im[f])+ \mu_{S}\big(\raisebox{-1pt}{$A\cap S^{|L|}$}\big)
        \qquad\forall\:  A\in \Borel\!\left(X_L\right)	
      \end{align}
      is a finite Radon measure on $\Borel\big(X_L\big)$.
    \end{enumerate}
    \endgroup
  \end{enumerate}
  \begin{beweis}
    \begin{enumerate}
    \item
      We first collect the following facts we have already proven during this section:
      \begin{enumerate}
      \item[(a)]
        \itspacecc
        The topology on $\qR$ induces the standard topologies on $\RR$ and $\RB$.
      \item[(b)]
        $U\in \T_F$ iff $\pi_L^{-1}(U)$ is open in $\qR$.
      \item[(c)]
	\vspace{-0.8pt}
        $W\subseteq \RR$ is open in $\qR$ iff $W$ is open in $\RR$.
      \item[(d)]
        \vspace{-1pt}
     	If $B\subseteq \RB$ is open, then there is $U\subseteq \im[f]$ such that $f^{-1}(U)\sqcup B$ is open in $\qR$.\footnote{By (e) this is equivalent to show that we find an open subset $W\subseteq \RR$ such that $W\sqcup B$ is open in $\qR$. But, this is clear if $B=\chih_l^{-1}(\OO)$ for some open subset $\OO\subseteq S^1$ and $l\in \RR$ (see Type 3 sets defined in Lemma and Remark \ref{remdefchris}). Since the sets of the form  $\chih_l^{-1}(\OO)$ provide a subbasis for the topology on $\RB$, the claim follows.}
      \item[(e)]
	\itspacecc	
	$f\colon \RR\rightarrow \im[f]$ is a homeomorphism.
      \item[(f)]
	\itspacecc	
        $\pih_L\colon \RB\rightarrow S^{|L|}$ is continuous and open.
      \end{enumerate}
      We start with the statements concerning the subspace topologies: 
      
      \vspace{1ex}
      
      $\boldsymbol{\im[f]}$:
      Let $U\subseteq \im[f]$. Then:
      
      \qquad\quad\hspace{9pt}
      \:$U$ is open w.r.t.\ the topology inherited from $X_L$

      \qquad$\Longleftrightarrow$ $\:\exists\: V\subseteq S^{|L|}$ such that $U\sqcup V$ is open in $X_L$

      \qquad$\Longleftrightarrow$ $\:\exists\: V\subseteq S^{|L|}$ such that $\pi_L^{-1}(U\sqcup V)$ is open in $\qR$\hspace*{\fill}{(b)}

      \qquad$\Longleftrightarrow$ $\:\exists\: V\subseteq S^{|L|}$ such that $f^{-1}(U)\sqcup \pih_L^{-1}(V)$ is open in $\qR$
      
      \qquad$\Longleftrightarrow$ $\:f^{-1}(U)$ is open in $\RR$\hspace*{\fill}{(c)}%\hspace*{\fill}{(a),(c)}

      \qquad$\Longleftrightarrow$ $\:U\in \T_f$  \hspace*{\fill}(e)
      
      \vspace{2ex}
      $\boldsymbol{S^{|L|}}$:
      Let $V\subseteq S^{|L|}$. Then:
      
      \qquad\quad\hspace{9pt}
      $V$ is open w.r.t.\ the topology inherited from $X_L$

      \qquad$\Longleftrightarrow$ $\:\exists\: U\subseteq \im[f]$ such that $U\sqcup V$ is open in $X_L$

      \qquad$\Longleftrightarrow$ $\:\exists\: U\subseteq \im[f]$ such that $\pi_L^{-1}(U\sqcup V)$ is open in $\qR$\hspace*{\fill}{(b)}

      \qquad$\Longleftrightarrow$ $\:\exists\: U\subseteq \im[f]$ such that $f^{-1}(U)\sqcup \pih_L^{-1}(V)$ is open in $\qR$

      \qquad$\Longleftrightarrow$ $\:\pih_L^{-1}(V)$ is open in $\RB$\hspace*{\fill}{(a),(d)}

      \qquad$\Longleftrightarrow$ $\:V\in \T_L$  \hspace*{\fill}(f) 			
      
      \vspace{1ex}
      Finally, observe that $\im[f]$ is open in $X_L$ because $\pi_L^{-1}(\im[f])=\RR$ is open in $\qR$. Then $U\subseteq \im[f]$ is open in $X_L$ iff $U$ is open w.r.t.\ the  topology on $\im[f]$ inherited from $X_L$. Since this topology equals $\T_f$, the claim follows.
    \item
      The spaces $X_L$ are compact by compactness of $\qR$ and continuity of $\pi_L$. For the Hausdorff property observe that $\T_F$ contains all sets of the following types:
      \begin{align*}
        \begin{array}{lcrclcl}
          \textbf{\textit{Type 1':}} && f(V) & \!\!\!\sqcup\!\!\! & \emptyset 
          && \text{with open $V \subseteq \RR$,} \\
          \textbf{\textit{Type 2':}} && f(K^c) & \!\!\!\sqcup\!\!\! & S^{|L|}
          && \text{with compact $K \subseteq \RR$,} \\
          \textbf{\textit{Type 3':}} && f\big(\chi_{l_i}^{-1}(\OO)\big) & \!\!\!\sqcup\!\!\! & \pr_i^{-1}(\OO)
          && \text{with $\OO\subseteq S^1$ open and $1\leq i\leq  k$.}
        \end{array}
      \end{align*} 
      Here $\pr_i\colon S^{|L|}\rightarrow S^1$, $(s_1,\dots,s_k)\mapsto s_i$ denotes the canonical projection. In fact,
      the preimage of a set of \textbf{\textit{Type m'}} under $\pi_L$ is a subset of $\qR$ of \textbf{\textit{Type m}}, cf.\ Lemma and Definition \ref{remdefchris}.  
      Then, by injectivity of $f$ the elements of $\im[f]$ are separated by sets of \textbf{\textit{Type 1'}}. Moreover, if $x\in \im[f]$ and $(s_1,\dots,s_k)\in S^{|L|}$, then we can choose a relatively compact neighbourhood $W$ of $f^{-1}(x)$ in $\RR$ and define $U:= f(W)$ and $V:=f\raisebox{1pt}{$\big($}\raisebox{-1pt}{${\ovl{W}\hspace{0.9pt}}^c$}\raisebox{1pt}{$\big)$}\sqcup S^{|L|}$. Finally, if $(s_1,\dots,s_k),(s'_1,\dots,s'_k)\in S^{|L|}$ are different elements, then $s_i\neq s'_i$ for some $1\leq i\leq k$. Then, for open neighbourhoods $\OO,\OO' \subseteq S^1$ of $s_i$ and $s_i'$, respectively, with $\OO \cap \OO' =\emptyset$ we have 	
      $\big[f\big(\chi_{l_i}^{-1}(\OO)\big) \sqcup \pr_i^{-1}(\OO)\big]\cap \big[f\big(\chi_{l_i}^{-1}(\OO')\big) \sqcup \pr_i^{-1}(\OO')\big]=\emptyset$.
    \item  
      We repeat the arguments from the proof of Lemma \ref{lemma:Radon}.
      
      If $U\subseteq X_L$ is open, then $U\cap \im[f]\in \T_f$ and $U\cap S^{|L|}\in \T_L$ 
      by Part \ref{lemma:reltop1}). This shows $U\in \Borel(\im[f])\sqcup\BTK$, i.e., $\Borel\left(X_L\right)\subseteq\Borel(\im[f])\sqcup\BTK$ as the right hand side is a $\sigma$-algebra. For the converse inclusion recall that $U\in \T_f$ iff $U$ is open in $X_L$, again by the first part, hence $\Borel(\im[f])\subseteq \Borel(X_L)$. Finally, if $A\subseteq S^{|L|}$ is closed, then $A$ is compact w.r.t.\ $\T_L$. This means that $A$ is compact w.r.t.\ the subspace topology inherited from $X_L$, implying that $A$ is compact as a subset of $X_L$. Then $A$ is closed by the Hausdorff property of $X_L$ so that $A\in \Borel\!\left(X_L\right)$, hence $\BTK\subseteq \Borel\!\left(X_L\right)$. Now,
      \begin{enumerate}
      \item[\textit{(a)}]
    	\itspacecc
        The measures $\mu|_{\Borel(\im[f])}$ and $\mu|_{\Borel(S^{|L|})}$ are well defined and obviously finite. Their inner regularities follow from the fact that subsets of $\im[f]$ and $S^{|L|}$ are compact w.r.t.\ $\T_f$ and $\T_L$, respectively, iff they are so w.r.t.\ the topology on $X_L$, just by by Part \ref{lemma:reltop1}).
      \item[\textit{(b)}]      
        \itspacecc
        If $\mu$ is defined by \eqref{eq:xh},  
        then $\mu$ is a finite Borel measure and its inner regularity follows by a simple $\epsilon\slash 2$ argument from the inner regularities of $\mu_{f}$ and $\mu_{S}$.
      \end{enumerate}
      % \endgroup
    \end{enumerate}
  \end{beweis}
\end{lemma}
Combining the Lemmata \ref{lemma:OpenMapp} and \ref{lemma:reltop} we obtain
\begin{proposition}
  \label{th:projlim}
  \begin{enumerate}
  \item
    $\qR$ is a projective limit of $\{X_L\}_{L\in I}$.
  \item
    A family $\{\mu_L\}_{L \in I}$ of measures $\mu_L$ on $X_L$ is a consistent family of normalized Radon measures w.r.t.\ $\{X_L\}_{L\in I}$ iff the following holds:
    \begingroup
    \setlength{\leftmarginii}{20pt}
    \begin{enumerate}
    \item
      There is $t\in [0,1]$ such that for each $L\in I$ and $A\in \Borel\!\left(X_L\right)$ we have
      \begin{align*}	
        \mu_L(A)=t\: \mu_f(A\cap \im[f])+ (1-t)\: \mu_{S,L}\big(\raisebox{-1pt}{$A\cap S^{|L|}$}\big)	
      \end{align*} 
      for $\mu_f$ and $\mu_{S,L}$ normalized\footnote{If $t$ equals $0$ or $1$, we allow $\mu_f=0$ or $\mu_{S,L}=0$, respectively.} Radon measure on $\im[f]$ and $S^{|L|}$, respectively. 
    \item
      For all $L,L'\in I$ with $L\leqZ L'$ we have $\pih^{L'}_{L}(\mu_{S,L'})=\mu_{S,L}$.
    \end{enumerate}
    \endgroup
  \end{enumerate}
  \begin{beweis}
    \begin{enumerate}
    \item
      The spaces $X_L$ are compact and Hausdorff by Lemma \ref{lemma:reltop}.\ref{lemma:reltop2}. 
      Moreover, each  $\pi_L$ is surjective by Lemma \ref{lemma:OpenMapp}.\ref{lemma:OpenMapp4}.
      If $L,L'\in I$ with $L\leqZ L'$, then continuity of the maps $\pi^{L'}_{L}$ is clear from
      $\pi^{L'}_{L}\cp \pi_{L'}=\pi_{L}$ which, in turn, is immediate from multiplicativity of the functions $\chi_l$. Finally, condition \ref{def:ProjLim3}) from
      Definition \ref{def:ProjLim} follows from injectivity of $f$ and the fact that the functions $\{\chi_l\}_{l\in \mathbb{R}}$ generate $\CAP(\RR)$.
    \item
      Let $\{\mu_L\}_{L\in I}$ be a consistent family of normalized Radon measures w.r.t.\ $\{X_L\}_{L\in I}$. Then Lemma \ref{lemma:reltop}.\ref{lemma:reltop3} shows that for each $L\in I$ we have
      \begin{align*}	
        \mu_L(A)= \mmu_{f,L}(A\cap \im[f])+ \mmu_{S,L}\big(\raisebox{-1pt}{$A\cap S^{|L|}$}\big)\qquad \forall\: A\in \Borel\!\left(X_L\right)
      \end{align*}
      for $\mmu_{f,L}$ and $\mmu_{S,L}$ finite Radon measures on $\im[f]$ and $S^{|L|}$, respectively. Then, consistency forces that $\mmu_{f,L}=\mmu_{f,L'}$ for all $L,L'\in I$. In fact, by Lemma \ref{lemma:normRM} there is a unique normalized Radon measure $\mu$ on $\qR$ for which $\mu_L=\pi_L(\mu)$ holds for all $L\in I$. Consequently, for each $A\in \Borel(\im[f])$ and all $L\in I$ we have	  
      \begin{align*}     	
        \mmu_{f,L}(A)=\mu_{L}(A)=\pi_L(\mu)(A)= \mu\big(f^{-1}(A)\big)=:\mmu_f(A).
      \end{align*}
      By the same arguments,\textit{(b)} follows from consistency of the measures $\{\mu_L\}_{L\in I}$. 
      Finally, if $t:= \mmu_f(\im[f])\in (0,1)$, then \textit{(a)} holds for $\mu_f:=\frac{1}{t}\mmu_f$ and $\mu_{S,L}:=\frac{1}{1-t}\mmu_{S,L}$ for $L\in I$.
      If $t=0$, we define $\mu_{S,L}:=\mmu_{S,L}$ for all $L\in I$ and if $t=1$, we define $\mu_f:=\mmu_f$.

      For the converse implication let
      $\{\mu_L\}_{L\in I}$ be a family of measures $\mu_L$ on $X_L$ such that \textit{(a)} and \textit{(b)} hold. Then, Lemma \ref{lemma:reltop}.\ref{lemma:reltop3} shows that each $\mu_L$ is a finite Radon measure, and obviously we have $\mu_L(X_L)=1$. 
      Finally, from \textit{(b)} for $A\in \Borel(X_L)$ we obtain 
      \begin{align*}
        \pi^{L'}_{L}(\mu_{L'})(A)&=\mu_{L'}\left(\pillstr^{-1}(A)\right)\\[-4pt]
        &= t\: \mu_f\left(\pillstr^{-1}(A)\cap \im[f]\right)+(1-t)\:\mu_{S,L'} \left(\pillstr^{-1}(A)\cap S^{|L'|}\right)\\[-4pt]
        &=t\: \mu_f\left(A\cap \im[f]\right)+ (1-t)\:\mu_{S,L'} \Big(\!\left(\raisebox{-0.1ex}{$\pih^{L'}_{L}$}\right)^{-1}\left(\raisebox{-0.1ex}{$A\cap S^{|L|}$}\right)\!\Big)\\[-4pt]
        &=t\: \mu_f\left(A\cap \im[f]\right)+(1-t)\:\pih^{L'}_{L}(\mu_{S,L'}) \big(\raisebox{-1pt}{$A\cap S^{|L|}$}\big)\\[-1pt]
        &=t\: \mu_f\left(A\cap \im[f]\right)+(1-t)\:\mu_{S,L} \big(\raisebox{-1pt}{$A\cap S^{|L|}$}\big)\\[-1pt]
      	&=\mu_{L}(A).
      \end{align*}
    \end{enumerate}		
  \end{beweis}
\end{proposition}

\subsection{Radon Measures on $\qR$} 
\label{subsec:CylMeas}
In this final subsection we use the results of the previous part in order to fix normalized Radon measures on $\qR$. Due to
Proposition \ref{th:projlim} 
this can be done as follows:
\begingroup
\setlength{\leftmargini}{20pt}
\begin{enumerate} 
\item[{\bf 1}]
  \itspace
  Determine a family of normalized Radon measures $\{\mu_{S,L}\}_{L\in I}$ on $S^{|L|}$ that fulfil condition \textit{(b)}.
\item[{\bf 2}]
  \itspace
  Fix an injective and nowhere vanishing element $f\in C_0(\RR)$ with suitable image together with a normalized Radon measure $\mu_f$ on $\im[f]$. 
\item[{\bf 3}]
  \itspace
  Adjust $t\in [0,1]$. 
\end{enumerate} 
\endgroup
\noindent 
In the following let $\lambda$ denote the Lebesgue measure on $\Borel(\RR)$. Moreover, for $B\in \Borel(\RR)$ and $\eta \colon B\rightarrow \RR$ measurable let  $\eta(\lambda):=\eta\big(\lambda|_{\Borel(B)}\big)$. %So, let us start with
% We start with

\vspace{10pt}
\noindent 
{\bf Step 1}
\newline
\vspace{-2.3ex}
\newline
We choose 
$\mu_{S,L}$ to be the Haar measure $\mu_{|L|}$ on $S^{|L|}$ because:
\begingroup
\setlength{\leftmargini}{20pt}
\begin{itemize}
\item
  \itspace
  This is canonical from the mathematical point of view, and these  
measures fulfil the required compatibility conditions as the next lemma shows. 
\item 
  \itspace
  This is in analogy to the case $\RB$ \cite{{oai:arXiv.org:0704.2397}}, where this choice
  results in the usual Haar measure on this space, see Remark and Definition \ref{rem:bohrproj}. 
\item
  \itspace
  These measures will suggest a natural choice of $f$ and $\mu_f$ in {\bf Step 2}.
\end{itemize}
\endgroup
\begin{lemma}
  \label{lemma:Projm}
  Let $\mu_f\colon \Borel(\im[f])\rightarrow [0,1]$ be a normalized Radon measure and $t\in[0,1]$. For each $L\in I$ let 
  \begin{align*}
    \mu_L(A):=  t\: \mu_f(A\cap \im[f]) +(1-t)\: \mu_{|L|}\raisebox{0pt}{$\big($}\raisebox{-1pt}{$A\cap S^{|L|}$}\big) \qquad \forall\: A\in \Borel(X_L). 
  \end{align*} 
  Then $\{\mu_L\}_{L\in I}$ is a consistent family of normalized Radon measures and the corresponding normalized Radon measure $\mu$ on $\qR$ is given by 
  \begin{align} 
    \label{eq:Radmeas}
    \mu(A)=t\: f^{-1}(\mu_f)(A\cap \mathbb{R}) +(1-t)\: \mu_{\mathrm{Bohr}}(A\cap \RB)\qquad \forall\:A\in \BRq.
  \end{align}
%  for $A\in \BRq$.
\end{lemma}
 %  \vspace{-10pt}
  \begin{beweis}
    Let $L\in I$, $A \in \Borel(X_L)$ and $\mu$ be defined by \eqref{eq:Radmeas}. Then
    \begin{align*}
      \pi_L(\mu)(A)&=t\: f^{-1}(\mu_f)\big(f^{-1}(A\cap \im[f])\big) + (1-t)\: \mu_{\mathrm{Bohr}}\big(\pih_L^{-1}\big(\raisebox{-1pt}{$A\cap S^{|L|}$}\big)\big)\\
      &= t\:\mu_f(A\cap \im[f])+ (1-t)\:\pih_L(\mu_{\mathrm{Bohr}})\big(\raisebox{-1pt}{$A\cap S^{|L|}$}\big).
    \end{align*}
    So, if we know that 
    $\pih_L(\mu_{\mathrm{Bohr}})=\mu_{|L|}$ holds for all $L\in I$, the claim follows. In fact, consistency of $\{\mu_L\}_{L\in I}$ then is automatically fulfilled because $\mu$ is a well-defined  normalized Radon measure. 
    Now, in order to show $\pih_L(\mu_{\mathrm{Bohr}})=\mu_{|L|}$, it suffices to show translation invariance of the normalized Radon measure $\pih_L(\mu_{\mathrm{Bohr}})$.
    For this, let $\tau \in S^{|L|}$. Then, by surjectivity of $\pih_L$ we find $\psi \in \RB$ with $\pih_L(\psi)=\tau$. Since $\pih_L$ is a homomorphism w.r.t.\ the group structure\footnote{Confer Subsection \ref{subsec:Bohrcomp}.} on $\RB$, for $A\subseteq S^{|L|}$ we have
    \begin{align*}
      \pih_L\left(\psi+ \pih_L^{-1}(A)\right)=\pih_L(\psi)\cdot \pih_L\left(\pih_L^{-1}(A)\right)=\tau \cdot A.
    \end{align*}
    Applying $\pih_L^{-1}$ to both sides gives $\psi+ \pih_L^{-1}(A)\subseteq \pih_L^{-1}(\tau \cdot A)$. For the opposite inclusion let $\psi'\in \pih_L^{-1}(\tau\cdot A)$. Then $\psi'- \psi \in \pih_L^{-1}(A)$ because 
    $\pih_L(\psi'- \psi)=\tau^{-1}\cdot \pih_L(\psi')\in A$. Consequently, $\psi'\in \psi+ \pih_L^{-1}(A)$, hence $\pih_L^{-1}(\tau\cdot A)\subseteq \psi+ \pih_L^{-1}(A)$, i.e., $\psi+ \pih_L^{-1}(A)=\pih_L^{-1}(\tau\cdot A)$. Then 
    \begin{align*}
      \pih_L(\mu_{\mathrm{Bohr}})(\tau\cdot A)&=\mu_{\mathrm{Bohr}}\left(\pih_L^{-1}(\tau\cdot A)\right)=\mu_{\mathrm{Bohr}}\left(\psi +\pih_L^{-1}(A)\right)\\
      &= \mu_{\mathrm{Bohr}}\left(\pih_L^{-1}(A)\right)=\pih_L(\mu_{\mathrm{Bohr}})(A)
    \end{align*}
    for all $A\in \BTK$. This shows that $\pih_L(\mu_{\mathrm{Bohr}})$ is translation invariant. 
  \end{beweis}
%\end{lemma}
\vspace{-2pt}
{\bf Step 2}
\newline
\vspace{-2.3ex}
\newline 
If $f,f'\in C_0(\mathbb{R})$ both are injective and vanish nowhere, then the respective projective structures from Definition \ref{def:ProjLimit} are equivalent in the sense that the corresponding spaces $X_L, X'_L$ are homeomorphic via the maps $\Omega_L\colon X_L\rightarrow X'_L$ defined by $\Omega_L|_{S^{|L|}}:=\id_{S^{|L|}}$ and $\Omega_L|_{\im[f]}:=f'\cp f^{-1}$. Moreover, if $\mu_f$ is a normalized Radon measure on  $\im[f]$, then $\mu_{f'}:=\big(f'\cp f^{-1}\big)(\mu_f)$ is a normalized Radon measure on $\im[f']$, and it is clear from \eqref{eq:Radmeas} that the corresponding Radon measures $\mu,\mu'$ on $\qR$
from Lemma \ref{lemma:Projm} coincide. All this makes sense because,
in contrast to $\CAP(\RR)$ where we have the canonical generators $\{\chi_l\}_{l\in \mathbb{R}}$, in $C_0(\mathbb{R})$ there is no distinguished nowhere vanishing, injective generator $f\in C_0(\RR)$. But, this also means that we can restrict to functions with a reasonable image such as the ``shifted'' circle $\Ts:=1+ S^1\backslash\{-1\}\subseteq \mathbb{C}$. In fact, here the analogy to $S^{|L|}$ suggests to use the Haar measure $\mu_1$ on $S^1$. So, in the following we will restrict to the elements of the subset
\begin{align*}
	\F:=\{f\in C_0(\RR)\:|\: \im[f]=\Ts\},
\end{align*}	
	 where for each $f\in \F$ we define $\mu_f:=\mus\colon \Borel\big(\Ts\big)\rightarrow [0,1]$. Here, 
\begin{align*}	
	 \mus(A):=\mu_1(A-1)=+_1(\mu_1)\qquad\forall\: A\in \Borel\big(\Ts\big)
\end{align*}	 
with $+_1\colon S^1\backslash\{-1\}\ni z\mapsto z+1\in \Ts$.
	  It follows that  
\begin{align}
  \label{eq:LebesgueHomeo}
  \big\{f^{-1}(\mu_f)\:\big|\: f\in \F\big\}=\big\{\rho(\muL)\:\big|\: \rho\in \HHH\big\}
\end{align}	
for \gls{HHH} the set of homeomorphisms $\rho\colon (0,1)\rightarrow \RR$.
\newline
\vspace{-8pt}
\newline
{\bf Proof of \eqref{eq:LebesgueHomeo}:} 
We consider the function $h\colon(0,1]\ni t\mapsto e^{\I\hspace{1pt} 2\pi[t-1\slash 2]} \in S^1$. Then
$\mu_1=h(\lambda)|_{\Borel(S^1)}$ and for $f\in \F$ we have $f^{-1}(\mu_f)= \rho(\lambda)$ for $\HHH\ni \rho:=f^{-1}\cp+_1\cp h|_{(0,1)}$. Conversely, if $\rho \in \HHH$, then $\rho(\lambda)=f^{-1}(\mus)$ for $\F\ni f:=+_1\cp h\cp \rho^{-1}$.\hspace*{\fill}{\scriptsize$\blacksquare$}
\newline
\vspace{-4pt}
\newline
So, if we restrict to projective structures arising from elements $f\in \F$, then Lemma \ref{lemma:Projm} and \eqref{eq:LebesgueHomeo} select the normalized Radon measures of the form
\begin{align}
  \label{eq:murhot}
  \mu_{\rho,t}(A):=t\:\rho(\muL)(A\cap \mathbb{R})+ (1-t)\:\mu_{\mathrm{Bohr}}(A\cap \RB)\qquad\forall\:A\in \BRq
\end{align}
for $\rho\colon (0,1)\rightarrow \mathbb{R}$ a homeomorphism and $t\in [0,1]$. 
\vspace{1ex}  
\newline
\vspace{-2ex}
\newline{\bf Step 3}
\newline 
\vspace{-2.3ex}
\newline
To adjust the parameter $t\in[0,1]$ we now take a look at the Hilbert spaces $\Hil_{\rho,t}:=\Lzw{\qR}{\mu_{\rho,t}}$. 
\begin{lemma}
  \label{lemma:techlemma}
  For $A\in \BRq$ let $\chi_A$ denote the corresponding characteristic function.
  \begin{enumerate}
  \item
    \label{lemma:techlemma1}
    If $\rho_1,\rho_2\colon (0,1)\rightarrow \RR$ are homeomorphisms and $t_1,t_2 \in (0,1)$, then 
    \begin{align*}
      \begin{split}
        \varphi \colon \Lzw{\qR}{\mu_{\rho_1,t_1}}&\rightarrow \Lzw{\qR}{\mu_{\rho_2, t_2}}\\
        \psi & \mapsto  \sqrt{\frac{t_1}{t_2}} \:(\chi_{\RR}\cdot \psi)\cp \big(\rho_1\cp \rho_2^{-1}\big) + \sqrt{\frac{(1-t_1)}{(1-t_2)}}\:\chi_{\RB}\cdot \psi
      \end{split}
    \end{align*}
    is an isometric isomorphism. The same is true for 
    \begin{align*}  
      &\varphi \colon \Hil_{\rho_1,1}\rightarrow \Hil_{\rho_2,1},\: \psi \mapsto (\chi_{\RR}\cdot \psi)\cp \big(\rho_1\cp \rho_2^{-1}\big),\\ 
      &\varphi \colon \Hil_{\rho_1,0}\rightarrow \Hil_{\rho_2,0},\: \psi \mapsto  \psi. 
    \end{align*}
  \item
    \label{lemma:techlemma2}
    If $t=1$, then $\Hil_{\rho,1}\cong \Lzw{\RR}{\rho(\muL)}\cong \Lzw{\RR}{\muL}$ for each $\rho\in H$. Here $\cong$ means canonically isometrically isomorphic.
  \end{enumerate}
  \end{lemma}
  \begin{proof}
    \begin{enumerate}
    \item
      This  is immediate from the general transformation formula.
    \item
      The first isomorphism is just because $\mu_{\rho,0}(\RB)=0$. Then, by the first part it suffices to specify the second isomorphism for the case that $\rho$ is a diffeomorphism. But, in this case we have $\rho(\lambda)=\frac{1}{|\dot\rho|}\lambda$, so that for the isomorphism
	\begin{align*}
		\varphi\colon \Lzw{\RR}{\rho(\muL)}\rightarrow \Lzw{\RR}{\muL},\quad \psi \mapsto  \frac{1}{\sqrt{|\dot \rho|}} \psi
	\end{align*}  we obtain 
      \begin{align*}
        \langle \varphi(\psi_1),\varphi(\psi_2)\rangle_\lambda&=\int_{\RR}  \psi_1 \ovl{\psi_2} \:\:\frac{1}{|\dot \rho|} \dd\muL= \int_{\RR}  \psi_1 \ovl{\psi_2} \:\: \dd\rho(\muL)
        =\langle \psi_1,\psi_2\rangle_{\rho(\lambda)}.
      \end{align*}
    \end{enumerate} 
  \end{proof}

\subsection{Summary}
\label{Finrem}
\begin{enumerate}
\item
  \label{Finrem4}
  In Subsection \ref{subsec:QuantvsRed} we have shown that quantization and reduction do not commute in homogeneous isotropic LQC, i.e., that 
  the inclusion $\ARRQw\cong \ARQw\subsetneq \AQRw$ is indeed proper in this case, see also Example \ref{ex:LQCInc}. 
\item
  \label{Finrem3}
  In Proposition \ref{lemma:bohrmassdichttrans} we have seen that the Haar measure on $\RB$ is uniquely determined by the condition that the translations w.r.t.\ the spectral extension $\Transl\colon \RR\times \RB \rightarrow \RB$ of the additive action \RPLUS$\colon \RR \times \RR \rightarrow \RR$, $(v,t)\mapsto v+t$ act as unitary operators on the respective Hilbert space of square integrable functions. In addition to that, we have shown that the one-parameter group $\{{\Transl_v\!}^*\}_{v\in \RR}$ of unitary operators ${\Transl_v\!}^*\colon \Lzw{\RB}{\muB}\rightarrow \Lzw{\RB}{\muB}$ is strongly continuous.
  Corollary \ref{cor:eindbohr} then states that the same unitality condition forces $\mu=\muB$ also for the space $\RR\sqcup\RB=\qR\cong \qRR=\ARRQw$, and that the respective family of unitary operators is strongly continuous as well. Consequently, if one wants to represent the exponentiated reduced fluxes (``momentum'' operators) by translations w.r.t.\ the respective spectral extension $\TTransw \colon \RR \times \qR \rightarrow \qR$ of \RPLUS$\colon \RR \times \RR \rightarrow \RR$, there is only the measure $\muB$ which can be used. So, following these lines, one ends up with the same kinematical Hilbert space as used in standard homogeneous isotropic LQC approach, namely $\Lzw{\RB}{\muB}$.
\item
  \label{Finrem1}
  In the last three subsections we have established a projective structure on $\qR\cong \qRR=\ARRQw$ in order to construct further normalized Radon measures thereon. Here, using Haar measures on tori we have derived the normalized Radon measures%$\mu_{\rho,t}$
  , cf. \eqref{eq:murhot} 
\begin{align*}
%  \label{eq:murhot}
  \mu_{\rho,t}(A):=t\:\rho(\muL)(A\cap \mathbb{R})+ (1-t)\:\mu_{\mathrm{Bohr}}(A\cap \RB)\qquad\forall\:A\in \BRq,	
\end{align*}
with $t\in [0,1]$, $\adif\colon (0,1)\rightarrow \RR$ a homeomorphism and $\lambda$ the restriction of the Lebesgue measure to $\Borel((0,1))$.  
  Then, 
  Lemma \ref{lemma:techlemma} shows that up to \emph{canonical} isometrical isomorphisms the parameters $\rho$ and $t$ give rise to the following three Hilbert spaces:
  \begingroup
  \setlength{\leftmarginii}{20pt} 
  \begin{enumerate}
  \item[1)]
    $\Hil_{\rho,1}\cong \Lzw{\RR}{\muL} \cong \Lzw{\RR}{\rho(\muL)}$ for all $\rho\in \text{\gls{HHH}}$\hspace*{\fill}{(Lemma \ref{lemma:techlemma}.\ref{lemma:techlemma2})}
  \item[2)] 
    $\Hil_{\rho,t}\cong L^2\big(\hspace{1pt}\raisebox{-0.1ex}{$\qR$},\mu_{\rho_0,t_0}\big)$ for all $\rho\in \HHH$, $t\in (0,1)$\hspace*{\fill}{(Lemma \ref{lemma:techlemma}.\ref{lemma:techlemma1})}
  \item[3)]
    $\Hil_{\rho,0}\cong \Lzw{\RB}{\mu_{\mathrm{Bohr}}}$ for all $\rho\in \HHH$  \hspace*{\fill}{($\RR$ is of measure zero)}
  \end{enumerate}
  \endgroup
  \noindent
  Here, the Hilbert spaces in $2)$ and $3)$ are isometrically isomorphic just because their Hilbert space dimensions coincide. In contrast to that, 
  the cases $1)$ and $3)$ cannot be isometrically isomorphic because $\Lzw{\RR}{\muL}$ is separable and $\Lzw{\RB}{\mu_{\mathrm{Bohr}}}$ is not so. 

  Anyhow, even if $L^2\big(\hspace{1pt}\raisebox{-0.1ex}{$\qR$},\mu_{\rho,t}\big)$ for $t\in(0,1)$ and $\Lzw{\RB}{\mu_{\mathrm{Bohr}}}$ are isometrically isomorphic, there may exist representations of the reduced holonomy-flux algebra (reduced algebra of observables) on the former space being not unitarily equivalent to the standard representation \cite{MathStrucLQG} on $\Lzw{\RB}{\mu_{\mathrm{Bohr}}}$. So, the next step towards physics might be to construct such representations on $L^2\big(\hspace{1pt}\raisebox{-0.1ex}{$\qR$},\mu_{\rho,t}\big)$. 
\end{enumerate}
In the previous sections we have discussed the problem of symmetry reduction in quantum gauge field theories, in particular, in the framework of loop quantum gravity. The problem of determining respective sets of invariant classical connections forming the reduced configuration spaces of the corresponding classical theories has been left open so far. This is the content of the final Section \ref{CHarinvconn}. There we prove a general characterization theorem for invariant connections on principal fibre bundles and, in particular, calculate the sets of invariant connections used in Subsection \ref{sec:inclrel} in order to show that (in the situations discussed there) quantization and reduction do not commute.

\section{A Characterization of Invariant Connections}
\label{CHarinvconn}
The set of connections on a principal fibre bundle $(P,\pi,M,S)$ is closed under pullback by automorphisms, and it is natural to search for connections that do not change under this operation. Especially, connections invariant under a Lie group  $(G,\Phi)$ of automorphisms are of particular interest as they reflect the symmetry of the whole group and, for this reason, find their applications in the symmetry reduction of (quantum) gauge field theories. \cite{MathStrucLQG, ChrisSymmLQG}
The first classification theorem for such connections was given by Wang \cite{Wang}, cf.\ Case \ref{th:wang}. This applies to the situation where the induced action $\varphi$ acts transitively on the base manifold and states that each point in the bundle gives rise to a bijection between the set of $\Phi$-invariant connections and certain linear maps $\psi\colon \mathfrak{g}\rightarrow \mathfrak{s}$. In \cite{HarSni} the authors generalize this to the situation where $\varphi$ admits only one orbit type. More precisely, they discuss a variation\footnote{Amongst others, they assume the $\varphi$-stabilizer of $\pi(p_0)$ to be the same for all $p_0\in P_0$.} of the case where the bundle admits a submanifold $P_0$ with $\pi(P_0)$ intersecting each $\varphi$-orbit in a unique point, see Case \ref{scase:OneSlice} and Example \ref{example:SCHSV}. 
Here, the $\Phi$-invariant connections are in bijection with such smooth maps $\psi\colon \mathfrak{g}\times P_0\rightarrow \mathfrak{s}$ for which the restrictions $\psi|_{\mathfrak{g}\times T_{p_0}P_0}$ are linear for all $p_0\in P_0$ and that fulfil additional consistency conditions.

Now, in the general case we consider \emph{$\Gg$-coverings} of $P$. These are families $\{P_\alpha\}_{\alpha\in I}$ of immersed submanifolds\footnote{At the moment assume that $P_\alpha\subseteq P$ is a subset which, at the same time, is a manifold such that the inclusion map $\iota_\alpha \colon P_\alpha \rightarrow P$ is an immersion. Here, we tacitly identify $T_{p_\alpha}P_\alpha$ with $\im[\dd_{p_\alpha}\iota_\alpha]$. Note that we do not require $P_\alpha$ to be an embedded submanifold of $P$. Details will be given in Convention \ref{conv:Submnfds}.} $P_\alpha$ of $P$ such that each $\varphi$-orbit has non-empty intersection with $\bigcup_{\alpha\in I}\pi(P_\alpha)$ and for which 
\begin{align*}
	T_{p}P = T_{p}P_\alpha + \dd_e\Phi_p(\mathfrak{g}) + Tv_{p}P 
\end{align*}
holds whenever $p\in P_\alpha$ for some $\alpha\in I$. Here, $Tv_{p}P\subseteq T_pP$ denotes the vertical tangent space at $p\in P$ and $e$ the identity in $G$. Observe that the intersection properties of the sets $\pi(P_\alpha)$ with the $\varphi$-orbits in the base manifold need not to be convenient in any sense. Here one might think of situations in which $\varphi$ admits dense orbits, or of the almost fibre transitive case, cf.\ Case \ref{scase:slicegleichredcluster}.

Let $\THA\colon (G\times S)\times P\rightarrow P$ be defined by $((g,s),p)\mapsto \Phi(g,p)\cdot s^{-1}$ for $(G,\Phi)$ a Lie group of automorphisms of $\PMS$. Then, the main result of this section can be stated as follows:
\begin{satz}
 % Let $\PMS$ be a principal fibre bundle and $(G,\Phi)$ a Lie group of automorphisms thereon. 
 Each $\Gg$-covering $\{P_\alpha\}_{\alpha\in I}$ of $P$ give rise to a bijection between the $\Phi$-invariant connections on $P$ and the families $\{\psi_\alpha\}_{\alpha\in I}$ of smooth maps $\psi_\alpha\colon \mathfrak{g}\times TP_\alpha \rightarrow \mathfrak{s}$ such that ${\psi_\alpha}|_{\mathfrak{g}\times T_{p_\alpha} P_\alpha}$ is linear for all $p_\alpha\in P_\alpha$ and that fulfil the following two (generalized Wang) conditions: 
  \begingroup 
  \setlength{\leftmargini}{20pt}
  \begin{itemize}
  \item
  \vspace{-1pt}
    $\wt{g}(p_\beta) + \vec{w}_{p_\beta}-\wt{s}(p_\beta)=\dd L_q\vec{w}_{p_\alpha}\quad \Longrightarrow\quad \psi_\beta(\vec{g},\vec{w}_{p_\beta})-\vec{s}=\text{\gls{QREP}}(q)\cp\psi_\alpha\big(\vec{0}_{\mathfrak{g}},\vec{w}_{p_\alpha}\big)$,
  \item
  	\vspace{-1pt}
    $\psi_\beta\big(\Add{q}(\vec{g}),\vec{0}_{p_\beta}\big)=\qrep(q)\cp \psi_\alpha\big(\vec{g},\vec{0}_{p_\alpha}\big)$.
  \end{itemize}
  \endgroup
  \vspace{-4pt}
  \noindent
  Here, $q\in G\times S$, $p_\alpha\in P_\alpha$, $p_\beta\in P_\beta$ with $p_\beta=q\cdot p_\alpha$ and $\vec{w}_{p_\alpha}\in T_{p_\alpha}P_\alpha$, $\vec{w}_{p_\beta}\in T_{p_\beta}P_\beta$. Moreover, $\wt{g}$ and $\wt{s}$ denote the fundamental vector fields that correspond to the elements $\vec{g}\in \mathfrak{g}$ and $\vec{s}\in \mathfrak{s}$, respectively, $\rho$ is the map from Definition \ref{def:Invconn} and $\Ad_q(\g):=\Ad_g(\g)$ for $q=(g,s)\in Q$. 
\end{satz}
\noindent
Using this theorem, the calculation of invariant connections reduces to identifying a $\Gg$-covering that makes the above conditions as easy as possible. Here, one has to 
find the balance between quantity and complexity of these conditions. Of course, the more submanifolds there are, the more conditions we have, so that usually it is convenient to use as few of them as possible. For instance, in the situation where $\varphi$ is transitive it suggests itself to choose a $\Gg$-covering that consists of one single point, which, in turn, has to be chosen appropriately. Also if there is some $m\in M$ contained in the closure of each $\varphi$-orbit, one single submanifold is sufficient, see Case \ref{scase:slicegleichredcluster} and Example \ref{ex:Bruhat}. The same example shows that sometimes pointwise\footnote{Here pointwise means to consider such elements $q\in G\times S$ that are 
  contained in the $\THA$-stabilizer of some fixed $p_\alpha\in P_\alpha$ for some $\alpha\in I$.} evaluation of the above conditions proves non-existence of $\Phi$-invariant connections.

In any case, one can use the inverse function theorem to construct a $\Gg$-covering $\{P_\alpha\}_{\alpha\in I}$ of $P$ such that the submanifolds $P_\alpha$ have minimal dimension in a certain sense, see Lemma \ref{lemma:suralpha} and Corollary \ref{cor:reductions}. This reproduces the description of connections by means of local 1-forms on $M$ provided that $G$ acts trivially or, more generally, via gauge transformations on $P$, see Case \ref{scase:GaugeTransf}. 

Finally, since orbit structures can depend very sensitively on the action or the group, one cannot expect to have a general concept for finding the $\Gg$-covering optimal for calculations.
Indeed, sometimes these calculations become easier if one uses coverings that seem less optimal at a first sight (as, e.g., if they have no minimal dimension, cf.\ calculations in Appendix \ref{subsec:IsotrConn}).

  In the first part, we will introduce the notion of a $\Gg$-covering, the central object of this section. In the second part, we prove the main theorem and deduce a slightly more general version of the result from \cite{HarSni}. In the last part, we will show how to construct $\Gg$-coverings to be used in special situations. In particular, we consider the (almost) fibre transitive case, trivial principal fibre bundles and Lie groups of gauge transformations. Along the way we give applications to loop quantum gravity.

\subsection{$\Phi$-Coverings}
\label{sec:phiCoverings}
We will start this subsection with some facts and conventions concerning submanifolds. Then, we provide the definition of a $\Gg$-covering and discuss some its properties.
\begin{convention}
  \label{conv:Submnfds}
  Let $M$ be a manifold.
  \begingroup
  \setlength{\leftmargini}{25pt}
  \begin{enumerate}
  \item
    \itspace
    A pair $(N,\tau_N)$ consisting of a manifold $N$ and an injective immersion $\tau\colon N\rightarrow M$ is called submanifold of $M$. 
  \item
    \itspace
    If $(N,\tau_N)$ is a submanifold of $M$, we tacitly identify $N$ and $TN$ with their images $\tau_N(N)\subseteq M$ and $\dd\tau_N(TN)\subseteq TM$, respectively.
    In particular, this means:
    \begingroup
    \setlength{\leftmarginii}{11pt}
    \begin{itemize}
    \item
      \itspace
      If $M'$ is a manifold and $\kappa\colon M\rightarrow M'$ a smooth map, then for $x\in N$ and $\vec{v}\in TN$ we write $\kappa(x)$ and $\dd\kappa(\vec{v})$ instead of $\kappa(\tau_N(x))$ and $\dd\kappa( \dd\tau(\vec{v}))$, respectively.
    \item
      If $\Psi\colon G\times M\rightarrow M$ is a left action of the Lie group $G$ and $(H,\tau_H)$ a submanifold of $G$, then the restriction of $\Psi$ to $H\times N$ is defined by
      \begin{align*}
        \Psi|_{H\times N}(h,x):= \Psi(\tau_H(h),\tau_N(x)) \qquad \forall\: (h,x)\in H\times N.  
      \end{align*}
    \item
      If $\w\colon TM\rightarrow V$ is a $V$-valued 1-form on $M$, then 
      \begin{align*}
        (\Psi^*\w)|_{TG\times TN}\:(\vec{m},\vec{v}):=(\Psi^*\w)(\vec{m},\dd\tau(\vec{v}))\qquad\quad\forall\: (\vec{m},\vec{v})\in TG\times TN.  
      \end{align*}
    \item	
      We will not explicitly refer to the maps $\tau_N$ and $\tau_H$ in the following. 
    \end{itemize}
    \endgroup
  \item
    \itspacec
    Open subsets $U\subseteq M$ are equipped with the canonical manifold structure making the inclusion map an embedding.
  \item
    \itspace
    If $L$ is a submanifold of $N$ and $N$ is a submanifold of $M$, we consider $L$ as a submanifold of $M$ in the canonical way.
    \hspace*{\fill}{$\Diamond$}
  \end{enumerate}
  \endgroup
\end{convention}	
\begin{definition}
  A submanifold $N\subseteq M$ is called $\Psi$-patch iff for each $x\in N$ there is an open neighbourhood $N'\subseteq N$ of $x$ and a submanifold $H$ of $G$ through $e$ such that the restriction $\Psi|_{H\times N'}$ is a diffeomorphism to an open subset $U\subseteq M$.  
\end{definition}
\begin{remark}
  \label{rem:patch}
  \begingroup
  \begin{enumerate}
  \item
    \itspacec
    It follows from the inverse function theorem and\footnote{The sums are not necessarily direct.}
    \begin{align*}
      \dd_{(e,x)}\Psi(\mathfrak{g}\times T_xN)= \dd_e\Psi_x(\mathfrak{g}) + \dd_x\Psi_e(T_xN)=  \dd_e\Psi_x(\mathfrak{g})+T_xN\qquad \forall\: x\in N
    \end{align*}
    that $N$ is a $\Psi$-patch iff for each $x\in N$ we have $T_xM=  \dd_e\Psi_x(\mathfrak{g})+T_xN $.\footnote{In fact, let $V\subseteq \dd_e\Psi_x(\mathfrak{g})$ be an algebraic complement of $T_xN$ in $T_xM$ and $V'\subseteq  \mathfrak{g}$ a linear subspace with $\dim[V']=\dim[V]$ and  $\dd_e\Psi_x(V')=V$. Then, we find a submanifold $H$ of $G$ through $e$ with $T_eH=V'$, so that $\dd_{(e,x)}\Psi\colon T_eH\times T_xN \rightarrow T_xM$ is bijective.}
  \item
    \itspacec
    Open subsets $U\subseteq M$ are always $\Psi$-patches. They are of maximal dimension, which, for instance, is necessary if there is a point in $U$ whose stabilizer equals $G$, see Lemma \ref{lemma:mindimslice}.1.
  \item
    \itspacec
    We allow zero-dimensional patches, i.e., $N=\{x\}$ for $x\in M$. Necessarily, then $\dd_e\Psi_x(\mathfrak{g})=T_xM$ and $\Psi|_{H\times N}=\Psi_x|_{H}$ for each submanifold $H$ of $G$.\hspace*{\fill}{{$\Diamond$}} 
  \end{enumerate}
  \endgroup
\end{remark}
\noindent
The second part of the next elementary lemma equals Lemma 2.1.1 in \cite{DuisKolk}.
\begin{lemma}
  \label{lemma:mindimslice}
  Let $(G,\Psi)$ be a Lie group that acts on the manifold $M$ and let $x\in M$. 
  \begin{enumerate}
  \item
    \label{item:s}  
    If $N$ is a $\Psi$-patch with $x\in N$, then $\dim[N]\geq \dim[M]-\dim[G]+\dim[G_x]$. 
  \item  	
    Let $V$ and $W$ be algebraic complements of $\dd_e\Psi_x(\mathfrak{g})$ in $T_xM$ and of $\mathfrak{g}_x$ in $\mathfrak{g}$, respectively. Then, there are submanifolds $N$ of $M$ through $x$ and $H$ of $G$ through $e$ such that $T_xN=V$ and $T_eH=W$. In particular, $N$ is a $\Psi$-patch and $\dim[N]=\dim[M]-\dim[G]+\dim[G_x]$.  
  \end{enumerate}
  \begin{proof}
    \begin{enumerate}
    \item
      By Remark \ref{rem:patch}.1 and since $\ker[\dd_e\Psi_x]=\mathfrak{g}_x$, we have
      \begin{align}
        \label{eq:leq}
        \dim[M]\leq \dim[\dd_e\Psi_x(\mathfrak{g})]+\dim[T_xN]=\dim[G]-\dim[G_x]+\dim[N].
      \end{align}
    \item
      Of course, we find submanifolds $N'$ of $M$ through $x$ and $H'$ of $G$ through $e$ such that $T_xN'=V$ and $T_eH'=W$. So, if $\vec{g}\in \mathfrak{g}$ and $\vec{v}_x\in T_xN'$, then $0=\dd_{(e,x)}\Psi(\vec{g},\vec{v}_x)=\dd_e\Psi_x(\vec{g})+\vec{v}_x$ implies $\dd_e\Psi_x(\vec{g})=0$ and $\vec{v}_x=0$. Hence, $\vec{g}\in \ker[\dd_e\Psi_x]=\mathfrak{g}_x$ so that\footnote{Recall that $\dd_{(e,x)}\Psi|_{T_eH'\times T_eN'}\colon \big(\raisebox{-1pt}{$\vec{h},\vec{v}_x$}\big)\mapsto \dd_{(e,x)}\Psi\big(\raisebox{-1pt}{$\dd_e\tau_H(\vec{h}),\dd_x\tau_N(\vec{v}_x)$}\big)$.} $\dd_{(e,x)}\Psi|_{T_eH'\times T_eN'}$ is injective. It is immediate from the definitions that this map is surjective so that by the inverse function theorem we find open neighbourhoods $N\subseteq N'$ of $x$ and $H\subseteq G$ of $e$ such that $\Psi|_{H\times N}$ is a diffeomorphism to an open subset $U\subseteq M$. Then $N$ is a $\Psi$-patch and since in \eqref{eq:leq} equality holds, also the last claim is clear.
    \end{enumerate}
  \end{proof}
\end{lemma}
\begin{definition}
  \label{def:pmappe}
  Let $(G,\Phi)$ be a Lie group of automorphisms of the principal fibre bundle $P$ and recall the actions $\varphi$ and $\THA$ defined by \eqref{eq:INDA} and \eqref{eq:THETA}, respectively. 
  A family of $\THA$-patches $\{P_\alpha\}_{\alpha\in I}$ is said to be a $\Gg$-covering of $P$ iff each $\varphi$-orbit intersects at least one of the sets $\pi(P_\alpha)$. 
\end{definition} 
\begin{remark}
  \label{bem:Psliceeigensch}
  \begin{enumerate}
  \item
    If $O\subseteq P$ is a $\THA$-patch, then Lemma \ref{lemma:mindimslice}.1 and \eqref{eq:staoQ} yield 
    \vspace{-1ex}
    \begin{align*}
      \dim[O]&\geq \dim[P]-\dim[Q]+\dim[Q_p]
      \stackrel{\eqref{eq:staoQ}}{=}\dim[M]-\dim[G]+\dim[G_{\pi(p)}].
    \end{align*}
  \item
    \itspace 
    It follows from Remark \ref{rem:patch}.1 and $\dd_e\THA_p(\mathfrak{q})=\dd_e\Phi_p(\mathfrak{g}) + Tv_pP$ that $O$ is a $\THA$-patch iff
    \begin{align}
      \label{eq:transv}
      T_pP=T_pO+ \dd_e\Phi_p(\mathfrak{g}) + Tv_pP\qquad \forall\: p\in O.
    \end{align}    
    As a consequence
    \begingroup
    \setlength{\leftmarginii}{13pt}
    \begin{itemize}
    \item
      \itspacec
      each $\Phi$-patch is a $\THA$-patch,
    \item
      $P$ is always a $\Gg$-covering by itself, and if $P=M\times S$ is trivial, then $M\times \{e\}$ is a $\Gg$-covering.
    \end{itemize}
    \endgroup
  \item
    \itspace
    If $N$ is a $\varphi$-patch and $s_0\colon N \rightarrow P$ a smooth section, i.e., $\pi\cp s_0 =\id_N$, then $O:=s_0(N)$ is a $\THA$-patch by Lemma \ref{lemma:suralpha}.2.\hspace*{\fill}{{$\Diamond$}} 
  \end{enumerate}
\end{remark}
\begin{lemma}
  \label{lemma:suralpha} 
  Let $(G,\Phi)$ be a Lie group of automorphisms of the principal fibre bundle $(P,\pi,M,S)$.
  \begin{enumerate}
  \item
    If $O\subseteq P$ is a $\THA$-patch, then for each $p\in O$ and $q\in Q$ the differential 
    $\dd_{(q,p)}\THA\colon  T_qQ\times T_{p}O\rightarrow T_{q\cdot p}P$
    is surjective.
  \item
    If $N$ is a $\varphi$-patch and $s_0\colon N \rightarrow P$  a smooth section, then $O:=s_0(N)$ is a $\THA$-patch.
  \end{enumerate}
  \begin{proof}
    \begin{enumerate}
    \item
      Since $O$ is a $\THA$-patch, the claim is clear for $q=e$. If $q$ is arbitrary, then for each $\kk_q\in T_qQ$ we find some $\vec{q}\in\mathfrak{q}$ such that $\kk_q=\dd L_q\vec{q}$. Consequently, for $\vec{w}_{p}\in T_{p}P$ we have
      \begin{align*}
        \dd_{(q,p)}\THA\left(\kk_q,\vec{w}_{p}\right)&=\dd_{(q,p)}\THA(\dd L_q\vec{q},\vec{w}_{p})=\dd_{p}L_q\left(\dd_{(e,p)}\THA(\vec{q},\vec{w}_{p})\right).
      \end{align*}
      So, since left translation w.r.t.\ $\THA$ is a diffeomorphism, $\dd_{p}L_q$ is surjective.
    \item
      First observe that $O$ is a submanifold of $P$ because $s_0$ is an injective immersion.
      By Remark \ref{bem:Psliceeigensch}.2 it suffices to show that
      \begin{align*}
        \dim\big[T_{s_0(x)}O + \dd_e\Phi_{s_0(x)}(\mathfrak{g})+ Tv_{s_0(x)}P\big]\geq \dim[\hspace{1pt}T_{s_0(x)}P\hspace{1pt}]\qquad \forall\: x\in N. 
      \end{align*}
      For this, let $x\in N$ and $V'\subseteq \mathfrak{g}$ be a linear subspace with $V'\oplus \mathfrak{g}_x$ and $T_xM=T_xN \oplus
\mathrm{d}_e\varphi_x(V')$. Then,   
      $T_{s_0(x)}O \oplus \dd_e\Phi_{s_0(x)}(V')\oplus Tv_{s_0(x)}P$ 
      because if $\dd_xs_0(\vec{v}_x)  +\dd_e\Phi_{s_0(x)}(\vec{g}\os')+ \vec{v}_v=0$ for $\vec{v}_x\in T_xN$, $\vec{g}\os'\in V'$ and $\vec{v}_v\in Tv_{s_0(x)}P$, then 
      \begin{align*}
        0=\dd_{s_0(x)}\pi \big(\dd_xs_0(\vec{v}_x)  +\textstyle\dd_e\Phi_{s_0(x)}(\vec{g}\os')+ \vec{v}_v\big)=\vec{v}_x \oplus \dd_e\varphi_x(\vec{g}\os')
      \end{align*}
      showing that $\vec{v}_x =0$ and $\mathrm{d}_e\phi_{x}(\vec{g}')=0$, hence $\vec{g}'=0$ by the choice of $V'$, i.e., $\vec{v}_v=0$ by assumption.
In particular, $\mathrm{d}_e\phi_{x}(\vec{g}')=0$ if $\mathrm{d}_e\Phi_{s_0(x)}(\vec{g}')=0$, hence $\dim[\mathrm{d}_e\Phi_{s_0(x)}(V')]\geq\dim[\mathrm{d}_e\varphi_x(V')]$, from which we obtain
      \begin{align*}
        \dim\big[T_{s_0(x)}O  + \dd_e\Phi_{s_0(x)}(\mathfrak{g})+ Tv_{s_0(x)}P\big]&\\
        & \hspace{-60pt}\geq \dim\big[T_{s_0(x)}O  \oplus \dd_e\Phi_{s_0(x)}(V')\oplus Tv_{s_0(x)}P\big] \\
        &\hspace{-60pt} =\dim[T_xN] + \dim[\dd_e\Phi_{s_0(x)}(V')] + \dim[S]\\
        &\hspace{-60pt} \geq\dim[T_xN] + \dim[\dd_e\varphi_x(V')] + \dim[S]\\
        &\hspace{-60pt} =\dim[P].
      \end{align*}
    \end{enumerate}
  \end{proof}
\end{lemma}

\subsection{Characterization of Invariant Connections}
\label{sec:mainth}
In this subsection, we will use $\Gg$-coverings $\{P_\alpha\}_{\alpha\in I}$ of the bundle $P$ in order to characterize the set of $\Phi$-invariant connections by families $\{\psi_\alpha\}_{\alpha\in I}$ of smooth maps $\psi_\alpha\colon \mathfrak{g}\times TP_\alpha \rightarrow \mathfrak{s}$ whose restrictions $\psi_\alpha|_{\mathfrak{g}\times T_{p_\alpha}P_\alpha }$ are linear and that fulfil two additional compatibility conditions. Here, we follow the lines of Wang's original approach, which basically means that we will generalize the proofs from \cite{Wang} to the non-transitive case.
We will proceed in two steps, the first one being performed in the next subsection. There, we show that a $\Phi$-invariant connection gives rise to a consistent family $\{\psi_\alpha\}_{\alpha\in I}$ of smooth maps as described above. We also discuss the situation in \cite{HarSni} in order to make the two conditions more intuitive.
Then, in Subsection \ref{subsec:Reconstruc}, we will verify that such families $\{\psi_\alpha\}_{\alpha\in I}$ glue together to a $\Phi$-invariant connection on $P$.
\subsubsection{Reduction of Invariant Connections}
\label{subsec:RedInvConn}
In the following, let $\{P_\alpha\}_{\alpha\in I}$ be a fixed $\Gg$-covering of $P$ and $\w$ a $\Gg$-invariant connection on $P$. We define $\w_\alpha:=(\THA^*\w)|_{TQ\times TP_\alpha}$ and $\psi_\alpha:=\w_\alpha|_{\mathfrak{g}\times TP_\alpha}$, and for $q'\in Q$ we let $\Co{q'}\colon Q\times P\rightarrow Q\times P$ denote the map $\Co{q'}(q,p):=\left(\Co{q'}(q),p\right)$ for $\Co{q'}\colon Q\mapsto Q$ the conjugation map w.r.t.\ $q'$. 
Finally, we let $\Ad_q(\g):=\Ad_g(\g)$ for $q=(g,s)\in Q$ and $\g\in \mg$ in the sequel.
\begin{lemma}
  \label{lemma:omegaalpha} 
  Let $q\in Q$, $p_\alpha\in P_\alpha$,  $p_\beta\in P_\beta$ with\footnote{More precisely, $\tau_{P_\beta}(p_\beta)=q\cdot \tau_{P_\alpha}(p_\alpha)$ by Convention \ref{conv:Submnfds}.} $p_\beta=q\cdot p_\alpha$ and $\vec{w}_{p_\alpha}\in T_{p_\alpha}P_\alpha$. Then  
  \begin{enumerate}
  \item
    $\w_\beta(\xii\:)=\qrep(q)\cp \w_\alpha(\vec{0}_\mathfrak{q},\vec{w}_{p_\alpha})$ for all $\xii\in TQ\times TP_\beta$ with $\dd\THA(\xii\:)=\dd L_q\vec{w}_{p_\alpha}$,
  \item 
    $\left(\Co{q}^*\w_\beta\right)\big(\raisebox{-0ex}{$\kk,\vec{0}_{p_\beta}$}\big)=\qrep(q)\cp \w_\alpha\big(\raisebox{-0ex}{$\kk,\vec{0}_{p_\alpha}$}\big)$ for all $\kk\in TQ$.
  \end{enumerate}
  \begin{proof}
    \begin{enumerate}
    \item
      Let $\xii\in  T_{q'}Q\times T_pP_\beta$ for $q'\in Q$. Then, since\footnote{See end of Subsection \ref{subsec:InvConn}.} $L_q^*\w = \qrep(q)\cp \w$ for each $q\in Q$ and $q'\cdot p= q\cdot p_\alpha =p_\beta$, we have
      \begin{align*} 
        \w_\beta(\xii\:)&=\w_{q'\cdot p}(\dd_{(q', p)}\THA (\xii\:))=\w_{p_\beta}(\dd L_q\vec{w}_{p_\alpha})
        =(L_q^*\w)_{p_\alpha}(\vec{w}_{p_\alpha})\\
        &=\qrep(q)\cp \w_{p_\alpha}(\vec{w}_{p_\alpha})=\qrep(q)\cp \w_{p_\alpha}\big(\dd_{(e,p_\alpha)}\THA (\vec{0}_\mathfrak{q},\vec{w}_{p_\alpha})\big)\\
        &=\qrep(q)\cp \w_{\alpha}\raisebox{1pt}{$\big($}\vspace{2pt}\vec{0}_\mathfrak{q},\vec{w}_{p_\alpha}\raisebox{1pt}{$\big)$}.
      \end{align*}
    \item
      For $\kk_{q'} \in T_{q'}Q$ let $\gamma \colon (-\epsilon,\epsilon)\rightarrow Q$ be smooth with $\dot\gamma(0)=\kk_{q'}$. Then
      \begin{align*}
        \big(\Co{q}^*\w_\beta \big)_{(q',p_\beta)}\big(\raisebox{-0ex}{$\kk_{q'},\vec{0}_{p_\beta}$}\big)&={\w_\beta}_{(\Co{q}(q'),p_\beta)}\big(\Add{q}(\kk_{q'}),\vec{0}_{p_\beta}\big)\\
        &=\w_{qq'q^{-1}q\cdot p_\alpha }\left(\dttB{t}{0}q\gamma(t) q^{-1}q\cdot p_\alpha\right)\\
        & =\left(L_q^*\w\right)_{q'\cdot p_\alpha }\left(\dttB{t}{0}\gamma(t)\cdot p_\alpha\right)\\
        & =\qrep(q)\cp \w_{q'\cdot p_\alpha}\left(\dd_{(q',p_\alpha)}\THA\left( \kk_{q'}\right)\right)\\
        &=\qrep(q)\cp {\w_\alpha}_{(q',p_\alpha)} \big(\kk_{q'},\vec{0}_{p_\alpha}\big).
      \end{align*}
    \end{enumerate}
  \end{proof}
\end{lemma}

\begin{corollary}
  \label{cor:psialpha}
  Let $q\in Q$, $p_\alpha\in P_\alpha$,  $p_\beta\in P_\beta$ with $p_\beta=q\cdot p_\alpha$ and $\vec{w}_{p_\alpha}\in T_{p_\alpha}P_\alpha$. Then for
  $\vec{w}_{p_\beta}\in T_{p_\beta}P_\beta$, $\vec{g}\in \mathfrak{g}$ and $\vec{s}\in \mathfrak{s}$ we have
  \begin{enumerate}
  \item[\textit{i.)}]
    $\wt{g}(p_\beta) + \vec{w}_{p_\beta}-\wt{s}(p_\beta)=\dd L_q\vec{w}_{p_\alpha}\quad \Longrightarrow\quad \psi_\beta(\vec{g},\vec{w}_{p_\beta})-\vec{s}=\qrep(q)\cp\psi_\alpha\big(\vec{0}_{\mathfrak{g}},\vec{w}_{p_\alpha}\big)$,
  \item[\textit{ii.)}]
    $\psi_\beta\big(\Add{q}(\vec{g}),\vec{0}_{p_\beta}\big)=\qrep(q)\cp \psi_\alpha\big(\vec{g},\vec{0}_{p_\alpha}\big)$.
  \end{enumerate}
  \begin{proof}
    \begin{enumerate}
    \item[\textit{i.)}]
      In general, for $\vec{w}_p\in T_pP$, $\vec{g}\in\mathfrak{g}$ and $\vec{s}\in\mathfrak{s}$ we have
      \begin{align}
        \label{eq:ThetaPhi}
        \dd_{(e,p)}\THA((\vec{g},\vec{s}),\vec{w}_p)=\dd_{(e,p)}\Phi(\vec{g},\vec{w}_p)-\wt{s}(p)=\wt{g}(p)+\vec{w}_p-\wt{s}(p)
      \end{align}
      and, since $\w$ is a connection, for $((\vec{g},\vec{s}),\vec{w}_{p_\alpha})\in  \mathfrak{q}\times TP_\alpha$ we obtain
      \begin{align}
        \label{eq:gugu}
        \begin{split}
          \w_\alpha((\vec{g},\vec{s}),\vec{w}_{p_\alpha})&=\w\big(\dd_{(e,p_\alpha)}\Phi(\vec{g},\vec{w}_{p_\alpha})-\wt{s}(p_\alpha)\big)\\
          &=\w\big(\dd_{(e,p_\alpha)}\Phi(\vec{g},\vec{w}_{p_\alpha})\big)-\vec{s}\\
          &=\w_\alpha\left(\vec{g},\vec{w}_{p_\alpha}\right)-\vec{s} =\psi_\alpha\left(\vec{g},\vec{w}_{p_\alpha}\right)-\vec{s}.
        \end{split}
      \end{align}
      Now, assume that $\dd_e\Phi_{p_\beta}(\vec{g}\hspace{1pt})+\vec{w}_{p_\beta}-\wt{s}(p)=\dd L_q\vec{w}_{p_\alpha}$. Then $\dd_{(e,p_\beta)}\THA((\vec{g},\vec{s}),\vec{w}_{p_\beta})=\dd L_q \vec{w}_{p_\alpha}$ by \eqref{eq:ThetaPhi} so that
      $\w_\beta((\vec{g},\vec{s}),\vec{w}_{p_\beta})=\qrep(q)\cp \w_\alpha\big(\vec{0}_{\mathfrak{g}},\vec{w}_{p_\alpha}\big)$  by Lemma \ref{lemma:omegaalpha}.1. Consequently, 
      \vspace{-8pt}
      \begin{align*}
        \psi_\beta\left(\vec{g},\vec{w}_{p_\beta}\right)-\vec{s}&\stackrel{\eqref{eq:gugu}}{=}\w_\beta((\vec{g},\vec{s}),\vec{w}_{p_\beta})\\ & \:\: =\qrep(q)\cp \w_\alpha\big(\vec{0}_\mathfrak{q},\vec{w}_{p_\alpha}\big)\stackrel{\eqref{eq:gugu}}{=}\qrep(q)\cp \psi_\alpha\big(\vec{0}_{\mathfrak{g}},\vec{w}_{p_\alpha}\big).
      \end{align*}
    \item[\textit{ii.)}]
      \vspace{-8pt}
      Lemma \ref{lemma:omegaalpha}.2 yields
      \begin{align*}
        \psi_\beta\big(\Add{q}(\vec{g}),\vec{0}_{p_\beta}\big)&=(\Co{q}^*\w_\beta)_{(e,p_\beta)}\big(\vec{g},\vec{0}_{p_\beta}\big)
        \\ &=\qrep(q)\cp (\w_\alpha)_{(e,p_\alpha)}\big(\vec{g},\vec{0}_{p_\alpha}\big)
        =\qrep(q)\cp \psi_\alpha\big(\vec{g},\vec{0}_{p_\alpha}\big).
      \end{align*}
    \end{enumerate}
  \end{proof}
\end{corollary} 

\begin{definition}
  A family $\{\psi_\alpha\}_{\alpha\in I}$ of smooth maps $\psi_\alpha\colon \mathfrak{g}\times TP_\alpha \rightarrow \mathfrak{s}$ which are linear in the sense that $\psi_\alpha|_{\mathfrak{g}\times T_{p_\alpha}P_\alpha }$ is linear for all $p_\alpha\in P_\alpha$ is called reduced connection w.r.t.\ $\{P_\alpha\}_{\alpha\in I}$ iff it fulfils the conditions \textit{i.)} and \textit{ii.)} from Corollary \ref{cor:psialpha}. 
\end{definition}
\begin{remark}
  \label{rem:Psialphaconderkl}
  \begingroup
  \setlength{\leftmargini}{18pt}
  \begin{enumerate}
  \item[{\bf 1)}]
    In particular, Corollary \ref{cor:psialpha}.\textit{i.)} encodes the following condition 
    \begingroup
    \setlength{\leftmarginii}{20pt}
    \begin{enumerate}
    \item[\textit{a.)}]
      For all $\beta \in I$, $(\vec{g},\vec{s})\in \mathfrak{q}$ and $\vec{w}_{p_\beta}\in T_{p_\beta}P_\beta$ we have
      \begin{align*}
        \wt{g}(p_\beta) +\vec{w}_{p_\beta}-\wt{s}(p_\beta)=0\quad \Longrightarrow\quad \psi_\beta(\vec{g},\vec{w}_{p_\beta})-\vec{s}=0.
      \end{align*}    
    \end{enumerate}
    \endgroup
  \item[{\bf 2)}]
    Assume that \textit{a.)} is true and let $q\in Q$, $p_\alpha\in P_\alpha$, $p_\beta \in P_\beta$ with $p_\beta=q\cdot p_\alpha$. Moreover, assume that we find elements $\vec{w}_{p_\alpha}\in T_{p_\alpha}P_\alpha$ and $((\vec{g},\vec{s}),\vec{w}_{p_\beta})\in \mathfrak{q}\times T_{p_\beta}P_\beta$ such that 
    \begin{align*}
      \dd_{(e,p_\beta)}\THA((\vec{g},\vec{s}),\vec{w}_{p_\beta})=\dd L_q\vec{w}_{p_\alpha}\quad \text{and}\quad \psi_\beta(\vec{g},\vec{w}_{p_\beta})-\vec{s}=\qrep(q)\cp\psi_\alpha(\vec{0}_{\mathfrak{g}},\vec{w}_{p_\alpha})
    \end{align*}
    holds.
    Then
    $\psi_\beta\big(\vec{g}\hspace{1pt}',\vec{w}'_{p_\beta}\big)-\vec{s}\hspace{1pt}'=\qrep(q)\cp\psi_\alpha\big(\vec{0}_{\mathfrak{g}},\vec{w}_{p_\alpha}\big)$ holds for each element\footnote{Observe that due to surjectivity of $\dd_{(e,p_\beta)}\Phi$ such elements always exist.} $\big((\vec{g}\hspace{1pt}',\vec{s}\hspace{1pt}'),\vec{w}'_{p_\beta}\big)\in \mathfrak{q}\times T_{p_\beta}P_\beta$ with  $\dd_{(e,p_\beta)}\THA\big((\vec{g}\hspace{1pt}',\vec{s}\hspace{1pt}'),\vec{w}'_{p_\beta}\big)=\dd L_q\vec{w}_{p_\alpha}$.
    In fact, we have 
    \begin{align*}
      \dd_{(e,p_\beta)}\THA\big((\vec{g}-\vec{g}\hspace{1pt}',\vec{s}-\vec{s}\hspace{1pt}'),\vec{w}_{p_\beta}-\vec{w}'_{p_\beta}\big)=0,
    \end{align*}	    
    so that by \eqref{eq:ThetaPhi} condition \textit{a.)}  gives
    \vspace{-4pt}
    \begin{align*}
      0 & \stackrel{\textit{a.)}}{=}\psi_\beta(\vec{g}-\vec{g}\hspace{1pt}',\vec{w}_{p_\beta}-\vec{w}'_{p_\beta})-(\vec{s}-\vec{s}\hspace{1pt}'))
      \hspace{2pt}= [\psi_\beta(\vec{g},\vec{w}_{p_\beta})-\vec{s}\hspace{2pt}]-[\psi_\beta(\vec{g}\hspace{1pt}',\vec{w}'_{p_\beta})-\vec{s}\hspace{1pt}']\\ 
      &= \qrep(q)\cp\psi_\alpha\big(\vec{0}_{\mathfrak{g}},\vec{w}_{p_\alpha}\big)-[\psi_\beta(\vec{g}\hspace{1pt}',\vec{w}'_{p_\beta})-\vec{s}\hspace{1pt}'].
    \end{align*}
  \item[{\bf 3)}]
    Assume that $\dd L_q \vec{w}_{p_\alpha}\in T_{p_\beta}P_\beta$ holds for all $q\in Q$, $p_\alpha\in P_\alpha$, $p_\beta \in P_\beta$ with $p_\beta=q\cdot p_\alpha$ and all $\vec{w}_{p_\alpha}\in T_{p_\alpha}P_\alpha$. Then, $\dd_{(e,p_\beta)}\THA\left(\dd L_q \vec{w}_{p_\alpha}\right)=\dd L_q \vec{w}_{p_\alpha}$, so that it follows from {\bf 2)} that in this case we can substitute 
    \textit{i.)} by \textit{a.)} and condition
    \begingroup
    \setlength{\leftmarginii}{20pt}
    \begin{enumerate}
    \item[\textit{b.)}]
      \itspacec
      Let $q\in Q$, $p_\alpha\in P_\alpha$, $p_\beta \in P_\beta$ with $p_\beta=q\cdot p_\alpha$. Then
      \begin{align*}
        \psi_\beta\big(\vec{0}_{\mathfrak{g}},\dd L_q\vec{w}_{p_\alpha}\big)=\qrep(q)\cp\psi_\alpha\big(\vec{0}_{\mathfrak{g}},\vec{w}_{p_\alpha}\big)\qquad\quad \forall\:\vec{w}_{p_\alpha}\in T_{p_\alpha}P_\alpha.
      \end{align*}
    \end{enumerate}   
    \endgroup 
    \noindent
    Now, \textit{b.)} looks similar to \textit{ii.)} and makes it plausible that the conditions \textit{i.)} and \textit{ii.)} from Corollary \ref{cor:psialpha} together encode the $\qrep$-invariance of the corresponding connection $\w$. 
    However, usually there is no reason for $\dd L_q \vec{w}_{p_\alpha}$ to be an element of $T_{p_\beta}P_\beta$. Even for $p_\alpha=p_\beta$ and $q\in Q_{p_\alpha}$ this is usually not true. So, typically there is no way to split up \textit{i.)} into parts whose meaning is more intuitive. \hspace*{\fill}{{$\Diamond$}}
  \end{enumerate}
  \endgroup
\end{remark}
\noindent
Remark \ref{rem:Psialphaconderkl} immediately proves
\begin{scase}[Gauge Fixing]
  \label{scase:OneSlice}
  Let $P_0$ be a $\THA$-patch of the bundle $P$ such that $\pi(P_0)$ intersects each $\varphi$-orbit in a unique point and $\dd L_q(T_pP_0)\subseteq T_pP_0$ for all $p\in P_0$ and all $q \in Q_p$. Then, a corresponding reduced connection consists of one single smooth map $\psi\colon \mathfrak{g}\times TP_0\rightarrow \mathfrak{s}$, and we have $p=q\cdot p'$ for $q\in Q$, $p,p'\in P_0$ iff $p=p'$ and $q\in Q_p$. So, by Remark \ref{rem:Psialphaconderkl}  the two conditions from Corollary \ref{cor:psialpha} are equivalent to:
  \newline
  \vspace{-10pt}
  \newline
  Let $p\in P_0$, $q=(h,\phi_p(h))\in Q_p$, $\vec{w}_p\in T_pP_0$ and $\vec{g}\in \mathfrak{g}$, $\vec{s}\in \mathfrak{s}$. Then
  \begin{enumerate}
  \item[\textit{i'.)}]
    $\wt{g}(p) +\vec{w}_{p}-\wt{s}(p)=0\quad \Longrightarrow\quad \psi(\vec{g},\vec{w}_{p})-\vec{s}=0$, 
  \item[\textit{ii'.)}]
    $\psi\big(\vec{0}_{\mathfrak{g}},\dd L_q\vec{w}_{p}\big)=\qrep(q)\cp\psi\big(\vec{0}_{\mathfrak{g}},\vec{w}_{p}\big)$,
  \item[\textit{iii'.)}]
    $\psi\big(\Add{h}(\vec{g}),\vec{0}_{p}\big)=\Add{\phi_p(h)}\cp\: \psi\big(\vec{g},\vec{0}_{p}\big)$.\hspace*{\fill}{{\scriptsize$\blacksquare$}} 
  \end{enumerate}
\end{scase} 
\noindent
The next example is a slight generalization of Theorem 2 in \cite{HarSni}. There the authors assume that $\varphi$ admits only one orbit type so that $\dim[G_x]=l$ for all $x\in M$. Then, they restrict to the situation where one finds a triple $(U_0,\tau_0,s_0)$ consisting of an open subset $U_0\subseteq \RR^k$ for $k=\dim[M]- [\hspace{1pt}\dim[G]- l\hspace{1pt}]$, an embedding $\tau_0\colon U_0 \rightarrow M$ and a smooth map $s_0\colon U_0\rightarrow P$ with $\pi\cp s_0=\tau_0$ and the addition property that $Q_p$ is the same for all $p\in \im[s_0]$. More precisely, they assume that $G_x$ and the structure group of the bundle are compact. Then they show the non-trivial fact that $s_0$ can be modified in such a way that in addition $Q_p$ is the same for all $p\in \im[s_0]$.

Observe that the authors forgot to require that $\im[\dd_x\tau_0] + \im\big[\dd_e\varphi_{\tau_0(x)}\big]=T_{\tau_0(x)}M$ holds for all $x\in U_0$, i.e., that $\tau_0(U_0)$ is a $\varphi$-patch (so that $s_0(U_0)$ is a $\THA$-patch). Indeed, Example \ref{ex:transinv}.2 shows that this additional condition is crucial. The next example is a slight modification of the result \cite{HarSni} in the sense that we do not assume $G_x$ and the structure group to be compact, but make the ad hoc requirement that $Q_p$ is the same for all $p\in P_0$.
\begin{example}[Harnad, Shnider, Vinet]
  \label{example:SCHSV}
  Let $P_0$ be a $\THA$-patch of the bundle $P$ such that $\pi(P_0)$ intersects each $\varphi$-orbit in a unique point. Moreover, assume that the $\THA$-stabilizer $L:=Q_{p}$ is the same for all $p\in P_0$. Then, it is clear from \eqref{eq:staoQ} that $H:=G_{\pi(p)}$ and
  $\phi:=\phi_{p}\colon H \rightarrow S$ are independent of the choice of $p\in P_0$. Finally, we require that
  \begin{align}
    \label{eq:dimP}
    \begin{split}
      \dim[P_0]&=\dim[M]-[\dim[G]-\dim[H]]
      =\dim[P]-[\dim[Q]-\dim[H]]
    \end{split}
  \end{align}
  holds. 
  Now, let $p\in P_0$ and $q=(h,\phi(h))\in Q_p$. Then, for $\vec{w}_{p}\in T_{p}P_0$ we have
  \begin{align*}
    \dd L_q\vec{w}_{p}&=\dttB{t}{0} \Phi(h,\gamma(t))\cdot \phi^{-1}_{p}(h)
    =\dttB{t}{0} [\gamma(t)\cdot \phi_{\gamma(t)}(h)]\cdot \phi^{-1}_{p}(h)\\
    &=\dttB{t}{0} [\gamma(t)\cdot \phi_{p}(h)]\cdot \phi^{-1}_{p}(h)=\vec{w}_{p},
  \end{align*}
  where $\gamma\colon (-\epsilon,\epsilon)\rightarrow P_0$ is some smooth curve with $\dot\gamma(0)=\vec{w}_{p}$. Consequently, $\dd L_q(T_pP_0)\subseteq T_pP_0$ so that we are in the situation of Case \ref{scase:OneSlice}. Here, \textit{ii'.)} now reads $\psi\big(\vec{0}_{\mathfrak{g}},\vec{w}_{p}\big)=\Add{\phi(h)}\cp\:\psi\big(\vec{0}_{\mathfrak{g}},\vec{w}_{p}\big)$ for all $h\in H$ and \textit{iii'.)} does not change. For \textit{i'.)}, observe that the Lie algebra $\mathfrak{l}$ of $L$ is contained in the kernel of $\dd_{(e,p_0)}\THA$.
  But $\dd_{(e,p_0)}\THA$ is surjective since $P_0$ is a $\THA$-patch (cf.\ Lemma \ref{lemma:suralpha}.1), so that
  \vspace{-1ex}
  \begin{align*}
    \dim\!\big[\!\ker\!\big[\dd_{(e,p_0)}\THA\big]\big]= \dim[Q] + \dim[P_0]-\dim[P]\stackrel{\eqref{eq:dimP}}{=}\dim[H],
  \end{align*}
  whereby $\dd_{(e,p_0)}\THA$ means the differential of the restriction of $\THA$ to $Q\times P_0$. 
  This shows $\ker[\dd_{(e,p)}\THA]=\mathfrak{l}$ for all $p\in P_0$. Altogether, it follows that a reduced connection w.r.t.\ $P_0$ is a smooth, linear\footnote{In the sense that $\psi|_{\mathfrak{g}\times T_pP_0}$ is linear for all $p\in P_0$.} map $\psi\colon \mathfrak{g}\times TP_0\rightarrow \mathfrak{s}$ which fulfils the following three conditions:
  \begingroup
  \setlength{\leftmargini}{30pt}
  \begin{itemize}
  \item[\textit{i''.)}]
    \vspace{-4pt}
    $\psi\big(\raisebox{-0.15ex}{$\vec{h},\vec{0}_{p}$}\big)=\dd_e\phi\big(\raisebox{-0.15ex}{$\vec{h}$}\:\big)\qquad \qquad\quad\hspace{10pt} \qquad \qquad\forall \:\vec{h}\in \mathfrak{h}, \forall\:p\in P_0$,\hspace*{\fill}(use \eqref{eq:ThetaPhi})
  \item[\textit{ii''.)}]
    \itspace
    $\psi\big(\vec{0}_{\mathfrak{g}},\vec{w}\big)=\Add{\phi(h)}\cp \:\psi\big(\vec{0}_{\mathfrak{g}},\vec{w}\big) \qquad\qquad\quad\hspace{6.1pt}  \forall\:h\in H, \forall\: \vec{w}\in TP_0$, 
  \item[\textit{iii''.)}]
    \itspace
    $\psi\big(\Add{h}(\vec{g}),\vec{0}_{p}\big)=\Add{\phi(h)}\cp\: \psi\big(\raisebox{-0.5pt}{$\vec{g},\vec{0}_{p}$}\big)\quad\qquad\hspace{2.5pt} \forall\: \vec{g}\in\mathfrak{g}, \forall\: h\in H, \forall\: p\in P_0$.
  \end{itemize}
  \endgroup
  \vspace{-4pt}
  \noindent
  Then, $\mu:= \psi|_{TP_0}$ and $A_{p_0}(\vec{g}):=\psi\big(\vec{g},\vec{0}_{p_0}\big)$ are the maps that are used for the characterization in Theorem 2 in \cite{HarSni}.\hspace*{\fill}{$\Diamond$}
\end{example}

\subsubsection{Reconstruction of Invariant Connections}
\label{subsec:Reconstruc}
Let $\{P_\alpha\}_{\alpha\in I}$ be a $\Gg$-covering of $P$. We now show that each respective reduced connection $\{\psi_\alpha\}_{\alpha\in I}$ gives rise to a unique $\Gg$-invariant connection on $P$. To this end, for each $\alpha\in I$ we define the maps 
$\lambda_\alpha\colon \mathfrak{q}\times TP_\alpha\rightarrow \mathfrak{s},
((\vec{g},\vec{s}),\vec{w})\mapsto\psi_\alpha(\vec{g},\vec{w})-\vec{s}$ 
and
\begin{align*}
  \w_\alpha\colon  TQ\times TP_\alpha &\rightarrow \mathfrak{s}\\
  \big(\kk_q,\vec{w}_{p_\alpha}\big)&\mapsto \qrep(q)\cp \lambda_\alpha\left(\dd L_{q^{-1}}\kk_q,\vec{w}_{p_\alpha}\right)\!,
\end{align*}
where $\kk_q\in T_qQ$ and $\vec{w}_{p_\alpha}\in T_{p_\alpha}P_\alpha$.

\begin{lemma}
  \label{lemma:lambda}
  Let $q\in Q$, $p_\alpha\in P_\alpha$, $p_\beta\in P_\beta$ with $p_\beta=q\cdot p_\alpha$ and $\vec{w}_{p_\alpha}\in T_{p_\alpha}P_\alpha$. Then
  \begin{enumerate}
  \item
    $\lambda_\beta(\xii\:)=\qrep(q)\cp \lambda_\alpha\big(\vec{0}_{\mathfrak{q}},\vec{w}_{p_\alpha}\big)$ for all $\xii\in \mathfrak{q}\times T_{p_\beta}P$ with $\dd\THA_{(e,p_\beta)}(\xii\:)=\dd L_q\vec{w}_{p_\alpha}$,
  \item
    $\lambda_\beta\big(\Add{q}(\vec{q}\:),\vec{0}_{p_\beta}\big)=\qrep(q)\cp \lambda_\alpha\big(\vec{q},\vec{0}_{p_\alpha}\big)$ for all $\vec{q}\in \mathfrak{q}$.
  \end{enumerate}
  For all $\alpha \in I$ we have
  \begin{enumerate}
  \item[3)]
    $\ker\!\big[\lambda_\alpha|_{\mathfrak{q}\times T_{p_\alpha}P_\alpha}\big]\subseteq \ker\!\big[\dd_{(e,p_\alpha)}\THA\big]$ for all $p_\alpha\in P_\alpha$,
  \item[4)]
    the map $\w_\alpha$ is the unique $\mathfrak{s}$-valued 1-form on $Q\times P_\alpha$ that extends $\lambda_\alpha$ and for which
    we have $L_q^*\w_\alpha=\qrep(q)\cp \w_\alpha$ for all $q\in Q$.
  \end{enumerate}
  \begin{proof}
    \begin{enumerate}
    \item
      Write $\xii=((\vec{g},\vec{s}\:),\vec{w}_{p_\beta})$ for $\vec{g}\in\mathfrak{g}$, $\vec{s}\in\mathfrak{s}$ and $\vec{w}_{p_\beta}\in T_{p_\beta}P_\beta$. Then 
      \vspace{-1ex}  
      \begin{align*}
        \wt{g}(p_\beta) +\vec{w}_{p_\beta}-\wt{s}(p_\beta)\stackrel{\eqref{eq:ThetaPhi}}{=}\dd\THA_{(e,p_\beta)}(\xii\:)=\dd L_q\vec{w}_{p_\alpha}
      \end{align*}
      so that from condition \textit{i.)} in Corollary \ref{cor:psialpha} we obtain
      \begin{align*}
        \lambda_\beta(\xii\:)=\psi_\beta(\vec{g},\vec{w}_{p_\beta})-\vec{s}=\qrep(q)\cp\psi_\alpha\big(\vec{0}_{\mathfrak{g}},\vec{w}_{p_\alpha}\big)=\qrep(q)\cp\lambda_\alpha\big(\vec{0}_{\mathfrak{q}}, \vec{w}_{p_\alpha}\big).
      \end{align*}
    \item
      Let $\vec{q}=(\vec{g},\vec{s}\:)$ for $\vec{g}\in\mathfrak{g}$ and $\vec{s}\in \mathfrak{s}$. Then by Corollary \ref{cor:psialpha}.\textit{ii.)} we have
      \begin{align*}
        \lambda_\beta\big(\!\Add{q}(\vec{q}),\vec{0}_{p_\beta}\big)
        &=\psi_\beta\big(\!\Add{q}(\vec{g}),\vec{0}_{p_\beta}\big)- \Add{q}(\vec{s}\:)\\
        &=\qrep(q)\cp [\:\psi_\alpha\big(\vec{g},\vec{0}_{p_\alpha}\big)- \vec{s}\:]
        =\qrep(q) \cp \lambda_\alpha\big(\raisebox{-0.1ex}{$\vec{q},\vec{0}_{p_\alpha}$}\big).
      \end{align*}
    \item
      This follows from the first part for $\alpha=\beta$, $q=e$ and $\vec{w}_{p_\alpha}=\vec{0}_{p_\alpha}$. 
    \item
      By definition we have $\w_\alpha|_{\mathfrak{q}\times TP_\alpha}=\lambda_\alpha$ and for the pullback property we calculate
      \begin{align*}
        \left(L_{q'}^*\w_\alpha\right)_{(q,p_\alpha)}\big(\raisebox{-0.1ex}{$\kk_q,\vec{w}_{p_\alpha}$}\big)&={\w_\alpha}_{(q'q,p_\alpha)}\big(\dd L_{q'}\kk_q,\vec{w}_{p_\alpha}\big)\\[-3pt]
        &=\qrep\left(q'q\right)\cp\lambda_\alpha\left(\dd L_{q^{-1}q'^{-1}}\dd L_{q'}\kk_q,\vec{w}_{p_\alpha}\right)\\[1pt]
        &=\qrep\left(q'\right)\cp\qrep(q)\cp \lambda_\alpha\left(\dd L_{q^{-1}}\kk_q,\vec{w}_{p_\alpha}\right)\\[1pt]
        &=\qrep\left(q'\right)\cp {\w_\alpha}_{(q,p_\alpha)}(\kk_q,\vec{w}_{p_\alpha}),
      \end{align*}
      where $q,q'\in Q$  and $\kk_q\in T_qQ$.
      For uniqueness let $\w$ be another $\mathfrak{s}$-valued 1-form on $Q\times P_\alpha$ whose restriction to $\mathfrak{q}\times TP_\alpha$ is $\lambda_\alpha$ and that fulfils $L_q^*\w=\qrep(q)\cp \w$ for all $q\in Q$. Then 
      \begin{align*}
        \w_{(q,p_\alpha)}\left(\kk_q,\vec{w}_{p_\alpha}\right)&={\w}_{(q,p_\alpha)}\left(\dd L_q\cp \dd L_{q^{-1}}\kk_q,\vec{w}_{p_\alpha}\right)
        =(L_q^*\w)_{(e,p_\alpha)}\left(\dd L_{q^{-1}}\kk_q,\vec{w}_{p_\alpha}\right)\\
        &=\qrep(q)\cp \w_{(e,p_\alpha)}(\dd L_{q^{-1}}\kk_q,\vec{w}_{p_\alpha})
        = \qrep(q)\cp \lambda_\alpha\left(\dd L_{q^{-1}}\kk_q,\vec{w}_{p_\alpha}\right)\\
        &=\w_\alpha(\dd L_{q^{-1}}\kk_q,\vec{w}_{p_\alpha}). 
      \end{align*}
      Finally, smoothness of $\w_\alpha$ is an easy consequence of smoothness of the maps 
      $\qrep$, $\lambda_\alpha$ and $\mu \colon TQ\rightarrow \mathfrak{q}$, $\kk_q \mapsto \dd L_{q^{-1}}\kk_q$ with $\kk_q\in T_qQ$. 
      For this, observe that $\mu=\dd\tau \cp \kappa$ for $\tau\colon Q\times Q\rightarrow Q$, $(q,q')\mapsto q^{-1}q'$ and $\kappa\colon TQ\rightarrow TQ\times TQ$, $\kk_q\mapsto \big(\vec{0}_q,\kk_q\big)$ for $\kk_q\in T_qQ$. 
    \end{enumerate}
  \end{proof}
\end{lemma}
\noindent
So far, we have shown that each reduced connection $\{\psi_\alpha\}_{\alpha \in I}$ gives rise to uniquely determined maps $\{\lambda_\alpha\}_{\alpha \in I}$ 
and $\{\omega_\alpha\}_{\alpha \in I}$. In the final step, we will construct a unique $\Phi$-invariant connection $\w$ out of the data $\{(P_\alpha,\lambda_\alpha)\}_{\alpha\in I}$. Here, uniqueness and smoothness of $\w$ will follow from uniqueness and smoothness of the maps $\w_\alpha$.
\begin{proposition}
  \label{prop:reconstr}
  There is one and only one $\mathfrak{s}$-valued 1-form $\w$ on $P$ with $\w_\alpha=(\THA^*\w)|_{TQ\times TP_\alpha}$ for all $\alpha \in I$. This 1-form $\w$ is a $\Phi$-invariant connection on $P$.
  \begin{proof}
    For uniqueness, we have to show that the values of such an $\w$ are uniquely determined by the maps $\w_\alpha$. To this end, let $p\in P$, $\alpha\in I$ and $p_\alpha\in P_\alpha$ such that $p=q\cdot p_\alpha$ for some $q\in Q$. By Lemma \ref{lemma:suralpha}.1 for $\vec{w}_p\in T_pP$ we find some $\xii\in T_qQ\times T_{p_\alpha}P_\alpha$ with $\vec{w}_p=\dd_{(q,p_\alpha)}\THA(\xii\:)$, so that uniqueness follows from
    \begin{align*}
      \w_p(\vec{w}_p)=\w_{q\cdot p_\alpha}\left(\dd_{(q,p_\alpha)}\THA(\xii\:)\right)
      =(\THA^*\w)_{(q,p_\alpha)}(\xii\:)
      =\w_\alpha(\xii\:).
    \end{align*} 
    For existence, let $\alpha\in I$ and $p_\alpha\in P_\alpha$. Due to surjectivity of $\dd_{(e,p_\alpha)}\THA$ and Lemma \ref{lemma:lambda}.3 there is a (unique) map $\widehat{\lambda}_{p_\alpha}\colon T_{p_\alpha}P\rightarrow \mathfrak{s}$ with
    \begin{align}
      \label{eq:lambda}
      \widehat{\lambda}_{p_\alpha}\!\cp \dd_{(e,p_\alpha)}\THA=\lambda_\alpha\big|_{\mathfrak{q}\times T_{p_\alpha}P_\alpha}.
    \end{align}
    Let $\widehat{\lambda}_\alpha \colon \bigsqcup_{p_\alpha\in P_\alpha} T_{p_\alpha} P\rightarrow \mathfrak{s}$
    denote the (unique) map whose restriction to $T_{p_\alpha}P$ is $\widehat{\lambda}_{p_\alpha}$ for each $p_\alpha\in P_\alpha$. Then 
    $\lambda_\alpha=\widehat{\lambda}_\alpha\cp \dd \THA|_{\mathfrak{q}\times TP_\alpha}$ and we construct the connection $\w$ as follows. For $p\in P$ we choose some $\alpha\in I$ and $(q,p_\alpha)\in Q\times P_\alpha$ such that $q\cdot p_\alpha=p$ and define 
    \begin{align}
      \label{eq:defomega}
      \w_p\big(\vec{w}_p\big):=\qrep(q)\cp  \widehat{\lambda}_\alpha\left(\dd L_{q^{-1}}\big(\vec{w}_p\big)\right)\qquad \forall \:\vec{w}_p\in T_pP.   
    \end{align} 
    We have to show that this depends neither on $\alpha\in I$ nor on the choice of $(q,p_\alpha)\in Q\times P_\alpha$.
    For this, let $p_\alpha\in P_\alpha$, $p_\beta \in P_\beta$ and $q\in Q$ with $p_\beta=q\cdot p_\alpha$. Then for $\vec{w}\in T_{p_\alpha}P$ we have 
    $\vec{w}=\dd\THA(\vec{q},\vec{w}_{p_\alpha})$ for some $(\vec{q},\vec{w}_{p_\alpha})\in \mathfrak{q}\times T_{p_\alpha}P_\alpha$, and since $\dd L_q\vec{w}_{p_\alpha}\in T_{p_\beta}P$, 
    there is $\xii\in \mathfrak{q}\times T_{p_\beta}P_\beta$ such that 
    $\dd_{(e,p_\beta)}\THA(\xii\:)=\dd L_q\vec{w}_{p_\alpha}$. It follows from the conditions 1) and 2) in Lemma \ref{lemma:lambda}
    that
    \begin{align}
      \label{eq:wohldefs}
      \begin{split}
        \widehat{\lambda}_\beta(\dd L_q\vec{w})&=\hspace{4.6pt}\widehat{\lambda}_\beta((\dd L_q\cp \dd\THA)(\vec{q},\vec{w}_{p_\alpha})) 
        =\widehat{\lambda}_\beta\big((\dd L_q\cp \dd\THA)\big(\vec{q},\vec{0}_{p_\alpha}\big)\big)+\widehat{\lambda}_\beta\big(\dd L_q \vec{w}_{p_\alpha}\big)\\[-2pt]
        &\hspace{-4.3pt}\stackrel{\eqref{eq:thirdstep}}{=}\widehat{\lambda}_\beta\cp \dd\THA \big( \Add{q}(\vec{q}),\vec{0}_{p_\beta}\big)+\widehat{\lambda}_\beta\cp \dd\THA(\xii\:)\\[-3pt]
        &\hspace{-4.3pt}\stackrel{\eqref{eq:lambda}}{=}\lambda_\beta\big(\Add{q}(\vec{q}),\vec{0}_{p_\beta}\big)+\lambda_\beta(\xii\:)\\[1pt]
        &=\hspace{4.6pt}\qrep(q)\cp \lambda_\alpha\big(\vec{q},\vec{0}_{p_\alpha}\big)+\qrep(q)\cp \lambda_\alpha\big(\vec{0}_{\mathfrak{q}},\vec{w}_{p_\alpha}\big)\\[1pt]
        &=\hspace{4.6pt}\qrep(q)\cp\lambda_\alpha(\vec{q},\vec{w}_{p_\alpha})
        =\qrep(q)\cp \widehat{\lambda}_\alpha\cp \dd\THA(\vec{q},\vec{w}_{p_\alpha})
        =\qrep(q)\cp \widehat{\lambda}_\alpha(\vec{w}\hspace{1pt}),
      \end{split}
    \end{align}
    where for the third equality we have used that
    \begin{align}
      \label{eq:thirdstep}
      \begin{split}
        \left(\dd L_q\cp \dd\THA\right)\big(\vec{q},\vec{0}_{p_\alpha}\big)&=\dttB{t}{0}\:q\cdot(\exp(t\vec{q}\:)\cdot p_\alpha)\\
        &=\dttB{t}{0}\:\Co{q}(\exp(t\vec{q}\:))\cdot p_\beta=\dd\THA\big(\Add{q}(\vec{q}\:),\vec{0}_{p_\beta}\big).
      \end{split}
    \end{align}
    Consequently, if $\wt{q}\cdot p_\beta=p$ with $(\wt{q},p_\beta)\in  Q\times P_\beta$ for some $\beta \in I$, then $p_\beta=(q^{-1}\wt{q}\:)^{-1}\cdot p_\alpha$ and well-definedness follows from
    \begin{align*}
      \qrep(\wt{q}\:)\cp \widehat{\lambda}_\beta\left(\dd L_{{\wt{q}}^{-1}}(\vec{w}_p)\right)&=\qrep(q)\cp\qrep\big(q^{-1}\wt{q}\:\big)\cp\widehat{\lambda}_\beta\left(\dd L_{(q^{-1}\wt{q}\:)^{-1}}\big( \dd L_{q^{-1}}\vec{w}_p\big)\right)\\
      &=\qrep(q)\cp \widehat{\lambda}_\alpha\big(\dd L_{q^{-1}}\vec{w}_p\hspace{1pt}\big),
    \end{align*}
    where the last step is due to \eqref{eq:wohldefs} with $\vec{w}=\dd L_{q^{-1}}\vec{w}_p\in T_{p_\alpha}P$.
    Next, we show that $\w$ fulfils the pullback property. For this let $(\kk,\vec{w}_{p_\alpha})\in T_qQ\times T_{p_\alpha}P_\alpha$. Then 
    \begin{align*}
      (\THA^*\w)\left(\kk_q,\vec{w}_{p_\alpha}\right)&=\w_{q\cdot p_\alpha}\left(\dd\THA(\kk_q,\vec{w}_{p_\alpha})\right)
      \stackrel{\eqref{eq:defomega}}{=}\qrep(q)\cp \widehat{\lambda}_\alpha\!\left(\dd L_{q^{-1}}\dd\THA(\kk_q,\vec{w}_{p_\alpha})\right)\\[-2pt]
      &=\qrep(q)\circ \widehat{\lambda}_\alpha\cp \dd\THA\left(\dd L_{q^{-1}} \kk_q,\vec{w}_{p_\alpha}\right)
      \stackrel{\eqref{eq:lambda}}{=}\qrep(q)\cp \lambda_\alpha\!\left(\dd L_{q^{-1}} \kk_q,\vec{w}_{p_\alpha}\right)\\[3pt]
      &=\w_\alpha(\kk_q,\vec{w}_{p_\alpha}).
    \end{align*}
    In the third step we have used that $L_{q^{-1}}\cp \THA= \THA(L_{q^{-1}}(\cdot),\cdot)$.
    Finally, we have to verify that $\w$ is a $\Phi$-invariant, smooth connection.
    For this let $p\in P$ and $(\wt{q},p_\alpha)\in Q\times P_\alpha$ with $p=\wt{q}\cdot p_\alpha$. Then for $q\in Q$ and $\vec{w}_p\in T_pP$ we have
    \begin{align*}
      \left(L_q^*\w\right)_{p}(\vec{w}_{p})&=\w_{q\cdot p}\left(\dd L_q \vec{w}_{p}\right)=\w_{(q \wt{q})\cdot p_\alpha}\left(\dd L_q \vec{w}_{p}\right)\\
      &=\qrep(q)\cp\qrep\left(\wt{q}\hspace{2pt}\right)\cp \widehat{\lambda}_\alpha\left(\dd L_{\wt{q}^{-1}}\vec{w}_{p}\right)
      =\qrep(q)\cp \w_{p}(\vec{w}_{p}),
    \end{align*}
    hence
    \begin{align*}
      R_s^*\w&=L_{\left(e,s^{-1}\right)}^*\w=\qrep\big(\big(e,s^{-1}\big)\big)\cp \w=\Add{s^{-1}}\cp \w,\\
      L_g^*\w& =L_{(g,e)}^*\w=\qrep((g,e))\cp \w=\w.
    \end{align*}
    So, it remains to show smoothness of $\w$ and that $\w_p(\widetilde{s}(p))=\vec{s}$ holds for all $p\in P$ and all $\vec{s}\in \mathfrak{s}$. For the second property, let $p=q\cdot p_\alpha$ for $(q,p_\alpha)\in Q\times P_\alpha$. Then $q=(g,s)$ for some $g\in G$ and $s\in S$ and we obtain
    \begin{align*}
      \w_{p}(\widetilde{s}(p))
      &=\qrep(q)\cp \widehat{\lambda}_\alpha\left(\dd L_{q^{-1}}\widetilde{s}(q\cdot p_\alpha)\right)\\
      &=\qrep(q)\cp \widehat{\lambda}_\alpha\left(\dttB{t}{0}\:  p_\alpha\cdot \left(\Co{s^{-1}}(\exp(t\vec{s}\:)\right)\right)\\
      &=\qrep(q)\cp\widehat{\lambda}_\alpha\big(\dd\THA\big(\Add{s^{-1}}(\vec{s}\hspace{2pt}),\vec{0}_{p_\alpha}\big)\big)\\
      &=\Add{s}\cp\: \lambda_\alpha\big(\Add{s^{-1}}(\vec{s}\hspace{2pt}),\vec{0}_{p_\alpha}\big)
      =\Add{s}\cp\Add{s^{-1}}(\vec{s}\hspace{2pt})=\vec{s}.
    \end{align*}
    For smoothness let $p_\alpha\in P_\alpha$ and choose a submanifold $Q'$ of $Q$ through $e$, an open neighbourhood $P'_\alpha\subseteq P_\alpha$ of $p_\alpha$ and an open subset $U\subseteq P$ such that the restriction 
    $\widehat{\THA}:=\THA|_{Q' \times P'_\alpha}$
    is a diffeomorphism to $U$. Then $p_\alpha \in U$ because $e\in Q'$, hence
    \begin{align*}
      \w|_U=\widehat{\THA}^{-1}\hspace{-1pt}\raisebox{2pt}{$^*\big[$}\widehat{\THA}\hspace{1pt}\raisebox{2pt}{$^*$} \w\raisebox{2pt}{$\big]$}=\widehat{\THA}^{-1}\hspace{-1pt}\raisebox{0pt}{$^*$}\big[(\THA^*\w)|_{TQ\times TP_\alpha}\big] =\widehat{\THA}^{-1}\hspace{-1pt}\raisebox{2pt}{$^*$}\w_\alpha.
    \end{align*}
    Since $\w_\alpha$ is smooth and $\widehat{\THA}$ is a diffeomorphism, $\w|_U$ is smooth as well. Finally, if $p=q\cdot p_\alpha$ for $q\in Q$, then $L_q(U)$ is an open neighbourhood of $p$ and
    \begin{align*}
      \w|_{L_q(U)}=\big(L_{q^{-1}}^*\!\left(L_q^*\hspace{1pt}\w\right)\!\big)\big|_{L_q(U)}= \qrep(q)\cp \big(L_{q^{-1}}^*\w\big)\big|_{L_q(U)} =\qrep(q)\cp L_{q^{-1}}^* \!\left(\w|_U\right)
    \end{align*}
    is smooth because $\w|_U$ and $L_{q^{-1}}$ are smooth.
  \end{proof} 
\end{proposition}
\noindent
Corollary \ref{cor:psialpha} and Proposition \ref{prop:reconstr} now prove
\begin{theorem}
  \label{th:InvConnes}
  Let $G$ be a Lie group of automorphisms of the principal fibre bundle $P$. Then for each $\Gg$-covering $\{P_\alpha\}_{\alpha\in I}$ of $P$ there is a bijection between the corresponding set of reduced connections and the $\Phi$-invariant connections on $P$.
  \begin{proof}
    Corollary \ref{cor:psialpha} and Proposition \ref{prop:reconstr}
  \end{proof}
\end{theorem}
\noindent
As already mentioned in the preliminary remarks to Example \ref{example:SCHSV}, the second part of the next example shows the importance of the transversality condition 
\begin{align*}
  \im[\dd_x\tau_0] + \im\big[\dd_e\varphi_{\tau_0(x)}\big]=T_{\tau_0(x)}M\qquad\forall\: x\in U_0
\end{align*}
for the formulation in \cite{HarSni}.
\begin{example}[(Semi-)homogeneous Connections]
  \label{ex:transinv}
  \begin{enumerate}
  \item
    Let $P=X \times S$ for an $n$-dimensional $\RR$-vector space $X$ and an arbitrary structure group $S$. 
    Moreover, let $G \subseteq X$ be a linear subspace of dimension $1\leq k\leq n$ acting via 
	\begin{align*}
	\Phi\colon G\times P \rightarrow P,\quad(g,(x,\sigma))\mapsto(g+x,\sigma).
	\end{align*}	    
    Let $W$ be an algebraic complement of $G$ in $X$ and 
    $P_0:=W\times \{e_S\}\subseteq P$. Then, $P_0$ is a $\Gg$-covering because $\THA\colon (G\times S)\times P_0\rightarrow P$ is a diffeomorphism and each $\varphi$-orbit intersects $W$ in a unique point. Consequently, identifying $G$ with its Lie algebra $\mg$, the $\Phi$-invariant connections on $P$ are in bijection with the smooth  maps $\psi\colon G\times TW\rightarrow \mathfrak{s}$ such that $\psi_w:=\psi|_{G \times T_wW}$ is linear for all $w\in W$. This is because the conditions \textit{i.)} and \textit{ii.)} from Corollary \ref{cor:psialpha} give no further restrictions in this case. It is straightforward to see that the $\Phi$-invariant connection that corresponds to $\psi$ is explicitly given by
    \begin{align}
      \label{eq:transinv}
      \w^\psi(\vec{v}_x,\vec{\sigma}_s)=\Add{s^{-1}}\!\cp\:\psi_{\pr_W(x)}\big(\pr_G(\vec{v}_x),\pr_W(\vec{v}_x)\big)+\dd L_{s^{-1}}(\vec{\sigma}_s)
    \end{align}
    for $(\vec{v}_x,\vec{\sigma}_s)\in T_{(x,s)}P$.
  \item
    In the situation of Part 1), let $X=\RR^2$, $G=\Span_\RR(\vec{e}_1)$, $W=\Span_\RR(\vec{e}_2)$ and $P_0=W\times \{e\}$. We fix $0\neq \vec{s}\in \mathfrak{s}$ and define $\psi\colon \mathfrak{g}\times TP_0 \rightarrow \mathfrak{s}$ by
    \begin{align*}    
      \psi_{y}(\lambda \cdot\vec{e}_1,\mu\cdot \vec{e}_2):= \mu \cdot f(y)\cdot\vec{s}\qquad \text{for}\qquad (\lambda \cdot\vec{e}_1,\mu \cdot\vec{e}_2)\in \mathfrak{g}\times T_{(y\cdot\vec{e}_{2},e)}P_0,
    \end{align*}    	
    where $f(0):=0$ and $f(y):=1\slash \sqrt[3]{y}$ for $y\neq 0$. Then, $\w^\psi$ defined by \eqref{eq:transinv} is smooth on $P':=Z\times S$ for $Z:= [\hspace{0.5pt} X\backslash \spann_\RR(\vec{e}_1)]$, but not smooth at $((x,0),e)$ because 
    \begin{align*}    
      \w^\psi_{((x,y),e)}\big(\big(\vec{0},\vec{e}_2\big),\vec{0}_{\mathfrak{s}}\big)=\psi_y\big(\vec{0},\vec{e}_2\big)=f(y)\cdot \vec{s}\qquad \forall\: y\in \RR. 
    \end{align*}
	In addition to that, there cannot exist any smooth invariant connection $\wt{\w}$ on $P$ which coincides on $P'$ with $\w^\psi$, just because $\lim_{y\rightarrow 0}f(y)\cdot \s$ does not exist. 
	    
    Now, let $U_0=\RR$, $\tau_0 \colon U_0 \rightarrow \RR^2$, $t\mapsto \big(t,t^3\big)$ and $s_0\colon t \mapsto (\tau_0(t),e)$. Then $(U_0,\tau_0,s_0)$ fulfils the conditions in \cite{HarSni}, but we have $\im[\dd_0\tau_0] + \im\hspace{-1pt}\big[\dd_e\varphi_{\tau_0(0)}\big] = \Span_\RR(\vec{e}_1)\neq T_0X =T_0\RR^2=\RR^2$.\footnote{Then     
      $\im[\dd_0s_0]+ \im[\dd_e\Phi_{s_0(x)}]+Tv_{s_0(0)}P=\Span_\RR(\vec{e}_1) \oplus Tv_{(0,e)}P\neq\RR^2 \oplus Tv_{(0,e)}P= T_{(0,e)}P$ so that $(U_0,s_0)$ is not a $\THA$-patch as it fails the transversality condition \eqref{eq:transv} from Remark \ref{bem:Psliceeigensch}.2.} 
      
      As a consequence, $\ovl{\psi}\colon \mathfrak{g}\times TU_0\rightarrow \mathfrak{s}$ defined by $\ovl{\psi}_t:=(\Phi^*\w^\psi)|_{\mathfrak{g}\times T_tU_0}$ is smooth because for $t\neq 0$ and $r\in T_tU_0 =\RR$ we have
    \begin{align*}
      \ovl{\psi}_{t}(\lambda\hspace{1pt} \vec{e}_1,r)&=\big(\Phi^*\w^\psi\big)\left(\lambda\hspace{1pt} \vec{e}_1,r\cdot \vec{e}_1 + 3t^2 r\cdot \vec{e}_2\right)
      =\w^\psi_{((t,t^3),e)}\!\left(\left(\lambda+r\right)\cdot \vec{e}_1 + 3t^2 r\cdot \vec{e}_2\right)\\
      &=\psi_{t^3}\!\left(\left(\lambda+r\right)\cdot \vec{e}_1,3t^2 r\cdot \vec{e}_2\right)= 3 t r\cdot \vec{s},
    \end{align*}
    as well as $\ovl{\psi}_{0}(\lambda\hspace{1pt} \vec{e}_1,r)=0$ if $t=0$. For the first step, keep in mind that 
	\begin{align*}    
    (\Phi^*\w^\psi)|_{\mathfrak{g}\times T_tU_0}(\vec{g},r)=(\Phi^*\w^\psi)(\vec{g},\dd_t s_0(r))
    \end{align*} 
    by Convention \ref{conv:Submnfds}.2.
    Then, the maps $\mu:= \ovl{\psi}|_{TU_0}$ and $A_{t_0}(\vec{g}):=\ovl{\psi}\big(\vec{g},\vec{0}_{t_0}\big)$ fulfil the conditions in Theorem 2 in \cite{HarSni} because $\ovl{\psi}$ fulfils the three algebraic conditions in Example \ref{example:SCHSV}, being trivial in this case because $H=\{e\}$. 
  
   Now, Theorem 2 in \cite{HarSni} states that $\ovl{\psi}$ can also be obtained by pullbacking and restricting (w.r.t.\ $s_0$) a unique smooth invariant connection $\wt{\w}$ on $P$ (instead of pullbacking and restricting $\w^\psi$). However, such a connection cannot exist as it necessarily has to coincide on $P'$ with $\w^\psi$.
 	
	In fact, let $U'_0:=\RR_{\neq 0}$ and $\tau_0'\colon U_0'\rightarrow Z$, $t\mapsto \big(t,t^3\big)$ as well as $s_0'\colon t\mapsto(\tau_0'(t),e)$ be defined as above. Then, $(U_0',s_0')$ is a $\THA$-patch as we have removed the point $0\in U_0$ for which transversality fails. Thus, the restriction of $\ovl{\psi}$ to $\mg\times U_0'$ corresponds to a unique smooth invariant connection $\w'$ on $P'$. This connection must be given by the restriction of $\w^\psi$ to $P'$ because pullbacking and restricting $\w^\psi$ w.r.t.\  $s_0'$  gives rise to the restriction $\ovl{\psi}|_{\mg \times TU_0'}$, just because $s_0'=s_0|_{U_0}$ holds. However, the same is true for $\wt{\w}$, so that both connections coincide on $P'$.                   
\hspace*{\fill}{{$\Diamond$}} 
  \end{enumerate}
\end{example} 

\subsection{Particular Cases and Applications}
\label{sec:PartCases}
In the first part of this Subsection we consider $\Gg$-coverings of $P$ that arise from the induced action $\varphi$ on the base manifold $M$ of $P$. Then we discuss the case where $\Phi$ acts via gauge transformations on $P$. This leads to a straightforward generalization of the description of connections by consistent families of local 1-forms on $M$. In the second part we discuss the (almost) fibre transitive case and deduce Wang's original theorem \cite{Wang} from Theorem \ref{th:InvConnes}. Finally, we consider the situation where $P$ is trivial and give examples in loop quantum gravity. 

\subsubsection{$\Gg$-Coverings and the Induced Action}
\label{subsec:GCovIndAct}
Let $(G,\Phi)$ be a Lie group of automorphisms of the principal fibre bundle $P$. According to Lemma \ref{lemma:mindimslice} for each $x\in M$ there is a $\varphi$-patch (with minimal dimension) $M_x$ with $x\in M$. Consequently, there is an open neighbourhood $M'_x\subseteq M_x$ of $x$ and a local section $s_x\colon U\rightarrow P$ such that $M'_x\subseteq U$ for an open neighbourhood $U\subseteq M$.
Let $I\subseteq M$ be a subset such that\footnote{It is always possible to choose $I=M$.} each $\varphi$-orbit intersects at least one of the sets $M_x$ for some $x\in I$. Then it is immediate from Lemma \ref{lemma:suralpha}.2  that $\{s_x(M'_x)\}_{x\in I}$ is a $\Gg$-covering of $P$. More generally, we have 
\begin{corollary}
  \label{cor:reductions}
  Let $\PMS$ be a principal fibre bundle and $(G,\Phi)$ a Lie group of automorphisms of $P$. Denote by $(M_\alpha,s_\alpha)_{\alpha\in  I}$ a family consisting of a collection of $\varphi$-patches $\{M_\alpha\}_{\alpha\in I}$ and smooth sections\footnote{This is that $\pi\cp s_\alpha= \id_{M_\alpha}$.} $s_\alpha\colon M_\alpha\rightarrow P$. Then the sets $P_\alpha:=s_\alpha(M_\alpha)$ are $\THA$-patches. They provide a $\Gg$-covering of $P$ iff each $\varphi$-orbit intersects at least one patch $M_\alpha$. 
  \begin{proof}
    This is immediate from Lemma \ref{lemma:suralpha}.2.
  \end{proof}
\end{corollary}
\noindent
We now consider the case where  $(G,\Phi)$ is a Lie group of gauge transformations of $P$, i.e., $\varphi_g=\id_M$ for all $g\in G$. Here, we show that Theorem \ref{th:InvConnes} can be seen as a generalization of the description of smooth connections by consistent families of local 1-forms on the base manifold $M$. For this, let $\{U_\alpha\}_{\alpha\in I}$ be an open covering of $M$ and $\{s_\alpha\}_{\alpha\in I}$ a family of smooth sections $s_\alpha\colon U_\alpha \rightarrow P$.
We define $U_{\alpha\beta}:=U_\alpha \cap U_\beta$ and consider the smooth maps $\delta_{\alpha\beta}\colon G\times U_{\alpha\beta}\rightarrow S$ determined by $s_\beta(x)=\Phi(g,s_\alpha(x))\cdot \delta_{\alpha\beta}(g,x)$, and for which we have $\delta_{\alpha\beta}(g,x)=\phi^{-1}_{s_\alpha(x)}(g) \cdot\delta_{\alpha\beta}(e,x)$. 
Finally, let $\mu_{\alpha\beta}(g,\vec{v}_x):= \dd L_{\delta^{-1}_{\alpha\beta}(g,x)}\cp \dd_x\delta_{\alpha\beta}(g,\cdot)(\vec{v}_x)$ for $\vec{v}_x\in T_xU_{\alpha\beta}$ and $g\in G$. Then we have
\begin{scase}[Lie Groups of Gauge Transformations]
  \label{scase:GaugeTransf}
  Let $(G,\Phi)$ be a Lie group of gauge transformations of the principal fibre bundle $(P,\pi,M,S)$. Then, the $\Phi$-invariant connections on $P$ are in bijection with the families $\{\chi_\alpha\}_{\alpha\in I}$ of $\mathfrak{s}$-valued 1-forms $\chi_\alpha\colon U_\alpha\rightarrow\mathfrak{s}$ for which we have
  \begin{align}
    \label{eq:consistgauge}
    \chi_\beta(\vec{v}_x)=\left(\Add{\delta_{\alpha\beta}(g,x)}\cp\: \chi_\alpha\right)(\vec{v}_x)+\mu_{\alpha\beta}(g,\vec{v}_x)\qquad \forall\: \vec{v}_x\in T_xU_{\alpha\beta},\forall\: g\in G.
  \end{align}
  \begin{proof}
    By Corollary \ref{cor:reductions} $\{s_\alpha(U_\alpha)\}_{\alpha\in I }$ is a $\Gg$-covering of $P$. So, let $\{\psi_\alpha\}_{\alpha\in I}$ be a reduced connection w.r.t.\ this covering. We first show that condition \textit{i.)} from Corollary \ref{cor:psialpha} implies 
    \begin{align}
      \label{eq:eq1}
      \psi_\beta\big(\raisebox{-0.5pt}{$\vec{g},\vec{0}_{p}$}\big) =\dd_e\phi_p(\vec{g}\hspace{1pt})\qquad \forall\:\vec{g}\in \mathfrak{g}, \forall\:p\in s_\beta(U).
    \end{align}	       
    For this
    observe that condition \textit{a.)} from Remark \ref{rem:Psialphaconderkl} means that for all $\beta\in I$, $p \in s_\beta(U_\beta)$, $\vec{w}_p\in T_ps_\beta(U_\beta)$ and $\vec{g}\in \mathfrak{g}$, $\vec{s}\in \mathfrak{s}$ we have 
    \begin{align*}
      \dd_e\Phi_p(\vec{g}\hspace{1pt}) + \vec{w}_p -\widetilde{s}(p)=0 \quad \Longrightarrow \quad\psi_\beta(\vec{g},\vec{w}_p)-\vec{s}=0.  
    \end{align*}
    Now, $T_ps_\beta(U_\beta)$  
    is complementary to $Tv_pP$ and $\im[\dd_e\Phi_p]\subseteq \ker[\dd_p\pi]$ so that \textit{a.)} is the same as 
    \begin{enumerate}
    \item[\textit{a'.)}]
      $\dd_e\Phi_p(\vec{g}\hspace{1pt})=\widetilde{s}(p)\quad \Longrightarrow\quad \psi_\beta\big(\raisebox{-0.5pt}{$\vec{g},\vec{0}_p$}\big)=\vec{s}$\quad for\: $\vec{g}\in \mathfrak{g}$, $\vec{s}\in \mathfrak{s}$ \:and all\: $p\in P_\beta$.    
    \end{enumerate}
    But, since $G_x=G$ for all $x\in M$, this just means\footnote{$\dd_e\Phi_p(\vec{g}\hspace{1pt})- \widetilde{s}(p)=0$ iff $(\vec{g},\vec{s}\hspace{1pt})\in \mathfrak{q}_p$ iff $\vec{s}=\dd_e\phi_p(\vec{g}\hspace{1pt})$.} $\psi_\beta\big(\raisebox{-0.5pt}{$\vec{g},\vec{0}_{p}$}\big) =\dd_e\phi_p(\vec{g}\hspace{1pt})$ for all $\vec{g}\in \mathfrak{g}$ and already implies condition \textit{ii.)} from Corollary \ref{cor:psialpha} as $\phi_p$ is a  Lie group homomorphism. Consequently, we can ignore this condition in the following. 
    Now, we have $p_\beta = q\cdot p_\alpha$ for $q\in Q$, $p_\alpha \in P_\alpha$, $p_\beta \in P_\beta$ iff 
    $\pi(p_\alpha)=\pi(p_\beta)=x\in U_{\alpha\beta}$ and $q=\big(g,\delta^{-1}_{\alpha\beta}(g,x)\big)$. Consequently, the left hand side of condition \textit{i.)} from Corollary \ref{cor:psialpha} reads
    \begin{align*}
      \wt{g}(s_\beta(x))+\dd_x s_\beta (\vec{v}_\beta)  - \wt{s}(s_\beta(x))= \big(\dd L_g\cp \dd R_{\delta_{\alpha\beta}(g,x)}\cp \dd_x s_\alpha\big) (\vec{v}_\alpha),
    \end{align*}
    \newline
    \vspace{-26pt}    
    \newline
    where $\vec{v}_\alpha, \vec{v}_\beta \in T_xM$ and $g\in G$.
    This is true for $\vec{v}_\alpha=\vec{v}_\beta=\vec{v}_x$, $\vec{g}=0$ and $\vec{s}=\mu_{\alpha\beta}(g,\vec{v}_x)$, which follows from
    \begin{align*}    
      \dd_x s_\beta (\vec{v}_\beta)&= \dd_x \Big[L_g\cp R_{\delta_{\alpha\beta}(g,\cdot)}\cp s_\alpha\Big] (\vec{v}_x)\\
      &=\dd L_g\left[\dd_{s_\alpha(x)}R \big(\dd_x\delta_{\alpha\beta}(g,\cdot)(\vec{v}_x)\big) + \dd R_{\delta_{\alpha\beta}(g,x)}(\dd_x s_\alpha(\vec{v}_x)) \right],\\[4pt]
      \wt{s}(s_\beta(x))&=\dttB{t}{0}L_g \cp R_{\delta_{\alpha\beta}(g,x)\cdot \exp(t\vec{s}\:)}(s_\alpha(x))\\[2pt]
      &=\dd L_g\left[\dd_{s_\alpha(x)}R \left(\dd L_{\delta_{\alpha\beta}(g,x)}(\vec{s}\:)\right)\right]
      =\dd L_g\left[\dd_{s_\alpha(x)}R \big(\dd_x\delta_{\alpha\beta}(g,\cdot)(\vec{v}_x)\big)\right].
    \end{align*}
    Consequently, by Corollary \ref{cor:psialpha}.\textit{i.)} and for 
	\begin{align*}    
    (\psi_\alpha \cp \dd_x s_{\alpha})(\vec{v}_x):=  \psi_\alpha\big(\vec{0}_{\mathfrak{g}},\dd_x s_\alpha(\vec{v}_x)\big) \qquad \forall\:\vec{v}_x\in T_xU_{\alpha\beta}.
	\end{align*}    
     we have
    \begin{align}
      \label{eq:gaugetraf}
      \psi_\beta\big(\vec{0}_{\mathfrak{g}},\dd_x s_\beta(\vec{v}_x)\big)=\big(\Add{\delta_{\alpha\beta}(g,x)} \cp\: \psi_\alpha \cp \dd_x s_{\alpha}\big)(\vec{v}_x)+\mu_{\alpha\beta}(g,\vec{v}_x)
    \end{align}
    for all $g\in G$ and all $\vec{v}_x\in T_xU_{\alpha\beta}$. 
    Due to part {\bf 2.)} in Remark \ref{rem:Psialphaconderkl} the condition \textit{i.)} from Corollary \ref{cor:psialpha} now gives no further restrictions, so that for $\chi_\beta:=\psi_\beta \cp \dd s_\beta$ we have
    \begin{align*}
      \psi_\beta(\vec{g}, \dd_xs_\beta(\vec{v}_x))=\dd_e\phi_{s_\beta(x)}(\vec{g})+ \chi_\beta(\vec{v}_x)\qquad \forall\:\vec{g}\in \mathfrak{g},\forall\:\vec{v}_x\in T_xM,\forall\: x\in U_\beta.  
    \end{align*}
    Then, $\psi_\beta$ is uniquely determined by $\chi_\beta$ for each $\beta\in I$, so that \eqref{eq:gaugetraf} yields the consistency condition \eqref{eq:consistgauge} for the maps $\{\chi_\alpha\}_{\alpha\in I}$.
  \end{proof} 
\end{scase}
\begin{example}[Trivial Action]
  If $G$ acts trivially, then for each $x\in U_{\alpha\beta}$ we have 
  \begin{align*}
    \delta_{\alpha\beta}(g,x)=\phi_{s_\alpha(x)}^{-1}(g)\cdot \delta_{\alpha\beta}(e,x)=\delta_{\alpha\beta}(e,x).
  \end{align*}
  Hence, $\delta_{\alpha\beta}$ is independent of $g\in G$ so that here Case \ref{scase:GaugeTransf} just reproduces the description of smooth connections by means of consistent families of local 1-forms on the base manifold $M$. \hspace*{\fill}$\Diamond$
\end{example}

\subsubsection{(Almost) Fibre Transitivity}
\label{subsec:AlmFTC}
In this subsection, we discuss the situation where $M$ admits an element which is contained in the closure of each $\varphi$-orbit. For instance, this holds for all $x\in M$ if each $\varphi$-orbit is dense in $M$ and, in particular, is true for fibre transitive actions.
\label{subsec:FIbtrcase}
\begin{scase}[Almost Fibre Transitivity]
  \label{scase:slicegleichredcluster}
  Let $x\in M$ be contained in the closure of each $\varphi$-orbit and let $p\in F_x$. Then, each $\THA$-patch $P_0\subseteq P$ with $p\in P_0$ is a $\Gg$-covering of $P$. Hence, the $\Gg$-invariant connections on $P$ are in bijection with the smooth maps $\psi\colon \mathfrak{g}\times TP_0\rightarrow \mathfrak{s}$ for which $\psi|_{\mathfrak{g}\times T_pP_0}$ is linear for all $p\in P_0$ and that fulfil the two conditions from Corollary \ref{cor:psialpha}.
  \begin{proof}
    It suffices to show that $\pi\left(P_0\right)$ intersects each $\varphi$-orbit $[o]$. Since $P_0$ is a $\THA$-patch, there is an open neighbourhood $P'\subseteq P_0$ of $p$ and a submanifold $ Q'$ of $Q$ through $(e_G,e_S)$ such that $\THA|_{Q' \times P'}$ is a diffeomorphism to an open subset $U\subseteq P$. Then $\pi(U)$ is an open neighbourhood of $\pi(p)$ and by assumption we have $[o]\cap \pi(U)\neq \emptyset$ for each $[o]\in M\slash G$. Consequently, for $[o]\in M\slash G$ we find $\wt{p}\in U$ with $\pi(\wt{p}\hspace{1.5pt})\in [o]$. Let $\wt{p}=\THA((g',s'),p')$ for $((g',s'),p')\in Q'\times P'$. Then
    \begin{align*}
      [o]\ni\pi(\wt{p}\hspace{1.5pt})=\pi\left(\Phi(g',p')\cdot s'\right)=\varphi(g',\pi(p'))\in [\pi(p')]
    \end{align*}  
    shows that $[o]=[\pi(p')]$, hence $\pi\left(P_0\right)\cap [o]\neq \emptyset$.  
  \end{proof}
\end{scase}
\noindent
The next example to Case \ref{scase:slicegleichredcluster} shows that evaluating the conditions \textit{i.)} and \textit{ii.)} from Corollary \ref{cor:psialpha} at one single point can be sufficient to verify non-existence of invariant connections.

\begin{example}[General linear group]
  \label{ex:Bruhat}
  \begingroup
  \setlength{\leftmargini}{15pt}
  \begin{itemize}
  \item
    Let $P:=GL(n,\mathbb{R})$ and $G=S=B \subseteq GL(n,\mathbb{R})$ the subgroup of upper triangular matrices. Moreover, let $S_n\subseteq GL(n,\mathbb{R})$ be the group of permutation matrices. 
    Then $P$ is a principal fibre bundle with base manifold $M:=P\slash S$, structure group $S$ and projection map $\pi\colon P\rightarrow M$, $p\mapsto [p]$. Moreover, $G$ acts via automorphisms on $P$ by $\Phi(g,p):=g\cdot p$ and we have the Bruhat decomposition
    \begin{align*}
      GL(n,\mathbb{R})=\bigsqcup_{w\in S_n}B w B.
    \end{align*}
    Then $M=\bigsqcup_{w\in S_n}G\cdot\pi(w)$, $G\cdot \pi(e)=\pi(e)$ and $\pi(e)\in \ovl{G\cdot\pi(w)}$ for all $w\in S_n$. Now, $\im[\dd_e\THA_e]=\mathfrak{g}$, since $\dd_e\THA_e(\vec{g})=\vec{g}$ for all $\vec{g}\in \mathfrak{g}$. Moreover, $\mathfrak{g}={\Span_\mathbb{R}}\{E_{ij} \:|\: 1\leq i\leq j\leq n\}$ so that $V:={\Span_\mathbb{R}}\{E_{ij} \:|\: 1\leq j<i\leq n\}$ is an algebraic complement of $\mathfrak{g}$ in $T_eP=\mathfrak{gl}(n,\mathbb{R})$. By Lemma \ref{lemma:mindimslice}.2 we find a patch $H\subseteq P$ through $e$ with $T_eH=V$ and due to Case \ref{scase:slicegleichredcluster} this is a $\Gg$-covering.
  \item
    \vspace{-6pt}
    A closer look at the point $e\in P$ shows that there cannot exist any $\Phi$-invariant connection on $GL(n,\RR)$. In fact, if $\psi\colon \mathfrak{g}\times TH \rightarrow \mathfrak{s}$ is a reduced connection w.r.t.\ $H$, then for $\vec{w}:=\vec{0}_e$ and $\vec{g}=\vec{s}$ we have
    \begin{align*}
      \wt{g}(e)+\vec{w} -\widetilde{s}(e)=\vec{g}+\vec{w}-\vec{s}=0.
    \end{align*}
    So, condition \textit{i.)} from Corollary \ref{cor:psialpha} gives $\psi\big(\vec{g},\vec{0}_e\big)-\vec{g}=0$, hence $\psi\big(\vec{g},\vec{0}_e\big)=\vec{g}$ for all $\vec{g}\in \mathfrak{g}$. Now $q\cdot e =e$ iff $q=(b,b)$ for some $b\in B$. Let 
    \begin{align*}
      V\ni\vec{h}:=E_{n1}\hspace{18pt} B\ni b:= \me + E_{1n}\hspace{18pt} \mathfrak{g}\ni  \vec{g}:=E_{11}- E_{1n}- E_{nn}.
    \end{align*} 
    Then, $\wt{g}(e)+\vec{h}=\vec{g}+\vec{h}=b\vec{h} b^{-1}=\dd L_q\vec{h}$,  
    so that condition \textit{i.)} yields 
    \begin{align*}
      \psi\big(\vec{g},\vec{h}\big)=\qrep(q)\cp  \psi\big(\vec{0}_{\mathfrak{g}},\vec{h}\big)=\Add{b}\cp \psi\big(\vec{0}_{\mathfrak{g}},\vec{h}\big),
    \end{align*}
    hence $\vec{g}+\left[\id-\Add{b}\right]\cp \psi\big(\vec{0}_{\mathfrak{g}},\vec{h}\big)=0$. But $(\vec{g}\hspace{1pt})_{11}=1$ and   
    \begin{align*}
      \big(\raisebox{-0.5pt}{$\psi\hvect-\Add{b}\cp\: \psi \hvect$}\big)_{11}=   \big(\raisebox{-1pt}{$\psi\hvect $}\big)_{11} -  \big(\raisebox{-0.5pt}{$\psi\hvect $}\big)_{11} =0
    \end{align*}
    so that $\psi$ cannot exist. \hspace*{\fill}{{$\Diamond$}} 
  \end{itemize}
  \endgroup
\end{example}
\begin{corollary}
  \label{cor:SliceFibTran}
  If $\Phi$ is fibre transitive, then $\{p\}$ is a $\Gg$-covering for all $p\in P$.
  \begin{proof}
    It suffices to show that $\{\pi(p)\}$ is a $\varphi$-patch, since then $\{p\}$ is a $\THA$-patch by Corollary \ref{cor:reductions} and a $\Gg$-covering by Case \ref{scase:slicegleichredcluster}. This, however, is clear from Remark \ref{rem:patch}.1. In fact, if $x:=\pi(p)$, then by general theory we know that $M$ is diffeomorphic to $G\slash G_{x}$ 
    via $\phi \colon [g]\mapsto \varphi(g,x)$ and that for each $[g]\in G\slash G_{x}$ we find an open neighbourhood $U\subseteq G\slash G_{x}$ of $[g]$ and a smooth section $s\colon U\rightarrow G$. Then surjectivity of $\dd_e\varphi_x$ is clear from surjectivity of $\dd_{[e]}\phi$ and 
    \begin{align*}    
      \dd_e\varphi_{x}\cp \dd_{[e]}s= \dd_{[e]}(\varphi_{x} \cp s)= \dd_{[e]}\varphi(s(\cdot),x)=\dd_{[e]}\phi,
    \end{align*} 
    showing that $T_xM=\dd_e\varphi_x(\mathfrak{g})$.
  \end{proof}
\end{corollary}
\noindent
Let $\wm$ be transitive and $p\in P$. Then $\{p\}$ is a $\Phi$-covering by Corollary \ref{cor:SliceFibTran} and  $T_p\{p\}$ is the zero vector space. Moreover, we have $p_\alpha=q\cdot p_\beta$ iff $p_\alpha=p_\beta=p$ and $q\in Q_p$. It follows that a reduced connection w.r.t.\ $\{p\}$ can be seen as a linear map $\psi\colon\mathfrak{g}\rightarrow \mathfrak{s}$ that fulfils the following two conditions.
\begingroup
\setlength{\leftmargini}{20pt}
\begin{itemize} 
\item
  $\dd_e\THA_p(\vec{g},\vec{s}\hspace{1pt})=0$\quad $\Longrightarrow$ \quad $\psi(\vec{g}\hspace{1pt})=\vec{s}$\qquad\quad for $\vec{g}\in \mathfrak{g}$, $\vec{s}\in \mathfrak{s}$,
\item
  \vspace{-2pt}
  $\psi\big(\Add{q}(\vec{g}\hspace{1pt})\big)=\qrep(q)\cp \psi(\vec{g}\hspace{1pt})$\qquad\qquad\quad\hspace{10.7pt} $\forall\: q\in Q_p$, $\forall\:\vec{g}\in \mathfrak{g}$.
\end{itemize}
\endgroup
\noindent
Since $\ker[d_e\THA_p]=\mathfrak{q}_p$, we have shown 
\begin{scase}[Hsien-Chung Wang, \cite{Wang}]
  \label{th:wang}
  Let $(G,\Phi)$ be a fibre transitive Lie group of automorphisms of $P$. Then for each $p\in P$ there is a bijection between the $\Phi$-invariant connections on $P$ and the linear maps $\psi\colon \mathfrak{g}\rightarrow \mathfrak{s}$ that fulfil 
  \begingroup
  \setlength{\leftmargini}{30pt}
  \begin{enumerate}
  \item[\textrm{a)}]
    $\psi\big(\raisebox{-1pt}{$\vec{h}$}\hspace{2pt}\big)=\dd_e\phi_p\big(\raisebox{-1pt}{$\vec{h}$}\hspace{2pt}\big)$ \quad\hspace{34.6pt} $\forall\: \vec{h}\in \mathfrak{g}_{\pi(p)}$,
  \item[\textrm{b)}]
    \vspace{-2pt}
    $\psi\cp\Add{h}=\Add{\phi_p(h)}\cp \:\psi$ \quad\hspace{10pt} $\forall\:h\in G_{\pi(p)}$.
  \end{enumerate}
  \endgroup
  \noindent
  This bijection is explicitly given by $\w\mapsto \Phi_p^*\w$.\hspace*{\fill}{{\scriptsize$\blacksquare$}} 
\end{scase}

\begin{example}
  \label{ex:eukl}
  \begingroup
  \setlength{\leftmargini}{15pt}
  \begin{itemize}
  \item \textbf{Homogeneous Connections}
    \newline
    \vspace{-13pt}
    \newline
    In the situation of Example \ref{ex:transinv} let $k=n$ and $X=\RR^n$. Then $\Phi$ is fibre transitive and for $p=(0,e)$ we have $G_{\pi(p)}=e$ and $\mathfrak{g}_{\pi(p)}=\{0\}$. Consequently, the reduced connections w.r.t.\ $\{p\}$ are just the linear maps $\psi\colon \RR^n \rightarrow \mathfrak{s}$, and the corresponding homogeneous connections are given by 
    \vspace{-4pt}
    \begin{align*}
      {\w^\psi}(\vec{v}_x,\vec{\sigma}_s)=\Add{s^{-1}}\cp\:\psi(\vec{v}_x)+\dd L_{s^{-1}}(\vec{\sigma}_s) \qquad\forall\:(\vec{v}_x,\vec{\sigma}_s)\in T_{(x,s)}P.
    \end{align*}
  \item 
    \vspace{-8pt}
    \textbf{Homogeneous Isotropic Connections}
    \newline
    \vspace{-13pt}
    \newline
    As already stated in Example \ref{ex:LQC}, the $\Pe$-invariant connections are of the form
    \vspace{-3pt}
    \begin{align*}
      \w^c(\vec{v}_x,\vec{\sigma}_x)= c \Add{s^{-1}}[\murs(\vec{v}_x)]+s^{-1}\vec{\sigma}_s \qquad \forall\:(\vec{v}_x,\vec{\sigma}_s)\in T_{(x,s)}P,
    \end{align*}   
    where $c$ runs over $\mathbb{R}$.
    \hspace*{\fill}{$\Diamond$}
  \end{itemize}
  \endgroup
\end{example}
\noindent
We close this subsection with a remark concerning the relations between sets of invariant connections that correspond to different lifts of the same Lie group action on the base manifold of a principal fibre bundle.
\begin{remark}
  \label{rem:liftuntersch} 
  Let $P$ be a principal fibre bundle and $\Phi,\Phi'\colon G\times P\rightarrow P$ be two Lie groups of automorphisms with $\varphi=\varphi'$. Then the respective sets of invariant connections can differ significantly. In fact, in the situation of the second part of Example \ref{ex:eukl} let
  $\Phi'((v,\sigma),(x,s)):=(v+ \varrho(\sigma)(x),s)$. Then $\varphi'=\varphi$ and Appendix \ref{subsec:DifferentLifts} shows that $\w_0(\vec{v}_x,\vec{\sigma}_s):=s^{-1}\vec{\sigma}_s$ for $(\vec{v}_x,\vec{\sigma}_s)\in T_{(x,s)}P$ is the only $\Phi'$-invariant connection on $P$.\hspace*{\fill}{{$\Diamond$}} 
\end{remark}
\subsubsection{Trivial Bundles -- Applications to LQG}
\label{sec:ApplTrivB}
In this final Subsection we determine the set of spherically symmetric connections on $\RR^3\times \SU$, to be used for the description of spherically symmetric gravitational systems in the framework of loop quantum gravity. To this end, we reformulate Theorem \ref{th:InvConnes} for trivial bundles. 

The spherically symmetric connections on $P=\RR^3\times \SU$ are such connections, invariant under the action $\Phi\colon \SU \times P\rightarrow P$, $(\sigma,(x,s))\mapsto (\sigma(x),\sigma s)$. Since $\Phi$ is not fibre transitive, we cannot use Case \ref{th:wang} for the necessary calculations. Moreover, it is not possible to apply the results from \cite{HarSni} (see Example \ref{example:SCHSV}) because the $\varphi$-stabilizer of $x=0$ equals $\SU$ whereas that of each $x\in \RR^3\backslash \{0\}$ is given by the maximal torus $H_x=\{\exp(t \murs(x) \:|\: t\in \RR)\}\subseteq \SU$. 
Of course, we could ignore the origin and consider the bundle $\RR^3\backslash \{0\}\times \SU$ together with the $\Gg$-covering $\{\lambda \cdot \vec{e}_1\:|\: \lambda \in \RR_{>0}\}$. This, however, is a different situation because an invariant connection on $\RR^3\backslash \{0\}\times \SU$ is not necessarily extendible to an invariant connection on $\RR^3\times \SU$. This is illustrated in (see also remarks following Example \ref{bsp:Rotats})
\begin{example}
  \label{ex:OnePoint}
  \begingroup
  \setlength{\leftmargini}{18pt}
  \begin{itemize}
  \item
    Let $S$ be a Lie group and $P=\RR^n\times S$. We consider the action $\Phi\colon \RR_{>0}\times P \rightarrow P$, $(\lambda,(x,s))\mapsto (\lambda x,s)$ and claim that the only $\Gg$-invariant connection is given by 
    \begin{align*}
      \w_0(\vec{v}_x,\vec{\sigma}_s):=\dd_sL_{s^{-1}}(\vec{\sigma}_s)\qquad\forall\: (\vec{v}_x,\vec{\sigma}_s)\in T_{(x,e)}P. 
    \end{align*}
    In fact, $P_\infty:=\RR^n\times \{e\}$ is a $\Gg$-covering of $P$ by Corollary \ref{cor:reductions} and it is straightforward to see that condition \textit{i.)} from Corollary \ref{cor:psialpha} is equivalent to the conditions \textit{a.)} and \textit{b.)} from Remark \ref{rem:Psialphaconderkl}.
    Let $\psi\colon \mathfrak{g}\times TP_\infty$ be a reduced connection w.r.t.\ $P_{\infty}$ and define $\psi_x:=\psi|_{\mathfrak{g}\times T_{(x,e)}}$.
    
    Since the exponential map $\exp\colon \mathfrak{g}\rightarrow \RR_{>0}$ is just given by $\mu \mapsto \mathrm{e}^{\mu}$ for $\mu \in \RR=\mathfrak{g}$, we have $\wt{g}((x,e))=\vec{g}\cdot x\in T_{(x,e)}P_\infty$ for $\vec{g}\in \mathfrak{g}$. Then for $\vec{w}:=-\vec{g}\cdot x \in T_{(x,e)}P_\infty$ from \textit{a.)} we obtain
    \begin{align}
      \label{eq:conda}    
      \psi_x\big(\vec{g},\vec{0}\big)=\psi_x\big(\vec{0}_{\mathfrak{g}},\vec{g}\cdot x\big)\qquad \forall\:\vec{g}\in \mathfrak{g},\forall\: x\in \RR^n.
    \end{align}
    In particular, $\psi_0\big(\vec{g},\vec{0}\big)=0$, and since $Q_{(0,e)}=\RR_{>0}\times \{e\}$, for $q=(\lambda,e)$ condition \textit{b.)} yields 
    \begin{align*}
      \lambda\:\psi_{0}\big(\vec{0}_{\mathfrak{g}},\vec{w}\big)=\psi_{0}\big(\vec{0}_{\mathfrak{g}},\lambda \vec{w}\big)\stackrel{\textit{b.)}}{=}\psi_{0}\big(\vec{0}_{\mathfrak{g}},\vec{w}\big)\qquad \forall\:\lambda> 0,\forall\:\vec{w} \in T_{(0,e)}P_\infty,
    \end{align*}
    hence $\psi_0=0$. 
    Analogously, for $x\neq 0$, $\vec{w} \in T_{(\lambda x,e)}P_\infty$, $\lambda>0$ and $q=(\lambda,e)$ we obtain
    \begin{align*}
      \lambda\hspace{1pt}\psi_{\lambda x}\big(\vec{0}_{\mathfrak{g}},\vec{w}\big)=\psi_{\lambda x}\big(\vec{0}_{\mathfrak{g}},\dd L_q(\vec{w})\big)\stackrel{\textit{b.)}}{=}\qrep(q)\cp \psi_{x}\big(\vec{0}_{\mathfrak{g}},\vec{w}\big)=
      \psi_{x}\big(\vec{0}_{\mathfrak{g}},\vec{w}\big),
    \end{align*}
    i.e., $\psi_{\lambda x}\big(\vec{0}_{\mathfrak{g}},\vec{w}\big)=\frac{1}{\lambda}\: \psi_{x}\big(\vec{0}_{\mathfrak{g}},\vec{w}\big)$. Here, in the second step, we have used the canonical identification of the linear spaces $T_{(x,e)}P_\infty$ and $T_{(\lambda x,e)}P_\infty$. Using the same identification, from continuity (smoothness) of $\psi$ and $\psi_{0}=0$ we obtain
    \begin{align*}
      0=\lim_{\lambda \rightarrow 0}\psi_{\lambda x}\big(\vec{0}_{\mathfrak{g}},\vec{w}\big)=\lim_{\lambda \rightarrow 0}\frac{1}{\lambda}\: \psi_{x}\big(\vec{0}_{\mathfrak{g}},\vec{w}\big)\qquad\forall\: x\in \RR^n,\forall\:\vec{w}\in T_{(x,e)}P_\infty 
    \end{align*}
    so that $\psi_{x}\big(\vec{0}_{\mathfrak{g}},\cdot\big)=0$ for all $x\in \RR^n$, hence $\psi=0$ by \eqref{eq:conda}.
    Finally, it is straightforward to see that $(\Phi^*\w_0)|_{\mathfrak{g}\times TP_\infty}=\psi=0$ holds.
  \item 
    \itspace
    Let $P'=\RR^n\backslash \{0\}\times S$ and $\Phi$ be defined as above. Then $K\times\{e\}$, for the unit-sphere $K:=\{x\in \RR^n\:|\: \|x\|=1\}$, is a $\Gg$-covering of $P'$ with the properties from 
    Example \ref{example:SCHSV}. Evaluating the corresponding conditions \textit{i''.)}, \textit{ii''.)}, \textit{iii''.)} immediately shows that the set of $\Gg$-invariant connections on $P'$ is in bijection with the smooth maps $\psi\colon \RR \times TK\rightarrow\mathfrak{s}$ for which $\psi|_{\RR \times T_kK}$ is linear for all $k\in K$. The corresponding invariant connections are given by
    \begin{align*}
      \w^\psi(\vec{v}_x,\vec{\sigma}_s)=\psi\left(\textstyle\frac{1}{\|x\|}\pr_{\|}(\vec{v}_x),\pr_{\perp}(\vec{v}_x)\right)+ s^{-1}\vec{\sigma}_s\qquad \forall\:(\vec{v}_x,\vec{\sigma}_s)\in T_{(x,s)}P'. 
    \end{align*}  
    Here, $\pr_{\|}$ denotes the projection onto the axis defined by $x\in \RR^n$ and $\pr_{\perp}$ the projection onto the corresponding orthogonal complement in $\RR^n$.\hspace*{\fill}{$\Diamond$}
  \end{itemize}
  \endgroup
\end{example}
\noindent
Also in the spherically symmetric case the $\varphi$-stabilizer of the origin has full dimension, and it turns out to be convenient (cf.\ Appendix \ref{subsec:IsotrConn}) to use the $\Gg$-covering $\RR^3\times \{e\}$ in this situation as well. 
Since the choice $P_\infty:=M\times\{e\}$ is always reasonable (cf.\ Lemma \ref{lemma:mindimslice}.1) if there is a point in the base manifold $M$ (of the trivial bundle $M\times S$) whose stabilizer is the whole group, we now adapt Theorem \ref{th:InvConnes} to this situation.
For this, we identify $T_xM$ with $T_{(x,e)}P_\infty$ for each $x\in M$ in the sequel.
\begin{scase}[Trivial Principal Fibre Bundles] 
  \label{scase:trivbundle}
  Let $(G,\Phi)$ be a Lie group of automorphisms of the trivial principal fibre bundle $P=M\times S$. Then the $\Phi$-invariant connections are in bijection with the smooth maps $\psi\colon \mathfrak{g}\times TM\rightarrow \mathfrak{s}$ for which $\psi|_{\mathfrak{g}\times T_xM}$ is linear for all $x\in M$ and that fulfil the following properties. 
  \newline
  \vspace{-8pt}
  \newline  
  Let $\psi^{\pm}\left(\vec{g},\vec{v}_y,\vec{s}\:\right):=\psi\left(\vec{g},\vec{v}_y\right)\pm\vec{s}$ for $((\vec{g},\vec{s}\:),\vec{v}_y)\in \mathfrak{q}\times T_yM$. Then for $q\in Q$, $x\in M$ with $q\cdot (x,e)=(y,e)\in M\times \{e\}$ and all  $((\vec{g},\vec{s}\:),\vec{v}_x)\in \mathfrak{q}\times T_xM$ we have
  \begin{enumerate}
  \item[\textrm{i.)}]
    $\wt{g}(x,e) + \vec{v}_x 
    -\vec{s}=0 \quad \Longrightarrow \quad \psi^{-}(\vec{g},\vec{v}_x,\vec{s}\hspace{1pt})=0$,
  \item[\textrm{ii.)}]
    $\psi^+(\dd L_q \vec{v}_x)=\qrep(q)\cp \psi\big(\vec{0}_{\mathfrak{g}},\vec{v}_x\big)$\hspace{35pt}  $\forall\:\vec{v}_x\in T_xM$,
  \item[\textrm{iii.)}]
    $\psi\big(\!\Add{q}(\vec{g}),\vec{0}_{y}\big)=\qrep(q)\cp \psi\big(\vec{g},\vec{0}_{x}\big)$ \:\hspace{19.5pt}  $\forall\:\vec{g}\in\mathfrak{g}$.
  \end{enumerate}
  \begin{proof}  
    The elementary proof can be found in Appendix \ref{subsec:TrivBund}.
  \end{proof}
\end{scase} 

\begin{example}[Spherically Symmetric Systems in LQG]
  \label{bsp:Rotats}
  Let $\varrho\colon \SU \rightarrow \SOD$ be the universal covering map and $\sigma(x):=\varrho(\sigma)(x)$ for $x\in \RR^3$. Moreover, let $\text{\gls{MURS}}\colon \RR \rightarrow \mathfrak{su}(2)$ be defined as in Convention \ref{conv:sutwo1}.\ref{conv:sutwo111}. We consider the action of $G=\SU$ on $P=\RR^3\times \SU$ defined by $\Phi(\sigma, (x,s)):=(\varrho(\sigma)(x),\sigma s)$. It is shown in Appendix \ref{subsec:IsotrConn} that the corresponding invariant connections are of the form
  \begin{align}
    \label{eq:rotinvconn}
    \begin{split}
      \w^{abc}(\vec{v}_x,\vec{\sigma}_s):= \Add{s^{-1}}\!\big[&a(x)\murs(\vec{v}_x)+ b(x)[\murs(x),\murs(\vec{v}_x)]%\\
      % &\hspace{14pt} 
      +c(x)[\murs(x),[\murs(x),\murs(\vec{v}_x)]]\big]+ s^{-1}\vec{\sigma}_s
    \end{split}
  \end{align}
  for $(\vec{v}_x,\vec{\sigma}_s)\in T_{(x,s)}P$ and with rotation invariant maps $a,b,c\colon \mathbb{R}^3\rightarrow \mathbb{R}$ for which the whole expression is a smooth connection.
  We claim that the functions $a,b,c$ can be assumed to be smooth as well. More precisely, we show that we can assume that
  \begin{align*}
    a(x)=f\big(\|x\|^2\big) \qquad\quad b(x)=g\big(\|x\|^2\big)\qquad\quad c(x)=h\big(\|x\|^2\big)
  \end{align*}
  holds for smooth functions $f,g,h\colon (-\epsilon,\infty)\rightarrow \RR$ with $\epsilon>0$. Then, each pullback of such a spherically symmetric connection by the global section $x\mapsto (x,e)$ can be written in the form 
  \begin{align*}
    \wt{\w}^{abc}(\vec{v}_x)=f'\big(\|x\|^2\big)\:\murs(\vec{v}_x)+g'\big(\|x\|^2\big)\:\murs(x\times \vec{v}_x) +h'\big(\|x\|^2\big)\:\murs\left(x\times (x\times \vec{v}_x)\right)
  \end{align*}
  for smooth functions $f',g',h'\colon (-\epsilon,\infty)\rightarrow \RR$ with $\epsilon>0$.
  \newline
  \vspace{-10pt}
  \newline
  {\bf Proof of the Claim.}
  \begingroup
  \setlength{\leftmargini}{20pt}
  \begin{enumerate}  
  \item[{\bf 1)}]
    \vspace{-6pt} 
    Smoothness of $\w^{abc}$ 
    implies smoothness of the real functions
    \begin{align*}
      a_{\vec{n}}(\lambda):= a(\lambda\vec{n})\qquad   b_{\vec{n}}(\lambda):=\lambda\: b(\lambda\vec{n})\qquad c_{\vec{n}}(\lambda):= \lambda^2 c(\lambda\vec{n})\qquad \forall\:\lambda\in \mathbb{R}
    \end{align*}
    for each $\vec{n}\in\mathbb{R}^3\backslash\{0\}$. In fact, $a_{\vec{n}}(\lambda)\cdot \murs(\vec{n})=\w^{abc}_{(\lambda \vec{n},e)}(\vec{n})$ is smooth, so that smoothness of $b_{\vec{n}}$ and $c_{\vec{n}}$ is immediate from smoothness of $\lambda \mapsto \w^{abc}_{(\lambda \vec{e}_1,e)}(\vec{e}_2)$.  
  \item[{\bf 2)}]
    \itspace
    Let $\vec{n}$ be fixed. Then
    $a_{\vec{n}}$ is even so that $a_{\vec{n}}(\lambda)=f\big(\lambda^2\big)$ for a smooth function $f\colon (-\epsilon_1,\infty)\rightarrow \mathbb{R}$, see \cite{HasslerWhitneyb}.  
    Moreover, $b_{\vec{n}}$ is smooth and odd so that $b_{\vec{n}}(\lambda) =\lambda\: g\big(\lambda^2\big)$ for some smooth function $g\colon (-\epsilon_2,\infty)\rightarrow \mathbb{R}$, again by \cite{HasslerWhitneyb}. Similarly, $c_{\vec{n}}(\lambda)= l(\lambda^2)$ for a smooth function $l\colon (-\epsilon_3,\infty)\rightarrow \mathbb{R}$. Since $\lambda \mapsto l(\lambda^2)$ is even and $l(0)=0$,  for $s\in \mathbb{N}_{>0}$ Taylor's formula yields 
    \begin{align*}
      l\big(x^2\big)&= a_1 x^2  +\dots + a_{s}x^{2s} + x^{2(s+1)}\phi(x)\\
      &=x^2 \big(a_1  +\dots + a_{s}x^{2s-2} + x^{2s}\phi(x)\big)=x^2L(x)
    \end{align*}
    with remainder term $\phi(x):=\frac{1}{(2s+1)!}\frac{1}{x^{2s+2}}\int_{0}^{x}(x-t)l^{(2s+2)}(t) \hspace{1.5pt} \dd t$ for $x\neq 0$ and $\phi(0):=l^{(2s+2)}(0)$. Now, $\phi$ is continuous by Theorem 1 in \cite{HasslerWhitneya} so that $L$ is continuous as well. But $x\mapsto x^2 L(x)$ is smooth so that Corollary 1 in \cite{HasslerWhitneya} shows that $L$ is smooth as well. Now, $L$ is even, hence $L(x)=h(x^2)$ for some smooth function $h\colon (-\epsilon_4,\infty )\rightarrow \mathbb{R}$. Then $c_{\vec{n}}(\lambda)=l\big(\lambda^2\big)=\lambda^2h\big(\lambda^2\big)$
    and for $x\neq 0$ we get
    \begin{align*}
      b(x)&=\|x\|\:b\left(\|x\|\frac{x}{\|x\|}\right)\frac{1}{\|x\|}=
      b_{\textstyle{\frac{x}{\|x\|}}}(\|x\|)\frac{1}{\|x\|}
      =g\left(\|x\|^2\right),\\
      c(x)&=\|x\|^2\:c\left(\|x\|\frac{x}{\|x\|}\right)\frac{1}{\|x\|^2}
      =c_{\textstyle\frac{x}{\|x\|}}\left(\|x\|\right)\frac{1}{\|x\|^2}
      =h\left(\|x\|^2\right).
    \end{align*}
    Moreover, for $x=0$ we have 
    \begin{align*}     
      b(x)[\murs(x),\murs(\vec{v}_x)]=\:&0=g\big(\|x\|^2\big)[\murs(x),\murs(\vec{v}_x)],\\
      c(x)[\murs(x),[\murs(x),\murs(\vec{v}_x)]\big]=\:&0=h(x)[\murs(x),[\murs(x),\murs(\vec{v}_x)]\big]
    \end{align*}
    so that we can assume $a(x)=f(\|x\|^2)$, $b(x)=g(\|x\|^2)$ and $c(x)=h(\|x\|^2)$ for the smooth functions $f,g,h\colon \left(-\min(\epsilon_1,\dots,\epsilon_4),\infty\right)\rightarrow \RR$. 
  \end{enumerate}
  \endgroup
  \noindent
  In particular, there are spherically symmetric connections on $\RR^3\backslash \{0\}\times \SU$ which cannot be extended to those on $P$. For instance, if $b=c=0$ and $a(x):=1\slash \|x\|$ for $x\in \RR^3\backslash \{0\}$, then $\w^{abc}$ cannot be extended smoothly to an invariant connection on $\RR^3\times \SU$ since elsewise $a_{\vec{n}}$ could be extended to a continuous (smooth) function on $\RR$. \hspace*{\fill}{$\Diamond$}
\end{example}

\subsection{Summary}
\label{invconnconclusion}
We conclude with a short review of the particular cases that follow from Theorem \ref{th:InvConnes}. For this let $(G,\Phi)$ be a Lie group of automorphisms of the principal fibre bundle $(P,\pi,M,S)$ and $\varphi$ the induced action on $M$.
\begingroup
\setlength{\leftmargini}{20pt}
\begin{itemize}
\item
 % \vspace{-4pt}
  If $P=M\times S$ is trivial, then $M\times \{e\}$ is a $\Gg$-covering of $P$. As we have demonstrated in the spherically symmetric and scale invariant case (cf.\ Examples \ref{ex:OnePoint} and \ref{bsp:Rotats}), this choice can be useful for calculations if there is a point in $M$ whose $\varphi$-stabilizer is the whole group $G$. 
\item
 % \vspace{-4pt}
  If there is an element $x\in M$ which is contained in the closure of each $\varphi$-orbit, 
  then each $\THA$-patch that contains some $p\in\pi^{-1}(x)$ is a $\Gg$-covering of $P$, see Example \ref{ex:Bruhat}. If $\varphi$ acts transitively on $M$, then for each $p\in P$ the zero-dimensional submanifold $\{p\}$ is a $\Gg$-covering of $P$ giving back Wang's original theorem, see Case \ref{th:wang} and Example \ref{ex:eukl}.
\item
  %\vspace{-4pt}		
  Let $\Phi$ act via gauge transformations on $P$. In this case, each open covering $\{U_\alpha\}_{\alpha\in I}$ of $M$ together with smooth sections $s_\alpha \colon U_\alpha \rightarrow P$ provides the $\Gg$-covering $\{s_\alpha(U_\alpha)\}_{\alpha\in I}$ of $P$. If $G$ acts trivially, this specializes to the usual description of smooth connections by means of consistent families of local 1-forms on the base manifold $M$.
\item
  %\vspace{-4pt}
  If we find a $\THA$-patch $P_0$ such that $\pi(P_0)$ intersects each $\varphi$-orbit in a unique point, then $P_0$ is a $\Gg$-covering. If in addition the stabilizer $Q_p$ does not depend on $p\in P_0$, then we are in the situation of \cite{HarSni}, see Example \ref{example:SCHSV}. 
\item
  %\vspace{-4pt}
  Assume that there is a collection of $\varphi$-orbits forming an open subset
  $U\subseteq M$. Then $O:=\pi^{-1}(U)$ is a principal fibre bundle and each $\Phi$-invariant connection on $P$ restricts to a $\Phi$-invariant connection on $O$. Conversely, if $U$ is in addition dense in $M$, then one can ask the question whether a $\Phi$-invariant connection on $O$ extends to a $\Phi$-invariant connection on $P$. Since such an extension is necessarily unique (continuity), $\varphi$-orbits not contained in $U$ can be seen as sources of obstructions for the extendibility of invariant connections on $O$ to $P$. Indeed, as the examples in Subsection \ref{sec:ApplTrivB} show, smoothness of these extension gives crucial restrictions. Moreover, by Example \ref{ex:Bruhat}, taking one additional orbit into account can shrink the number of invariant connections to zero.  
  Of particular interest, in this context, is the case where $G$ is compact as then the orbits of principal type always form a dense and open subset $U$ of $M$ on which the situation of \cite{HarSni} always holds locally \cite{RumSchmBuch}. This gives rise to a canonical $\Phi$-covering of $O$ consisting of convenient patches. So, using the present characterization theorem, there is a realistic chance to get some general classification results in the compact case. These can be used, e.g., to extend the framework of the foundational LQG reduction paper \cite{BojoKa}.
\end{itemize}
\endgroup
\noindent
As Corollary \ref{cor:reductions} shows, in the general situation one can always construct $\Gg$-coverings of $P$ from families of $\varphi$-patches in $M$. In particular, the first three cases arise in this way.

\section{Conclusions and Outlook}
\begingroup
\setlength{\leftmargini}{15pt}
\begin{itemize}
\item
	In Subsection \ref{sec:ModifreeSeg}, we have considered the situation where the  action $\wm$ induced on the base manifold $M$ is analytic and pointwise proper. We have shown that then the following decomposition of the set $\Paw$ (of embedded analytic curves in $M$) holds 
	\begin{align*}
		\Paw=\Pags\sqcup \Pacs\sqcup \Pafns\sqcup \Pafs,
	\end{align*}
	and that each of these subsets is invariant under decomposition and inversion of its elements. So, by Proposition \ref{rem:euklrem2b} we have 
	\begin{align*}
    \AQRw  \cong  \AQRInd{\mg}\times \AQRInd{\mathrm{CNL}}   
 \times  \AQRFNS\times \AQRInd{\mathrm{FS}}
  \end{align*}	
	provided, of course, that each of the above sets of curves is non-empty.\footnote{Elsewise, we just remove the respective factor in the above product.} Recall that
	\begingroup
\setlength{\leftmarginii}{12pt}
	\begin{itemize}
	\item[$\triangleright$]
	 $\Paf=\Pafns\sqcup \Pafs$ is the set of embedded analytic curves that contain a segment $\delta$ for which $\wm_g\cp \delta=\delta$ holds whenever $\im[\wm_g\cp\delta]\cap \im[\delta]$ is infinite. Here, $\Pafns$ consists of all such curves whose stabilizer (a well-defined quantity in this context) is trivial, so that $\Pafs$ denotes the set of all free curves for which this is not the case. Obviously,  $\Pafs=\emptyset$ holds if $\wm$ is free.
	\item[$\triangleright$]
	$\Pac=\Pags\sqcup \Pacs$ (set of continuously generated curves) consists of all embedded analytic curves which are not free. Here, $\Pags$ consists of such curves which are generated by the Lie algebra of symmetry group and $\Pacs$ is just its complement in $\Pac$.  At this point, it is the set $\Pacs$ which makes it hard to define measures on the full space $\A_\w$. However, as we have seen in the last two parts of Proposition \ref{prop:freeseg}, $\Pacs=\emptyset$  holds if $\wm$ admits only normal stabilizers and is transitive or proper.
	\end{itemize}
	\endgroup
	So, for $\wm$ non-trivial we have:
\renewcommand*{\arraystretch}{1.2}	
	\begin{center}
  \begin{tabular}{c|c|c|c|c}
    $\wm$       & $\Pags$ & $\Pacs$ & $\Pafns$ & $\Pafs$\\[3pt] \hline 
      free &  many& ? & ?& $\emptyset$\\
    transitive $+$ normal stabilizers   &  many & $\emptyset$ & ?& ?\\
      \hspace{10pt} proper $+$ normal stabilizers  &  many & $\emptyset$ & ?& ?
  \end{tabular}
  	\end{center}		
	Thus, if $\wm$ is proper and free such as in (semi-)homogeneous LQC (see Example \ref{ex:LQC}), even $\Paw=\Pags\sqcup \Pafns$, hence $\AQRw  \cong  \AQRInd{\mg}   
 \times  \AQRFNS$ holds. 
 
 In Section \ref{sec:MOQRCS}, we have constructed normalized Radon measures $\mLAS$ and $\mFNS$ on $\AQRInd{\mg}$ (for $S=\SU$) and $\AQRFNS$ (for $S$ compact and connected), respectively. 
	This means that we have the normalized Radon measure $\mLAS\times \mFNS$ on $\AQRw \cong \AQRInd{\mg}\times\AQRFNS$ whenever $\Paw=\Pags\sqcup \Pafns$ holds and $S=\SU$. In particular, this provides us with a normalized Radon measure on $\AQRw$ in \mbox{(semi-)homogeneous} LQC. 

Unfortunately, in homogeneous isotropic and spherically symmetric LQC the $\wm$-stabilizers are not normal subgroups. So, we do not know whether $\Pacs=\emptyset$  holds in these cases. More precisely, there we have
		\begin{center}
  \begin{tabular}{c|c|c|c|c}
     LQC    & $\Pags$ & $\Pacs$ & $\Pafns$ & $\Pafs$\\[3pt] \hline 
     homogeneous 	  &  many & $\emptyset$ & many& $\emptyset$\\ 
        semi-homogeneous &  many & $\emptyset$ & many & $\emptyset$\\ 
    homogeneous isotropic  &  many & ? & many & $\emptyset$\\
    spherically symmetric   &  many & ?& many& linear curves  through origin.
  \end{tabular}
  	\end{center}
Hence, in order to obtain a normalized Radon measure on $\AQRw$, in the last two cases we first had to calculate the set $\Pacs$ by hand. 
For this observe that in the spherically symmetric case 
it is possible to construct a measure on $\AQRInd{\mathrm{FS}}$ by hand, see Remark \ref{ex:Fullmeas}.\ref{ex:Fullmeas2}.
Consequently, there is still some serious demand for concepts allowing to determine the set $\Pacs$ also in the general case. 
\item
 Since we have constructed normalized Radon measures on quantum-reduced configuration spaces in LQG, we now can start to reduce the holonomy-flux algebra and try to define reasonable representations on the respective Hilbert spaces of square integrable functions. Then, aiming at some uniqueness results for these representations similar to that one has for the full theory \cite{UniqLew, ChUn}, we should investigate the invariance properties of the constructed measures in detail. 
 Here, the case of (semi-)homogeneous LQC, where we have the Radon measure $\mLAS\times \mFNS$ on the full quantum-reduced configuration space  $\AQRw \cong \AQRInd{\mg}\times\AQRFNS$, may serve as a working example for this. 
In addition to that, one also should search for such representations on the Hilbert spaces 
we have constructed in Section \ref{sec:HomIsoCo} for the space $\RR\sqcup \RB$.  
These representations then can be compared (w.r.t.\ unitary equivalence) with the standard representation on the standard kinematical Hilbert space $\Lzw{\RB}{\muB}$ of homogeneous isotropic LQC. 
Complementary to that, here one might use Proposition \ref{lemma:bohrmassdichttrans} and Corollary \ref{cor:eindbohr} in order to prove some kind of uniqueness statement for this standard representation. 
\item 
	Even if the characterization theorem \ref{th:InvConnes} looks quite technical at a first sight, its flexibility makes it a powerful tool for explicit calculations. It simplifies significantly in several situations and provides us, e.g., with the Cases \ref{scase:OneSlice}, \ref{scase:GaugeTransf}, \ref{scase:slicegleichredcluster}, \ref{th:wang} and \ref{scase:trivbundle}. Aiming at some classification results, here the next step will be to investigate such special situations in more detail. 
	Predestinated for this is the case where $G$ is compact as there
the orbits of principal type always form a dense open subset of $U$ of $M$, defining a dense open subbundle $O$ of $P$ for which the situation of \cite{HarSni} always holds locally \cite{RumSchmBuch}. 
This provides us with a $\Phi$-covering of $O$ which consists of very convenient patches. So, using the present characterization theorem there is a realistic chance to get some general classification results which can be used to extend the framework of the foundational LQG reduction paper \cite{BojoKa}.	
\item
  In gauge field theories, instead of $\Con$ usually the quotient $\Con \slash \GAG$ is considered as the physically relevant configuration space.  Here, $\GAG$ denotes the set of gauge transformations\footnote{These are all automorphisms $\sigma$ of $P$ with $\pi\cp \sigma =\pi$.} on the underlying principal fibre bundle and we have $\w\sim_{\GAG} \w'$ for $\w,\w'\in \Con$ iff there is $\sigma\in \GAG$ such that $\w'=\sigma^*\w$ holds. 
  
   Similarly, if $(G,\Phi)$ is a Lie group of automorphisms of $P$, instead of $\AR$ also the set 
  \begin{align*}
    \Con_{\red,\GAG}=\{\w\in \Con\:|\: \forall\: g\in G\:\: \exists\:\sigma\in \GAG : \Phi_g^*\w=\sigma^*\w\} 
  \end{align*} 
    of connections invariant up to gauge transformations is considered as reduced configuration space.     
 However, in contrast to the set $\AR$, for $\Con_{\red,\GAG}$ no general characterization theorems seem to exist so far, so that it is potentially harder to compute this space. So, instead of quantizing $\Con_{\red,\GAG}$ (which would also be a classical reduction) one can use the lifted action $\specw$ in order to define ``invariance up to gauge'' directly on the quantum level. 
  Hence, instead of the space $\AQR\cong \IHOM$ one can consider the space $\RedGauge$ consisting of all 
  elements $\homm\in \HOM$ for which we have that for each $g\in G$ there is a generalized gauge transformation\footnote{This just means that $\pi\cp \sigma =\pi$ and $\sigma(p\cdot s)=\sigma(p)\cdot s$ holds for all $\in S$.} $\sigma\colon P\rightarrow P$ with $\specw_g(\homm)=\sigma(\homm)$, i.e., 
  \begin{align*}
    \specw_g(\homm)(\gamma)(p)=(\sigma \cp \homm(\gamma))(\sigma^{-1}(p))\qquad \forall\: \gamma\in \Pa, \forall\:p\in F_{\gamma(a)}
  \end{align*} 
  for $\dom[\gamma]=[a,b]$. Then,  $\IHOM \subseteq \RedGauge$ holds and the concepts of the Sections \ref{susec:LieALgGenC} and \ref{sec:MOQRCS} seem adaptable to the ``up to gauge''-case. Indeed, we rather expect technical than conceptual difficulties at this point. 
  \item
  An alternative construction of the Ashtekar-Lewandowski measure has been presented in \cite{Tlas}. There, the author uses a so-called nicely shrinking net of open neighbourhoods (thickenings) of the closed subset $\HOM\subseteq \mathrm{Maps}(\Pa,\IsoF)$ in order to define the Ashtekar-Lewandowski measure on $\HOM$.\footnote{The author considers the situation where $S$ is compact, connected and semisimple. Moreover, he assumes that $\Pa$ consists of  piecewise smooth and immersive loops with fixed base point, i.e., piecewise smooth and immersive curves whose end points equal a fixed point in the base manifold. The arguments, however, also go through if $\Pa=\Paw$ and $S$ is just compact and connected.} Here, the space $\mathrm{Maps}(\Pa,\IsoF)$ (of all maps $\Pa\rightarrow \IsoF$), equipped with the topology defined in analogy to that one in  Definition \ref{def:indepref}.\ref{conv:muetc}, is homeomorphic to the Tychonoff product $S^{|\Pa|}$.  
  
   More generally, for $X$ a compact Hausdorff space carrying a normalized Radon measure $\mu$  and $C\subseteq X$ a closed subspace, one can define a positive and continuous linear functional 
  	$\III\colon C(C)\rightarrow \mathbb{C}$, i.e., a finite Radon measure on $C$  
  as follows: (see also \cite{Tlas})
	\begingroup
	\setlength{\leftmarginii}{20pt}
		\begin{itemize}
		\item[i)]
			Let $\{C_\lambda\}_{\lambda \in \Lambda}$ be a net of neighbourhoods of $C$ such that for each neighbourhood $U$ of $C$ we find $\lambda_0\in \Lambda$ with $C_\lambda\subseteq U$ for all $\lambda\geq \lambda_0$. Such a net is called nicely shrinking. 
		\item[ii)]
			For each $f\in C(C)$ let $\ovl{f}\in C(X)$ be an extension (apply Tietze extension theorem) for which $\big\|\wt{f}\big\|_\infty=\|f\|_\infty$ holds, and define
	\begin{align*}			
			f_\lambda:=\frac{1}{\mu(C_\lambda)}\int_{C_\lambda}\wt{f}\: \dd\mu\qquad \forall\: \lambda\in \Lambda. 
			\end{align*}
		\item[$\triangleright$]
			Now, choose $\{C_\lambda\}_{\lambda \in \Lambda}$ in such a way that $\III(f):=\lim_\lambda f_\lambda$ exists for all $f$ contained in a suitable dense subset $\mathfrak{D}\subseteq C(C)$. It is straightforward to see \cite{Tlas} that then $\III(f)$ does not depend on the explicit extension $\wt{f}$ of $f$. 		
			Moreover, it follows that $\III\colon \mathfrak{D}\rightarrow \CCC$ is positive and bounded by 1, hence extends by continuity to a positive (and continuous) linear functional on $C(X)$. 
		\end{itemize}
		\endgroup   
		Now, the measure used in \cite{Tlas} for $S^{|\Pa|}$ is just the Radon product (cf.\ Lemma and Definition \ref{def:ProductMa}.\ref{def:ProductMa3}) of copies of the Haar measure on $S$, and the net $\{C_\lambda\}_{\lambda \in \Lambda}$ is constructed in a very natural way. 		
	In particular, in view of the complications (non-trivial restrictions to the images of the  projection maps) arising from the invariance and inclusion properties of the reduced spaces  discussed in this work, the above ``thickening approach'' seems to be predestinated  for providing a general notion of a reduced Ashtekar-Lewandowski measure on these spaces. Indeed, it even might help to drop the restriction to the structure group $\SU$ (and tori) we have made in Subsection \ref{sec:ConSp} in order to define the measure on $\AQRInd{\mg}$. The developments of this thesis then should help to find the correct definitions, e.g., by requiring that, when restricting to the set $\Pafns$, we get back the measure constructed in Subsection \ref{sec:FreeM}. 
Moreover, the net of thickenings constructed in \cite{Tlas} for $\HOM$ even seems generalizable to the space\footnote{This space is indeed closed as one easily deduces from compactness of the structure group $S$.} $\RedGaugew$. This is basically because this space (under mild assumptions) seems to admit sufficiently many elements in order to make the relevant maps surjective.
\end{itemize} 
\endgroup   

\appendix
\section*{Appendix}

\section{Appendix to Preliminaries}

\subsection{Projective Structures and Radon Measures}
In this section we adapt the standard facts on projective structures to our definitions from Subsection \ref{subsec:ProjStruc}.
\begin{Lemma} 
  \label{lemma:equivalence}
  Let $X$ be a projective limit of $\{X_\alpha\}_{\alpha \in I}$ w.r.t.\ the maps $\pi_\alpha \colon X\rightarrow X_\alpha$ for $\alpha\in I$ and $\pi^{\alpha_2}_{\alpha_1}\colon X_{\alpha_2}\rightarrow X_{\alpha_1}$ for $\alpha_1,\alpha_2 \in I$ with $\alpha_2 \geq \alpha_1$. Then $X$ is homeomorphic to 
  \begin{align} 
  \label{eq:projlimprod} 
    \widehat{X}:=\left\{\hx \in \prod_{\alpha\in I}X_\alpha\: \:\Bigg|\:\: \pi_{\alpha_1}^{\alpha_2}(x_{\alpha_2})=x_{\alpha_1}\:\: \forall\:\alpha_2\geq \alpha_1\right\}
  \end{align}  
  equipped with the subspace topology inherited from the Tychonoff topology on $\prod_{\alpha\in I}X_\alpha$. 
  \begin{proof}
    The map $\eta\colon X\rightarrow \widehat{X}$, $x\mapsto \{\pi_\alpha(x)\}_{\alpha\in I}$ is well-defined by Definition \ref{def:ProjLim}.\ref{def:ProjLim2}. Moreover, $\eta$ is continuous as it is so as a map from $X$ to $\prod_{\alpha\in I}X_\alpha$. This is  because $\pr_\alpha \cp \eta=\pi_\alpha$ is continuous for all projection maps $\pr_\alpha\colon\prod_{\alpha\in I}X_\alpha \rightarrow X_\alpha$ just by Definition \ref{def:ProjLim}.\ref{def:ProjLim1}. 
    Then, $\eta$ is a homeomorphism if it is bijective because $X$ is compact and $\widehat{X}$ is Hausdorff. 
    Injectivity of $\eta$ is immediate from Definition \ref{def:ProjLim}.\ref{def:ProjLim3}. For surjectivity, assume that $\{x_\alpha\}_{\alpha\in I}=\hx\in \widehat{X}$ with $\hx\notin \im[\eta]$, i.e.,
    $\bigcap_{\alpha\in I}\pi_\alpha^{-1}(x_\alpha)=\eta^{-1}(\hx)=\emptyset$. By continuity of $\pi_\alpha$, the sets $\pi_\alpha^{-1}(x_\alpha)\subseteq X$ are closed, hence compact by compactness of $X$. Consequently, there are finitely many $\alpha_1,\dots,\alpha_k \in I$ such that $\pi_{\alpha_1}^{-1}(x_{\alpha_1})\cap {\dots} \cap \pi_{\alpha_k}^{-1}(x_{\alpha_k})=\emptyset$. By directedness of $I$, we find some $\alpha\in I$ such that $\alpha_j\leq \alpha$ for all $1\leq j\leq k$, hence
    \begin{align}
      \label{eq:projlimmm}    
      \{x_{\alpha_j}\}= \big(\pi_{\alpha_j}^{\alpha}\cp \pi_\alpha\big)\left(\pi_\alpha^{-1}(x_\alpha)\right)= \pi_{\alpha_j}\left(\pi_\alpha^{-1}(x_\alpha)\right)\quad  \text{for all}\quad 1\leq j\leq k
    \end{align}
    because $\pi_\alpha^{-1}(x_\alpha)$ is non-empty ($\pi_\alpha$ is surjective) and $\pi^{\alpha}_{\alpha_j}(x_\alpha)=x_{\alpha_j}$ for all $1\leq j\leq k$. Applying $\pi_{\alpha_j}^{-1}$ to both sides of \eqref{eq:projlimmm} gives $\pi_{\alpha_j}^{-1}(x_{\alpha_j})\supseteq\pi_\alpha^{-1}(x_\alpha)$ for all $1\leq j\leq k$, which contradicts that $\pi_{\alpha_1}^{-1}(x_{\alpha_1})\cap \dots \cap \pi_{\alpha_k}^{-1}(x_{\alpha_k})=\emptyset$ holds.  
  \end{proof}
\end{Lemma}

\begin{lemma}
  \label{lemma:ConstMeas}
  Let $X$ and $\{X_\alpha\}_{\alpha\in I}$ be as in Definition \ref{def:ProjLim}. Then the normalized Radon measures on $X$ are in bijection with the consistent families of normalized Radon measures on $\{X_\alpha\}_{\alpha\in I}$.
  \begin{proof}
    If $\mu$ is a normalized Radon measure on $X$, then it is straightforward to see that $\{\pi_\alpha(\mu)\}_{\alpha\in I}$ is a consistent family of normalized Radon measures on $\{X_\alpha\}_{\alpha\in I}$. For the converse statement define $\mathrm{Cyl}(X):=\bigcup_{\alpha\in I}\pi_\alpha^*(C(X_\alpha))\subseteq C(X)$. Then, $\mathrm{Cyl}(X)$ is closed under involution, separates the points in $X$ and vanishes nowhere. Moreover, $\mathrm{Cyl}(X)$ is closed under addition, since
    \begin{align}
      \label{eq:addd}
      f\cp \pi_{\alpha_1} + g\cp \pi_{\alpha_2}&=\left(f \cp\pi^{\alpha_3}_{\alpha_1}\right) \cp \pi_{\alpha_3}+ \left(g \cp\pi^{\alpha_3}_{\alpha_2}\right)\cp \pi_{\alpha_3}\in \mathrm{Cyl}(X),
    \end{align}
    where $\alpha_1,\alpha_2,\alpha_3\in I$ with $\alpha_1,\alpha_2\leq \alpha_3$.
    It follows in the same way that $\mathrm{Cyl}(X)$ is closed under multiplication.
    By Stone-Weierstrass theorem $\mathrm{Cyl}(X)$ is a dense $^*$-subalgebra of $C(X)$ and the map
    $\III\colon \mathrm{Cyl}(X)\rightarrow \mathbb{C}$, $f\cp \pi_\alpha \mapsto \int_{X_\alpha}f\:\dd\mu_\alpha$
    is well-defined, linear and continuous w.r.t.\ supremum norm on $C(X)$. In fact, if $f\cp \pi_{\alpha_1}= g\cp \pi_{\alpha_2}$, then $(f \cp\pi^{\alpha_3}_{\alpha_1}) \cp \pi_{\alpha_3}= (g \cp\pi^{\alpha_3}_{\alpha_2})\cp \pi_{\alpha_3}$ if $\alpha_1,\alpha_2\leq \alpha_3$. Hence, $f \cp\pi^{\alpha_3}_{\alpha_1}=g \cp\pi^{\alpha_3}_{\alpha_2}$ by surjectivity of $\pi_{\alpha_3}$ so that the transformation formula yields
    \begin{align*}
      \III(f\cp \pi_{\alpha_1})&=\int_{X_{\alpha_1}}f\: \dd\mu_{\alpha_1} = \int_{X_{\alpha_3}}f\cp \pi_{\alpha_1}^{\alpha_3}\: \dd\mu_{\alpha_3}\\
      &= \int_{X_{\alpha_3}}g\cp \pi_{\alpha_2}^{\alpha_3}\: \dd\mu_{\alpha_3}=\int_{X_{\alpha_2}}g\: \dd\mu_{\alpha_2}=\III(g\cp \pi_{\alpha_2})
    \end{align*}
    showing well-definedness of $\III$. Linearity follows from \eqref{eq:addd} and for continuity observe that 
	\begin{align*}
		|\III(f\cp \pi_\alpha)|\leq\|f\|_\infty=\|f\cp \pi_\alpha \|_\infty,
	\end{align*}	    
    by surjectivity of $\pi_\alpha$. Since $\III$ is linear and continuous, it extends to a continuous functional on $C(X)$ which, by Riesz-Markov theorem (see, e.g., 2.5 Satz in \cite[Chap.VIII, \S 2]{Elstrodt}), defines a finite Radon measure $\mu$ on $\Borel(X)$. Then 
    $\mu(X)=\III(1)=1$ and for each $f\in C(X_\alpha)$ we have
    \begin{align*}
      \int_{X_\alpha}f\:d(\pi_\alpha^*\mu)=\int_X (f\cp \pi_\alpha) \:\dd\mu=\III(f\cp \pi_\alpha)=\int_{X_\alpha}f\:\dd\mu_\alpha
    \end{align*}
    so that $\pi_\alpha^*\mu=\mu_\alpha$, again by Riesz-Markov theorem. Finally, if $\mu'$ is a further finite Radon measure with $\pi_\alpha^*\mu'=\mu_\alpha$ for all $\alpha\in I$, then $\III'\colon C(X)\rightarrow \mathbb{C}$, $f\mapsto \int_{X}f\:\dd\mu'$ is continuous and $\III|_{\mathrm{Cyl}(X)}=\III'|_{\mathrm{Cyl}(X)}$ by the transformation formula. Consequently, $\III=\III'$ by denseness of $\mathrm{Cyl}(X)$ in $C(X)$ so that $\mu=\mu'$.
  \end{proof}
\end{lemma}

\section{Appendix to Special Mathematical Background}
\subsection{Homogeneous Isotropic Connections}
\label{subsec:InvIsoHomWang}
Assume that we are in the situation of Example \ref{ex:LQC}, i.e., 
$P=\RR^3\times \SU$ and $G:=\Ge=\Gee$. We show that the connections of the form
\begin{align*}
  \w^c_{(x,s)}(\vec{v}_x,\vec{\sigma}_s)= c\: \Ad(s^{-1})[\murs(\vec{v}_x)]+s^{-1}\vec{\sigma}_s \qquad (\vec{v}_x,\vec{\sigma}_s)\in T_{(x,s)}P 
\end{align*} 
are exactly the $\Pe$-invariant ones.
To verify their $\Pe$-invariance let $(x,s)\in P$, $(v,\sigma)\in G$ and $(\vec{v}_x,\vec{\sigma}_s)\in T_{(x,s)}P$. Then $ \dd_{(x,s)}L_{(v,\sigma)}(\vec{v}_x,\vec{\sigma}_s)= (\uberll{\sigma}{\vec{v}_x},\sigma \vec{\sigma}_s)$ so that
\begin{align*}
  (L_{(v,\sigma)}^*\w^{c})_{(x,s)}(\vec{v}_x,\vec{\sigma}_s)&=\w^{c}_{(v+\uberll{\sigma}{x},\sigma s)}(\dd_{(x,s)}L_{(v,\sigma)}(\vec{v}_x,\vec{\sigma}_s))\\
  &=c\: \Ad(s^{-1}\sigma^{-1})\left[\murs\cp \uberll{\sigma}{\vec{v}_x}\right]+ s^{-1}\sigma^{-1}\sigma\:\vec{\sigma}_s\\
  &=c\: \Ad(s^{-1})\cp \Ad(\sigma^{-1})\cp \Ad(\sigma)\cp \murs(\vec{v}_x) + s^{-1}\vec{\sigma}_s\\
  &=c\: \Ad(s^{-1})\left[\murs(\vec{v}_x)\right]+s^{-1}\vec{\sigma}_s\\
  &=\w^{c}_{(x,s)}(\vec{v}_x,\vec{\sigma}_s).
\end{align*}
It remains to show that each $\Pe$-invariant connection equals $\w^c$ for some $c\in\mathbb{R}$. For this we compute $\Phi_p^*\w^c$ for $p=(0,e)$ and show that there are no other linear maps $\psi\colon \mathbb{R}^3\times \mathfrak{su}(2)\rightarrow \mathfrak{su}(2)$ that fulfil the two conditions in Wang's theorem \cite{Wang}, see Case \ref{th:wang}. 
For the first step we calculate
\begin{align*}
  \psi^c(\vec{v},\vec{s})&:=\big(\Phi_{(0,e)}^* \w^c\big)(\vec{v},\vec{s})
  =\w^c_p\left(\dt{t}{0}\Phi\left((t\vec{v},\exp(t\vec{s})),(0,e)\right)\right)\\
  &=\w^c_p\left(\dt{t}{0}(t\vec{v},\exp(t\vec{s}))\right)
  =c\:\murs(\vec{v})+\vec{s}.
\end{align*}
For the second one observe that $\pi(p)=0$ and $G_{0}=\{0\} \times \SU$. Then for $j=(0,\sigma)\in G_{0}$ we have $j\cdot p=(0,\sigma) (0,e)=(0, \sigma)=p\cdot \sigma$, hence $\phi_p(j)=\sigma$. 
Consequently, if $\psi\colon \mathbb{R}^3\times \mathfrak{su}(2)\rightarrow \mathfrak{su}(2)$ is a linear map as in Wang's theorem, then,
together with condition \textit{i.)}, this gives $\psi(\vec{s})=\dd_e\phi_p(\vec{s})=\vec{s}$. So, it remains to show that $\psi(\vec{v})=c\:\murs(\vec{v})$ for some $c\in \mathbb{R}$ and each $\vec{v}\in\mathbb{R}^3$. Due to condition \textit{ii.)}, we have
$\psi(\vec{v})=\Ad(\sigma^{-1})\cp \psi\cp\Ad(\sigma)(\vec{v})$ for each $\sigma\in \SU$. Moreover, 
\begin{align*}
  \Ad(\sigma)(\vec{v})&=\dt{t}{0}(0,\sigma)\cdot_\delta(t\vec{v},e)\cdot_\delta(0,\sigma)^{-1} =\dt{t}{0}(\uberll{\sigma}{t\vec{v}},\sigma)\cdot_\delta(0,\sigma^{-1})\\ &=\dt{t}{0}(\uberll{\sigma}{t\vec{v}},\sigma\sigma^{-1})=(\uberll{\sigma}{\vec{v}},0),
\end{align*}
hence $\psi(\vec{v})=\Ad(\sigma)\cp \psi\cp \uberll{\sigma^{-1}}{\vec{v}}$ for all $\vec{v}\in \mathbb{R}^3$ and all $\sigma\in \SU$. 
Then for $\sigma_t:=\exp(t\vec{s})$ with $\vec{s}\in \mathfrak{su}(2)$ it follows from linearity of $\psi$ that
\begin{align*}
  0&=\dt{t}{0}\psi(\vec{v})
  =\dt{t}{0}\Ad(\sigma_t^{-1})\cp \psi(\uberll{\sigma_t}{\vec{v}})\\
  &=\dt{t}{0}\sigma^{-1}_t (\psi\cp \murs^{-1})\left(\sigma_t\:\murs(\vec{v})\:\sigma^{-1}_t\right) \sigma_t\\
  &\stackrel{\text{lin.}}{=}-\vec{s}\: \psi(\vec{v})+ (\psi\cp \murs^{-1})\left(\vec{s}\:\murs(\vec{v})-\murs(\vec{v})\:\vec{s}\right)  +\psi(\vec{v})\:\vec{s}.
\end{align*}
This is $[\vec{s},\psi(\vec{v})]=(\psi\cp \murs^{-1})\left([\vec{s},\murs(\vec{v})]\right)$ for all $\vec{v}\in \mathbb{R}^3$ and all $\vec{s}\in \mathfrak{su}(2)$ so that for $1\leq i,j,k\leq 3$ we obtain 
\begin{align*}
  [\tau_i,\psi(\vec{e}_j)]=(\psi\cp \murs^{-1})([\tau_i,\tau_j])=2\epsilon_{ijk}(\psi\cp \murs^{-1})(\tau_k)=2\epsilon_{ijk}\psi(\vec{e}_k).
\end{align*} 
This forces $\psi(\vec{v})=c\:\murs(\vec{v})$ for some $c\in\mathbb{R}$ and all $\vec{v}\in \mathbb{R}^3$,
hence
\begin{align*}
  \psi(\vec{v},\vec{s})=\psi(\vec{v})+\psi(\vec{s})=c\:\murs(\vec{v})+\vec{s}=\psi^c(\vec{v},\vec{s}). 
\end{align*}

\subsection{Abelian Group Structures and Quasi-Characters}
\label{app:GrStrOnSp} 
\begin{proposition}[{\bf Proof of Proposition \ref{prop:Specgroup}} ]
  \label{prop:SpecgroupA} 
  \begin{enumerate}
  \item
  \label{eq:gg}
  Let $X$ be a set and $\aA\subseteq B(X)$ a unital $C^*$-algebra. Then, the families of quasi-characters are in bijection with the continuous abelian group structures on $\X$. 
  \item
  \label{eq:ggg}
  If $X$ carries an abelian group structure, then a continuous abelian group structure on $\X$ is compatible in the sense that
  \begin{align*} 
  	\iota_X(x)+  \iota_X(y)=\iota_X(x + y)\qquad \forall\:x,y\in X_\aA
  \end{align*}  	
  	iff the respective family $\uU$ consists of characters.  
  \end{enumerate}
  \begin{proof} 
	\begingroup
	\setlength{\leftmargini}{15pt}
	\begin{itemize}
	\item  
    First, assume that $\X=\Spec(\aA)$ carries an abelian group structure continuous w.r.t.\ the Gelfand topology and let $\DG$ denote the dual of $\X$. Since the elements of $\uU_0:=\DG\subseteq C(\X)$ separate the point in $\X$, we have $\ovl{\uU_0}=C(\X)$ by the Stone-Weierstrass theorem. Moreover, the elements of $\uU_0$ are linearly independent because $\int_{\X}\chi\: \dd \muH=0$ iff $\chi\neq 1$. Since the Gelfand transform is an isometric $^*$-isomorphism, we have 
    \begingroup
	\setlength{\leftmarginii}{20pt}
	\begin{itemize}
	\item
		 $\uU:=\GT^{-1}(\uU_0)$ is closed under pointwise multiplication and complex conjugation,
	\item
		$\ovl{\uU}=\aA$,
	\item
		the elements of $\uU$ are linearly independent.
	\end{itemize}
	\endgroup
	\noindent	    
   Moreover, if $f\in \uU$ and $x\in X$, then $f(x)=\iota_X(x)(f)=\GT(f)(\iota_X(x))\in S^1$, so that it remains to show the property \ref{def:quasichar4}) from Definition \ref{def:quasichar}. For this, let $x,y\in X$ and $f\in \uU$. Then we have
    \begin{align}
      \label{eq:netplus}
      f(x)\cdot f(y)&=\GT(f)(\iota_X(x))\cdot\GT(f)(\iota_X(y))=\GT(f)(\iota_X(x) + \iota_X(y))=(\iota_X(x) + \iota_X(y))(f),
    \end{align}
    whereby the third step is due to $\GT(f)\in \Gamma$.
    Since $\iota_X(X)\subseteq \X$ is dense, we find a net $\{x_\alpha\}_{\alpha \in I}$ with $\iota_X(x_\alpha)\rightarrow (\iota_X(x) + \iota_X(y))$ so that
    \begin{align*}
      f(x)\cdot f(y)\stackrel{\eqref{eq:netplus}}{=}\lim_\alpha \iota_X(x_\alpha)(f)=\lim_\alpha f(x_\alpha).
    \end{align*} 
    Moreover, there is a net $\{\iota_X(e_\alpha)\}_{\alpha \in J}\longrightarrow e\in \X$, hence 
	\begin{align*}    
    	\lim_\alpha f(e_\alpha)=\lim_\alpha \GT(f)(\iota_X(e_\alpha))=\GT(f)(e)=1
    \end{align*}
    because $\GT(f)\in \DG$.
    If $X$ is an abelian group with $\iota_X(x+y)=\iota_X(x)+ \iota_X(y)$ for all $x,y\in X$, then 
	\begin{align*}    
    	f(x+ y)=\iota_X(x+y)(f)=\GT(f)(\iota_X(x+y))=\GT(f)(\iota_X(y))+\GT(f)(\iota_X(y))=f(x)\cdot f(y).
    \end{align*}
	 \item
    Let $\uU=\{f_\alpha\}_{\alpha\in I}= \aA$ be a set of quasi-characters. We define the group structure on $\X$ as follows.  
    Let $\psi_1,\psi_2,\psi \in \X$ and $\Gen$ the $^*$-algebra generated by $\uU$. For
    $f:=\textstyle\sum_{i=1}^n\beta_i f_{\alpha_i}\in \Gen$  we define
    \begin{align*}
      (\psi_1+\psi_2)(f):= \textstyle\sum_{i=1}^n \beta_i\:\psi_1(f_{\alpha_i})\psi_2(f_{\alpha_i})\qquad \psi^{-1}(f):=\textstyle\sum_{i=1}^n \beta_i\:\ovl{\psi_1(f_{\alpha_i})}\qquad e(f):=\sum_{i=1}^n\beta_i.
    \end{align*}
    These maps are well defined and linear just because the elements $\{f_\alpha\}_{\alpha\in I}$ are linearly independent. Moreover, they are multiplicative because $f_\alpha f_\beta\in \uU$ if $f_\alpha, f_\beta\in \uU$ and since $\psi_1,\psi_2,\psi$ are multiplicative. It follows from $f_\alpha\ovl{f}_\alpha=1$ that $\psi+\psi^{-1}=e$, and it is straightforward to see that $+$ is associative. 
    
    The crucial part now is to extend $\psi_1+\psi_2$, $\psi^{-1}$ and $e$ to $\aA$. Since these maps are linear, for this it suffices to show their continuities. The algebraic properties of these maps then also hold on $\aA$ just by continuity. 
    By denseness of $\iota_X(X)\subseteq \X$, for each $\epsilon>0$ we find $x_\epsilon,y_\epsilon\in X$ such that for $i=1,\dots,n$ and $C:=2\cdot n\cdot \mathrm{max}(|\beta_1|,\dots,|\beta_n|)$ we have
    \begin{align*}
      |\psi_1(f_{\alpha_i})-\iota_X(x_{\epsilon})(f_{\alpha_i})|\leq \epsilon/C \qquad\text{as well as}\qquad
      |\psi_2(f_{\alpha_i})-\iota_X(y_{\epsilon})(f_{\alpha_i})|\leq \epsilon/C. 
    \end{align*}	
    It follows that $\left|(\psi_1 + \psi_2)\left(\textstyle\sum_{i=1}^n\beta_i\: f_{\alpha_i}\right)\right|\leq \textbf{A}+\underbrace{\left|\textstyle\sum_{i=1}^n\beta_i\: \iota_X(x_\epsilon)(f_{\alpha_i})\iota_X(y_\epsilon)(f_{\alpha_i})\right|}_{\textbf{B}}$ for
    \begin{align*}	
      \textbf{A}&=\left|\textstyle\sum_{i=1}^n\beta_i \big[\psi_1(f_{\alpha_i})\psi_2(f_{\alpha_i})-\iota_X(x_\epsilon)(f_{\alpha_i})\iota_X(y_\epsilon)(f_{\alpha_i})\right]\big|\\
      & \leq\textstyle\sum_{i=1}^n|\beta_i|\:\big[ |\psi_1(f_{\alpha_i})|\: |\psi_2(f_{\alpha_i})-\iota_X(y_\epsilon)(f_{\alpha_i})| + |\iota_X(y_\epsilon)(f_{\alpha_i})|\:|\psi_1(f_{\alpha_i})-\iota_X(x_\epsilon)(f_{\alpha_i})|\big]
      \\ &=  \textstyle\sum_{i=1}^n|\beta_i|\:|\psi_2(f_{\alpha_i})-\iota_X(y_\epsilon)(f_{\alpha_i})|+ \textstyle\sum_{i=1}^n|\beta_i|\: |\psi_1(f_{\alpha_i})-\iota_X(x_\epsilon)(f_{\alpha_i})|\leq \epsilon,
    \end{align*}
    where the second last step is clear from $\im[f_{\alpha_i}]\subseteq S^1$ for $i=1,\dots ,n$.
    For the summand $\textbf{B}$ observe that we find a sequence $\{z_n\}_{n\in\mathbb{N}}\subseteq X$ with
    \begin{align*}
      f_{\alpha_i}(x_\epsilon)f_{\alpha_i}(y_\epsilon)=\lim_k f_{\alpha_i}(z_k)\qquad\forall\:1\leq i\leq k,
    \end{align*}
    so that this summand reads
    \begin{align*}
      \textbf{B}=\left|\textstyle\sum_{i=1}^n\beta_i\: \iota_X(x_\epsilon)(f_{\alpha_i})\iota_X(y_\epsilon)(f_{\alpha_i})\right|&
      =\left|\textstyle\sum_{i=1}^n\beta_i\: \lim_k f_{\alpha_i}(z_k)\right|
      =\lim_k \left|\textstyle\sum_{i=1}^n\beta_i\: f_{\alpha_i}(z_k)\right|\leq  \|f\|_\infty.
    \end{align*}
    This shows $|(\psi_1+\psi_2)(f)|\leq \|f\|_\infty +\epsilon$ for each $\epsilon>0$, hence $|(\psi_1 + \psi_2)(f)|\leq \|f\|_\infty$. Consequently, $\psi_1+\psi_2$ is continuous on $\Gen$.
    For $\psi^{-1}$ we choose $x_\epsilon$ as above, but now for $\ovl{f}_{\alpha_i}$ instead of $f_{\alpha_i}$, and conclude in the same way that $|\psi^{-1}(f)|\leq \|f\|_\infty$ for $f\in \Gen$. For continuity of $e$, we choose 
    a sequence $\{e_k\}_{k\in \mathbb{N}}\subseteq X$ with $\lim_k f_{\alpha_i}(e_k)=1$ for all $1\leq i\leq n$. Then
    \begin{align*}
    \left|e(f)\right|=\left|\textstyle\sum_{i=1}^n\beta_i\right|=\left|\left(\textstyle\sum_{i=1}^n\beta_i\: \lim_k f_{\alpha_i}(e_k)\right)\right|
    = \lim_k \left|\textstyle\sum_{i=1}^n\beta_i\: f_{\alpha_i}(e_k)\right| 
    \leq \left\|f\right\|_\infty
    \end{align*}
    showing continuity of $e$.

    It remains to show continuity of the group structure. For this, keep in mind that a map into $\X$ is continuous iff its composition with each $f_\alpha\in \uU$ is continuous. This is because $\aA$ is generated by these elements. So, let $\{\psi_{1}^{\kappa},\psi_2^\kappa\}_{\kappa\in J}\rightarrow (\psi_1,\psi_2)$ be a converging net. Then $\{\psi_i^\kappa\}_{\kappa\in J}\rightarrow \psi_i$ for $i=1,2$, and for $f_\alpha\in \uU$ fixed we find $\kappa_\epsilon\in J$ such that $\|\psi_i-\psi_i^\kappa\|_{f_\alpha}\leq \frac{\epsilon}{2}$ for $i=1,2$ and  all $\kappa\geq \kappa_0$. Hence, 
    \begin{align*}
      \|\psi_1 + \psi_2 -\psi_1^\kappa + \psi_2^\kappa\|_{f_\alpha}&=|\psi_1(f_\alpha)\psi_2(f_\alpha)-\psi_1^\kappa(f_\alpha)\psi_2^\kappa(f_\alpha)|
      \\ & \leq|\psi_1(f_\alpha)|\:|\psi_2(f_\alpha)-\psi_2^\kappa(f_\alpha)|+|\psi_2^\kappa(f_\alpha)|\:|\psi_1(f_\alpha)-\psi_1^\kappa(f_\alpha)|
      \\ &=  |\psi_2(f_\alpha)-\psi_2^\kappa(f_\alpha)|+|\psi_1(f_\alpha)-\psi_1^\kappa(f_\alpha)|\leq \epsilon.
    \end{align*}
    Continuity of inversion is straightforward now.  
    
    If $X$ is an abelian group such that the elements in $\uU$ are characters, then 
    \begin{align*}
      (\iota_X(x)+  \iota_X(y))(f_\alpha)=\iota_X(x)(f_\alpha)\iota_X(y)(f_\alpha)=f_\alpha(x + y)=\iota_X(x+y)(f_\alpha),
    \end{align*}
    hence $\iota_X(x)+  \iota_X(y)=\iota_X(x+y)$ as $\Gen\subseteq \aA$ is dense.
 	 \item
    We have to show that the established correspondence is one-to-one:
    
     Let $\uU$ be a set of quasi-characters that corresponds to a continuous group structure $(+,^{-1})$ on $\X$. Then, for $(+',^{-1'})$ the continuous abelian group structure induced by $\uU$, and $f_\alpha\in \uU$, we have 
	\begin{align*}   
    	(\psi_1+'\psi_2)(f_\alpha)=\psi_1(f_\alpha)\psi_2(f_\alpha)=\GT(f_\alpha)(\psi_1)\GT(f_\alpha)(\psi_2)=\GT(f_\alpha)(\psi_1+\psi_2)=(\psi_1+\psi_2)(f_\alpha)
	\end{align*}    	
    because $\GT(f_\alpha)\in \DG$ by construction. As above, here $\Gamma$ denotes the dual of $\X$ (w.r.t.\ $(+,^{-1})$). Since $\psi^{-1'}(f_\alpha)=\psi(\ovl{f}_\alpha)=\ovl{\GT(f_\alpha)}(\psi)=\GT(f_\alpha)(\psi^{-1})$, the group structures $(+,^{-1})$ and $(+',^{-1'})$ coincide. Here, the last equality is due to the fact that $\GT(f_\alpha)$ is a character w.r.t.\ $(+,^{-1})$.
    
    Let $\uU'$ be the family of quasi-characters that corresponds to the group structure on $\X$ induced by the family of quasi-characters $\uU$. Then, by definition, $\GT(\uU')=\DG$ is the dual group of $\X$. Obviously, $\GT(\uU)\subseteq \DG$, and since $\ovl{\uU}=\aA$, for $\chi\in \DG\backslash \GT(\uU)$ we find a sequence $\Gen\supseteq \{h_n\}_{n\in \mathbb{N}}\rightarrow \chi$ with $\Gen$ the $^*$-algebra generated by $\GT(\uU)$. Then 
	\begin{align*} 
    	1=\int_{\X}1\:\dd \muH =\lim_n\int_{\X}\ovl{\chi}\cdot h_n\:\dd \muH=\lim_n 0=0
	\end{align*}    	
     shows $\GT(\uU)=\Gamma=\GT(\uU')$, hence $\uU=\uU'$.
    \end{itemize}
    \endgroup
  \end{proof}
\end{proposition}

\section{Appendix to Spectral Extensions of Group Actions}

\subsection{Invariant Generalized Connections}
\begin{lemma}
  \label{lemma:separatingA}
  The map $\iota_\Con\colon \Con \rightarrow \A$ is injective if for each $\vec{v}\in TM$ there is $\gamma\in \Pa$ and $s\in\dom[\gamma]=[a,b]$, such that $\dot\gamma(s)=\vec{v}$ and $\gamma|_{[a,t]}\in \Pa$ for all $t\in (a,s]$ or $\gamma|_{[t,b]}\in \Pa$ for all $t\in [s,b)$. 
  \begin{proof}
    By Lemma \ref{lemma:dicht}.\ref{lemma:dicht1} it suffices to show that the respective cylindrical functions separate the points in $\Con$. Since $\rho$ is injective, it suffices to show that a connection $\w$ is uniquely determined by the values $\{h_\gamma^\nu(\w)\}_{\gamma\in \Pa}$ for $\nu=\{\nu_y\}_{y\in M}\subseteq P$ fixed with $\nu_y\in F_y$ for all $y\in M$. Moreover, 
    since $\w$ is a connection, it even suffices to show that $\w|_{T_{\nu_y}P}$ is uniquely determined by this family for all $y\in M$. Now, $\w|_{Tv_{\nu_y}P}$ is determined by the connection property of $\w$. Thus,  
    we only have to show that for $x\in M$ fixed and $p:=\nu_x$, there is an algebraic complement $H_{p}$ of $Tv_{p}P$ in $T_{p}P$ such that $\omega|_{H_{p}}$ is determined by the values $\{h^\nu_\gamma(\w)\}_{\gamma\in \Pa}$, or, due to \eqref{eq:indep}, by any other family $\{h^{\nu'}_\gamma(\w)\}_{\gamma\in \Pa}$ with $\nu'=\{\nu'_y\}_{y\in M}\subseteq P$ another collection with $\nu'_y\in F_y$ for all $y\in M$. For this, let $(U_0,\phi_0)$ be a bundle chart of $P$ with $\phi_0(p)=(x,e)$, and define (observe that $\nu'_x=\nu_x=p$)
    \begin{align*}
      \nu'_y:=\left\{
	\begin{array}{ll}
          \phi_0^{-1}(y,e)  & \mbox{if } y \in U_0\\
          \nu_{y} & \mbox{if } y \notin U_0.
        \end{array}\right.
    \end{align*}
    Then, $H_p:=\dd_{x}\phi_0^{-1}(T_{x}U_0)$ is an algebraic complement of $Tv_{p}P$ in $T_{p}P$ for which we have to show that $\w|_{H_p}$ is determined by the values $\{h^{\nu'}_\gamma(\w)\}_{\gamma\in \Pa}$. 
    
    For this, fix $\vec{v}\in H_p$ and choose $\gamma \in \Pa$ with $\dot\gamma(s)=\dd_{p}\phi_0(\vec{v})$. Moreover, let $a=\tau_0<{\dots}<\tau_k=b$ be a decomposition of $\gamma$ such that $\gamma_i:=\gamma|_{[\tau_i,\tau_{i+1}]}$ is of class $C^k$.
    \begingroup
\setlength{\leftmargini}{14pt}
\begin{itemize}
\item
    First, assume that $s\in (a,b]$ with $\gamma|_{[a,t]}\in \Pa$ for all $t\in (a,s]$. We find $0\leq i\leq k-1$ with $s\in (\tau_i,\tau_{i+1}]$ and denote by  
   $\delta_q^\w$ the horizontal lift of $\gamma_i$ w.r.t.\ $\w$ in $q\in F_{\gamma(\tau_i)}$. Then, for each $t\in (\tau_i,\tau_{i+1}]$ 
    the restriction $\delta_q^\w|_{[\tau_i,t]}$ is the horizontal lift of $\gamma_i|_{[\tau_i,t]}$ w.r.t\ $\w$ in $q$. It follows from horizontality of $\delta^\w_q$ and the connection properties of $\w$ that 
    \begin{align*}
    %\label{eq:abbbabab}
      \dttB{t}{s}[\pr_2\cp \phi_0]\left(\delta^\w_{q}(t)\right)&=-\dd R_{[\pr_2\cp \phi_0]\left(\delta^\w_q(s)\right)}\cp \w\left(\dt{t}{s}\phi_0^{-1}(\gamma_i(t),e)\right).
    \end{align*}
    Consequently, for $\gamma_t:=\gamma|_{[a,t]}$ and $q:=\parall{\gamma_{i-1}}{\w}\cp\dots\cp\parall{\gamma_{0}}{\w}\big(\nu'_{\gamma(a)}\big)$ we have
    \begin{align*}
      \dt{t}{s}h_{\gamma_t}^{\nu'}(\w)&=\dt{t}{s}[\pr_2\cp \phi_0]\big(\delta^\w_{q}(t)\big)\\
      &=-\dd R_{[\pr_2\cp \phi_0]\left(\delta^\w_{q}(s)\right)}\cp \w\left(\dt{t}{s}\phi_0^{-1}(\gamma_i(t),e)\right)\\
      &=-\dd R_{h^{\nu'}_{\gamma_s}(\w)}\cp \w(\vec{v}),
    \end{align*}
    so that $\omega(\vec{v})=-\frac{\dd}{\dd t}\big|_{t=s}h_{\gamma_t}^{\nu'}(\w)\cdot h^{\nu'}_{\gamma_s}(\w)^{-1}$ is completely determined by the family $\{h^{\nu'}_\gamma(\w)\}_{\gamma\in \Pa}$. 
    \item
    Second, if $s\in [a,b)$ with $\gamma|_{[t,b]}\in \Pa$ for all $t\in [s,b)$, by the first part we have
    \begin{align*}
    	\textstyle\omega(-\vec{v})=-\frac{\dd}{\dd t}\big|_{t=u}h_{\ovl{\gamma}_t}^{\nu'}(\w)\cdot h^{\nu'}_{\ovl{\gamma}_{u}}(\w)^{-1}
    \end{align*}
    for $\ovl{\gamma}$ the inverse of $\gamma$ and $u:=b+a-s$ so that $\ovl{\gamma}(u)=\gamma(s)$ and 
    $\dot{\ovl{\gamma}}(u)=-\dot\gamma(s)$. Thus, the claim is clear because the occurring parallel transports are determined by that along the curves $\gamma|_{[t,b]}$ for $t\geq s$. This is clear from   the homomorphism properties of parallel transports and that, up to parametrization, $\ovl{\gamma}_t$ is the inverse of some curve $\gamma|_{[t',b]}$ for some $t'>s$. 
    \end{itemize}
    \endgroup
  \end{proof} 
\end{lemma}

\begin{lemma} 
  \label{lemma:CylSpecActionvorbA}
  If $\Pa$ is $\Phi$-invariant, then $\PaC$ is $\cw$-invariant.
  \begin{proof}
    Let $\Pa\ni \gamma\colon [a,b]\rightarrow M$, $\w\in \Con$ and $g\in G$. Moreover, denote by $\gamma_p^\w$ the horizontal lift of $\gamma$ w.r.t.\ $\w$ in $p\in F_{\gamma(a)}$ and define
    \begin{align*}
      \vg:=\wm_g\cp \gamma,\qquad\vg_p^\w:= \Phi_g\cp \gamma_p^\w,\qquad \vw:=\cw(g,\w), \qquad \vp:=\Phi(g,p).
    \end{align*}
    Then $\vg_p^\w(a)=\vp$, $\pi\cp \vg_p^\w(t)=\pi \cp \Phi\big(g,\gamma_p^\w(t)\big)=\wm(g,\gamma(t))=\vg(t)$ and
    \begin{align*}
      \vw_{\vg_p^\w(t)}\big({\dot{\vg}}{}^\w_p(t)\big)=\big(\Phi_{g^{-1}}^*\w\big)_{\vg_p^\w(t)}\big({\dot{\vg}}{}^\w_p(t)\big)=\w_{\gamma_p^\w(t)}\big(\dot{\gamma}_p^\w(t)\big)=0.
    \end{align*}
    This shows that $\vg_p^\w(t)$ is the horizontal lift of $\vg$ w.r.t.\ $\vw$ in $\vp$, hence
    \begin{align*}
      \parall{\vg}{\vw}(\vp)=\vg_p^\w(b)=\Phi_g\big(\parall{\gamma}{\w}(p)\big).
    \end{align*}
    Then, if we substitute $p$ by $p'=\Phi_{g^{-1}}(p)$ and $\gamma$ by $\gamma':=\wm_{g^{-1}}\cp \gamma$, this gives
    \begin{align}
      \label{eq:patra} 
      \parall{\gamma}{\vw}(p)=\Phi_g\big(\parall{\gamma'}{\w}(p')\big).
    \end{align}
    Now, let $\nu=\{\nu_x\}_{x\in M}\subseteq P$ with $\nu_x\in F_x$ for all $x\in M$ and define $\psi_x(p):=\diff(\nu_x,p)$. %as in Definition \ref{def:Connectionss}. 
    Then for $p:=\nu_{\gamma(a)}$ and $\gamma'$ as above we obtain
    from the morphism properties of the maps $\parall{\gamma'}{\w}$, $\Phi_g$, $\psi_x$ that
    \begin{align}
      \label{eq:trafogenerators}
      \begin{split}
        \big(\cw_g^*h_\gamma^\nu\big)(\w)&=h_\gamma^\nu(\cw(g,\w))=\big[\psi_{\gamma(b)}\cp \parall{\gamma}{\vw}\big](\nu_{\gamma(a)})\\
        &=\big[\psi_{\gamma(b)}\cp \Phi_g \cp\parall{\gamma'}{\w}\big]\left(\Phi_{g^{-1}}(\nu_{\gamma(a)})\right)\\
        &=\big[\psi_{\gamma(b)}\cp \Phi_g\cp \parall{\gamma'}{\w}\big](\nu_{\gamma'(a)})\cdot\psi_{\gamma'(a)}\big(\Phi_{g^{-1}}(\nu_{\gamma(a)})\big)\\
        &=\psi_{\gamma(b)}\left[\Phi_g(\nu_{\gamma'(b)})\cdot\psi_{\gamma'(b)}\left(\parall{\gamma'}{\w}\left(\nu_{\gamma'(a)}\right)\right)\right]\cdot
        \psi_{\gamma'(a)}\big(\Phi_{g^{-1}}(\nu_{\gamma(a)})\big)\\
        &=\psi_{\gamma(b)}\left(\Phi_g(\nu_{\gamma'(b)})\right)\cdot\psi_{\gamma'(b)}\left(\parall{\gamma'}{\w}\left(\nu_{\gamma'(a)}\right)\right)\cdot\psi_{\gamma'(a)}\big(\Phi_{g^{-1}}(\nu_{\gamma(a)})\big)\\
        &=\underbrace{\psi_{\gamma(b)}\left(\Phi_g(\nu_{\gamma'(b)})\right)}_{\delta_1}\cdot\: h^\nu_{\gamma'}(\w)\cdot\underbrace{\psi_{\gamma'(a)}\big(\Phi_{g^{-1}}(\nu_{\gamma(a)})\big)}_{\delta_2}.
      \end{split}
    \end{align}
    Consequently, for the generators $f \cp h_\gamma^\nu$ of $\PaC$ with $f\in C(S)$ we have
    \begin{align*}
      \cw^*_g\left(f\cp h_\gamma^\nu\right)=(f\cp L_{\delta_1}\cp R_{\delta_2}) \cp h_{\gamma'}^\nu \in \PaC, %=\sum_{p,q} \rho_{ip}(\delta_1)\rho_{qj}(\delta_2)\: \rho_{pq}\cp h_{\gamma'}^\nu \in \PaC
    \end{align*}
    hence $\cw_g^*(\cC)\subseteq \PaC$ for $\cC\subseteq  \PaC$ the dense unital $^*$-subalgebra generated by the functions $f\cp h_\gamma^\nu$. For this, observe that $\cw^*_g1=1$ for all $g\in G$. Finally, for $f\in \PaC$ let $\cC\supseteq \{f_n\}_{n\in \mathbb{N}}\rightarrow f$ be a converging sequence. Then $\cw_g^*f_n\rightarrow \cw_g^*f$, since $\cw_g^*$ is an isometry, hence $\cw_g^*f\in \PaC$.
  \end{proof}
\end{lemma}

\section{Appendix to Modification of Invariant Homomorphisms}

\subsection{Analytic and Lie Algebra Generated Curves}
\label{subsec:AnalytCurvesA}
 
\begin{lemma}
  \label{lemma:analytCurvesIndepetc}
  \begin{enumerate}
  \item
    \label{lemma:analytCurvesIndepetc1}
	 Let $\gamma_i\colon (a_i,b_i)\rightarrow M$ be analytic embeddings and $x$ an accumulation point of $\im[\gamma_1]\cap \im[\gamma_2]$. Then $\gamma_1(I_1)=\gamma_2(I_2)$ for open intervals $I_i\subseteq (a_i,b_i)$ with $x\in  \gamma_i(I_i)$ for $i=1,2$.     
  \item
    \label{lemma:analytCurvesIndepetc3}
   If $\dim[S]\geq 1$, then 
		$\:\gamma_1\csim \gamma_2\:\:\: \Longleftrightarrow\:\:\: \gamma_1\isim \gamma_2 \:\:\: \Longleftrightarrow\:\:\: \gamma_1 \psim \gamma_2$.
  \item
    \label{lemma:analytCurvesIndepetc5}
    Let $S$ be connected. If $\Pas\subseteq \Paw$ is closed under decomposition and inversion, then $\Pas$ is independent.
  \end{enumerate}
  \begin{proof} 
    \begin{enumerate}
    \item
    See Lemma \ref{lemma:BasicAnalyt}.\ref{lemma:BasicAnalyt1}.
    \item
    See Lemma \ref{lemma:BasicAnalyt}.\ref{lemma:BasicAnalyt2}.
    \item
      It follows from Proposition A.1 in \cite{ParallTranspInWebs} that a set $\{\delta_1,\dots,\delta_n\}\subseteq \Paw$ is independent if $\im[\delta_i]\cap \im[\delta_j]$ is finite for all $1\leq i\neq j\leq n$. Consequently, it suffices to show that each collection of the form $\{\gamma , \sigma_1,\dots,\sigma_l\} \subseteq \Pas$, where  $\im[\sigma_i]\cap \im[\sigma_j]$ is finite for all $1\leq i\neq j\leq l$, admits a refinement $\{\delta_1,\dots,\delta_n\}\subseteq\Pas$, such that $\im[\delta_i]\cap \im[\delta_j]$ is finite for all $1\leq i\neq j\leq n$ as well. 

      In fact, if  $\{\gamma_1,\dots,\gamma_d\}$ is given, we can apply this to $\gamma:=\gamma_1$ and $\sigma_1=\gamma_2$ in order to obtain a refinement of $\{\gamma_1,\gamma_2\}$ with the above intersection property. We now argue by induction.  For this, let $1<l< d$ and 
      $\{\delta_1,\dots,\delta_n\}$ be a refinement of $\{\gamma_1,\dots, \gamma_l\}$ such that $\im[\delta_i]\cap\im[\delta_j]$ is finite for all $1\leq i\neq j\leq n$. By assumption, for $\gamma:=\gamma_{l+1}$ and $\sigma_i:=\delta_i$ for $i=1,\dots,n$ there is a refinement $\{\delta'_1,\dots ,\delta'_{n'}\}$ of $\{\gamma_{l+1},\delta_1,\dots \delta_n\}$ such that $\im[\delta'_i]\cap\im[\delta'_j]$ is finite for all $1\leq i \neq j\leq n'$. Using \ref{lemma:analytCurvesIndepetc3}), 
      it is clear that $\{\delta'_1,\dots \delta'_{n'}\}$ is a refinement of $\{\gamma_{l+1},\gamma_1,\dots,\gamma_{l}\}$ as well. 
      
      Now, if $\{\gamma, \sigma_1,\dots,\sigma_l\} \subseteq \Pas$ is as above, for each pair $(\gamma,\sigma_q)$ we denote by          
      $\{L_q^{p}\}_{1\leq p\leq m_q}$ and $\{J^p_q\}_{1\leq p\leq m_q}$ the corresponding        
      overlapping intervals from Lemma \ref{lemma:BasicAnalyt}.\ref{Basanalyt} in $\dom[\gamma]=[a,b]$ and $\dom[\sigma_q]$ for $q=1,\dots,l$, respectively. We find $a=\tau_0< {\dots} <\tau_n=b$ such that each $L_q^p$ occurs exactly once as an interval $[\tau_{i(q,p)},\tau_{i(q,p)+1}]$. For this, observe that 
	\begin{align*}      
      L_{q}^p\cap L_{q'}^{p'}=\emptyset\:\:\:\text{ if }\:\:q=q'\:\:\:\text{ and }\:\:\:p\neq p', 
	\end{align*}      
      and only can contain start and end points if $q\neq q'$. This is because $\im[\sigma_i]\cap \im[\sigma_j]$ was assumed to be finite for $1\leq i\neq j\leq l$. Then, $\gamma|_{L_{q}^p}\isim \big[\sigma_q|_{J_{q}^p}\big]^{\pm 1}$ so that for 
      \begin{align*}      
        I:=\{0\leq i\leq n-1\:|\: i\neq i(q,p) \text{ for all }1\leq q\leq l,\: 1\leq p\leq m_q\}
      \end{align*}      
      the families $\{\gamma|_{[\tau_i,\tau_{i+1}]}\}_{i \in I}$ and $\{\sigma_q|_{J_q^p}\}_{1\leq p\leq m_q}$ for $1\leq q\leq l$ provide a refinement in $\Pas$ of $\{\gamma, \sigma_1,\dots,\sigma_l\}$ which has  the desired intersection properties.
    \end{enumerate}
  \end{proof} 
\end{lemma} 

\subsection{Inclusion Relations}
\label{app:incl} 

{\bf Proof of Proposition \ref{prop:incl}.\ref{prop:incl2}:}
\begin{enumerate}
\item[\ref{prop:incl2})]
  By assumption, we find an $\Add{G_x}$-invariant linear subspace $V\neq \{0\}$ of $\mg$ with $V\oplus \mathfrak{g}_x\subsetneq \mg$, and a non-trivial $\Add{G_x}^p$-equivariant linear map $L\colon V \rightarrow \ms$. We choose $\g_3\in V\backslash \ker[L]$, $\vec{g}_1\in \mathfrak{g}\backslash\left[ V\oplus \mathfrak{g}_x\right]$ and define $\g_2:=\g_3+\g_1$. Then, $\g_2\in \mg\backslash\mg_x$ by definition, and we have $\g_3\nsim_x \g_1,\g_2$. In fact, we even have 
\begin{align}
	\label{eq:inped}
	  \g'\nsim_x \g_1,\g_2\qquad \forall\: g'\in V
\end{align}  
 since elsewise it would follow from Lemma \ref{lemma:sim}.\ref{lemma:sim4} that      
  \begin{align*}  
  \lambda \g_1 + \Add{h}(\g')\in \mg_x\qquad \text{or}\qquad \lambda (\g_3+\g_1) + \Add{h}(\g')\in \mg_x 
	\end{align*}  
   for some $\lambda\neq 0$ and $h\in G_x$. Then, by $\Add{G_x}$-invariance of $V$, in both cases we would have that $\g_1\in V\oplus \mg_x$, which contradicts the choice of $\g_1$. 
   
   Now, for $\mu\in \RR$ let $\homm_\mu$ denote the map \eqref{eq:homdef} from Proposition \ref{th:invhomm}.\ref{th:invhomm2}  that corresponds to the linear map $\mu L$ and arises by modifying some $\homm'$ along $\g_3$. %\ref{th:invhomm}.\ref{th:invhomm2}.   
  Recall the open subsets \eqref{eq:opensets}, and define $\gamma_i:=\gamma_{\g_i}^x|_{[0,l]}$ for $i=1,2$ as well as $\gamma_3:=\gamma_{\g_3}^x|_{[0,l]}$ for some fixed $l>0$ with $l<\tau_{\g},\tau_{\g_1},\tau_{\g_2}$. 
  
  We are going to show that there is $\mu\in \RR$ and $U\subseteq S$ open such that
	\begin{align}
	\label{eq:dfdffdf}
		\kappa_{\mg'}(\iota_\Con(\w))\in U^{p,p}_{\gamma_1,\gamma_2}(\homm_\mu)\qquad\Longrightarrow \qquad \kappa_{\mg'}(\iota_\Con(\w))\notin U^{p}_{\gamma_3}(\homm_\mu),
	\end{align}	  
  	hence $\kappa_{\mg'}(\iota_\Con(\AR))\cap U^{p,p,p}_{\gamma_1,\gamma_2,\gamma_3}(\homm_\mu)=\emptyset$. From this, it follows that 
	\begin{align*}
		\homm_\mu \notin \kappa_{\mg'}\big(\ARQ\big)=\kappa_{\mg'}\Big(\ovl{\iota_\Con(\AR)}\Big),
	\end{align*}	  	
  	 hence $\AQR\ni\kappa_{\mg'}^{-1}(\homm_\mu)\notin \ARQ$, which shows the claim. Observe that the neighbourhood $U^{p,p}_{\gamma_1,\gamma_2}(\homm_\mu)$ on the left hand side of \eqref{eq:dfdffdf} does not depend on $\mu\in \RR$ just because the values $\epsilon_\mu(\gamma_i)$ for $i=1,2$ do not so.
	
	  \vspace{4pt}
	{\bf Step 1:} 
	
		  \vspace{-2pt}
  Let $U$ be a neighbourhood of $e$ in $S$ and $a_i\in S$ for $i=1,2$ the unique element with $\homm'(\gamma_i)(p)=\Phi_p(\exp(l\g_i))\cdot a_i$.  
   Then, for $\w\in \AR$ we have
	\begin{align}
	\label{eq:rfrtrt}
		\kappa_{\mg'}(\iota_\Con(\w))\in U^{p,p}_{\gamma_1,\gamma_2}(\homm_\mu)\qquad \Longrightarrow\qquad  \exp(-\I\he l \w(\wt{g_i}(p)))\in a_i\cdot U\:\text{ for }\:i=1,2.
	\end{align}	  

  In fact, since $\homm_{\mu}(\gamma_i)=\homm'(\gamma_i)$ for $i=1,2$, we have
  \begin{align*}
  	\kappa_{\mg'}(\iota_\Con(\w))\in U^{p,p}_{\gamma_1,\gamma_2}(\homm_\mu)\qquad \Longleftrightarrow\qquad \kappa_{\mg'}(\iota_\Con(\w))\in U^{p,p}_{\gamma_1,\gamma_2}(\homm') 
  \end{align*} 
  Consequently, $\kappa_{\mg'}(\iota_\Con(\w))\in U^{p,p}_{\gamma_1,\gamma_2}(\homm_\mu)$ implies %$\iota(\w)\in U^{p,p}_{\gamma_1,\gamma_2}(\homm')$, which means
  \begin{align*}
    \Phi_p(\exp(l \g_i))\cdot \exp(-\I\he l\w(\wt{g}_i(p)))\stackrel{\eqref{eq:trivpar}}{=}\kappa_{\mg'}(\iota_\Con(\w))(\gamma_i)(p)\in  \homm'(\gamma_i)(p)\cdot U\qquad \text{for }i=1,2.
  \end{align*}
  This is equivalent to $\exp(-\I\he l \w(\wt{g_i}(p)))\in a_i\cdot U$ for $i=1,2$.  

  \vspace{4pt}
  {\bf Step 2:}
  
  		\vspace{-2pt}
  Let  $U:=\exp(\hspace{1pt}\I (-\epsilon,\epsilon))$ for $\epsilon < \frac{\pi}{4}$, and $t_i \in [0,2\pi)$ such that $a_i=\exp(\hspace{1pt}\I\hspace{1pt} t_i)$ for $i=1,2$. Then, from \eqref{eq:rfrtrt} and $\g_3=\g_1-\g_2$, we see that $\kappa_{\mg'}(\iota_\Con(\w))\in U^{p,p}_{\gamma_1,\gamma_2}(\homm_\mu)$ implies  
  \begin{align*}
    l\w(\wt{g}(p))=-l\w(\wt{g}_1(p))- [-l\w(\wt{g}_2(p))]\in \bigsqcup_{n\in \mathbb{Z}} 2\pi n + [t_1-t_2] + (-2\epsilon, 2\epsilon),
  \end{align*}
  just because $\exp(-\I l\w(\wt{g}_1(p))- [-\I\he l\w(\wt{g}_2(p))])\in \exp( \I\he2\pi n + \I\he[t_1-t_2])\cdot\exp(\I\he(-2\epsilon, 2\epsilon))$.
  Now, 
\begin{align*}  
  \kappa_{\mg'}(\iota_\Con(\w))(\gamma_3)(p)=\Phi_p(\exp(l\g))\cdot \exp(-\I\he l\w(\wt{g}(p)))
  \end{align*}
    by \eqref{eq:trivpar}, hence
  \begin{align*}      
    \kappa_{\mg'}(\iota_\Con(\w))(\gamma_3)(p)\in \Phi_p(\exp(l\g))\cdot\exp(\I [t_2-t_1]) \cdot \exp(\I (-2\epsilon, 2\epsilon)). 
  \end{align*}       
  Since $L(\g)\in \ms=\RR$, for $\mu=\frac{t_2-t_1 + \pi}{l L(\g)}$ we have just by definition of $\homm_\mu$ that
  \begin{align*} 
    \homm_{\mu}(\gamma_3)(p)\cdot U\stackrel{\eqref{eq:defmodh}}{=}\Phi_p(\exp(l\g))\cdot \exp(\I [t_2-t_1+\pi]) \cdot \exp(\I (-\epsilon, \epsilon)),
  \end{align*}
  hence $\kappa_{\mg'}(\iota_\Con(\w))\notin U_{\gamma_3}^p(\homm_\mu)$ 
  by the choice of $\epsilon$. 
\hspace*{\fill}{\scriptsize$\blacksquare$}
\end{enumerate}

\section{Appendix to Lie Algebra Generated Configuration Spaces}
\label{sec:ConSpAA}
\subsection{Lemma and Definition \ref{def:topo}}
\label{app:Bohrmodl}
{\bf Well-definedness of \eqref{eq:pip}:} 
  We have to show that $\pip_p(\homm)\in \Eq_p$ for each $\homm \in \IHOMLAS$. For this, let $\g\in\mg\backslash\mg_x$ be fixed.
  We first show that \eqref{eq:Equi0}, \eqref{eq:Equi} hold for such $\lambda\neq 0$, $l>0$ with $|\lambda|l<\tau_\g$ and such $l,l'>0$ with $l,l',l+l'<\tau_\g$ as well as equivariance for all $0<l<\tau_\g$. Then, an easy refinement argument shows that the value 
	\begin{align*}  
  	\pi_p(\g,l,\homm):=\prod_{i=1}^k\pi_p(\g,l_i,\homm)\quad \text{for}\quad 0<l_1,\dots,l_k<\tau \quad\text{with}\quad l_1+\dots+l_k=l
	\end{align*}  
  does not depend on the decomposition of $l$. From this, then it is immediate that \eqref{eq:Equi0}, \eqref{eq:Equi} as well as equivariance also hold in the general case. For instance, if $\lambda \neq 0$, we have 
	\begin{align*}  
  	\pip_p(\homm)(\lambda\cdot\g,l_1+\dots+l_k)&=\pi_p(\homm)(\lambda\cdot \g,l_k)\cdot \cdot \pip_p(\homm)(\lambda\cdot \g,l_1)\\
  	&=\pip_p(\homm)(\g,|\lambda|l_k)^{\sign(\lambda)}\cdot {\dots}\cdot \pip_p(\homm)(\g,|\lambda|l_1)^{\sign(\lambda)}\\
  	&= \pip_p(\homm)(\g,|\lambda|[l_1+\dots+l_k])^{\sign(\lambda)}.
  \end{align*}
  Now, let $g:=\exp(l\g)$, $g':=\exp(l'\g)$ and $h:=\exp([l+l']\cdot\g)$. Then we have
  {\allowdisplaybreaks
    \begin{align*}
      \pip_p(\homm)(\g,l+l')&=\Delta\left(\Phi_{\exp((l+l')\g)}(p),\homm\big(\gamma_\g^{x}|_{[0,l+l']}\big)(p)\right)\\ 
      &=\Delta\left(\Phi_h(p),\homm\big(\gag|_{[l',l+l']}\big)\left(\homm\left(\gag|_{[0,l']}\right)(p) \right)\right)\\  
      &=\Delta\left(\Phi_h(p),\homm\big(\wm_{\exp(l'\g)}\cp\gag|_{[0,l]}\big)\left(\homm\left(\gag|_{[0,l']}\right)(p) \right)\right)\\   
      &=\Delta\left(\Phi_h(p),\Phi_{g'}\cp\homm\big(\gag|_{[0,l]}\big)\left(\Phi_{{g'}^{-1}}\cp\homm\left(\gag|_{[0,l']}\right)(p) \right)\right)\\   
      &=\Delta\big(\Phi_{g}(p),\homm\big(\gag|_{[0,l]}\big)\big(p\cdot \Delta\big(p,\Phi_{{g'}^{-1}}\cp\homm\left(\gag|_{[0,l']}\right)(p)\big)\big)\big)\\
      &=\Delta\big(\Phi_{g}(p),\homm\big(\gag|_{[0,l]}\big)\big(p\big)\big)\cdot \Delta\big(p,\Phi_{{g'}^{-1}}\cp\homm\left(\gag|_{[0,l']}\right)(p)\big)\\
      &=\pip_p(\homm)(\g,l)\cdot \Delta\big(\Phi_{g'}(p),\homm\left(\gag|_{[0,l']}\right)(p)\big)\\   
      &=\pip_p(\homm)(\g,l)\cdot\pip_p(\homm)(\g,l')                 
    \end{align*}}For equivariance, let $\g':=\Add{h}(\g)$. Then $x':=\wm_{h^{-1}}(x)=x$, so that $\Delta(p\cdot s,p'\cdot s)=\alpha_{s^{-1}}(\Delta(p,p'))$ shows
  \begin{align*}
    \pip_p(\homm)\left(\Add{h}(\g),l\right)
    &=\Delta\left(\Phi_{h\exp(l\g)h^{-1}}(p),\homm\left(\wm_h\cp\gamma_{\g}^{x'}|_{[0,l]}\right)(p)\right)\\
    &=\Delta\Big(\Phi_{h\exp(l\g)}\left(p\cdot \fiba_p\left(h^{-1}\right)\right),\Phi_h\cp\homm\left(\gamma_{\g}^{x}|_{[0,l]}\Big)(\Phi_{h^{-1}}(p))\right)\\
    &=\alpha_{\fiba_p(h)}\cp\Delta\Big(\Phi_h\cp \Phi_{\exp(l\g)}\left(p\right),\Phi_h\cp\homm\left(\gamma_{\g}^{x}|_{[0,l]}\Big)\left(p\right)\right)\\
    &=\alpha_{\fiba_p(h)}\cp\Delta\Big(\Phi_{\exp(l\g)}\left(p\right),\homm\left(\gamma_{\g}^{x}|_{[0,l]}\Big)\left(p\right)\right)\\
    &=\alpha_{\fiba_p(h)}\cp\pip_p(\homm)(\g,l).
  \end{align*}
  Finally, if $\lambda>0$, then $\gamma^x_{\lambda\g}|_{[0,l]}\csim\gamma^x_{\g}|_{[0,\lambda l]}$, and we obtain
  \begin{align*}
    \pip_p(\homm)(\lambda\cdot \g,l)
    &=\Delta\left(\Phi_{\exp(\lambda l\g)}(p),\homm\big(\gamma_{\lambda\g}^{x}|_{[0,l]}\big)(p)\right)\\
    &=\Delta\left(\Phi_{\exp(\lambda l\g)}(p),\homm\big(\gamma_{\g}^{x}|_{[0,\lambda l]}\big)(p)\right)=\pip_p(\homm)( \g,\lambda l).
  \end{align*}
  Moreover, since $\wm_{\exp(l\g)}\cp \gamma^x_{-\g}|_{[0,l]}\csim \big[\gamma^x_{\g}|_{[0, l]}\big]^{-1}$, we have
  \begin{align*} 
    \pip_p(\homm)(- \g,l)
    &=\Delta\left(\Phi_{\exp(-l\g)}(p),\homm\big(\gamma_{-\g}^{x}|_{[0,l]}\big)(p)\right)\\
    &=\Delta\left(\Phi_{\exp(-l\g)}(p),\homm\Big(\wm_{\exp(-l\g)}\cp\big[\gamma_{\g}^{x}|_{[0,l]}\big]^{-1}\Big)(p)\right)\\
    &=\Delta\left(\Phi_{\exp(-l\g)}(p),\Phi_{\exp(-l\g)}\cp\homm\Big(\big[\gamma_{\g}^{x}|_{[0,l]}\big]^{-1}\Big)(\Phi_{\exp(l\g)}(p))\right)\\
    &=\Delta\left(p,\homm\Big(\big[\gamma_{\g}^{x}|_{[0,l]}\big]^{-1}\Big)(\Phi_{\exp(l\g)}(p))\right).
  \end{align*}
  Then,
  \begin{align*}
    \homm\Big(\big[\gamma_{\g}^{x}|_{[0,l]}\big]^{-1}\Big)\left(\Phi_{\exp(l\g)}(p)\right)\cdot \pi_p(\g,l,\homm)&=\homm\Big(\big[\gamma_{\g}^{x}|_{[0,l]}\big]^{-1}\Big)\left(\Phi_{\exp(l\g)}(p)\cdot \pi_p(\g,l,\homm)\right)\\
    &=\homm\Big(\big[\gamma_{\g}^{x}|_{[0,l]}\big]^{-1}\Big)\left(\homm\big(\gamma_{\g}^{x}|_{[0,l]}\big)(p)\right)=p
  \end{align*}
  shows $\pip_p(\homm)(- \g,l)\cdot \pip_p(\homm)(\g,l)=\me$, hence $\pip_p(\homm)(- \g,l)=\pip_p(\homm)(\g,l)^{-1}$.
  \vspace{8pt}  

  \noindent
  For the continuity statement let $\IHOMLA\supseteq \{\homm_\iota\}_{\iota\in I}\rightarrow \homm\in \IHOMLA$ be a converging net, $U\subseteq \SU$ open, $\g_1,\dots,\g_k \in \mg\backslash\mg_x$ and $l_1,\dots,l_k>0$. Let $\gamma_i:=\gamma^x_{\g_i}\big|_{[0,l_i]}$ for $i=1,\dots,k$.
  By Lemma \ref{lemma:homeotopo}.\ref{it:ddd} we find $\iota_0\in I$ such that for all $\iota\geq \iota_0$ we have
  \begin{align*} 
    \homm_\iota  \in\: & U^{p,\dots,p}_{\gamma_1,\dots,\gamma_k}(\homm)\\
    &\hspace{5pt} = \left\{\homm'\in \IHOMLA\:|\: \homm'(\gamma_i)(p)\in  \homm(\gamma_i)(p)\cdot U\quad \text{ for } i=1,\dots,k \right\}\\
    &\hspace{5pt}=\left\{\homm'\in \IHOMLA\:|\: \Delta\left(\homm(\gamma_i)(p), \homm'(\gamma_i)(p)\right) \in U \quad \text{ for } i=1,\dots,k \right\}\\
    &\hspace{5pt}=\left\{\homm'\in \IHOMLA\:|\: \pip_p(\homm')(\g_i,l_i) \in \pip_p(\homm)(\g_i,l_i)\cdot U\quad \text{ for } i=1,\dots,k  \right\}\\
    &\hspace{5pt}=\left\{\homm'\in \IHOMLA\:|\: \pip_p(\homm')\in U_{\g_1,l_1}(\pip_p(\homm))\cap\dots \cap U_{\g_k,l_k}(\pip_p(\homm))\right\},
  \end{align*}
  where the third step is due to
  \begin{align*}
    \Delta\big(\homm(\gamma_i)(p), \homm'(\gamma_i)(p)\big)&=\Delta\big(\homm(\gamma_i)(p), \Phi_{\exp(l\g)}(p)\big)\cdot \Delta\big(\Phi_{\exp(l\g)}(p), \homm'(\gamma_i)(p)\big)\\
    &=\pip_p(\homm)\big(\g_i,l_i\big)^{-1}\cdot \pip_p(\homm')\big(\g_i,l_i\big).
  \end{align*}
  This shows $\pip_p(\homm_\iota)\rightarrow \pip_p(\homm)$, hence continuity of $\pip_p$.  \hspace*{\fill}{\scriptsize$\blacksquare$}
\vspace{20pt}

	\noindent
	{\bf Proof of \eqref{eq:verkn}:} 
  The claim follows from
  \begin{align*} 
    \pi_{g\cdot p\cdot s}(\Add{h}(\g),l,\homm)&=\Delta\!\left(\Phi_{\exp(l\Add{h}(\g))}(\Phi_g(p)) \cdot s),\homm\!\left(\gamma_{\Add{h}(\g)}^{\varphi_g(x)}\big|_{[0,l]}\right)(\Phi_g(p)\cdot s)\right)\\
    &=\alpha_{s^{-1}}\cp \Delta\!\left(\Phi_{h\exp(l\g)h^{-1}}(g\cdot p),\homm\!\left(\wm_h\cp \gamma^{\wm_g(x)}_{\g}\big|_{[0,l]}\right)(\Phi_g(p))\right)\\
    &=\alpha_{s^{-1}}\cp \Delta\!\left(\Phi_{h}\cp \Phi_{\exp(l\g)}(g\cdot p)\cdot \fiba_{g\cdot p}(h)^{-1},\Phi_h\cp \homm\!\left(\gamma^{\wm_g(x)}_{\g}\big|_{[0,l]}\right)(\Phi_{h^{-1}}(g\cdot p))\right)\\
    &=\alpha_{(s^{-1}\cdot \fiba_{g\cdot p}(h))}\cp \Delta\!\left(\Phi_{g}\cp \Phi_{\exp(l\Add{g^{-1}}(\g))}(p) ,\homm\!\left(\varphi_{g}\cp \gamma^{x}_{\Add{g^{-1}}(\g)}\big|_{[0,l]}\right)(g\cdot p)\right)\\
    &=\alpha_{(s^{-1}\cdot \fiba_{g\cdot p}(h))}\cp \Delta\!\left(\Phi_{g}\cp \Phi_{\exp(l\Add{g^{-1}}(\g))}(p) ,\Phi_g\cp \homm\!\left(\gamma^{x}_{\Add{g^{-1}}(\g)}\big|_{[0,l]}\right)( p)\right)\\
    &=\alpha_{(s^{-1}\cdot \fiba_{g\cdot p}(h))}\cp \Delta\!\left(\Phi_{\exp(l\Add{g^{-1}}(\g))}(p) ,\homm\!\left(\gamma^{x}_{\Add{g^{-1}}(\g)}\big|_{[0,l]}\right)( p)\right)\\
    &=\alpha_{(s^{-1}\cdot \fiba_{g\cdot p}(h))}\cp \pi_{p}\big(\Add{g^{-1}}(\g),l,\homm\big).
    % \Delta\!\left(\Phi_{\exp(l\Add{g^{-1}}(\g))}(p) ,\homm\left(\gamma^{x}_{\Add{g^{-1}}(\g)}\big|_{[0,l]}\right)( p)\right)
  \end{align*}
  for $0<l<\tau_\g$.
  \hspace*{\fill}{\scriptsize$\blacksquare$}

\subsection{Lemma \ref{prop:Bohrmod2}}
\label{app:Bohrmod}
{\bf Proof of Lemma \ref{prop:Bohrmod2}.\ref{prop:Bohrmod23}:}
\begin{enumerate}
\item[\ref{prop:Bohrmod23})]
  We always have $0\in J(\g,p)$ because $\Psi(\lambda\cdot \g):=\me$ is a well-defined element of $\Eq_p$. Now, assume that the first two cases do not hold and choose $\Psi_0,\Psi_1\in Y_\g^p$ with 
\begin{align*}  
  0\neq \beta_0:=\beta(\Psi_0)\neq\beta_1:=\beta(\Psi_1)\neq 0.
  \end{align*}
  % We define $\beta_0:=\beta(\g,x,\homm_0)$, $\s_0:=\s_{\beta_0}$ and find a 
  Let $\phi\in \Per$ denote the unique\footnote{This follows by the arguments as in  the proof of Lemma and Convention \ref{lemconv:RBMOD}.\ref{prop:Bohrmod21}.} element with 
	\begin{align*}
  \Psi_0(\lambda\cdot \g)=\exp(\phi(\lambda)\cdot \s_{\beta_0})\qquad \forall\: \lambda>0, 
	\end{align*}  
  and choose $\lambda_0,\lambda_1>0$ such that $\Psi_0(\lambda_0\cdot\g),\Psi_1(\lambda_1\cdot\g)\neq \pm \me$ holds. Then, for $h\in G^x_{[\g]}$ we find $\mu\in \{-1,1\}$ with
  \begin{align*}
    \alpha_{\fiba_p(h)}\cp \Psi_0(\lambda\cdot \g)=\Psi_0(\lambda \cdot  \g)^{\mu}\qquad \forall\: \lambda\in \RR.
  \end{align*}
  Evaluating this for $\lambda=\lambda_0$ and using Lemma \ref{lemma:torus}.\ref{lemma:torus2},  
  we see that one of the following situations holds:
  \begingroup
  \setlength{\leftmarginii}{22pt}
  \begin{enumerate}
  \item[1.)]
  \vspace{-3pt}
    $\mu=1$ and $\fiba_p(h)\in H_{\beta_0}$,
  \item[2.)]
  \vspace{2pt}
    $\mu=-1$ and $\fiba_p(h)=\exp\big(\textstyle\frac{\pi}{2}\murs(\vec{m})\big)$ for some $\mm\in \RR^3$ orthogonal to $\murs^{-1}(\s_{\beta_0})$ with $\|\mm\|=1$ and uniquely determined up to a sign. 	 
    	
    	In particular, $\alpha_h(s)=s^{-1}$ holds for $\pm \me\neq s\in H_{\vec{n}'}$ iff $\langle\vec{m},\vec{n}'\rangle=0$. 
  \end{enumerate}
  \endgroup
  We now distinguish between the following two cases:
  
  \vspace{4pt}
  {\bf Case A:}  
  For each  $h\in G^x_{[\g]}$ Situation 1.) holds. 
  Then,
   \begin{align*}
    \alpha_{\fiba_p(h)}\cp \Psi_1(\lambda_1\cdot \g)=\Psi_1(\lambda_1\cdot \Ad_{h}(\g))=\Psi_1(\lambda_1\cdot \g) %\qquad\forall\: l>0 
  \end{align*}
  shows $\fiba_p(h)\in H_{\beta_0}\cap H_{\beta_1}=\{-\me,\me\}$, so that 
  for each $\beta\in \Sp\ms$ the map
  \begin{align}
  \label{eq:modhshshs}
  	\Psi'(\lambda\cdot \g):= \exp\!\big(\!\sign(\lambda)\phi(|\lambda|)\cdot\s_\beta\big)\qquad\forall\:\lambda\in \RR
  \end{align}
is obviously equivariant, hence contained in $Y_\g^p$. This shows that $J(\g,p)$ is of Type {\rm 4)}.
  
  \vspace{4pt}
  {\bf Case B:} We find $h\in G^x_{[\g]}$ such that Situation 2.) holds,  hence $\Ad_h(\g)=-\g$ and $h\neq \pm \me$. %, i.e. with 
  Using equivariance, we obtain
  \begin{align*}
    \alpha_{\fiba_p(h)}\cp \Psi_1(\lambda_1\cdot \g)=\Psi_1(\lambda_1\cdot\Ad_{h}(\g))=\Psi_1(\lambda_1\cdot\g)^{-1}, 
  \end{align*}
  hence $\langle \murs^{-1}(\s_{\beta_1}),\vec{m}\rangle =0$ because 
  $\Psi_1(\lambda_1\cdot \g)\neq \pm\me$ just by the choice of $\lambda_1$. 
  %, or  
  Thus,
  \begin{align*}  
    J(\g,p)\subseteq  \{0\}\sqcup \textstyle\bigcup_{\vec{n}\in \RR^3\backslash\{0\} \colon\langle\vec{n},\vec{m}\rangle=0}\:[\hspace{1.5pt}\murs(\vec{n})\hspace{0.5pt}].
  \end{align*}
  so that it remains to show that even equality holds: 
  \begingroup
  \setlength{\leftmarginii}{15pt}
  \begin{itemize}
  \item
    Let $h\neq h'\in G_{[\g]}^p$ be a further element for which Situation 2.) holds.     
    Then, the above arguments show
    \begin{align*}
      J(\g,p)\backslash \{0\}\subseteq \bigcup_{\vec{n}\in \RR^3\backslash\{0\} \colon\langle\vec{n},\vec{m}\rangle=0}[\hspace{1.5pt}\murs(\vec{n})\hspace{0.5pt}]\quad\cap\quad  \bigcup_{\vec{n}\in \RR^3\backslash\{0\}  \colon\langle\vec{n},\vec{m}'\rangle=0}[\hspace{1.5pt}\murs(\vec{n})\hspace{0.5pt}] ,
    \end{align*}
    so that $ J(\g,p)$ would be of Type {\rm 2)} if $\vec{m}$ and $\vec{m}'$ were linearly independent. Since this contradicts the assumption, %by assumption $|J(\g,p)|>2$, it follows that 
    we have $\fiba_p(h')=\pm\exp\big(\frac{\pi}{2}\murs(\vec{m})\big)$. 
  \item
    Let $h\neq h'\in G_{[\g]}^p$ be an element for which Situation 1.) holds. 
    Then, as in Case A, we conclude that 
    $\fiba_p(h')\in H_{\beta_0}\cap H_{\beta_1}=\{\me,-\me\}$.
  \end{itemize}
  \endgroup
  Now, to show that $J(\g,p)$ is of Type {\bf 3)}, let $\vec{n}\in \RR^3\backslash \{0\}$ with $\langle\vec{n},\vec{m}\rangle=0$ and $\beta:=[\hspace{1.5pt}\murs(\vec{n})\hspace{0.5pt}] \in \Sp\ms$. We define $\Psi'$ by \eqref{eq:modhshshs}. This map is equivariant because for each $h_0\in G^x_{[\g]}$ we either have $\phi_p(h_0)=\pm \me$ and $\Ad_{h_0}(\g)=\g$ or $\phi_p(h_0)=\pm \exp\big(\frac{\pi}{2}\vec{m}\big)$ $\Ad_{h_0}(\g)=-\g$. In fact, let $\lambda \in \RR$. Then, in the first case we have
  	\begin{align*}
          \alpha_{\fiba_p(h_0)}\cp \Psi'(\lambda\cdot \g)
          &=\alpha_{\fiba_p(h_0)}\cp\exp\!\big(\!\sign(\lambda)\phi(|\lambda|)\cdot\s_{\beta}\big)
          = \exp\!\big(\!\sign(\lambda)\phi(|\lambda|)\cdot\s_{\beta}\big)\\
          &=\Psi'(\lambda\cdot \g)=(\Psi'\cp \Ad_h)(\lambda\cdot\g),
        \end{align*}
        and in the second one
	\begin{align*}
          \alpha_{\fiba_p(h_0)}\cp \Psi'(\lambda\cdot \g)
          &=\alpha_{\fiba_p(h_0)}\cp\exp\!\big(\!\sign(\lambda)\phi(|\lambda|)\cdot\s_{\beta}\big)
          = \exp\!\big(\!\sign(\lambda)\phi(|\lambda|)\cdot\s_{\beta}\big)^{-1}\\
          &=\Psi'(-\lambda\cdot \g)=(\Psi'\cp \Ad_h)(\lambda\cdot\g)
        \end{align*}
   holds.
  \hspace*{\fill}{\scriptsize$\blacksquare$}
\end{enumerate}

\subsection{Completeness and Independence of the Family in Example \ref{ex:cosmoliealgmaasse}}
\label{subsec:Aclacula}
% consider the elements of the form $(\vec{v},\s)$ for $\s\in E_0\backslash \{\lambda \mu\:|\: \lambda\neq 0\}$ where $E_0 \subseteq SU(2)$ a fixed plane through the origin.
Recall that $x=0$, that $\vv, \vv_{\perp}$ are orthogonal, $\s_0=\murs(\vv)$, $\s_{\perp}=\murs(\vv_\perp)$, and that
\begin{align*} 
    E_\geq&:=\{\lambda_1 \s_0 +\lambda_2 \s_\perp\: |\: \lambda_1 \in \RR,\lambda_2 \geq 0\}.
  \end{align*}
Then, for each $\g \in \mg\backslash \mg_x= [\RR^3\backslash \{0\}] \times \su$ we find 
\begin{align*}
\s\in E_{\geq 0}\qquad \qquad 
%Then, each $\Ad'$-orbit (cf.\ Remark \ref{lemremmgohnermgx}) in $\pr_\mg(\mg\backslash \mg_x)$ %contains an element of $\{\vv\}\times E_\geq$, i.e., for each $\g\in \mg\backslash \mg_x$ 
h=(0,\sigma)\in G_x=\{0\}\times \SU \qquad\qquad \lambda\neq 0,
\end{align*}
 such that $\g=\lambda\cdot \Ad_h((\vv,\s))$ holds, hence $\{\vv\}\times E_\geq$ is complete by \eqref{eq:compladorb}.
\begingroup
\setlength{\leftmargini}{10pt}
\begin{itemize} 
\item
In fact, since $\Add{(0,\sigma)}((\vec{v},\s))=(\varrho(\sigma)(\vec{v}),\Add{\sigma}(\s))$ holds  for all $\sigma\in \SU$, running over $\sigma\in H_{\vv}$ generates the set $\{\vv\}\times \su$.
\item
Thus, running over $\SU$ gives $S^2\times \su$ as $\|\vv\|=1$, and since each element of $[\RR^3\backslash \{0\}] \times \su$ equals, up to some non-zero factor, some element of $S^2\times \su$, the statement is clear.
\end{itemize}
\endgroup
\noindent
We now have to show that replacing $E_\geq$ by $E_0$ does not change this fact, that $E_0$ consists of stable elements, and that it is independent. For this, we first have to determine the Lie algebra generated curves. To this end,
  \begingroup
  \setlength{\leftmargini}{20pt}
 \begin{enumerate}
 \item[a.)]
 \vspace{-2pt}
We take a closer look at the group %exponential map of %$G=\RR^3\rtimes_{\varrho} \SU$. For this, let 
$\wt{G}:=\RR^3 \rtimes \SOD$ whose exponential map is given by\footnote{One easily verifies that $\dttB{t}{s}\wt{\exp}(t\cdot (\vv,\rr))=\dd L_{\wt{\exp}(s\cdot (\vv,\rr))} (\vv,\rr)$ holds.}
\begin{align*}
\wt{\exp}\big((\vv,\rr)\big)=\big(A(\rr)(\vv),\exp(\rr)\big)\qquad\text{for}\qquad A(\rr):=\textstyle\sum^\infty_{k=1}\frac{1}{k!}{\rr\:}^{k-1}
\end{align*}		
with $(\vv,\rr)\in \RR^3\times \mathfrak{so}(3)$ and $\exp(\rr)$ the matrix exponential in $\SOD$. We let
\begin{align*}
	\wt{\wm}((v,R),x):=v + R(x)\qquad\forall\:(v,R)\in \RR^3\times \SOD,\:x\in \RR^3
\end{align*}
and define $\wt{\varrho}\colon G\rightarrow \wt{G}$, $(v,\sigma)\mapsto (v,\varrho(\sigma))$. Then, $\wt{\varrho}$ is a Lie group homomorphism and
\begin{align}
	\label{eq:sfhfs}
 	\wm_y=\wt{\wm}_y\cp \wt{\varrho}\qquad\forall\:y\in M=\RR^3
\end{align}
holds for $\wm_y$ the induced action that corresponds to the action of $\Gee$ on $P$. 
Thus, for $\ovl{\exp}$ the exponential map in $\RR^3\times \SU$ and $\rr:=\dd_e\varrho(\s)$, we have
\begin{align}
\label{eq:liealgbsphi}
\begin{split}
  \gamma^x_{(\vv,\s)}(t)& =(\wm_x\cp\ovl{\exp})\big((t\cdot\vv,t\cdot\s)\big)\stackrel{\eqref{eq:sfhfs}}{=}(\wt{\wm}_x\cp\exp)\big(\dd_e\wt{\varrho}(t\cdot\vv,t\cdot\s)\big)\\
  &=(\wt{\wm}_x\cp\wt{\exp})\big((t\cdot\vv,t\cdot \dd_e\varrho(\s))\big)
  =A(t\cdot \rr)(t\cdot \vv)=  
  \textstyle\sum^\infty_{k=1}\frac{t^k}{k!}{\rr\:}^{k-1}\vv,
\end{split}
\end{align}
hence $\dot\gamma^x_{(\vv,\s)}(t)=\exp(t\cdot \rr)(\vv)$. In particular, since $\dd_e\varrho(\lambda\cdot \s_0)(\vv)=0$, \eqref{eq:liealgbsphi} shows that 
\begin{align*}
	\gamma^x_{(\vv,\lambda\cdot \s_0)}(t)= t\cdot \vv \qquad \forall\: \lambda\in \RR,
\end{align*}  
hence that $\{\vv\}\times E_0$ is complete as well.
 \item[b.)]
 Let $\s,\s'\in \su$ and $\sigma\in \SU$ be such that (we write $\sigma(\vv)$ instead of $\varrho(\sigma)(\vv)$ in the following)
 \begin{align}
 \label{eq:xdfdffd}
 	\gamma^x_{(\vv,\s)}|_{[0,l]}= \gamma^x_{\pm(\sigma(\vv),\s')}\cp\adif|_{[0,l]}
 \end{align}
 holds for some diffeomorphism $\adif$ with $\dot\adif>0$. Then, $\sigma(\vv)=\pm \vv$, as well as $\s'=\pm\s$ or $\s,\s'\in \spann_\RR(\s_0)$. 
 
 In fact, combining $\dot\gamma^x_{(\vv,\s)}(t)=\exp(t\cdot \rr)(\vv)$ ($\rr=\dd_e\varrho(\s)$) with \eqref{eq:xdfdffd}, for $\rr':=\dd_e\varrho(\s')$ we obtain 
\begin{align*}
  \exp(t\cdot\rr)(\vv)=\dot\gamma^x_{(\vv,\s)}(t)\stackrel{\eqref{eq:xdfdffd}}{=}\dot\adif(t)\cdot  \dot\gamma^x_{\pm(\sigma(\vv),\s')}(\adif(t))= \exp(\pm \adif(t)\cdot \rr')(\pm\dot\adif(0)\cdot\sigma(\vv)). 
\end{align*}
Since $\|\vv\|=1$ and $\|\exp(t\cdot\rr)(\vv)\|=1$ for all $t\in[0,l]$, we have $\dot\adif=1$, hence $\adif(t)= t$ for all $t\in [0,l]$. Evaluating this for $t=0$, we see that $\sigma(\vv)=\pm \vv$ holds, hence
\begin{align}
\label{eq:dgd}
 	\exp(t\cdot\rr)(\vv)=\exp(\pm t\cdot\rr')(\vv)\qquad\forall\: t\in \RR.
\end{align}
Since the images of the maps on both hand sides of \eqref{eq:dgd} are circles which traverse through $\vv$ in the planes orthogonal to $\rr$ and $\rr'$, respectively, we conclude that one of the following situations holds:
\begingroup
\setlength{\leftmarginii}{12pt}
\begin{itemize} 
\item
	$\rr=\lambda\cdot \rr'$ for $\lambda\neq 0$ 
\item
	$\rr,\rr'\in \spann_\RR(\vv)$\qquad $\Longleftrightarrow$\qquad $\s,\s'\in \spann_\RR(\s_0)$
\end{itemize}
\endgroup
\noindent
In the second case, we have done. In the first case, we deduce that $\rr=\pm\rr'$, hence $\s=\pm \s'$ holds because the tangent vectors of the above curves at $t=0$ have to coincide, and are given by $\rr\times \vv$ and $\pm  \rr'\times \vv$, respectively. This is because
\begin{align*}
	\|\rr\times \vv\|=\|\rr\|\cdot\|\vv\|\cdot|\sin(\theta)|\qquad\text{and}\qquad \|\rr'\times \vv\|=\|\rr'\|\cdot\|\vv\|\cdot|\sin(\theta')|, 
\end{align*}
whereby the angles $\theta$ and $\theta'$ between the respective vectors coincide modulo $\pi$.
\hspace*{\fill} $\dagger$
 \end{enumerate}
 \endgroup
 \noindent
Now, for stability, let \eqref{eq:xdfdffd} hold for $\s'=\Ad_\sigma(\s)$ with $\s\in E_0$. Then, b.) shows\footnote{Observe that $\s \in \spann_\RR(\s_0)$ already means that $\s=0$, hence $\s'=0$ holds.} 
\begin{align*}
\Ad_{(0,\sigma)}((\vv,\s))=(\sigma(\vv),\Ad_\sigma(\s))=\pm(\vv,\s),
\end{align*}
 hence $(0,\sigma)\in G^x_{[(\vv,\s)]}$. This shows stability of $(\vv,\s)$.
 \vspace{5pt}
 
 \noindent 
 For independence, assume that $(\vv,\s)\xsim (\vv,\s'')$ holds for $\s,\s''\in E_0$. Then, by \eqref{eq:simconfestfuss} in Lemma \ref{lemma:completee}.\ref{it:completee1} the equation \eqref{eq:xdfdffd} holds for $\s'=\Ad_\sigma(\s'')$ for some $\sigma\in \SU$. Hence, 
$(\sigma(\vv),\s)=\pm(\vv,\Ad_\sigma(\s''))$ by b.) because $\s,\Ad_\sigma(\s'')\in \spann_\RR(\s_0)$ already implies that both elements are zero. 
\begingroup
\setlength{\leftmargini}{15pt}
\begin{itemize}
\item
\vspace{-2pt}
If $\sigma(\vv)=\vv$, then $\sigma \in H_{\vv}=H_{\s_0}$ so that $\s=\Ad_\sigma(\s'')$ implies $\s=\s''$ just because $\s,\s''\in E_0$.
\item 
If $\sigma(\vv)=-\vv$, then $\sigma= \pm \exp(\textstyle\frac{\pi}{2} \murs(\mm))$ for $\|\mm\|=1$ with $\mm\perp\vv$. Since $\s=0$ iff $\s''=0$, we can assume that $\s,\s''\neq 0$ holds because elsewise the statement is clear. Then, the claim follows if we can show that either $\vv_\perp = \pm \mm$ holds or that $\vv_\perp$ and $\mm$ are orthogonal.  

In fact, if $\vv_\perp$ and $\mm$ are orthogonal, then we have 
\begin{align*}
	\Add{\sigma}(\s'')=-\s''\qquad \Longrightarrow \qquad -\s=\Add{\sigma}(\s'')=-\s''\qquad \Longrightarrow \qquad\s''=\s.
\end{align*}
 Moreover, if $\vv_\perp = \pm \mm$ holds, then $\Ad_\sigma(\s'')\in E_0$, hence $\s=0$ since elsewise $\Ad_\sigma(\s'')=-\s \notin E_0$ would give a contradiction as well.  

Now, to show that one of the above cases holds, let $\ww:=\murs^{-1}(\s)$ and denote by $D(\mm)$ the rotation in $\RR^3$ by $\pi$ around $\mm$. Since $0\neq \s\in E_0$, we have  $\ww=\lambda \cdot \vv + \mu \cdot \vv_\perp$ for some $\mu> 0$. Then, $D(\mm)(\ww)$ must be contained in $\spann_\RR(\vv,\vv_\perp)$ because $\Ad_\sigma(\s'')=-\s$ and $\s \in E_0$ holds. Then, since $D(\mm)(\ww)=-\lambda\cdot \vv + \mu \cdot D(\mm)(\vv_\perp)$, the element $D(\mm)(\vv_\perp)$ must be contained in $\spann_\RR(\vv,\vv_\perp)$  as well. However, $\vv_\perp = \langle \vv_\perp,\mm\rangle \cdot \mm + \langle \vv_\perp,\vv \times \mm \rangle \cdot \vv \times \mm$, so that
\begin{align*}
D(\mm)(\vv_\perp)=\langle \vv_\perp,\mm\rangle \cdot \mm - \langle \vv_\perp,\vv \times \mm \rangle \cdot \vv \times \mm
\end{align*}
can only be contained in $\spann_\RR(\vv,\vv_\perp)$ if 
\begin{align*}
	0&= \langle D(\mm)(\vv_\perp) , (\vv\times \vv_\perp) \rangle\\
	 &=  \langle \vv_\perp,\mm\rangle \cdot \langle \mm, \vv\times \vv_\perp\rangle - \langle \vv_\perp,\vv \times \mm \rangle \cdot \langle \vv \times \mm , \vv\times \vv_\perp\rangle\\
	 &= \det(\mm,\vv,\vv_\perp)\cdot \langle \mm,\vv_\perp\rangle - \det(\vv_\perp,\vv,\mm)\cdot \langle \mm,\vv_\perp\rangle\\
	 &= 2\cdot \det(\mm,\vv,\vv_\perp) \cdot \langle \mm,\vv_\perp\rangle.
\end{align*}
Here, in the third step, we have applied $\langle \vec{a}\times \vec{b}, \vec{c}\times \vec{d}\rangle=\langle \vec{a}, \vec{c}\rangle \cdot \langle \vec{b}, \vec{d}\rangle- \langle \vec{b}, \vec{c}\rangle \cdot \langle \vec{a}, \vec{d}\rangle$ to the second summand. The above equality now shows that either $\vv_\perp = \pm \mm$ holds or that $\mm$ is orthogonal to $\vv_\perp$.
\end{itemize}
\endgroup

\section{Appendix to Homogeneous Isotropic LQC}
\label{sec:homgeniso}

\subsection{Proof of the Lemma in Subsection \ref{subsec:Motivation}}
\label{sec:ProofOfLemmaImCirc}
{\bf Proof of Lemma \ref{lemma:BildCirc}:}
\begin{enumerate} 
\item[\ref{lemma:BildCirc2})]
  It suffices to show the claim for $\RR\subseteq \qR$. Now, $\pi_{\tau,r}\cp \iota_\RR$ is continuous by \eqref{eq:matrixentr}, so that
  $\pi'_{\tau,r}(\RR)=(\pi_{\tau,r}\cp \iota)(\RR)\subseteq \SU$ is connected.
  Since each proper Lie subalgebra of $\mathfrak{su}(2)$ is of dimension $1$ and since $\SU$ is connected, each proper Lie subgroup of $\SU$ is of dimension $1$ as well. Let $H\subseteq\SU$ be such a proper Lie subgroup with $\pi'_{\tau,r}(\RR)\subseteq H$. Then $\mathfrak{h}=\spann_{\mathbb{R}}(\vec{s}\hspace{2pt})$ for some $0\neq\vec{s}\in\mathfrak{su}(2)$. Then $H_{\s}$ is the unique connected Lie subgroup of $H$ with Lie algebra $\mathfrak{h}$, i.e., the component of $\me$ in $H$. 
  But $\me=\pi'_{\tau,r}(0_\RR)\in \pi'_{\tau,r}(\RR)\cap H$, hence $\pi'_{\tau,r}(\RR)\subseteq H_{\vec{s}}$ by connectedness of $\pi_{\tau,r}(\RR)$. This, however, cannot be true as \eqref{eq:patrall} shows. 
\item[\ref{lemma:BildCirc3})]
  Let $\x\in \RB \subseteq \qR$. Then 	
  \begin{align*}
    \pi'_{\tau,r}(\x)
    %=\pi_{\tau,r}(\xi(\x))=\exp\left(\textstyle\frac{\tau}{2}\tau_3\right)^{-1}\cdot\big(\Omega\cp \kappa\cp \ovl{i^*_\AR}\big)(\xi(\x))(\gamma_{\tau,r})\\
   % & \!\!\stackrel{\eqref{eq:hilfseq}}{=}\exp\left(\textstyle\frac{\tau}{2}\tau_3\right)^{-1}\cdot\big(\big(\ovl{i^*_\AR}\cp  \xi\big)(\ovl{x})\big([h_{\gamma_{\tau,r}}^\nu]_{ij}\big)\big)_{ij}\\
    &\!\!\stackrel{\eqref{eq:pilrtau}}{=}\exp\left(\textstyle\frac{\tau}{2}\tau_3\right)^{-1}\cdot\big(\xi(\ovl{x})\big([h_{\gamma_{\tau,r}}^\nu]_{ij}|_\AR\big)\big)_{ij}
    =\big(\xi(\ovl{x})\big([\exp\left(\textstyle\frac{\tau}{2}\tau_3\right)^{-1}\cdot h_{\gamma_{\tau,r}}^\nu]_{ij}|_\AR\big)\big)_{ij}\\
    % 
    % &=\exp\left(\textstyle\frac{\tau}{2}\tau_3\right)^{-1}\cdot\big(\xi(\ovl{x})\big([\pr_2\cp \parall{\gamma_{\tau,r}}{\RnA} ]_{ij} \big)\big)_{ij}\\ 
    % 
    &=\begin{pmatrix} \xi(\x)\left(c\mapsto \cos(\pc \tau)+\frac{\I}{2\pc}\sin(\pc \tau)\right) &\xi(\x)\left(c\mapsto\frac{cr}{\pc} \sin(\pc \tau)\right) \\ \xi(\x)\left(c\mapsto-\frac{cr}{\pc}\sin(\pc \tau)\right) & \xi(\x)\left(c\mapsto\cos(\pc \tau)-\frac{\I}{2\pc}\sin(\pc \tau)\right)  \end{pmatrix}\\
    &=\begin{pmatrix} \xi(\x)\left(c\mapsto \cos(c r\tau)\right) &\xi(\x)\left(c\mapsto\sin(c r \tau)\right) \\ \xi(\x)\left(c\mapsto-\sin(c r \tau)\right) & \xi(\x)\left(c\mapsto\cos(cr \tau)\right)  \end{pmatrix}\\
    & = \big(\ovl{x}\big(c\mapsto \exp(-c r\tau \cdot \tau_2)_{ij}\big) \big)_{ij}\in H_{\tau_2},
  \end{align*}
  where the second step is due to multiplicativity of $\xi(\x)$ and the third one follows from \eqref{eq:matrixentr}. The fourth step is due to the fact that 
  the (unique) decompositions of the matrix entries  
  \begin{align}
    \label{eq:entries}
    a\colon c\mapsto \cos(\pc \tau)+\textstyle\frac{\I}{2\pc}\sin(\pc \tau)\qquad\quad \text{and}\qquad\quad
    b\colon c\mapsto \frac{cr}{\pc} \sin(\pc \tau)
  \end{align}
  in \eqref{eq:matrixentr} into direct sums of the form $C_{0}(\mathbb{R})\oplus \CAP(\RR)$ are given by
  \begin{align}
    \label{eq:cappluco}
    a(c)&= \left(\cos(\pc \tau)+\textstyle\frac{\I}{2\pc}\sin(\pc \tau)- \cos(cr\tau)\right) + \cos(cr\tau)\\
    \label{eq:cappluco2}
    b(c)&=\left(\textstyle\frac{cr}{\pc} \sin(\pc \tau)-\sin(cr\tau)\right) + \sin(cr\tau).
  \end{align} 
  The last step is due to  
  closedness of $H_{\tau_2}$ and that we find a net $\{c_\alpha\}_{\alpha\in I}\subseteq \RR$ such that for each $1\leq i,j\leq 2$ we have 
  \begin{align*}      
    \ovl{x}\big(c\mapsto \exp(c r\tau \cdot\tau_2)_{ij}\big)= \lim_\alpha \exp\!\big(c_\alpha r\tau\cdot \tau_2\big)_{ij}. 
  \end{align*}
  Obviously, $\pi'_{\tau,r}|_{\RB}\colon \RB\rightarrow H_{\tau_2}$ is surjective. For the intersection statement observe that by \eqref{eq:matrixentr} and \eqref{eq:expSU2} we have $s\in \pi'_{\tau,r}(\mathbb{R})\cap H_{\tau_2}$ iff $\sin(\pc\tau)=0$, hence $\cos(\pc\tau)=\pm 1$, i.e., $s=\pm\me$. 
\item[\ref{lemma:BildCirc1})]
  Let $\mu_0$ denote the Haar measure on $\SU$.
  By \ref{lemma:BildCirc3}) we have $\mu_0\big(\pi'_{\tau,r}\big(\qR\big)\big)= \mu_0(\pi'_{\tau,r}(\RR))+\mu_0\big(H_{\tau_2}\big)$ and \eqref{eq:matrixentr} shows that $\pi'_{\tau,r}|_\RR$ is an immersion. In fact, the derivatives of the functions \eqref{eq:entries} are given by
  \begin{align*}
    \dot a(c)&= -\textstyle\frac{cr^2\tau}{\pc}\sin(\pc\tau)+ \textstyle\frac{\I c r^2}{2\pc^2}\left[\cos(\pc \tau)\tau-\textstyle\frac{1}{\pc}\sin(\pc\tau)\right]\\
    \dot b(c)&=\textstyle\frac{r}{\pc} \sin(\pc \tau) - \textstyle\frac{c^2r^3}{\pc^3} \sin(\pc \tau) +     \textstyle\frac{c^2r^3\tau}{\pc^2} \cos(\pc \tau),
  \end{align*} 
  so that we have $\dot b(0)= 2 r \sin\!\big(\textstyle\frac{\tau}{2}\big)\neq 0$ because $0<\tau<2\pi$ and $\dot a(c)\neq 0$ if $c\neq 0$. 
  Consequently, 
  $\pi'_{\tau,r}(\RR)$ can be decomposed into countably many $1$-dimensional embedded submanifolds each  of measure zero w.r.t.\ the Haar measure on $\SU$. Hence, $\mu_0(\pi'_{\tau,r}(\RR))=0$ and, obviously we have $\mu_0\big(H_{\tau_2}\big)=0$ as well.  
\item[\ref{lemma:BildCirc4})]
Let $x\neq y$ with $\pi'_{\tau,r}(x)=\pi'_{\tau,r}(y)$. If $x=-y$, a closer look at the entry $(\pi_{\tau,r})_{12}$ ($b$ in \eqref{eq:entries}) shows that $\sin(\beta_x \tau)=0=\sin(\beta_y\tau)$ holds, hence $x=a_n$ and $y=a_{-n}$ for some $n\in \ZZ_{\neq 0}$. 
If $|x|\neq |y|$, then $\beta_x\neq \beta_y$, so that a closer look at
the entry $(\pi_{\tau,r})_{11}$ ($a$ in \eqref{eq:entries}) shows\footnote{Consider the graph of the curve $\RR_{\geq 0}\ni \beta\mapsto\cos(\beta \tau)\vec{e}_1+ \frac{1}{2\beta}\sin(\beta \tau)\vec{e}_2\in \RR^2$. Then, compare its self intersection points with the zeroes of the function 
  $b\colon c\mapsto \frac{cr}{\beta_c}\sin(\beta_c \tau)$.} 
   that  
  either 
  $\tau\beta_x, \tau\beta_y \in \{2\pi n\:|\: n\in \mathbb{N}_{\geq 1}\}$ or $\tau\beta_x, \tau \beta_y\in \{(2n-1)\pi\:|\: n\in \mathbb{N}_{\geq 1}\}$ holds. 
  Thus, $x=a_n$ and $y=a_m$ for some $m,n\in \ZZ_{\neq 0}$, 
	from which    
  the first part follows. 
  The merging property is immediate from the formulas \eqref{eq:cappluco} and \eqref{eq:cappluco2}. \hspace*{\fill}{\scriptsize$\blacksquare$}
\end{enumerate}

\section{Appendix to A Characterization of Invariant Connections}

\subsection{The Proof of the Case in Subsection \ref{sec:ApplTrivB}}
\label{subsec:TrivBund}

\noindent  
{\bf Proof of Case \ref{scase:trivbundle}.}
The only patch is $M\times\{e\}$ so that a reduced connection is a smooth map $\psi\colon  \mathfrak{g}\times TM \rightarrow \mathfrak{s}$ with the claimed linearity property and that fulfils the two conditions from Corollary \ref{cor:psialpha}. Obviously, \textit{ii.)} and \textrm{iii.)} are equivalent. Moreover, \textrm{i.)} follows from \textit{i.)} for $p_\alpha=p_\beta=(x,e)$, $q=(e,e)$, $\vec{w}_{p_\beta}=\vec{v}_x$ and $\vec{w}_{p_\alpha}=\vec{0}_{(x,e)}$, see also \textit{a.)} in Remark \ref{rem:Psialphaconderkl}. To obtain \textrm{ii.)} 
let $\vec{v}_x\in T_xM$, $q\in Q$ and $q\cdot (x,e)=(y,e)$. Then 
$\dd L_q\vec{v}_x=(\vec{v}_y ,-\vec{s})$ for elements $\vec{v}_y\in T_yM$ and $\vec{s}\in \mathfrak{s}$ so that 
\begin{align*}
  \psi^+(\dd L_q\vec{v}_x)&=\psi^+(\vec{v}_y ,-\vec{s})=\psi\big(\vec{0}_{\mathfrak{g}},\vec{v}_y\big)-\vec{s}\stackrel{\textit{i.)}}{=}\qrep(q)\cp \psi\big(\vec{0}_{\mathfrak{g}},\vec{v}_x\big).  
\end{align*}
It remains to show that \textrm{i.)} and \textrm{ii.)} imply \textit{i.)}.
To this end, let $(y,e)=q\cdot (x,e)$ for $x,y\in M$ and $q\in Q$. Then \textit{i.)} reads
\begin{align*}
  \wt{g}(y,e)+ \vec{v}_{y}-\vec{s}=\dd L_q\vec{v}_{x}\quad\Longrightarrow\quad  \psi^-(\vec{g},\vec{v}_{y},\vec{s})=\qrep(q)\cp\psi\big(\vec{0}_{\mathfrak{g}},\vec{v}_{x}\big),
\end{align*}
where $\vec{v}_{x}\in T_xM$,$\vec{v}_{y}\in T_yM$, $\vec{s}\in \mathfrak{s}$ and $\vec{g}\in \mathfrak{g}$. Let $\dd L_q\vec{v}_x=(\vec{v}_y ,-\vec{s})$ be as above. If \textrm{ii.)} is true, then it is clear from 
\begin{align*}
  \psi^{-}(\vec{v}_y,\vec{s})=\psi^+(\dd L_q\vec{v}_x)\stackrel{\textrm{ii.)}}{=} \qrep(q)\cp \psi\big(\vec{0}_{\mathfrak{g}},\vec{v}_x\big)
\end{align*}
that \textit{i.)} is true for $\big(\!\big(\vec{0}_{\mathfrak{g}},\vec{s}),\vec{v}_y\big)$, i.e.,
\begin{align*}
  \vec{0}_{\mathfrak{g}}+\vec{v}_y-\vec{s}=\dd L_q\vec{v}_{x}\quad\Longrightarrow\quad \psi\big(\vec{0}_{\mathfrak{g}},\vec{v}_y\big)-\vec{s}=\qrep(q)\cp \psi\big(\vec{0}_{\mathfrak{g}},\vec{v}_x\big). 
\end{align*}
Due to \textrm{i)} and the linearity properties of $\psi$, the condition \textit{i.)} then is also true for each other element $((\vec{g}\hspace{1pt}',\vec{s}\hspace{1pt}'),\vec{v}\hspace{1pt}'_y)\in \mathfrak{q}\times T_yM$ with $\wt{g}'(y,e)+ \vec{v}'_{y}-\vec{s}'=\dd L_q\vec{v}_{x}$. \hspace*{\fill}{\scriptsize $\blacksquare$}

\subsection{A Result used in the End of Section \ref{sec:PartCases}}
\label{subsec:DifferentLifts}
We consider the fibre transitive action $\Phi'\colon \Ge\times P\rightarrow P$ that is defined by $\Phi'((v,\sigma),(x,s)):=(v+ \sigma(x),s)$ and claim that the connection
\begin{align*} 
  \w_0(\vec{v}_x,\vec{\sigma}_s)=s^{-1}\vec{\sigma}_s \qquad \forall\: (\vec{v}_x,\vec{\sigma}_s)
  \in T_{(x,s)}P
\end{align*}  
is the only $\Phi'$-invariant one.
For this, observe that the stabilizer of $x=0$ w.r.t.\ $\varphi'$ is given by $\SU$ and $\phi'_{(0,e)}(\sigma)=e$ for all $\sigma\in \SU$.
We apply Wang's theorem to $p=(0,e)$. Then condition \textrm{a)} yields $\psi(\vec{s}\hspace{2pt})=0$ for all $\vec{s}\in \mathfrak{su}(2)$ and \textrm{b)} now reads $\psi\cp \Add{\sigma}=\psi$ for all $\sigma\in \SU$. Consequently, for $\vec{v}\in \mathbb{R}^3\subseteq \mathfrak{e}=\RR^3 \times \mathfrak{su}(2)$ we obtain
\begin{align*}
  0=\dttB{t}{0}\psi(\vec{v})&=\dttB{t}{0} \psi\cp \Add{\exp(t\vec{s})}(\vec{v})\\
  &=\psi\left(\dttB{t}{0}\varrho(\exp(t\vec{s}))(\vec{v})\right)=\psi \cp \murs^{-1}([\vec{s},\murs(\vec{v})])
\end{align*}
for all $\vec{s}\in \mathfrak{su}(2)$ just by linearity of $\psi$. This gives 
\begin{align*}  
  0=\psi\big(\murs^{-1}([\tau_i,\murs(\vec{e}_j)])\big)=\psi\big(\murs^{-1}([\tau_i,\tau_j])\big)=2 \epsilon_{ijk}\psi(\vec{e}_k),
\end{align*}
hence $\psi=0={\Phi'_{p}}\hspace{-2pt}^*\w_0$. 

\subsection{Spherically Symmetric Connections}
\label{subsec:IsotrConn}
We consider the action $\Phi$ of $\SU$ on $P$ defined by $\Phi(\sigma, (x,s)):=(\sigma(x),\sigma s)$ and show that the corresponding invariant connections are given by (see \eqref{eq:rotinvconn} in Example \ref{bsp:Rotats})
\begin{align*}
  \begin{split}
    \w^{abc}(\vec{v}_x,\vec{\sigma}_s):= \Add{s^{-1}}\!\big[&a(x)\murs(\vec{v}_x)+ b(x)[\murs(x),\murs(\vec{v}_x)]
    +c(x)[\murs(x),[\murs(x),\murs(\vec{v}_x)]]\big]+ s^{-1}\vec{\sigma}_s
  \end{split}
\end{align*}
with rotation invariant maps $a,b,c\colon \mathbb{R}^3\rightarrow \mathbb{R}$ for which the whole expression is a smooth connection. 
Now, a straightforward calculation shows that each $\w^{abc}$ is $\Phi$-invariant so that it remains to verify that each $\Phi$-invariant connection is of the upper form. To this end, we reduce the connections $\w^{abc}$ w.r.t.\ $P_\infty=\mathbb{R}^3\times \{e\}$ and show that each map $\psi$ as in Case \ref{scase:trivbundle} can be obtained in this way. For this, let $\vec{g}\in \mathfrak{g}$, $p=(x,e)\in P_\infty$ and $\gamma_x\colon (-\epsilon,\epsilon)\rightarrow M$ be a smooth curve with $\dot\gamma_x(0)=\vec{v}_x\in T_xM\subseteq T_pP_\infty$. Then 
\begin{align}  
  \label{eq:theta}
  \begin{split}
    \dd_{(e,p)}\Phi(\vec{g},\vec{v}_x)
    &=\left(\dttB{t}{0}\murs^{-1}\left(\exp(t\vec{g})\murs(\gamma_x(t))\exp(t\vec{g})^{-1}\right),\exp(t\vec{g})\right)\\
    &=\left(\murs^{-1}\left([\vec{g},\murs(x)]\right)+\vec{v}_x,\vec{g}\right).
  \end{split}
\end{align}
This equals $\vec{s}$ iff $\vec{g}=\vec{s}$ and $\vec{v}_x=\murs^{-1}([\murs(x),\vec{g}])$. Consequently, for the reduced connection $\psi^{abc}$ that corresponds to $\w^{abc}$ we obtain
\begin{align*}
  \psi^{abc}(\vec{g},\vec{v}_x)
  =&\big(\Phi^*\w^{abc}\big)_{(e,p)}(\vec{g},\vec{v}_x)\\
  =&\:\w^{abc}_{p}\left(\murs^{-1}\left([\vec{g},\murs(x)]+\murs(\vec{v}_x)\right),\vec{g}\right)\\
  =&\:a(x)\big[[\vec{g},\murs(x)]+\murs(\vec{v}_x)\big]+ b(x)\big[[\murs(x),[\vec{g},\murs(x)]]+[\murs(x),\murs(\vec{v}_x)]\big]\\
  &+c(x)\big[[\murs(x),[\murs(x),[\vec{g},\murs(x)]]]+[\murs(x),[\murs(x),\murs(\vec{v}_x)]]\big] +\vec{g}.
\end{align*}
Now, assume that $\psi$ is as in Case \ref{scase:trivbundle}.
Then for $q\in Q$ and $p\in P_\infty$ we have $q\cdot p\in P_\infty$ iff $q=(\sigma,\sigma)$ for some $\sigma\in \SU$ and $p=(x,e)$ for some $x\in M$. Consequently, $q\cdot p=(\sigma(x),e)$ as well as $\dd L_q(\vec{v}_x)=\sigma(\vec{v}_x)$ for all $\vec{v}_x\in T_xM$ so that \textrm{ii.)} gives
\begin{align*}
  \psi\big(\vec{0}_{\mathfrak{g}},\sigma(\vec{v}_x)\big)=\psi^+(\dd L_q(\vec{v}_x))=\Add{\sigma}\cp\: \psi\big(\vec{0}_{\mathfrak{g}},\vec{v}_x\big), 
\end{align*}
hence
\begin{align}
  \label{eq:condrei}
  \psi\big(\vec{0}_{\mathfrak{g}},\vec{v}_x\big)=\Add{\sigma^{-1}}\cp\: \psi\big(\vec{0}_{\mathfrak{g}},\sigma(\vec{v}_x)\big)\qquad\forall\: \vec{v}_x\in T_xM.
\end{align}
If $x\neq 0$, then for $\sigma_t:=\exp(t\murs(x))$ we have $\sigma_t(x)=x$ and $\sigma_t(\vec{v}_x)\in T_{x}M$ for all $t\in \RR$. Then linearity of $\psi_x:=\psi|_{\mathfrak{g}\times T_{(x,e)}P_\infty}$ yields 
\vspace{-5pt}
\begin{align*}
  0&=\dttB{t}{0}\psi\big(\vec{0}_{\mathfrak{g}},\vec{v}_x\big)
  \stackrel{\eqref{eq:condrei}}{=}\dttB{t}{0}\Add{\sigma_t^{-1}}\cp\: \psi_x\big(\vec{0}_{\mathfrak{g}},\sigma_t(\vec{v}_x)\big)\\[3pt]
  &=\dttB{t}{0}\sigma^{-1}_t \big(\psi_x\cp \murs^{-1}\big)\left(\sigma_t\:\murs(\vec{v}_x)\:\sigma^{-1}_t\right) \sigma_t\\[3pt]
  &\stackrel{\text{lin.}}{=}-\murs(x)\: \psi_x\big(\vec{0}_{\mathfrak{g}},\vec{v}_x\big)+ \big(\psi_x\cp \murs^{-1}\big)\left[\murs(x)\murs(\vec{v}_x)-\murs(\vec{v}_x)\murs(x)\right]  +\psi_x\big(\vec{0}_{\mathfrak{g}},\vec{v}_x\big)\murs(x),
\end{align*}
hence $\big[\murs(x),\psi_x\big(\vec{0}_{\mathfrak{g}},\vec{v}_x\big)\big]=\big(\psi_x\cp \murs^{-1}\big)\left([\murs(x),\murs(\vec{v}_x)]\right)$. For $x=\lambda \vec{e}_1\neq 0$ and $\kappa_j:=\psi\big(\vec{0}_{\mathfrak{g}},\vec{v}_x\big)$ with $\vec{v}_x=\vec{e}_j$ this reads
\begin{align*}
  \left[\tau_1,\kappa_j \right]=\big(\psi_x\cp\murs^{-1}\big)([\tau_1,\tau_j])=\big(\psi_x\cp\murs^{-1}\big)(2\epsilon_{1jk}\tau_k)
  =2\epsilon_{1jk}\psi_x\big(\vec{0}_{\mathfrak{g}},\vec{e}_k\big)=2\epsilon_{1jk}\kappa_k.
\end{align*}
From these relations it follows that 
\begin{align*}
  \kappa_1=r(\lambda)\:\tau_1\qquad
  \kappa_2=s(\lambda)\:\tau_2+ t(\lambda)\:\tau_3\qquad
  \kappa_3=s(\lambda)\:\tau_3- t(\lambda)\:\tau_2
\end{align*}
for real constants $r(\lambda),s(\lambda),t(\lambda)$ depending on $\lambda\in\RR\backslash\{0\}$. 
Then for $x=\lambda \vec{e}_1$ and
\begin{align*}
  a(\lambda\vec{e}_1):=r(\lambda)\qquad b(\lambda \vec{e}_1):=\frac{t(\lambda)}{2\lambda}\qquad c(\lambda\vec{e}_1):=\frac{r(\lambda)-s(\lambda)}{4\lambda^2}
\end{align*}
linearity of $\psi_x$ yields that
\begin{align*}
  \psi\big(\vec{0}_{\mathfrak{g}},\vec{v}_x\big)=a(x) \murs(\vec{v}_x)+ b(x)[\murs(x),\murs(\vec{v}_x)]+c(x)[\murs(x),[\murs(x),\murs(\vec{v}_x)]].
\end{align*}
Now, if $x\neq 0$ is arbitrary, then $x=\sigma(\lambda\vec{e}_1)$ for some $\sigma\in \SU$ and $\lambda > 0$. So, $(\sigma,\sigma)\cdot (\lambda \vec{e}_1,e)=(x,e)$ and if we consider $\sigma^{-1}(\vec{v}_x)$ as an element of $T_{(\lambda\vec{e_1},e)}P_\infty$, then \textrm{ii.)} yields
\begin{align*}
  \psi\big(\vec{0}_{\mathfrak{g}},\vec{v}_x\big)&= \psi^+(\vec{v}_x)
  =\psi^+\big(\dd L_{(\sigma,\sigma)}\big(\sigma^{-1}(\vec{v}_x)\big)\big)\\[-5pt]
  &\hspace{-0.9pt}\stackrel{\textrm{ii.)}}{=}\Add{\sigma}\cp\: \psi^+\big(\sigma^{-1}(\vec{v}_x)\big)=\Add{\sigma}\cp \:\psi\big(\vec{0}_{\mathfrak{g}},\sigma^{-1}(\vec{v}_x)\big)\\
  &=a(\lambda\vec{e}_1)\murs(\vec{v}_x)+b(\lambda\vec{e}_1)\left[\murs(x),\murs(\vec{v}_x)\right]+c(\lambda\vec{e}_1) [\murs(x),[\murs(x),\murs(\vec{v}_x)]]. 
\end{align*}
\noindent    
For $x=0$ we have $\sigma(x)=x$ for all $\sigma\in \SU$ and analogous to the case $x\neq 0$ but now for $\sigma_t:=\exp(t\vec{g})$ with $\vec{g}\in \mathfrak{g}$ we obtain from \eqref{eq:condrei} that 
\begin{align*}
  \big[\vec{g},\psi_0\big(\vec{0}_{\mathfrak{g}},\vec{v}_0\big)\big]=\big(\psi_0\cp \murs^{-1}\big)\left([\vec{g},\murs(\vec{v}_0)]\right)\qquad\forall\:\vec{g}\in\mathfrak{su}(2), \forall\: \vec{v}_0\in T_0M.
\end{align*}
This gives $\big[\tau_i,\psi_0\big(\vec{0}_{\mathfrak{g}},\vec{e}_j\big)\big]=2\epsilon_{ijk}\psi_0\big(\vec{0}_{\mathfrak{g}},\vec{e}_k\big)$ and forces $\psi_0(\vec{v}_0)=a(0)\murs(\vec{v}_0)$ for all $\vec{v}_0\in T_{(0,e)}P_\infty$ where $a(0)\in \mathbb{R}$ is some constant. Together, this shows
\begin{align*}
  \psi\big(\vec{0}_{\mathfrak{g}},\vec{v}_x\big)=a(x)\murs(\vec{v}_x)+b(x)[\murs(x),\murs(\vec{v}_x)]+c(x)[\murs(x),[\murs(x),\murs(\vec{v}_x)]]
\end{align*} 
with functions $a,b,c$ that depend on $\|x\|$ in such a way that the whole expression is smooth. 
Finally, to determine $\psi\big(\vec{g},\vec{0}_x\big)$ for $\vec{g}\in \mathfrak{su}(2)=\mathfrak{g}$, we consider 
$\murs^{-1}([\murs(x),\vec{g}])$ as an element of $T_{(x,e)}P_\infty$. Then by \eqref{eq:theta} we obtain  from \textrm{i.)} that $\psi\big(\vec{g},\murs^{-1}([\murs(x),\vec{g}])\big)-\vec{g}=0$, hence
\begin{align*} 
  \psi\big(\vec{g},\vec{0}_x\big)&=\vec{g}-\psi\big(\vec{0}_{\mathfrak{g}},\murs^{-1}([\murs(x),\vec{g}])\big)\\
  &=\vec{g}-a(x)[\murs(x),\vec{g}]-b(x)[\murs(x),[\murs(x),\vec{g}]]-c(x)[\murs(x),[\murs(x),[\murs(x),\vec{g}]]]\\
  &=a(x)[\vec{g},\murs(x)]+b(x)[\murs(x),[\vec{g},\murs(x)]]+c(x)[\murs(x),[\murs(x),[\murs(x),\vec{g}]]]+\vec{g}.
\end{align*}

\newpage
\section*{List of Corrections:}
\begingroup
\setlength{\leftmargini}{12pt}
\begin{itemize}
\item
Second line in Subsection \ref{sec:notations}: 

``a smooth map'' has been replaced by ``differentiable''.
\item
Third line on page 15: 

$\{\Psi(g_\alpha,x_\alpha)\}_{\alpha\in _I}\subseteq M$\quad has been replaced by\quad $\{\Psi(g_\alpha,x_\alpha)\}_{\alpha\in I}\subseteq M$.
\item
Lemma \ref{lemma:dicht}.\ref{lemma:dicht4}: 

$f(\aA_X)$\quad has been replaced by\quad $\rho(\aA_X)$.
\item
Proof of Lemma \ref{lemma:leftactionsvsautos}.\ref{lemma:leftactionsvsautos2}: 

$|\ah(g)(a)-\ah(g_\alpha)(a)|$\quad has been replaced by\quad $\|\ah(g)(a)-\ah(g_\alpha)(a)\|_\aA$.
\item
Equation \eqref{eq:Absch}:

$h^\nu_{\gamma_{\lambda \vec{e}_2}}$\quad has been replaced by\quad $h^\nu_{\gamma_{\lambda}}$.
\item
Fifth line in Subsection \ref{susec:LieALgGenC}:

``properties .'' has been replaced by ``properties.''.
\item
Lemma \ref{lemma:BasicAnalyt}.\ref{lemma:BasicAnalyt3}: 

The assumption $\dim[S]\geq 1$ was added. 
\item
Proof of Lemma \ref{lemma:BasicAnalyt}.\ref{lemma:BasicAnalyt1}: 

$k=2,3$\quad has been replaced by\quad $k=2,\dots,\dim[M]$.
\item
Proof of Lemma \ref{lemma:BasicAnalyt}.\ref{lemma:BasicAnalyt2}:  

$\RR^3$\quad has been replaced by\quad $M$.
\item
Proof of Lemma \ref{lemma:BasicAnalyt}.\ref{lemma:BasicAnalyt2}:  

In the beginning of the last paragraph, 

``let $S$ be connected'' has been replaced by ``let $\dim[S]\geq 1$''.
\item
Second equation in Definition \ref{def:freeSegg}:

$g\sim_\delta g'$\quad has been replaced by\quad $g \sim_\gamma g'$.
\item
Second line on Page 96:

``in fact'' has been replaced by ``In fact''.
\item
Equation \eqref{eq:RadonMeasures}

$A\cap \Borel(\RR)$ and $A\cap \Borel(\RB)$\quad has been replaced by\quad $A\cap \RR$ and $A\cap \RB$,\quad respectively.
\item
Proof of Lemma \ref{lemma:WeierErzeuger}.\ref{lemma:WeierErzeuger2}:

$\ovl{f}\colon \RR\sqcup \{\infty\}$\quad has been replaced by\quad $\ovl{f}\colon \RR\sqcup \{\infty\} \rightarrow \CCC$.
\item
Page 132: 

Before the first point,

``$\pi'_l\cp\TT_1=\pi'_{\tau,r}$ or $\pi'_{\tau,r}\cp \TT_2= \pi'_l$''\quad has been replaced by\quad ``$\TT_1\cp \pi'_{\tau,r}=\pi'_l$ or $\TT_2\cp \pi'_l=\pi'_{\tau,r}$''.
\item
Page 132:

In the first point,

``Consequently, there cannot exist a transition map $\TT_1\colon \im[\pi'_l]\rightarrow \im[\pi'_{\tau,r}]$''

{\bf has been replaced by}

``Consequently, there cannot exist a transition map $\TT_2\colon \im[\pi'_l]\rightarrow \im[\pi'_{\tau,r}]$''.
\item
Page 132: 

In the second point,

$\TT_2\colon \im[\pi'_{l}]\rightarrow \im[\pi'_{\tau,r}]$ we would have  
  \begin{align*}
    \TT_2(\me)=\left(\TT_2\cp\pi'_{\tau,r}\right)(a_{2n})=\pi'_l( a_{2n}) \qquad\forall \: n\in \ZZ_{\neq 0}.
  \end{align*}
  {\bf has been replaced by}
  
  $\TT_1\colon  \im[\pi'_{\tau,r}]\rightarrow \im[\pi'_{l}]$ we would have  
  \begin{align*}
    \TT_1(\me)=\left(\TT_1\cp\pi'_{\tau,r}\right)(a_{2n})=\pi'_l( a_{2n}) \qquad\forall \: n\in \ZZ_{\neq 0}.
  \end{align*}
\item
First line on Page 133:

``Lemma \ref{lemma:BildCirc}.\ref{lemma:BildCirc3}'' has been replaced by ``Lemma \ref{lemma:WeierErzeuger}.\ref{lemma:WeierErzeuger3})''; and the same in the first line in 2).
\item
Proof of Lemma \ref{lemma:OpenMapp}.\ref{lemma:OpenMapp4}:

Continuity of $\pih_L$ is clear from 
      \begin{align*}
        \pih_L^{-1}(A_1,\dots,A_k)= \chih_{l_1,m_1}^{-1}(A_1)\cap\dots \cap \chih_{l_k,m_k}^{-1}(A_k)\qquad\forall\:A_1,\dots A_k \subseteq S^1,
      \end{align*}
      
{\bf has been replaced by}

Continuity of $\pih_L$ is clear from 
      \begin{align*}
        \pih_L^{-1}(A_1,\dots,A_k)= \chih_{l_1}^{-1}(A_1)\cap\dots \cap \chih_{l_k}^{-1}(A_k)\qquad\forall\:A_1,\dots A_k \subseteq S^1,
      \end{align*}
\item
Proof of Lemma \ref{lemma:reltop}.\ref{lemma:reltop2}:

``Lemma and Remark \ref{remdefchris}'' has been replaced by ``Lemma and Definition \ref{remdefchris}''.
\item
Proof of Lemma \ref{lemma:reltop}.\ref{lemma:reltop3}:

``w.r.t.\ subspace topology'' has been replaced by ``w.r.t.\ the subspace topology''.
\item
Proof of Lemma \ref{lemma:suralpha}.2):

``For this, let $x\in N$ and $V'\subseteq \mathfrak{g}$ be a linear subspace such that $T_xM=T_xN \oplus \dd_e\varphi_x(V')$. Then   
      $T_{s_0(x)}O \oplus \dd_e\Phi_{s_0(x)}(V')\oplus Tv_{s_0(x)}P$ 
      because if $\dd_xs_0(\vec{v}_x)  +\dd_e\Phi_{s_0(x)}(\vec{g}\os')+ \vec{v}_v=0$ for $\vec{v}_x\in T_xN$, $\vec{g}\os'\in V'$ and $\vec{v}_v\in Tv_{s_0(x)}P$, then 
      \begin{align*}
        0=\dd_{s_0(x)}\pi \big(\dd_xs_0(\vec{v}_x)  +\textstyle\dd_e\Phi_{s_0(x)}(\vec{g}\os')+ \vec{v}_v\big)=\vec{v}_x + \dd_e\varphi_x(\vec{g}\os')
      \end{align*}
      so that $\vec{v}_x =0$ and $\vec{g}\os'=0$ by assumption, hence $\vec{v}_v=0$. In particular, this shows $\dim[\dd_e\Phi_{s_0(x)}(\g')=0$ iff $\g'=0$, hence 
      $\dim[\dd_e\Phi_{s_0(x)}(V')]\geq\dim[\dd_e\varphi_x(V')]$, and we obtain''

{\bf has been replaced by}

``For this, let $x\in N$ and $V'\subseteq \mathfrak{g}$ be a linear subspace with $V'\oplus \mathfrak{g}_x$ and $T_xM=T_xN \oplus
\mathrm{d}_e\varphi_x(V')$. Then,   
      $T_{s_0(x)}O \oplus \dd_e\Phi_{s_0(x)}(V')\oplus Tv_{s_0(x)}P$ 
      because if $\dd_xs_0(\vec{v}_x)  +\dd_e\Phi_{s_0(x)}(\vec{g}\os')+ \vec{v}_v=0$ for $\vec{v}_x\in T_xN$, $\vec{g}\os'\in V'$ and $\vec{v}_v\in Tv_{s_0(x)}P$, then 
      \begin{align*}
        0=\dd_{s_0(x)}\pi \big(\dd_xs_0(\vec{v}_x)  +\textstyle\dd_e\Phi_{s_0(x)}(\vec{g}\os')+ \vec{v}_v\big)=\vec{v}_x \oplus \dd_e\varphi_x(\vec{g}\os')
      \end{align*}
      showing that $\vec{v}_x =0$ and $\mathrm{d}_e\phi_{x}(\vec{g}')=0$, hence $\vec{g}'=0$ by the choice of $V'$, i.e., $\vec{v}_v=0$ by assumption.
In particular, $\mathrm{d}_e\phi_{x}(\vec{g}')=0$ if $\mathrm{d}_e\Phi_{s_0(x)}(\vec{g}')=0$, hence $\dim[\mathrm{d}_e\Phi_{s_0(x)}(V')]\geq\dim[\mathrm{d}_e\varphi_x(V')]$, from which we obtain''.
\item
Proof of Lemma \ref{lemma:omegaalpha}.1):

In the second line of the first equation, 

$\qrep(q)\cp \w_{p_\alpha}\big(\dd_{(e,p_\alpha)}\THA (\vec{0}_\mathfrak{q}\vec{w}_{p_\alpha})\big)$\quad 
has been replaced by\quad 
$\qrep(q)\cp \w_{p_\alpha}\big(\dd_{(e,p_\alpha)}\THA (\vec{0}_\mathfrak{q},\vec{w}_{p_\alpha})\big)$.
\item
Fifth line in Subsection \ref{subsec:GCovIndAct}: 

``the sets $M_x$. Then'' 
has been replaced by 
``the sets $M_x$ for some $x\in I$. Then''.
\item
Example \ref{ex:OnePoint}:

In the second paragraph in the first point, $\lambda\in \RR$ has been replaced by $\mu\in \RR$. 
\item
Equation \eqref{eq:projlimmm}:

The ``.'' was removed.
\item
Proof of Lemma \ref{lemma:ConstMeas}:

In the third line from below,

``finite Radon measure with $\pi_\alpha^*\mu=\mu_\alpha$''

{\bf has been replaced by}

``finite Radon measure with $\pi_\alpha^*\mu'=\mu_\alpha$''.
\item
Proof of Lemma \ref{lemma:analytCurvesIndepetc}.\ref{lemma:analytCurvesIndepetc5}:

In the last line of the first paragraph, ``such that'' has been replaced by ``, such that''.
\item
References:

\cite{Elstrodt}: ``Ma{\ss}- und Integrationstheorie.'' instead of ``Ma{\ss} und Integrationstheorie.''.

\cite{RudinFourier}: ``Fourier Analysis on Groups.'' instead of ``Representations of Compact Lie Groups.''.
\end{itemize}
\endgroup

%Activate the two lines below to obtain the symbol list.
%\clearpage
%\printglossary[title=Symbols,toctitle=Symbols,style=longragged3col] 
\end{document}